\documentclass[twocolumn,trackchanges]{aastex62}

\usepackage{graphicx}
\usepackage{subfigure}
\usepackage{rotating}
\usepackage{float}
\usepackage{booktabs}
\usepackage{color}

\accepted{2018 September 19}
\submitjournal{ApJ}

\shorttitle{BHB stars in the outer bulge}
\shortauthors{Montenegro et al.}

\begin{document}

\title{\textsc{VVV Survey of Blue Horizontal-Branch Stars in the Bulge-Halo Transition Region of the Milky Way}}

\author{Katherine Montenegro}
\affiliation{Departamento de Ciencias F\'isicas, Facultad de Ciencias Exactas, Universidad Andr\'es Bello, Av. Fern\'andez Concha 700, Las Condes, Santiago, Chile.}
\affiliation{Instituto Milenio de Astrof\'isica, Santiago, Chile.}

\author{Dante Minniti}
\affiliation{Departamento de Ciencias F\'isicas, Facultad de Ciencias Exactas, Universidad Andr\'es Bello, Av. Fern\'andez Concha 700, Las Condes, Santiago, Chile.}
\affiliation{Instituto Milenio de Astrof\'isica, Santiago, Chile.}
\affiliation{Vatican Observatory, V-00120 Citt\`a del Vaticano, Vatican City State, Europe.}

\author{Javier Alonso-Garc\'ia}
\affiliation{Unidad de Astronom\'ia, Facultad de Cs. B\'asicas, Universidad de Antofagasta, Av. U. de Antofagasta 02800, Antofagasta, Chile.}
\affiliation{Instituto Milenio de Astrof\'isica, Santiago, Chile.}

\author{Maren Hempel}
\affiliation{Instituto de Astrof\'isica, Pontificia Universidad Cat\'olica de Chile, Av. Vicu\~na Mackenna 4860, Santiago, Chile.}

\author{Roberto K. Saito}
\affiliation{Departamento de F\'isica, Universidade Federal de Santa Catarina, Trinidade 88040-900, Florian\'opolis, SC, Brazil.}

\author{Timothy C. Beers}
\affiliation{Department of Physics and JINA Center for the Evolution of the Elements, University of Notre Dame, Notre Dame, IN 46556, USA.}

\author{David Brown}
\affiliation{Vatican Observatory, V-00120 Citt\`a del Vaticano, Vatican City State, Europe.}

\begin{abstract}
We characterize the population of blue horizontal-branch (BHB) stars in
the bulge-halo transition region of the Milky Way using the VISTA
Variables in the V\'ia L\'actea (VVV) ESO Public Survey data. The
selection of BHB stars is made using the globular cluster M22 as a
reference standard, and constructing color-magnitude and color-color
diagrams with specific cuts in the $ZYJHK_s$ near-infrared (IR)
passbands. A total of 12,554 BHB stars were detected, in a region within
$-10.0^{\circ} \leq \ell \leq 10.2^{\circ}$ and $-10.2^{\circ} \leq b
\leq -8.0^{\circ}$. We provide accurate coordinates and near-IR
photometry for this sample of BHB stars. We searched for over-densities
of stars with sizes similar to those of known globular clusters and
stellar streams. By comparing real data with Monte Carlo simulations, we
conclude that the few over-densities detected are of low significance.
We also constructed $K_s$-band light curves for the BHB stars to study
their variability. Taking an average of 52 epochs to calculate periods
and amplitudes, we identify hundreds of candidate eclipsing binaries and
a dozen pulsating stars. Finally, we made some comparisons with results
obtained in a previous study for RR Lyrae variable stars in this same
region.
\end{abstract}

\vspace{1cm}

\section{Introduction} \label{intro}

The VISTA Variables in the V\'ia L\'actea (VVV) is an ESO Public Survey
that is targeting the central parts of the Galaxy, covering an area of
$\sim$ 540 square degrees \citep{minniti}. Its main goal is to study our
galaxy's bulge and southern disk, in order to reveal the corresponding
3D structure through variable stars. VVV combines high-resolution
($\sim 0.34"$ pixel$^{-1}$), deep ($K_s \gtrsim 18$ mag), near-IR
photometry in five bands ($ZYJHK_s$), thus alleviating the problem of
high interstellar dust extinction and better resolving the high
stellar density regions, allowing us to unveil the stellar populations
in this complex area \citep{saito2012b}. In addition, multi-epoch
observations allow us to construct $K_s$-band light curves to
study different variable stars, enabling the construction of a 3D map of
the surveyed region \citep{gran2016}.

Old metal-poor stars, e.g., RR Lyrae and blue horizontal-branch (BHB)
stars, are ideal for tracing old regions of the Milky Way ($> 5$ Gyr)
and constructing their associated number-density maps. They are also
very good standard candles, which enable accurate determination of their
distances \citep{clewley}. However, BHB stars are $\sim$ 10
times more numerous than RR Lyrae stars. Considering these advantages, as well as the accuracy and depth of the photometric data obtained by the VVV
Survey, we can use BHB stars to study the (expected) most ancient part
of the Galaxy, such as the bulge-halo transition region. We define this
region loosely, at projected Galactocentric distances $\rm 1 \: kpc <
R_G < 2 \: kpc$. This very special place contains precious information
about the assembly history of the Milky Way.

In the Hertzsprung-Russell color-magnitude diagrams (CMDs) of globular
clusters, the horizontal branch is populated by stars which have evolved
past the main sequence and red giant branch stage. Such stars are now
burning helium in their cores and hydrogen in shells outside the core
\citep{ruhland}. Amongst these HB stars, RR Lyrae are found in the
instability strip of the horizontal branch, and BHB stars are found to
the left of (bluer than) the RR Lyrae along the HB. BHB stars are old,
metal-poor Population II objects, found in both globular clusters and in
the Galactic halo. Their masses are typically within the range of $0.5
M_{\odot} - 1.0 M_{\odot}$. They are a very useful tool to trace the
Milky Way's structure, due to their intrinsic brightness and
distinctive colors, as well as being readily identifiable because of
their spectral features, e.g., a strong Balmer jump, their strong and
deep Balmer lines (narrower than main-sequence A-type stars, and deeper
than, e.g., subdwarf B-type stars), and their lack of strong metallic
features/lines \citep{smith}.

BHB stars in the halo and thick-disk systems of the Galaxy have been
studied for decades. Early catalogs (\citealt{pier1982}, and references
therein; \citealt{beers1988, beers1996, beers2007};
\citealt{christlieb}), based primarily on identification from
objective-prism surveys, have been used to constrain the kinematics of
the halo and metal-weak thick disk of the Galaxy (e.g., \citealt{thom};
\citealt{brown2008}; \citealt{sommer}), and provided the first
identification of a likely age gradient in the halo of the Milky Way
\citep{preston}. Later, much larger catalogs of
spectroscopically-confirmed BHB stars from the Sloan Digital Sky Survey
(SDSS, \citealt{york}; \citealt{yanny}) were used to study the rotation
curve and mass of the Galaxy out to 60 kpc \citep{xue}, and to explore
the age structure of the halo \citep{santucci}. Some 130,000
photometrically selected BHB candidates from SDSS were used by
\cite{carollo} to obtain the first accurate estimate of the age gradient
in the Galactic halo based on field stars, and to confirm the existence
of a so-called ``ancient chronographic sphere" of old stars in the halo
(first suggested by \citealt{santucci}), extending from immediately
outside the bulge, past the solar vicinity, out to $\sim 15$ kpc from
the Galactic center. Due to the selection criteria employed in
the above-mentioned studies, none of these samples extended into the bulge
region.

Bulge BHB stars have also been studied in the past, both photometrically
and spectroscopically (e.g., \citealt{peterson}; \citealt{zoccali};
\citealt{terndrup}; \citealt{busso}; \citealt{koch}), firmly
establishing that the Galactic bulge, in spite of being predominantly
metal-rich, still contains old and metal-poor BHB stars. However, these
previous studies were limited to a relative handful of stars. One of the
goals of the present work is to assemble a massive catalog of bulge BHB
stars, complementing the existing halo- and disk-system catalogs, in
order to enable a suite of follow-up studies. Therefore, in this paper
we make a census of these old BHB stars in the bulge-halo transition
region, using the VVV near-IR data. We also use this new catalog to search
for new globular clusters, and for variable stars such as eclipsing
binaries and RR Lyrae.

\section{The VVV Data} \label{Sources}
\subsection{Observations}

\begin{figure}
	\centering
	\includegraphics[scale=0.3]{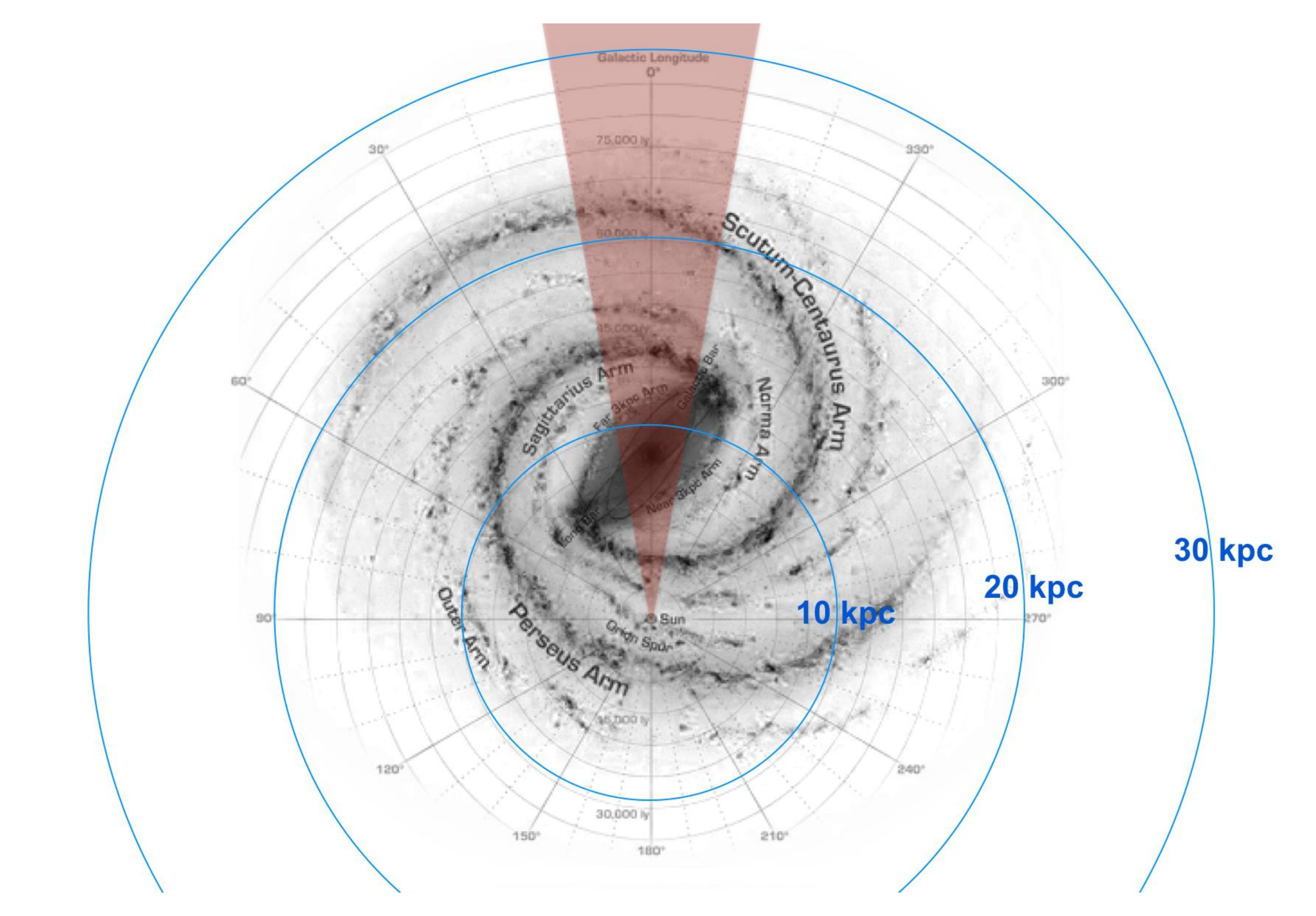}
	\caption{Artistic representation of the Milky Way, seen face-on, based on the Spitzer satellite data (Credits: R. Hurt/Spitzer Space Observatory/NASA). Each concentric circle indicates the distance to the Sun. The shaded region corresponds to the area covered by the VVV survey used in this study.}
	\label{fig:ourposition}
\end{figure}

The VVV is a public near-IR survey of the inner Milky Way using the
$ZYJHK_s$ passbands (\citealt{minniti}; \citealt{catelan};
\citealt{saito2012a}), whose coordinates have sub-arcsecond
accuracy \citep{smith2017}, a necessary requirement in this high
stellar-density region. The VVV observations were taken with the
near-IR camera VIRCAM at the Visible and Infrared Survey Telescope for
Astronomy (VISTA) located at ESO Cerro Paranal Observatory
\citep{emerson}. VISTA is a 4-m telescope that has been optimized for
the near-IR. The VISTA Infrared Camera (VIRCAM) has 16 near-IR detectors
with a scale of 0.34$\arcsec$/pixel, arranged in a 4$\times$4 pattern
with significant gaps between them. The observing sequence consists of 6
individual exposures, spatially shifted in an mosaic pattern, and later
combined into a single image, which we refer to as a ``tile". Each tile
covers approximately 1.1$\times$1.5 square degrees on the sky. The VVV
bulge observations comprise 196 tiles in total, covering a 307 square
degree field of view, within $-10.0^{\circ} \leq \ell \leq
10.2^{\circ}$, and $-10.2^{\circ} \leq b \leq 5.0^{\circ}$. Figure
\ref{fig:ourposition} shows a schematic of the bulge region observed by
the VVV survey.

The observations are made using the $ZYJHK_s$ filters, and consist of
two epochs in $ZYJH$, taken at the beginning and the end of the survey,
in years 2010 and 2015, respectively, plus multiple epochs in the
$K_s$-band. The $K_s$-band variability campaign was carried out between
2010 and 2016, obtaining a total of about 75 observations per field in
the VVV bulge area.

\begin{figure}
	\centering
	\includegraphics[scale=0.29]{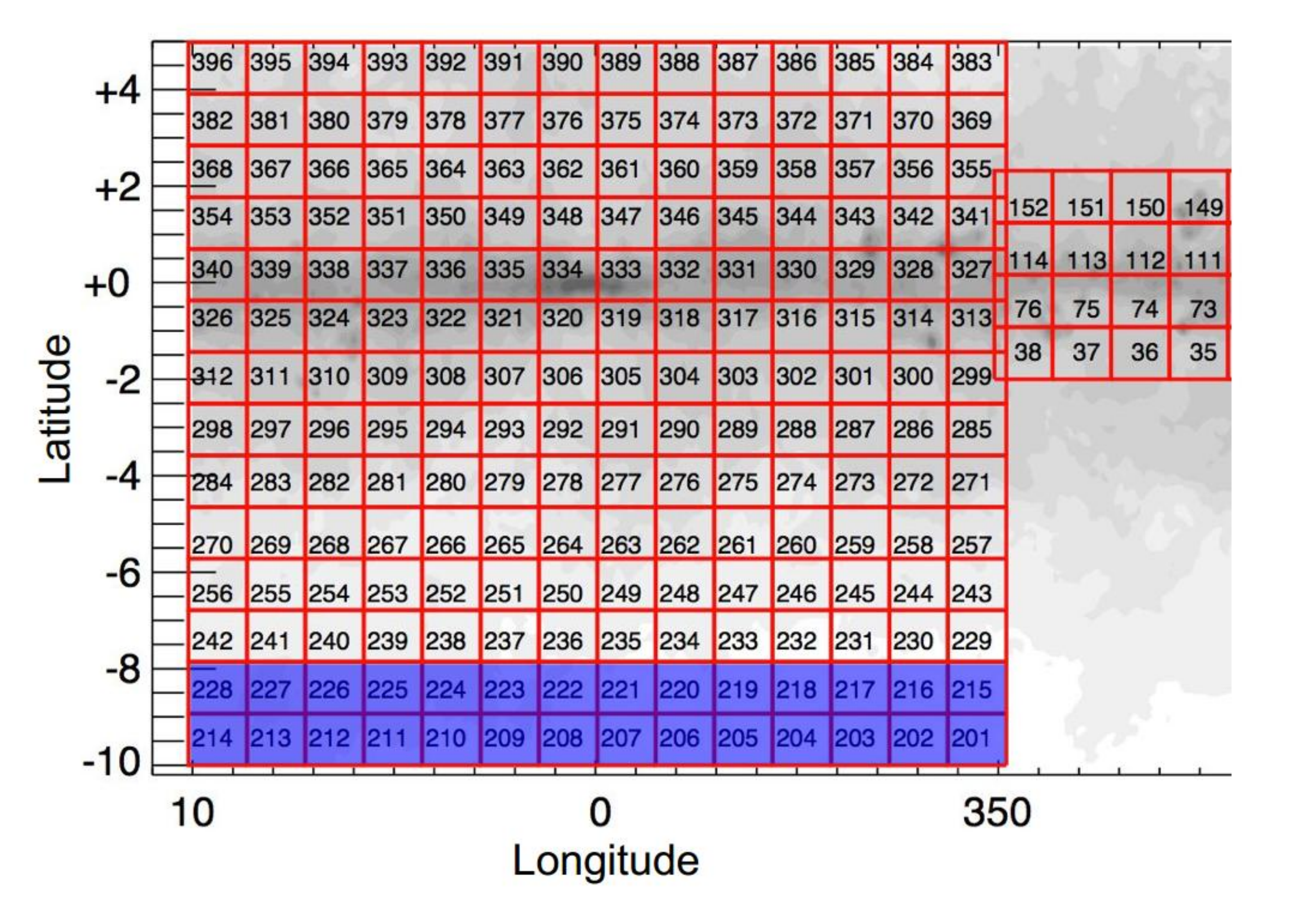}
	\caption{VVV Bulge area in Galactic coordinates (measured in degrees). Tiles used in this work (b201-b228) are colored in blue, and are located in the bulge-halo transition region of the Milky Way.}
	\label{fig:tiles}
\end{figure}

The data reduction, calibration, and aperture photometry were carried out
by the Cambridge Astronomical Survey Unit (CASU; \citealt{irwin}). The
CASU photometry is tied to the 2MASS system (\citealt{cutri};
\citealt{hodgkin}), and is made publicly available at the VISTA Science
Archive (VSA, \citealt{hambly}; \citealt{cross}). In addition, we have
performed single-epoch PSF photometry in the $ZYJHK_s$ bands for the
entire VVV bulge dataset using DoPhot \citep{alonsogarcia}, in order to
obtain deeper and more accurate photometry, adequate for the selection
of BHB stars. In this work we analyze the outermost southern VVV bulge
tiles, from b201 to b228 (Figure \ref{fig:tiles}), covering an area
within $10.0^{\circ} \leq \ell \leq 10.2^{\circ}$, and $-10.2^{\circ}
\leq b \leq -8.0^{\circ}$, for a total of about 44 square degrees,
comprising VVV tiles b201 to b228. This area is sufficiently far from
the Galactic plane that extinction and crowding do not present severe
limitations.

\subsection{Selection of BHB Stars}
In order to select BHB stars in the bulge, one can fit theoretical
isochrones to their CMDs (e.g., \citealt{brown2004}). However, we
preferred an empirical selection approach, since we needed to consider
the presence of complex stellar populations as well. For that reason we
use the globular cluster M22 (NGC~6656) as a standard sample, as it
comprises an old stellar population with a prominent horizontal branch.
This cluster was observed in the near-IR by the VVV in the same way,
with the same instrumental setup, sharing the biases and constraints of
our BHB star sample.

\begin{figure}
	\centering
    \includegraphics[scale=0.27]{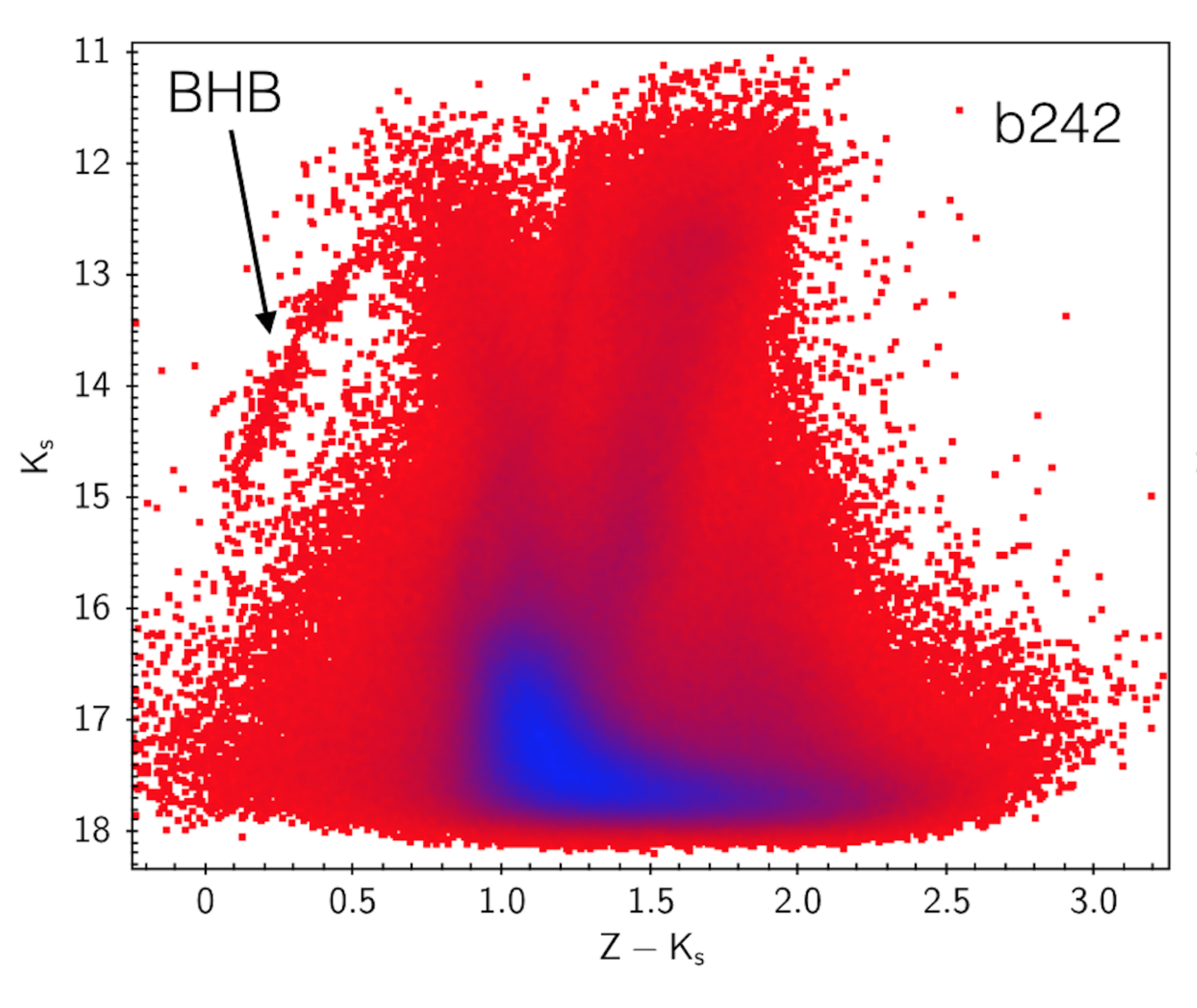}
    \includegraphics[scale=0.27]{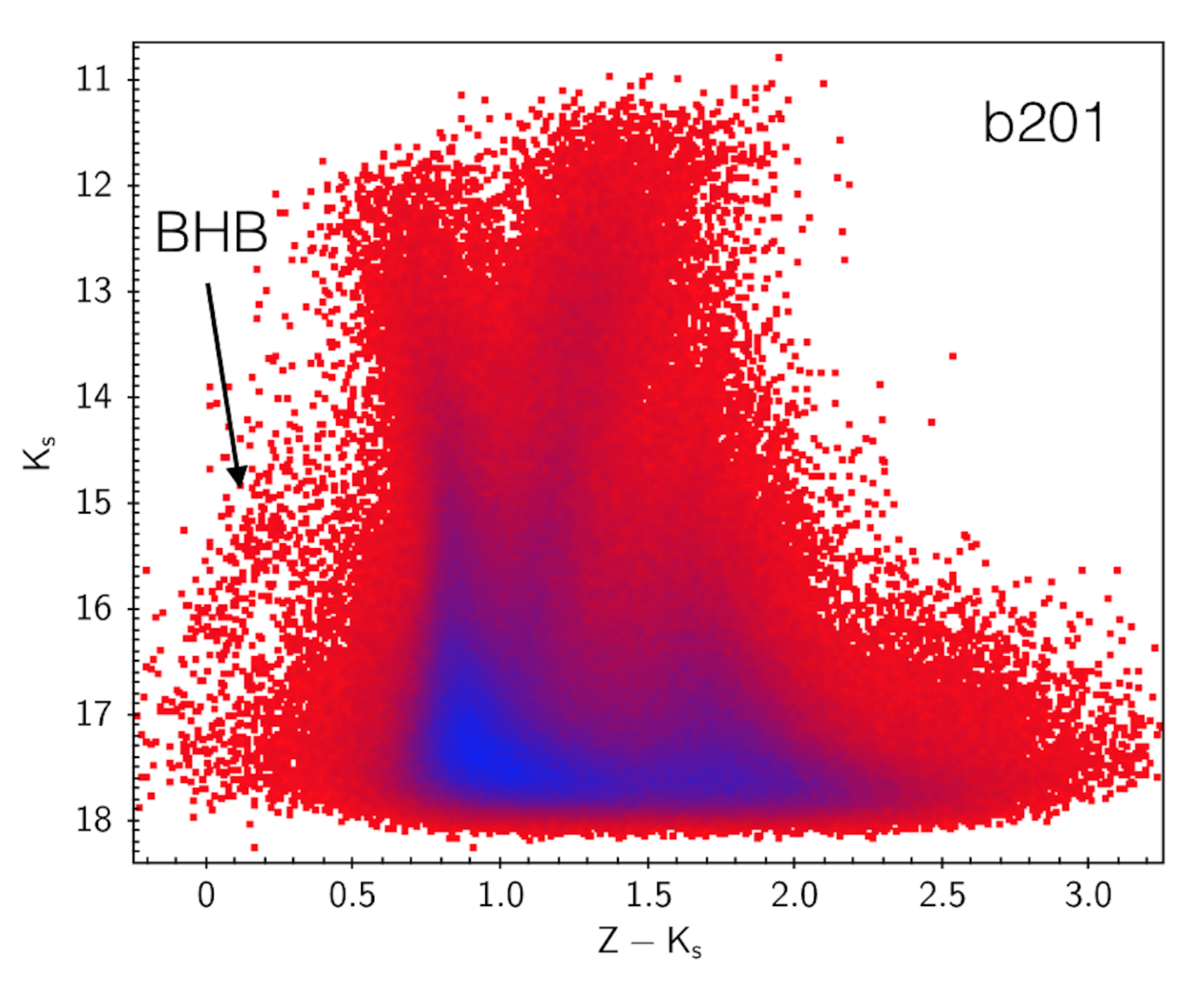}
	\caption{{\bf Top panel:} CMD of the sources in tile b242, including the stars from globular cluster M22. BHB stars in M22 are clearly distinct at the bluest colors and brightest magnitudes. {\bf Bottom panel:} CMD of the sources in tile b201, one of the fields included in our study. Bulge BHB stars are located at similar colors for M22, but about two magnitudes fainter.}
	\label{fig:photometry}
\end{figure}

Given the known distance difference, the bulge BHB stars would be
located in the CMD about two magnitudes fainter than the M22 BHB stars.
The bulge BHB stars can be clearly seen in the deep VVV near-IR CMDs
(Figure \ref{fig:photometry}).

\begin{figure}
	\centering
	\includegraphics[scale=0.28]{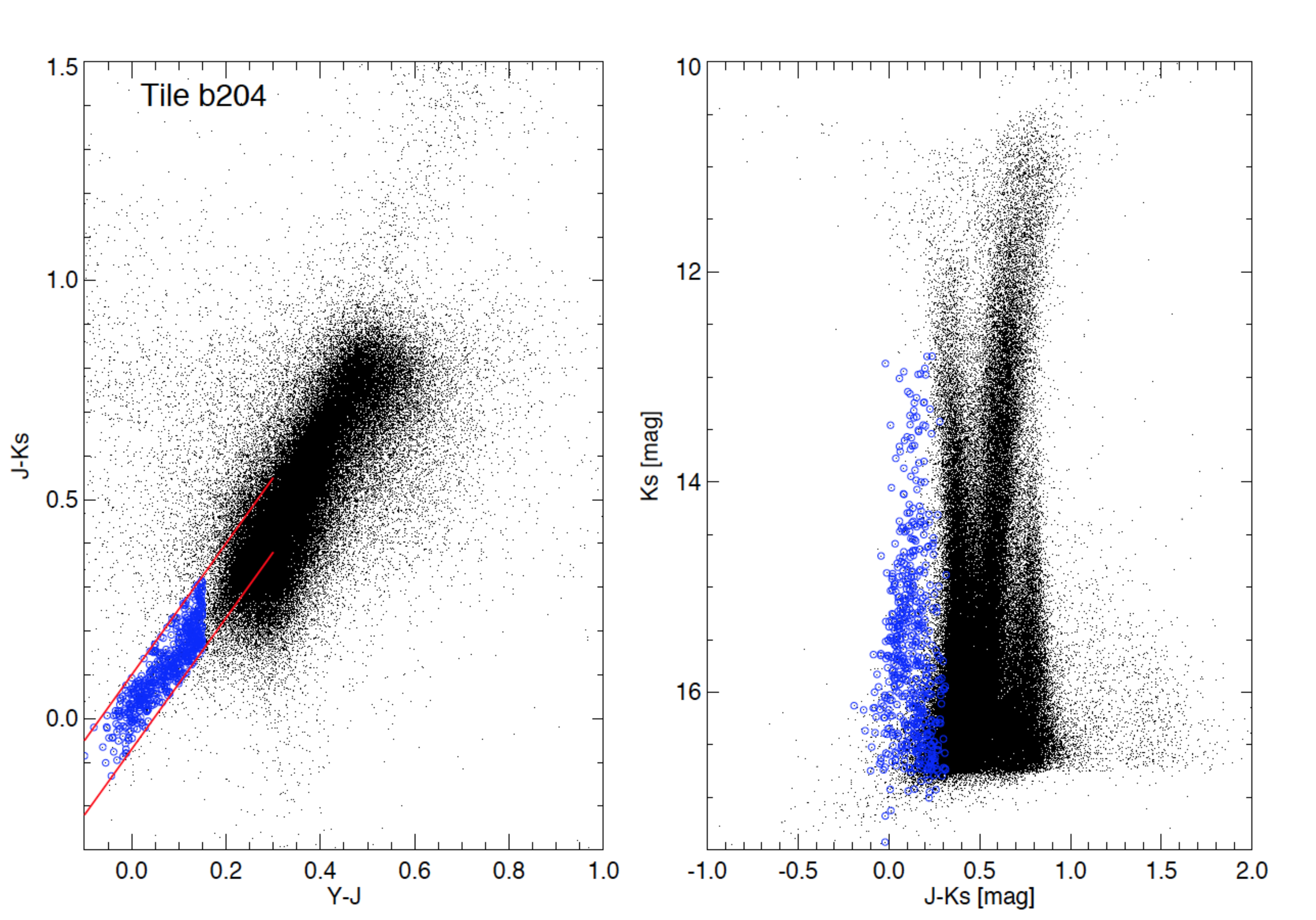}
	\includegraphics[scale=0.28]{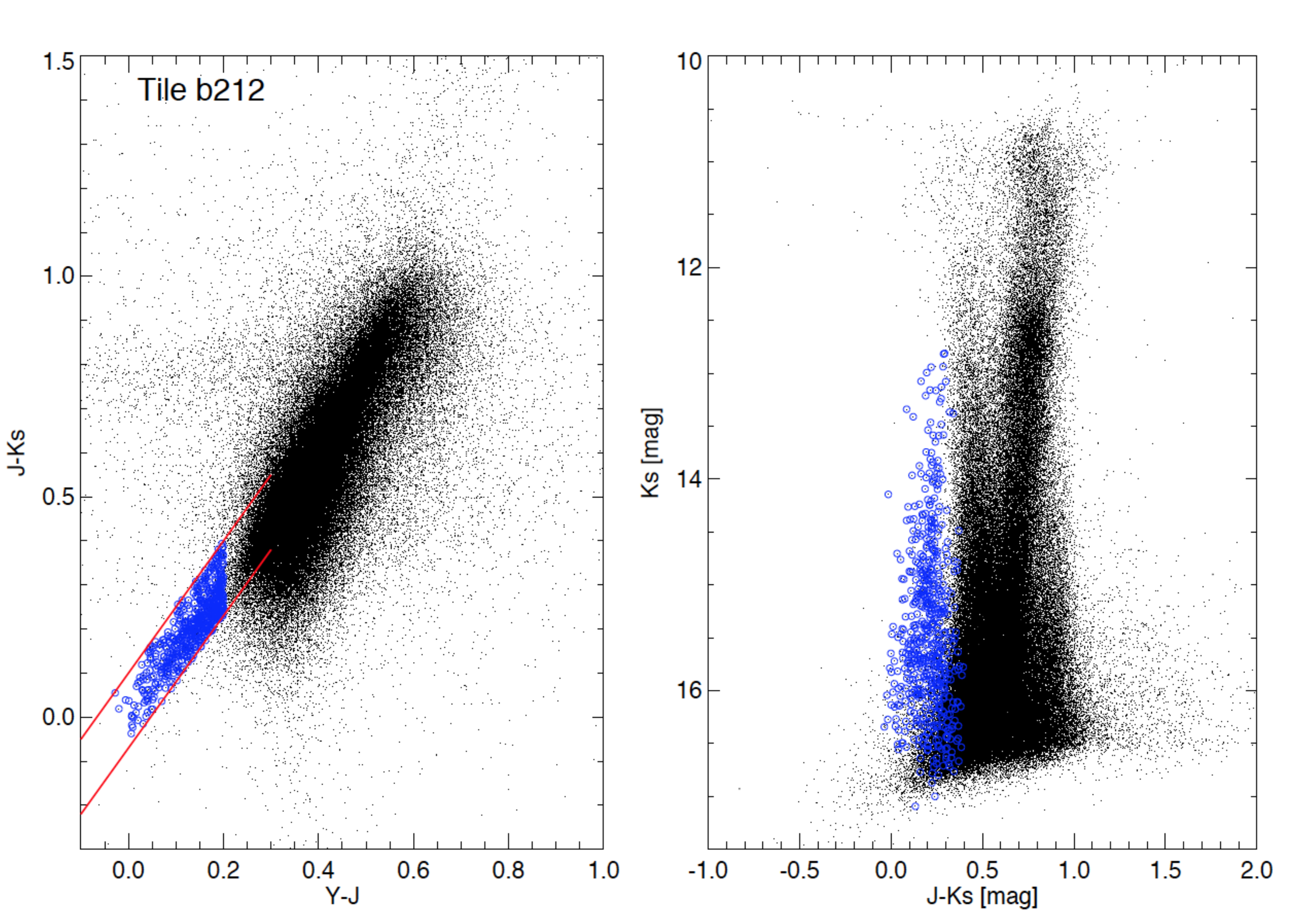}
	\caption{{\bf Top panels:} Color-color and color-magnitude diagrams for the tile b204, with selected BHB stars marked in blue and color cuts of $Y-J < 0.15$ and $J-K_s < 0.35$. {\bf Bottom panels:} Color-color and color-magnitude diagrams for the reddened tile b212, with selected BHB stars marked in blue and color cuts in $Y-J < 0.20$ and $J-K_s < 0.45$.}
	\label{fig:selection}
\end{figure}

We used the $Y-J$ vs. $J-K_s$ color-color diagram to select bulge
BHB stars, demanding $Y-J < 0.15$ and $J-K_s < 0.35$ (Figure
\ref{fig:selection}). These cuts are applied in most fields, and only
relaxed to $Y-J < 0.20$ and $J-K_s < 0.45$ in the most reddened fields,
e.g. b212, which are contaminated by RGB stars from the Sgr dwarf
galaxy.

\begin{figure}[h!]
	\centering
	\includegraphics[scale=0.26]{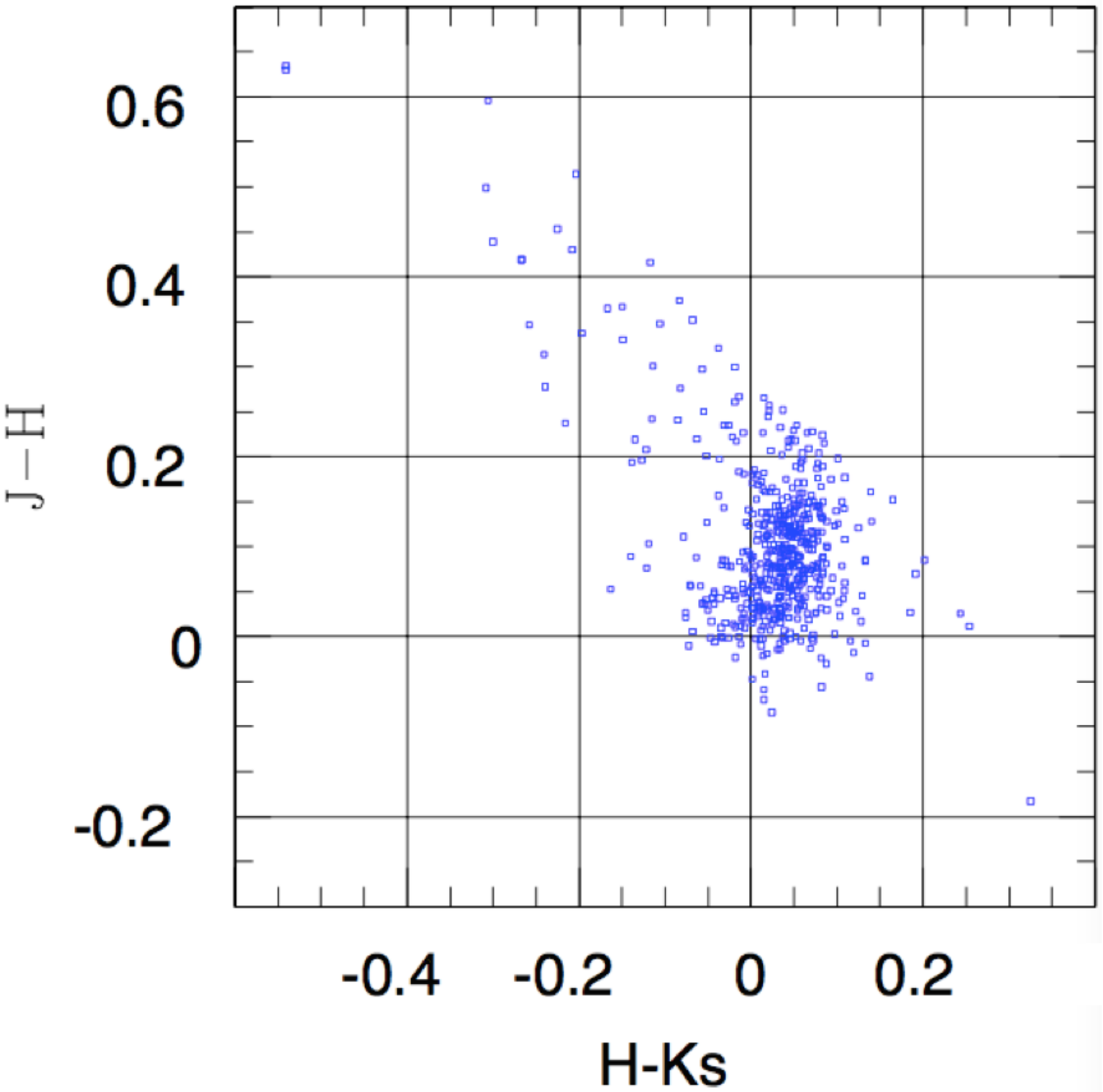}
    \includegraphics[scale=0.26]{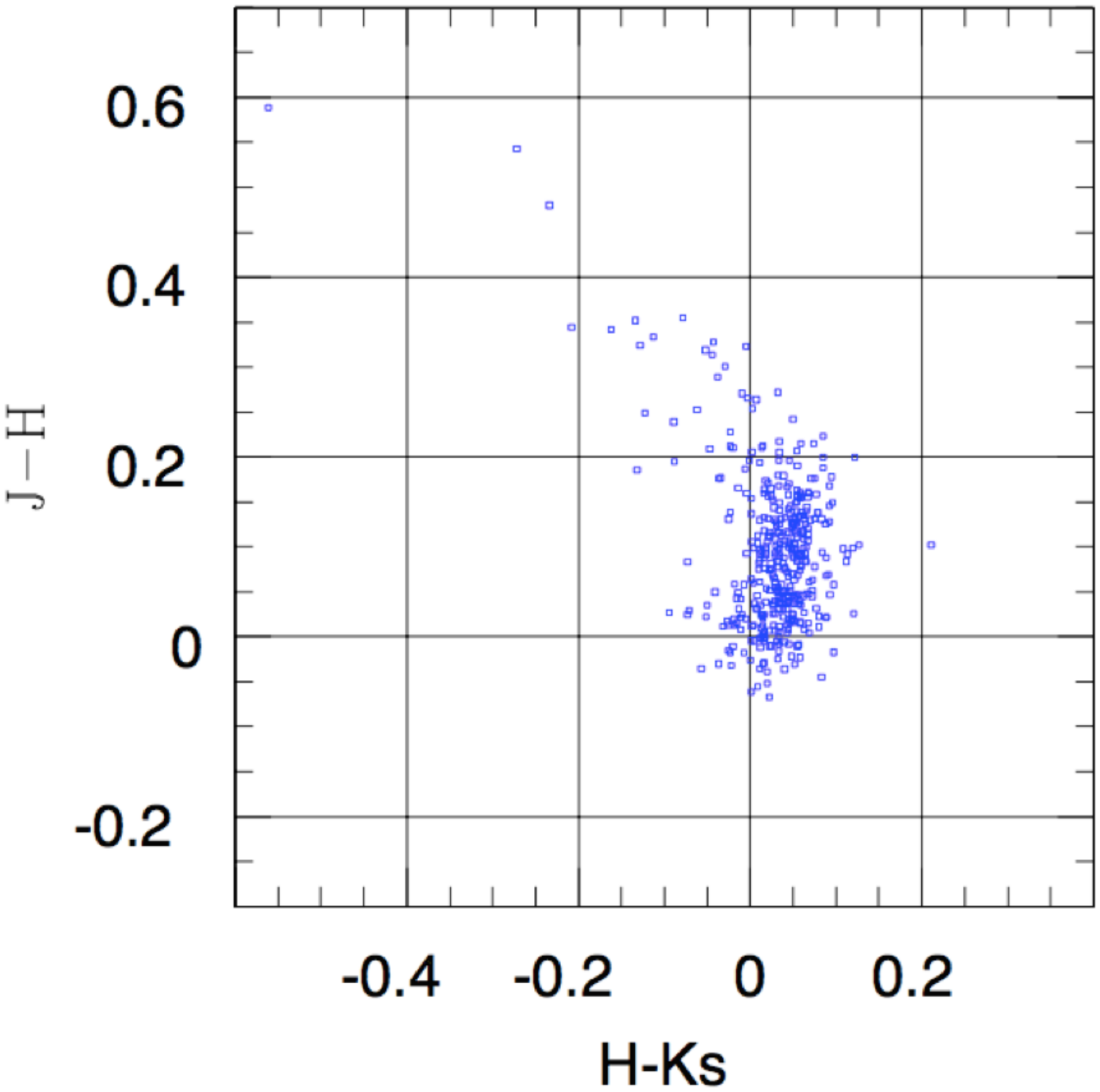}
	\caption{Color-color diagrams of selected BHB stars for tiles b207 (top panel) and b201 (bottom panel), according to the color cut suggested by \cite{brown2004}.}
	\label{fig:color}
\end{figure}

At the faintest magnitudes there is contamination from other A-type
stars (e.g. \citealt{brown2004} ; \citealt{clewley}). For
example, some bulge blue straggler stars in particular share a similar
location in the CMDs. Also, in these fields there is some contamination
from the main sequence of the Sgr dSph galaxy. In order to take the
contamination from A-type stars into account, we added another near-IR
color cut: $J-H < 0.10$. This is because, according to
\citet{brown2004}, the BHB stars have: $-0.20 < (J-H)_0 < 0.10$ and
$-0.10 < (H-K)_0 < 0.10$ (Figure \ref{fig:color}), with $J-H$ being a
good indicator to discard A-type stars, since they tend to have redder
$(J-H)_0$ colors. Our final sample contains 12,554 bulge BHB star
candidates.

\section{A Search for New Globular Clusters}

\citet{minniti2017a, minniti2017b, minniti2017c} recently reported the discovery 
of dozens of new globular cluster candidates in the Galactic bulge. BHB
stars are well known representative of old and metal-poor populations,
often found in old globular clusters. Single Stellar Population (SSP)
models tell us that, in a globular cluster, for every BHB star there are
hundreds of red giant branch and thousands of main-sequence stars (even
though recently it has been found that not all GCs are single stellar
populations, see e.g. \citealt{piotto}).

\begin{figure*}
	\centering
	\includegraphics[scale=0.6]{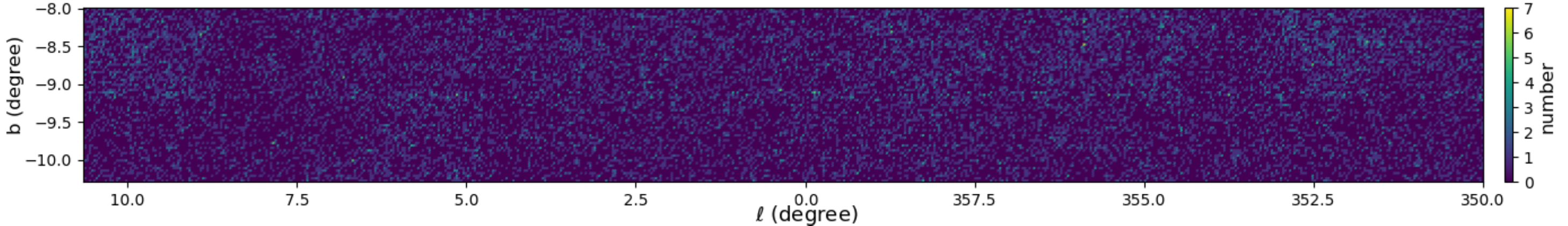}
	\includegraphics[scale=0.6]{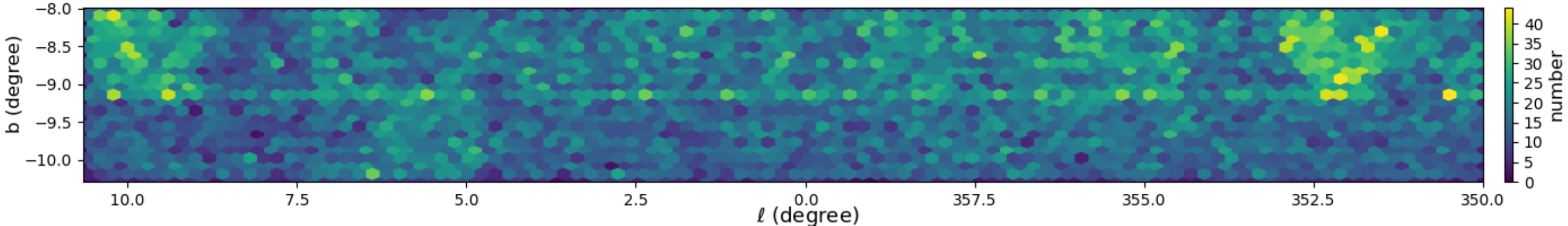}
        \caption{{\bf Top panel:} Density plot for the 28 tiles, using a
bin size = 2$\arcmin$. In the clearest areas it is possible to note 97 over-densities, mostly of low significance. {\bf Bottom panel:} An alternative density plot, with a bin size = 12$\arcmin$. In the
yellow areas it is possible to note 59 over-densities with a
greater number of objects per bin, and two possible streams. Targets
located in the overlap regions between tiles are accounted for twice in this
plot, thus producing the artificial, horizontal over-density stripe seen
crossing the plot at b $\sim 9.1^\circ$.} \label{fig:alltiles}
\end{figure*}

In this work we aim to detect over-densities of BHB stars in
the selected 28 fields, since they may lead to the discovery of
heretofore unknown globular clusters, and complement other stellar
population studies. To accomplish this, we generated density maps of BHB
candidates (Figure \ref{fig:alltiles}) with different bin sizes. We also
generated random homogeneous samples of 12,554 points (equivalent to the
total number of BHB stars in our catalog), distributed across the 28
tiles. The results of these Monte Carlo simulations were plotted,
together with the BHB stars of the sample. The objective was to analyze
if there were over-densities that could correspond to previously
undiscovered globular clusters or streams. We experimented with a number
of different bin searches. Ultimately, from a density plot with
hexagonal bins and a color gradient, we obtained a first-pass
identification of possible over-densities. The size of each bin was
chosen to be the average size that the typical globular clusters in our
galaxy would have at the distance of the Galactic bulge.

According to \cite{Chen10}, a typical Milky Way globular cluster has a
size of 5 pc $\equiv$ 2$\arcmin$ on the sky. Figure \ref{fig:alltiles}
(top panel) shows the panorama for the 28 tiles studied; some
over-densities, mostly of low significance ($<3 \sigma$), are seen. Then
we performed a new inspection, with a bin size of 12$\arcmin$, which is
a very large size for a globular cluster ($1^\circ$ could host 5
globular clusters). In Figure \ref{fig:alltiles} (bottom panel) we can
see some over-densities, but with a larger number of objects per bin
than in the previous simulation. Also, we can see two probable streams
close to the coordinates $\ell = 10^\circ$ and $\ell = 353^\circ$. Even
though these over-densities are more significant, the result is unlikely
to be successful in identifying any new clusters, since such a large
globular cluster probably would have been discovered already. In
addition, the streams we identify could be due to inhomogeneities in the
reddening, as the BHB search is extremely sensitive to reddening
variations. However, we point out that the stream at $\ell = 10^\circ$
could be associated with tidal tails from the globular cluster M22,
located in tile b242, just outside this map, and which contains a rich
population of BHB stars.

\begin{figure}
	\centering
	\includegraphics[scale=0.45]{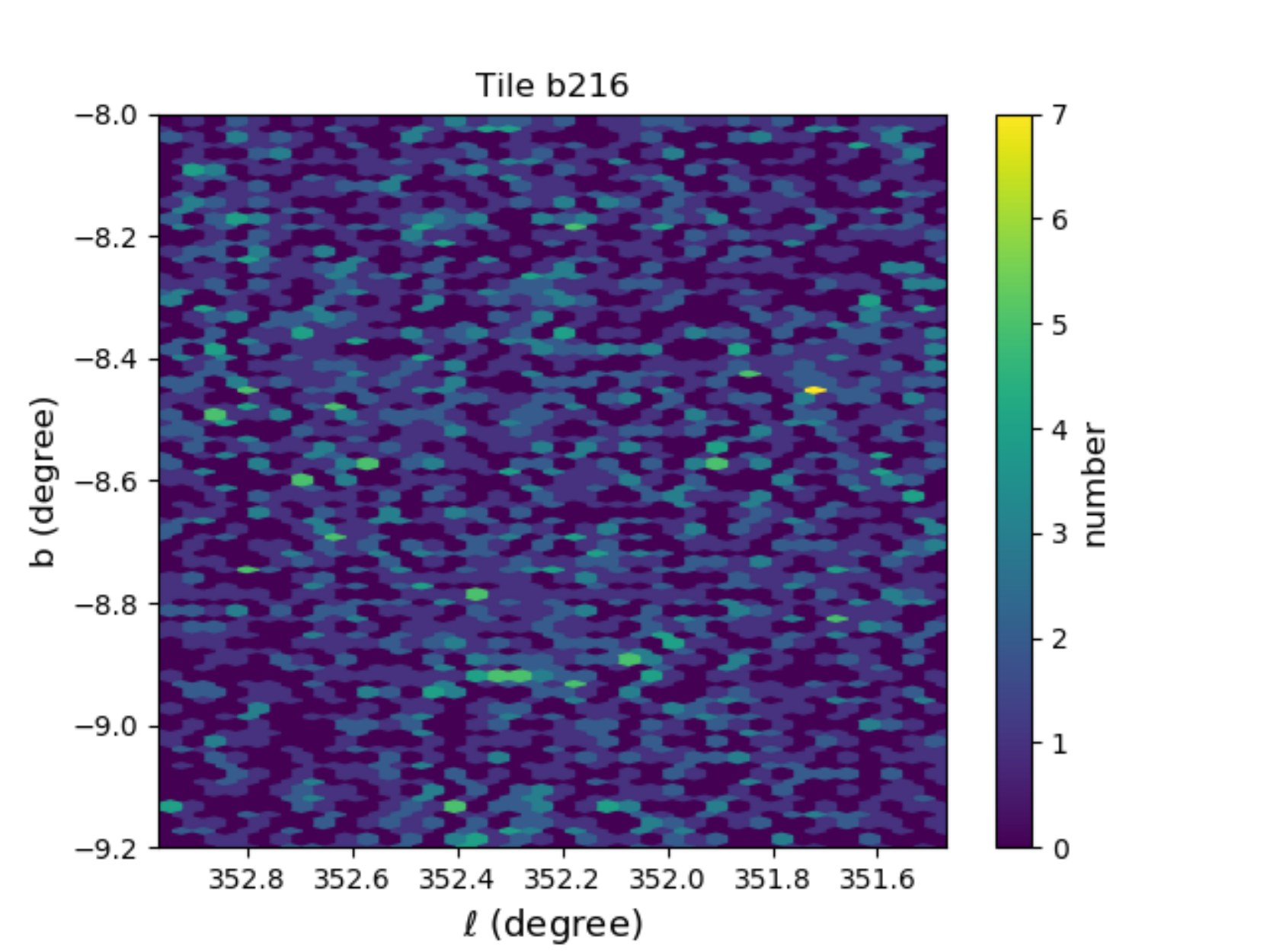}
	\includegraphics[scale=0.45]{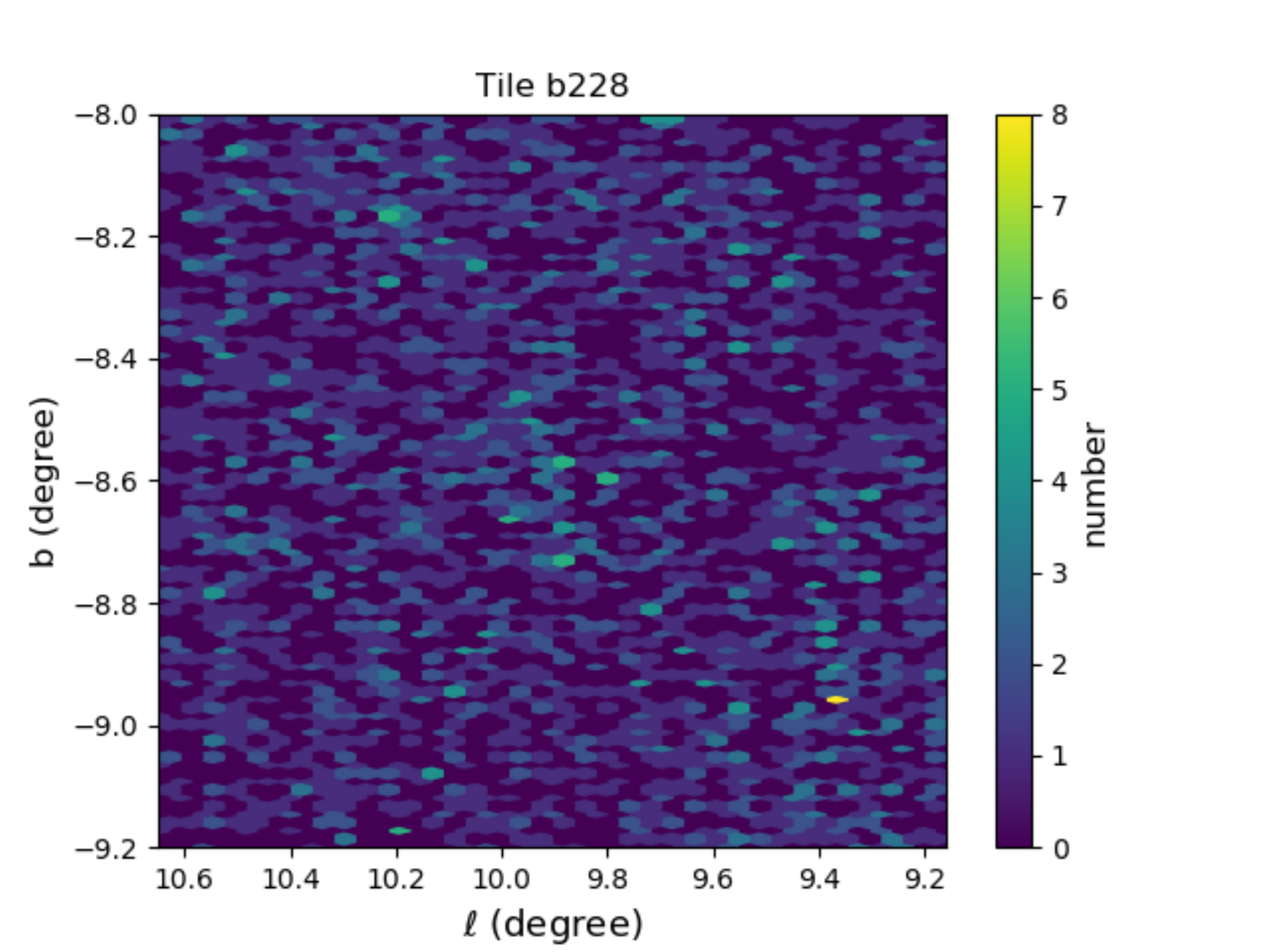}
	\caption{Some interesting over-densities corresponding to the first simulation (bin size of 2').}
	\label{fig:sim1}
\end{figure}

\begin{figure}
	\centering
	\includegraphics[scale=0.45]{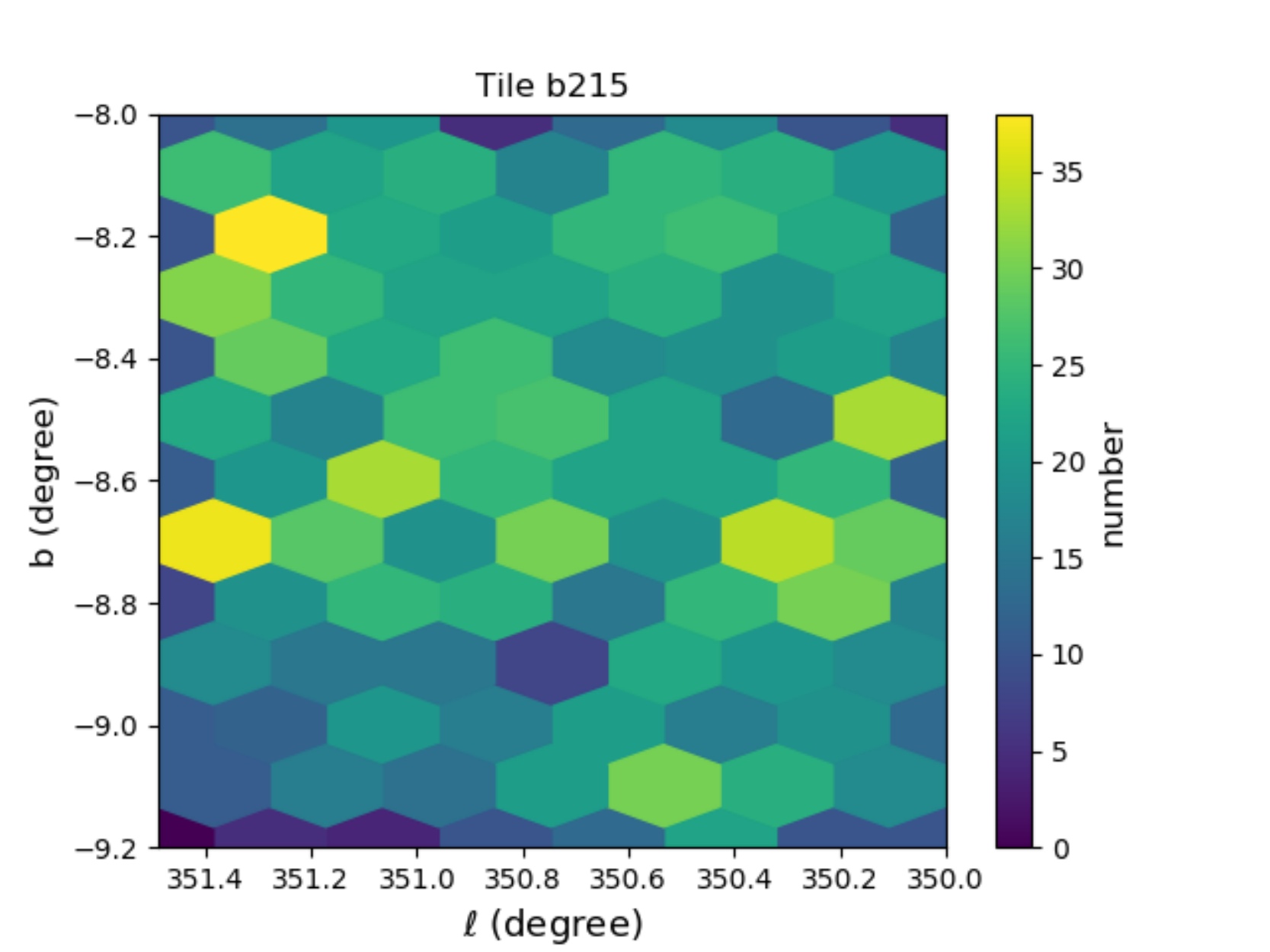}
	\includegraphics[scale=0.45]{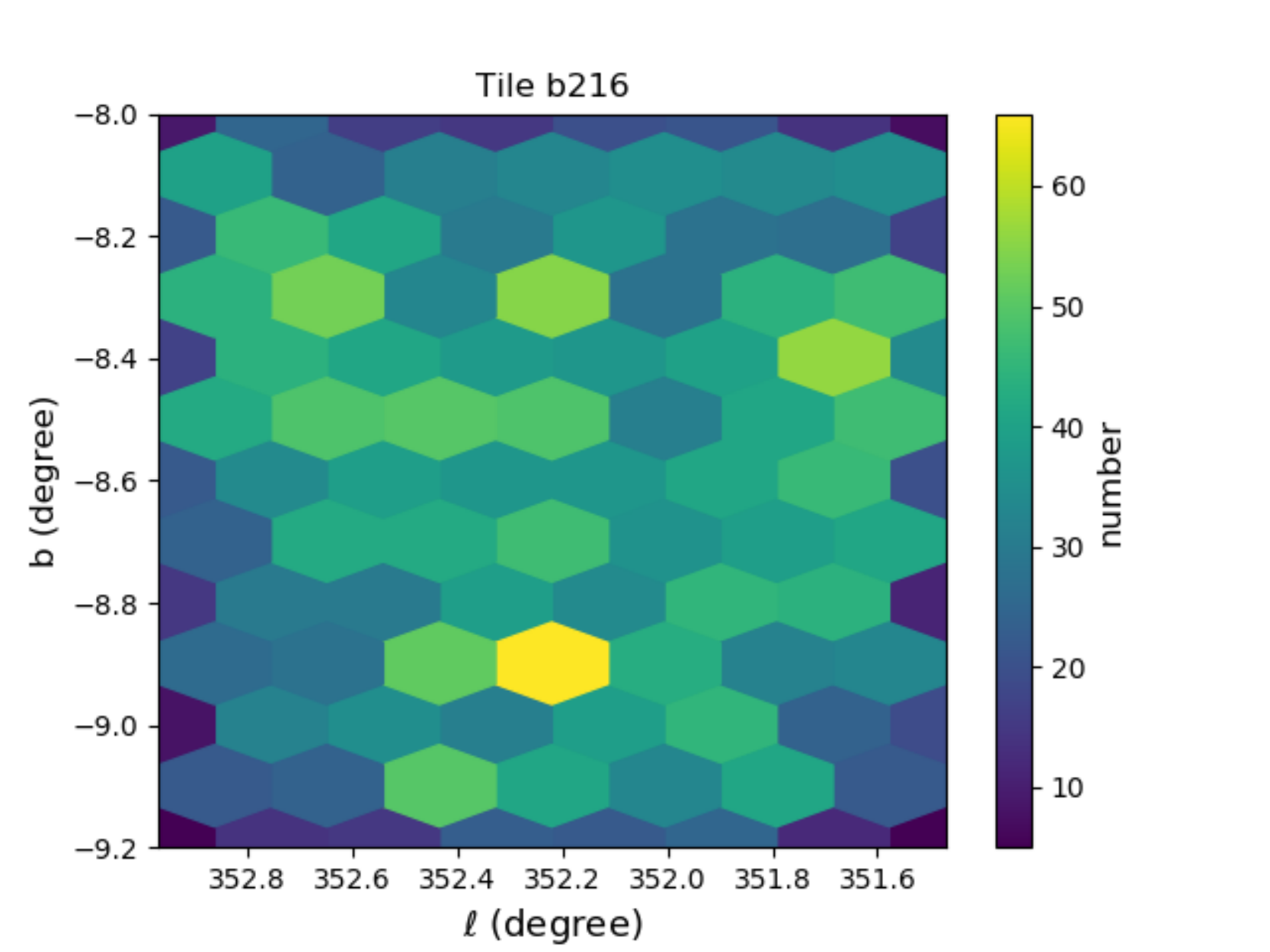}
	\includegraphics[scale=0.45]{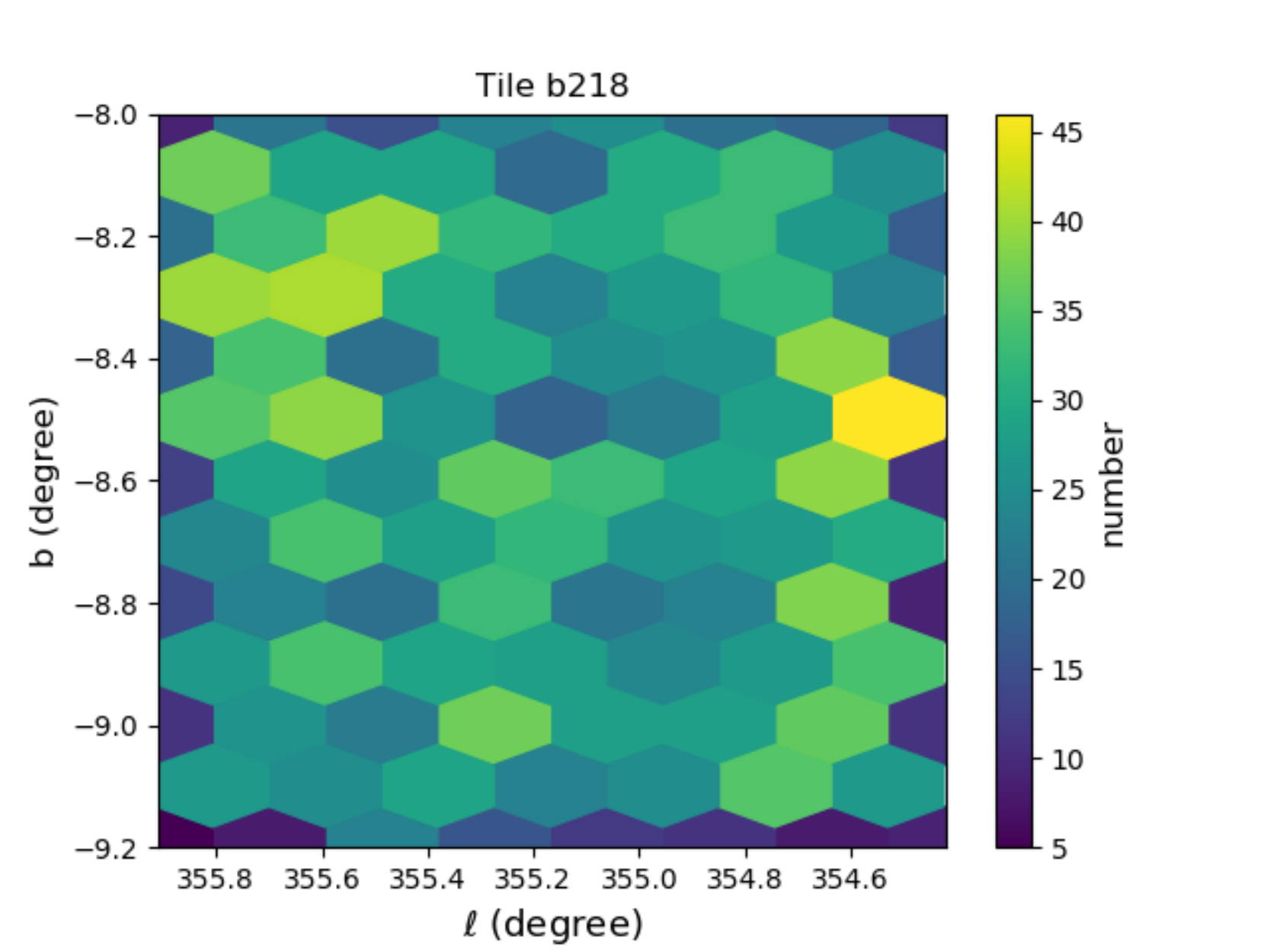}
	\includegraphics[scale=0.45]{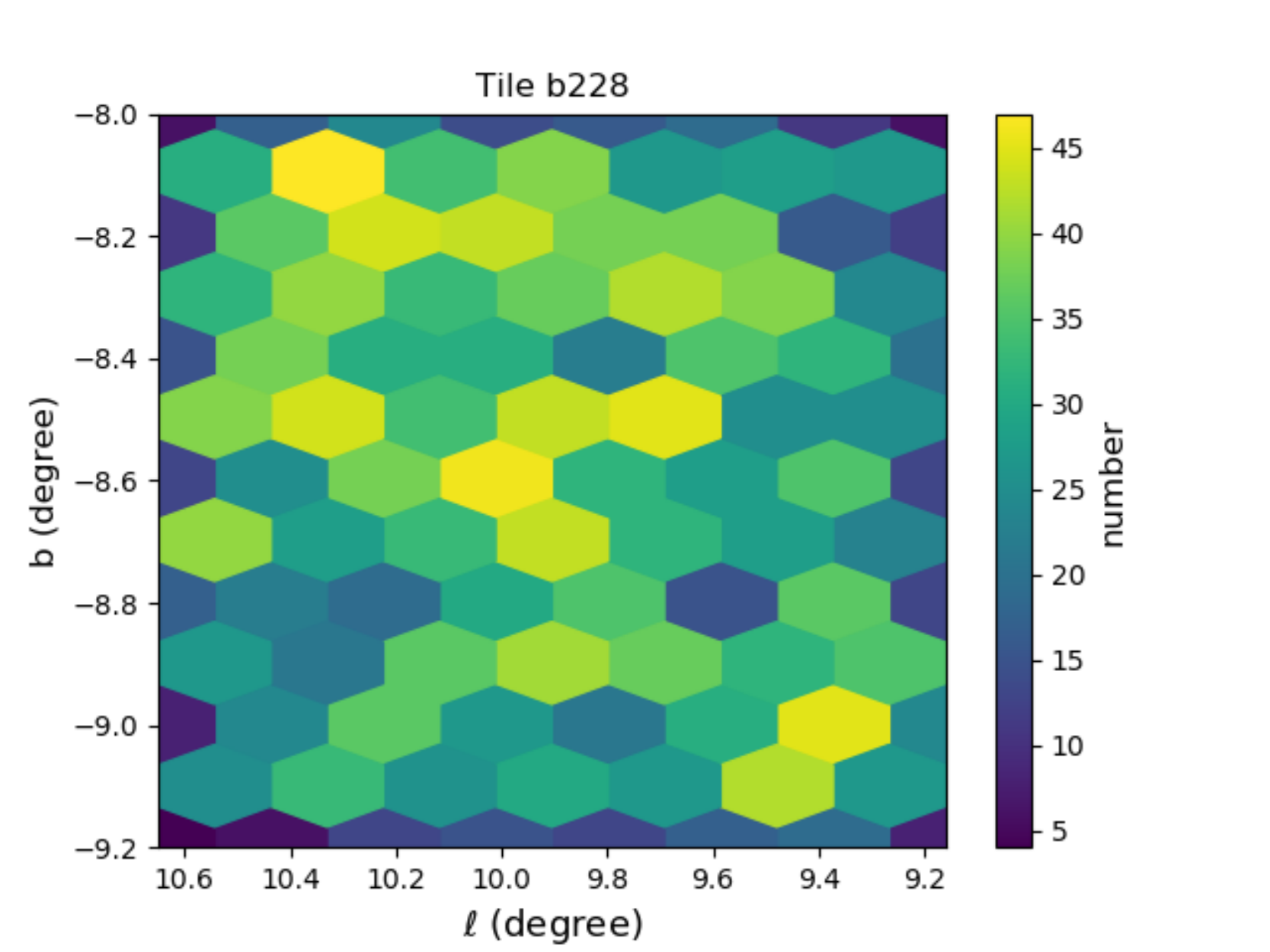}
	\caption{Some interesting over-densities corresponding to the second binning exercise (bin size of 12$\arcmin$).}
	\label{fig:sim2}
\end{figure}

In order to study these over-densities more clearly, Monte Carlo
simulations were performed for each tile, using as the sample the number
of stars contained in each tile, plus the real stars in each of them.
The most interesting results of each simulation are shown in Figures
\ref{fig:sim1} and \ref{fig:sim2}. This simple exercise shows that we
have to report a negative result; even though there may be some
over-densities, these are probably not due to previously unknown
globular clusters. These over-densities could correspond to streams,
which will be addressed in future work.

\section{light curves}
\subsection{Variability}

We search for variability by constructing light curves of all the BHB
stars in our sample. In order to do this, additional data were needed,
which were downloaded from the VSA (VISTA Science Archive)
\footnote{\url{http://horus.roe.ac.uk/vsa/}} hosted on the Royal
Observatory of Edinburgh website. Then, a cross-match was performed
using the VSA and the original data available for each star. For each
object there are on average 52 epochs; each magnitude has its respective
error. Typical errors range from 0.01 to 0.06 mag at the
brightest and faintest magnitudes of our sample, respectively.

The tiling pattern produces overlap regions between the tiles,
corresponding to about 7\% of the total VVV Survey area
\citep{saito2012b}. In these regions the data for each source are
combined, thus producing light curves with double, or even four times
more, data-points. In our sample 80 objects are in the overlap regions,
with light curves reaching +100 epochs in total. In order to
account for spuriously large photometric errors, we apply sigma clipping
by eliminating the points are more than $3\sigma$ away from the mean of
the light curve.

\begin{figure}
 \centering
 \includegraphics[scale=0.5]{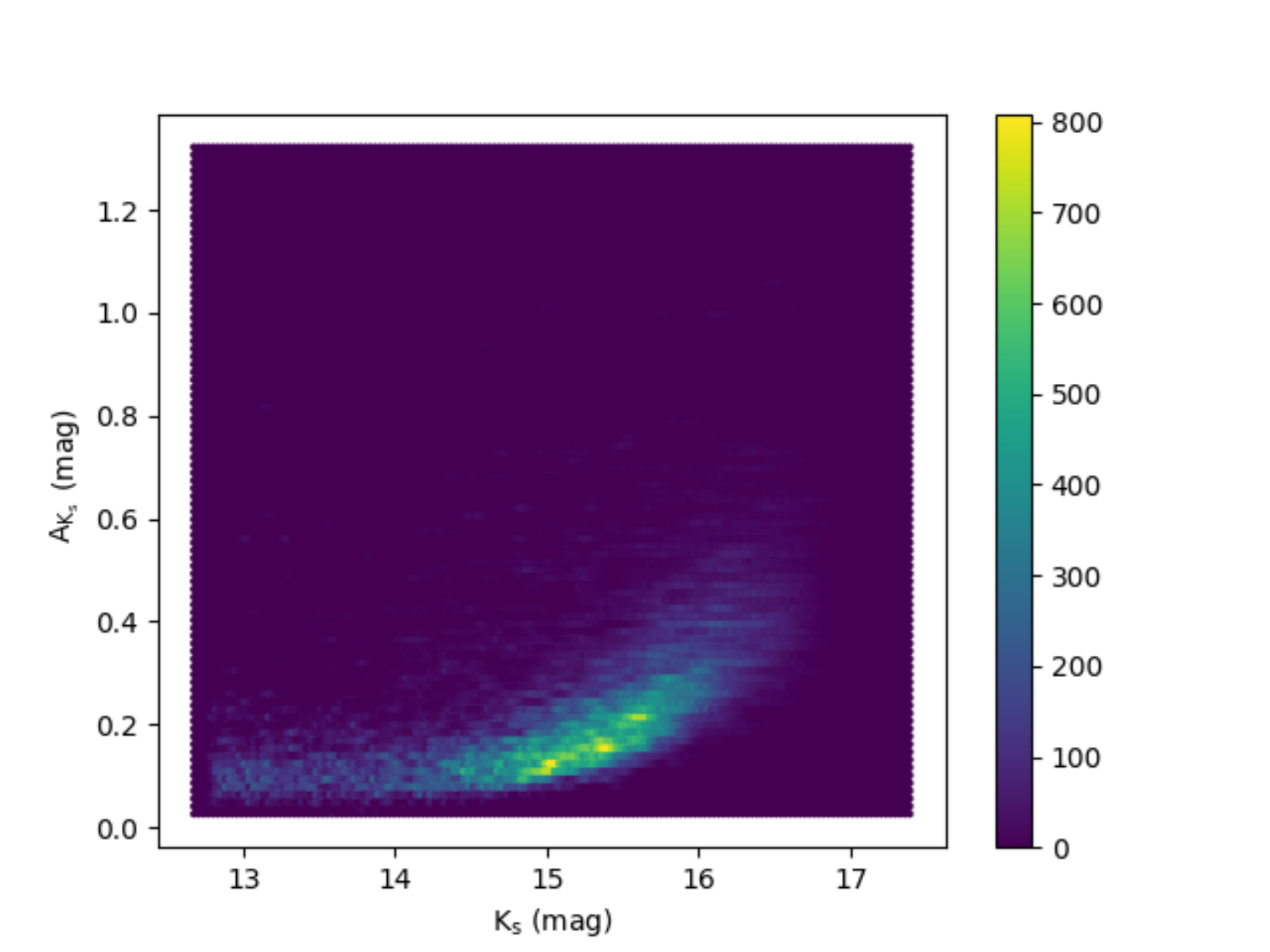} 
 \includegraphics[scale=0.5]{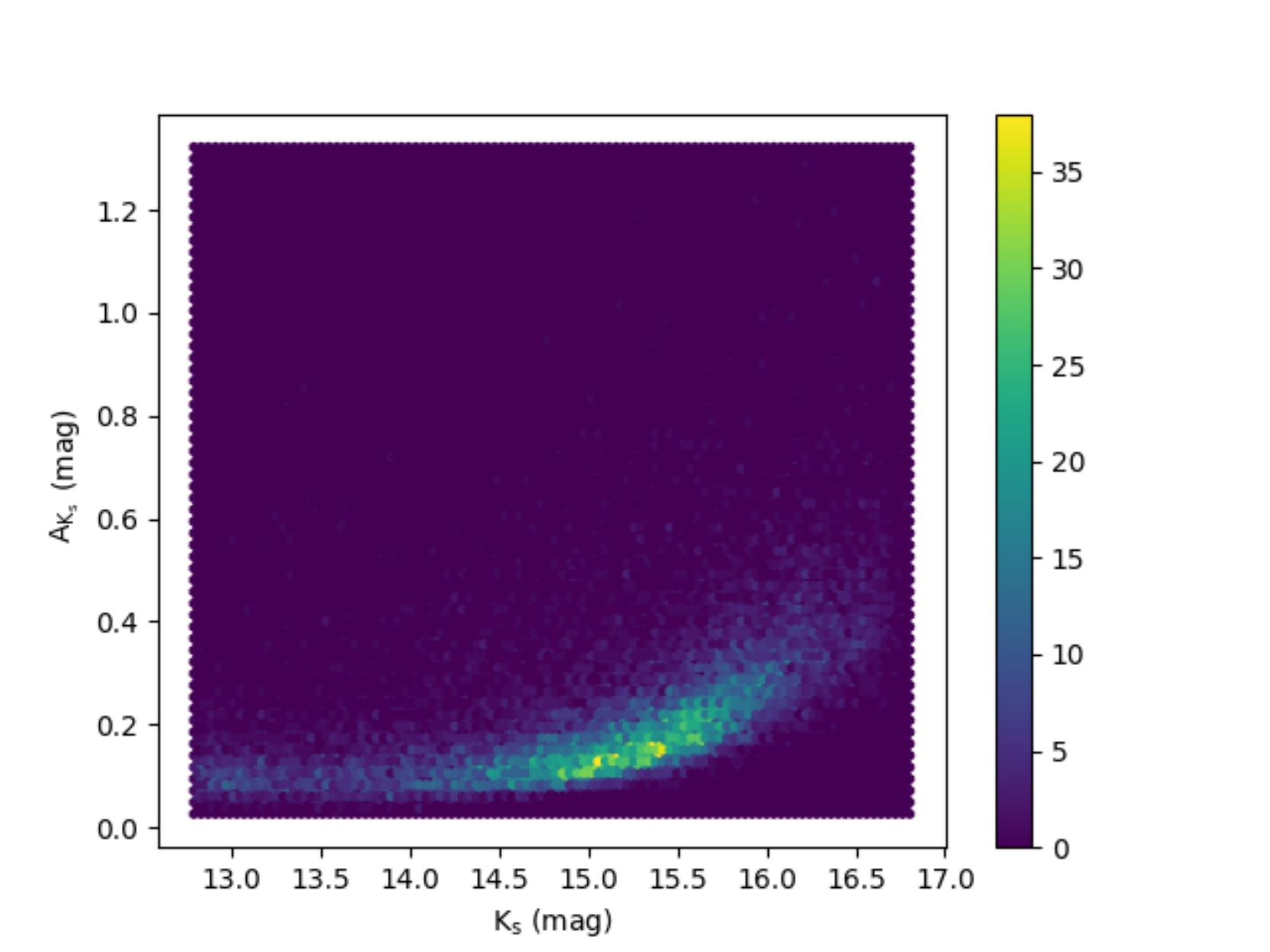}	
 \caption{{\bf Top panel:} Amplitude vs. $K_s$ magnitude for the total
 sample of BHB stars. This plot allows us to determine a cut of minimum
 amplitude. {\bf Bottom panel:} Amplitude of the light curve vs mean
 $K_s$ magnitude after cleaning the light curve with our sigma
 clipping constraint, and applying the cut at Amp $K_s$ = 0.15 mag.}
 \label{fig:variab}
\end{figure}

With this information, plots of amplitude versus magnitude were made
(Figure \ref{fig:variab} top panel), in order to find a suitable cut for
the minimum amplitude that could yield a cleaner and easier to study
light curve. A cut of minimum amplitude is set at Amp ${K_s}$ = 0.15 mag
for the light curves of interest, after a visual inspection of Figure
\ref{fig:variab} (top panel). 

Finally, Figure \ref{fig:variab} (bottom panel) shows the magnitudes of
the selected BHB stars and their amplitudes, after cleaning the light
curves, and applying the cut in amplitude. A total of 7,665
light curves were obtained using this criterion, which corresponds to
61\% of the original sample and whose magnitudes are within the range
$12.68 \leq K_s \leq 16.80$. The remaining 39\% of the sample are BHB
stars that do not exhibit variability, which are beyond the scope of this work, but these candidate stars are still being studied and will be presented in a future paper.

\subsection{Period Searches}
The Lomb-Scargle Periodogram (\citealt{lomb}; \citealt{scargle}) is a
statistical tool used to detect periodic signals in unevenly spaced
observations; it works by making a search for the maximum peak in
frequency of the light curve data. This estimating function was
used to determine the period of our BHB stars' phased light curves.

\begin{figure}[h!]
 \centering
 \includegraphics[scale=0.34]{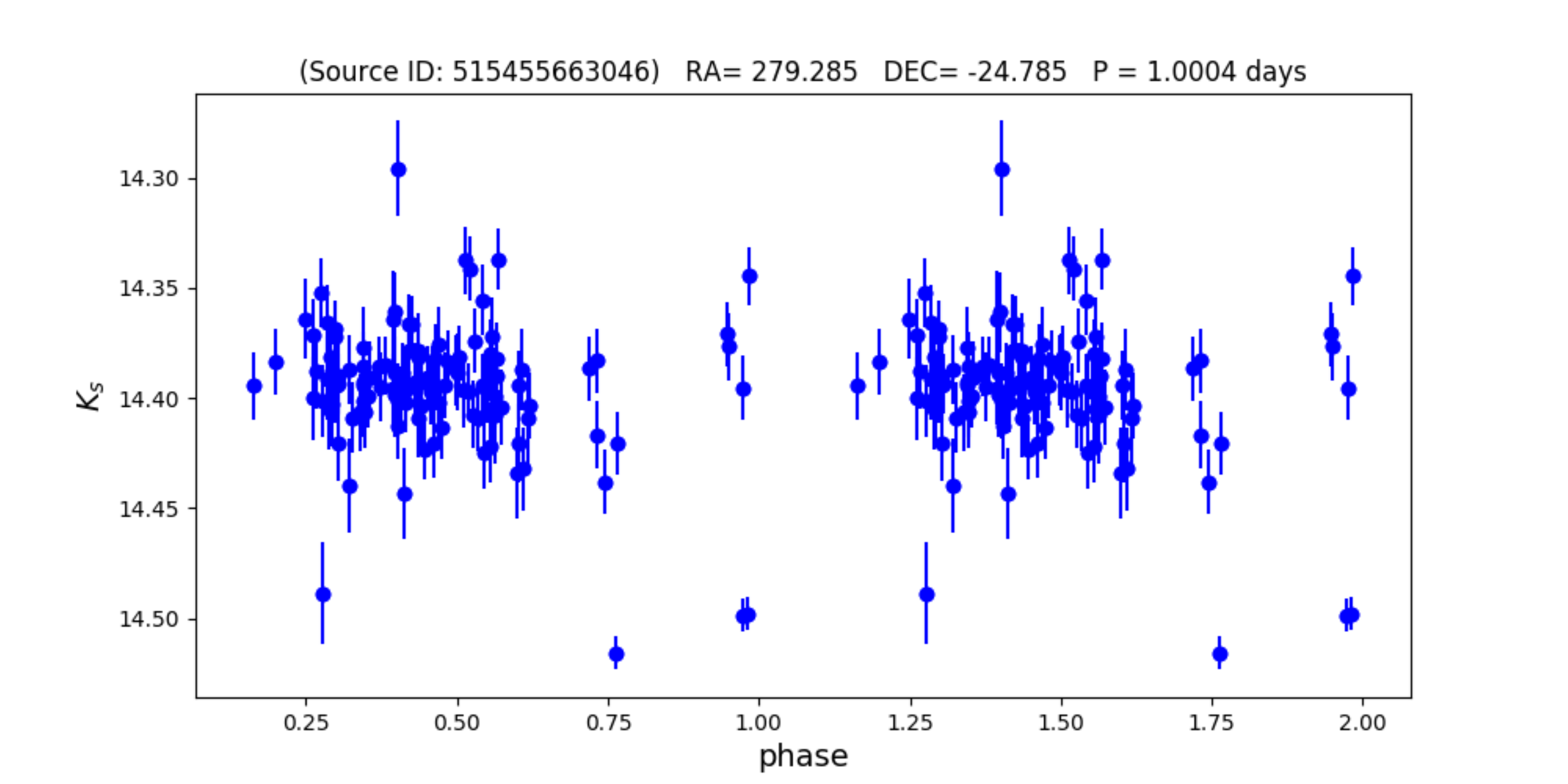}	
 \caption{Example of a BHB star light curve with a daily aliased period.}
 \label{fig:alias}
\end{figure}

An alias is a false period that contaminates the periodograms.
In particular, daily aliases occur when the interval between
observations matches the day/night cycle, half or some other multiple of
this (\citealt{scargle}; \citealt{baluev}). Observing the values
obtained for the periods and the light curves of our sample of BHB
stars, we find that 1,721 of them correspond to daily aliased
periods between 0.1 and 3.$\overline{3}$ days, being the most common alias of one day. These light curves, like the one
shown in Figure \ref{fig:alias}, are left out of our analysis.

\begin{figure}[h!]
 \centering
 \includegraphics[scale=0.34]{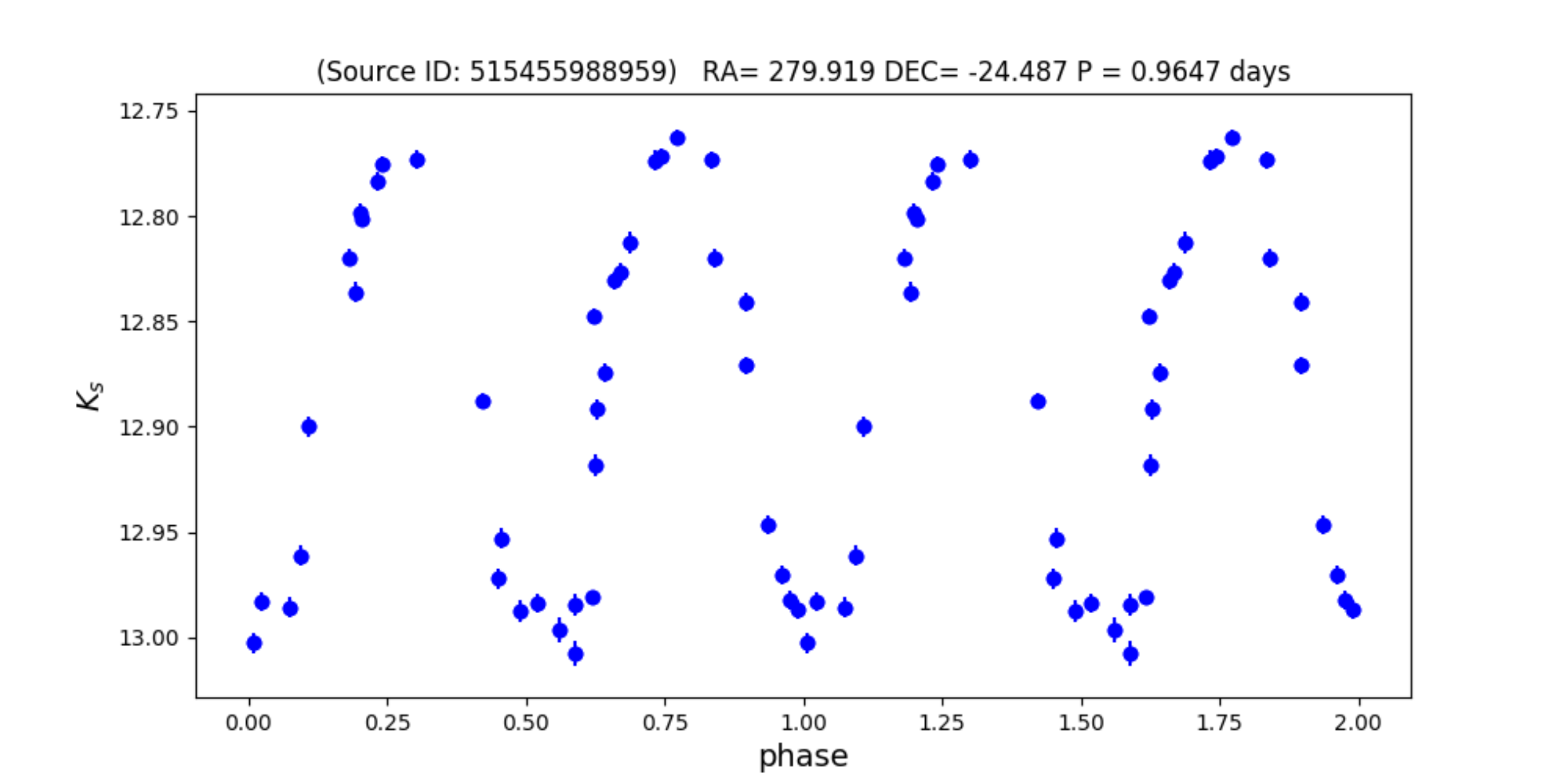}
 \includegraphics[scale=0.34]{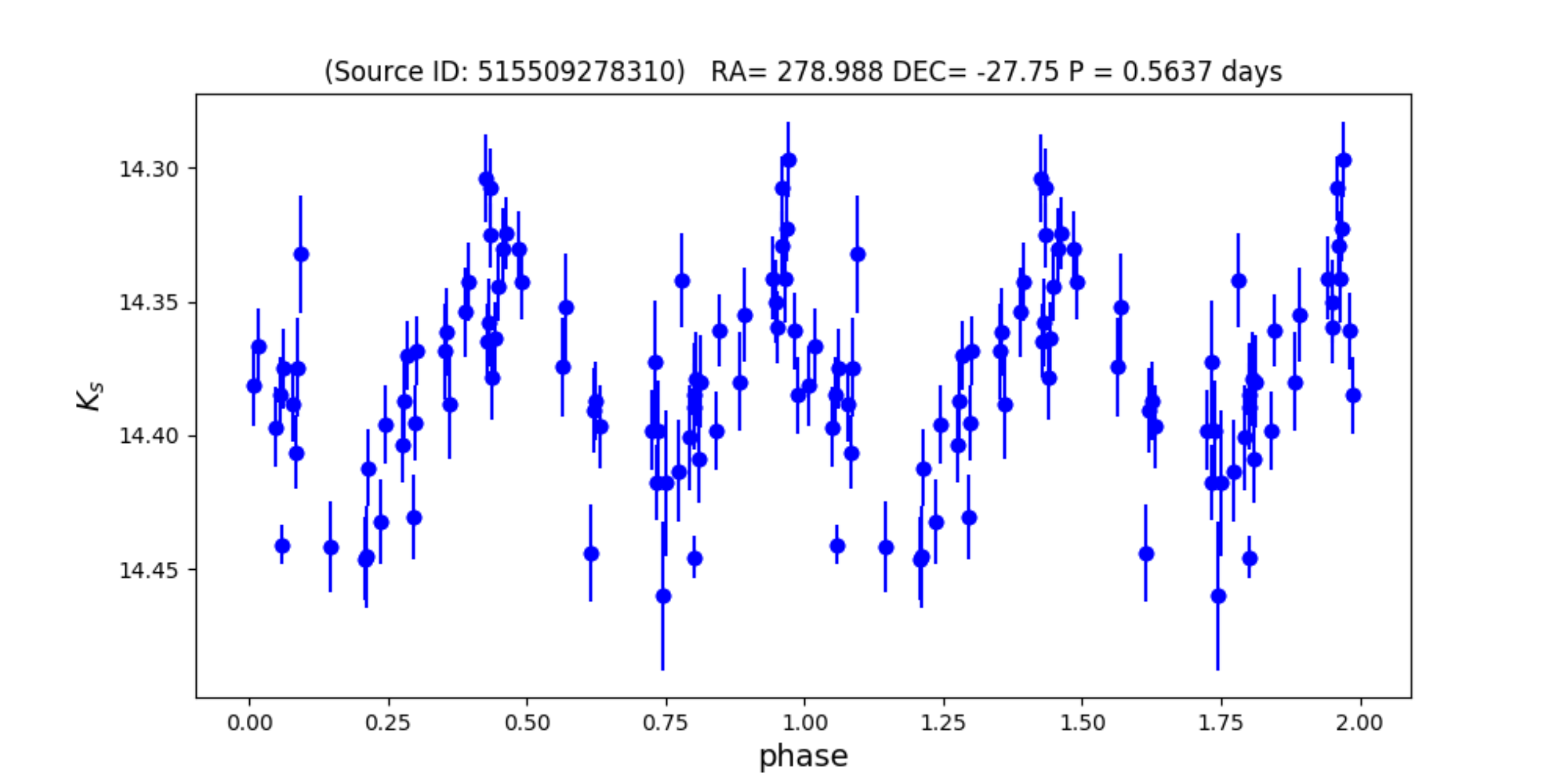}	
 \caption{{\bf Top panel:} light curve of a BHB star classified as a
 $1^{st}$ category eclipsing binary. {\bf Bottom panel:} light curve of
 a BHB star classified as a $2^{nd}$ category eclipsing binary.}
 \label{fig:eb}
\end{figure}

With a total of 5,944 good-quality light curves, we carried out a visual
inspection at least three times, to determine what kind of
variable stars they were. Following the models exhibited by
\cite{prsa2011}, we detected a total of 336 eclipsing binaries (Figure
\ref{fig:eb}). We classified 232 of them as first category (Figure
\ref{fig:eb} top panel), due to their error bars are on average 0.02 mag and the other 104 sources were classified as second
category, because they exhibited error bars with an average of 0.04 mag (Figure \ref{fig:eb} bottom panel). Regardless,
all of them exhibit light curves that are sufficiently clear
to distinguish them from other variable stars. The remaining
5,608 variable stars remain unclassified, until more epochs become
available in their light curves.

\begin{figure}[h!]
 \centering
 \includegraphics[scale=0.30]{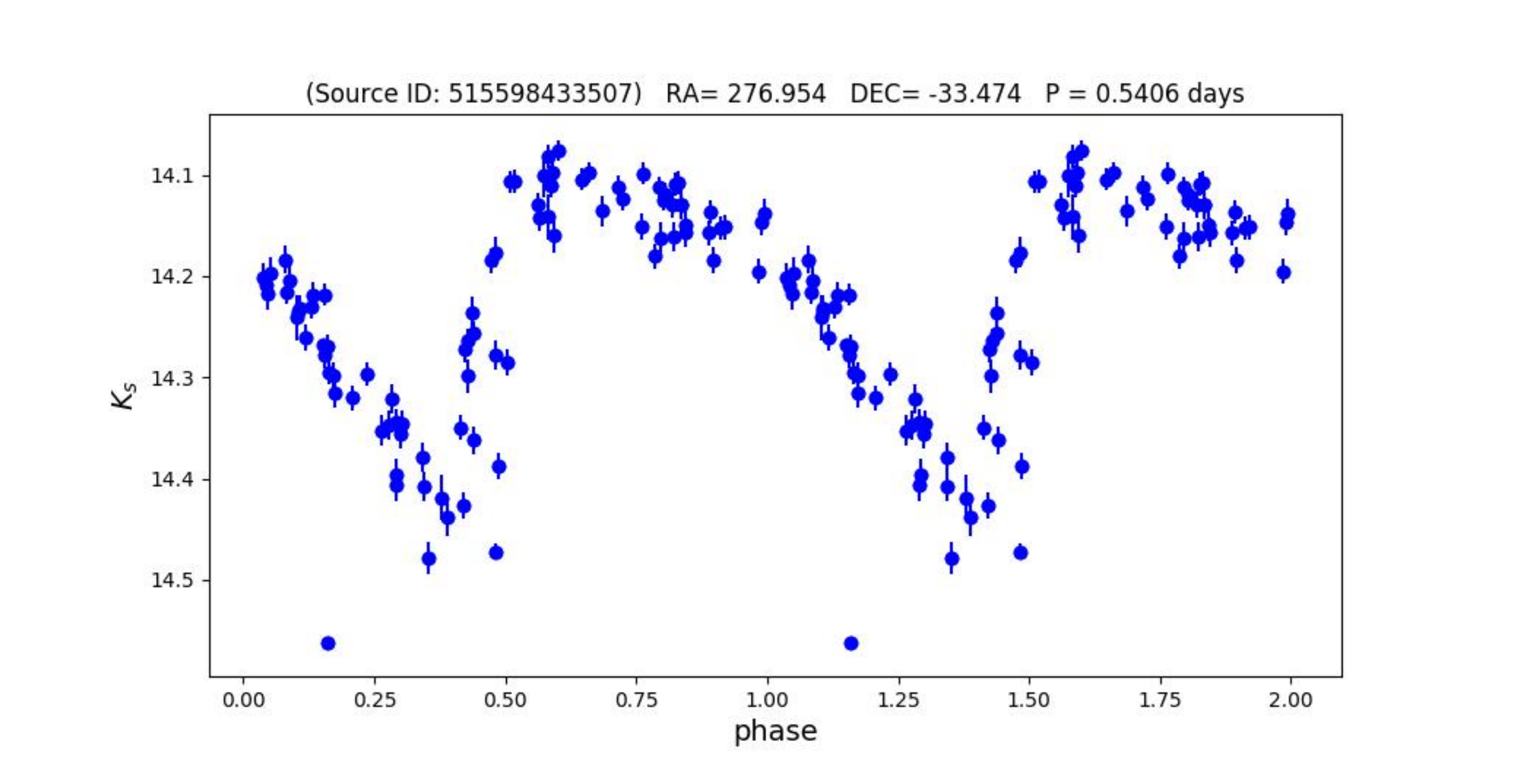}	
 \caption{RR Lyrae star found in our sample.}
 \label{fig:rrlyra}
\end{figure}

We also found 12 RR Lyrae stars, such as that shown in Figure
\ref{fig:rrlyra}; all are listed in Table \ref{table:rr}. Seven of these
RR Lyrae were independently discovered by \cite{gran2016}, using a
different selection method that does not consider the color of the stars
as in this work. The remaining 5 RR Lyrae have bluer colors ($J-K_s$
$\sim 0.27$). Considering the Hess diagram presented by \cite{gran2016}, our 5
RR Lyrae have colors within the color range of their stars.

\subsection{Classification of Eclipsing Binaries}
An eclipsing variable is a binary system with its orbital plane oriented
edge-on towards the Earth, in such a way that eclipses and transits can
occur. According to the General Catalogue of Variable Stars
\footnote{\url{http://www.sai.msu.su/gcvs/gcvs/}} \citep{samus2017},
eclipsing binaries can be classified into three groups, depending on the
shape of their light curves: EA (Algols), EB ($\beta$ Lyrae), and EW (W
Ursae Majoris).

\begin{figure}[h!]
 \centering
 \includegraphics[scale=0.55]{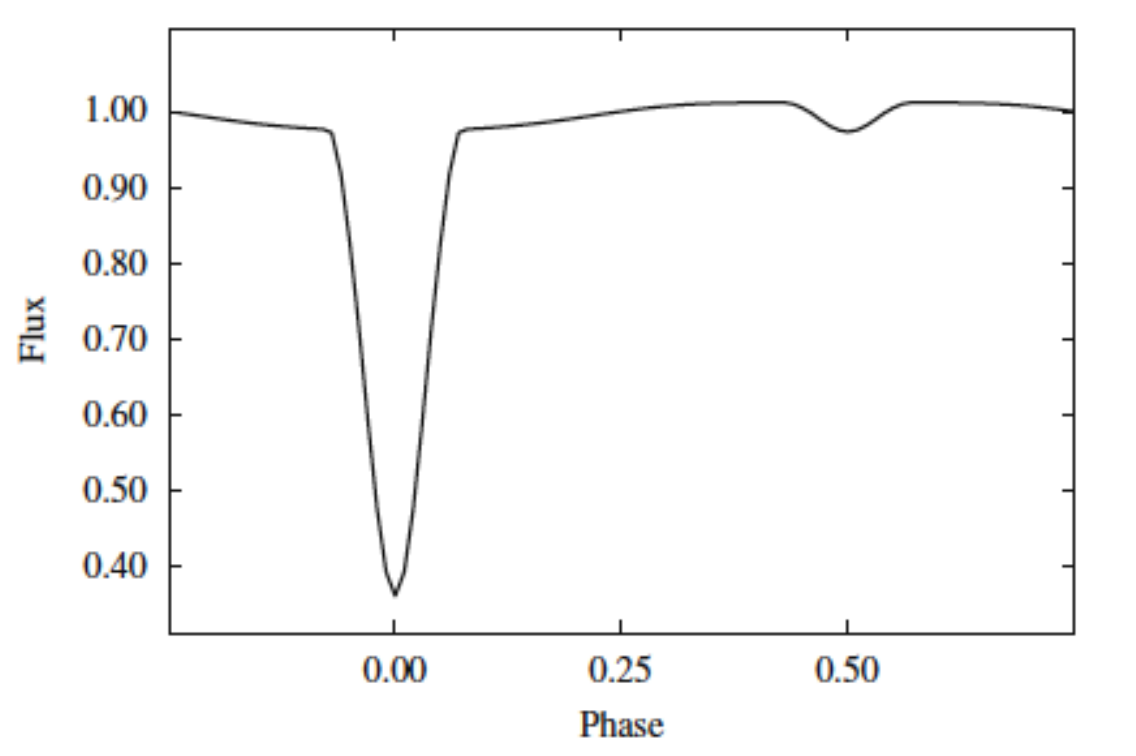}
 \includegraphics[scale=0.33]{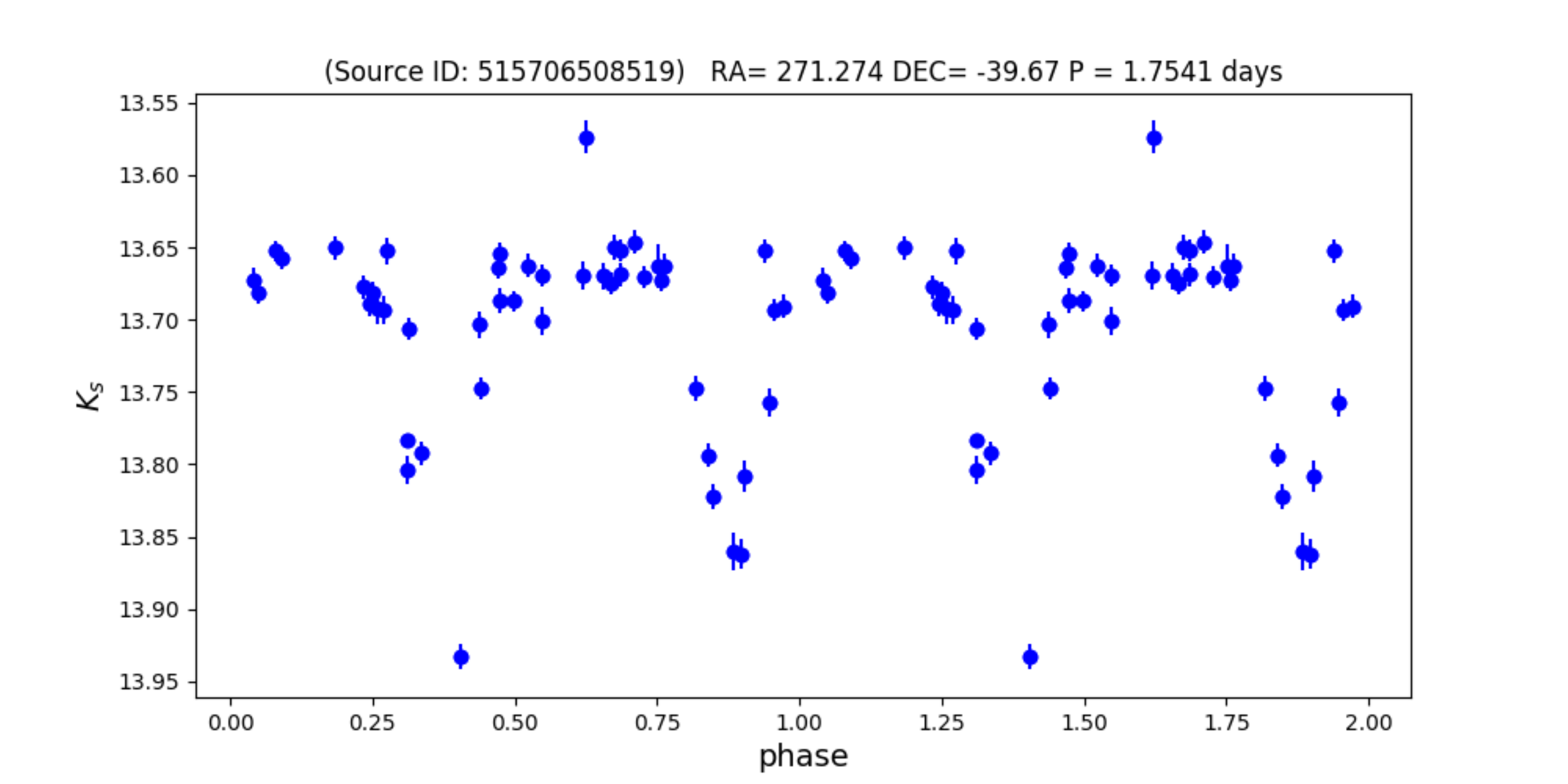}	
 \caption{EA eclipsing binaries. A synthetic light curve (obtained from \citealt{kallrath}) in the top panel can be compared with one of our EA light curves shown in the bottom panel.}
 \label{fig:3eba}
\end{figure}

\begin{figure}[h!]
 \centering
 \includegraphics[scale=0.55]{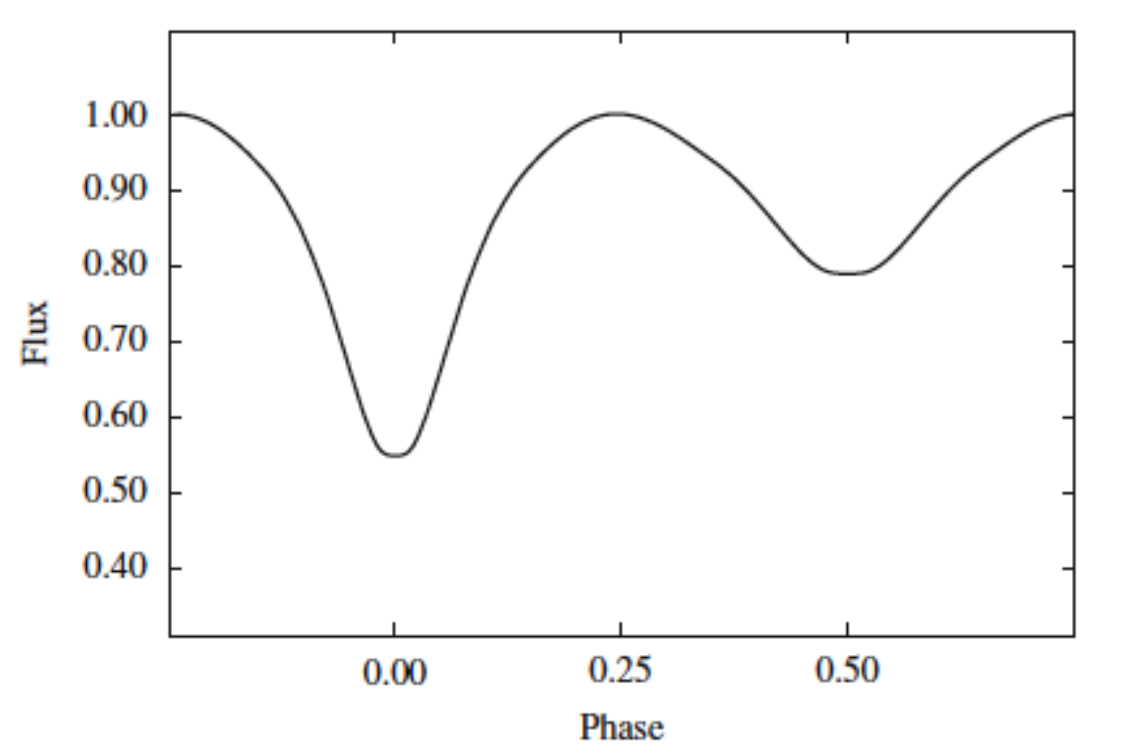}
 \includegraphics[scale=0.33]{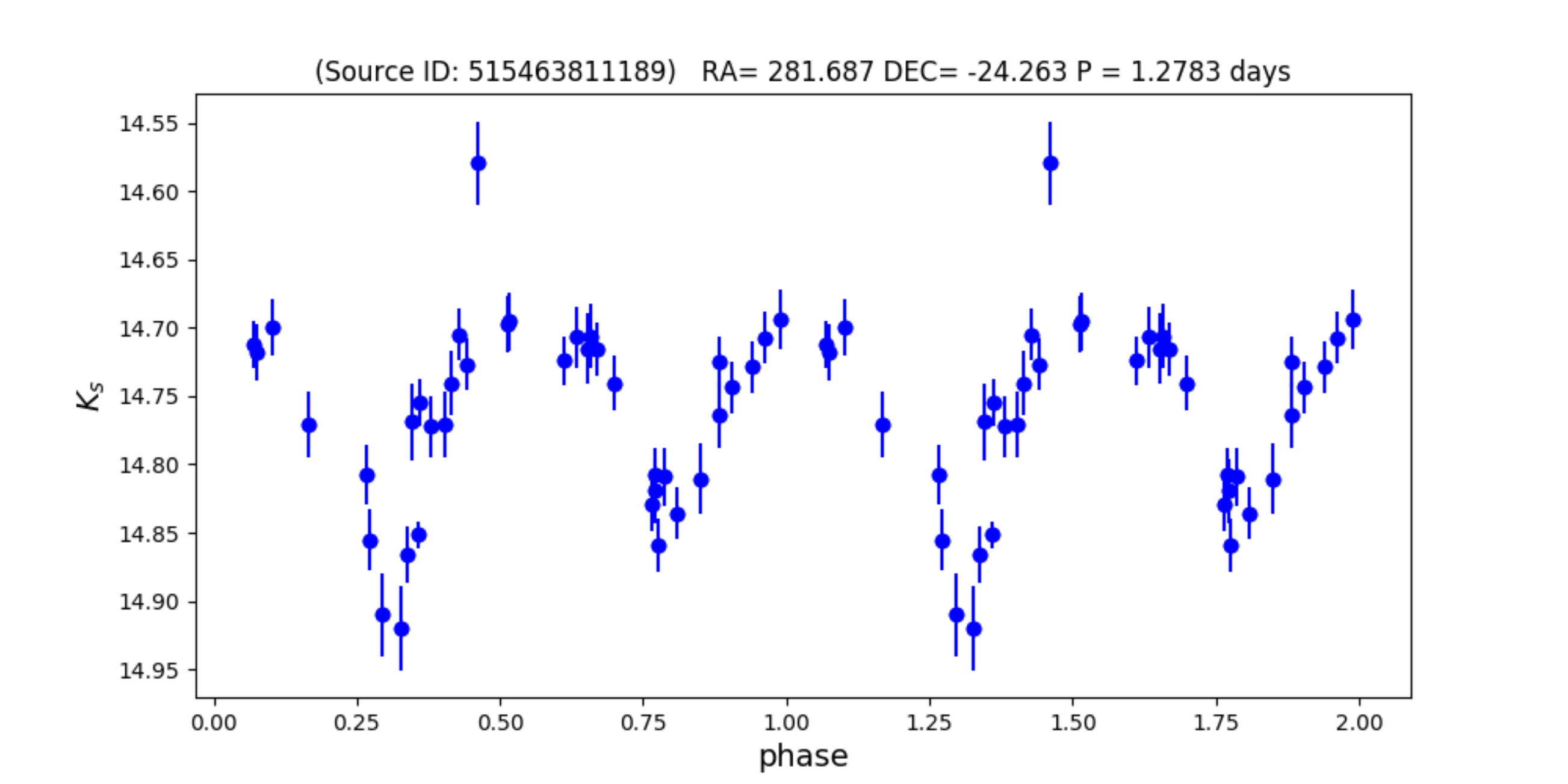}		
 \caption{EB eclipsing binaries. A synthetic light curve (obtained from \citealt{kallrath}) in the top panel can be compared with one of our EB light curves in the bottom panel.}
 \label{fig:3ebb}
\end{figure}

\begin{figure}[h!]
 \centering	
 \includegraphics[scale=0.55]{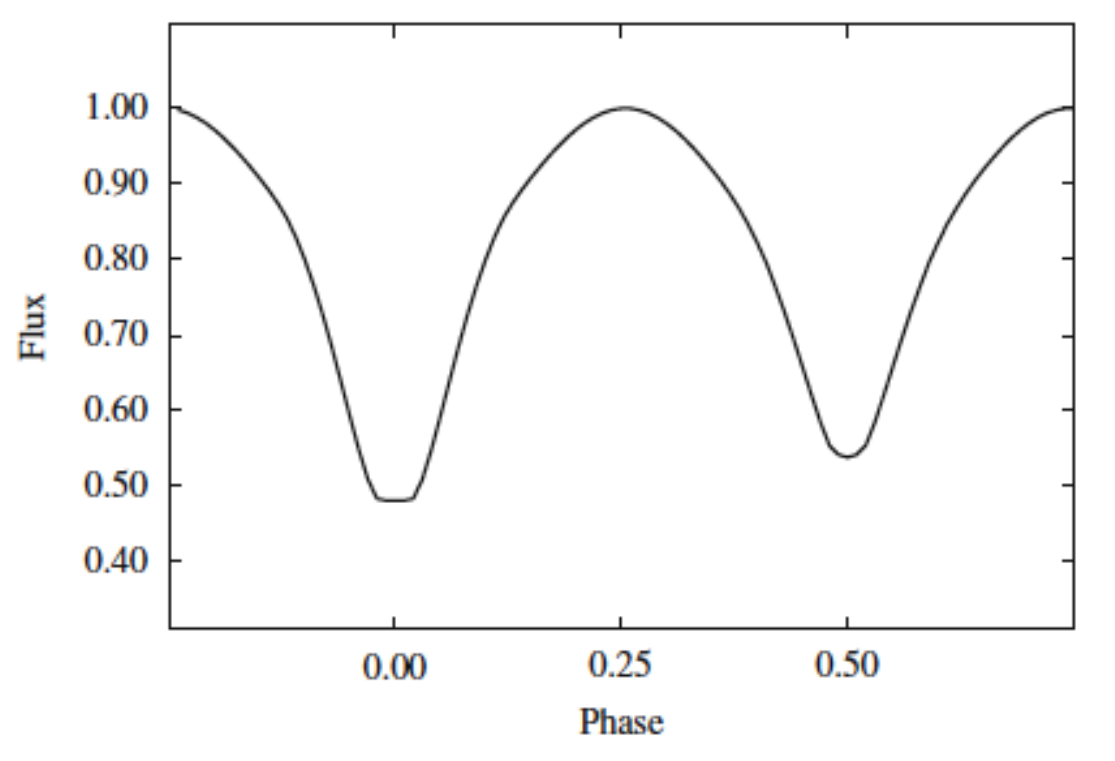}
 \includegraphics[scale=0.33]{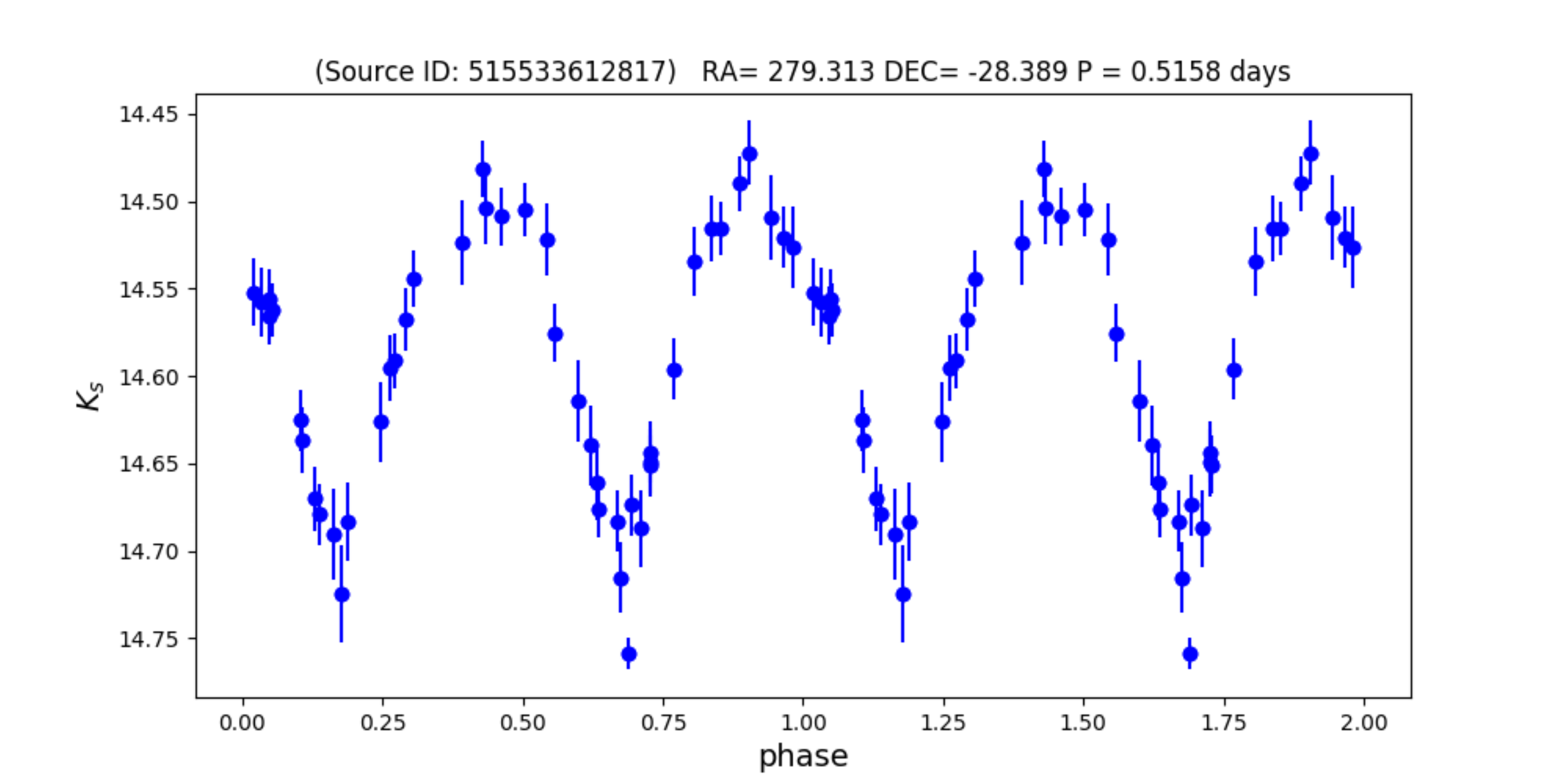}	
 \caption{EW eclipsing binaries. A synthetic light curve (obtained from \citealt{kallrath}) in the top panel can be compared with one of our EW light curves in the lower panel.}
 \label{fig:3ebc}
\end{figure}

From visual inspection, we classified a total of 42 EA eclipsing
binaries, since these are different from the other two types,
mainly due to the flat portion of their light curves. Then, to
distinguish between the other two types, EB and EW, a deeper analysis is
made, considering their periods and amplitudes in addition to visual
inspection of the light curve shape, thus obtaining 119 EB type and 175
EW type eclipsing binaries in our sample of BHB candidates
in the outer bulge. Tables \ref{table:1cat} and \ref{table:2cat} present
the characteristics (coordinates, periods, amplitudes, and magnitudes) of
each of these candidate BHB eclipsing binary stars. Likewise, Figures
\ref{fig:bhb1} - \ref{fig:bhb10} and \ref{fig:bhbB1} - \ref{fig:bhbB5}
show the light curves of $1^{st}$ and $2^{nd}$ category examples
of these 336 stars, respectively. 

The region of the CMD occupied by the BHB stars is rich in variable
stars, including eclipsing binaries, as recently shown by the Gaia
satellite \citep{eyer}. In particular, shorter period eclipsing binaries
predominate in our sample because they are easier to detect than longer
period ones. As expected, the EWs are more concentrated towards
shorter periods than the EAs and EBs.

\subsection{Period-Amplitude Diagram}

\cite{bailey} classified RR Lyrae stars by their periods and amplitudes of variation.
Even now, the Period-Amplitude diagram (a.k.a. the Bailey diagram) is
quite useful for the classification and study of the intrinsic
properties of variable stars.

\begin{figure}[h!]
 \centering
 \includegraphics[scale=0.28]{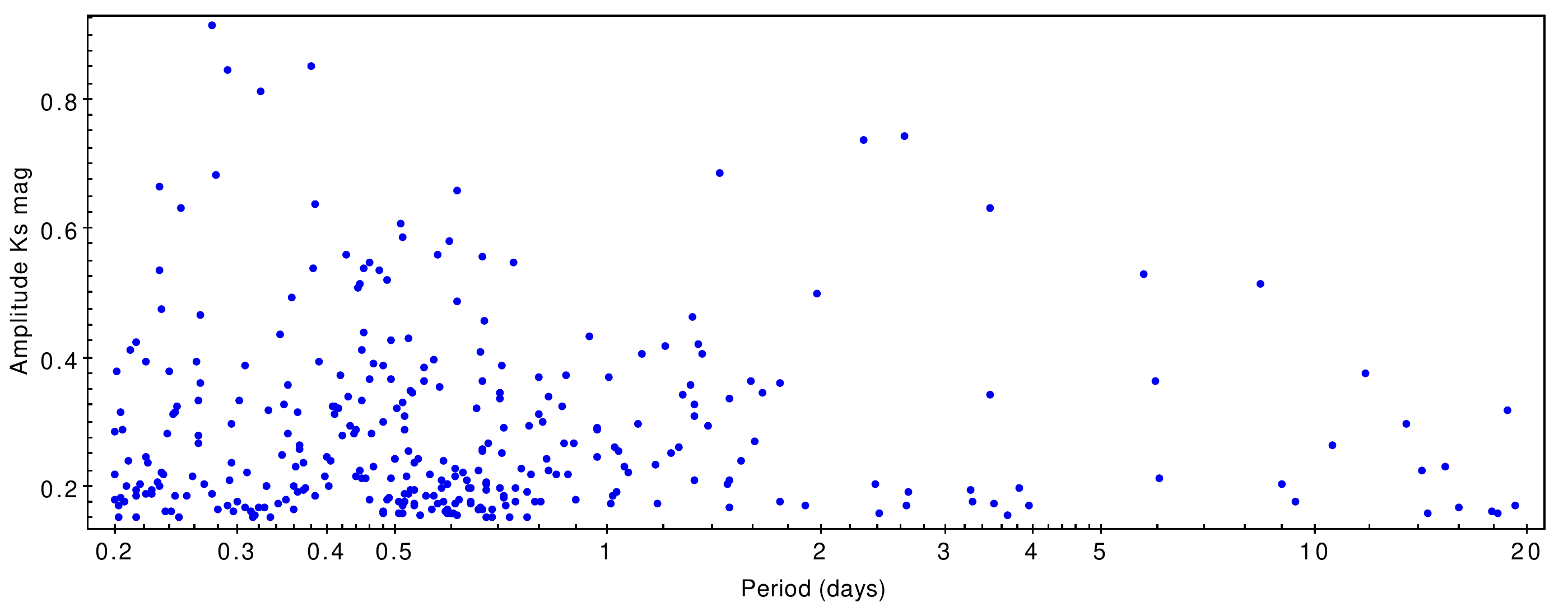}
 \includegraphics[scale=0.27]{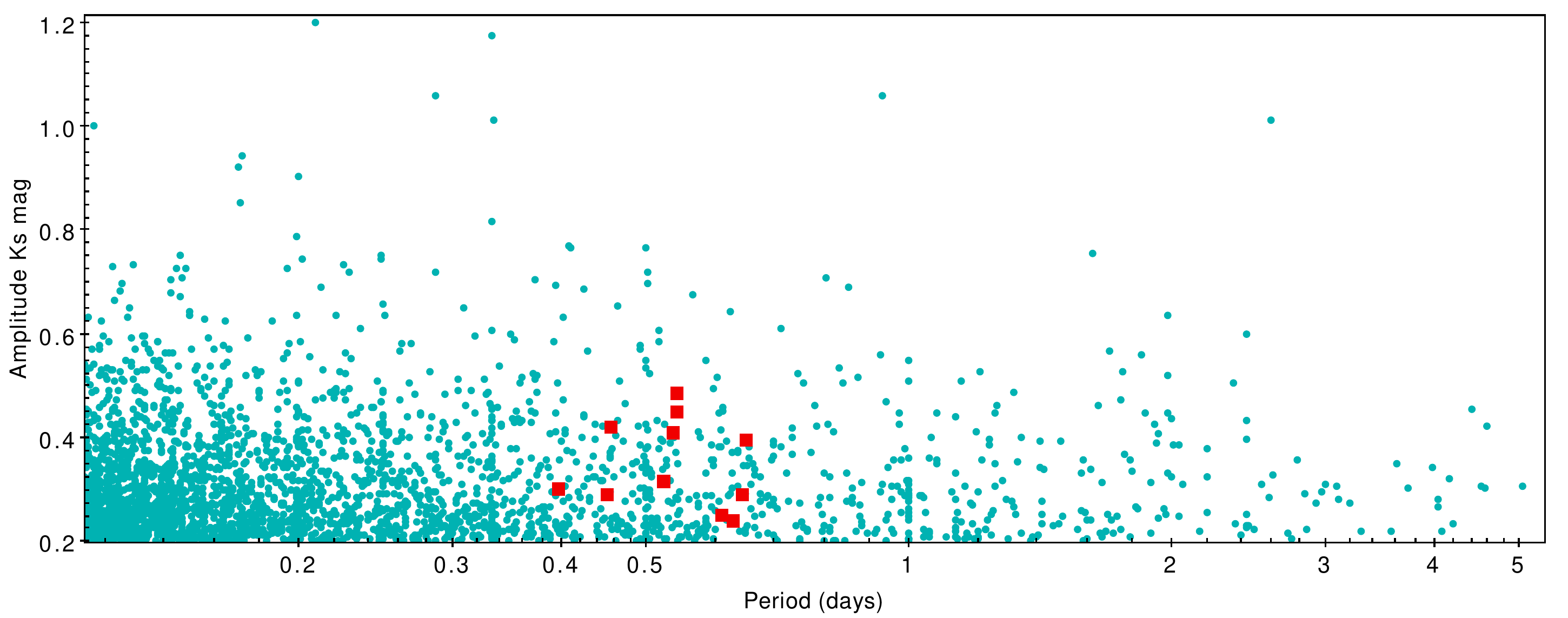}
 \caption{{\bf Top panel:} Period-Amplitude diagram of eclipsing binaries from our sample. {\bf Bottom panel:} Period-Amplitude diagram of the remaining variables. Turquoise points correspond to BHB stars and red squares are our 12 RR Lyrae.}
 \label{fig:bailey}
\end{figure}

From inspection of the top panel of Figure \ref{fig:bailey}, one
can see that the periods of
our eclipsing binaries span between $0.2 < P < 19.2$ days, and their
amplitudes are in the range $0.15 < \rm Amp \: K_s < 0.91$ mag.
Eclipsing binaries, corresponding to 2.7\% of the BHB stars original
sample, are more homogeneously distributed than the remaining variable stars in the sample, which are strongly concentrated
at low amplitudes and short periods. The average period is 1.12 days for
eclipsing binaries, which is a similar value to that obtained by
\cite{rozyczka} for 29 BHB stars that correspond to eclipsing binaries
in the field of the globular cluster M22, which have an average
period of 1.86 days. 

According to the Gaia DR2 variable star color-magnitude diagram
given by \cite{eyer} (their Figure 4), there are eclipsing binaries present in the
horizontal branch. For this reason, we decided to perform a cross-match
between our sample of BHB eclipsing binaries and the objects detected by
Gaia DR2 \footnote{\url{http://vizier.u-strasbg.fr/viz-bin/VizieR-3}} in
this area of the Milky Way, considering only sources which were
classified as variables. We found 100 matches in total, corresponding to
about one third of our eclipsing binaries sample, which we consider a
good confirmation of our results. We expect an increase in the number of
matches when Gaia completes the classification of variable and
non-variable objects in the area studied here.

From inspection of the bottom panel of Figure \ref{fig:bailey},
one can see that the periods
of the remaining variables are in the range $0.1 < P < 8.5$
days, and their amplitudes span between $0.15 < \rm Amp \: K_s < 1.32$
mag. These variables stars, corresponding to 44.7\% of the BHB candidates in the original sample, reach the highest amplitudes and
periods, almost as long as the eclipsing binaries; an average
period of 0.53 days. As expected, the 12 RR Lyrae found in our sample
are included within the ranges of this part of the sample. In addition,
in our sample, most of the long-period variables have smaller
amplitudes, while short-period variables have larger amplitudes, on
average. 

In spite of their small numbers, our RR Lyrae are located in the Oosterhoff type I 
region of the Bailey diagram, different from the M22 RR Lyrae that are
Oosterhoff type II. This is in agreement with the previous
results of \cite{gran2016}, who analyzed the Bailey diagram for a
thousand bulge RR Lyrae discovered by the VVV survey.

\section{Results}

We have studied an area of 40 square degrees on the sky, in order to
characterize the population of BHB stars in the direction of the
Galactic bulge using VVV data. This area corresponds to 28 tiles of
$1.1\times1.5$ square degrees each, in the bulge-halo transition region
of the Milky Way. We discovered a total of 12,554 BHB candidate stars
using a strict color selection, taking the BHB members of the old and
metal-poor globular cluster M22 as a reference. The present bulge BHB
catalog is used here to search for new globular clusters and variable
stars.

Monte Carlo simulations were performed with the goal of detecting
over-densities that could be undiscovered globular clusters, covering
the total area first and then each tile individually. The first
simulations used a bin size of 2$\arcmin$ -- some over-densities were
discovered, but all of them were of low significance. The second
simulation used an extreme bin size of 12$\arcmin$ -- some of the
corresponding over-densities reveal two possible streams, whose analysis
will be a part of future work.

We also examined the variability of the BHB sample, plotting light
curves having an average of 52 epochs. By choosing as a minimum
amplitude $\rm Amp \: {K_s}$ = 0.15 mag, we found 7,665 variable BHB
candidates with mean magnitudes between $12.68 \leq K_s \leq 16.80$.
Their mean magnitudes are consistent with the majority of them being
located at the distance of the Galactic bulge, $d\approx 8$ kpc. Their periods were determined using the Lomb-Scargle periodogram, 
yielding an average of 1.12 days for eclipsing binaries and 0.53 days
for the remaining variables. A total of 1,721 variable stars with aliased periods was detected, which were discarded. We
performed a visual inspection of the remaining sample of 5,944 BHB
stars, finding 336 good-quality eclipsing binary candidates, with
periods up to P = 19.2 days. These were further sub-classified as
detached, semi-contact, and contact binaries (12.5 \% EA, 35.4 \% EB,
52.1 \% EW). A total of 12 RR Lyrae were detected, including 7 of them
that were independently discovered by \cite{gran2016}. From a
period-amplitude diagram we found that the amplitude range for the remaining variables is $0.15 \leq \rm Amp \: {K_s} \lesssim
1.32$ mag and $0.15 \leq \rm Amp \: {K_s} \lesssim 0.91$ for eclipsing
binaries, with a period range $0.1 < P < 8.5$ days for the
remaining variables, and $0.2 < P < 19.2$ for eclipsing binaries.

Finally, comparing our sample of BHB stars with the study of RR Lyrae
stars in the same region by \cite{gran2016}, we see that the Galactic
bulge BHB population is more than an order of magnitude larger than that
of the RR Lyrae ($\sim 14 \times$).

\acknowledgments
The authors gratefully acknowledge the careful review of this
manuscript by an anonymous referee, which contributed to clarification of
our results.  We also gratefully acknowledge the use of data from the ESO Public Survey
program ID 179.B-2002 taken with the VISTA telescope, and data products
from the Cambridge Astronomical Survey Unit (CASU). Support for the
authors is provided by the BASAL Center for Astrophysics and Associated
Technologies (CATA) through grant PFB-06, and the Ministry for the
Economy, Development, and Tourism, Programa Iniciativa Cient\'ifica
Milenio through grant IC120009, awarded to the Millennium Institute of
Astrophysics (MAS). D.M. acknowledges support from FONDECYT Regular
grant No. 1170121. DM is also grateful for the hospitality of the
Vatican Observatory. T.C.B. acknowledges partial support from grant PHY
14-30152 (Physics Frontier Center/JINA-CEE), awarded by the U.S.
National Science Foundation. T.C.B. is also grateful for funding from
the Luksburg Foundation, which supported his travel to Chile in August,
2017, and enabled him to participate in this collaboration effort.
J.A.-G. acknowledges support from FONDECYT Iniciation grant 11150916.

\appendix
 
\startlongtable
\begin{deluxetable}{c|c|c|c|c|c|c|c|c|c|c} 
\tabletypesize{\scriptsize} 
\centering 
\tablecaption{Complete Parameters for the First Category Eclipsing Binaries \label{table:1cat}}       
\tablehead{
\colhead{VVV ID} & \colhead{RA(J2000)} & \colhead{DEC(J2000)} & \colhead{Period} & \colhead{$A_{K_s}$} &\colhead{Z} & \colhead{Y} & \colhead{J} & \colhead{H} & \colhead{$K_{\rm{s}}$} & \colhead{Class}} 
\startdata 
	VVV 515455705782 & 18:41:02.6 & $-$25:08:27.6 & 3.2937 & 0.18 & 14.75 & 14.93 & 14.51 & 14.24 & 14.12 & EB\\ 
	VVV 515455735619 & 18:37:32.4 & $-$24:42:03.6 & 14.4274 & 0.16 & 15.20 & 15.38 & 15.04 & 14.93 & 14.82 & EB\\ 
	VVV 515455749525 & 18:40:50.8 & $-$25:02:27.6 & 1.0277 & 0.19 & 15.67 & 15.80 & 15.53 & 15.35 & 15.27 & EB\\  
	VVV 515455859192 & 18:38:27.6 & $-$24:35:16.8 & 0.7135 & 0.29 & 14.74 & 14.92 & 14.52 & 14.29 & 14.17 & EB\\   
	VVV 515455988959 & 18:39:40.5 & $-$24:29:13.2 & 0.9647 & 0.25 & 13.47 & 13.66 & 13.23 & 12.98 & 12.85 & EW\\
	VVV 515456105530 & 18:41:39.6 & $-$24:29:24.0 & 1.0014 & 0.37 & 15.13 & 15.31 & 14.94 & 14.75 & 14.63 & EA\\
	VVV 515456121023 & 18:41:49.6 & $-$24:28:48.0 & 0.7075 & 0.39 & 15.45 & 15.61 & 15.25 & 15.01 & 14.92 & EB\\   
    VVV 515456130441 & 18:42:31.9 & $-$24:32:24.0 & 0.6329 & 0.21 & 14.23 & 14.40 & 14.09 & 13.88 & 13.80 & EB\\
\enddata 
\tablenotetext{a}{The complete parameters: VVV ID, right ascension (J2000), declination (J2000), period (days), amplitude of $K_s$ magnitude, \emph{ZYKH$K_s$} magnitudes and type of eclipsing binary.}
\end{deluxetable}

\startlongtable
\begin{deluxetable*}{c|c|c|c|c|c|c|c|c|c|c} 
\tabletypesize{\scriptsize} 
\centering 
\tablecaption{Complete Parameters for the Second Category of Eclipsing Binaries\label{table:2cat}}       
\tablehead{
\colhead{VVV ID} & \colhead{RA(J2000)} & \colhead{DEC(J2000)} & \colhead{Period} & \colhead{$A_{K_s}$} &\colhead{Z} & \colhead{Y} & \colhead{J} & \colhead{H} & \colhead{$K_{\rm{s}}$} & \colhead{Class}} 
\startdata 
	VVV 515455684844 & 18:39:58.3 & $-$25:03:36.0 & 0.1542 & 0.22 & 15.91 & 16.06 & 15.69 & 15.55 & 15.42 & EB\\
	VVV 515455868670 & 18:40:48.9 & $-$24:49:37.2 & 0.1043 & 0.20 & 15.95 & 16.10 & 15.75 & 15.55 & 15.52 & EW\\
	VVV 515455933928 & 18:39:48.9 & $-$24:35:56.4 & 0.1157 & 0.21 & 16.20 & 16.37 & 15.98 & 15.72 & 15.76 & EW\\
	VVV 515456085853 & 18:38:39.1 & $-$24:11:38.4 & 0.2052 & 0.31 & 15.41 & 15.59 & 15.16 & 14.90 & 14.79 & EA\\
	VVV 515456440625 & 18:39:44.6 & $-$23:38:24.0 & 0.1997 & 0.25 & 15.34 & 15.52 & 15.09 & 14.85 & 14.74 & EW\\
	VVV 515463134581 & 18:43:55.4 & $-$25:19:58.8 & 0.1374 & 0.19 & 15.61 & 15.78 & 15.43 & 15.32 & 15.20 & EW\\
	VVV	515463225981 & 18:42:07.9 & $-$24:57:36.0 & 0.1826 & 0.26 & 15.32 & 15.46 & 15.19 & 15.09 & 15.05 & EA\\ 
	VVV 515478514260 & 18:36:37.4 & $-$26:13:01.2 & 0.1662 & 0.15 & 15.51 & 15.69 & 15.36 & 15.21 & 15.16 & EB\\
\enddata  
\tablenotetext{a}{The complete parameters: VVV ID, right ascension (J2000), declination (J2000), period (days), amplitude of $K_s$ magnitude, \emph{ZYKH$K_s$} magnitudes and type of eclipsing binary.}
\end{deluxetable*}

\startlongtable
\begin{deluxetable*}{c|c|c|c|c|c|c|c|c|c|c}
\tabletypesize{\scriptsize} 
\centering
\tablecaption{RR Lyrae Stars in our Sample. \label{table:rr}} 
\tablehead{
\colhead{VVV ID} & \colhead{VVV ID \citet{gran2016}} & \colhead{RA(J2000)} & \colhead{DEC(J2000)} & \colhead{Period} & \colhead{Amp $K_s$} &\colhead{Z} & \colhead{Y} & \colhead{J} & \colhead{H} & \colhead{$K_{\rm{s}}$}} 
\startdata 
VVV 515706039923 & J180039.53-400902.4 & 18:00:39.6 & $-$40:09:03.6 & 0.6497 & 0.40 & 14.14 & 14.32 & 14.00 & 13.79 & 13.73\\
VVV 515673239493 & J180949.72-364838.9 & 18:09:49.6 & $-$36:48:39.6 & 0.5354 & 0.41 & 14.01 & 14.18 & 13.90 & 13.76 & 13.68\\
VVV 515672602244 & J180958.42-380307.7 & 18:09:58.3 & $-$38:03:07.2 & 0.5230 & 0.32 & 14.76 & 14.86 & 14.61 & 14.48 & 14.41\\
VVV 515698929724 & J181011.78-384805.0 & 18:10:11.7 & $-$38:48:03.6 & 0.6101 & 0.25 & 14.63 & 14.74 & 14.49 & 14.32 & 14.22\\ 
VVV 515656519651 & J181418.96-352242.8 & 18:14:18.9 & $-$35:22:44.4 & 0.3964 & 0.30 & 14.49 & 14.64 & 14.35 & 14.17 & 14.08\\
VVV 515598433507 & J182748.94-332826.0 & 18:27:48.9 & $-$33:28:26.4 & 0.5406 & 0.49 & 14.83 & 15.02 & 14.69 & 14.47 & 14.39\\ 
VVV 515533139402 & J183758.88-290835.4 & 18:37:58.8 & $-$29:08:34.8 & 0.5395 & 0.45 & 15.04 & 15.13 & 14.85 & 14.58 & 14.50\\ 
VVV 515705907121 & & 17:59:59.2 & $-$40:21:00.0 & 0.6415 & 0.29 & 14.85 & 14.97 & 14.70 & 14.51 & 14.40\\
\enddata 
\tablenotetext{a}{The complete parameters: VVV ID, right ascension (J2000), declination (J2000), period (days), amplitude of $K_s$ magnitude and \emph{ZYKH$K_s$} magnitudes.}
\end{deluxetable*}

\begin{figure*}[htb]
\centering
  \begin{tabular}{@{}ccc@{}}    
   \includegraphics[width=.33\textwidth]{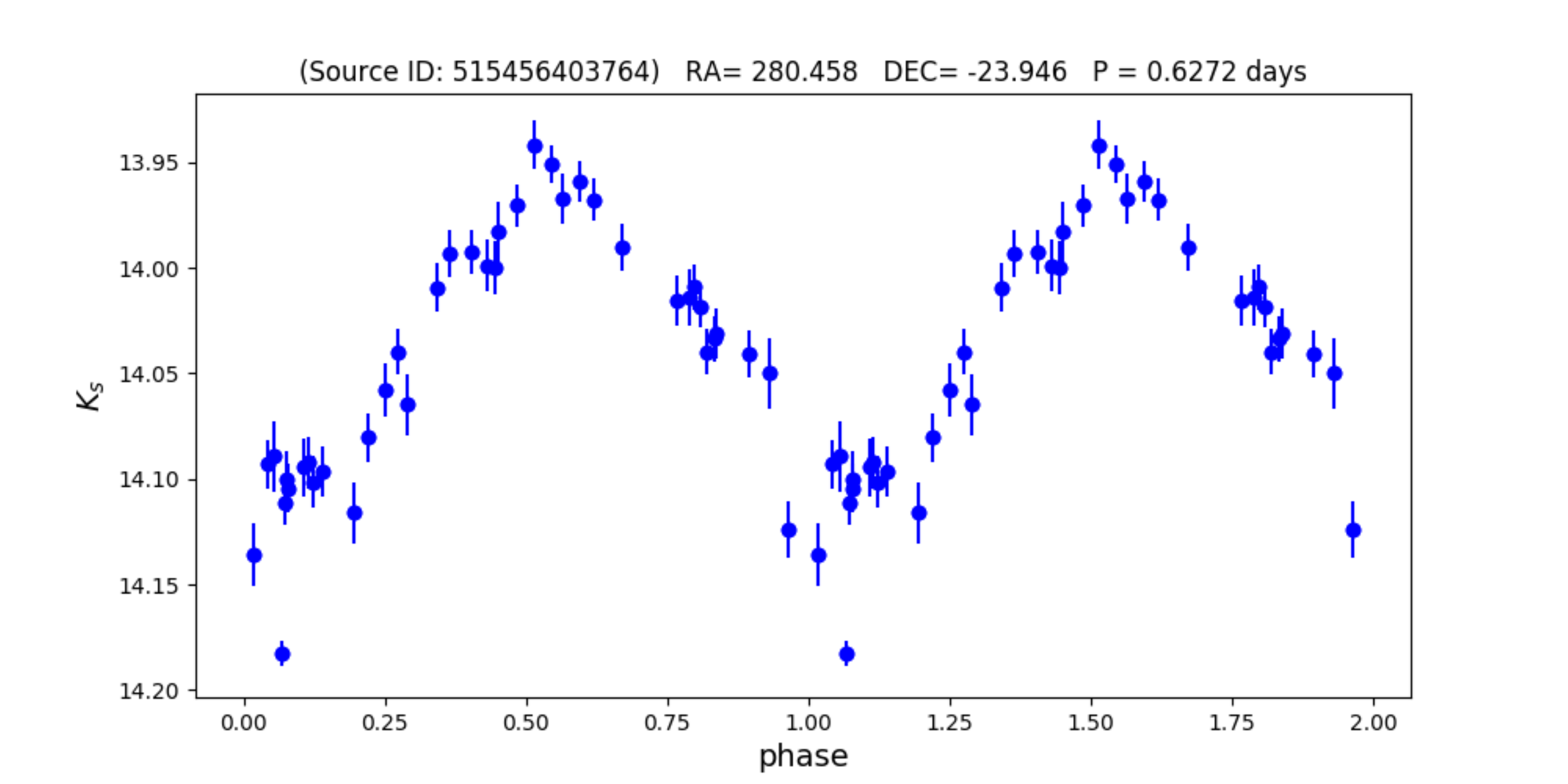} &  
   \includegraphics[width=.33\textwidth]{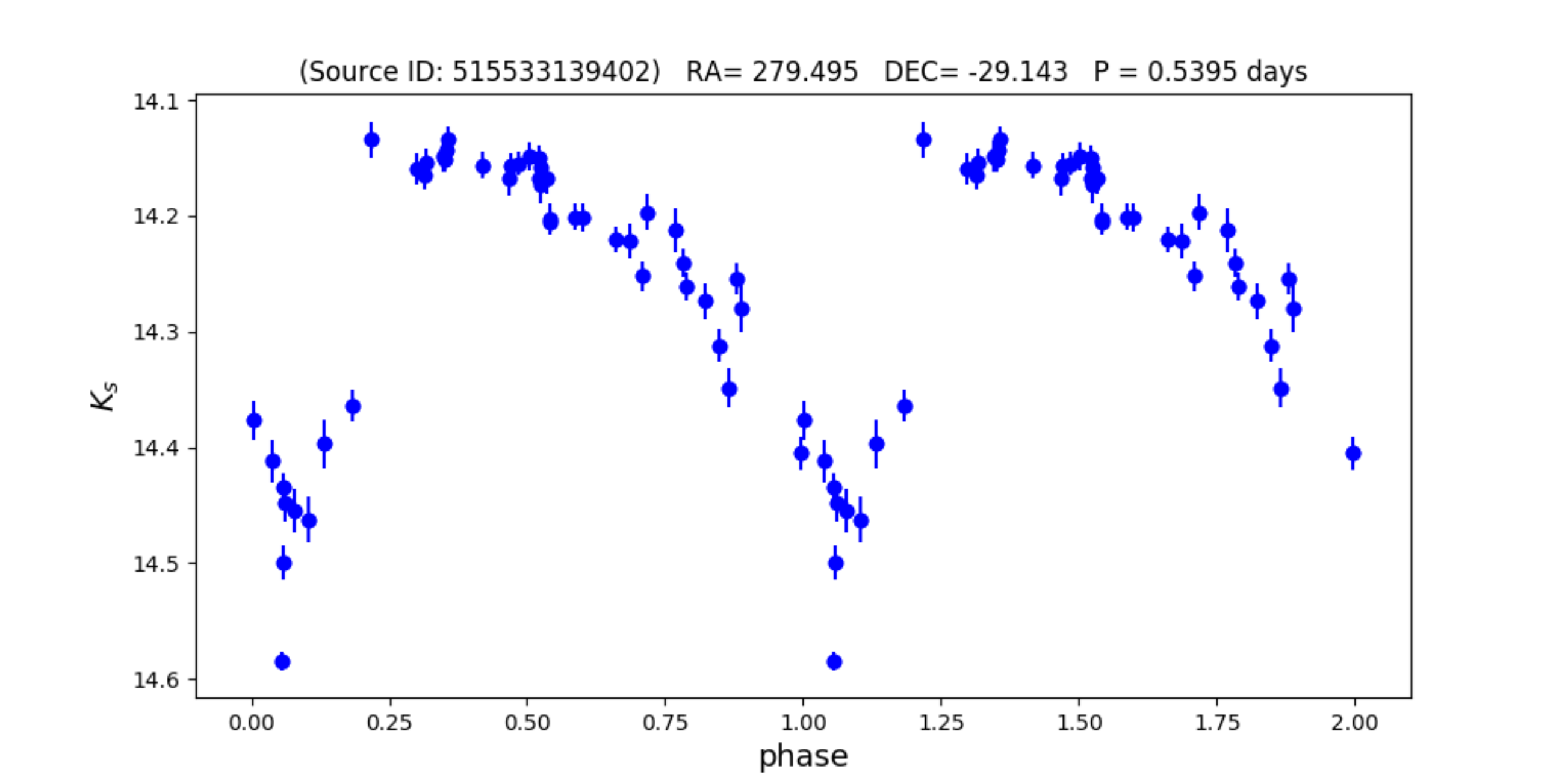} &  
   \includegraphics[width=.33\textwidth]{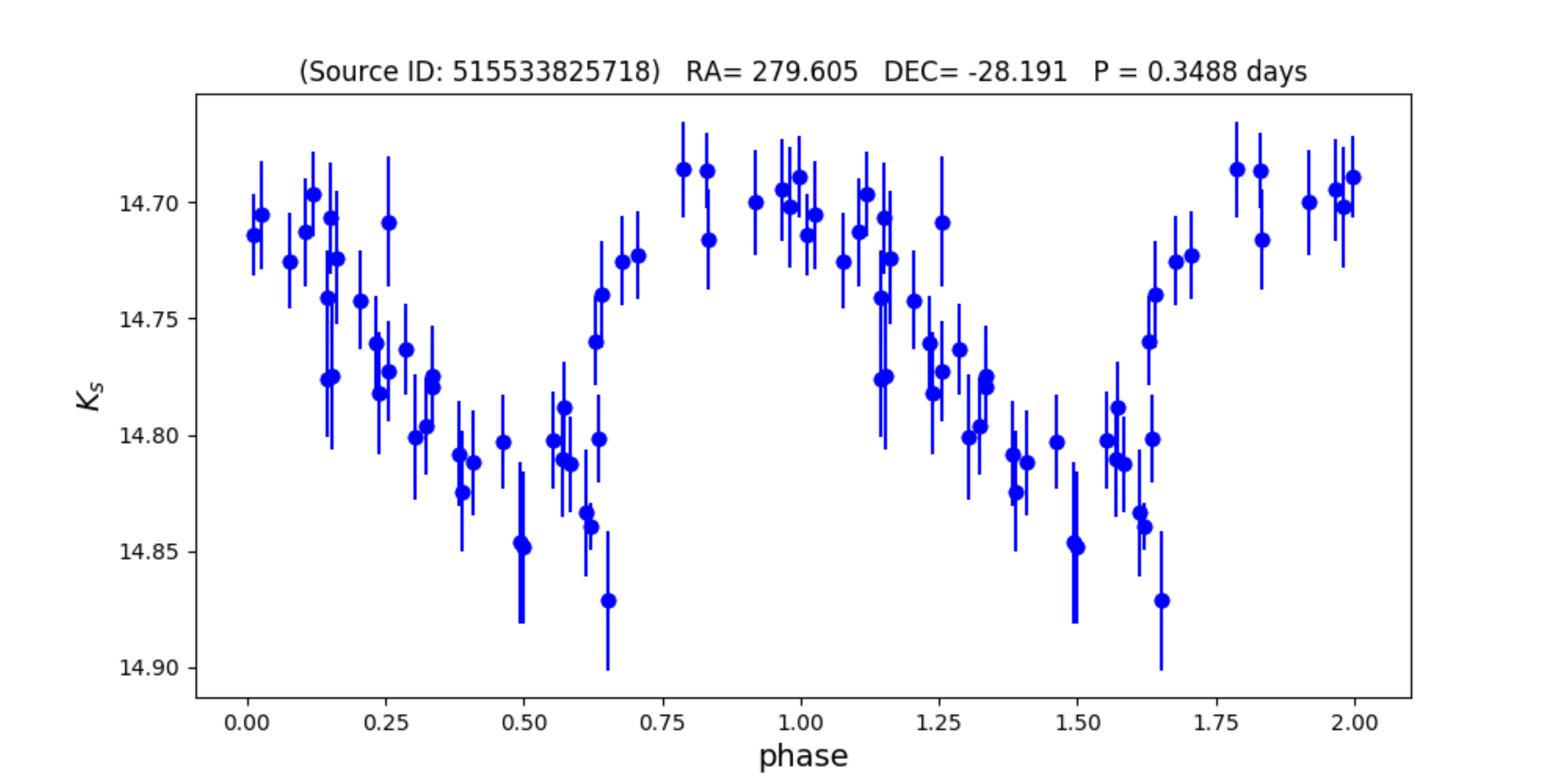} \\ 
   \includegraphics[width=.33\textwidth]{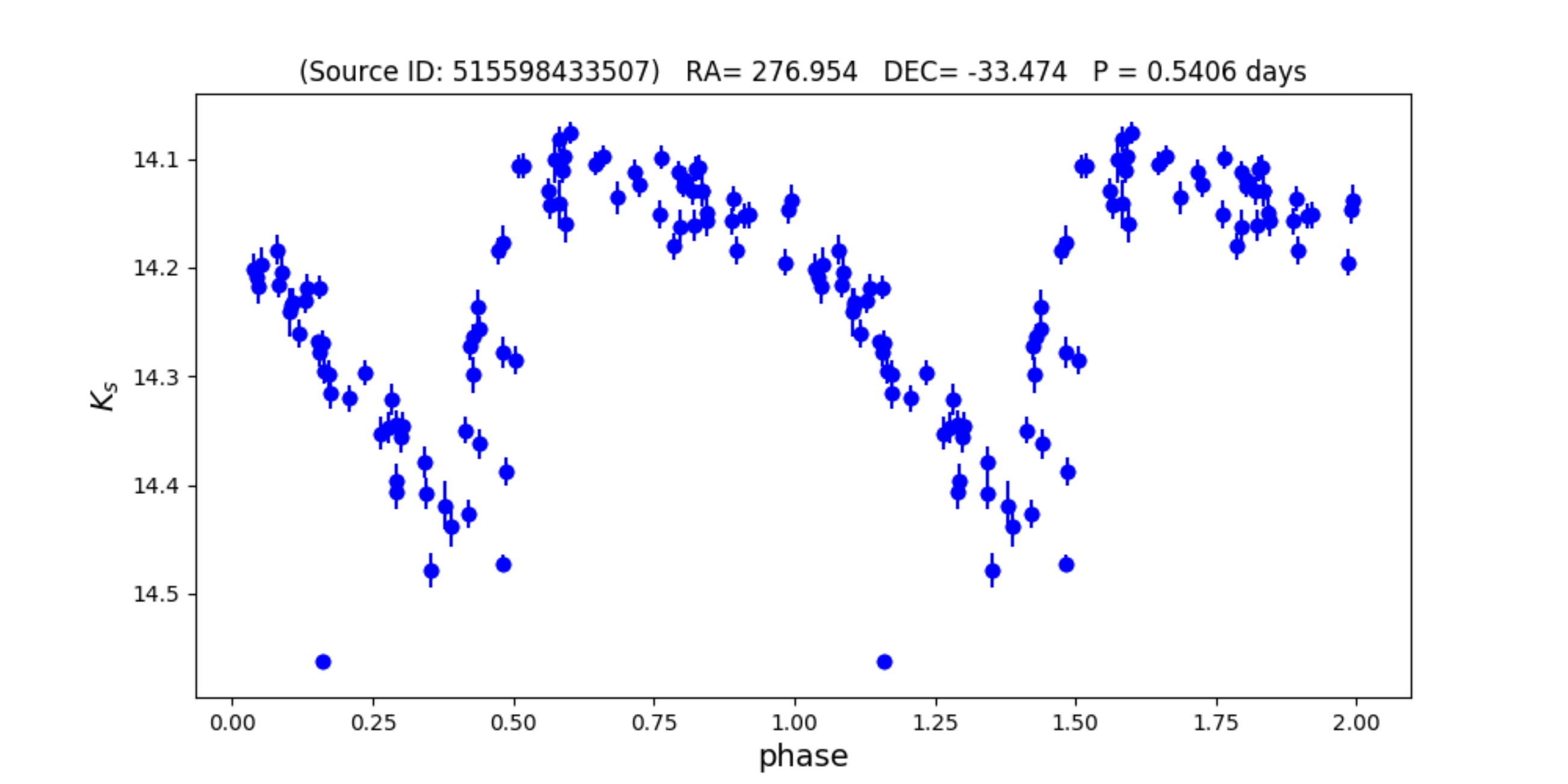} &
   \includegraphics[width=.33\textwidth]{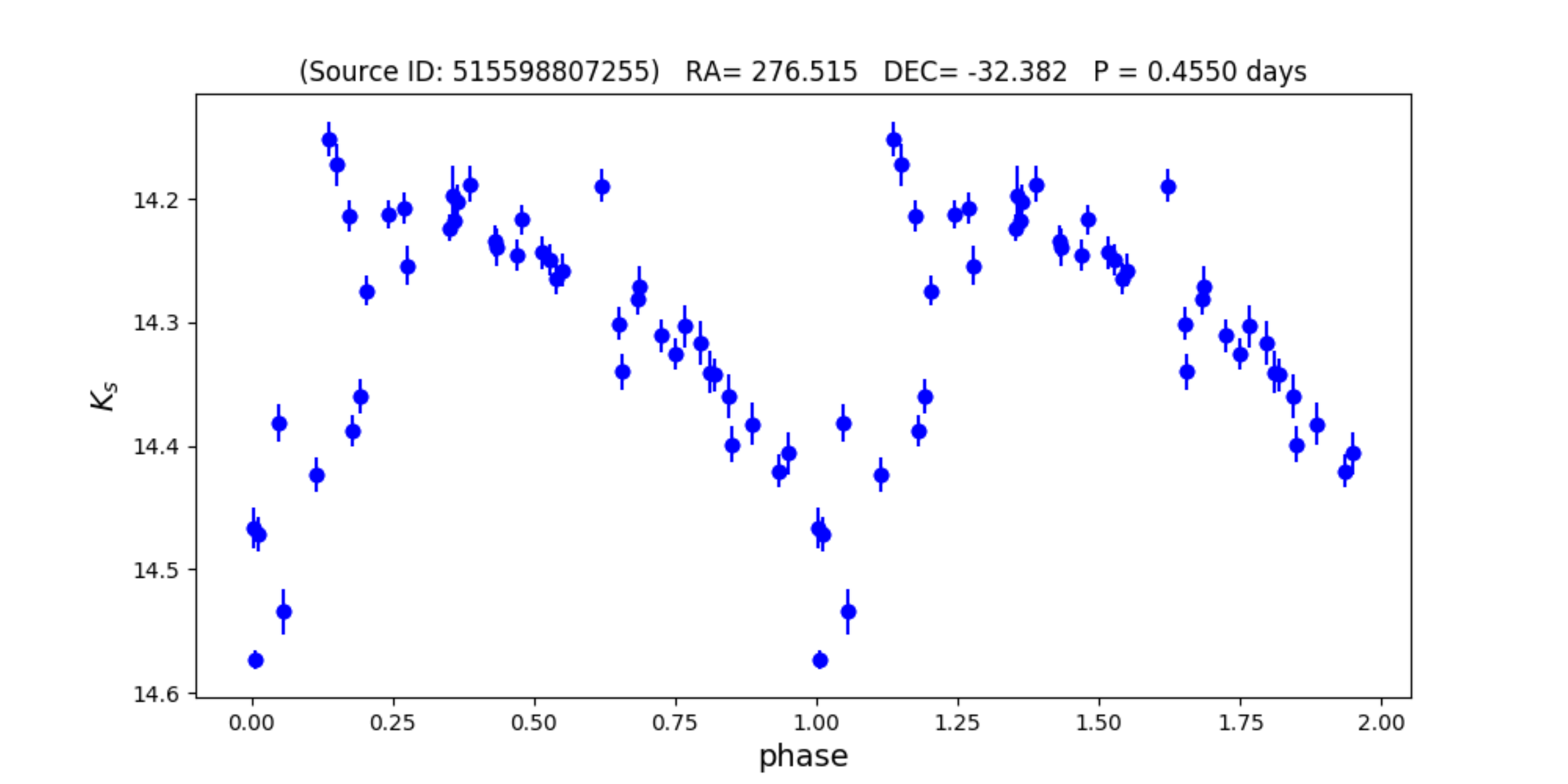} &
   \includegraphics[width=.33\textwidth]{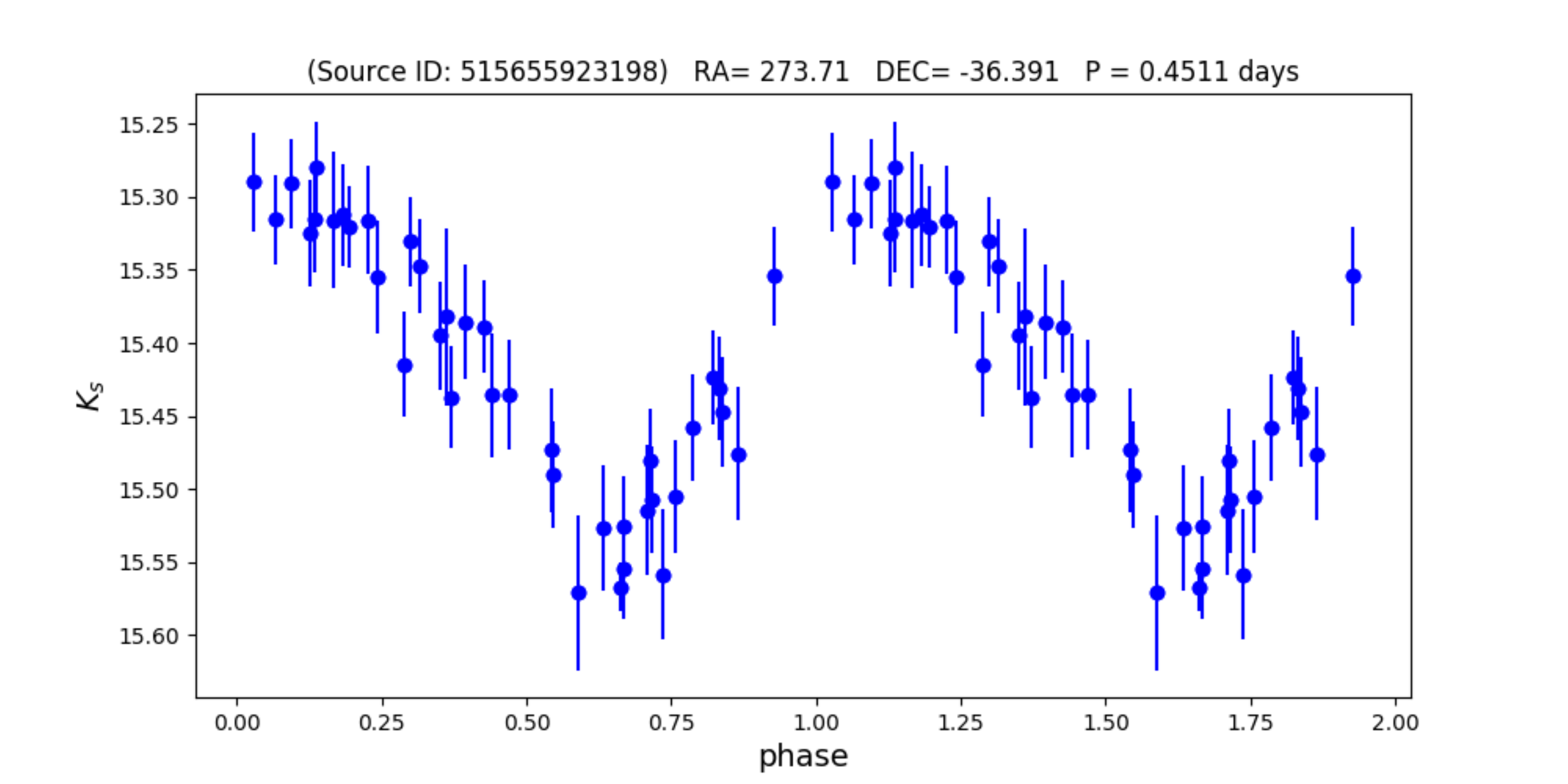} \\
   \includegraphics[width=.33\textwidth]{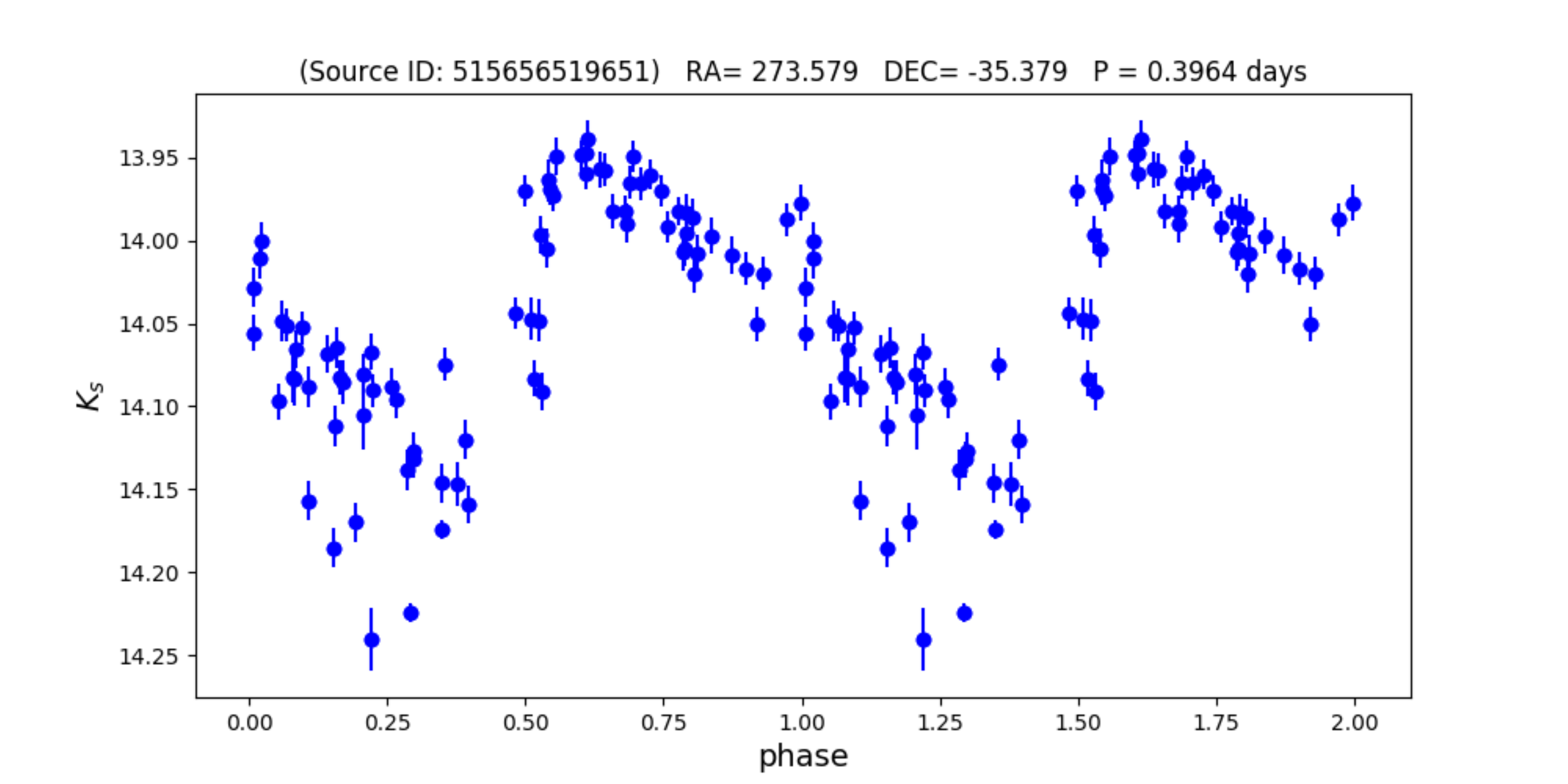} &
   \includegraphics[width=.33\textwidth]{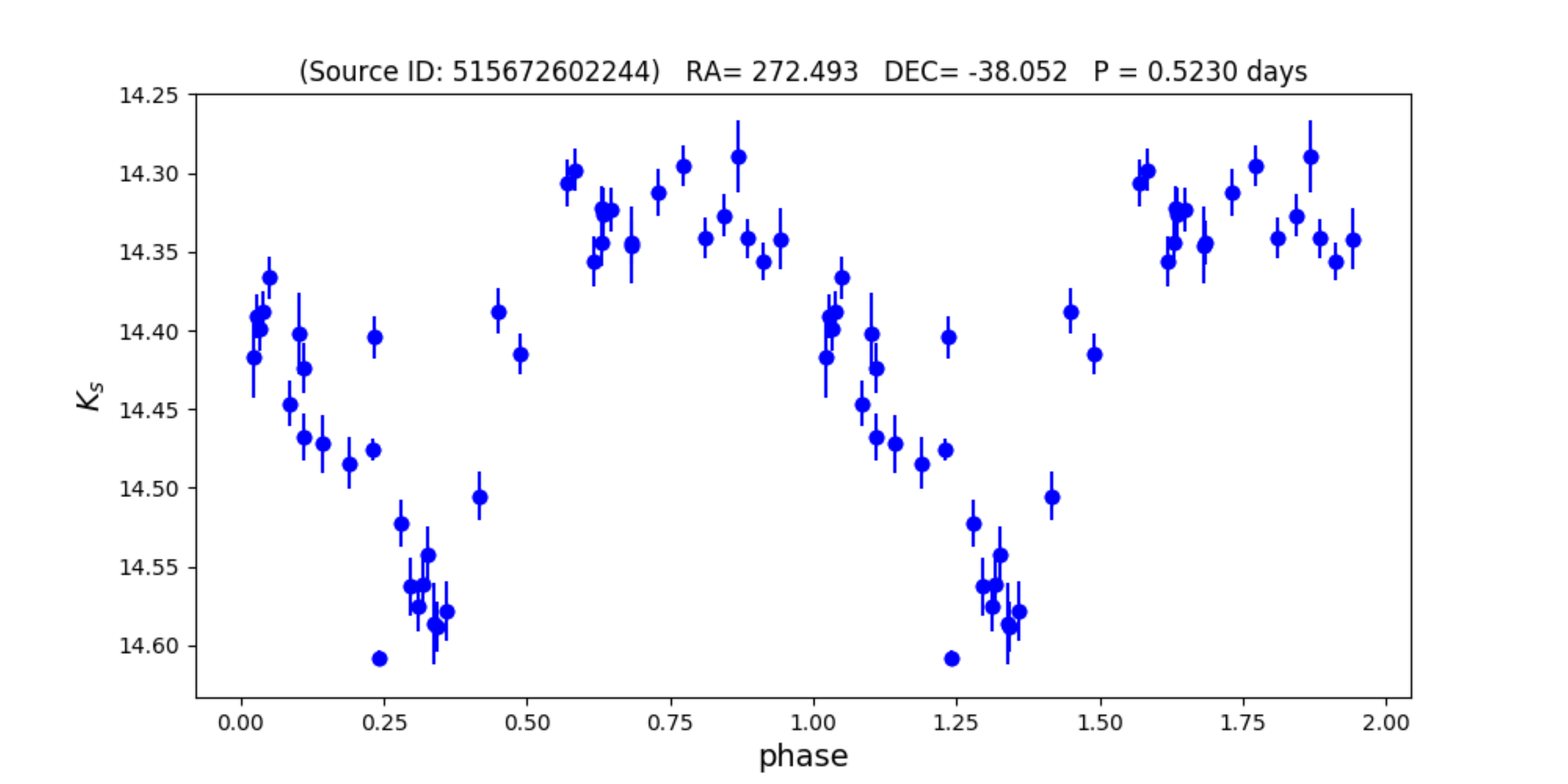} &
   \includegraphics[width=.33\textwidth]{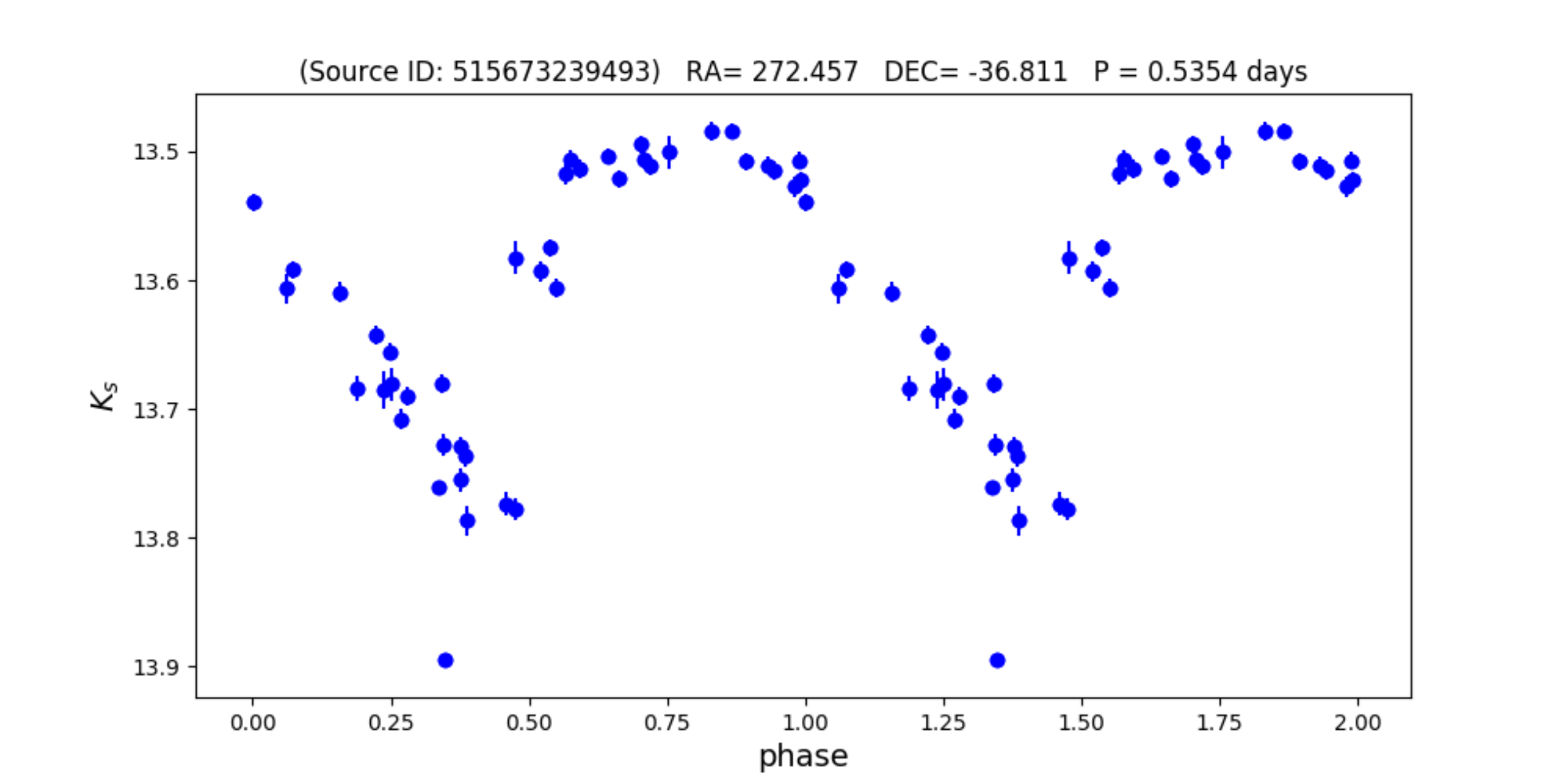} \\
   \includegraphics[width=.33\textwidth]{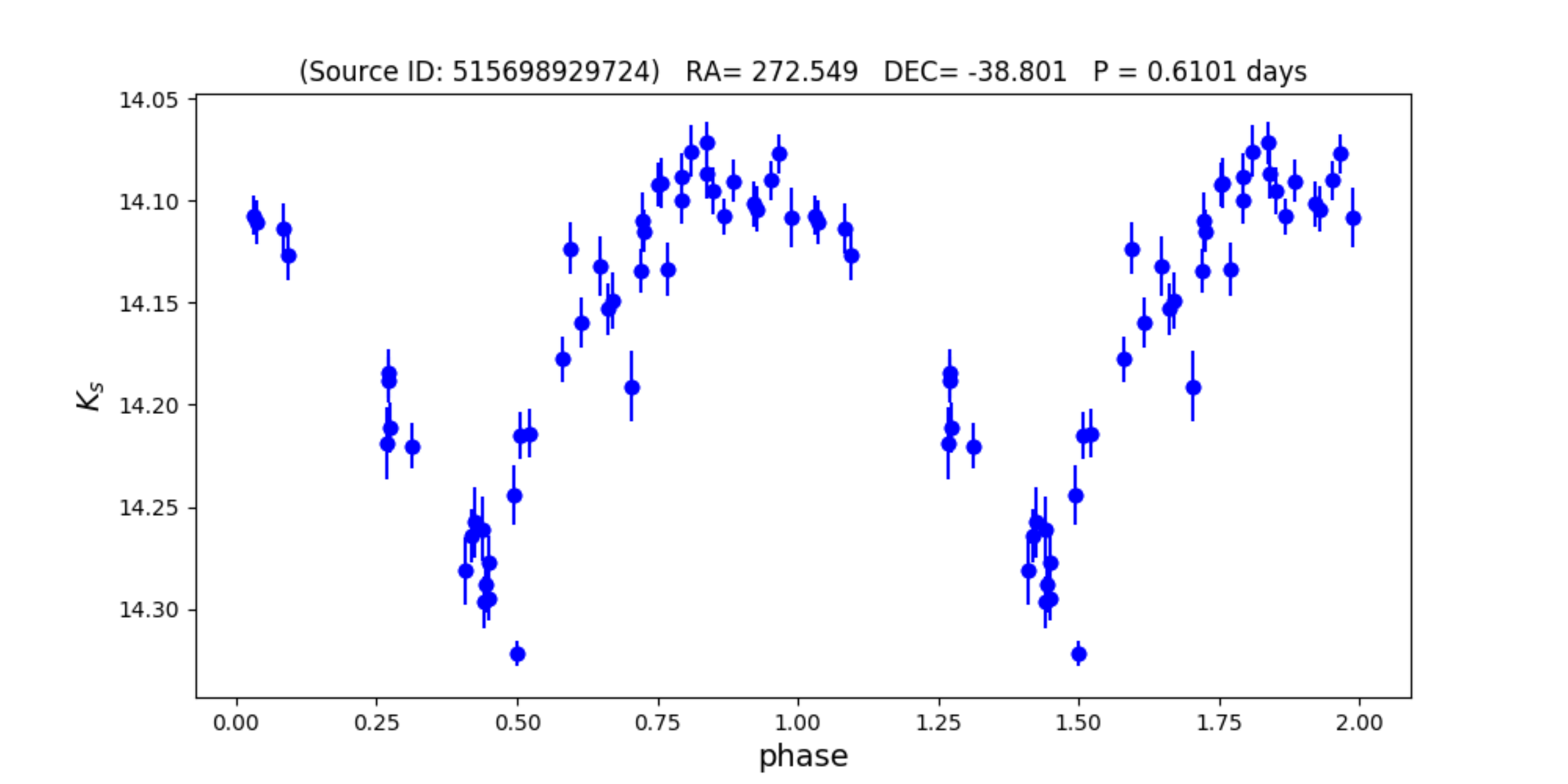} &
   \includegraphics[width=.33\textwidth]{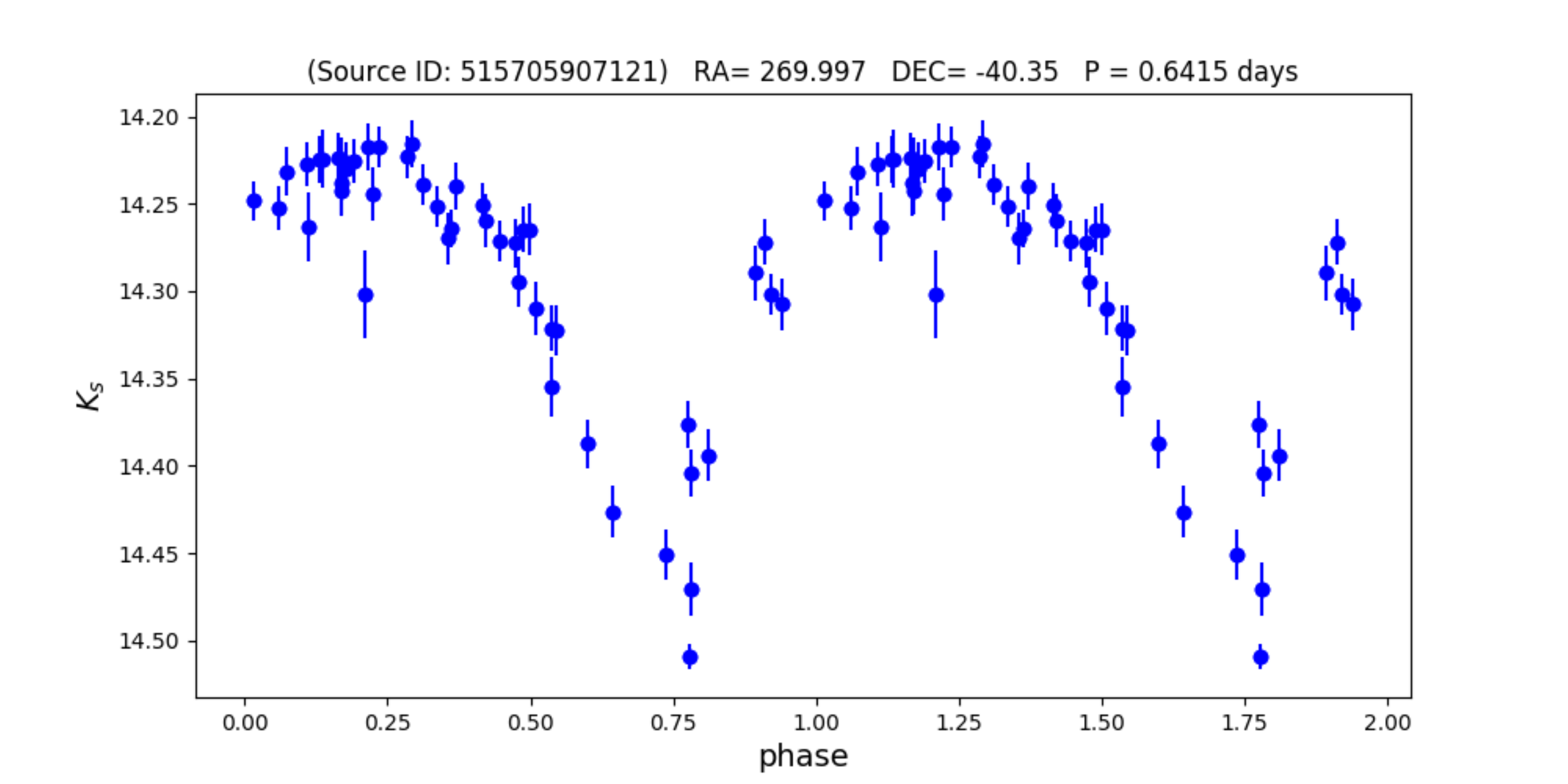} &
   \includegraphics[width=.33\textwidth]{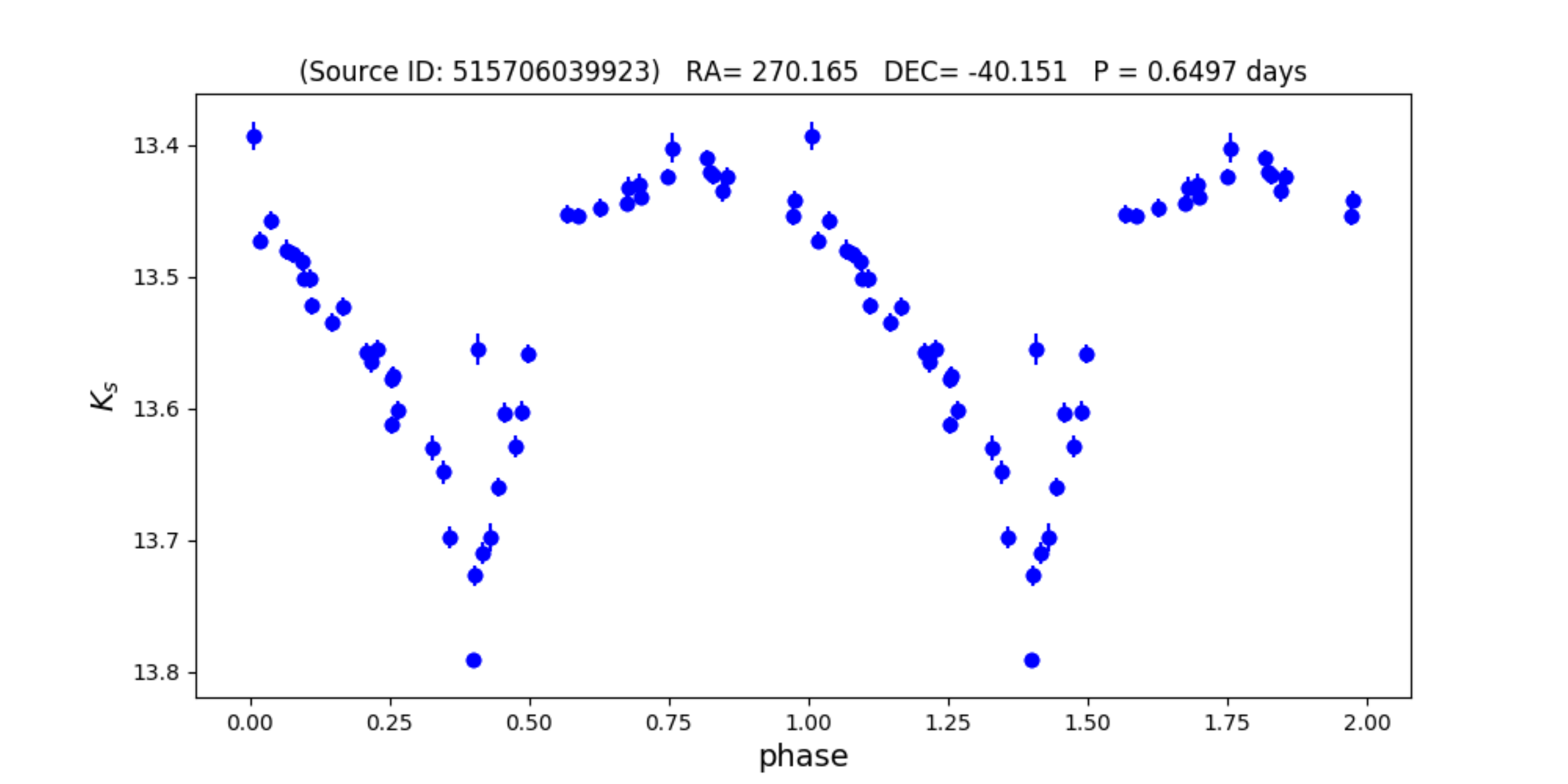} \\
  \end{tabular} 
  \caption{RR Lyrae light curves found in our sample.}
  \label{fig:rr1}
\end{figure*}

\begin{figure*}[htb]
\centering
  \begin{tabular}{@{}ccc@{}}
    \includegraphics[width=.33\textwidth]{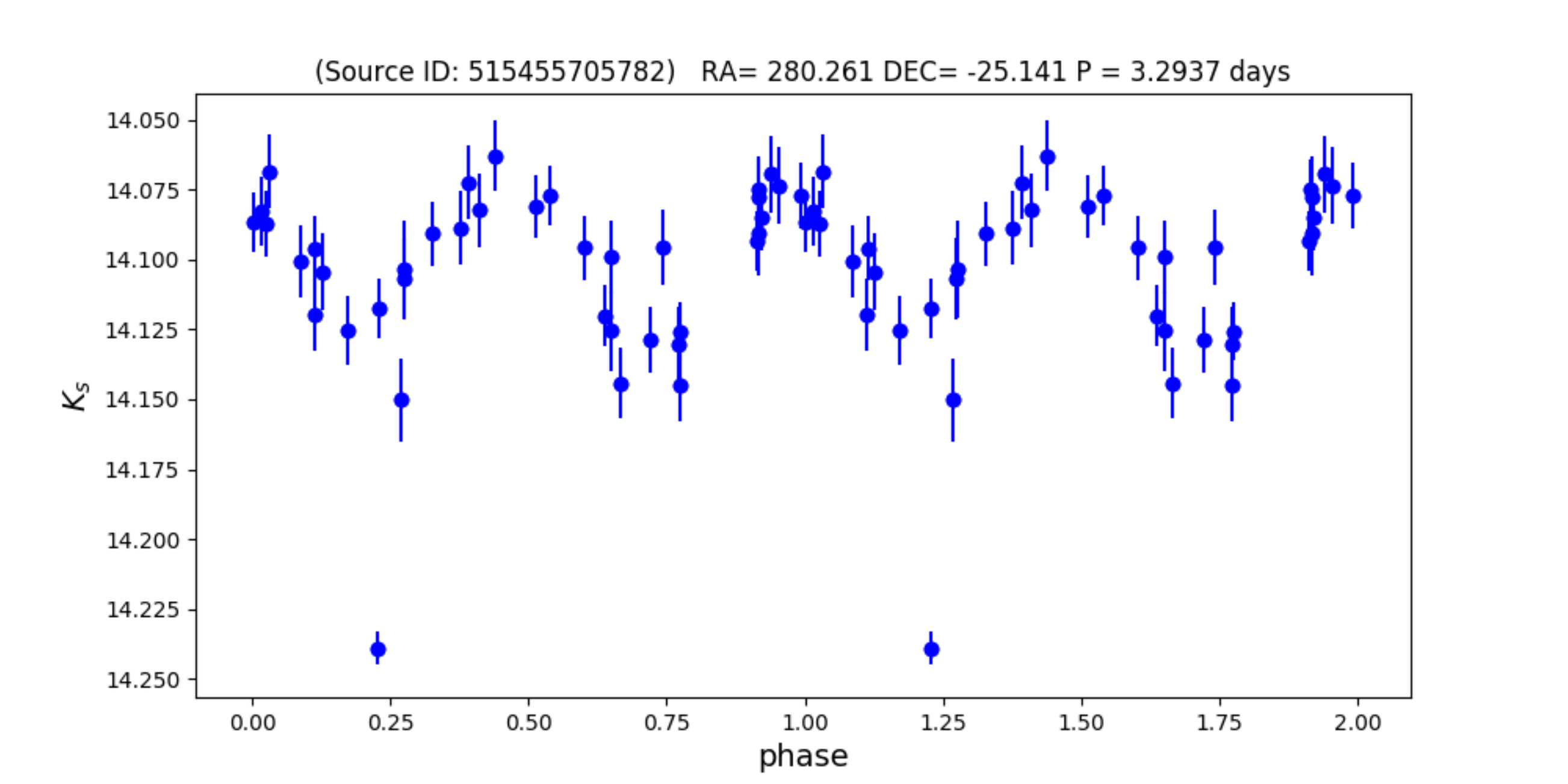} &
    \includegraphics[width=.33\textwidth]{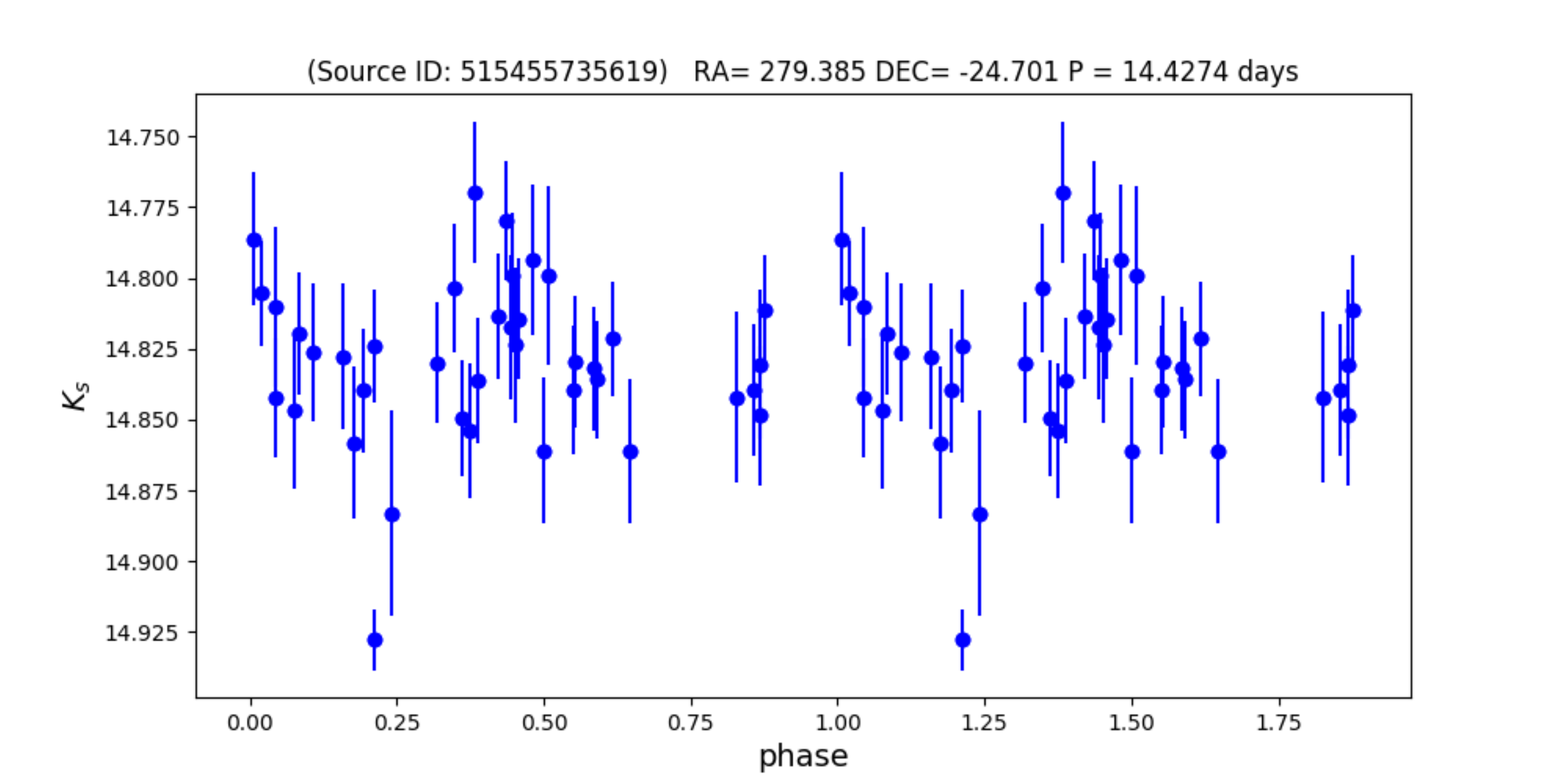} & 
    \includegraphics[width=.33\textwidth]{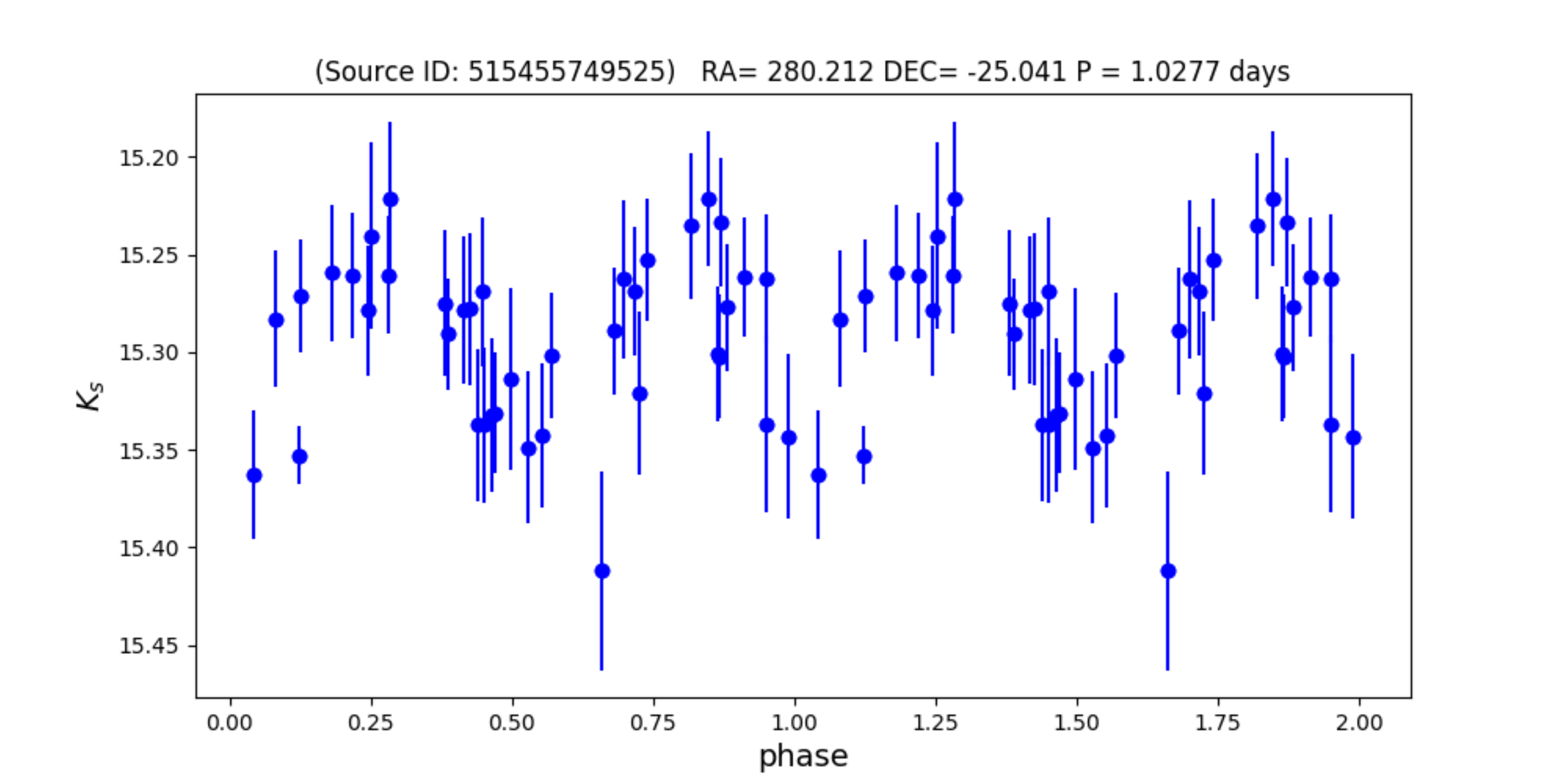} \\ 
    \includegraphics[width=.33\textwidth]{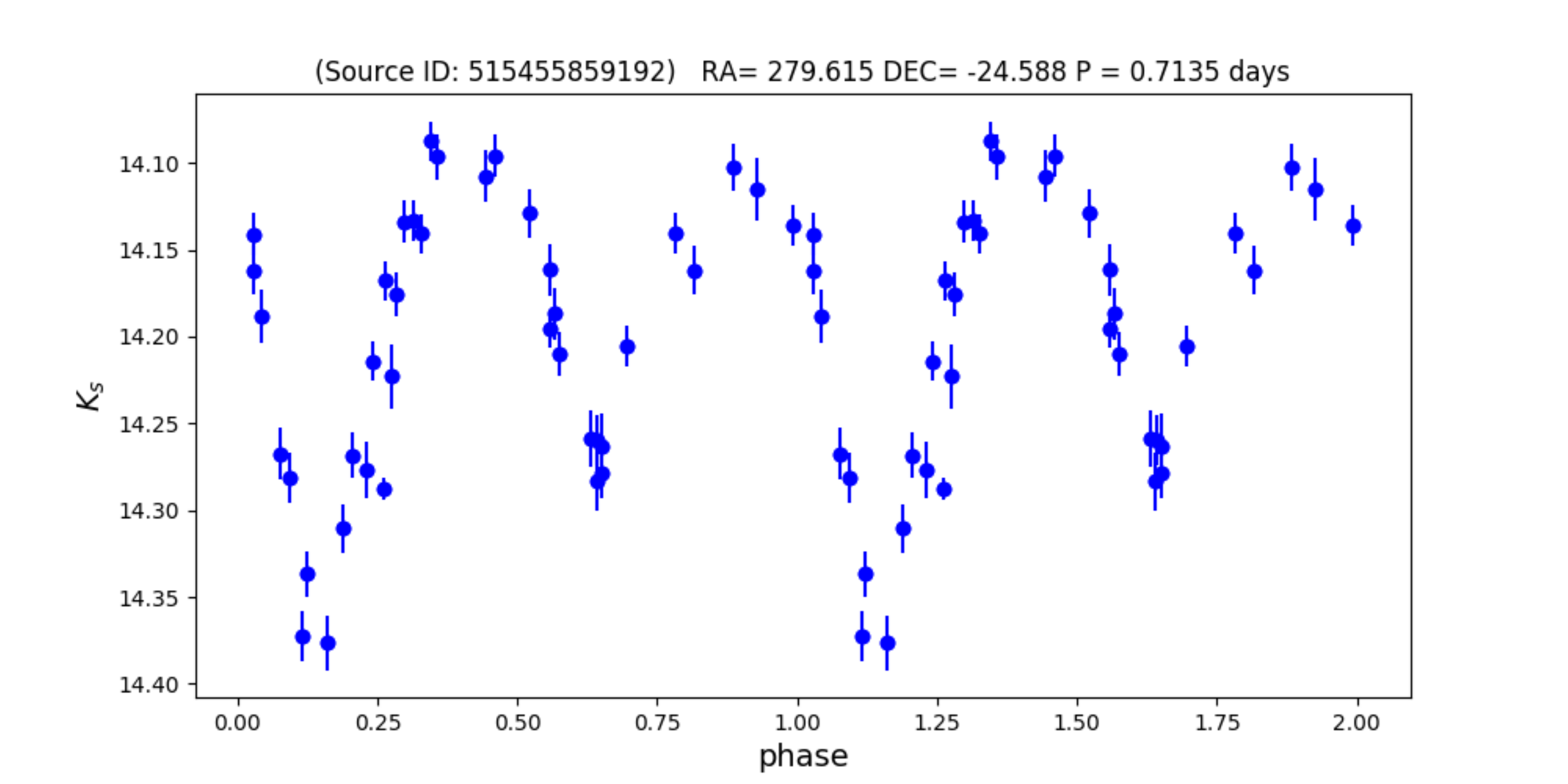} &
    \includegraphics[width=.33\textwidth]{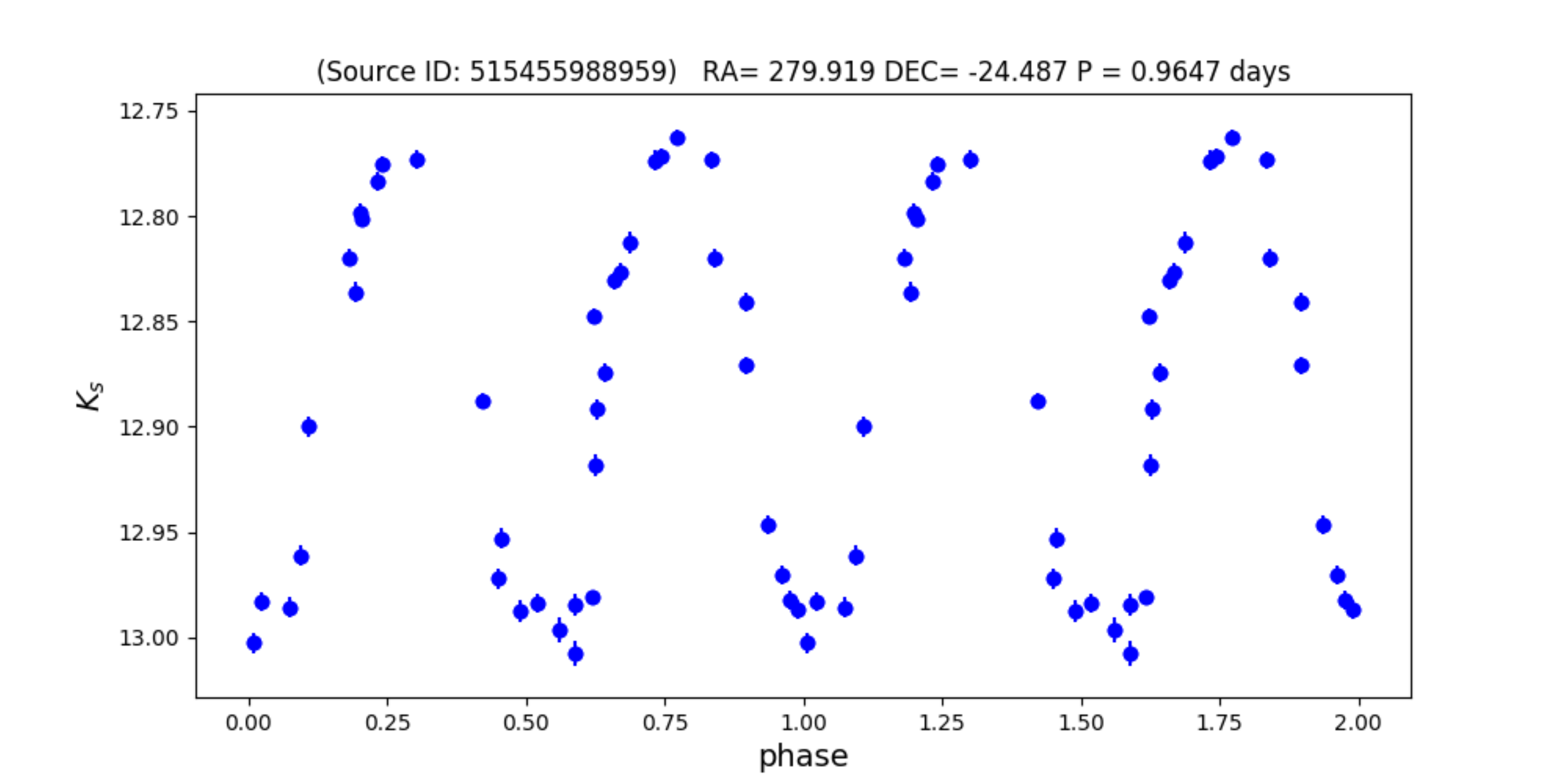} &
    \includegraphics[width=.33\textwidth]{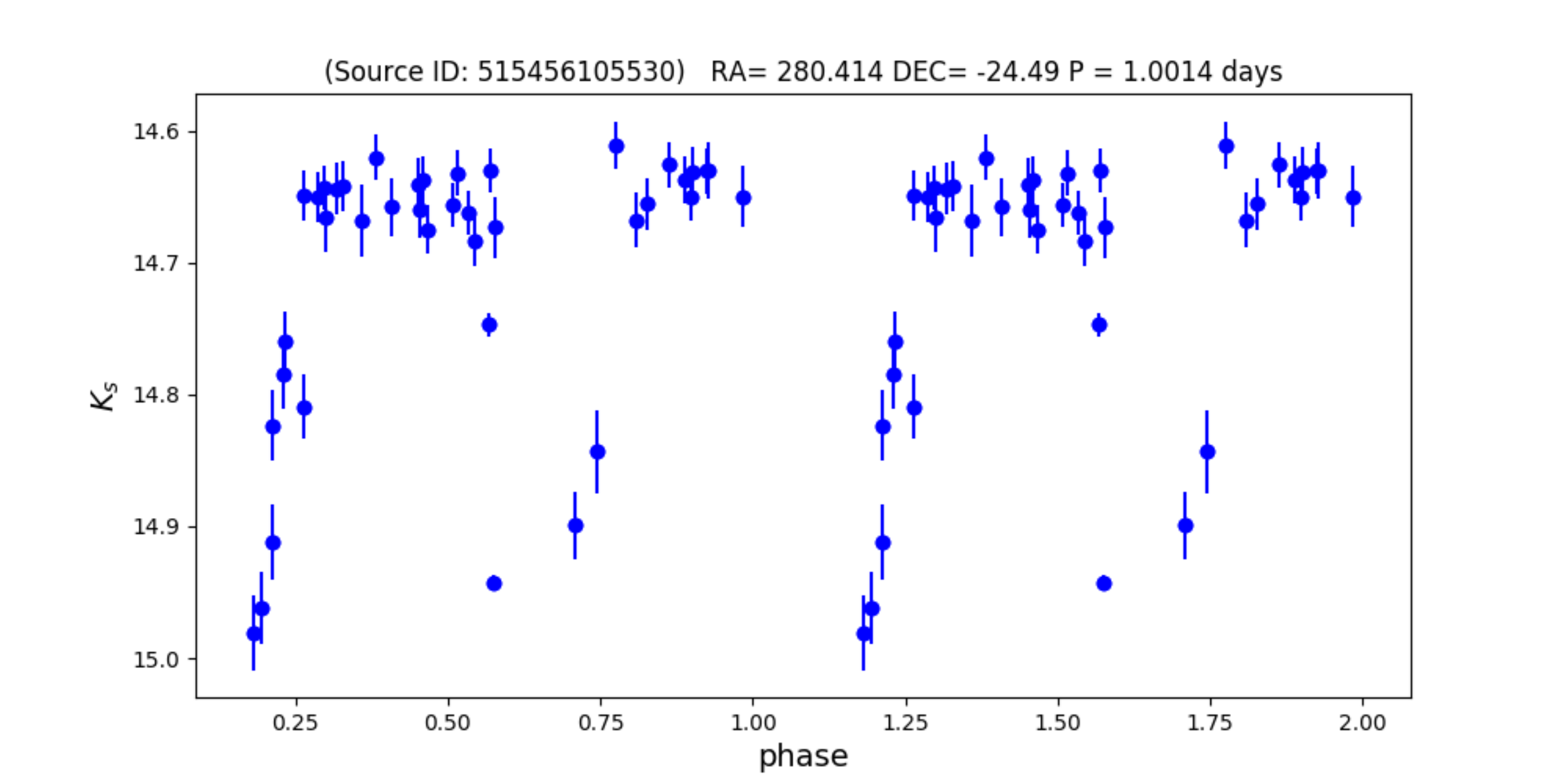} \\
    \includegraphics[width=.33\textwidth]{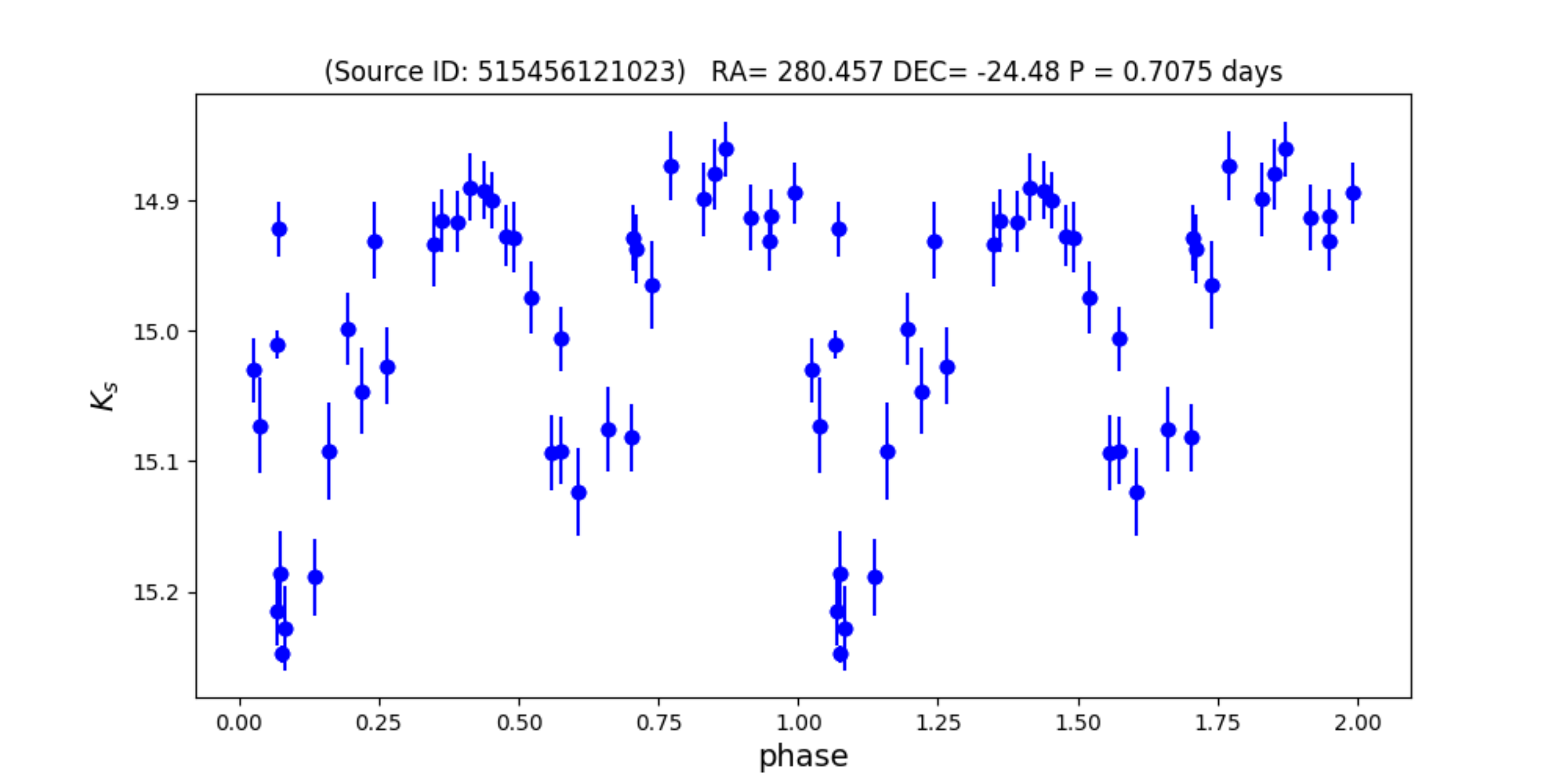} &
    \includegraphics[width=.33\textwidth]{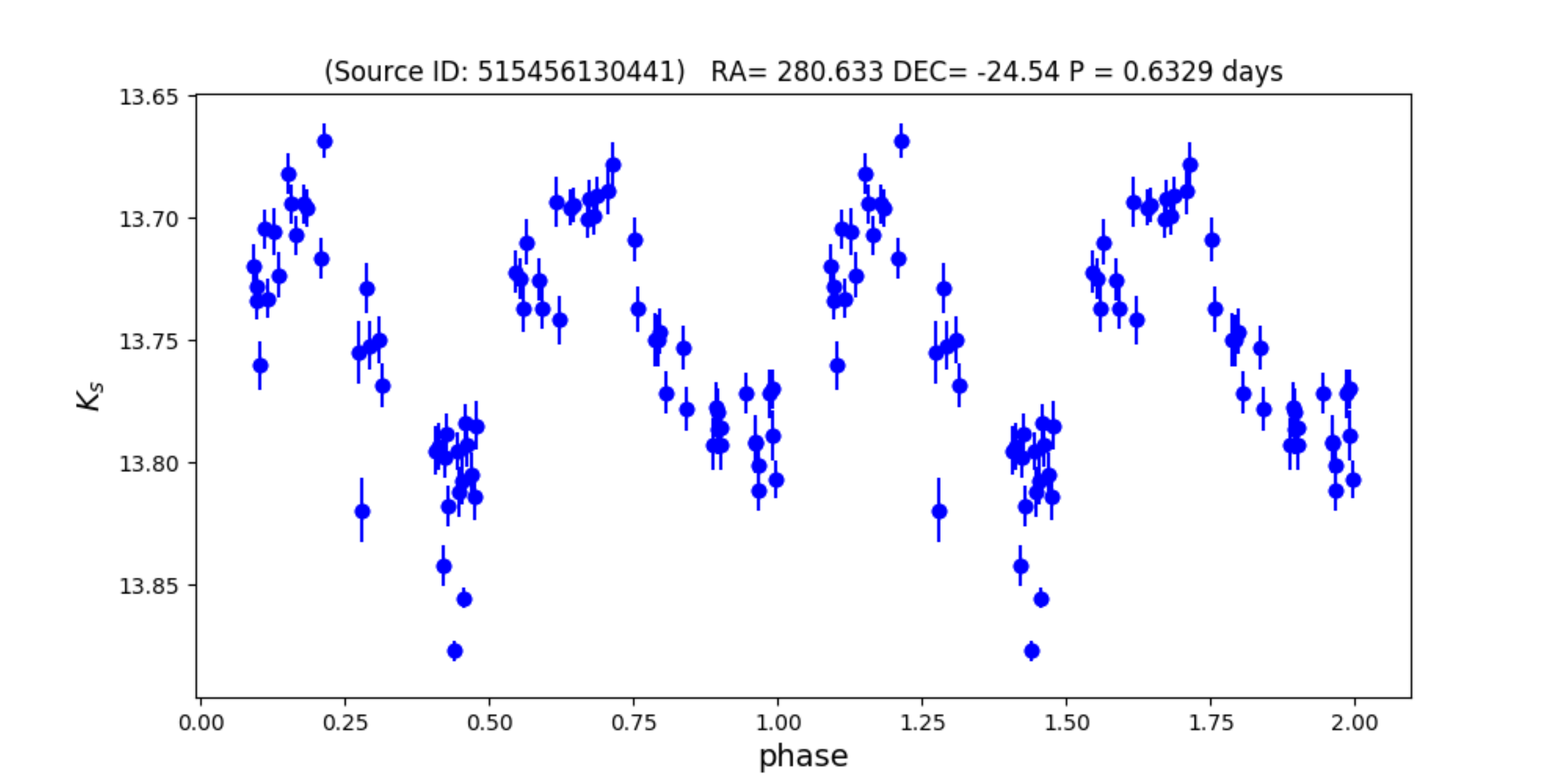} &
    \includegraphics[width=.33\textwidth]{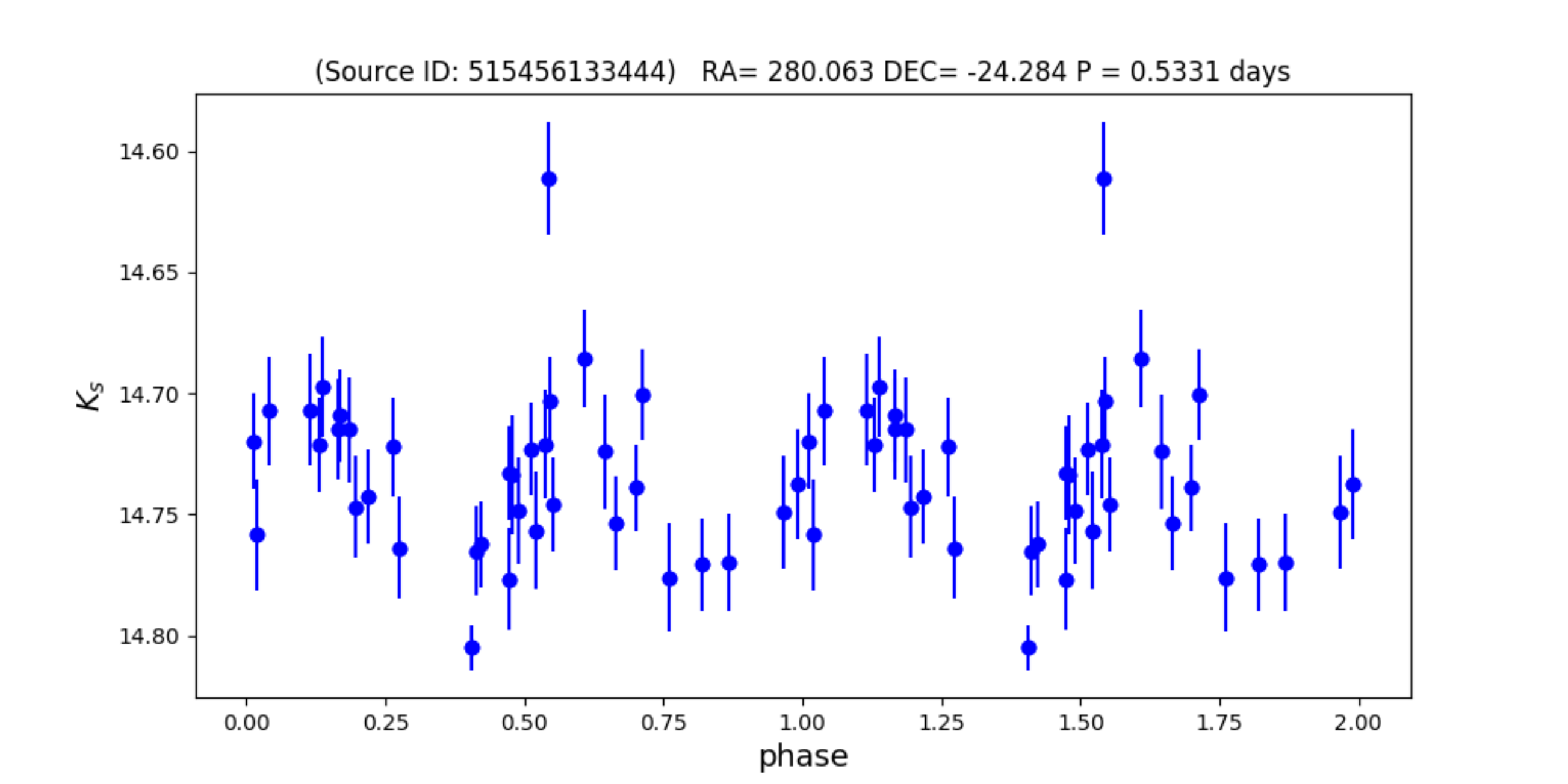} \\
    \includegraphics[width=.33\textwidth]{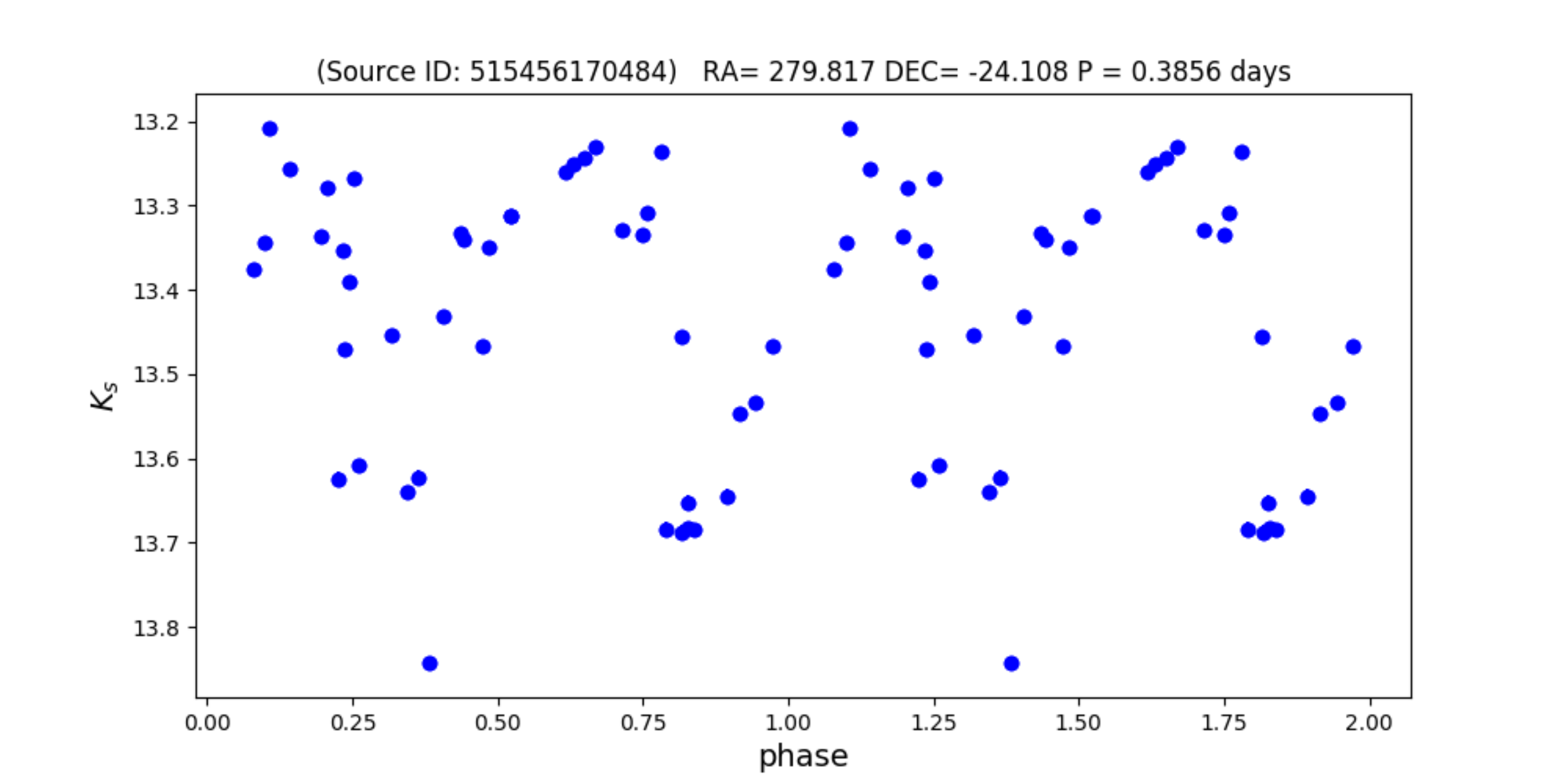} &
    \includegraphics[width=.33\textwidth]{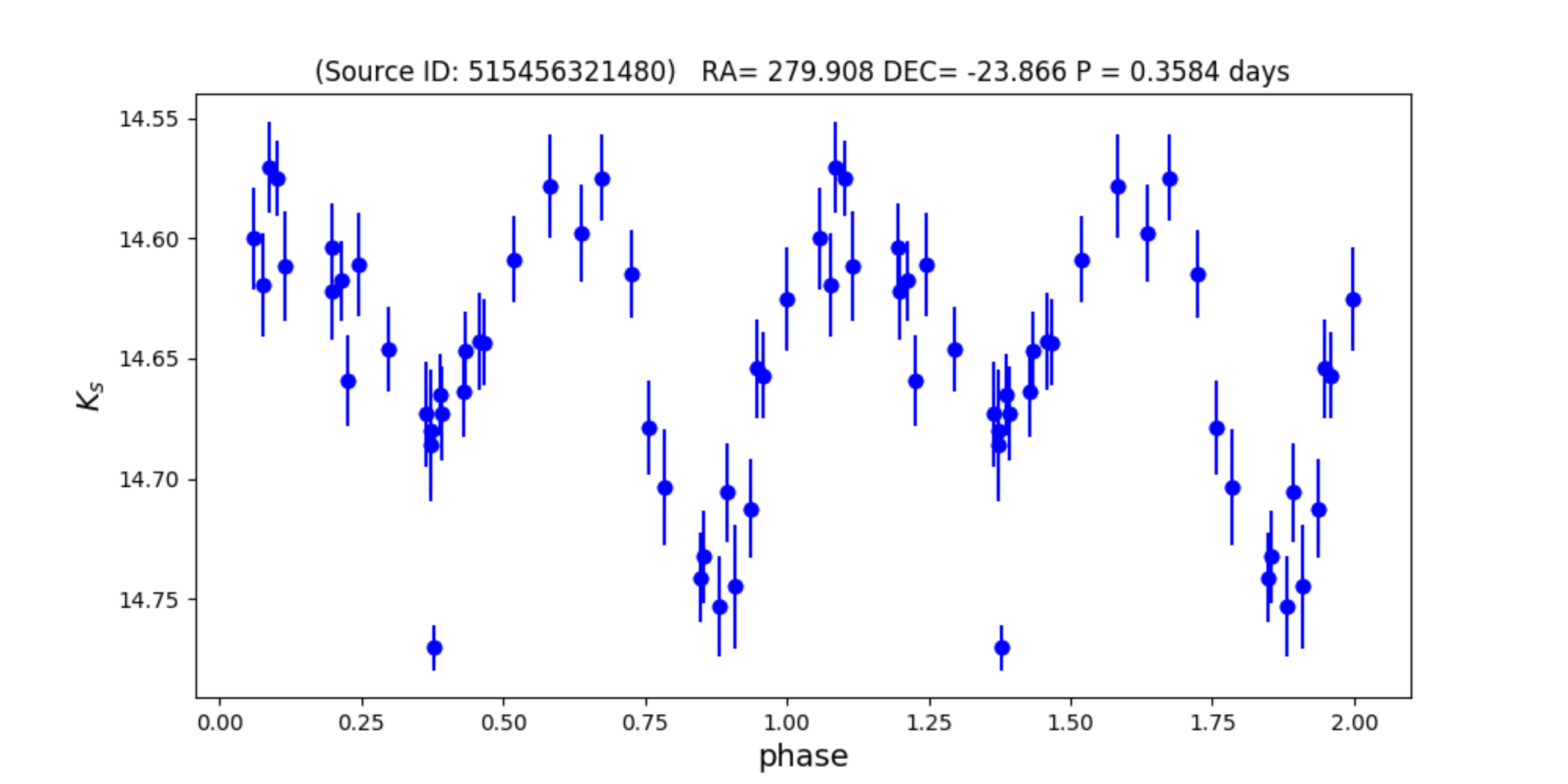} &
    \includegraphics[width=.33\textwidth]{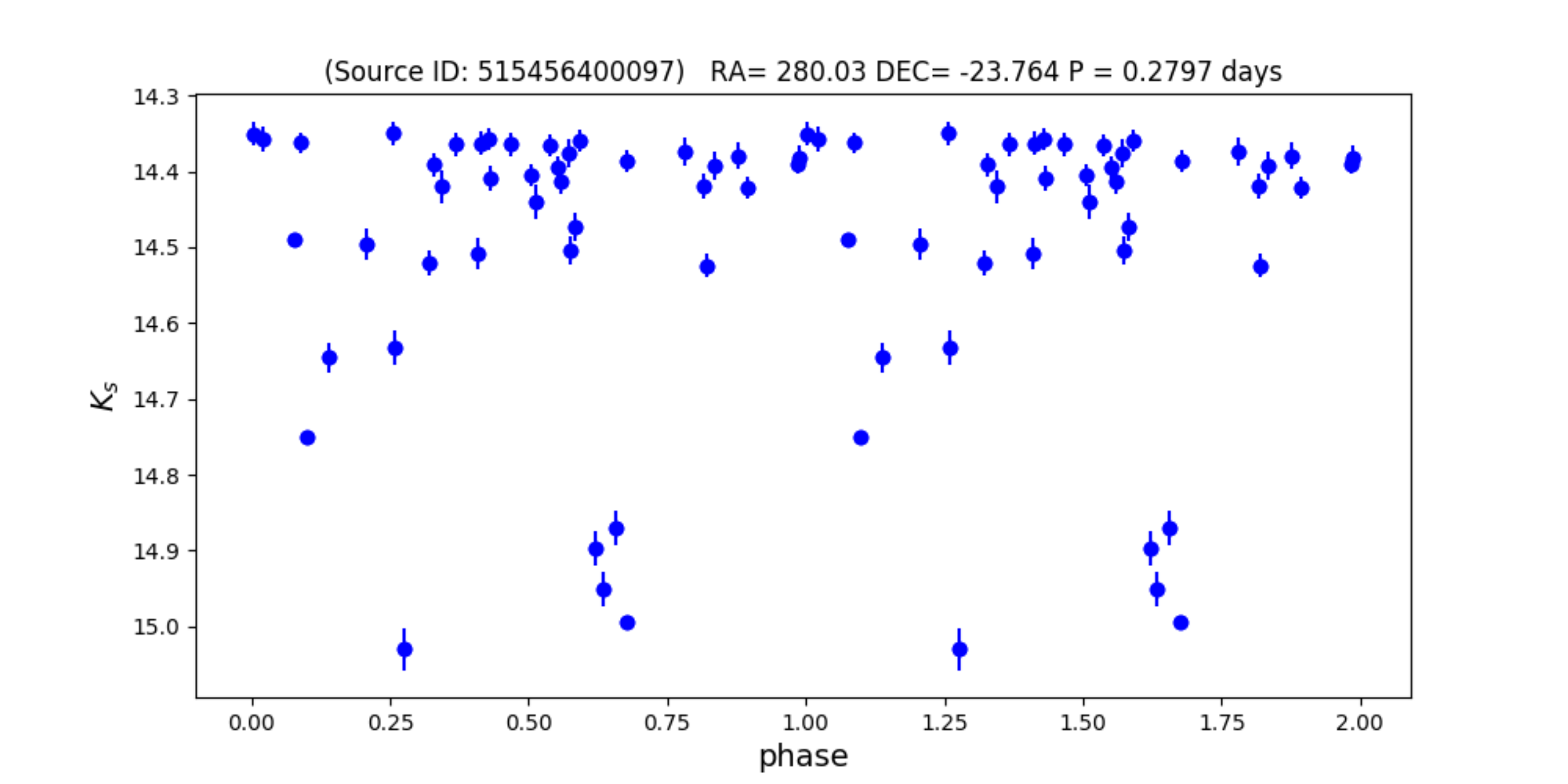} \\
    \includegraphics[width=.33\textwidth]{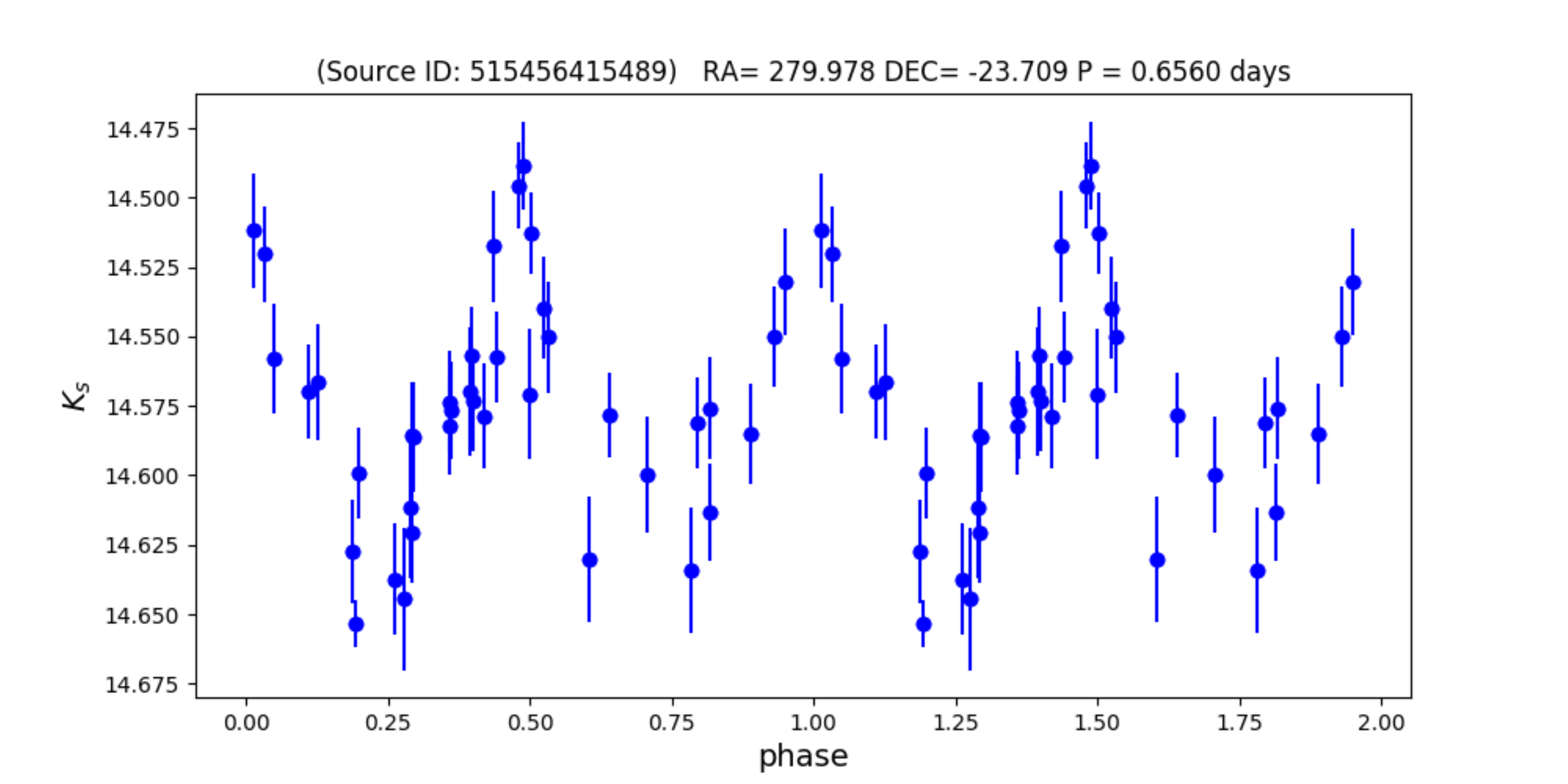} &
    \includegraphics[width=.33\textwidth]{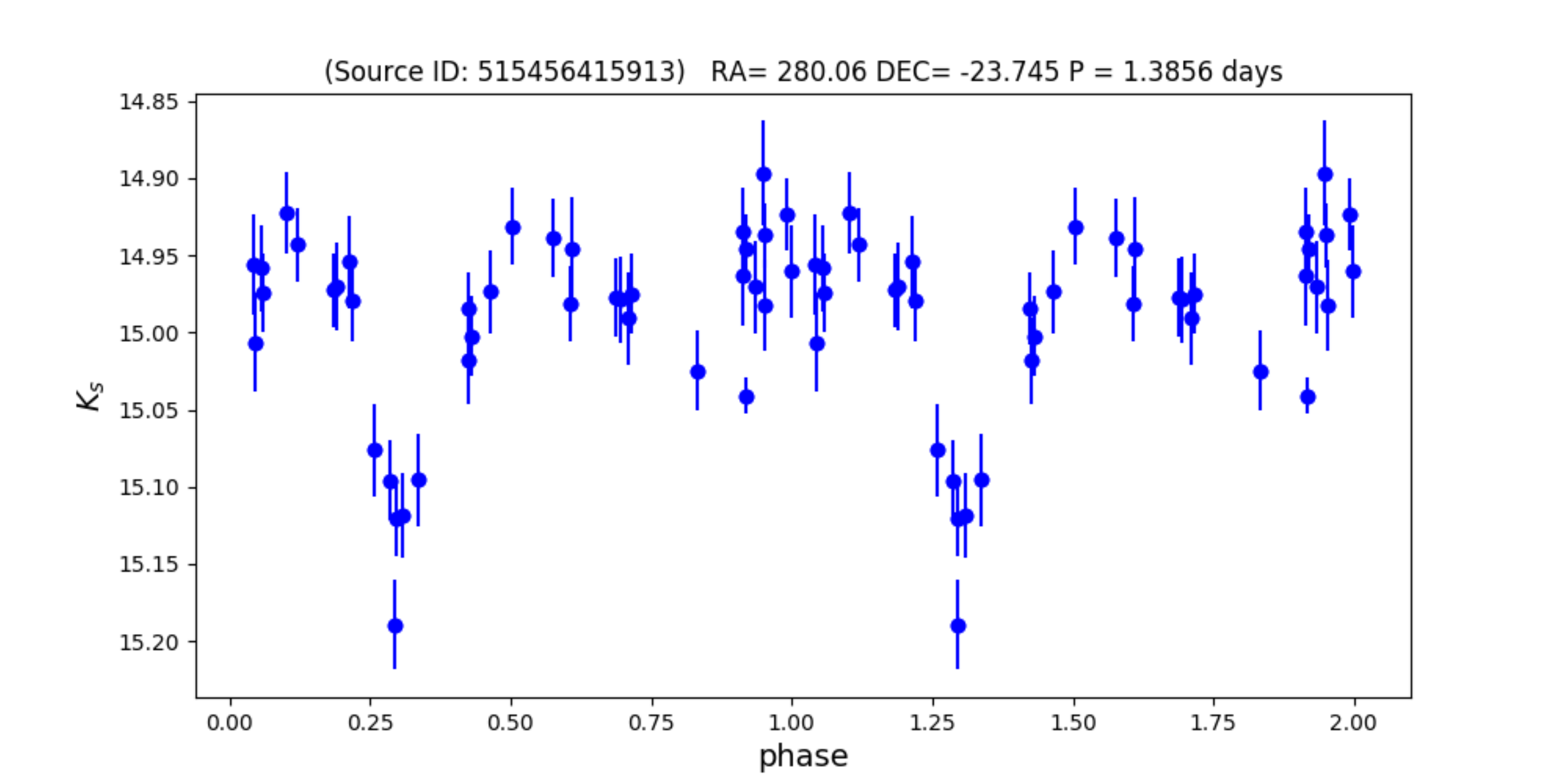} &
    \includegraphics[width=.33\textwidth]{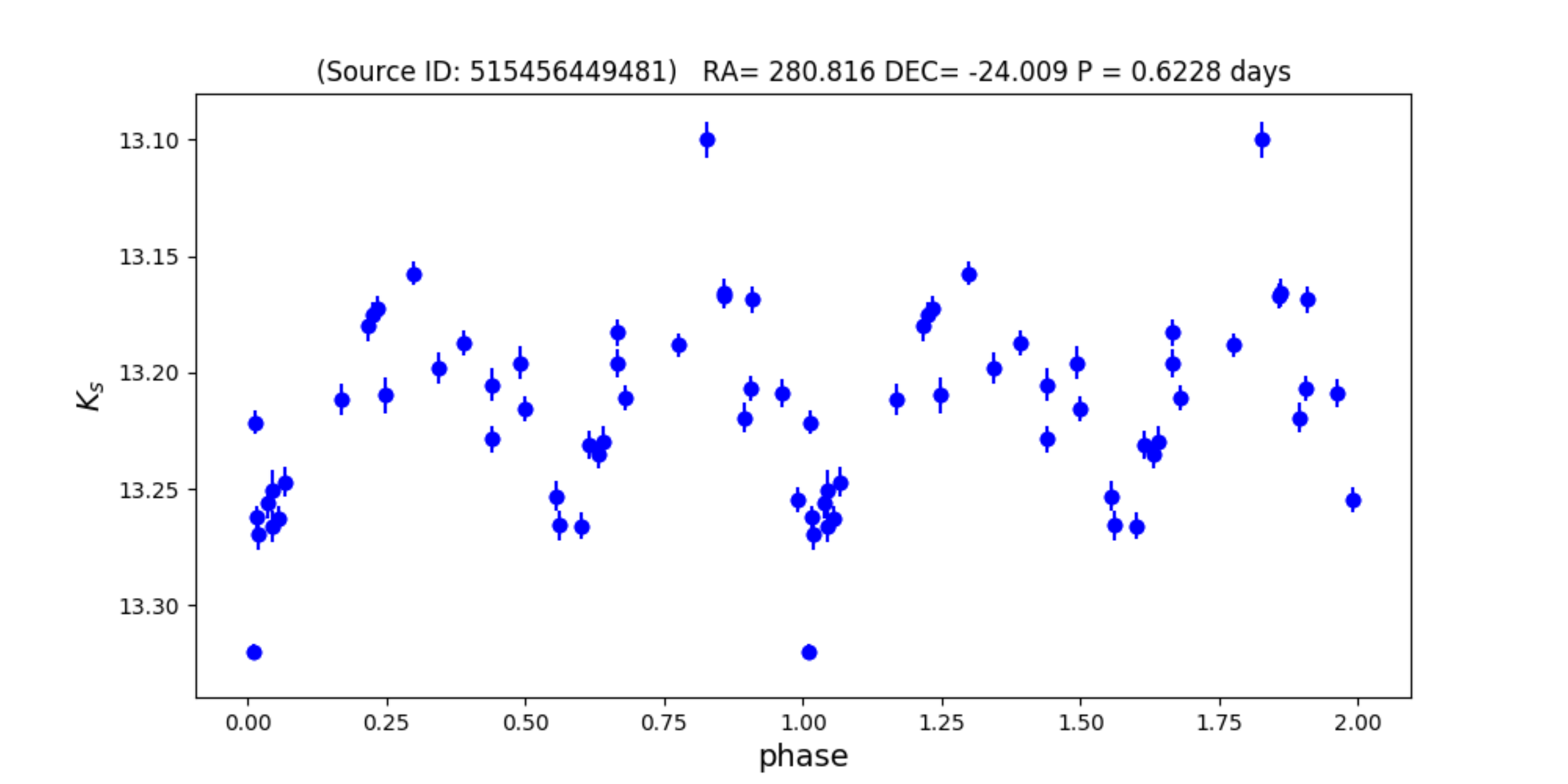} \\
    \includegraphics[width=.33\textwidth]{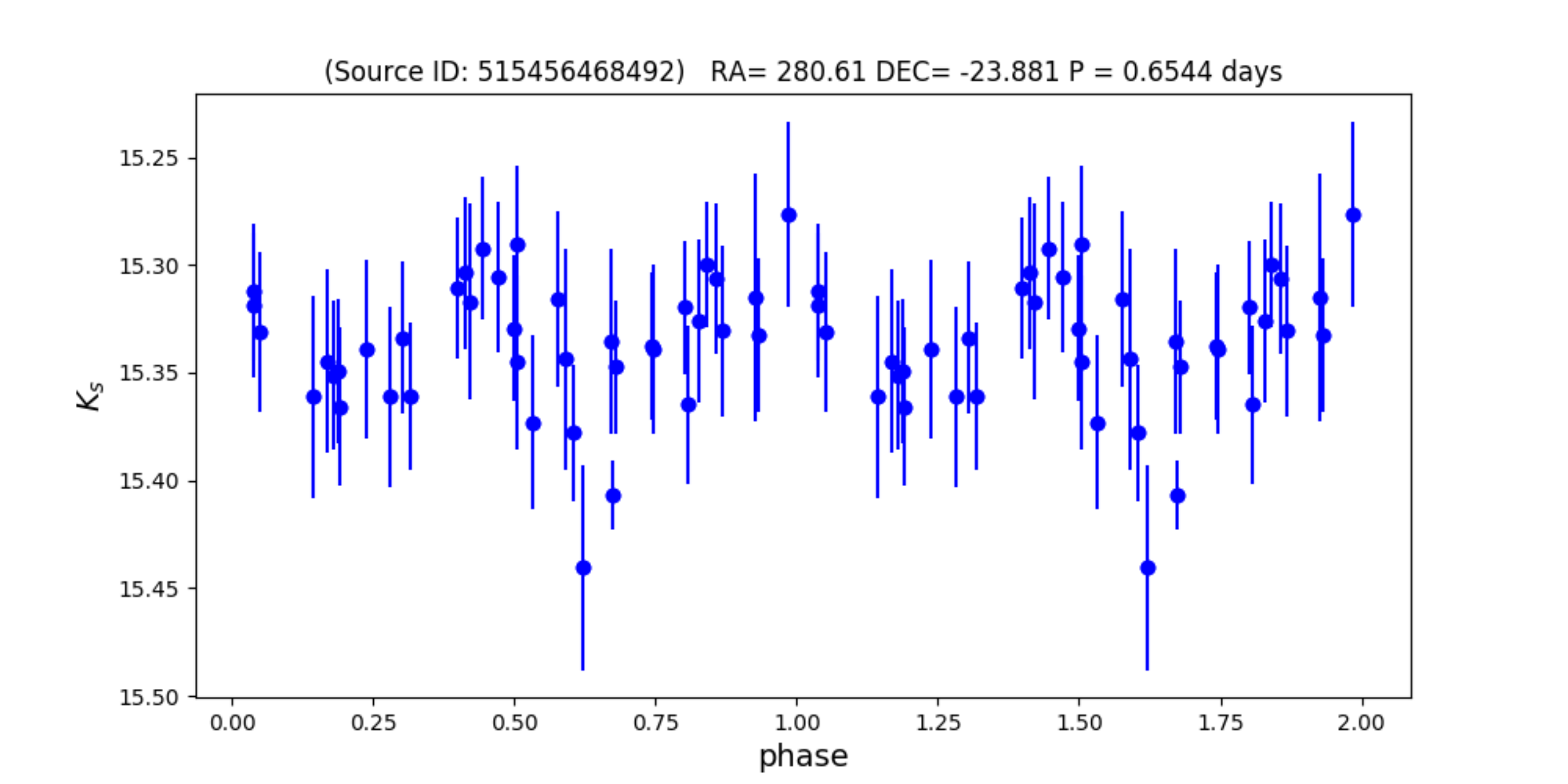} &
    \includegraphics[width=.33\textwidth]{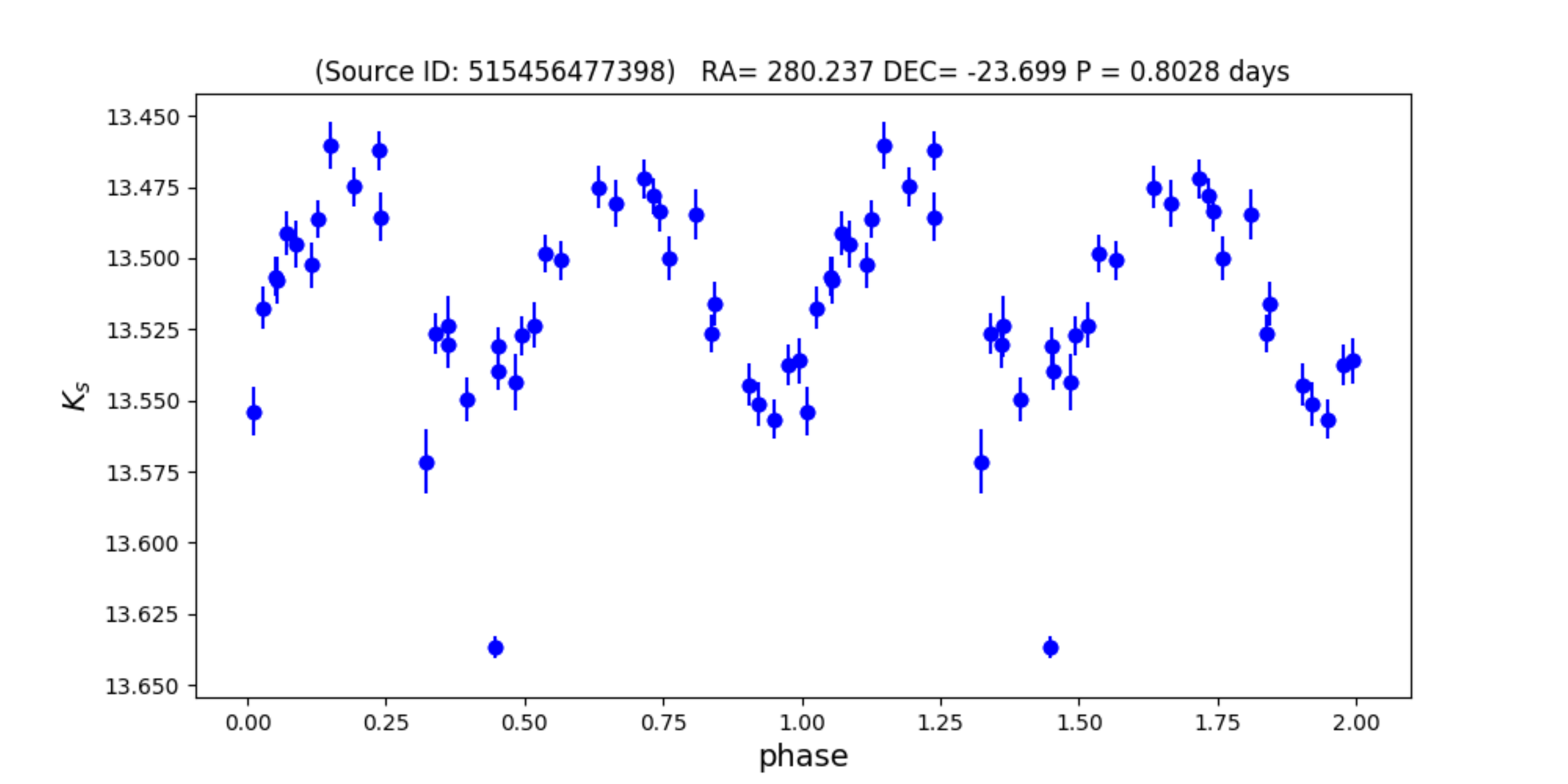} &
    \includegraphics[width=.33\textwidth]{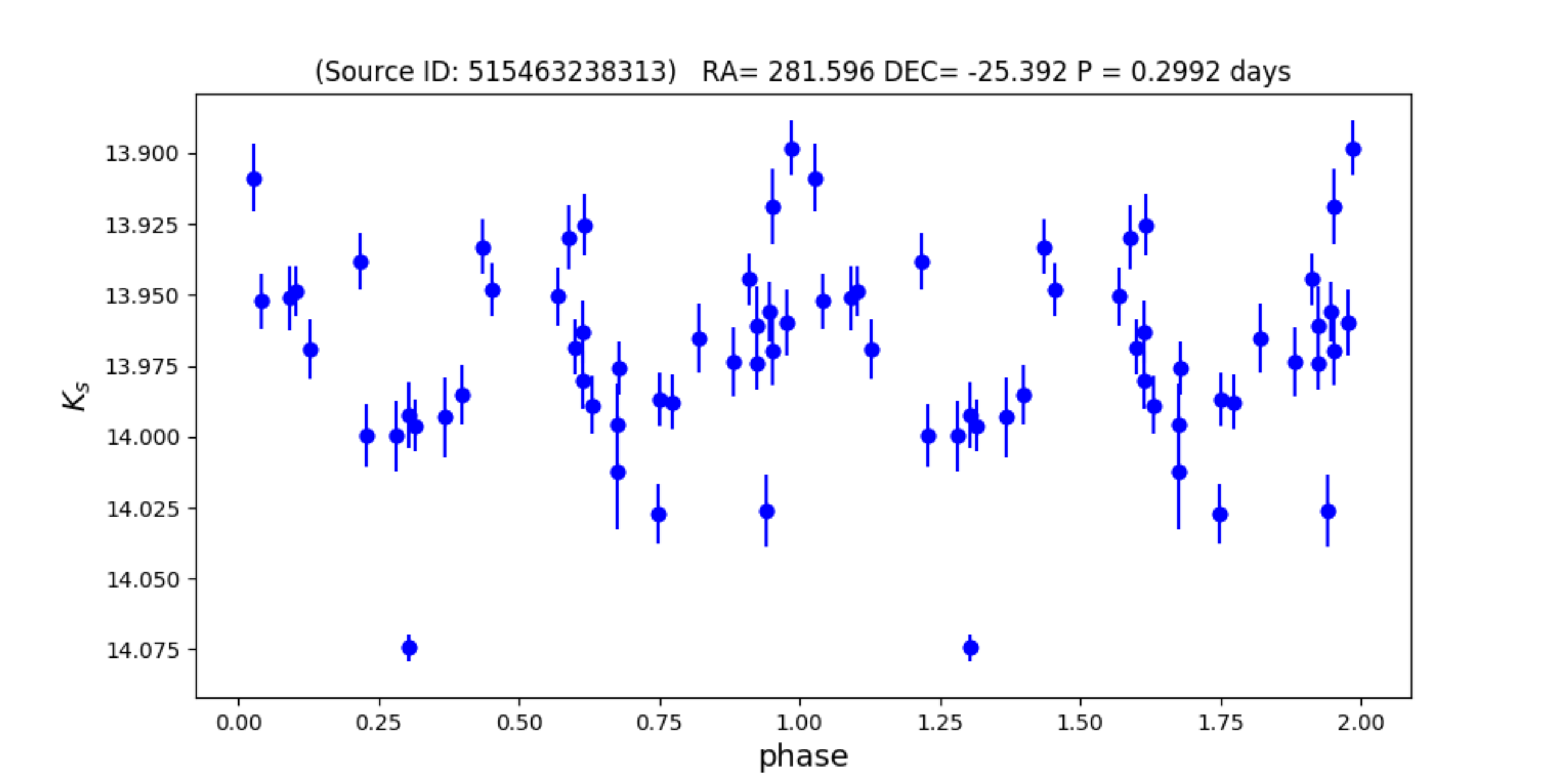} \\
    \includegraphics[width=.33\textwidth]{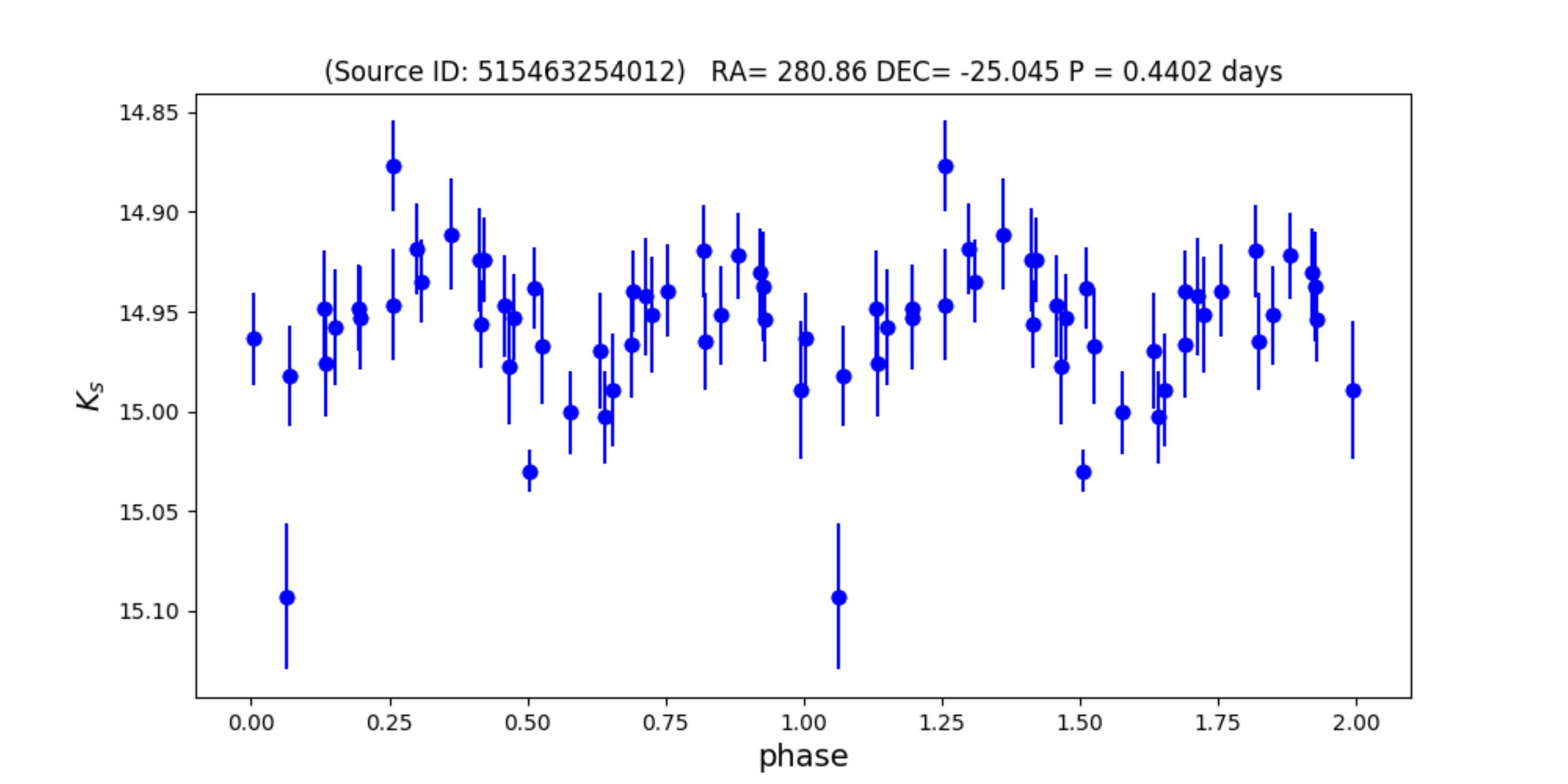} &
    \includegraphics[width=.33\textwidth]{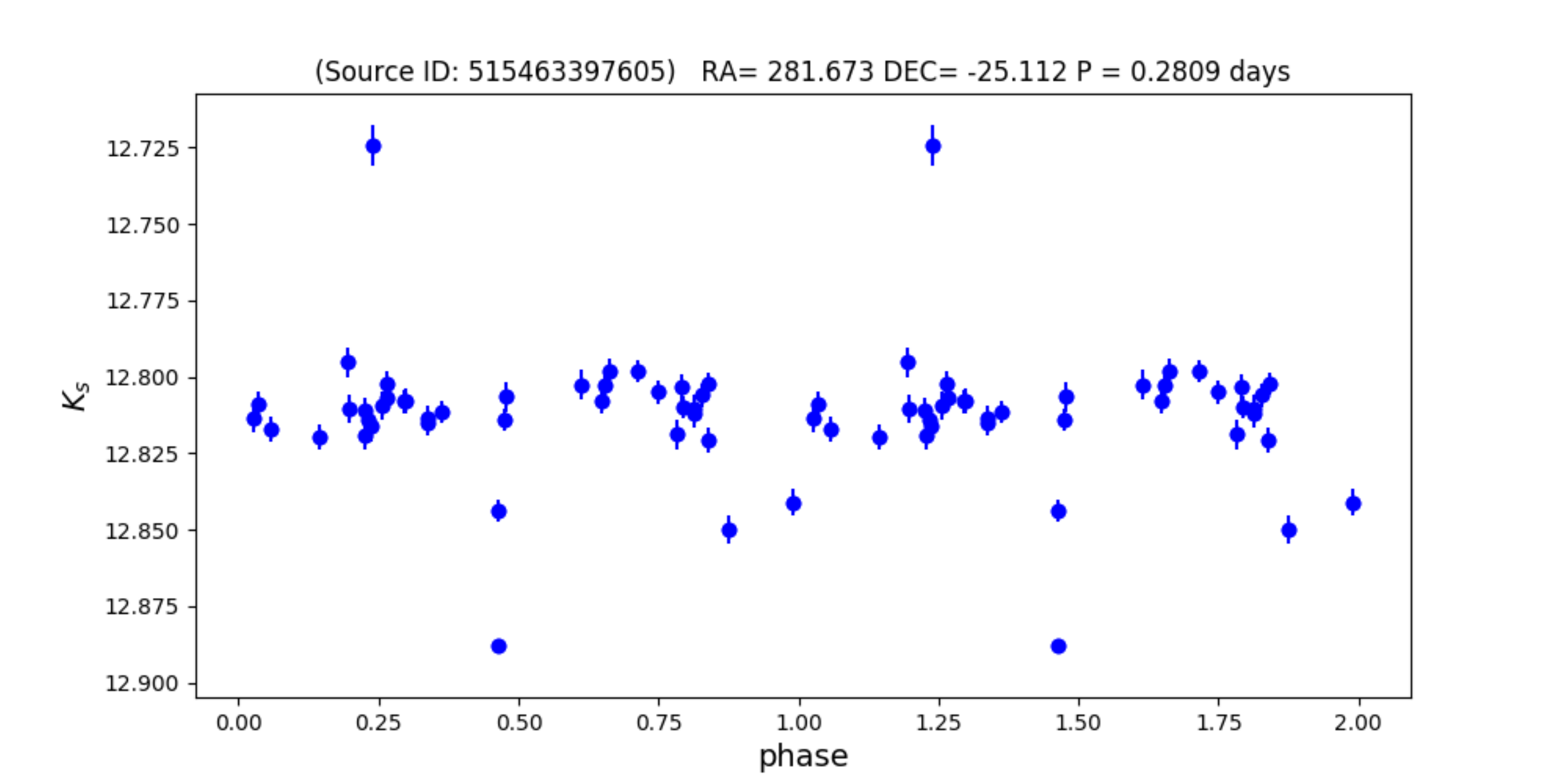} &
    \includegraphics[width=.33\textwidth]{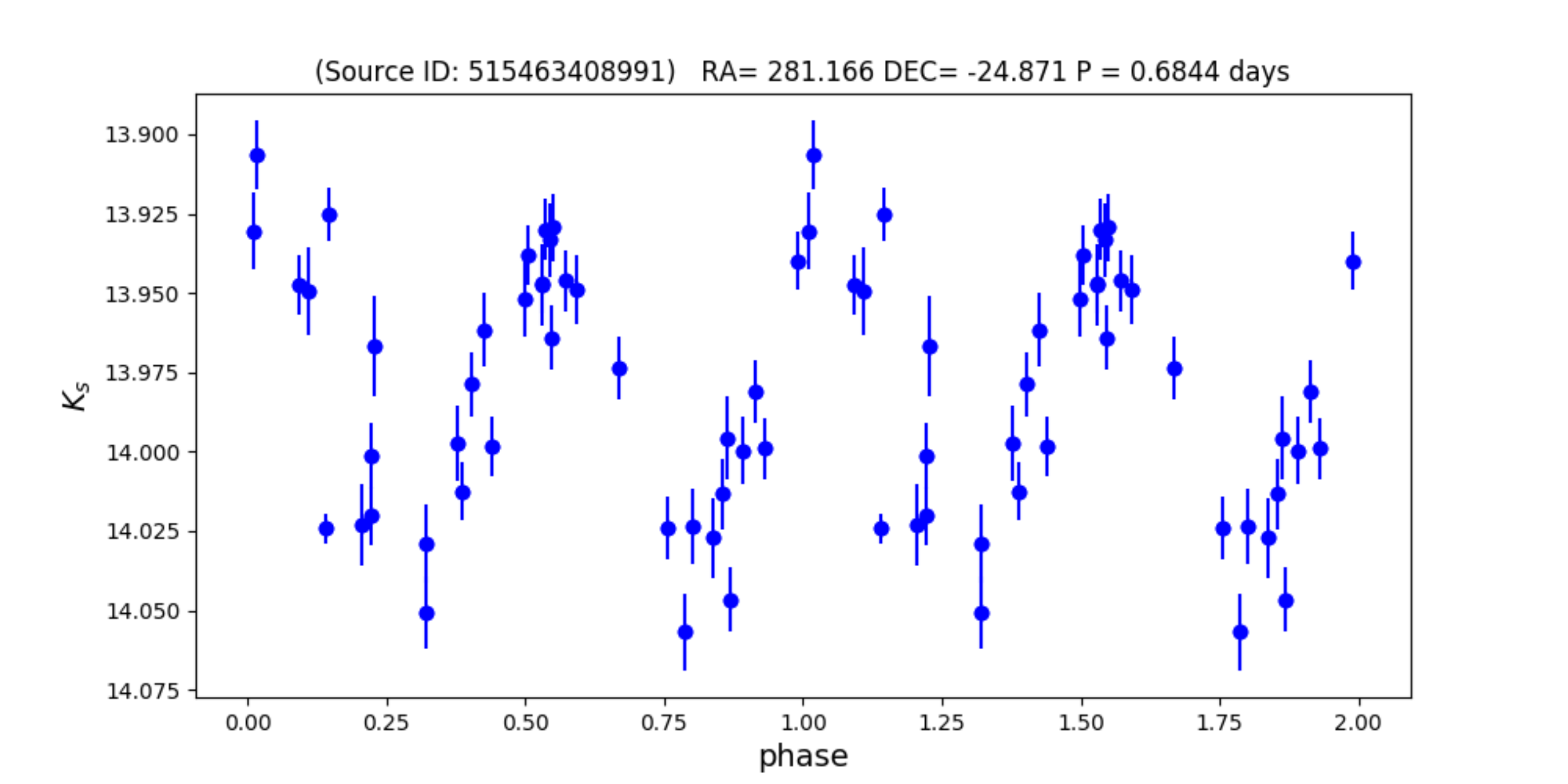} \\
    \includegraphics[width=.33\textwidth]{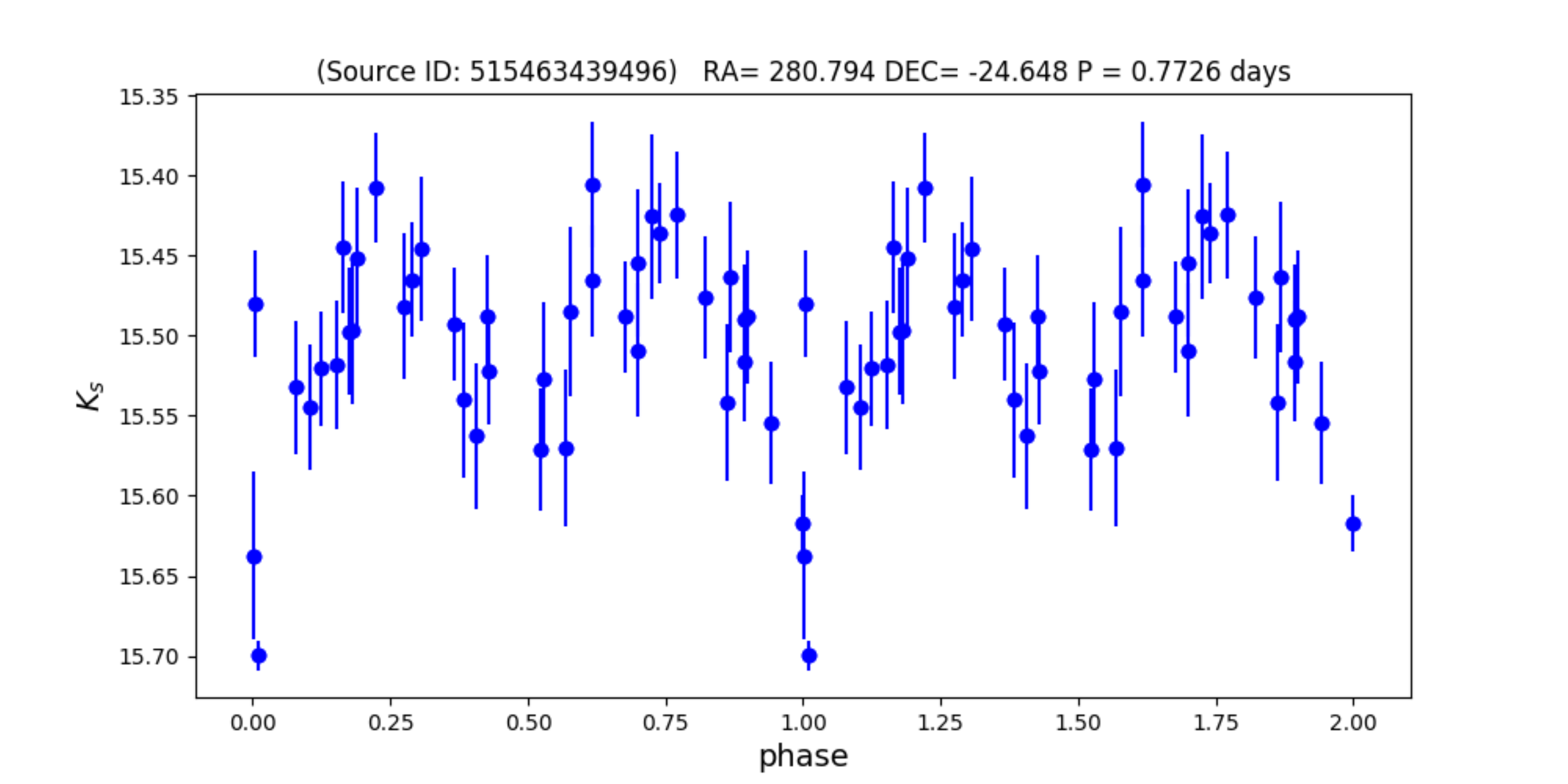} &
    \includegraphics[width=.33\textwidth]{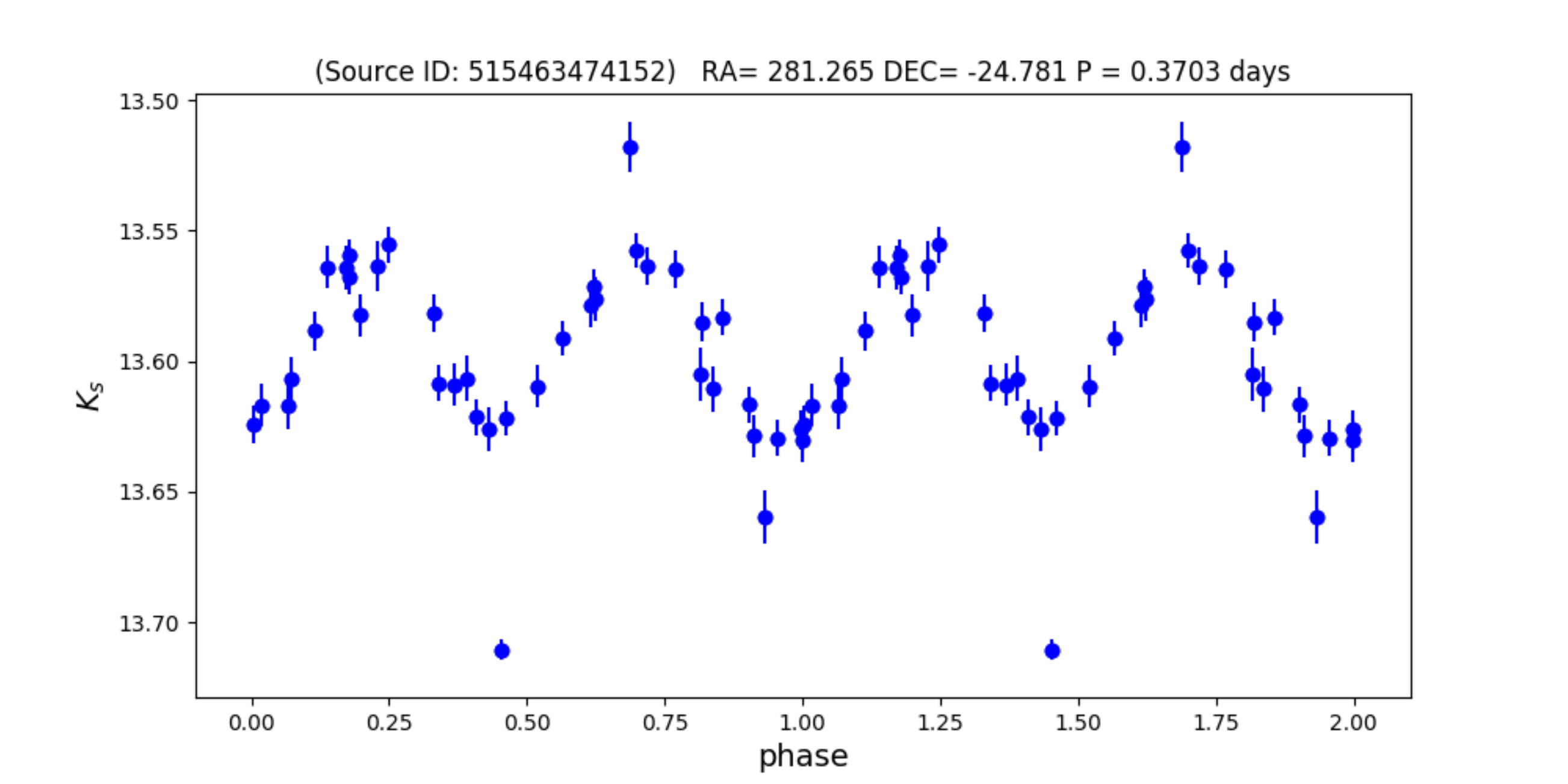} &
    \includegraphics[width=.33\textwidth]{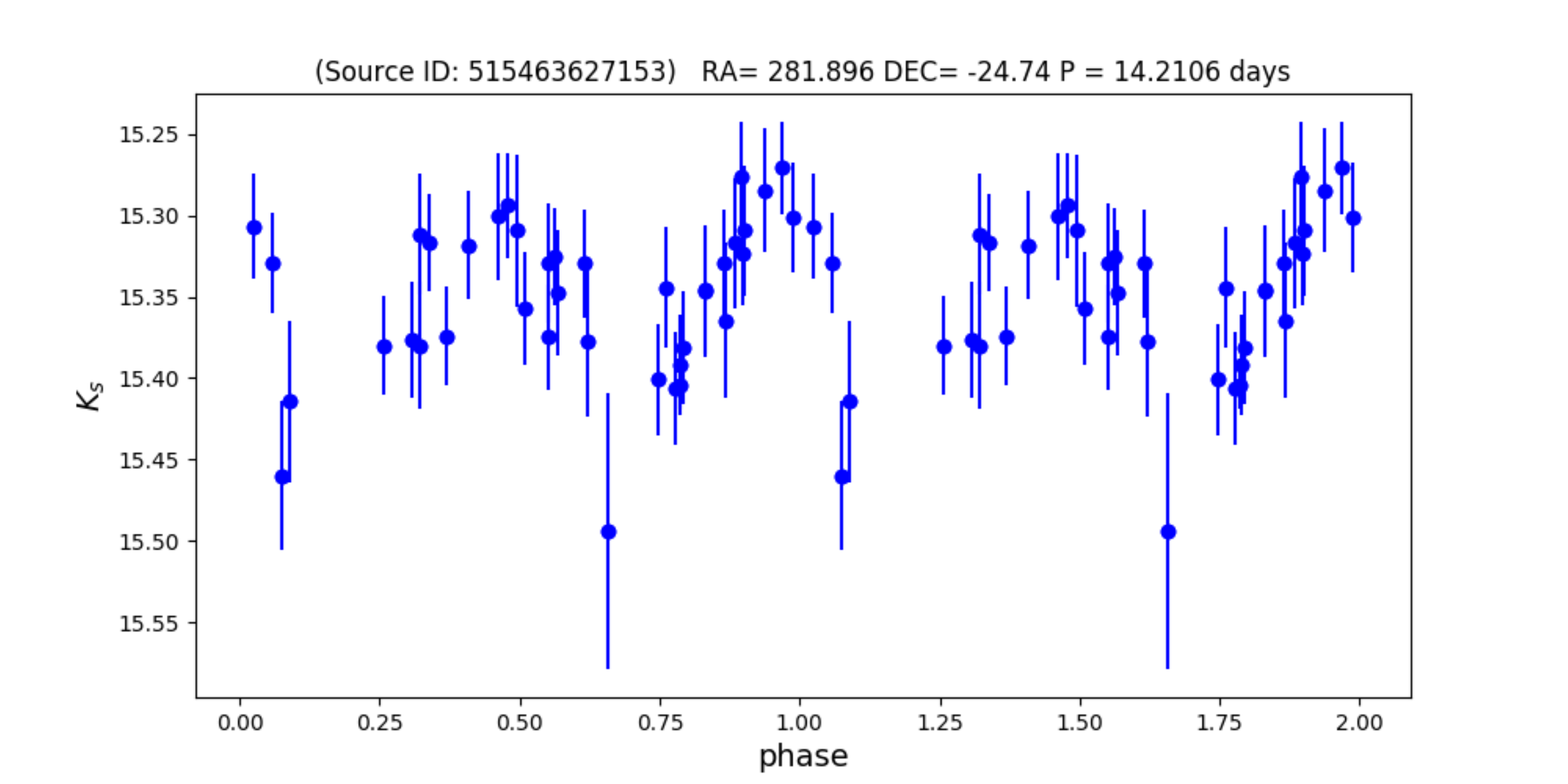} \\ 
  \end{tabular}
  \caption{BHB stars light curves corresponding to $1^{st}$ category eclipsing binaries.}
  \label{fig:bhb1}
\end{figure*}

\begin{figure*}[htb]
\centering
  \begin{tabular}{@{}ccc@{}}      
    \includegraphics[width=.33\textwidth]{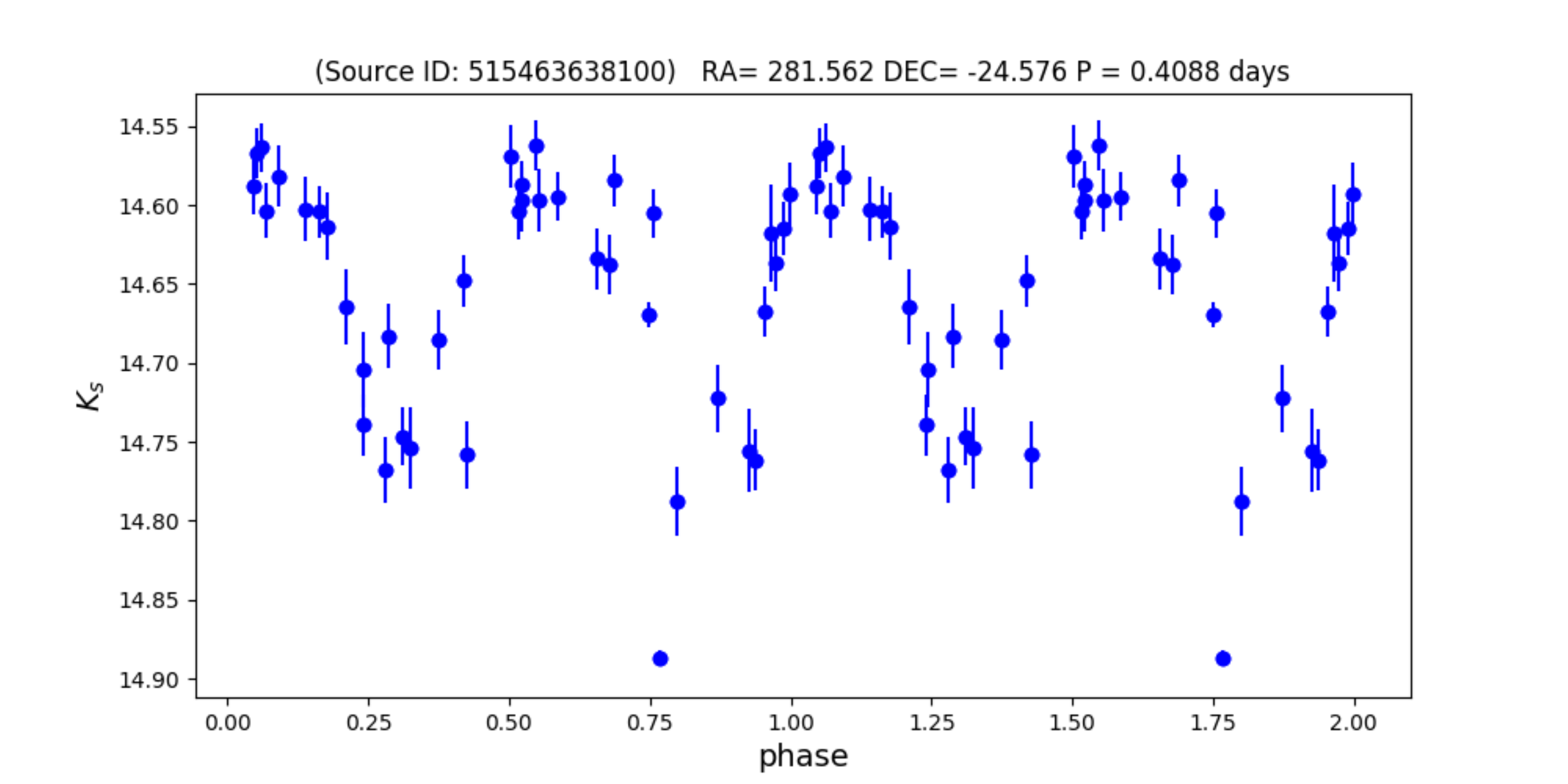} &
    \includegraphics[width=.33\textwidth]{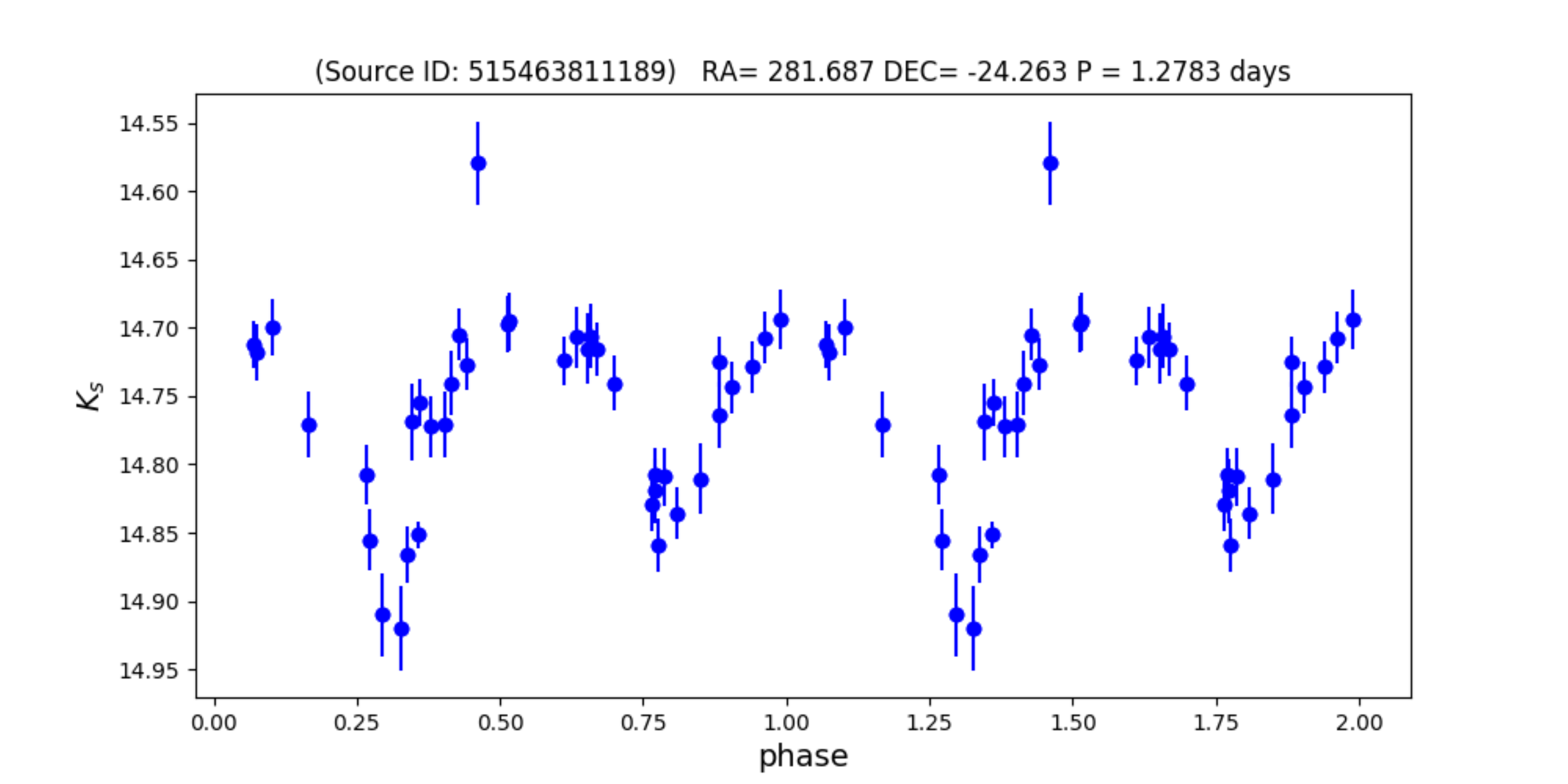} & 
    \includegraphics[width=.33\textwidth]{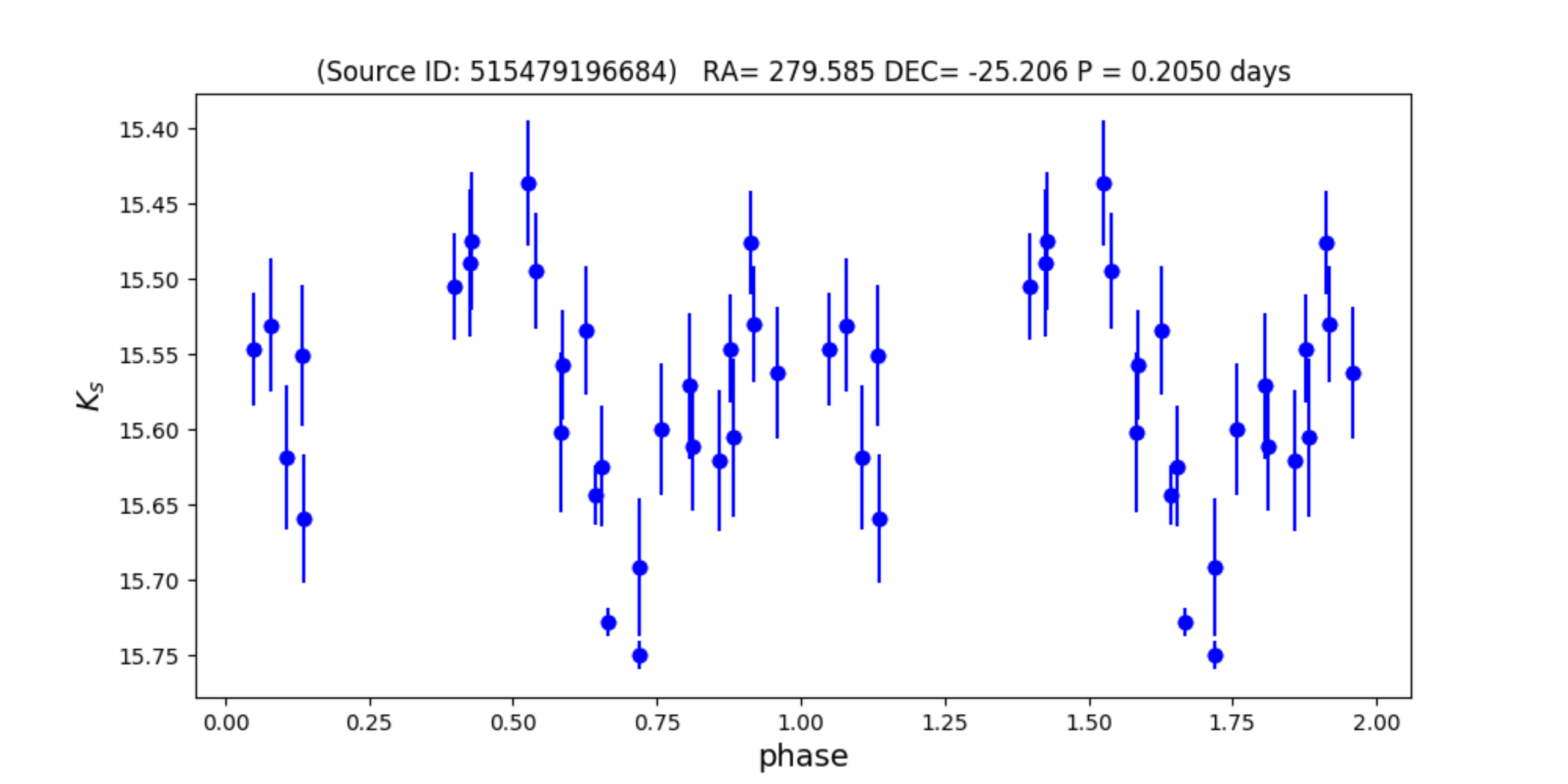} \\
    \includegraphics[width=.33\textwidth]{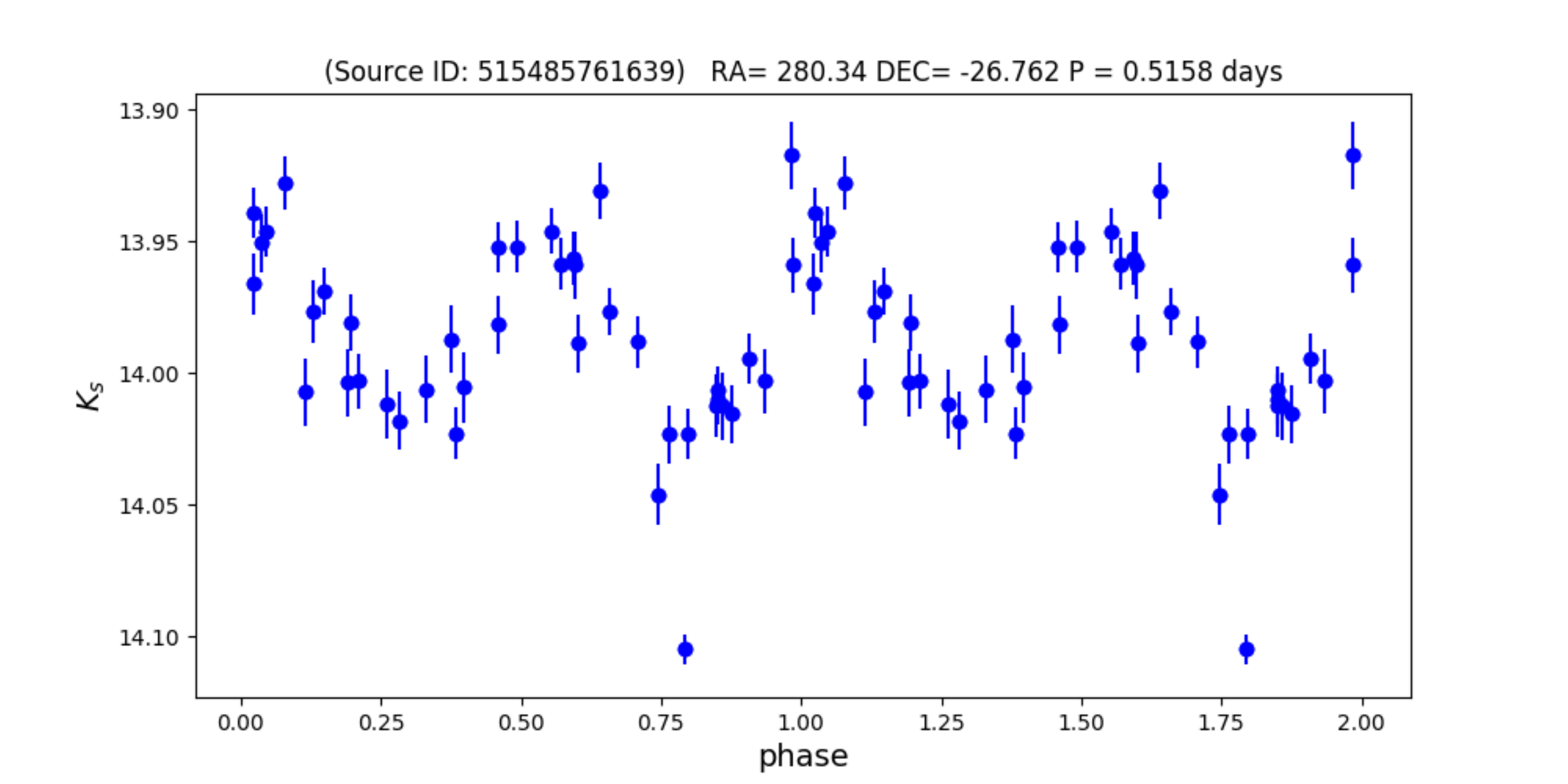} &  
    \includegraphics[width=.33\textwidth]{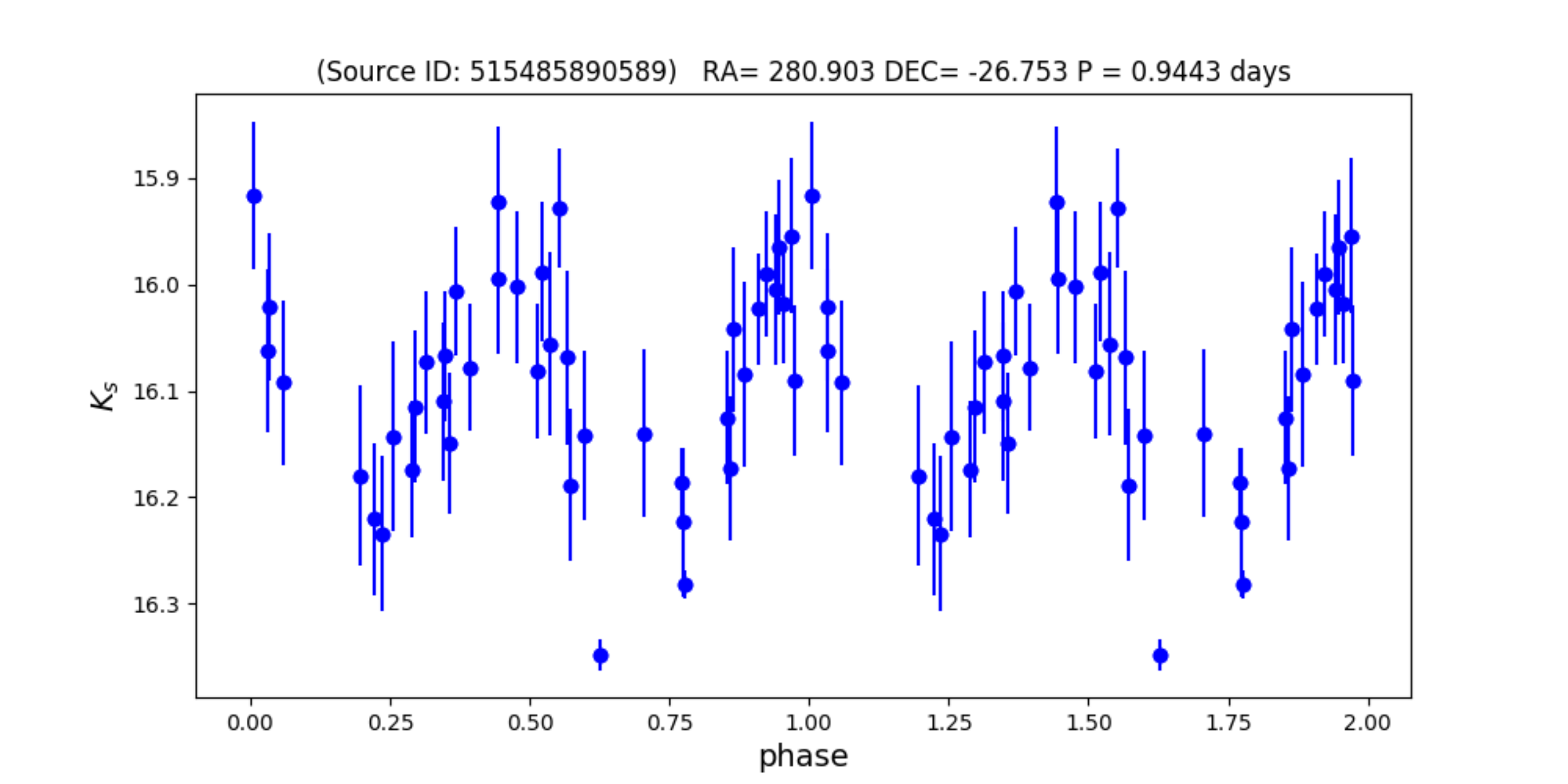} & 
    \includegraphics[width=.33\textwidth]{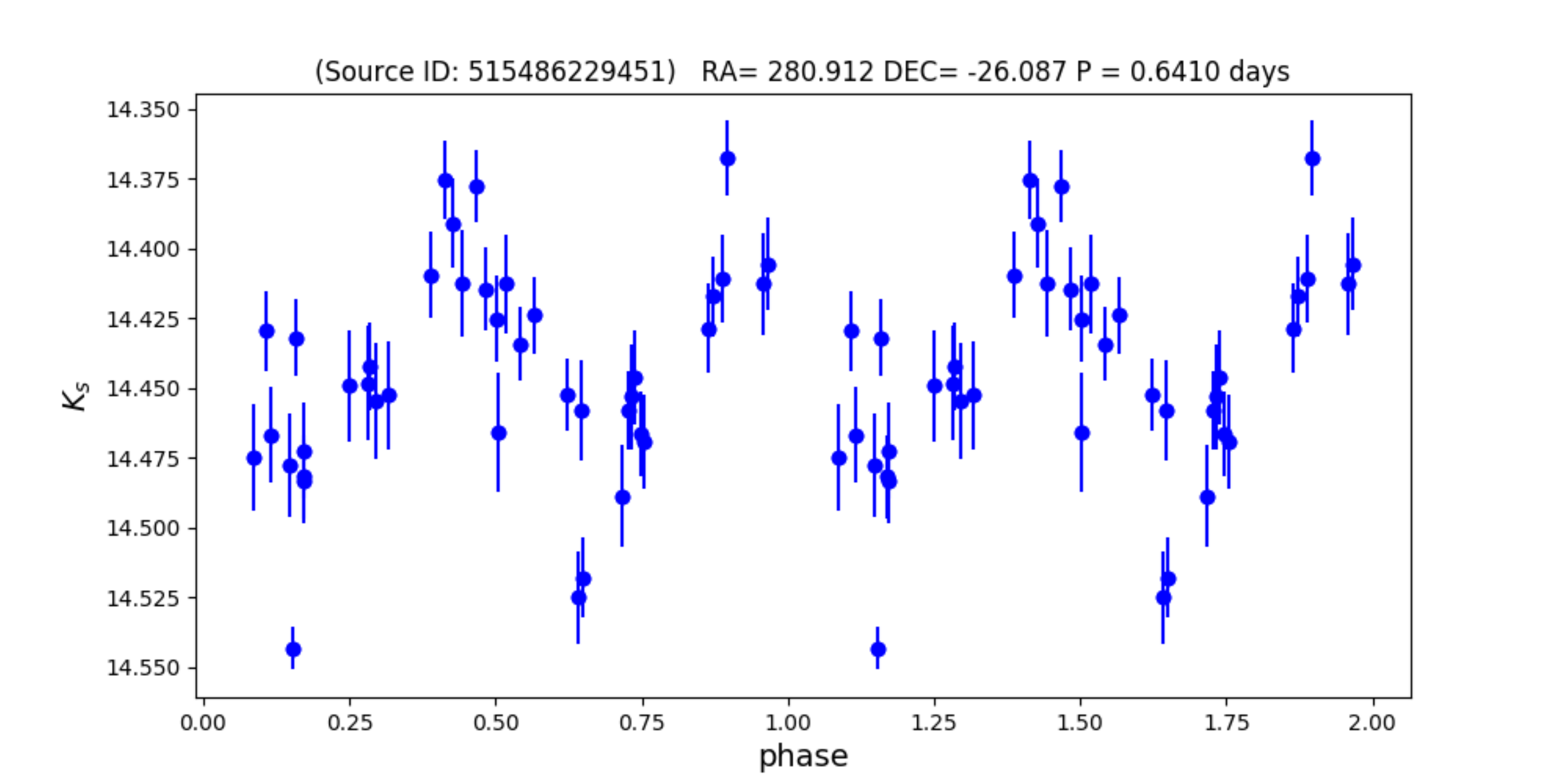} \\
    \includegraphics[width=.33\textwidth]{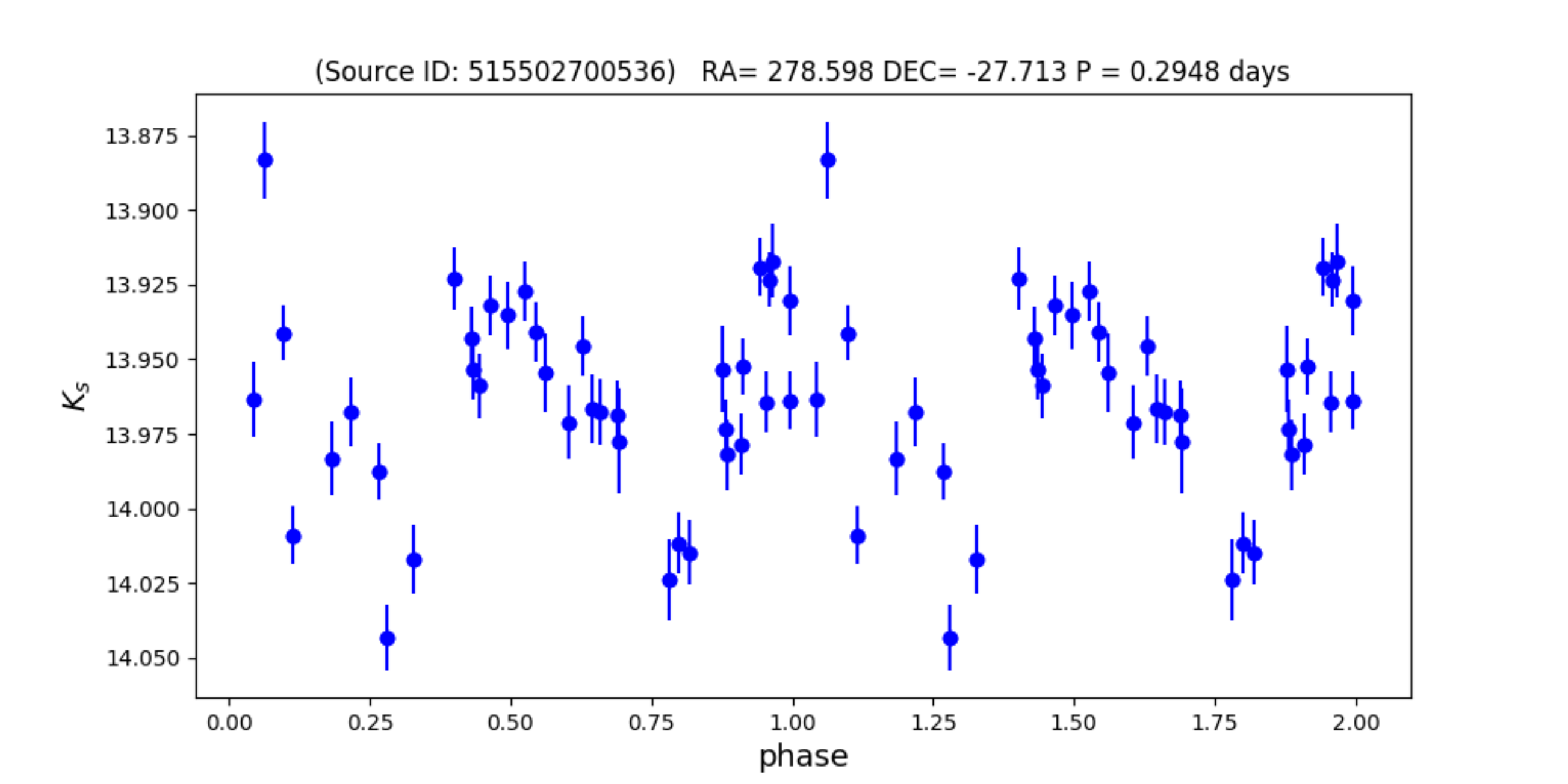} &
    \includegraphics[width=.33\textwidth]{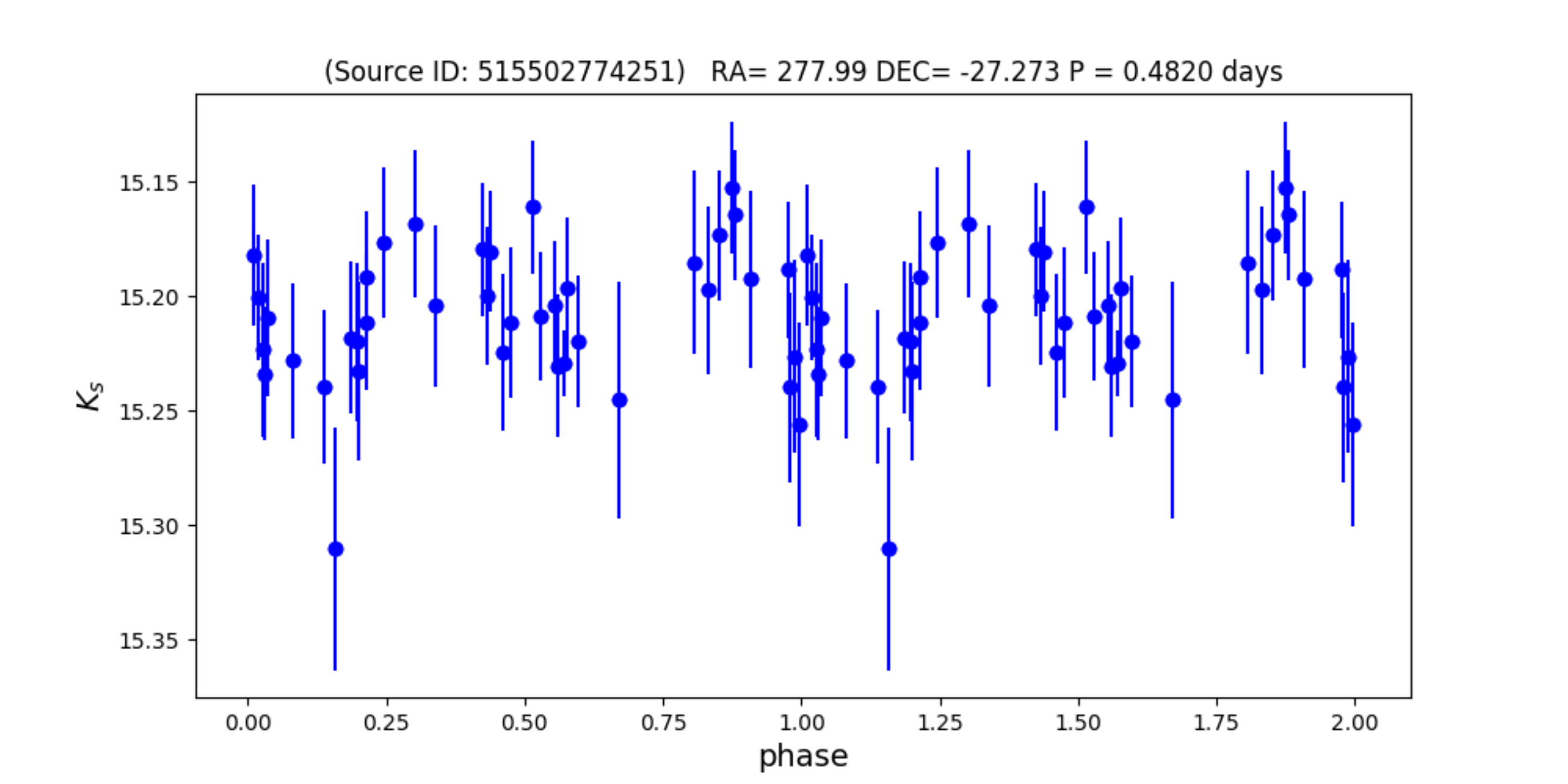} &
    \includegraphics[width=.33\textwidth]{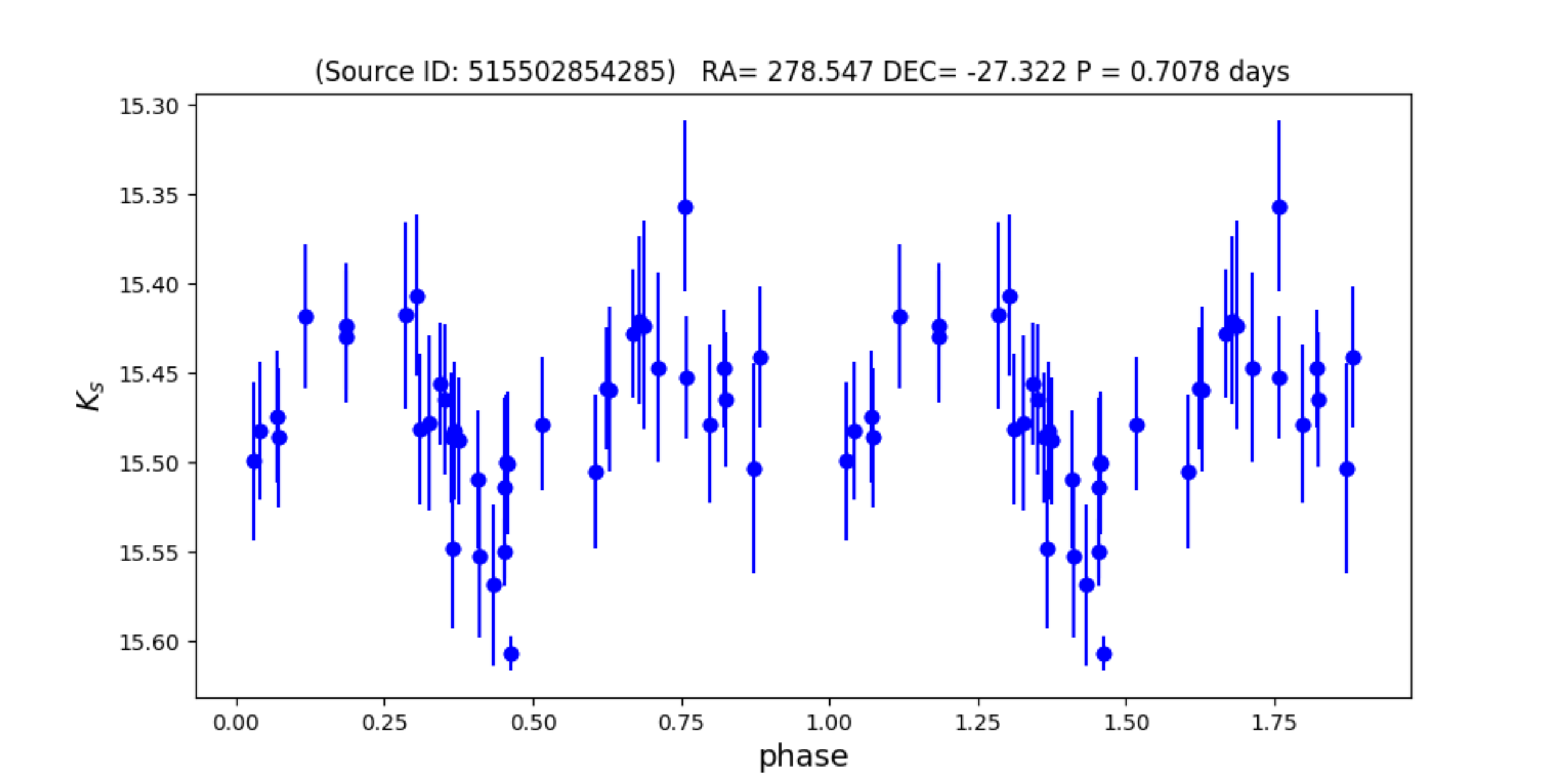} \\
    \includegraphics[width=.33\textwidth]{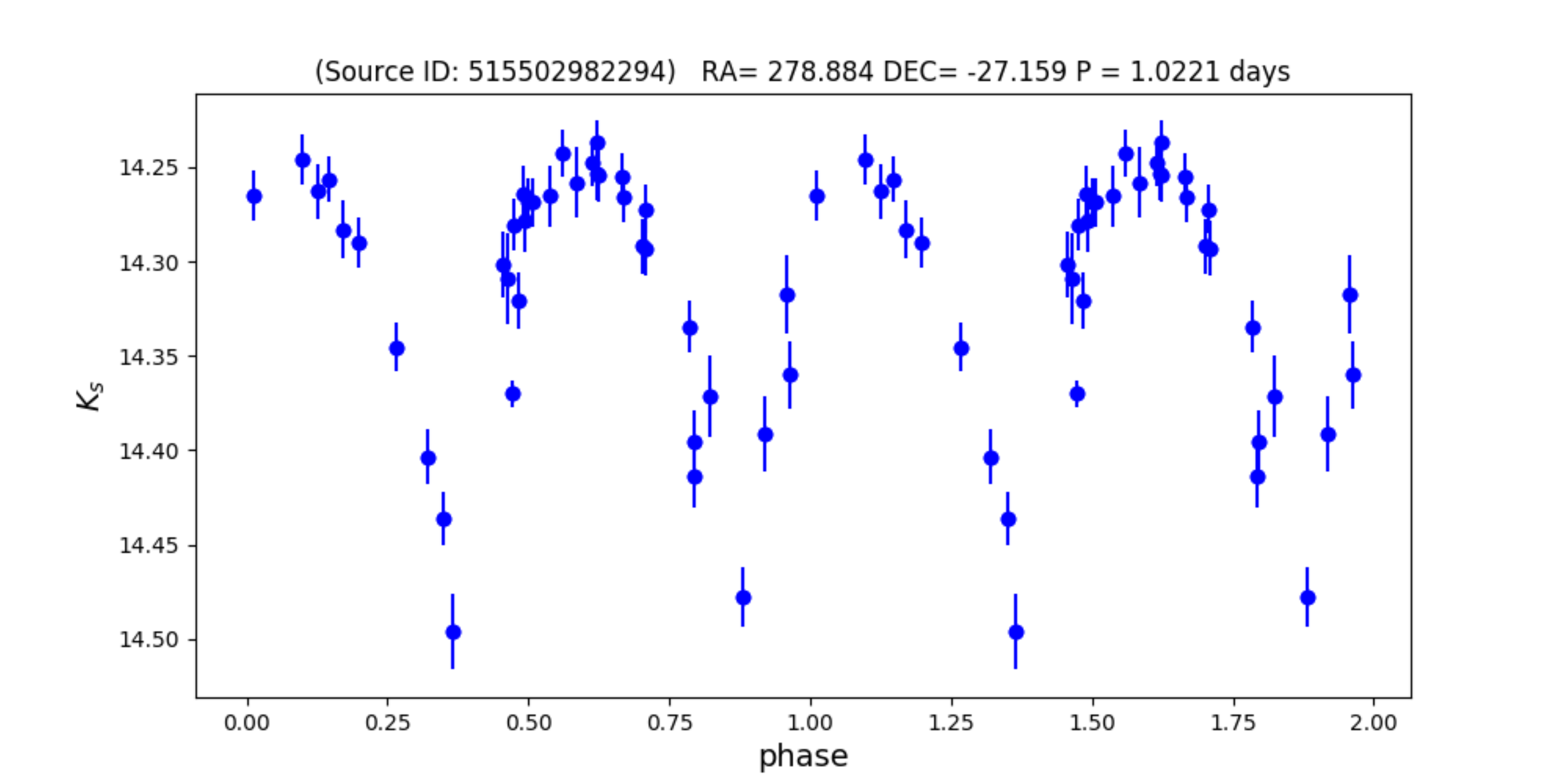} &
    \includegraphics[width=.33\textwidth]{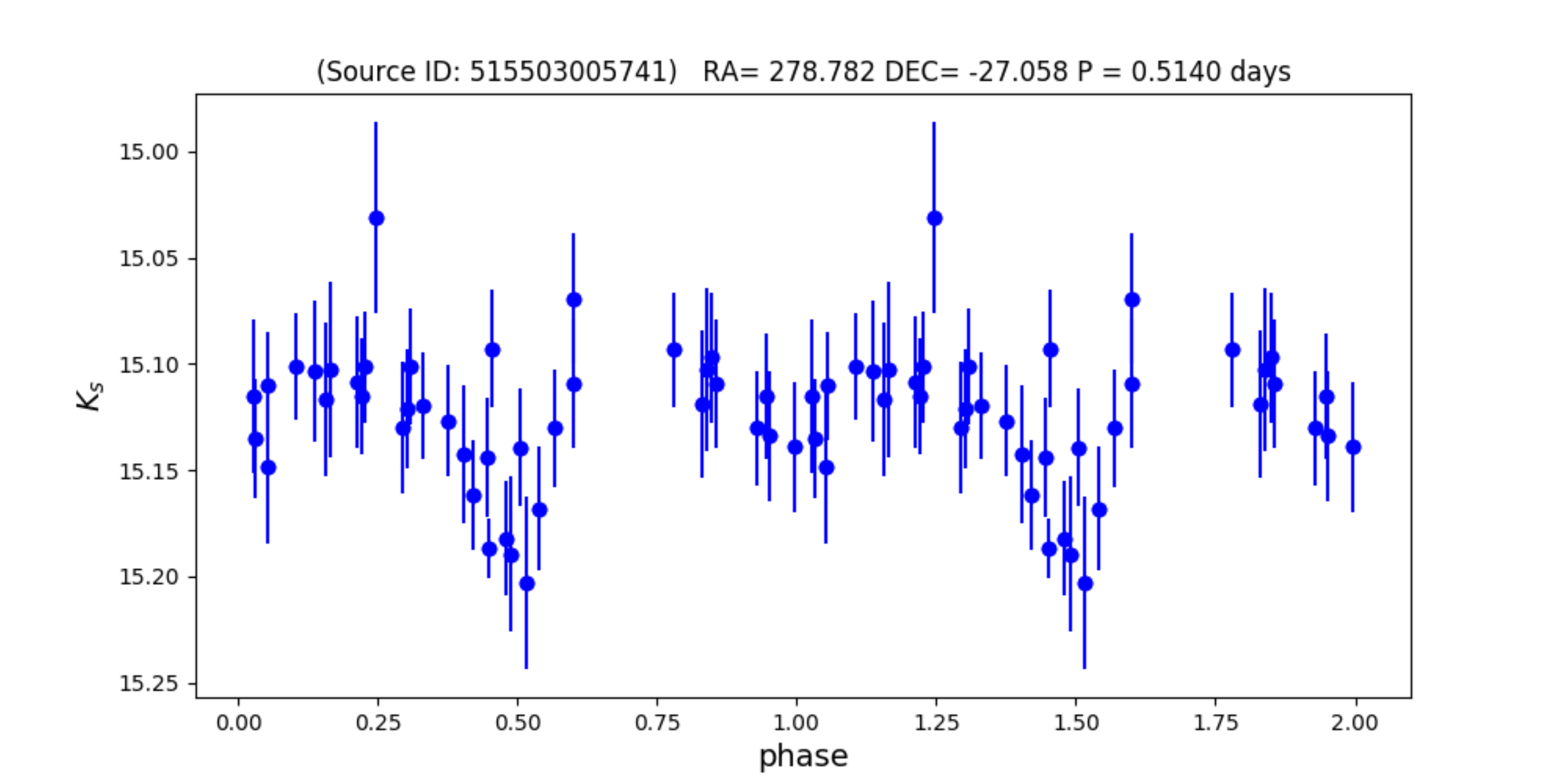} &
    \includegraphics[width=.33\textwidth]{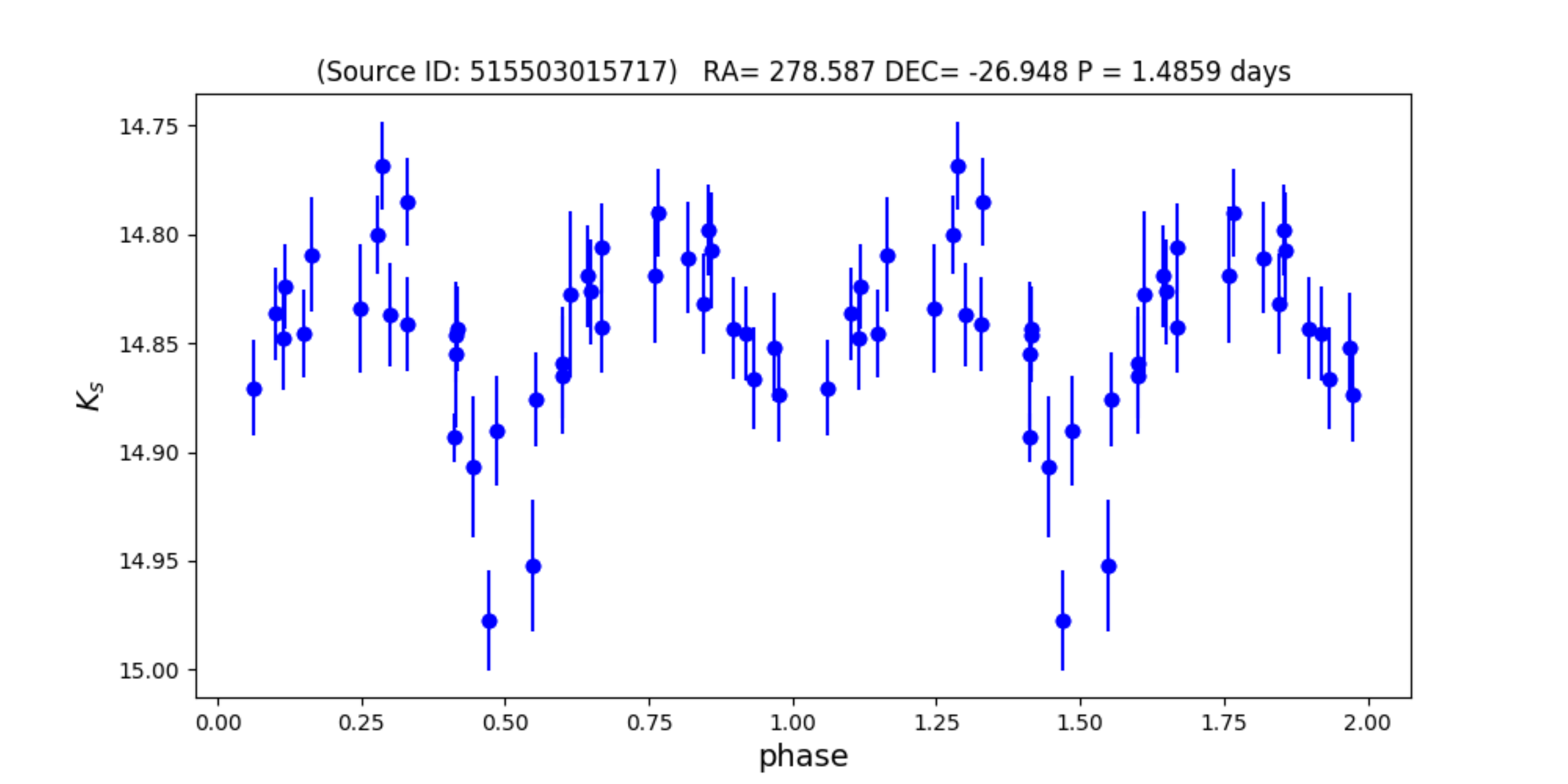} \\
    \includegraphics[width=.33\textwidth]{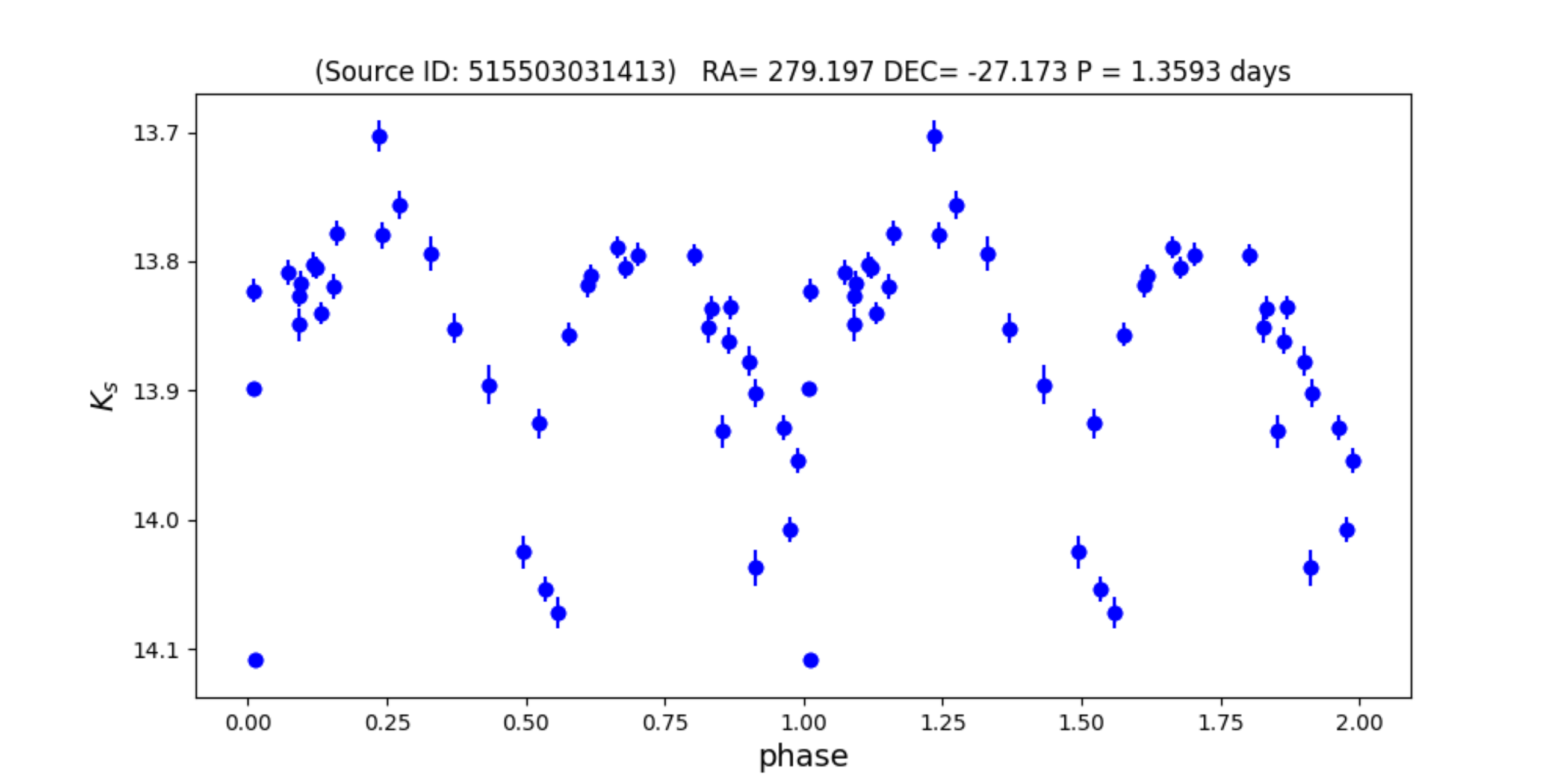} &
    \includegraphics[width=.33\textwidth]{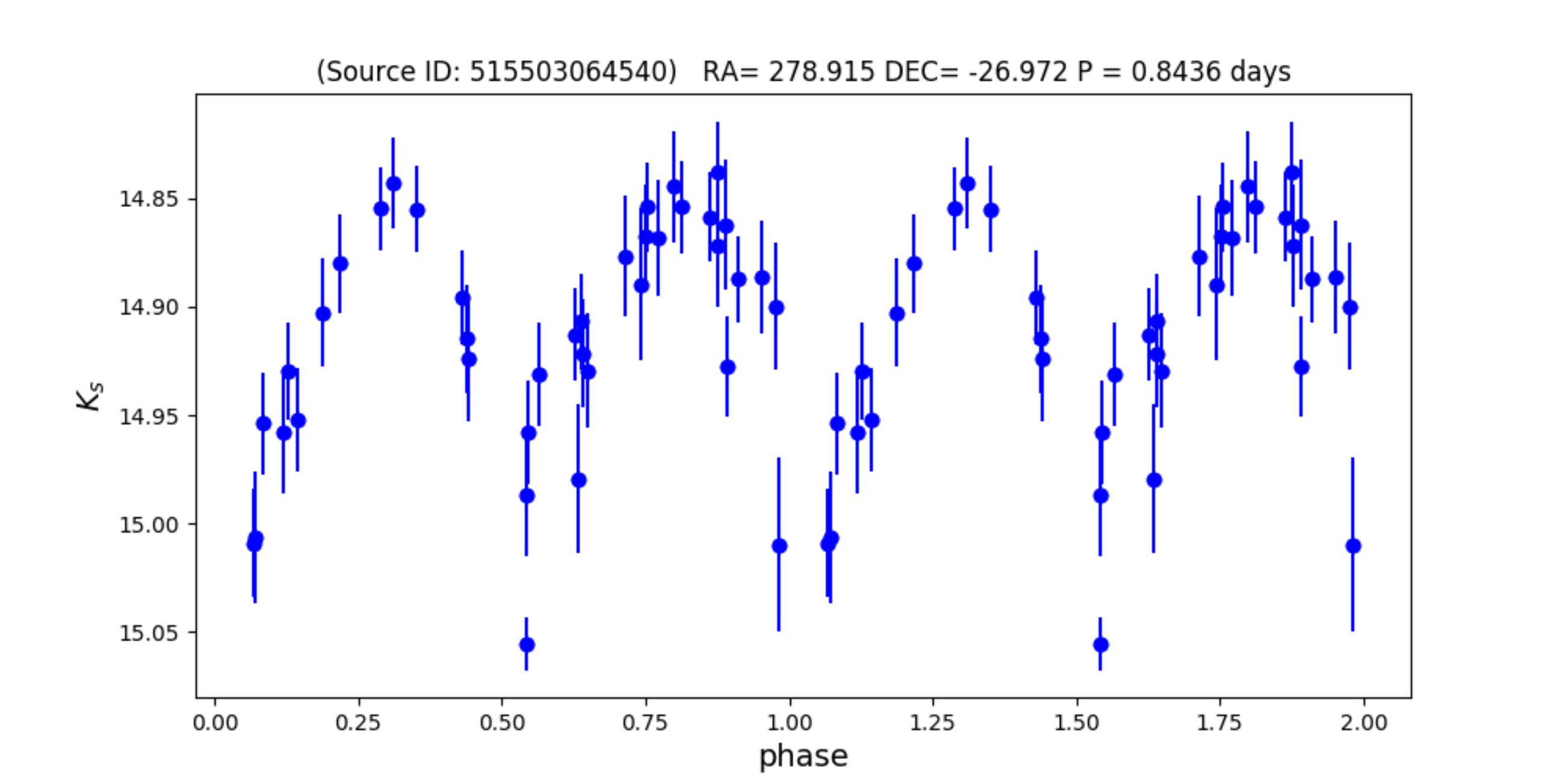} &
    \includegraphics[width=.33\textwidth]{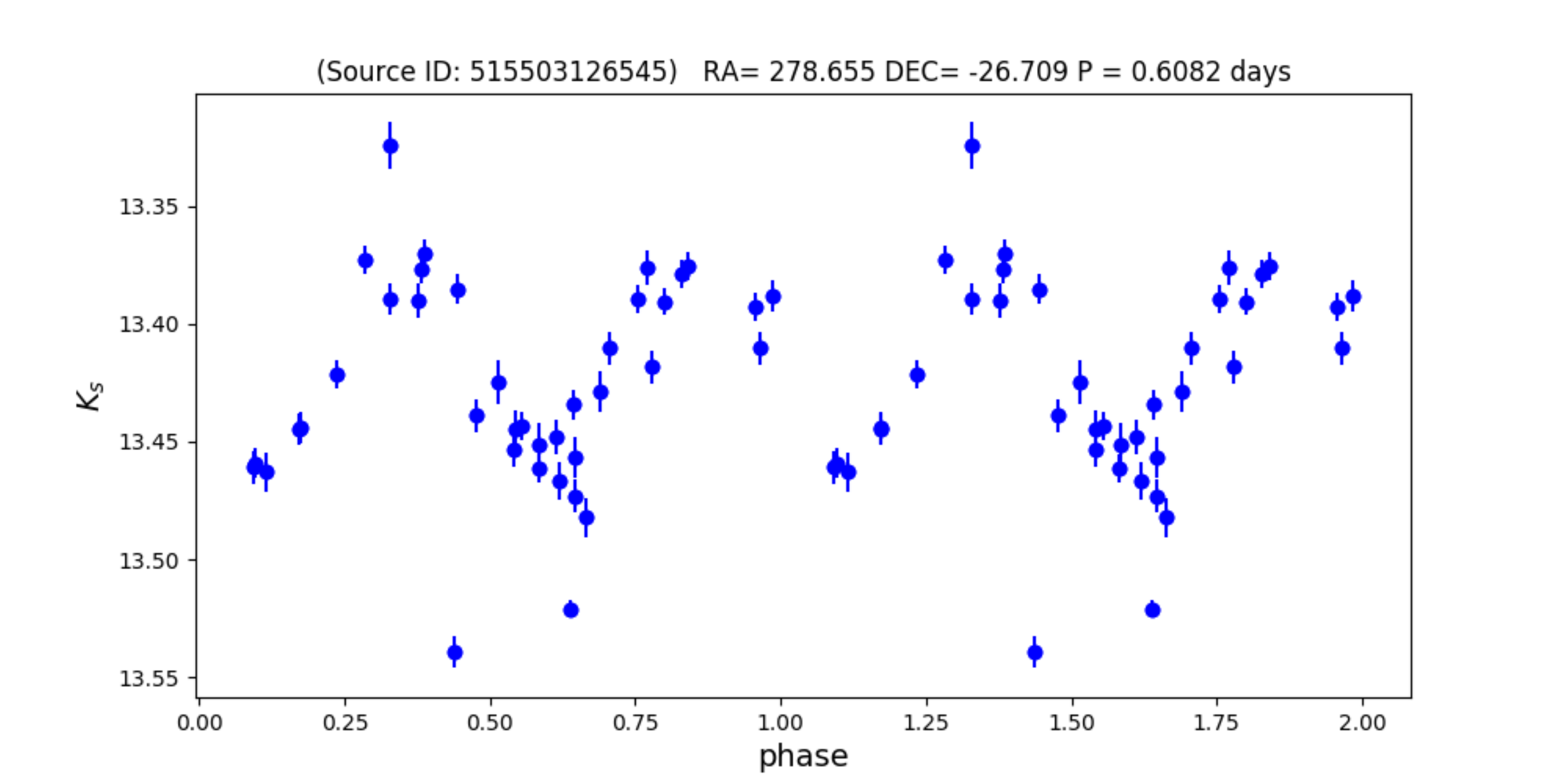} \\
    \includegraphics[width=.33\textwidth]{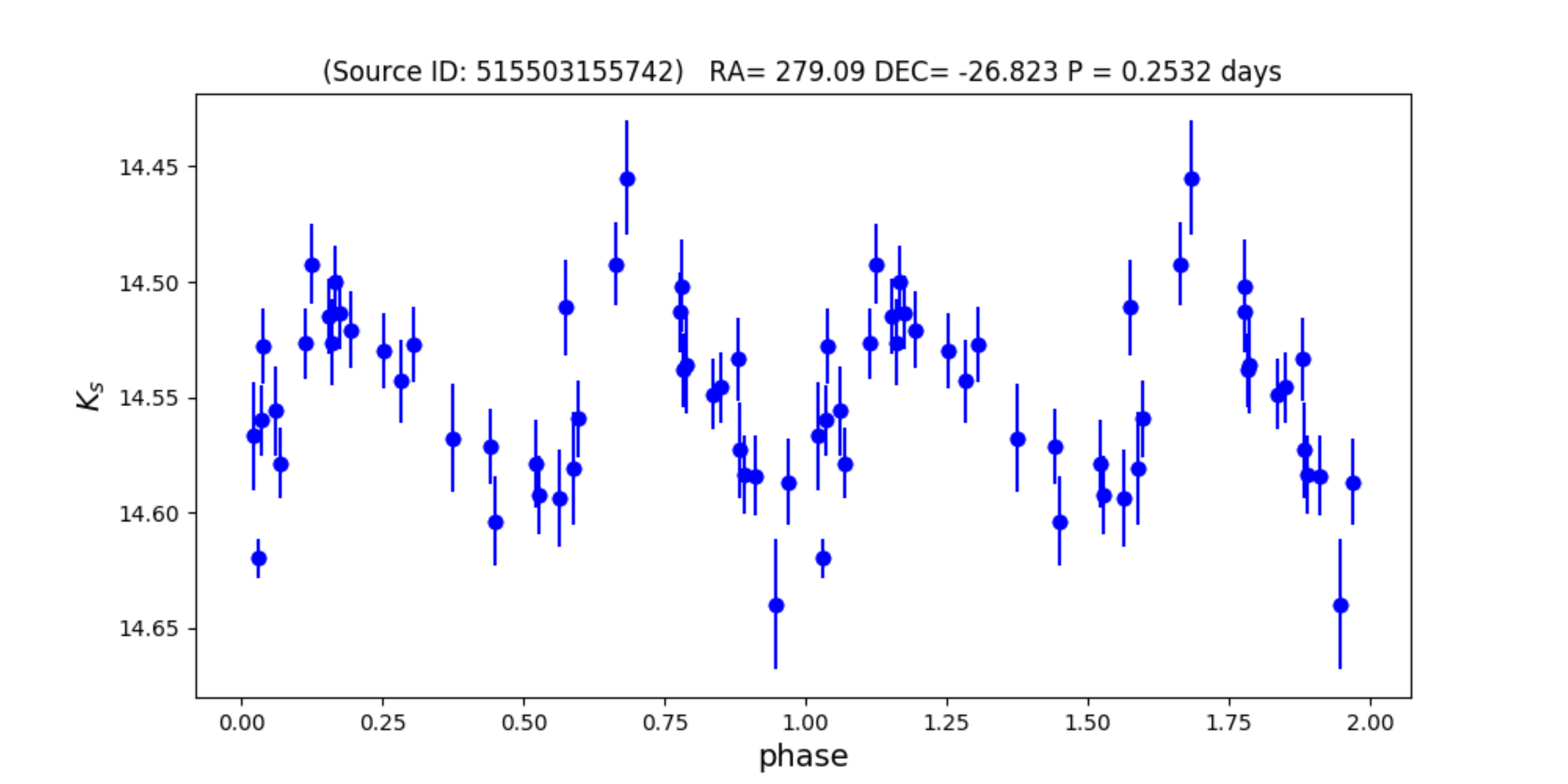} &
    \includegraphics[width=.33\textwidth]{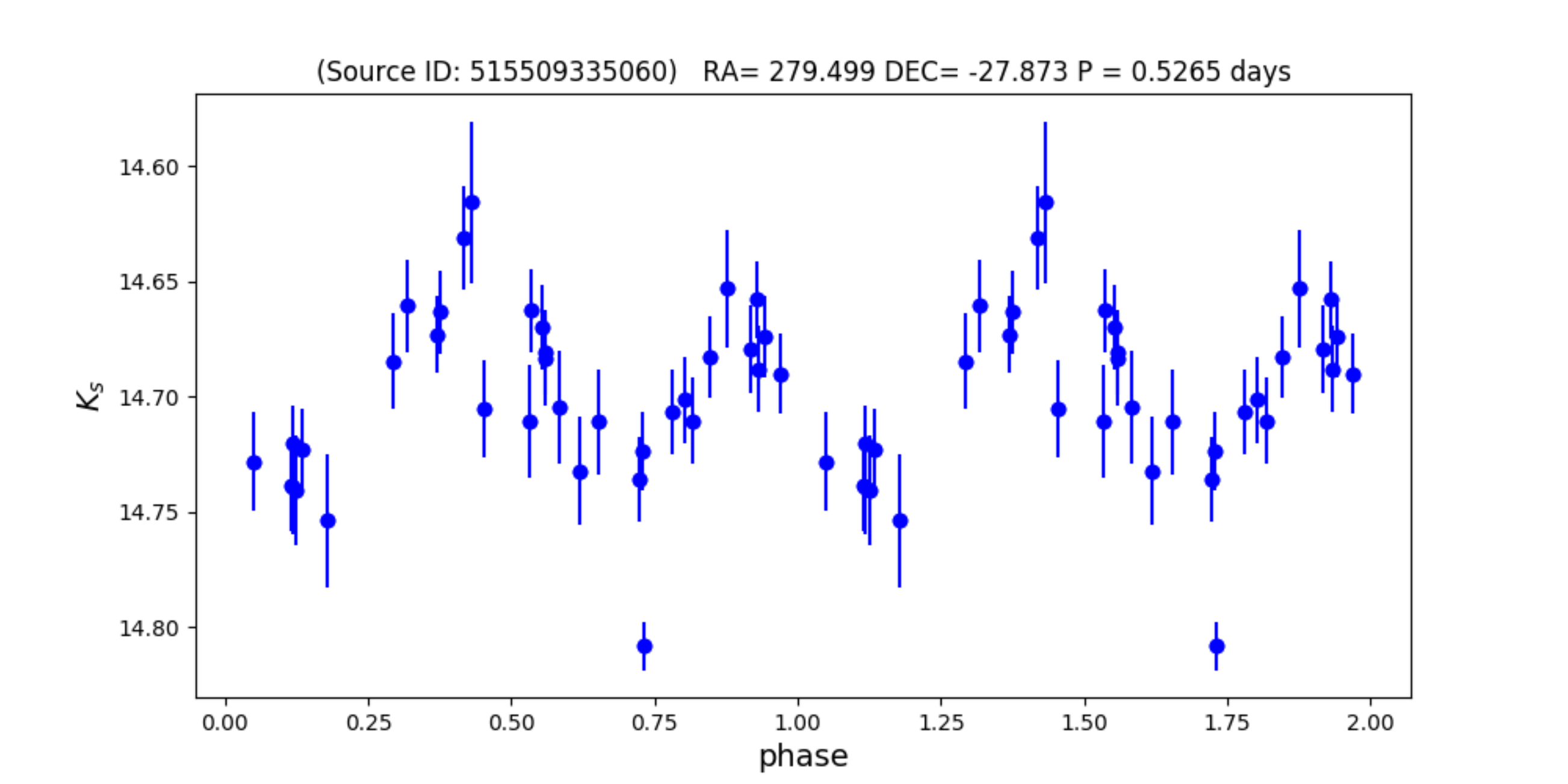} &
    \includegraphics[width=.33\textwidth]{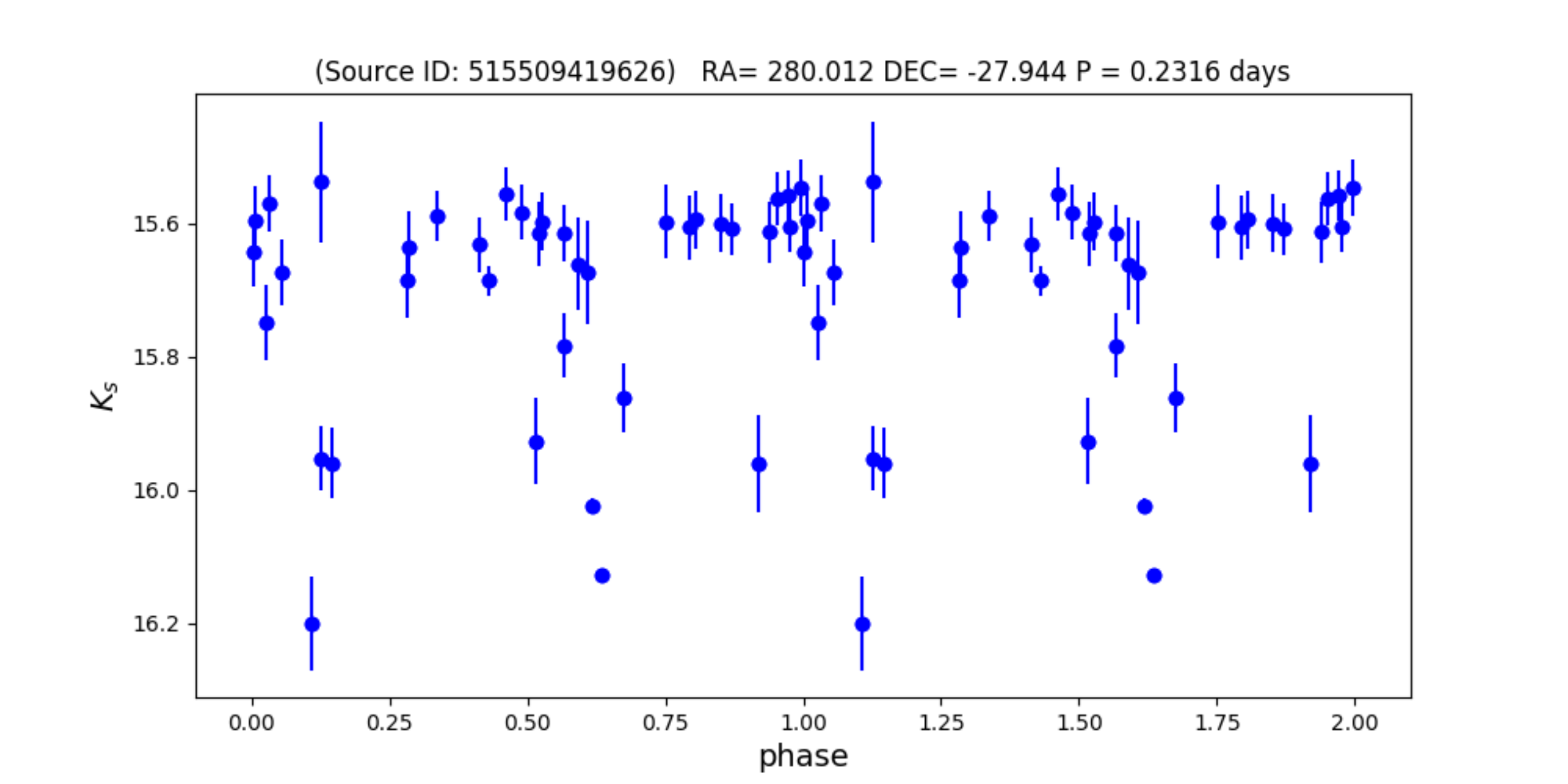} \\
    \includegraphics[width=.33\textwidth]{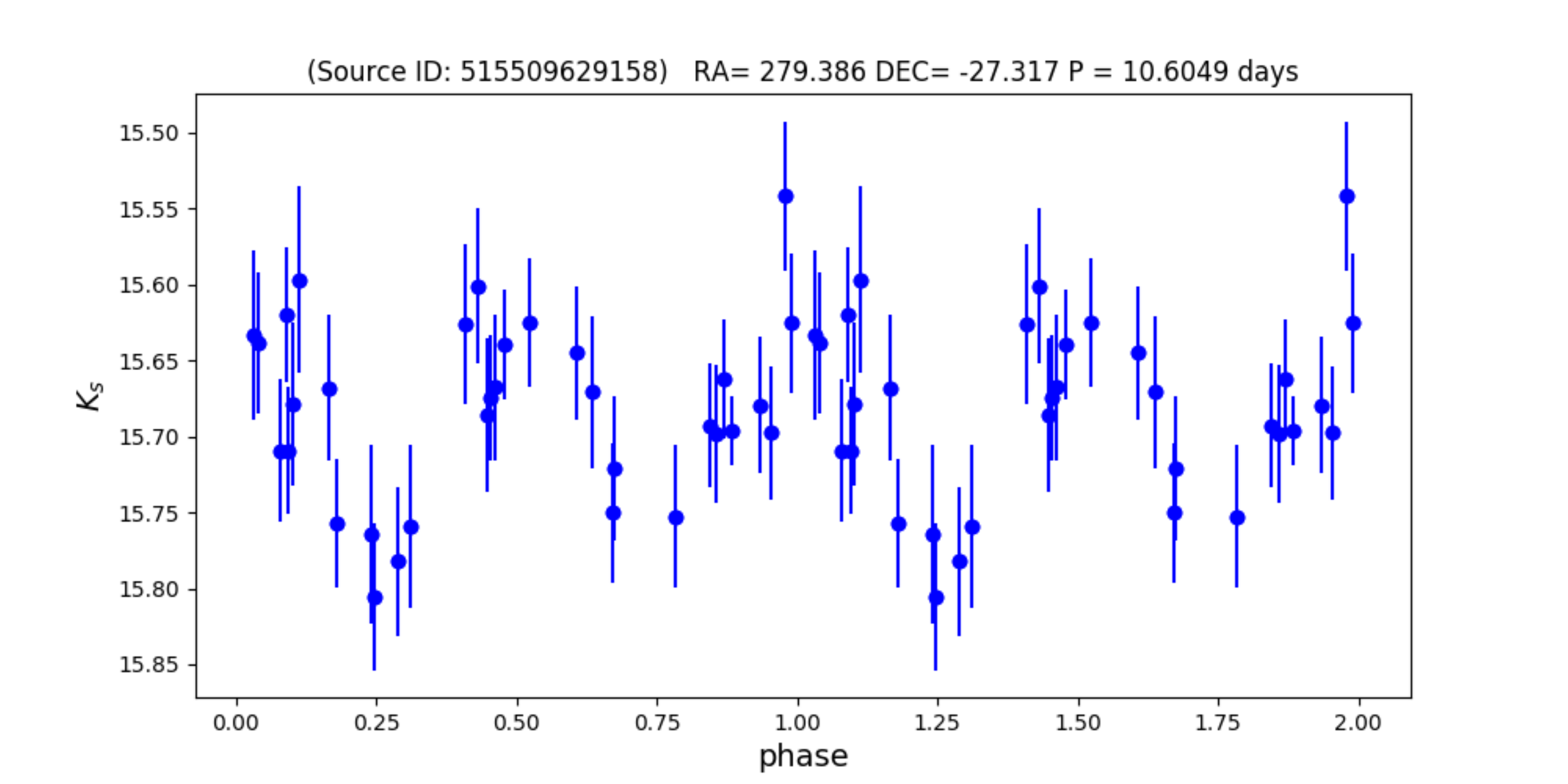} &
    \includegraphics[width=.33\textwidth]{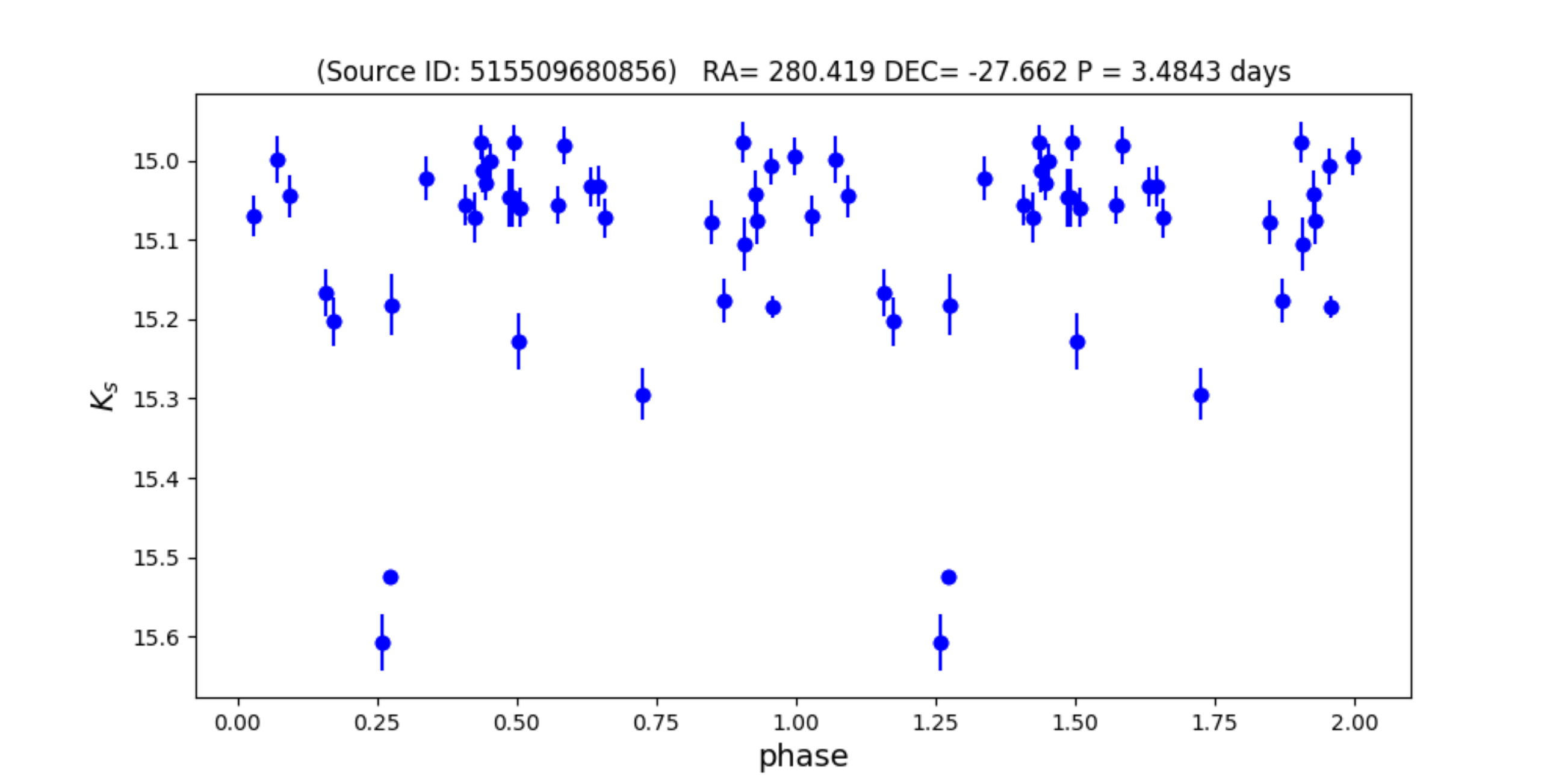} &
    \includegraphics[width=.33\textwidth]{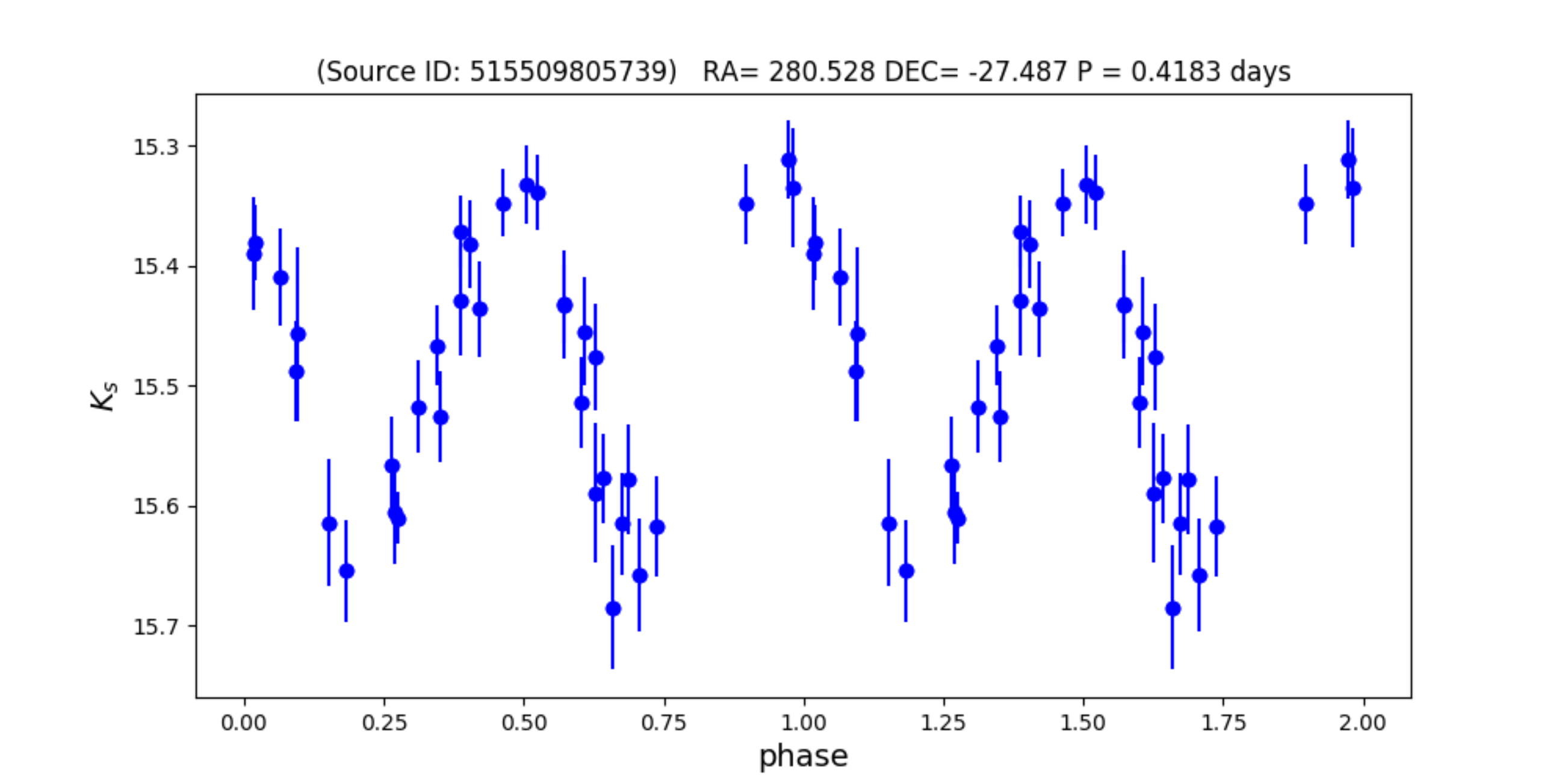} \\
    \includegraphics[width=.33\textwidth]{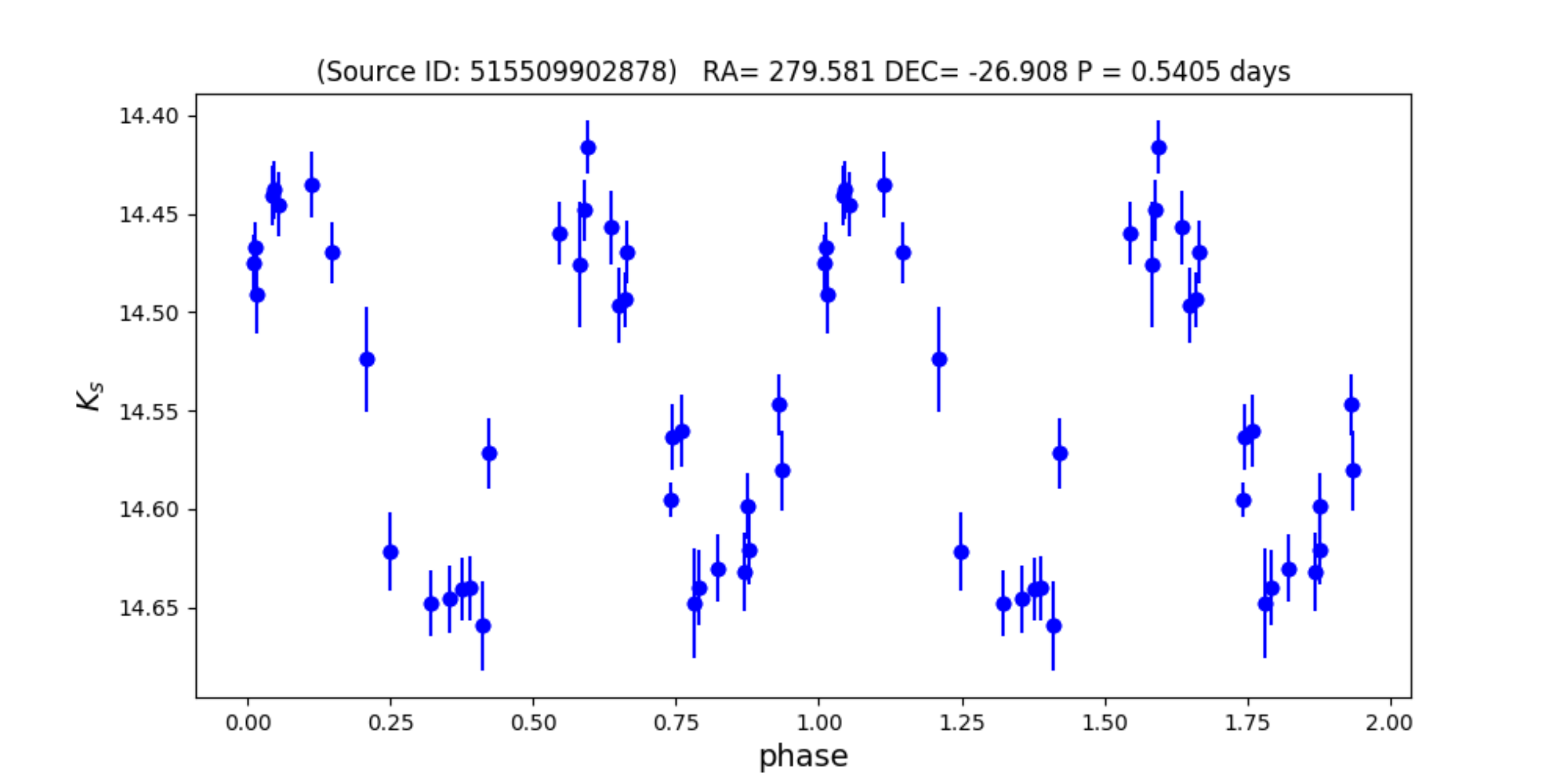} &
    \includegraphics[width=.33\textwidth]{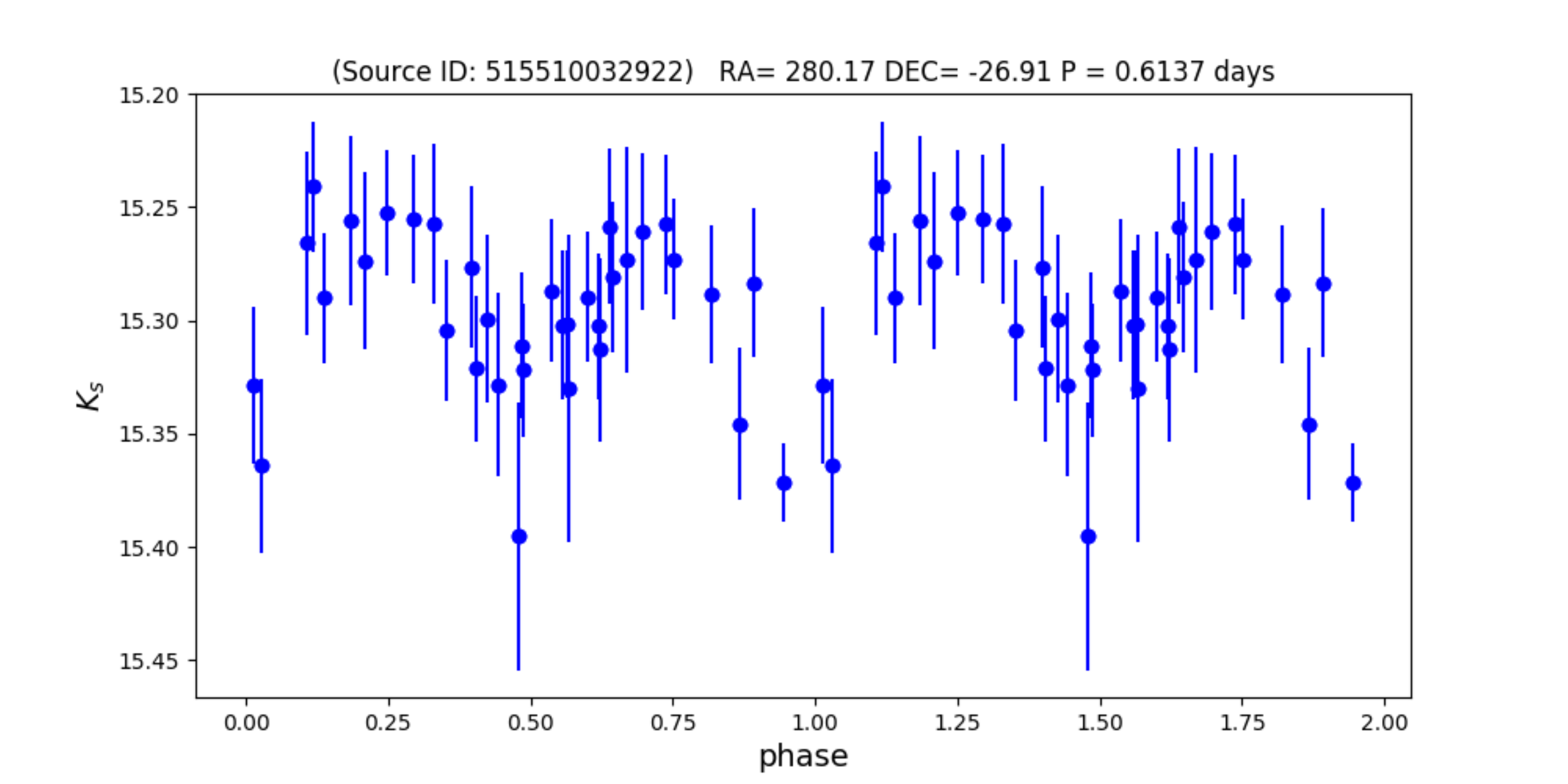} &
    \includegraphics[width=.33\textwidth]{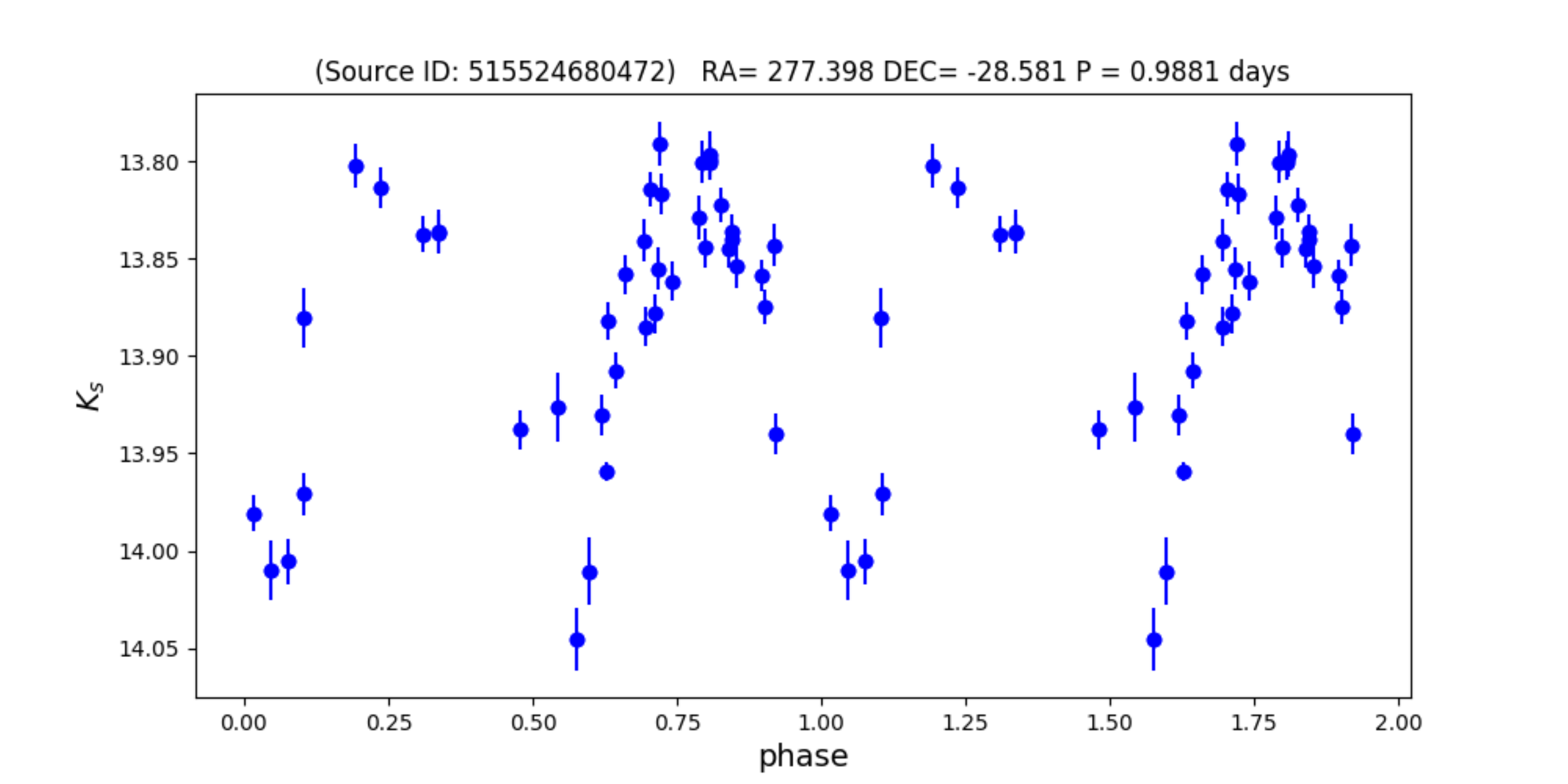} \\
  \end{tabular}
  \caption{BHB stars light curves corresponding to $1^{st}$ category eclipsing binaries.}
  \label{fig:bhb2}
\end{figure*}

\begin{figure*}[htb]
\centering
  \begin{tabular}{@{}ccc@{}}    
    \includegraphics[width=.33\textwidth]{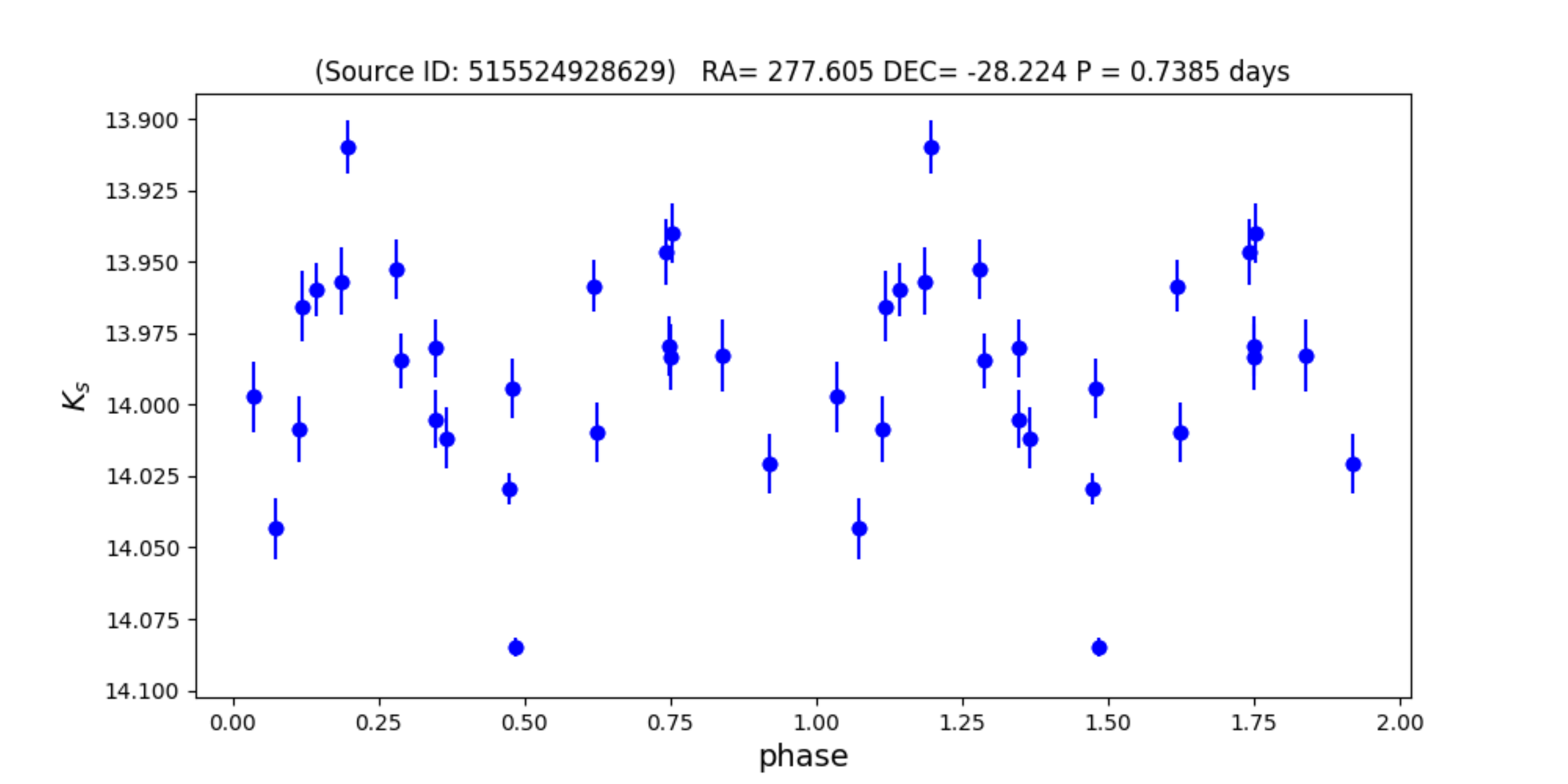} &
    \includegraphics[width=.33\textwidth]{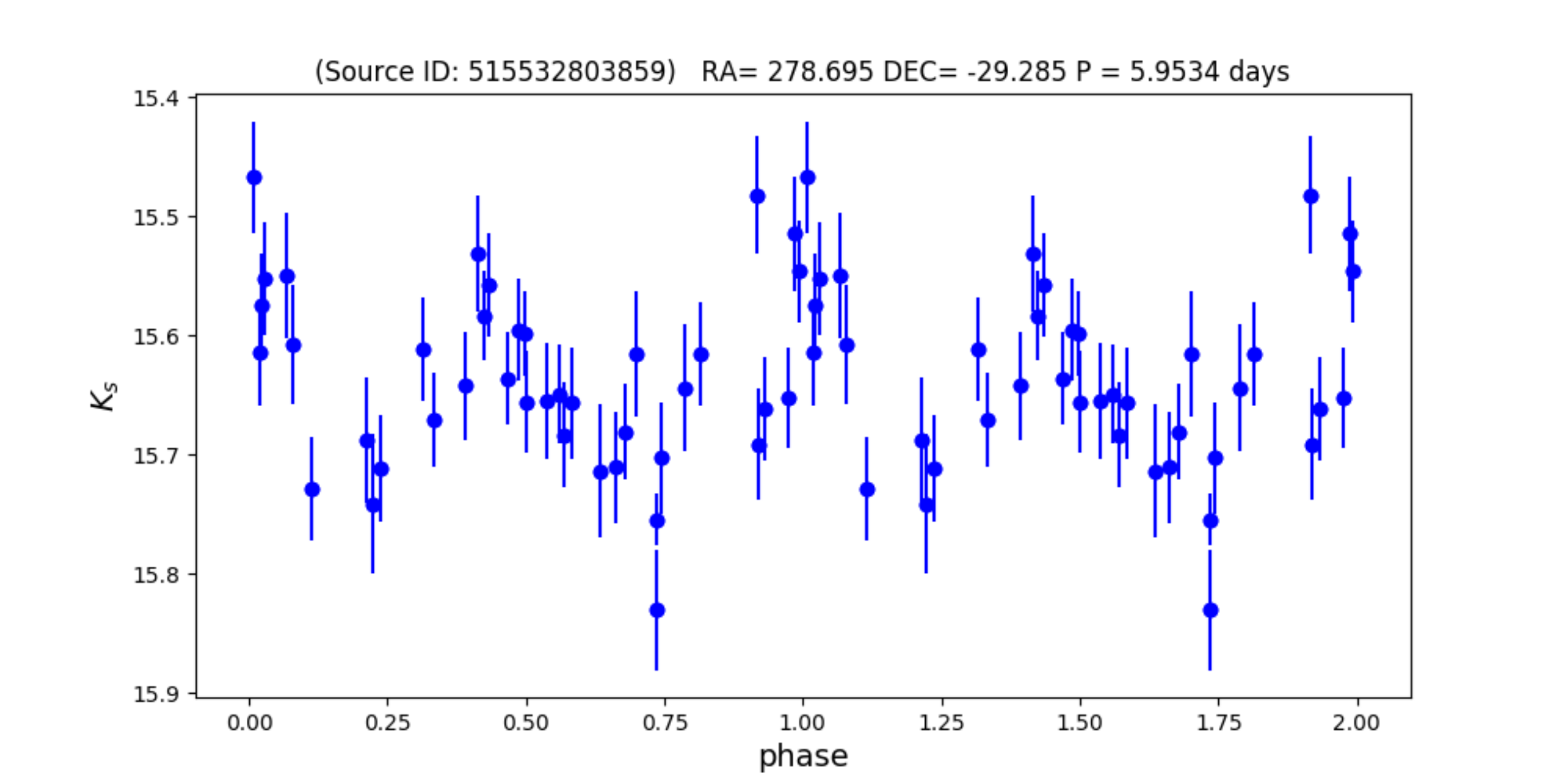} &
    \includegraphics[width=.33\textwidth]{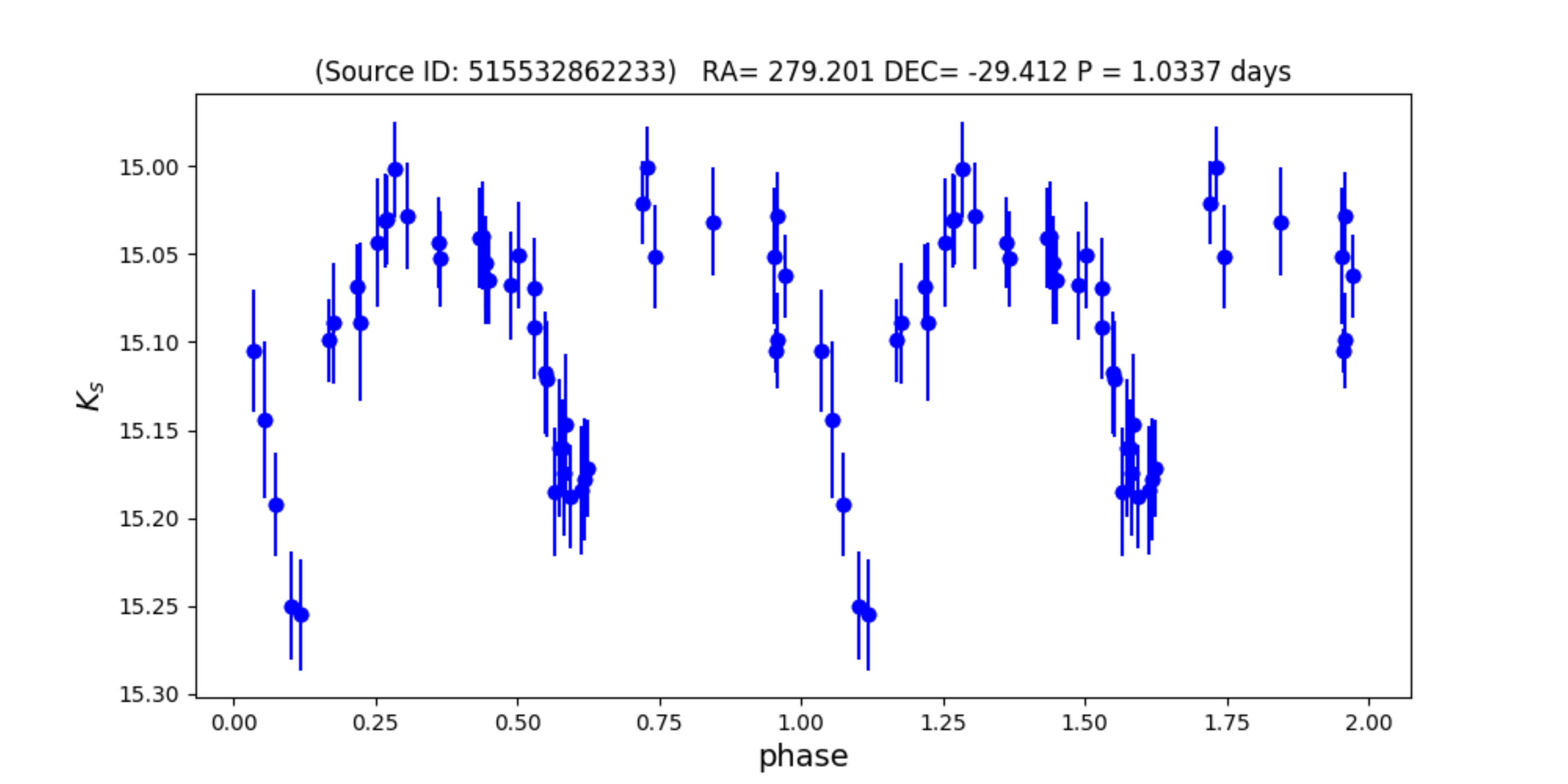} \\
    \includegraphics[width=.33\textwidth]{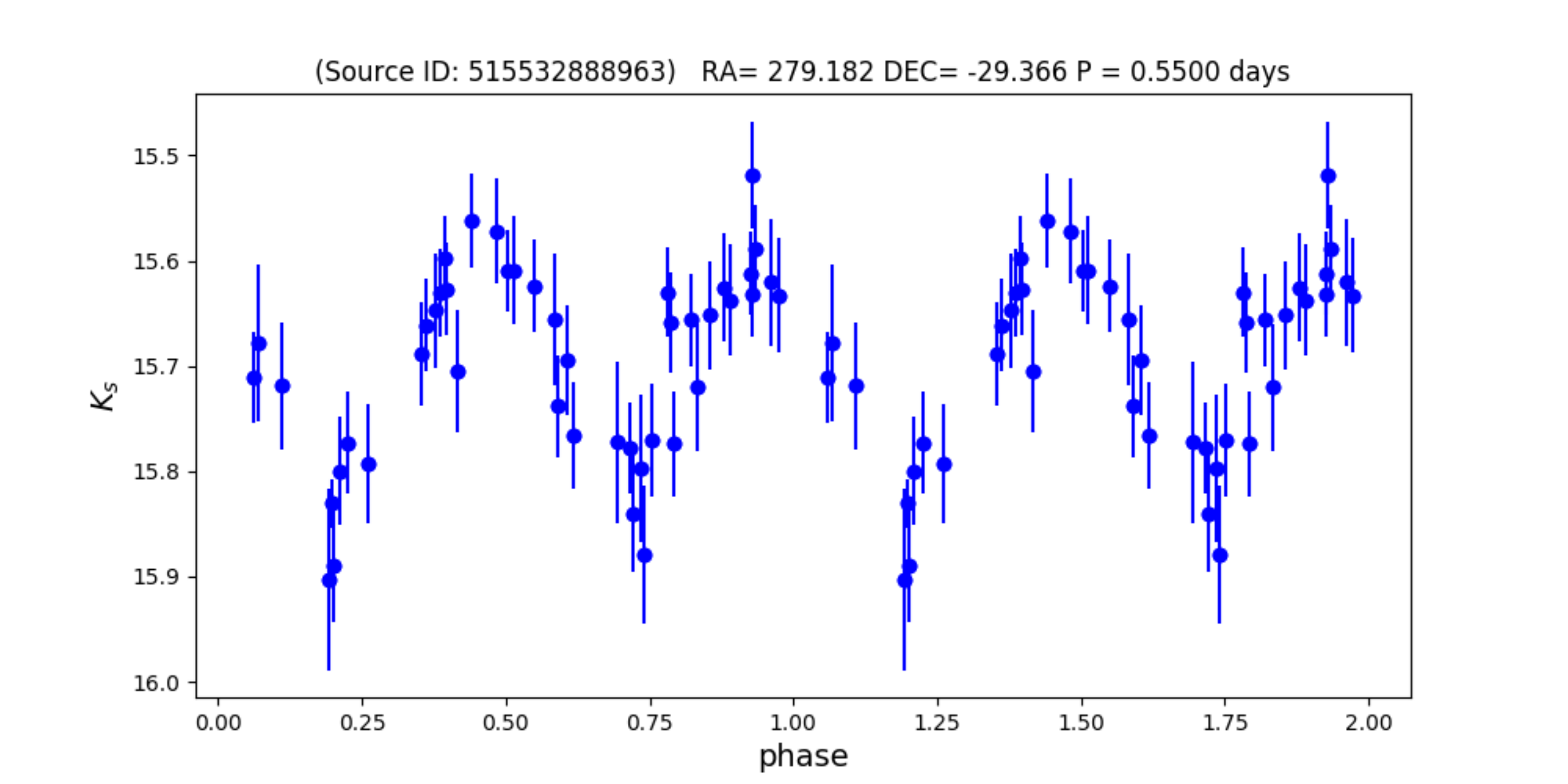} &
    \includegraphics[width=.33\textwidth]{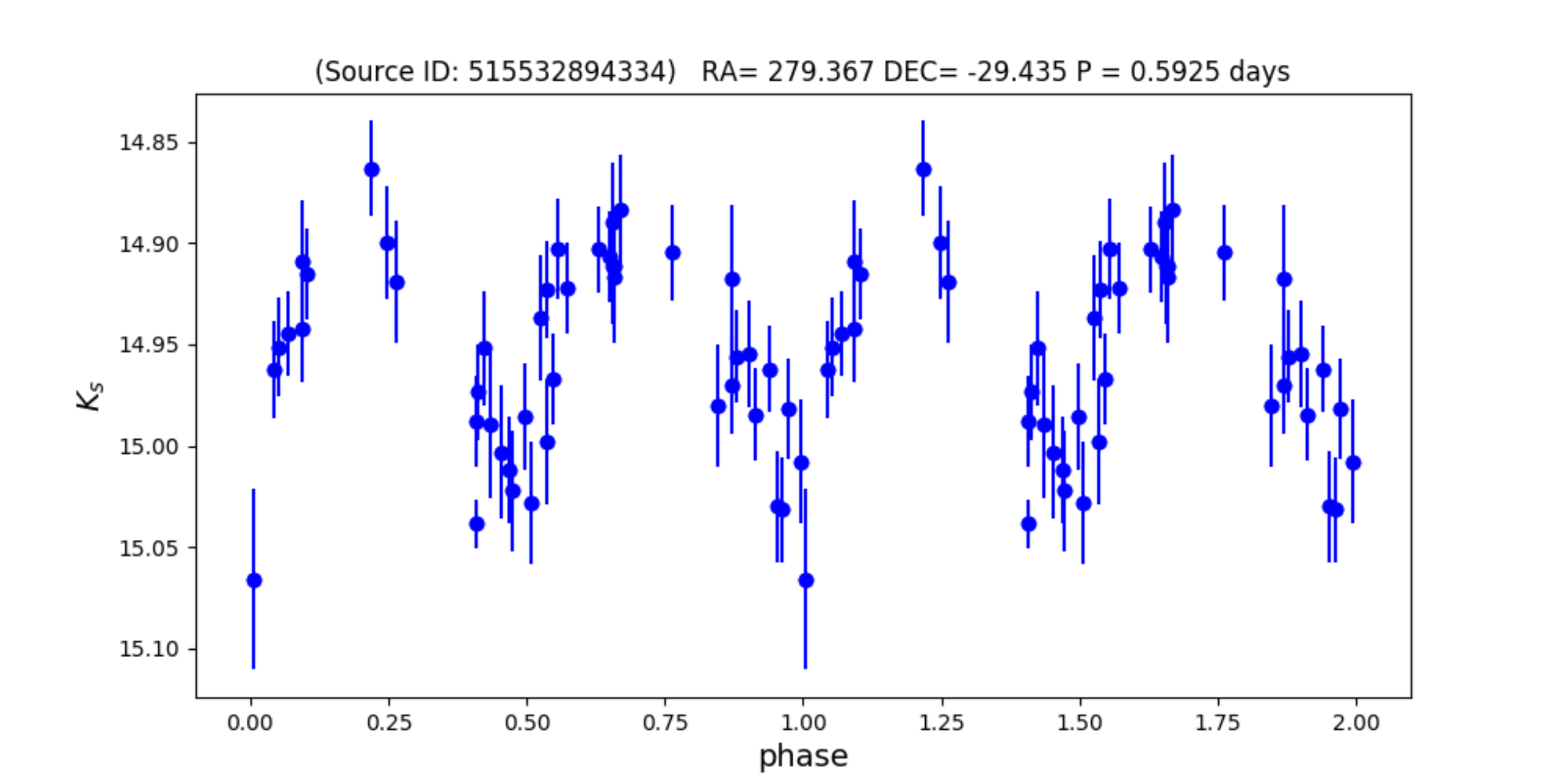} &
    \includegraphics[width=.33\textwidth]{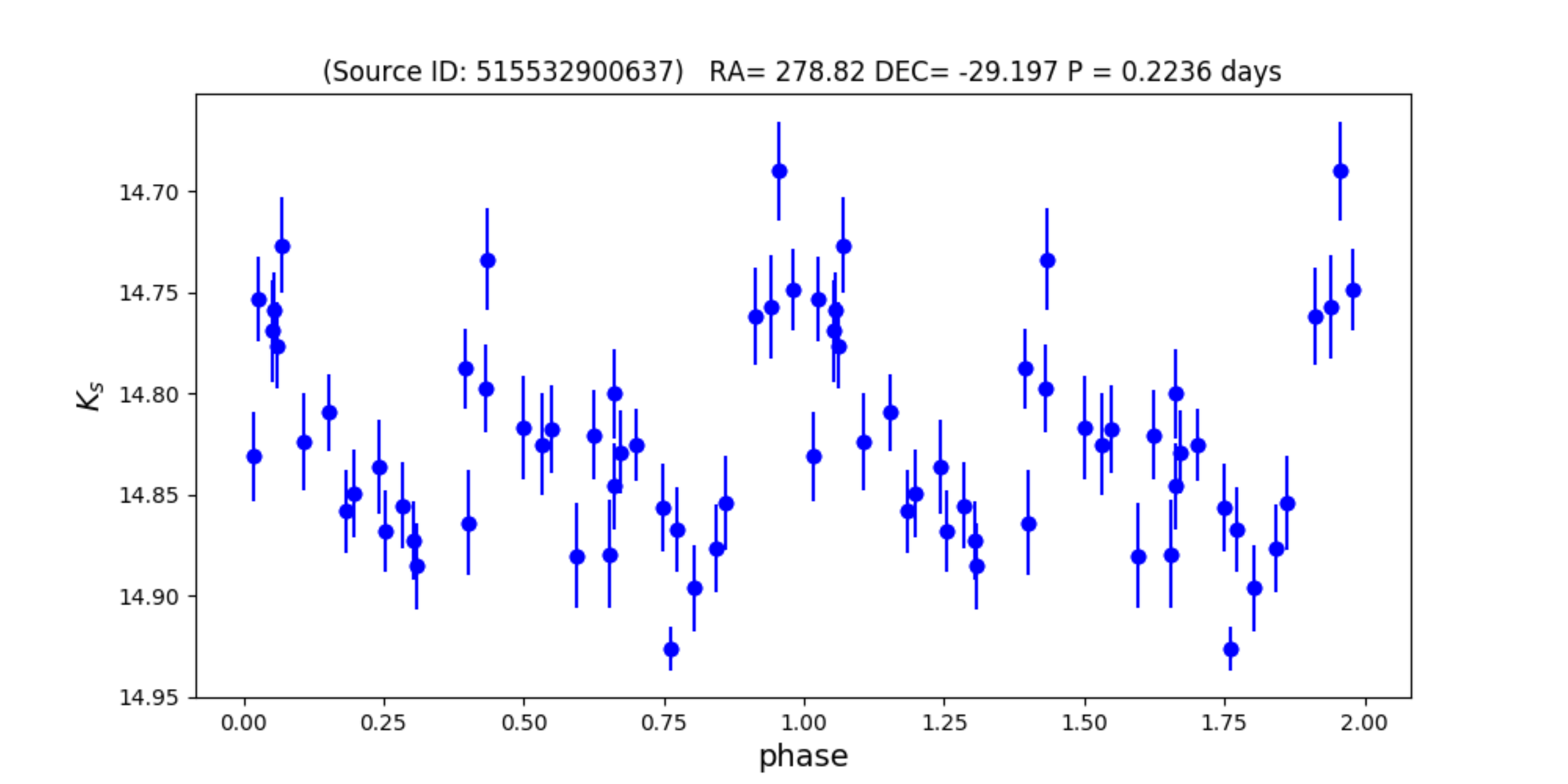} \\
    \includegraphics[width=.33\textwidth]{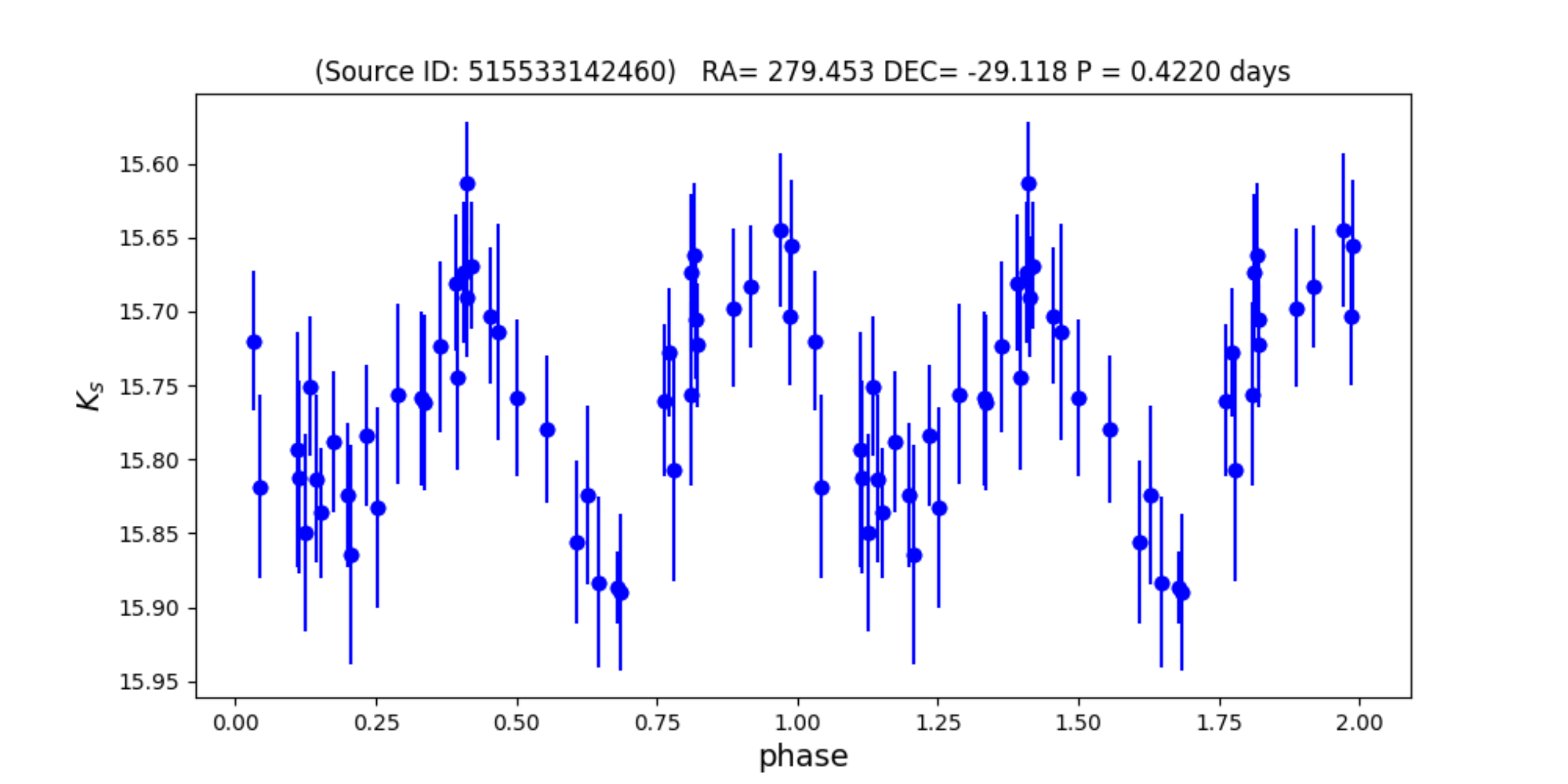} &
    \includegraphics[width=.33\textwidth]{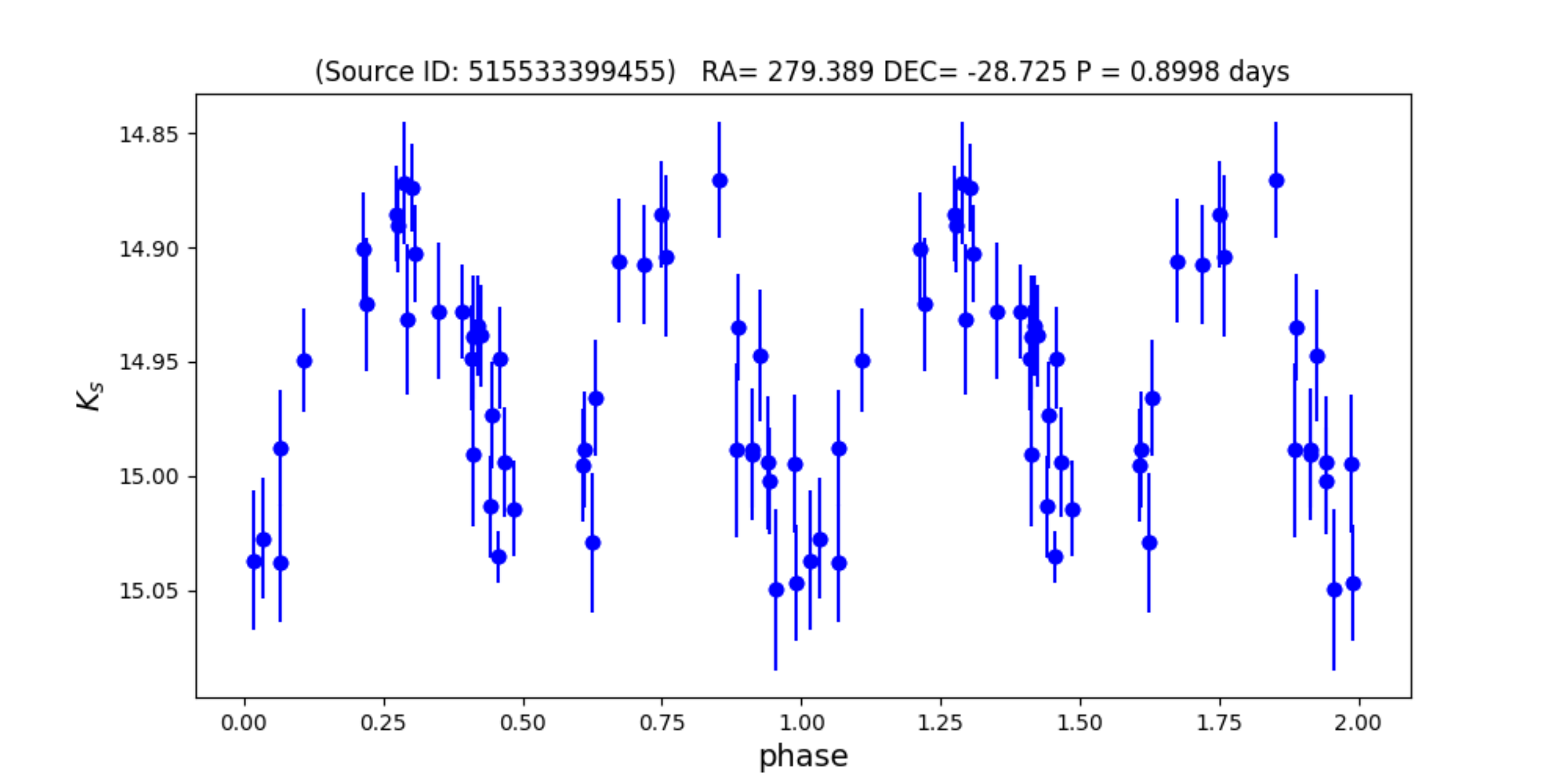} &
    \includegraphics[width=.33\textwidth]{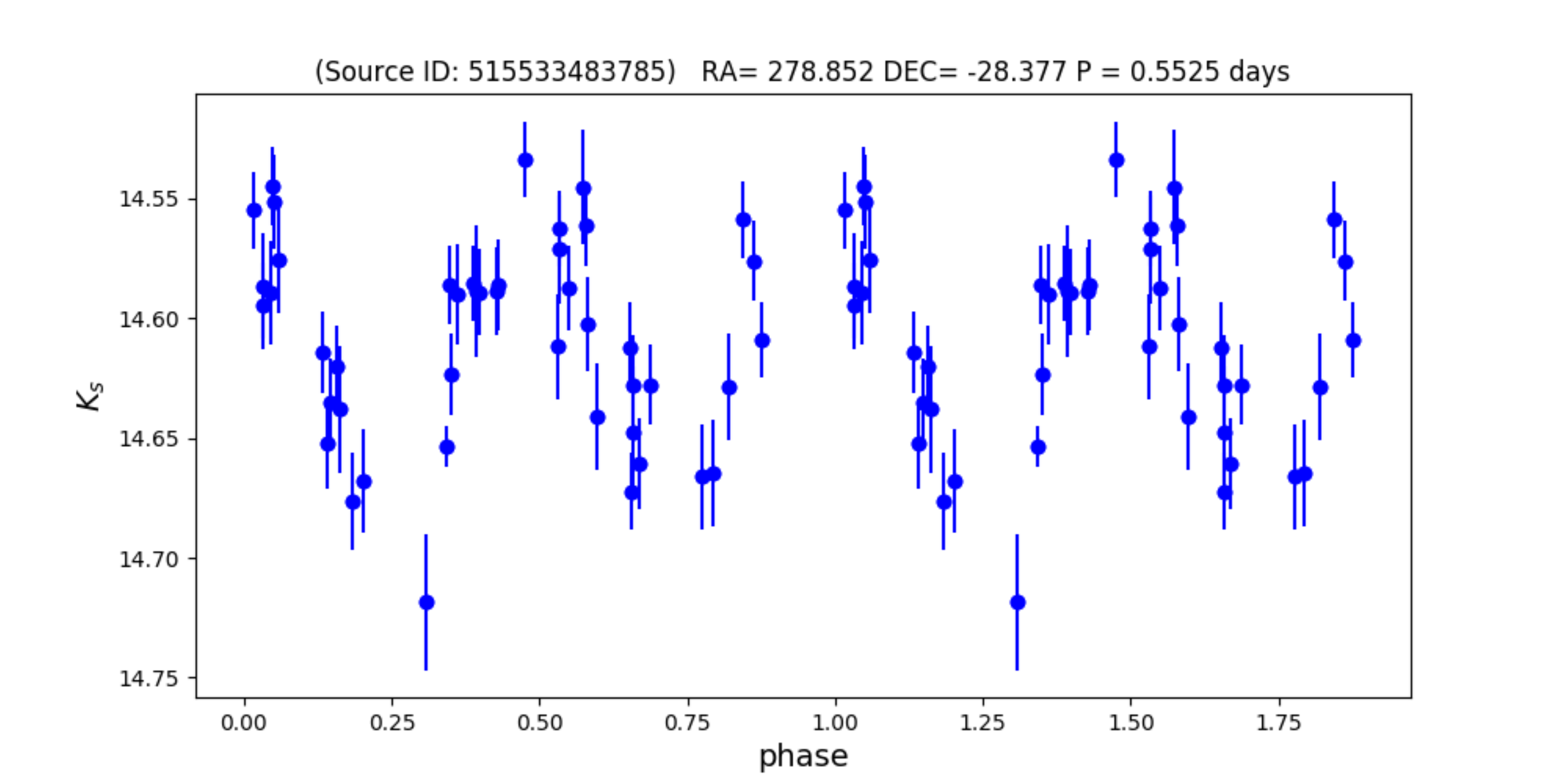} \\    
    \includegraphics[width=.33\textwidth]{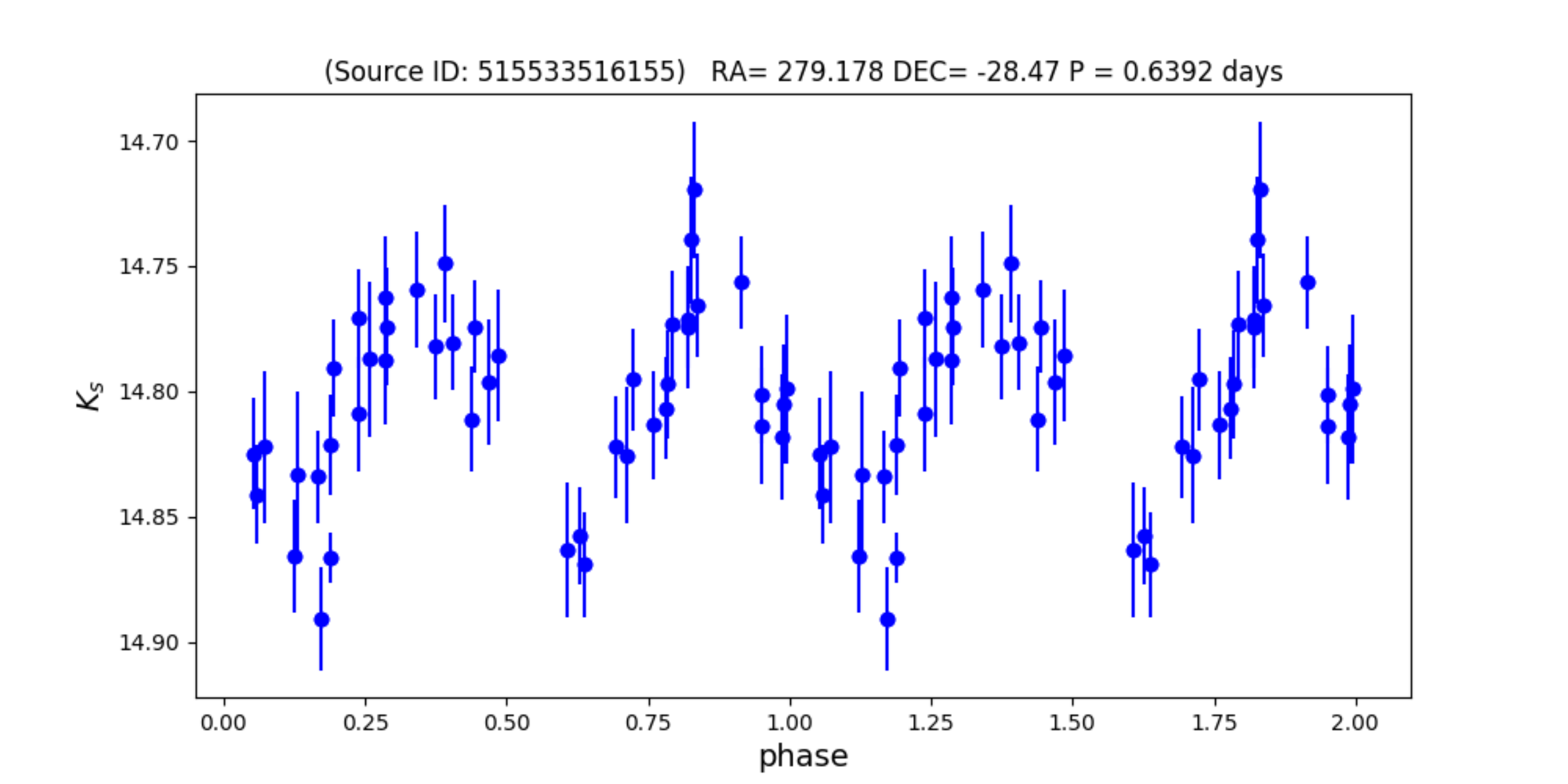} &
    \includegraphics[width=.33\textwidth]{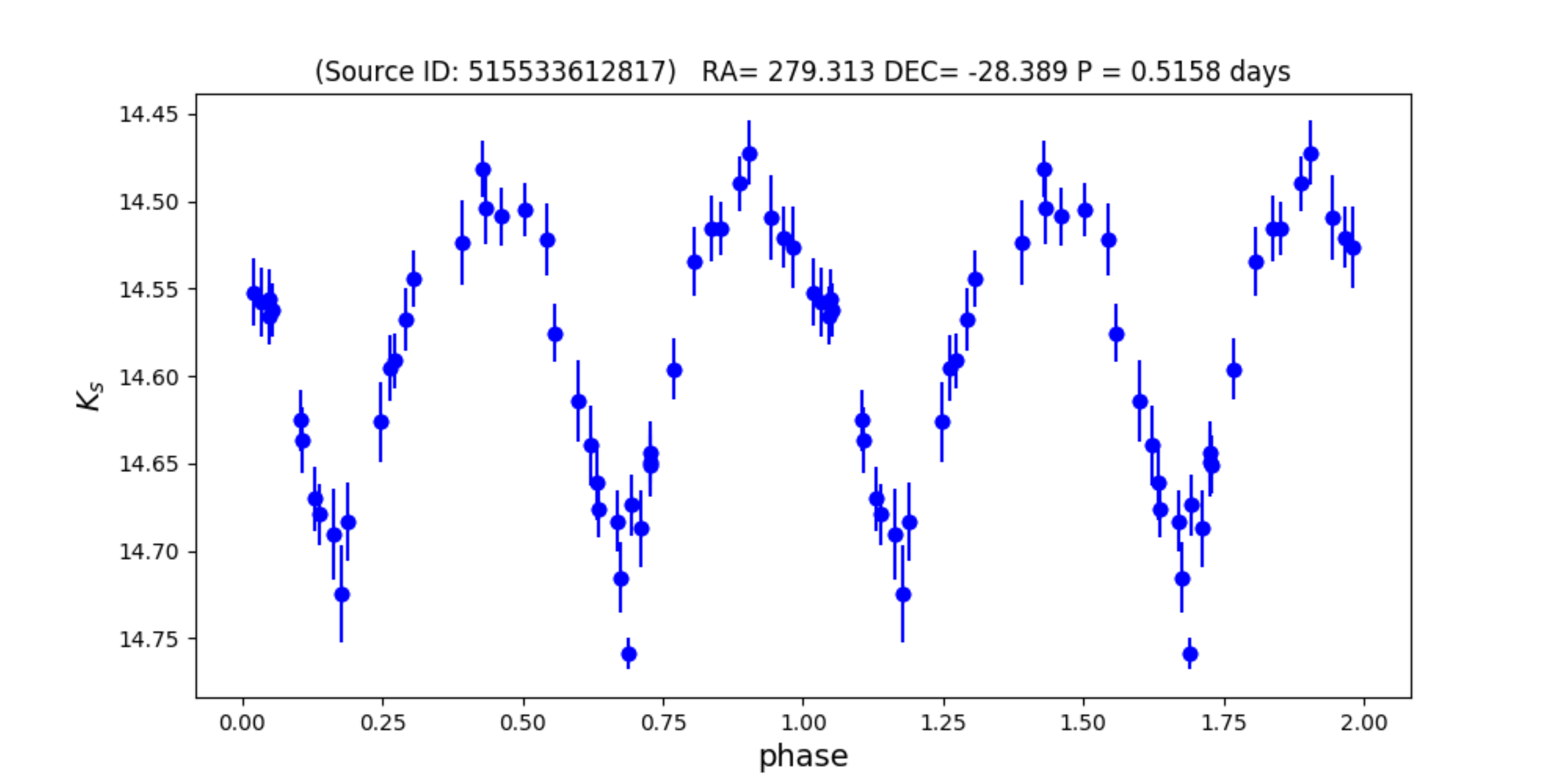} &
    \includegraphics[width=.33\textwidth]{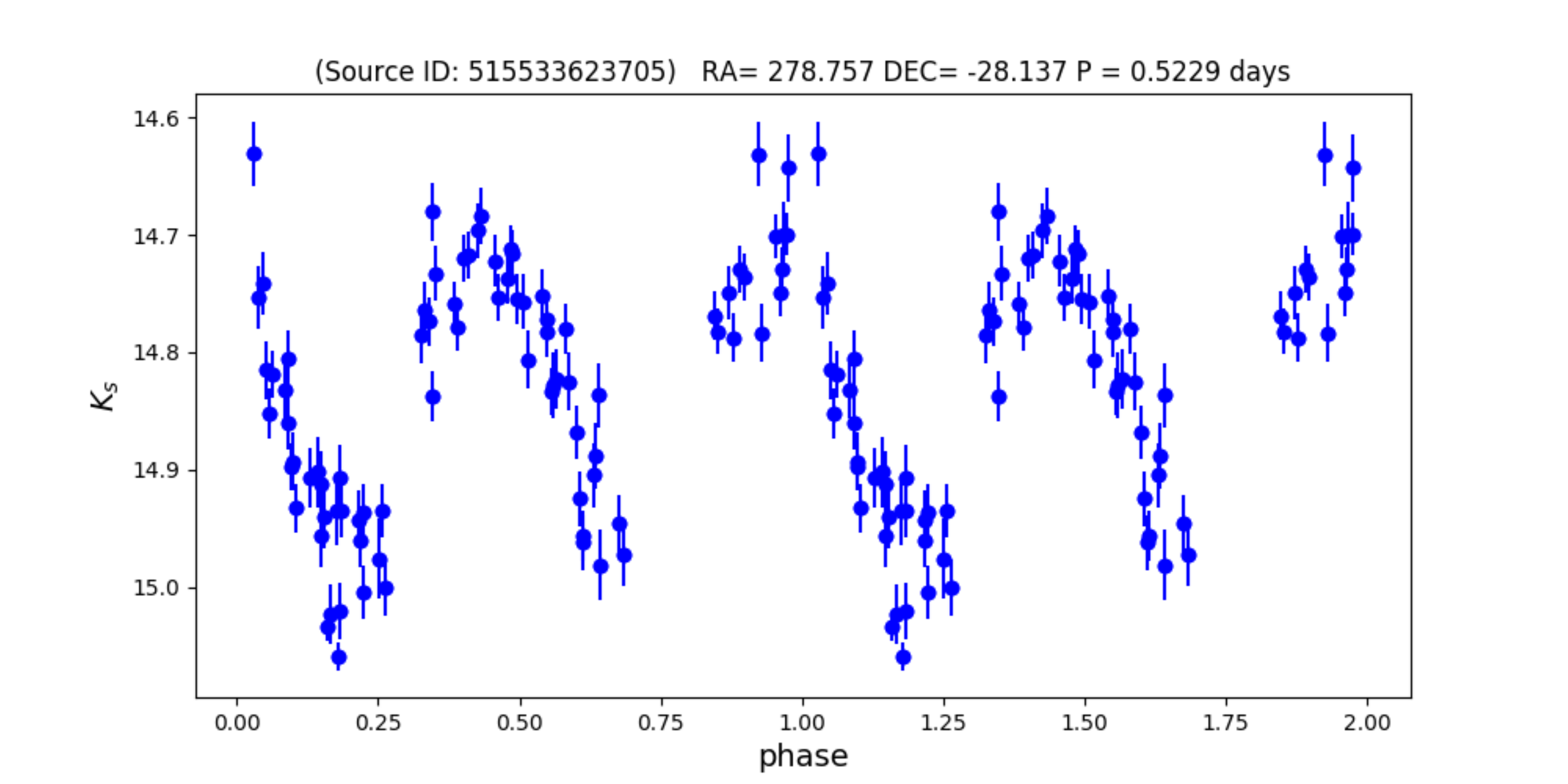} \\
    \includegraphics[width=.33\textwidth]{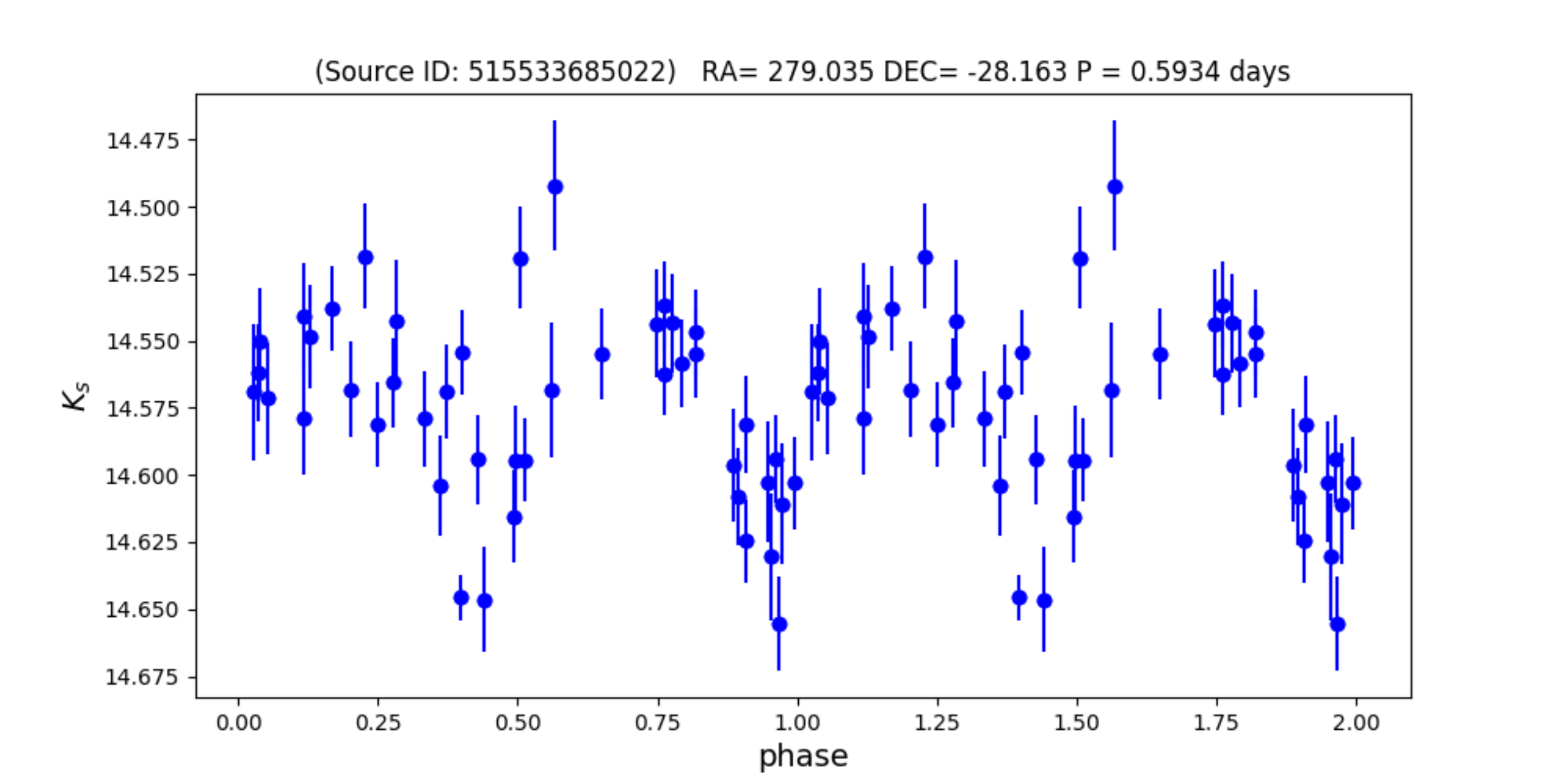} &
    \includegraphics[width=.33\textwidth]{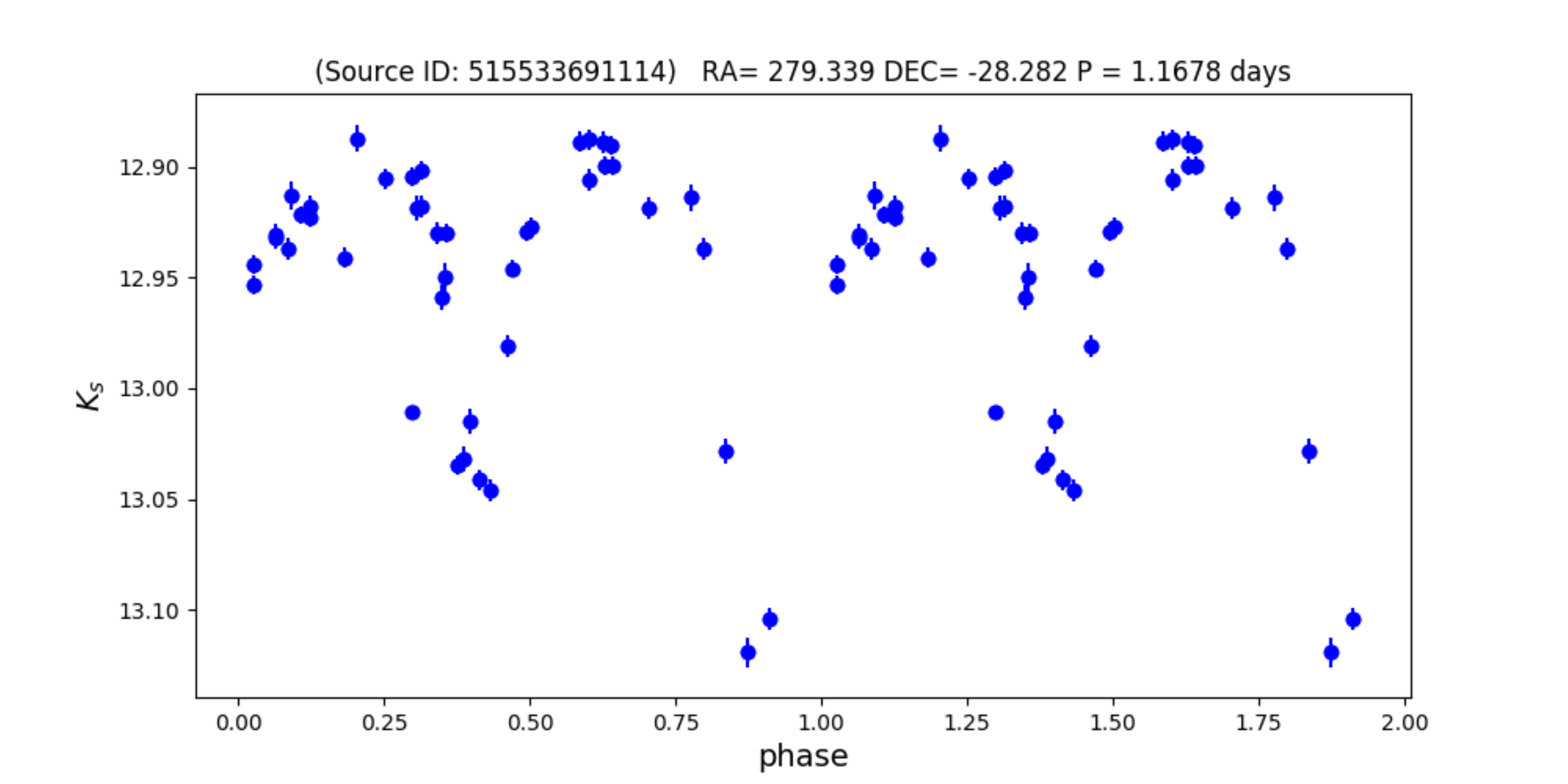} &
    \includegraphics[width=.33\textwidth]{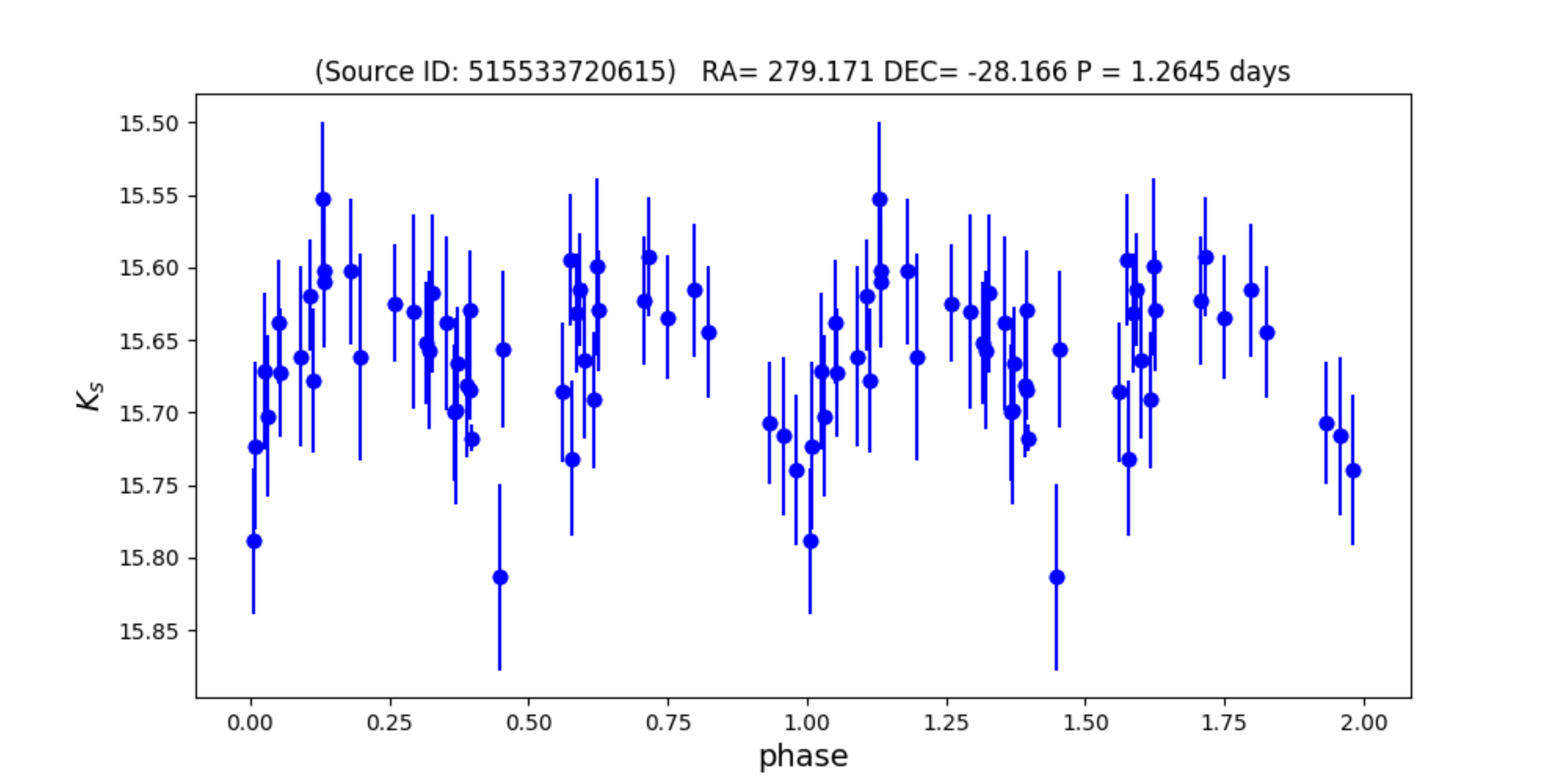} \\ 
    \includegraphics[width=.33\textwidth]{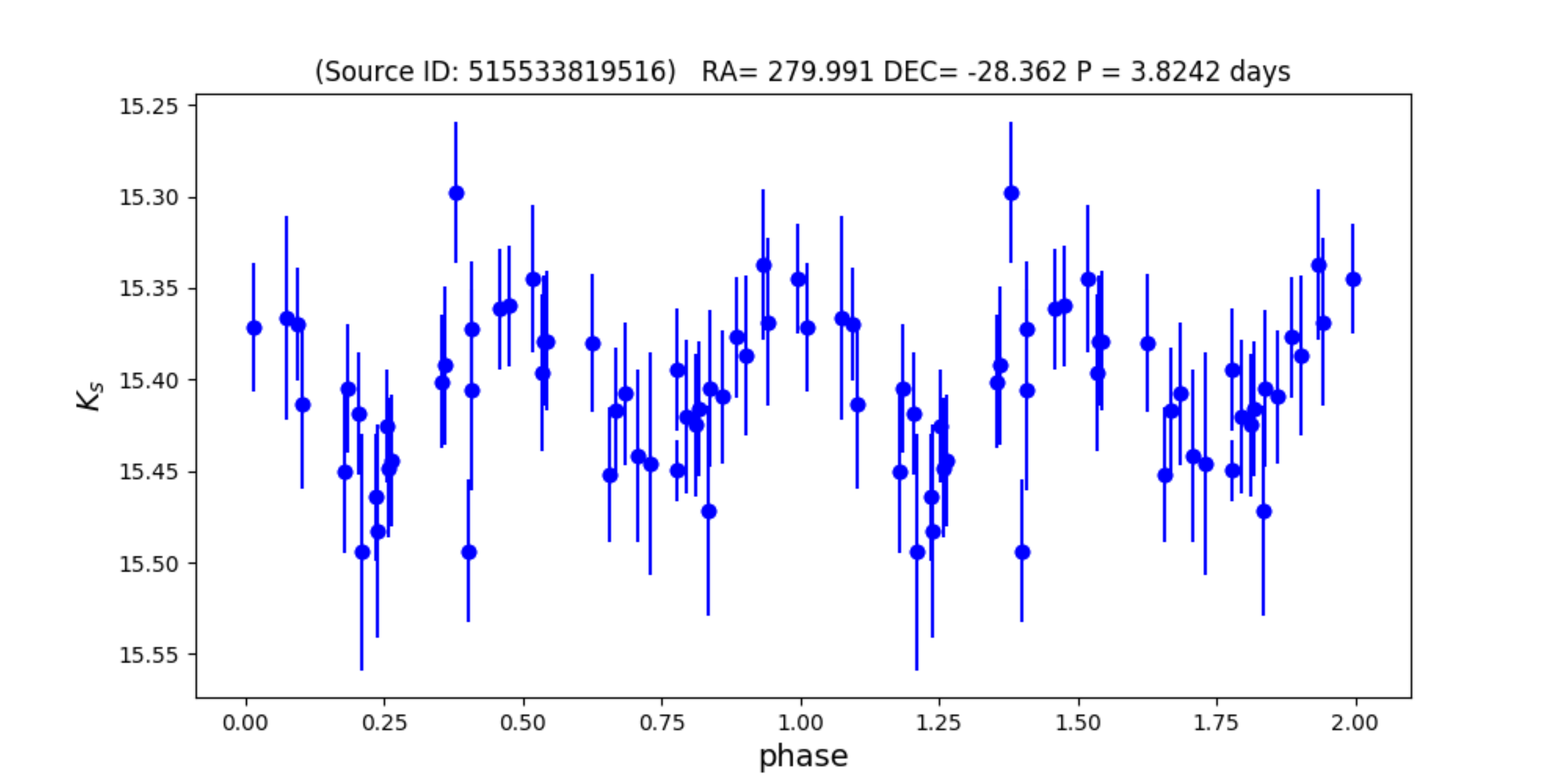} &
    \includegraphics[width=.33\textwidth]{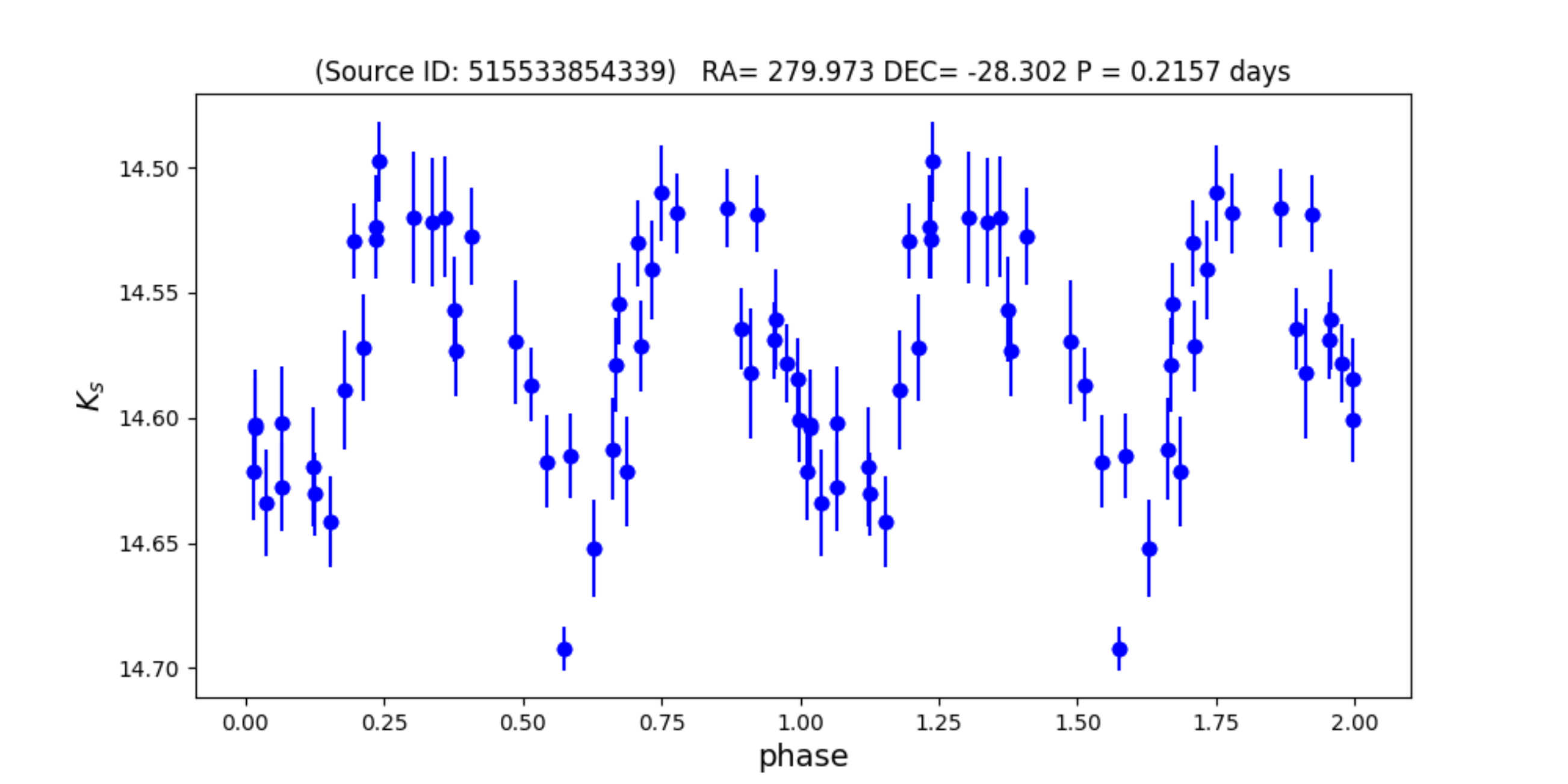} &
    \includegraphics[width=.33\textwidth]{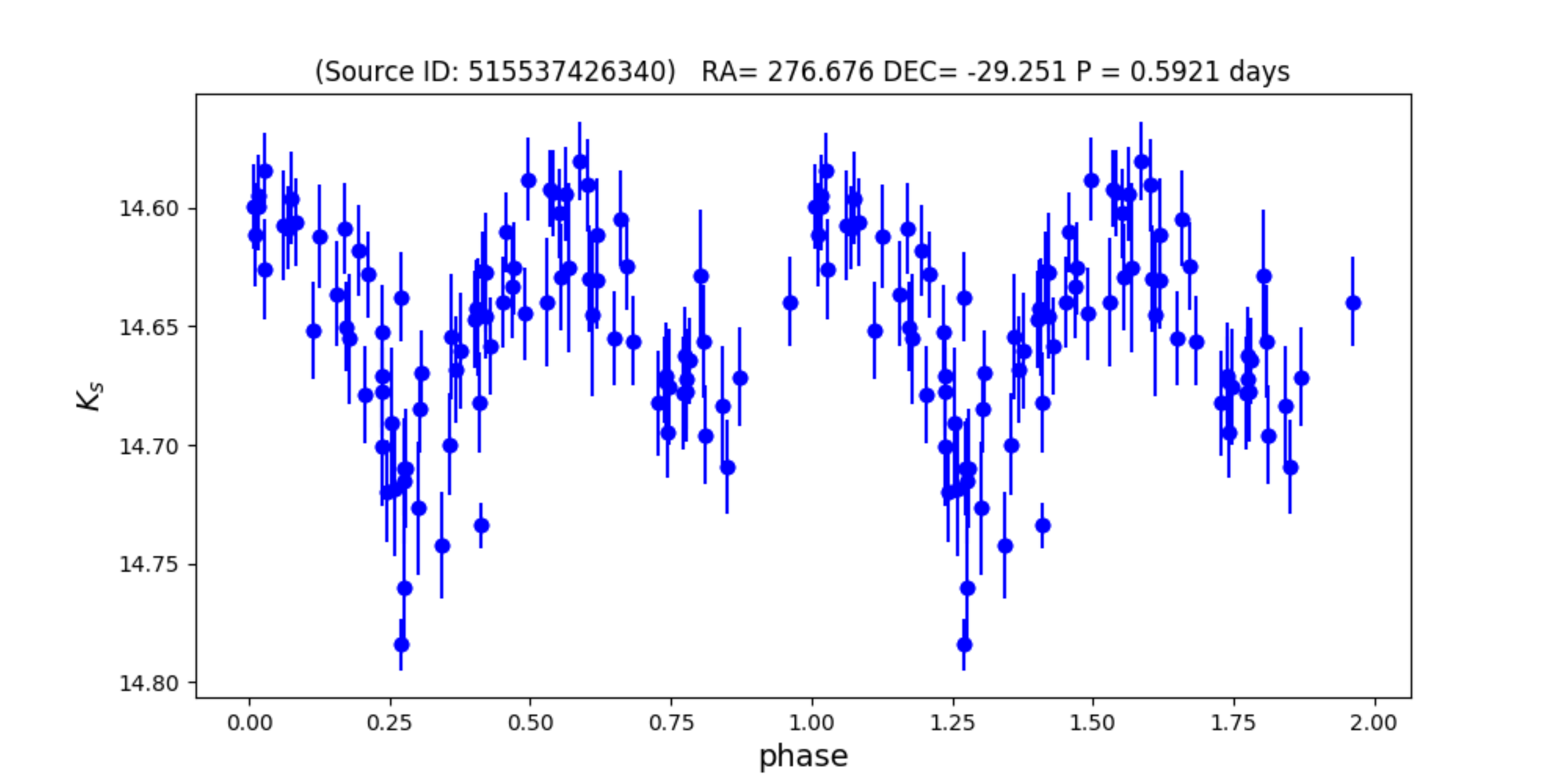} \\
    \includegraphics[width=.33\textwidth]{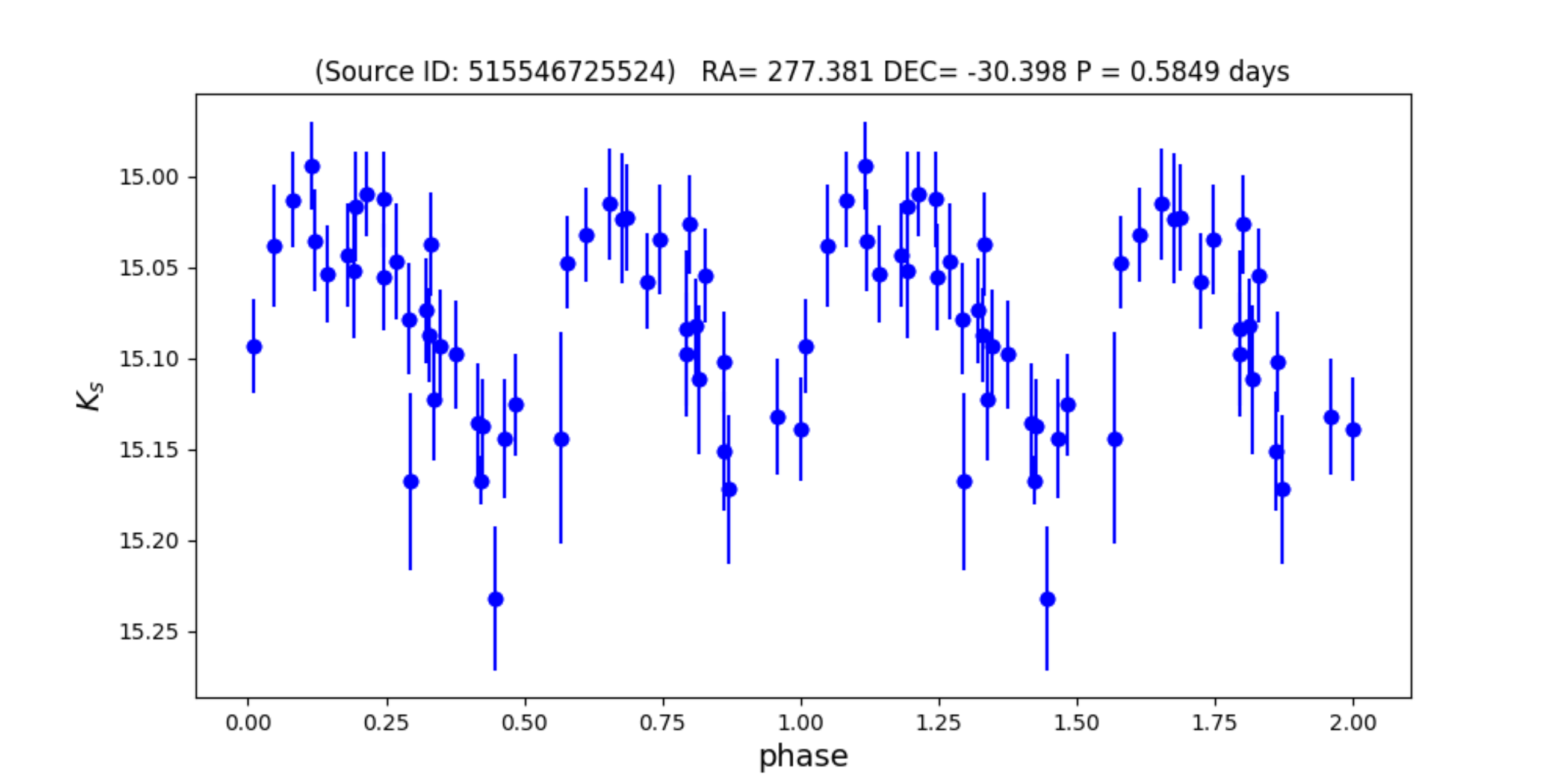} &
    \includegraphics[width=.33\textwidth]{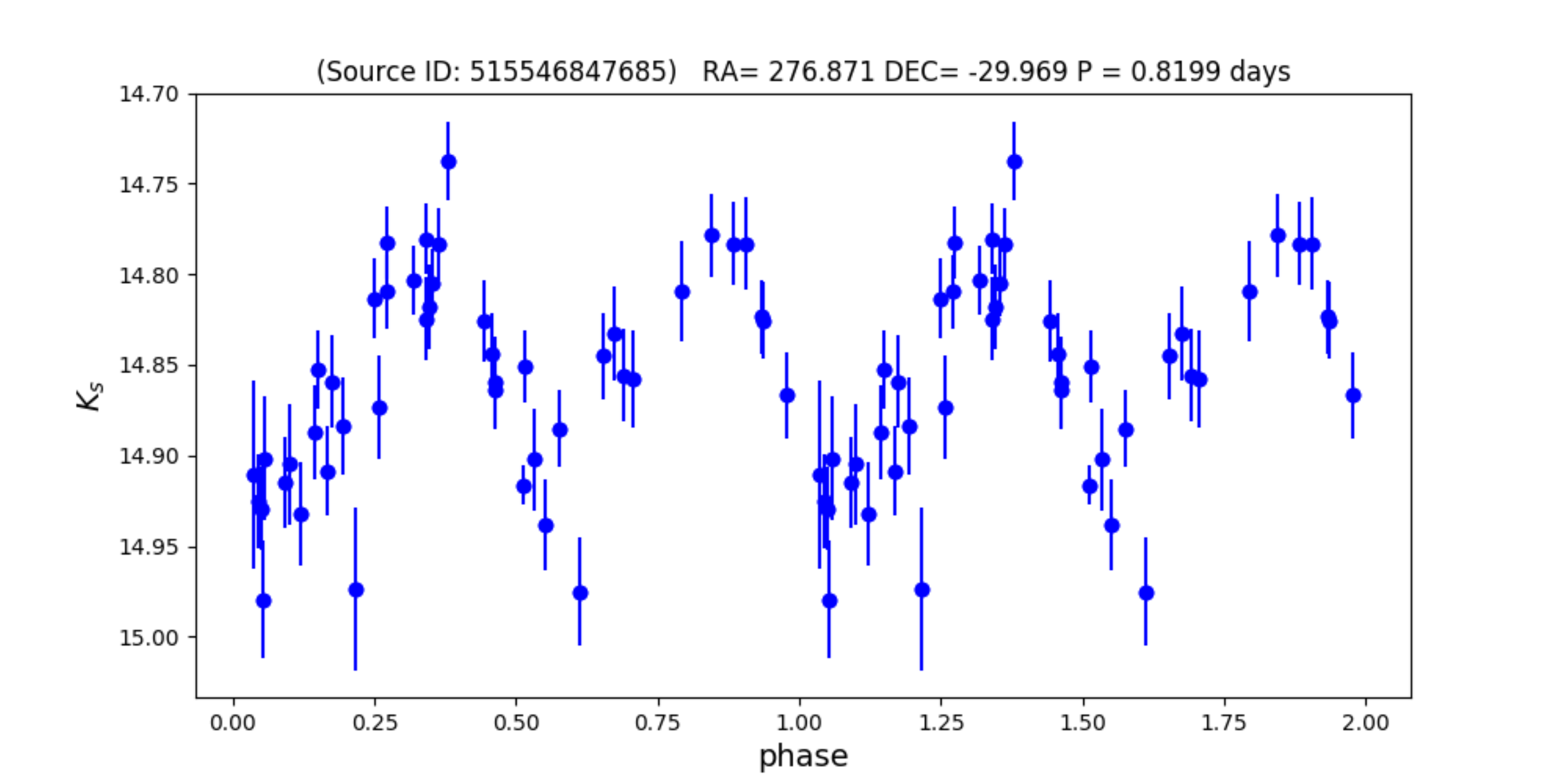} &
    \includegraphics[width=.33\textwidth]{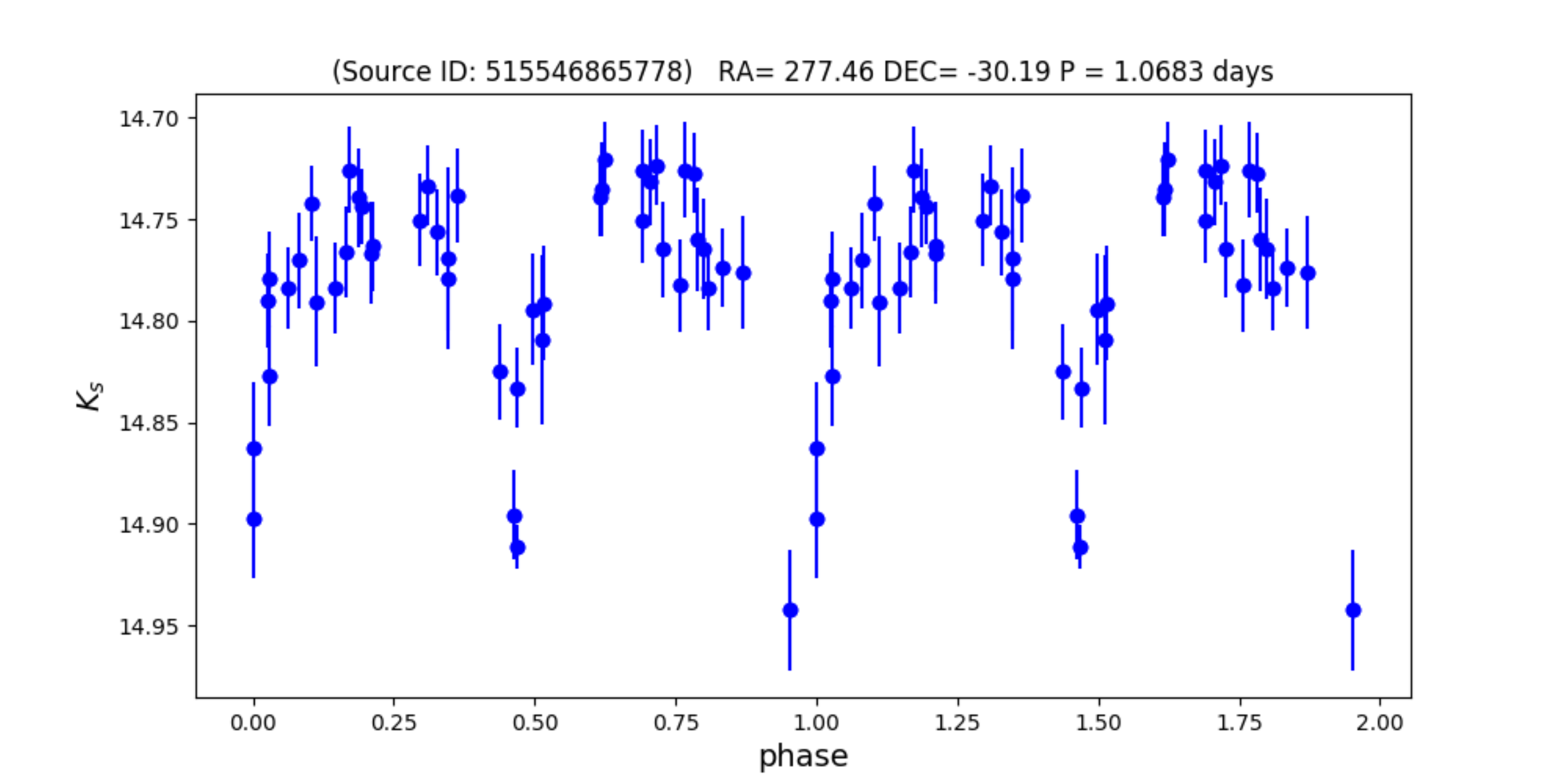} \\  
    \includegraphics[width=.33\textwidth]{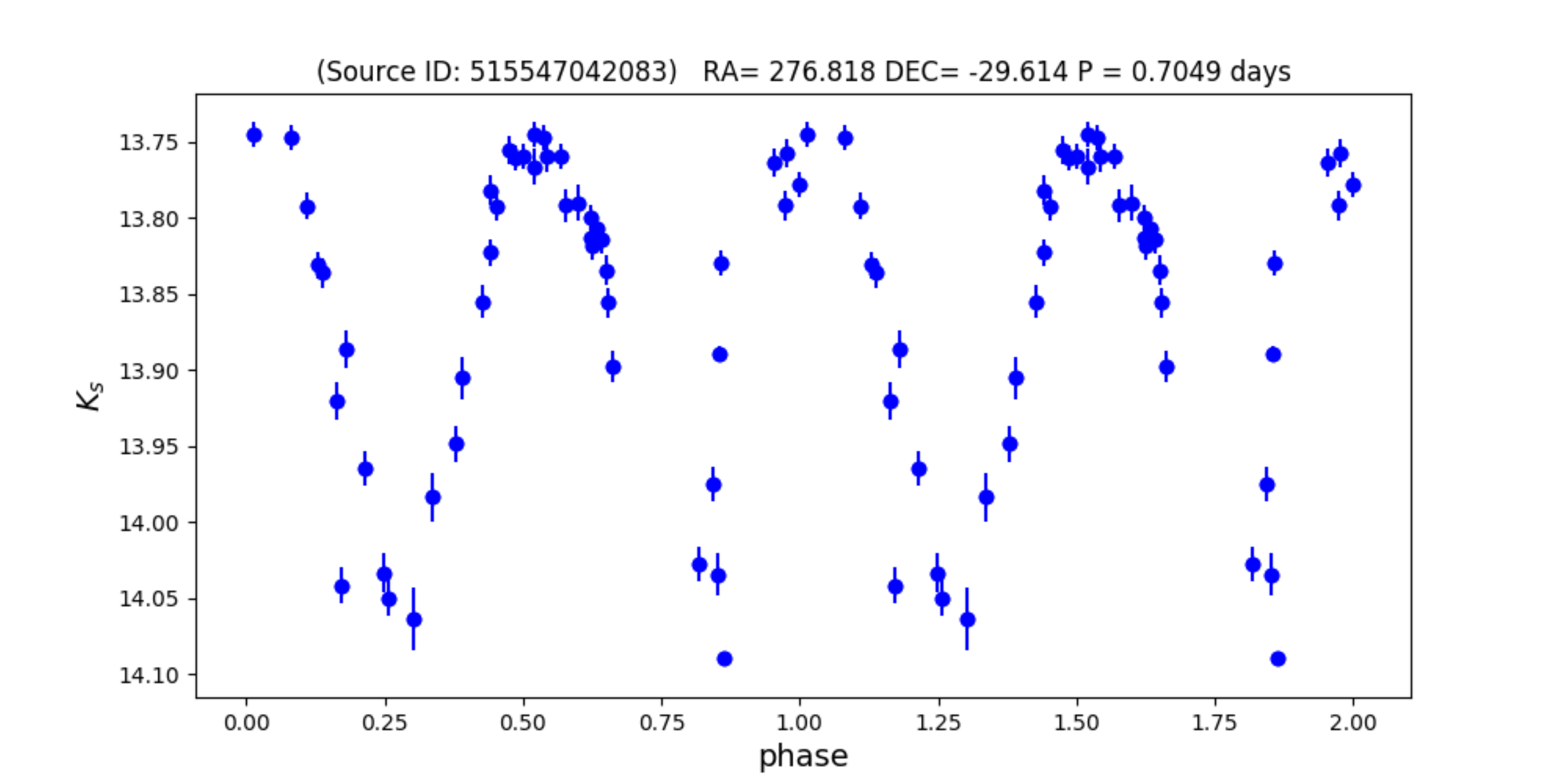} &  
    \includegraphics[width=.33\textwidth]{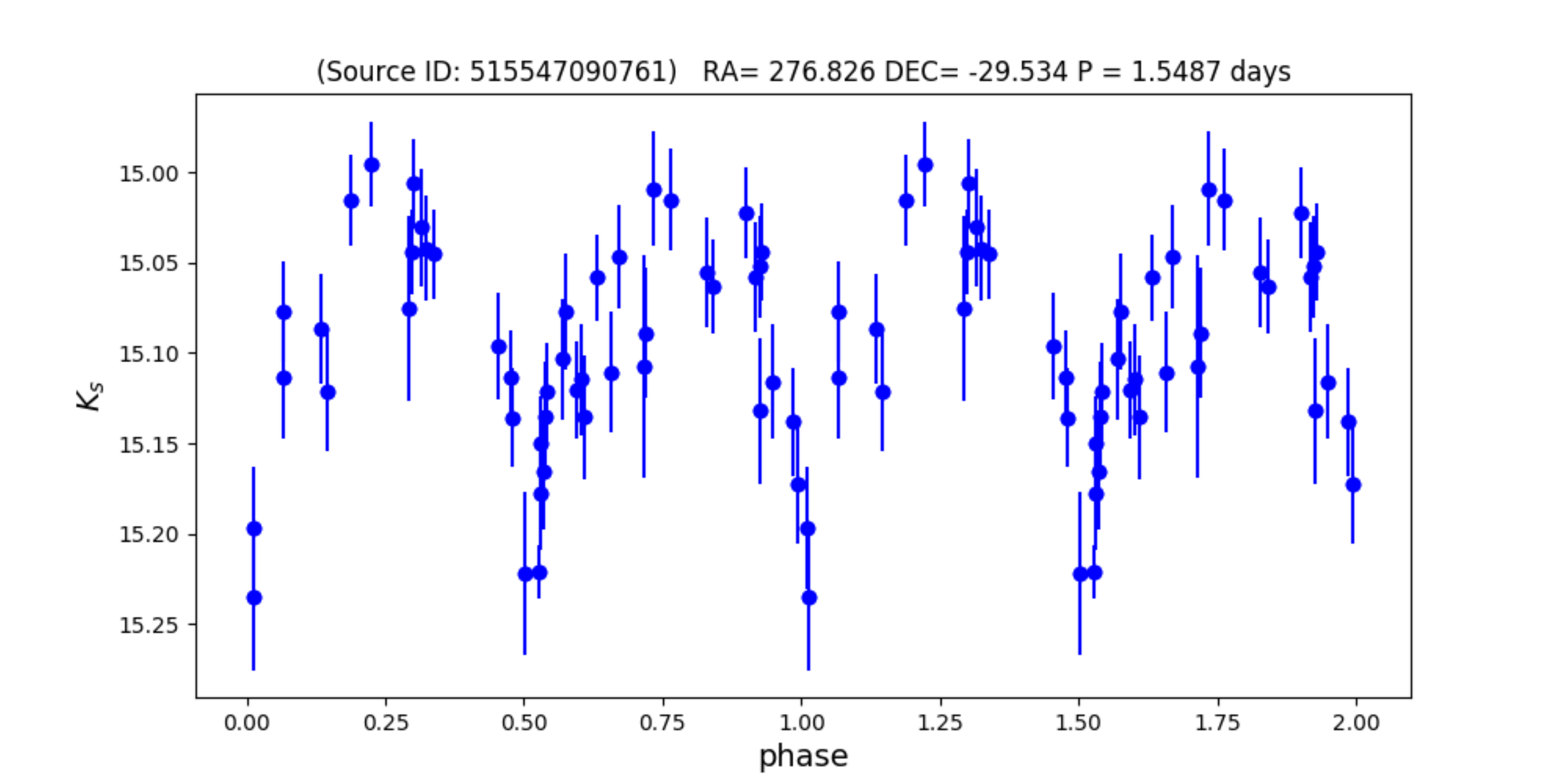} &
    \includegraphics[width=.33\textwidth]{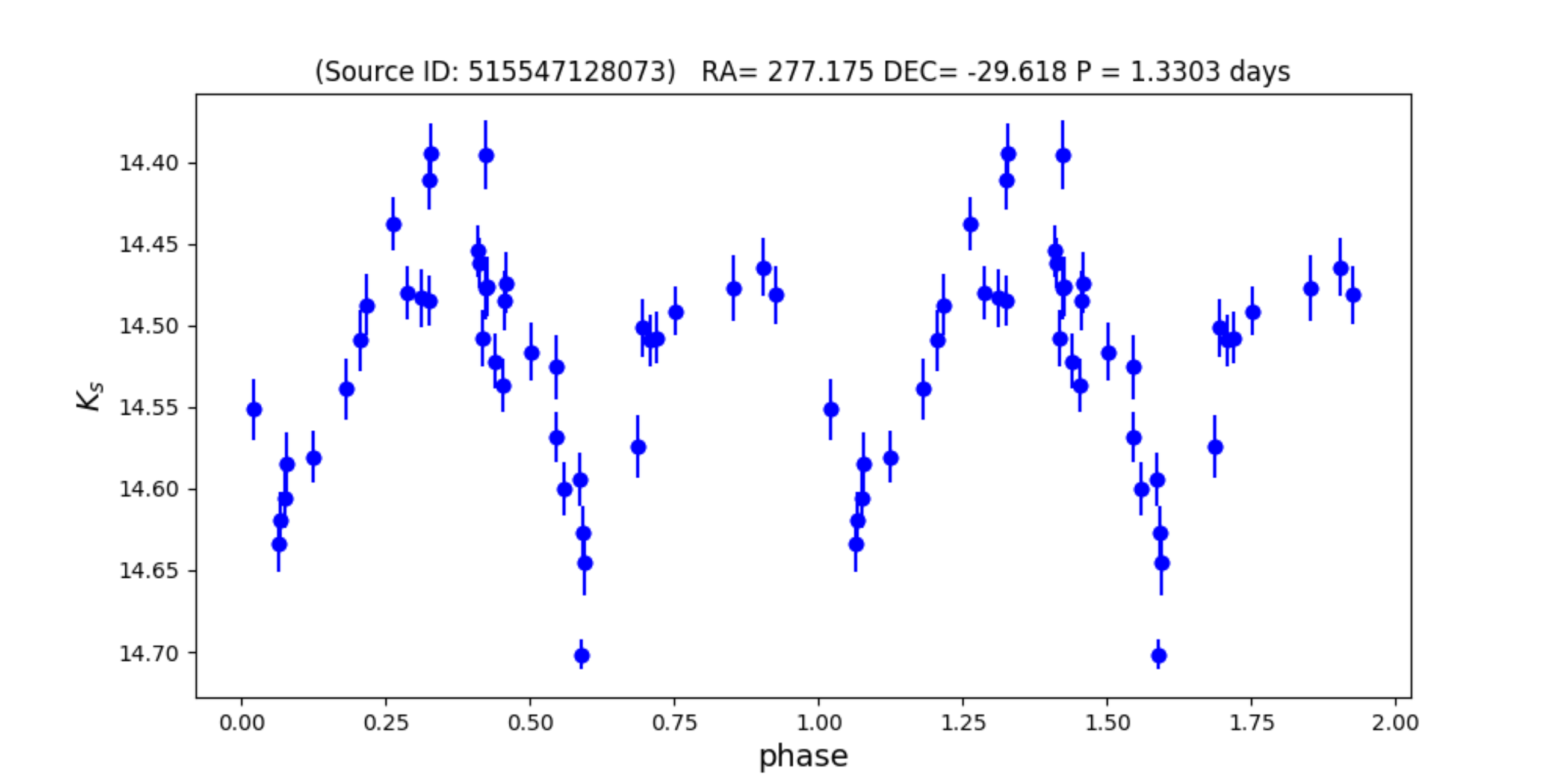} \\  
  \end{tabular}
  \caption{BHB stars light curves corresponding to $1^{st}$ category eclipsing binaries.}
  \label{fig:bhb3}
\end{figure*}

\begin{figure*}[htb]
\centering
  \begin{tabular}{@{}ccc@{}}   
    \includegraphics[width=.33\textwidth]{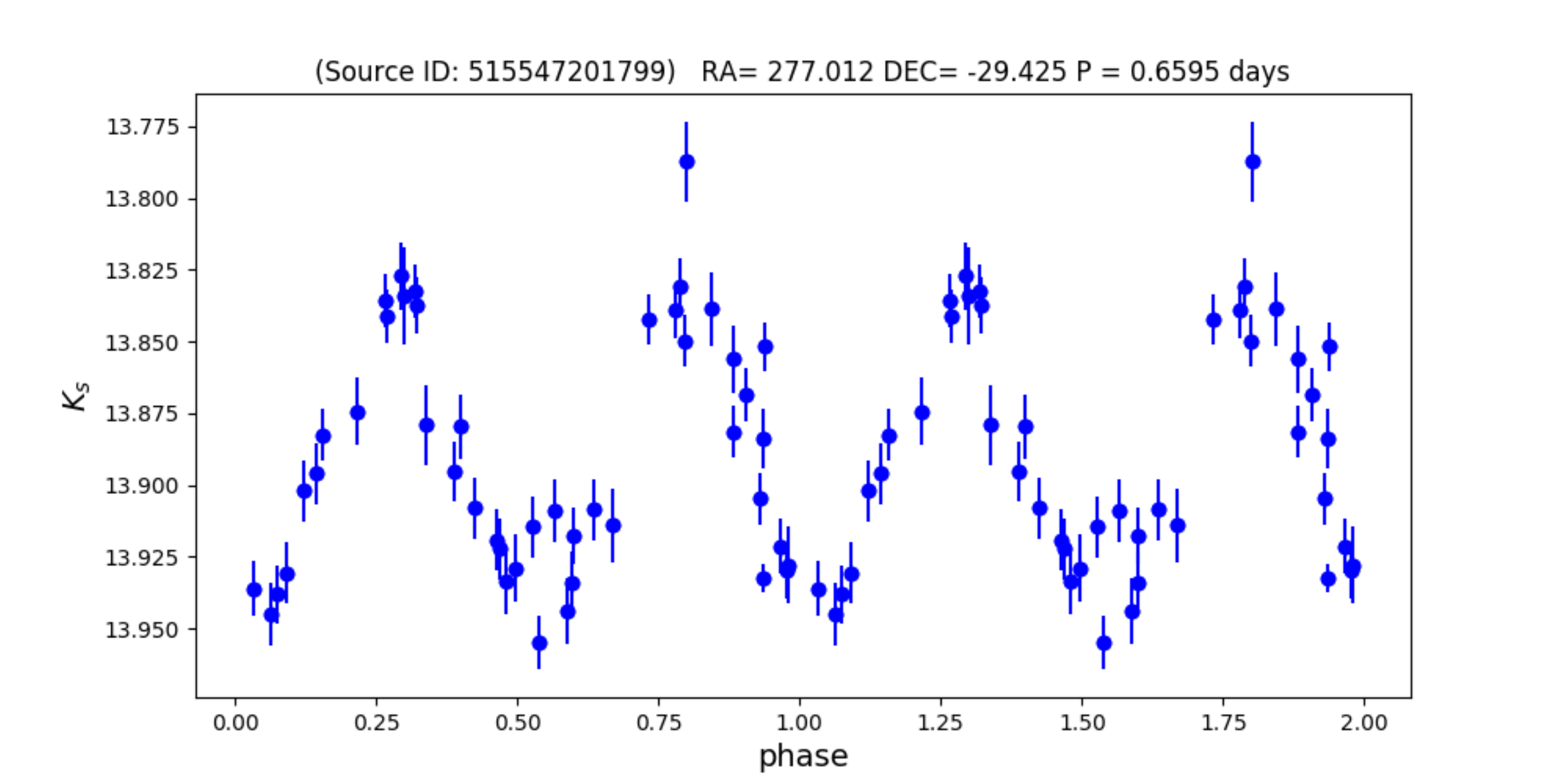} &
    \includegraphics[width=.33\textwidth]{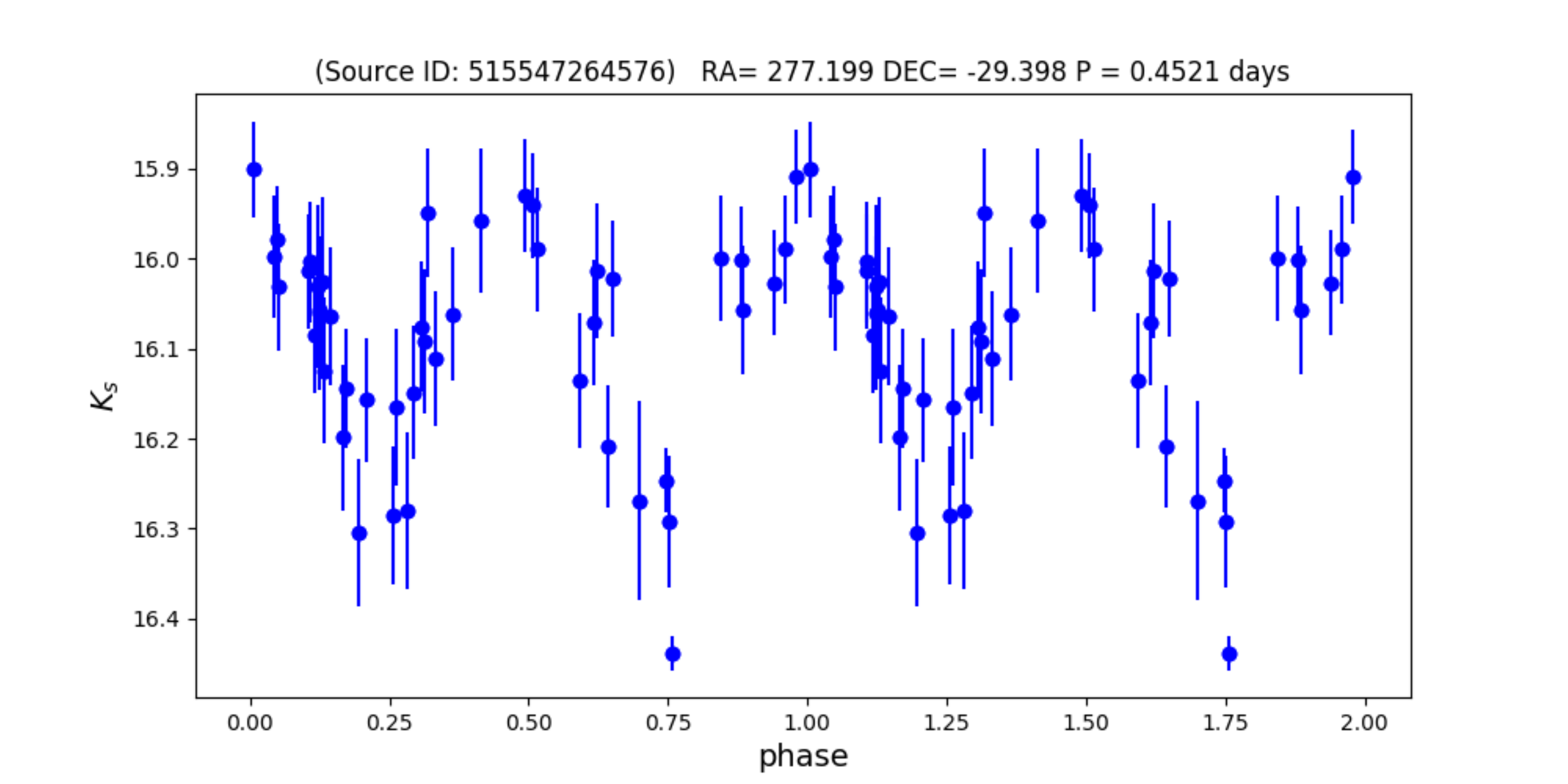} &
    \includegraphics[width=.33\textwidth]{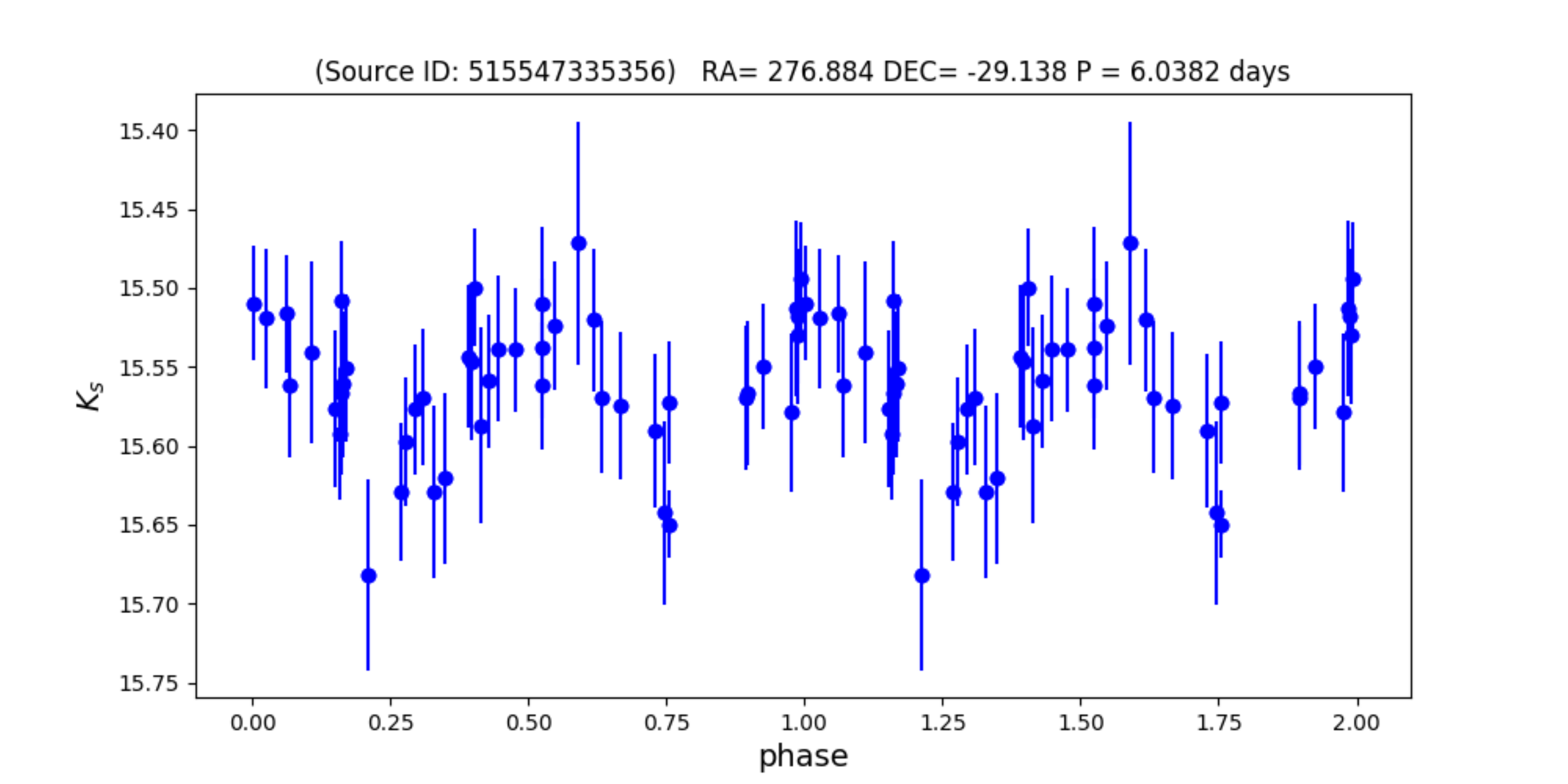} \\
    \includegraphics[width=.33\textwidth]{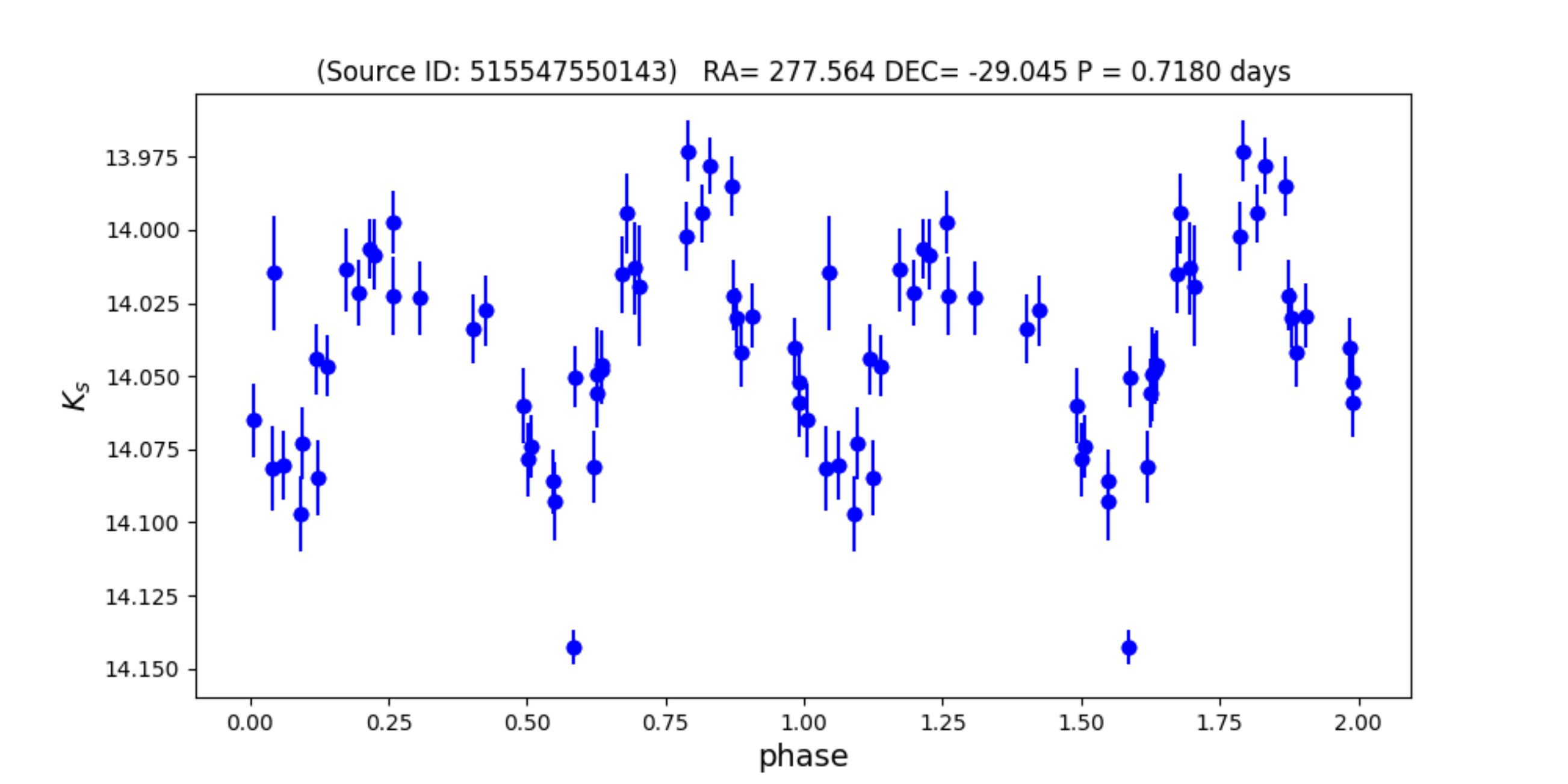} &
    \includegraphics[width=.33\textwidth]{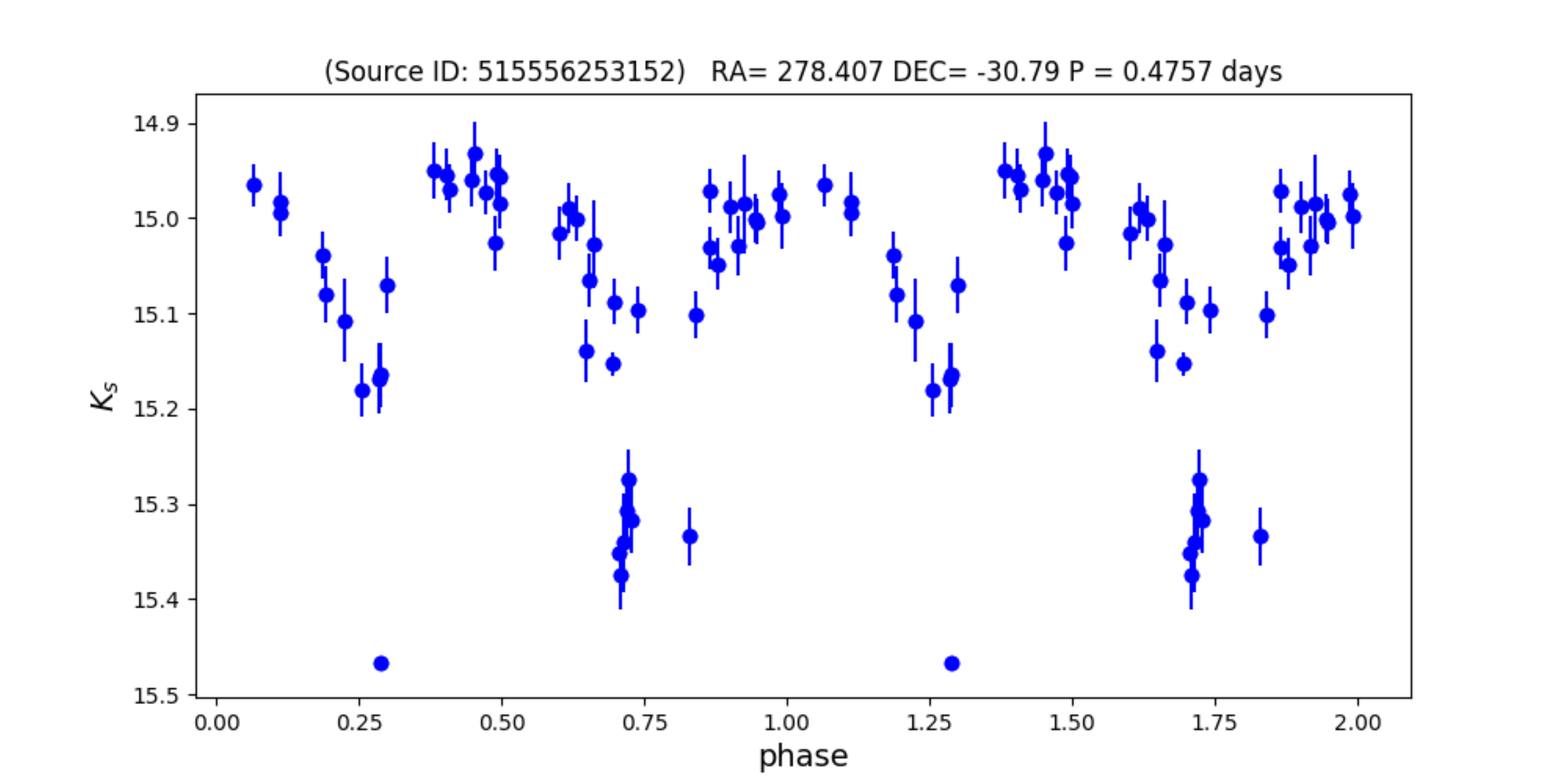} &
    \includegraphics[width=.33\textwidth]{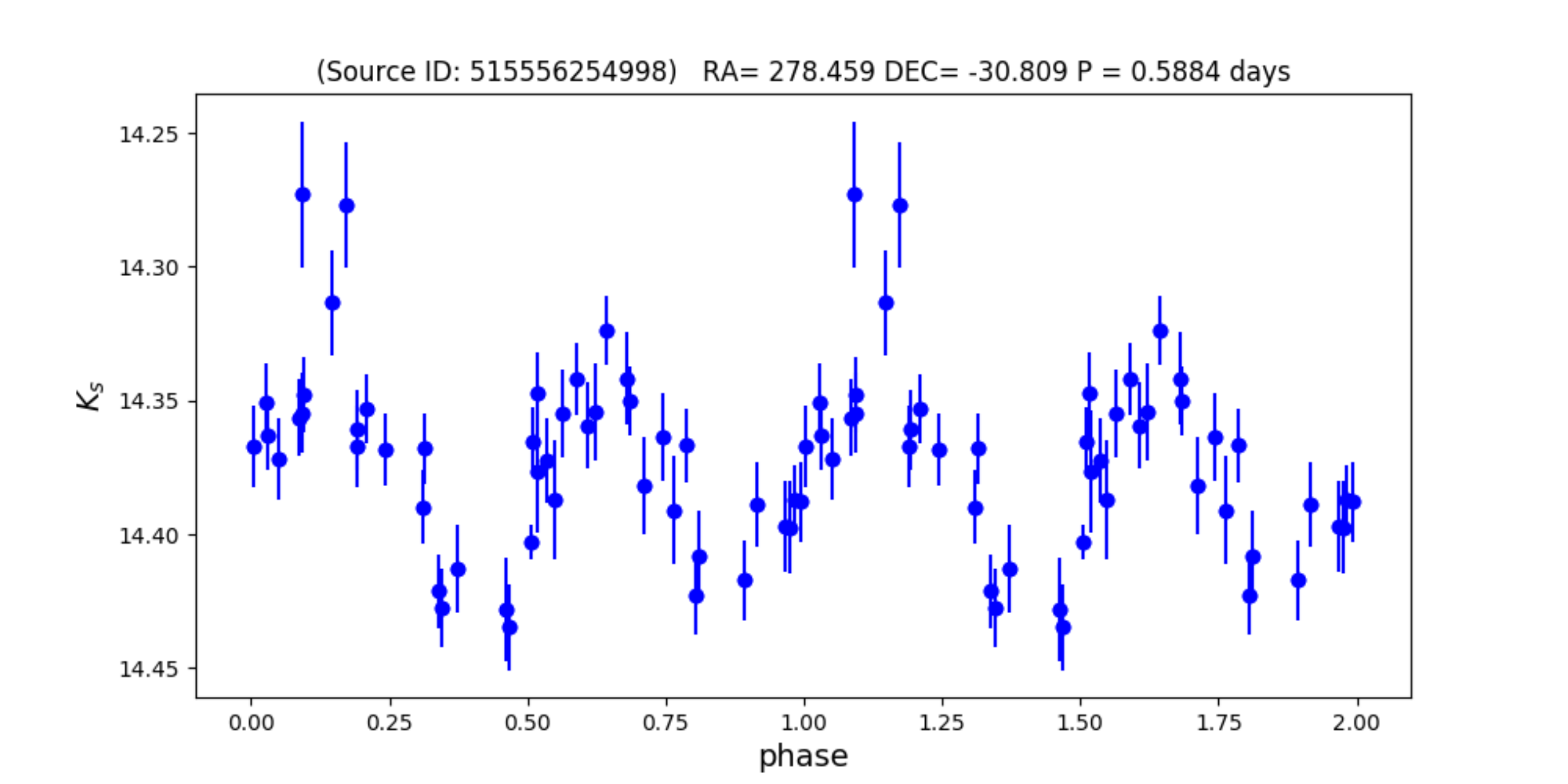} \\
    \includegraphics[width=.33\textwidth]{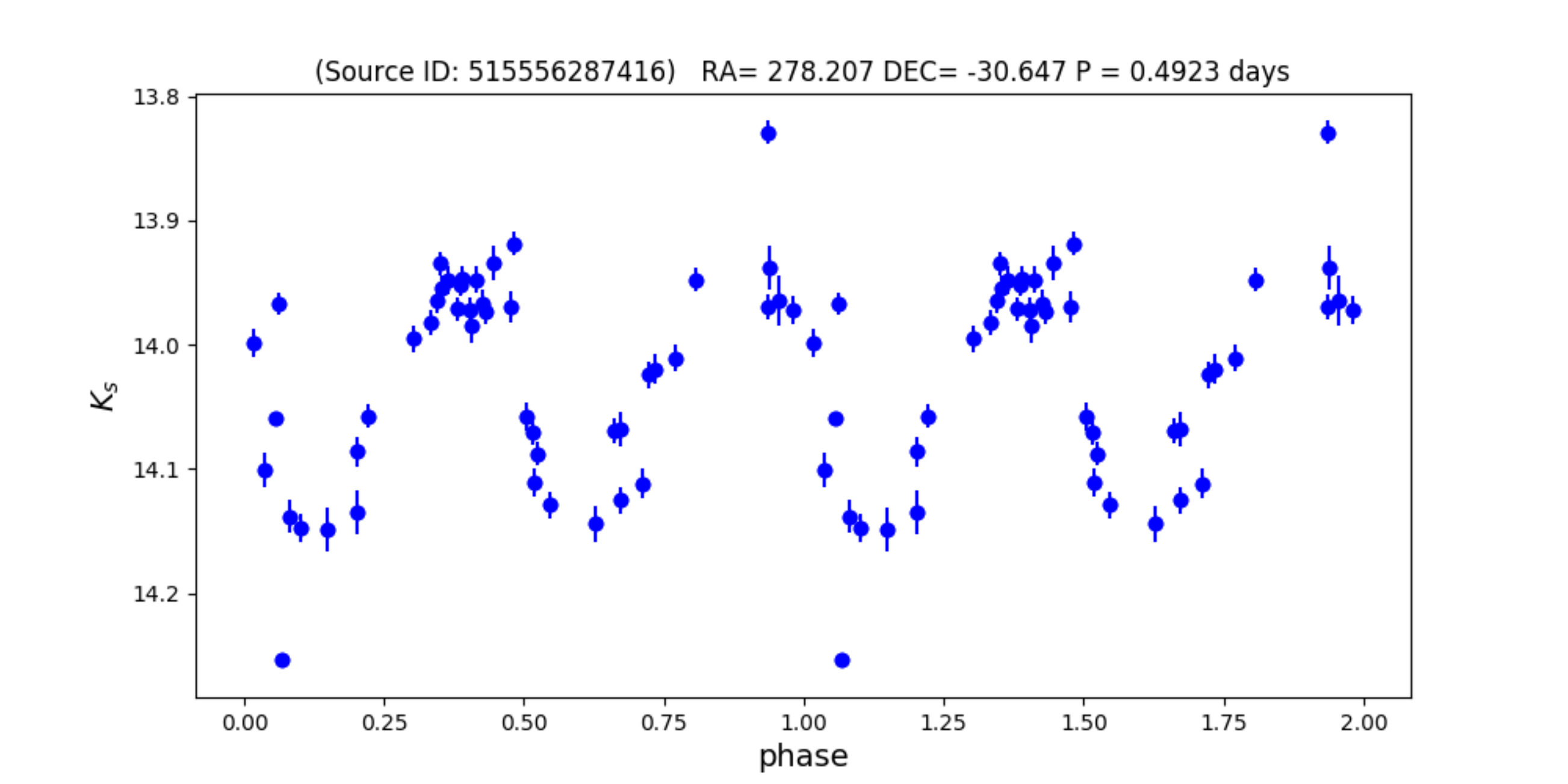} &
    \includegraphics[width=.33\textwidth]{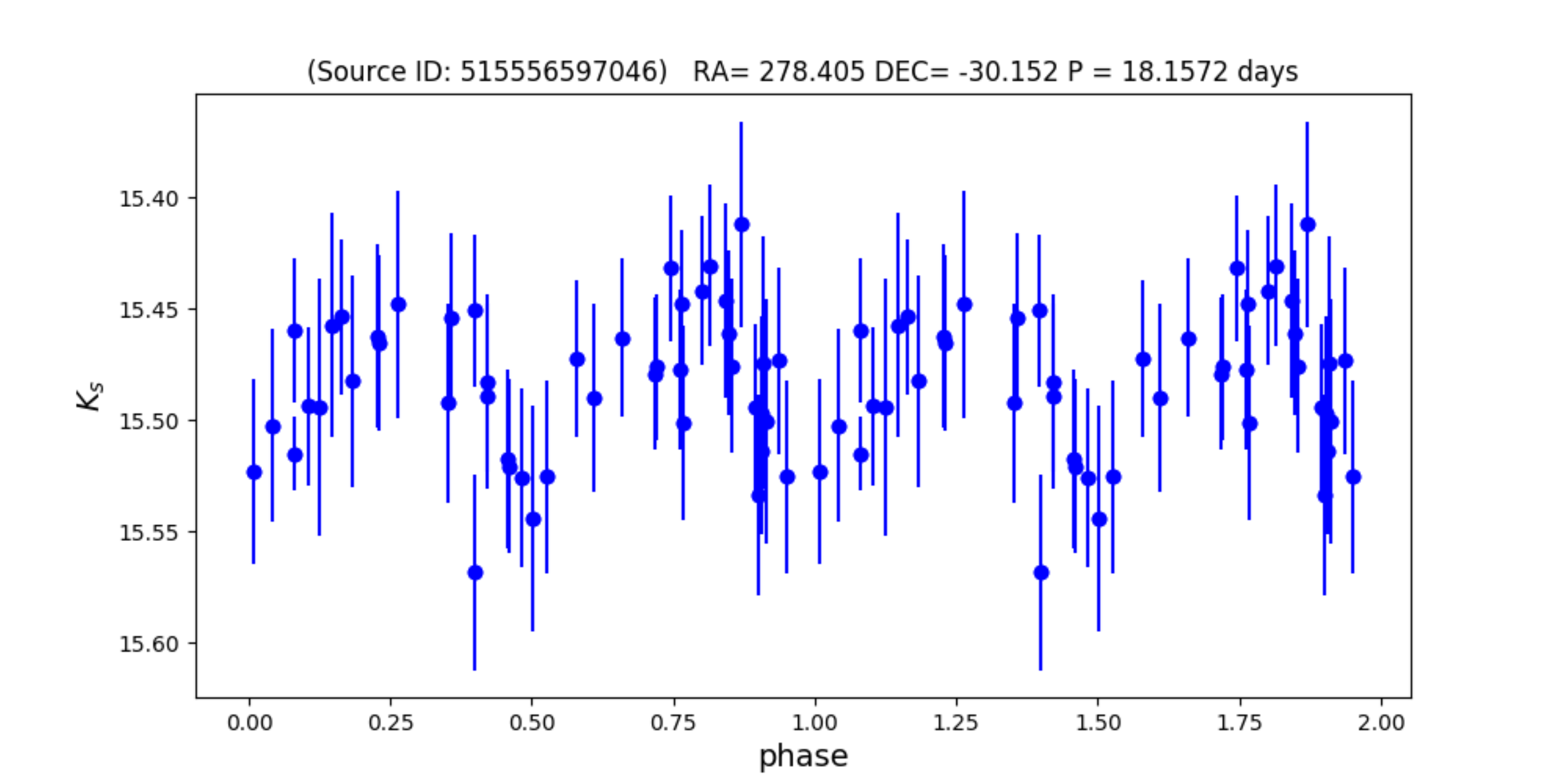} &
    \includegraphics[width=.33\textwidth]{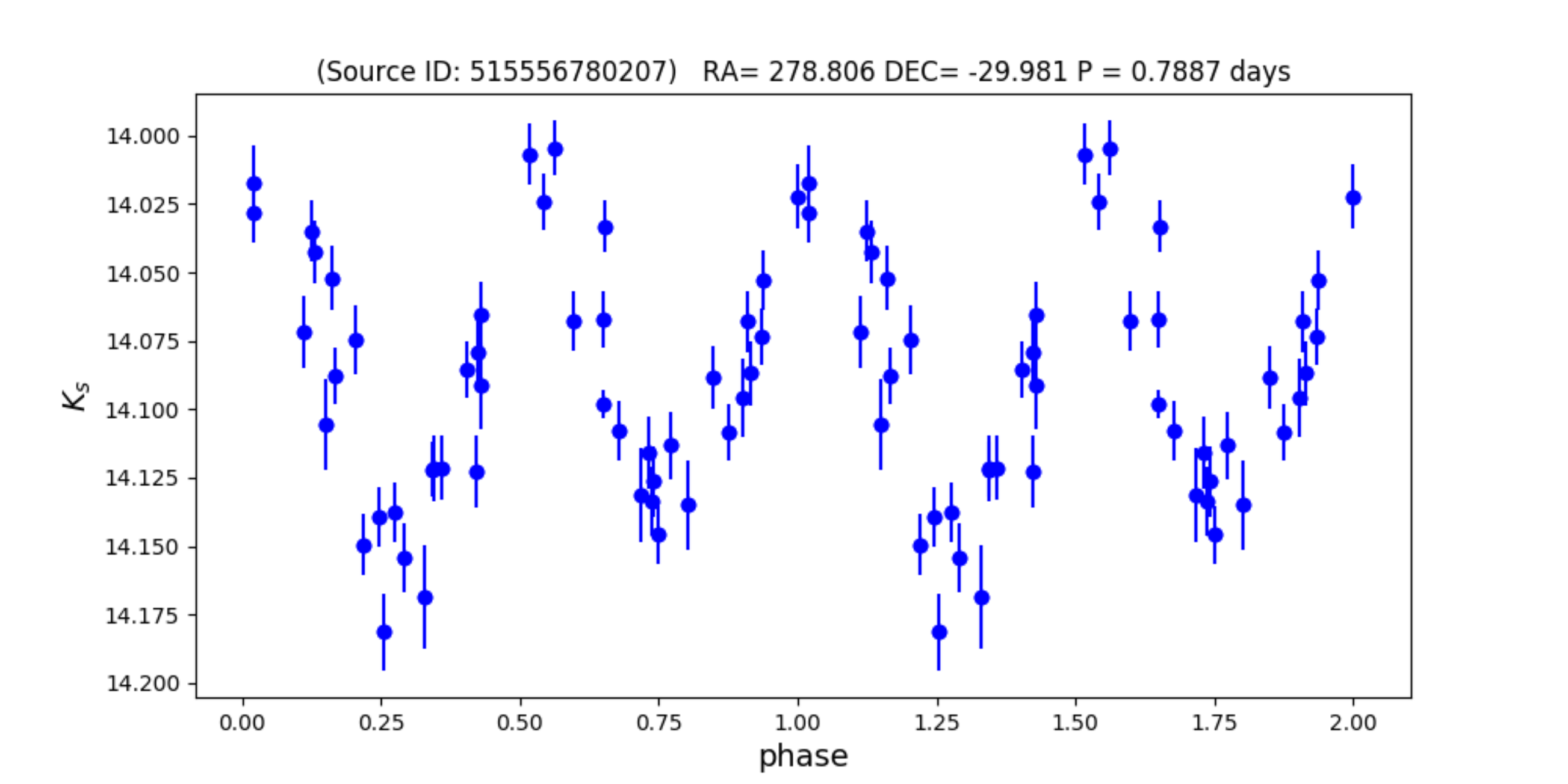} \\
    \includegraphics[width=.33\textwidth]{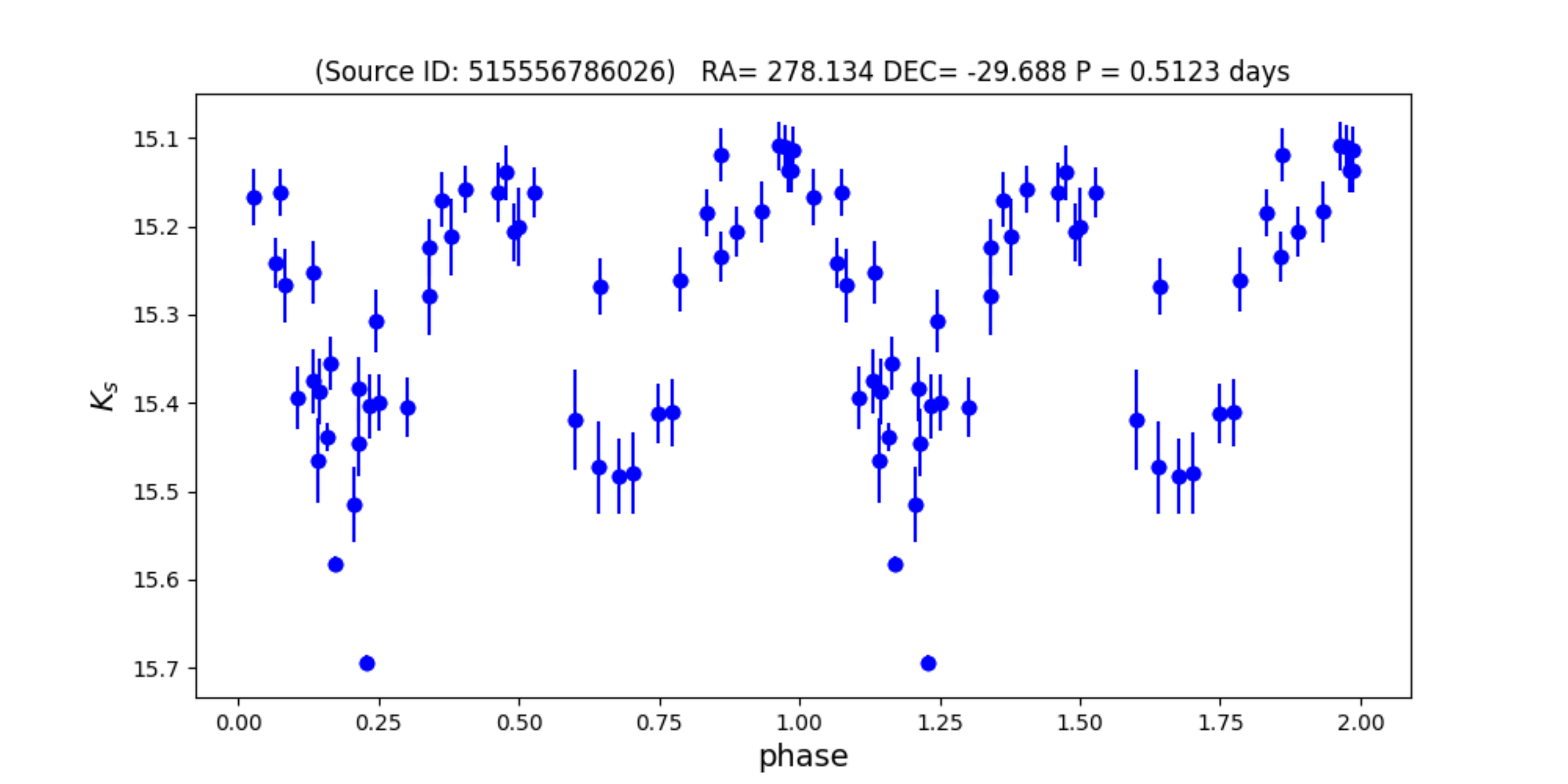} &
    \includegraphics[width=.33\textwidth]{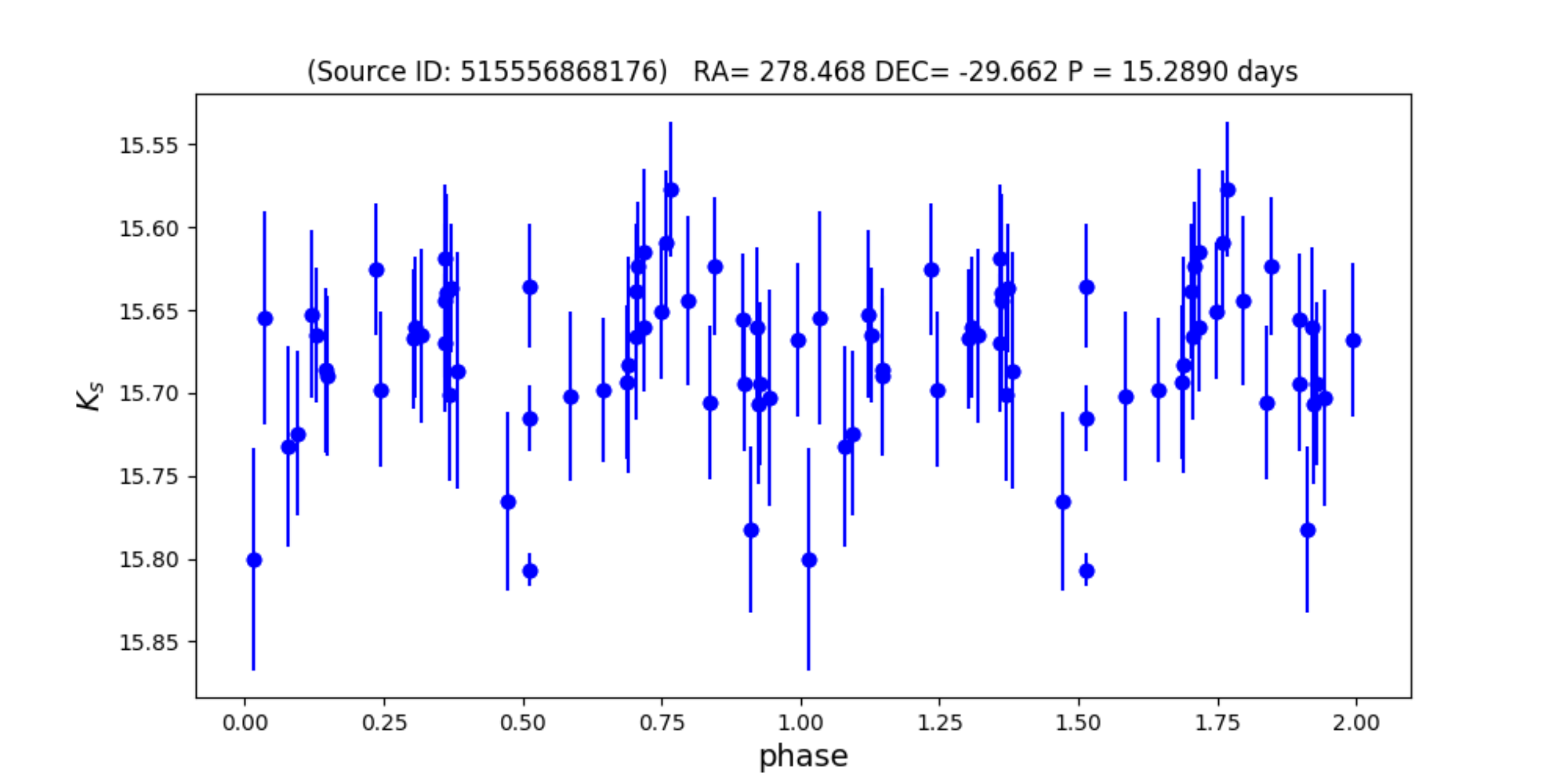} & 
    \includegraphics[width=.33\textwidth]{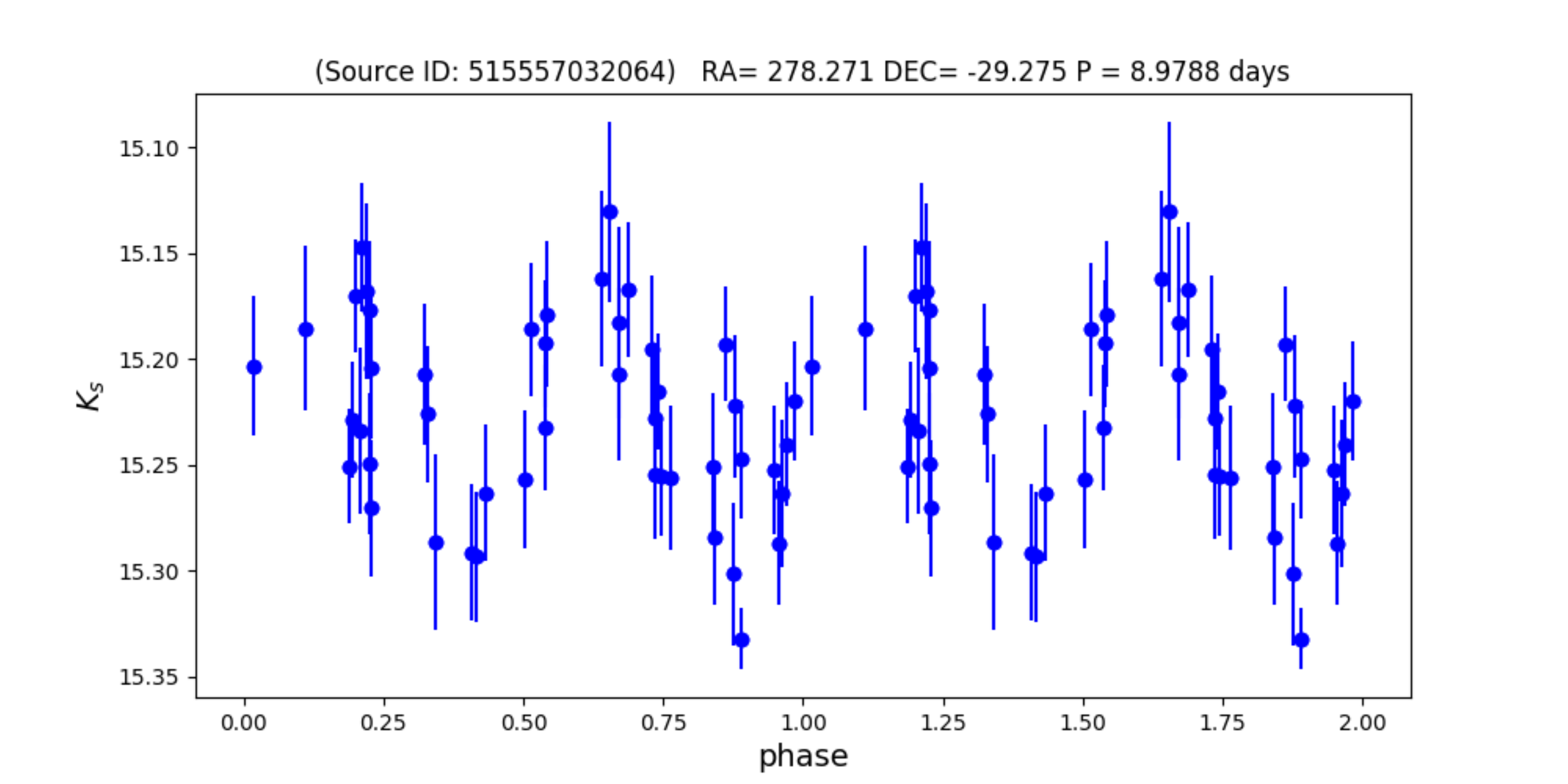} \\
    \includegraphics[width=.33\textwidth]{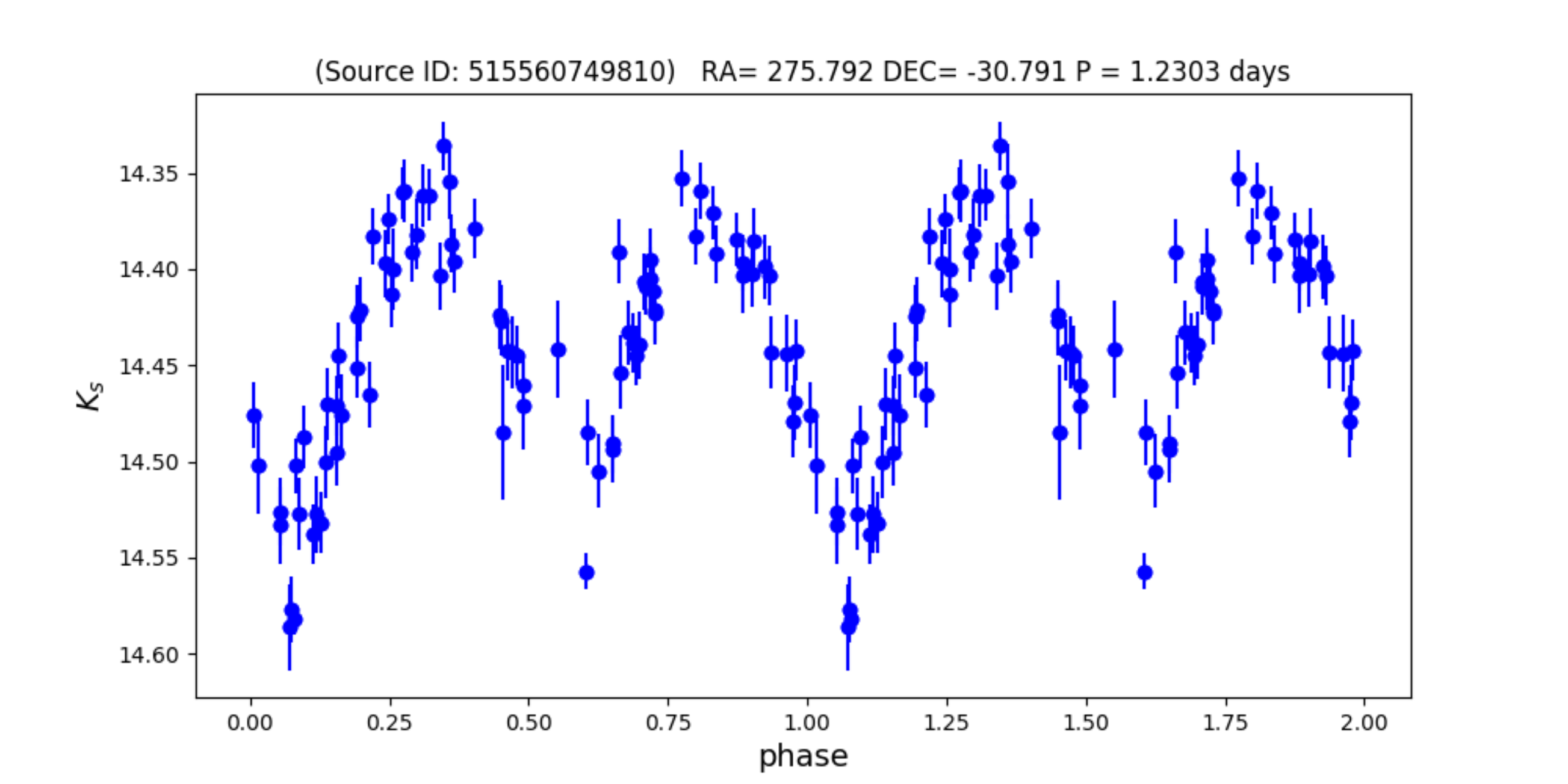} &
    \includegraphics[width=.33\textwidth]{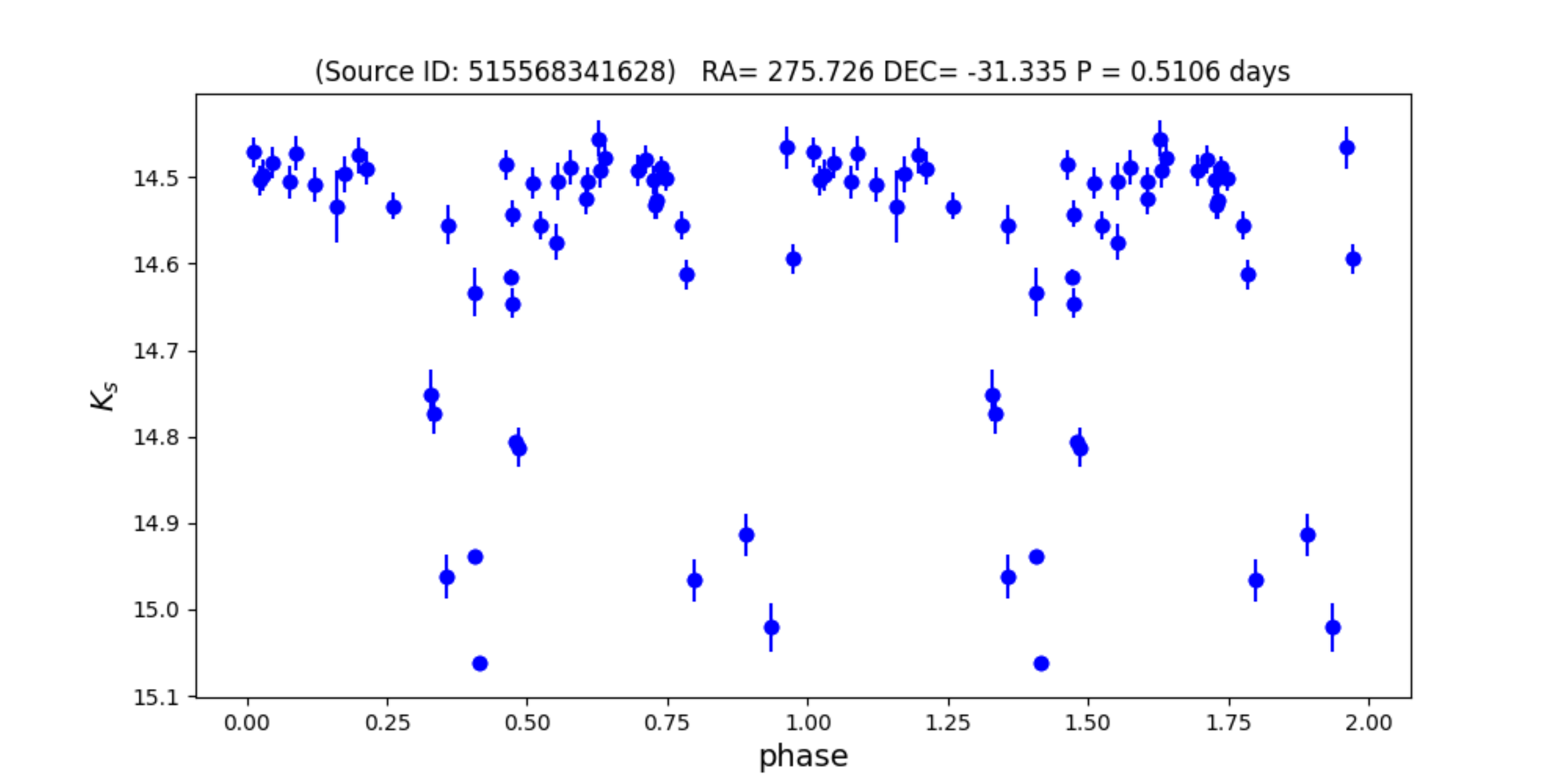} &    
    \includegraphics[width=.33\textwidth]{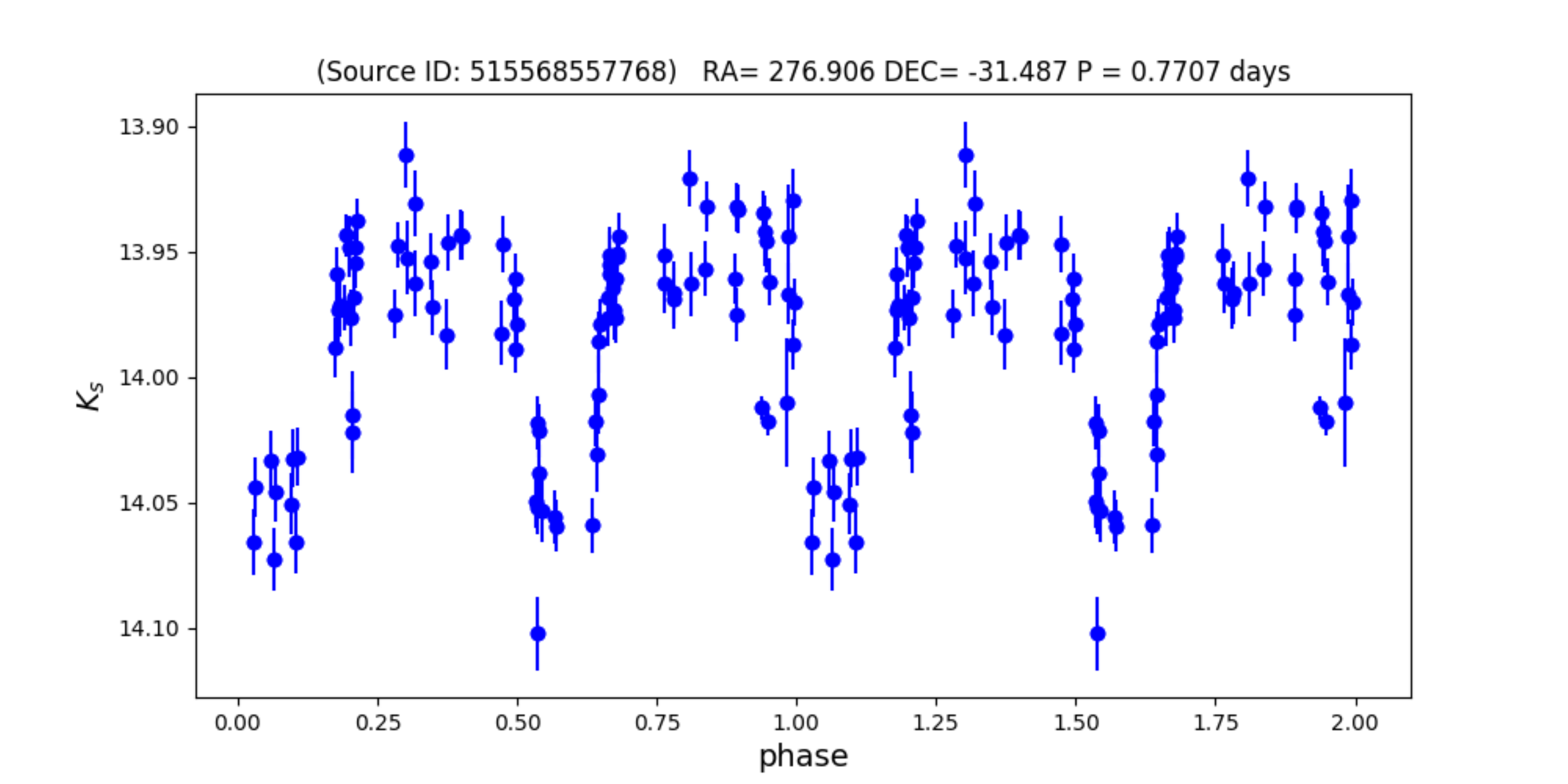} \\
    \includegraphics[width=.33\textwidth]{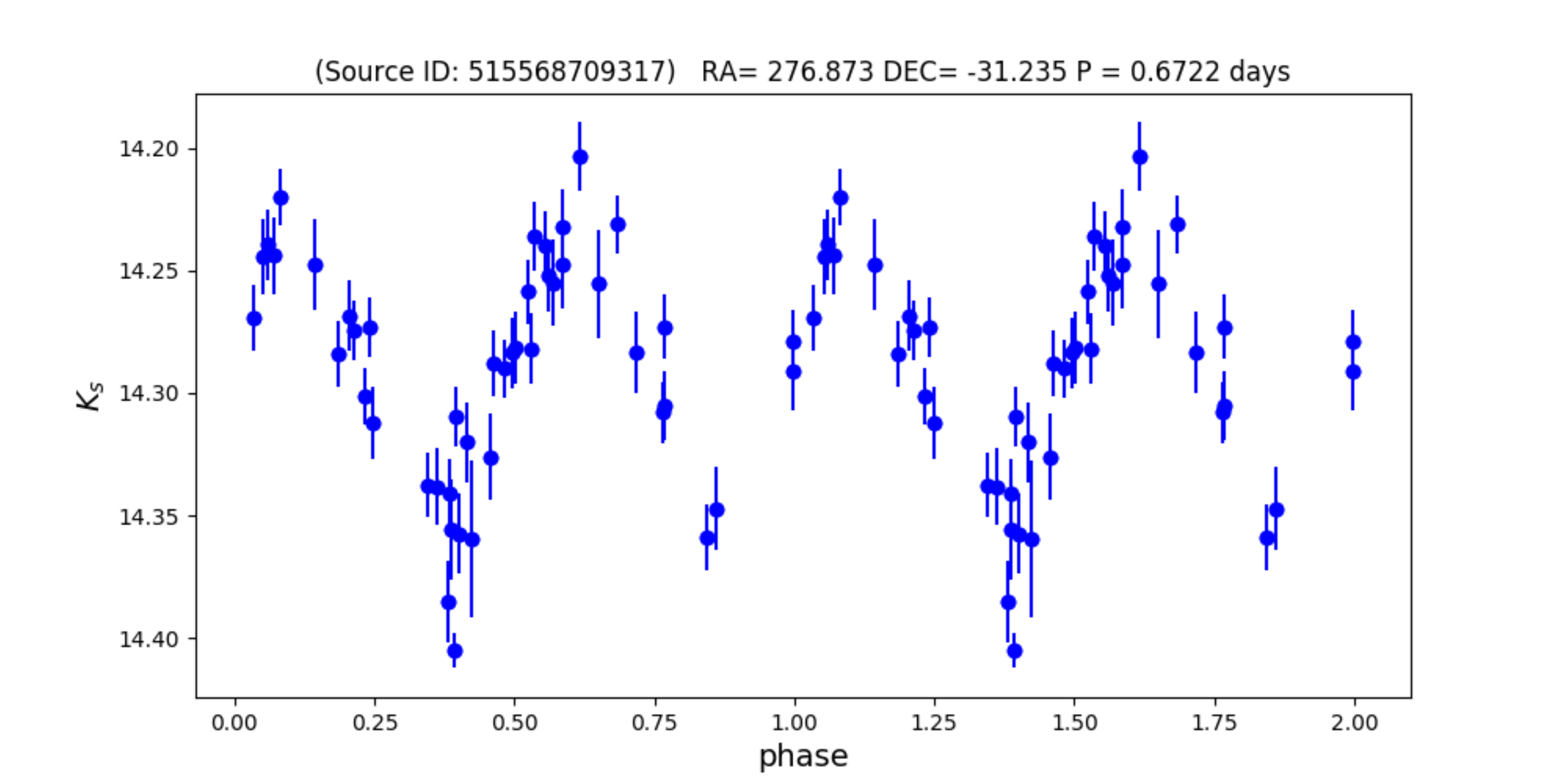} &
    \includegraphics[width=.33\textwidth]{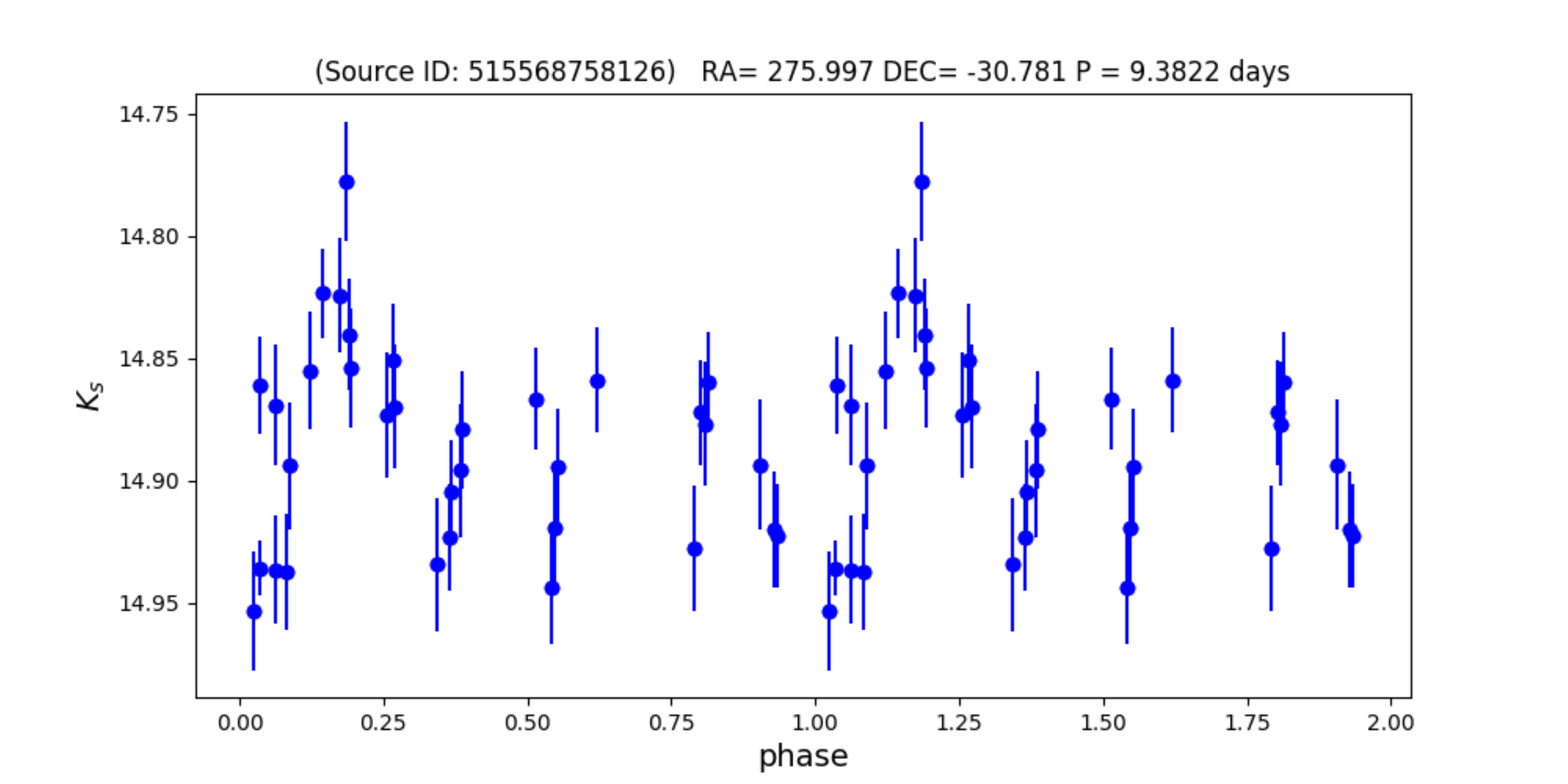} &
    \includegraphics[width=.33\textwidth]{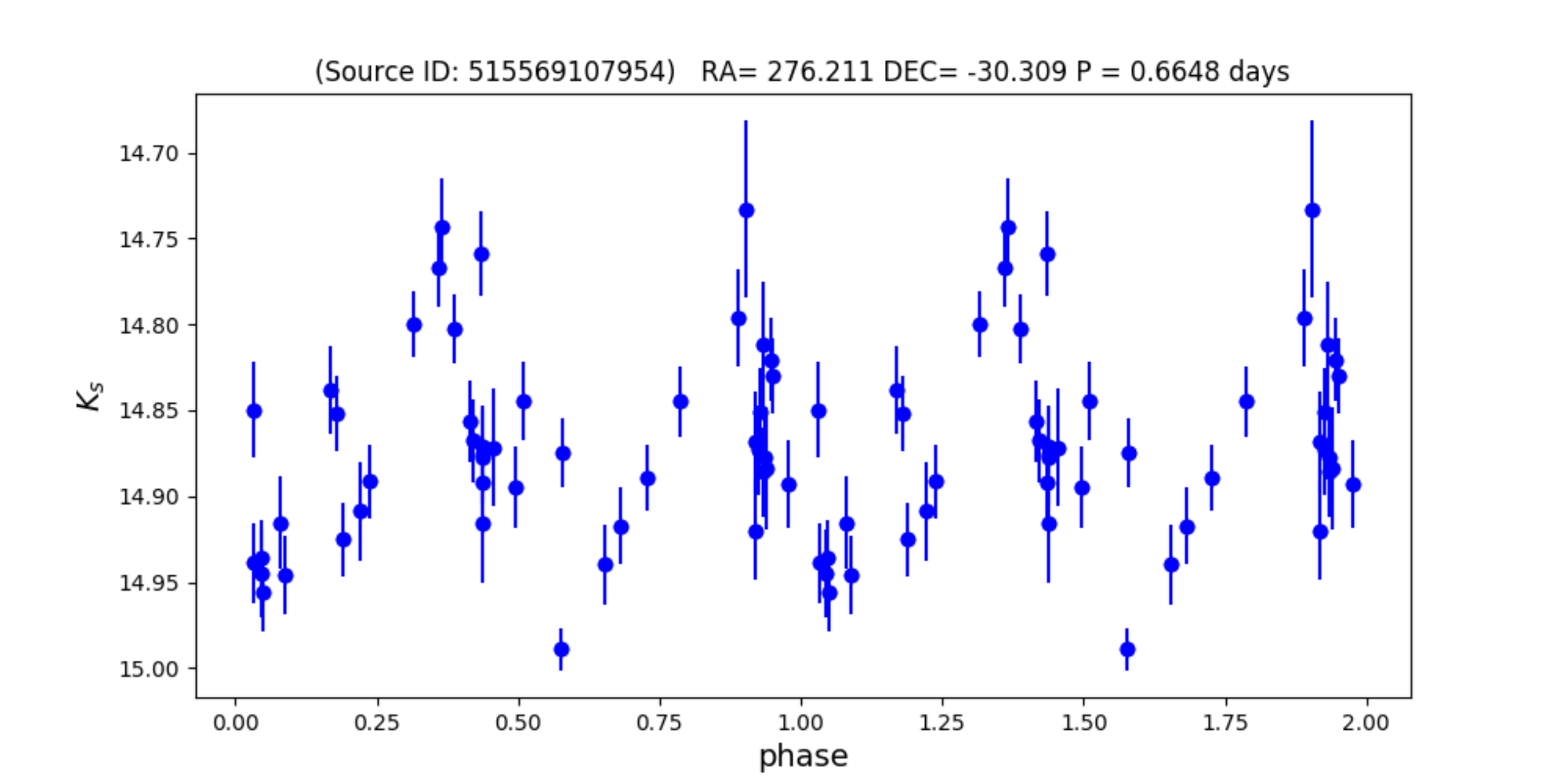} \\
    \includegraphics[width=.33\textwidth]{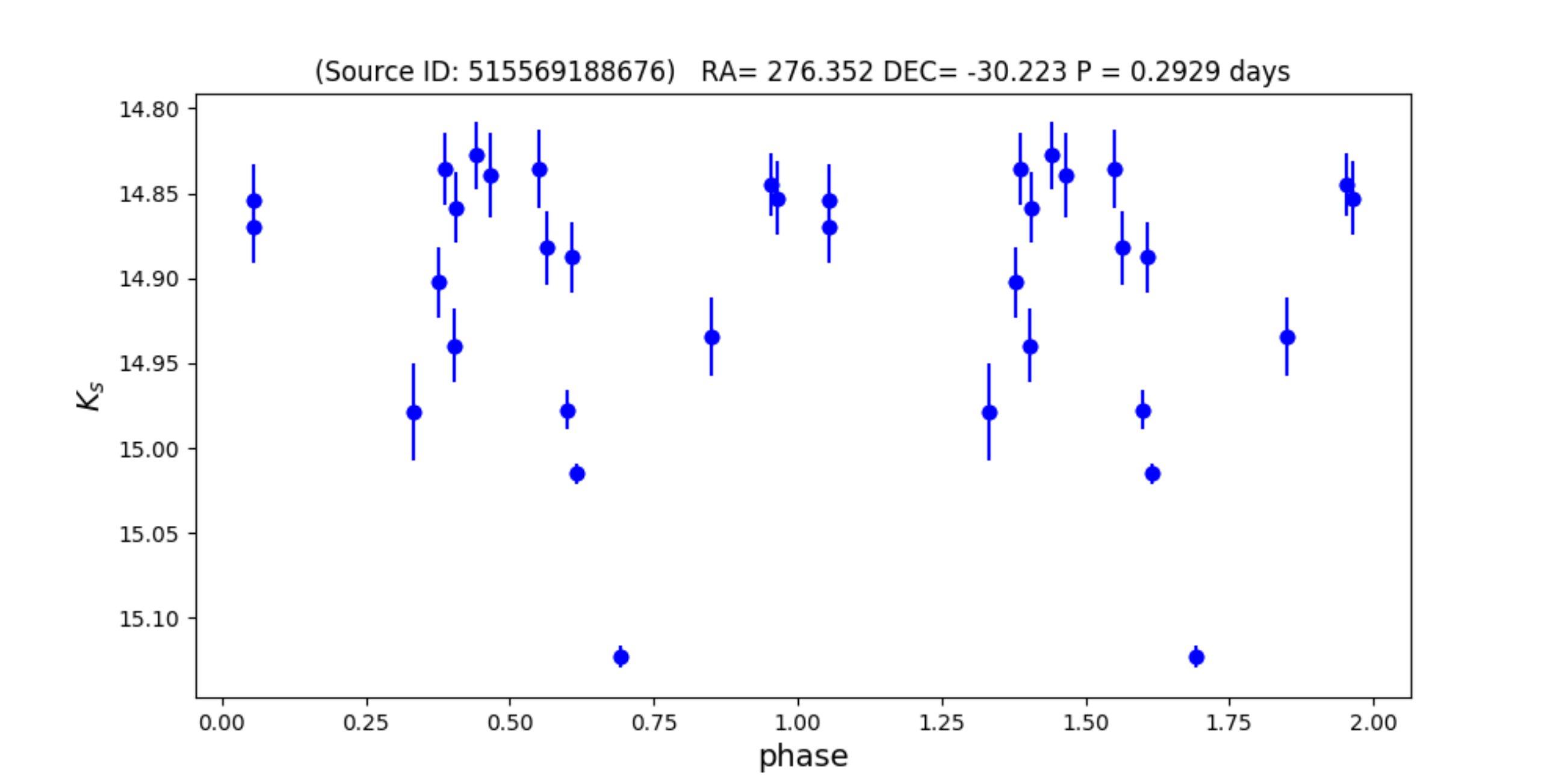} &
    \includegraphics[width=.33\textwidth]{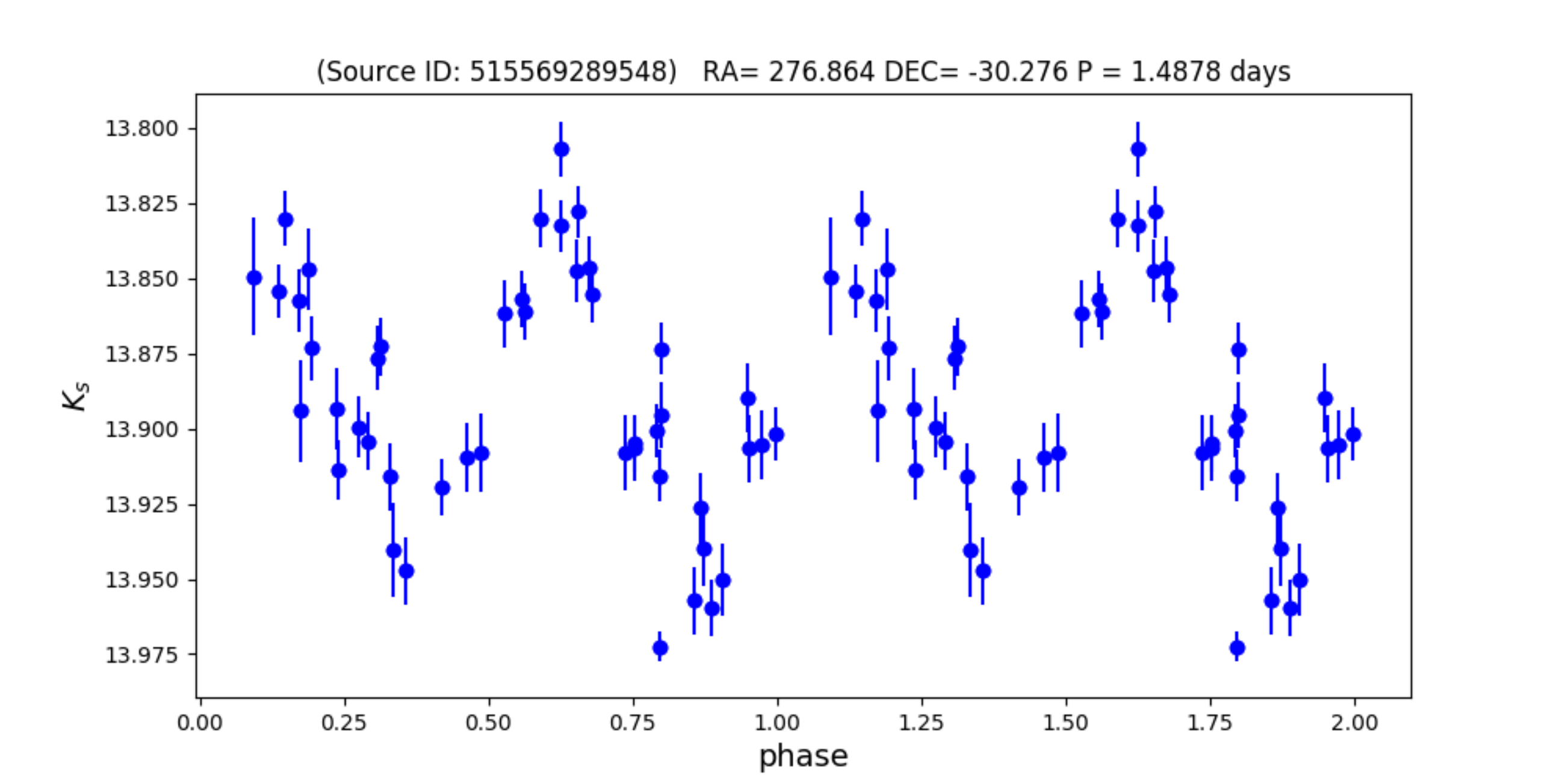} &
    \includegraphics[width=.33\textwidth]{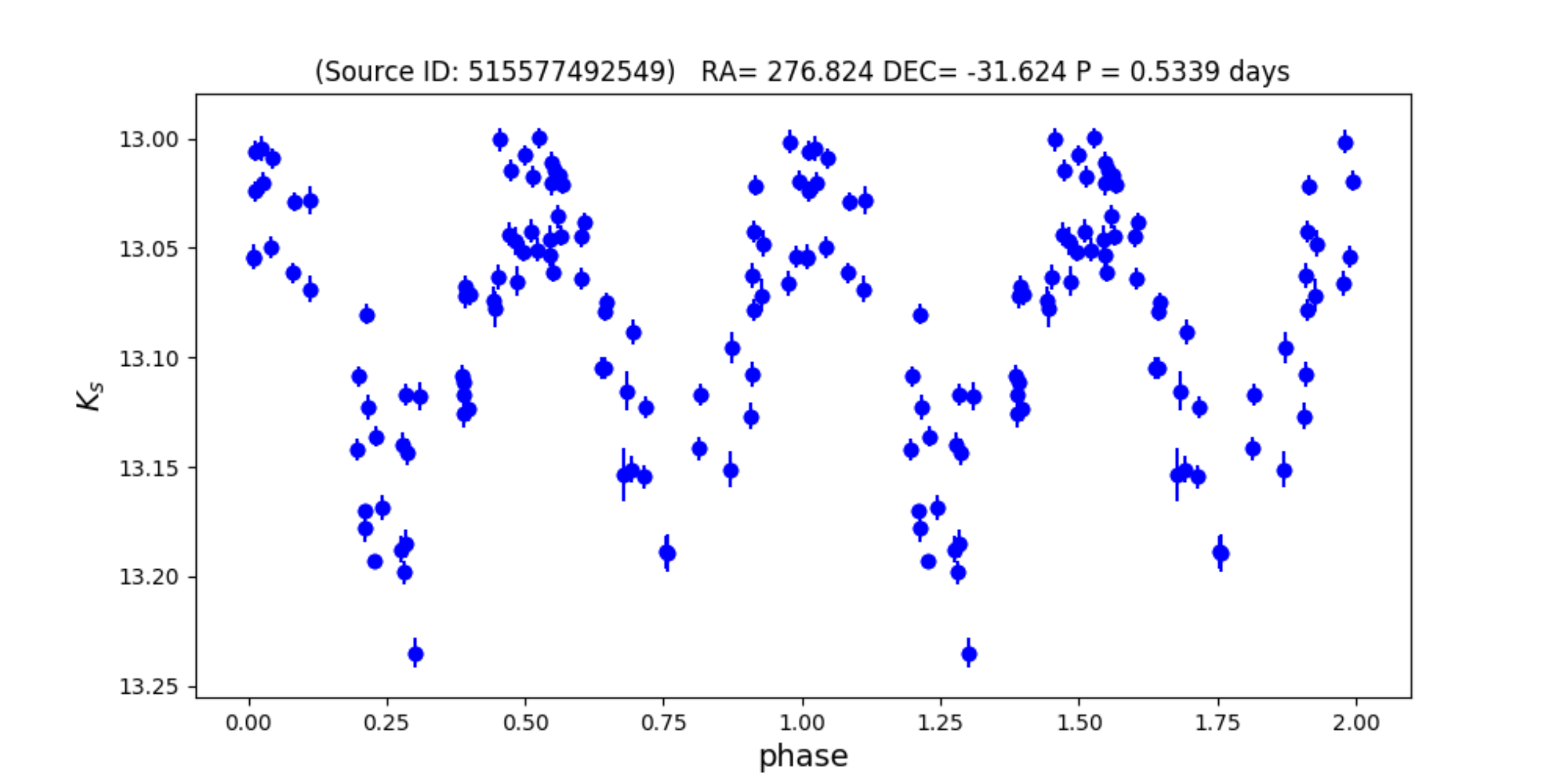} \\
    \includegraphics[width=.33\textwidth]{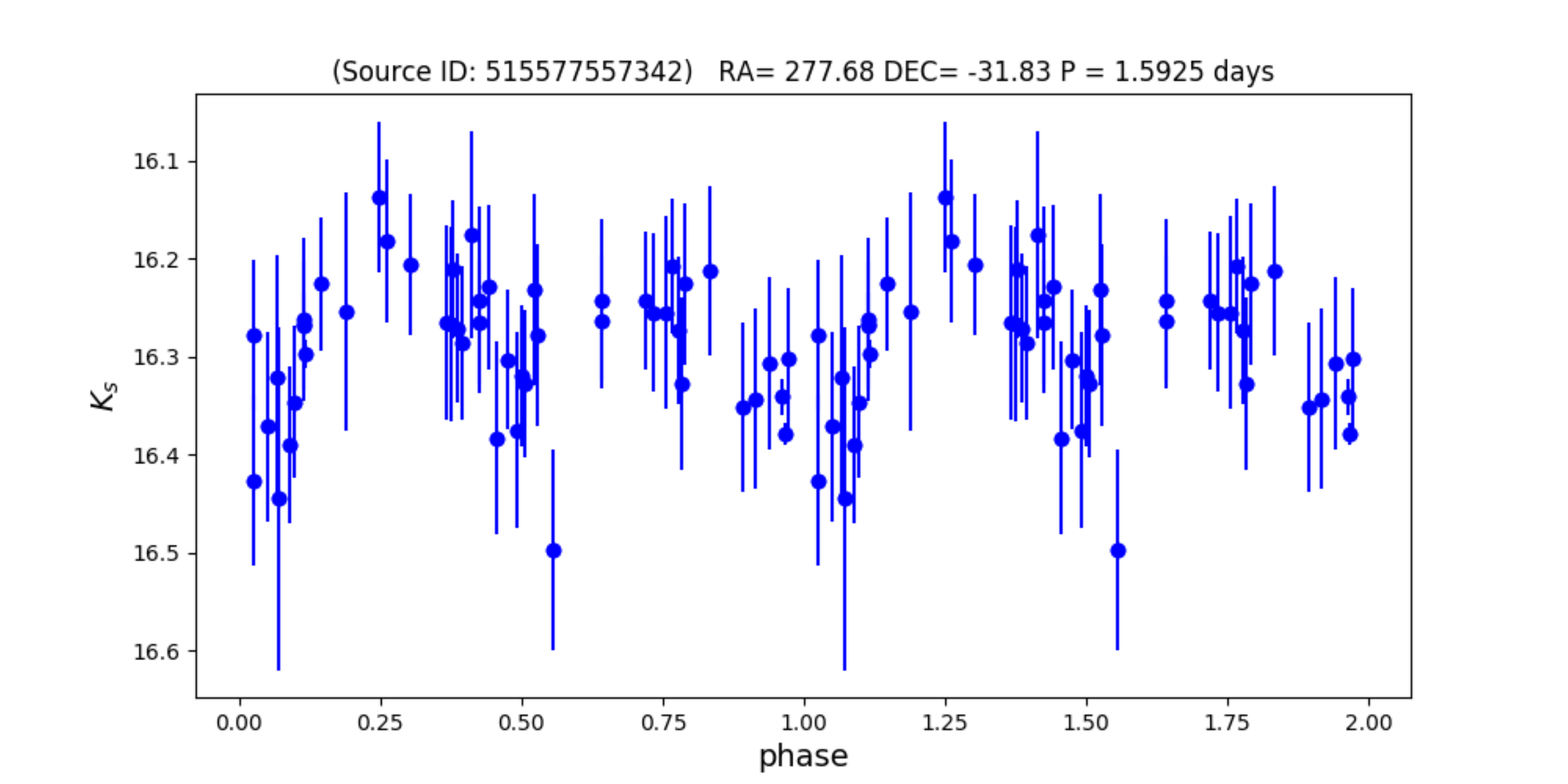} &
    \includegraphics[width=.33\textwidth]{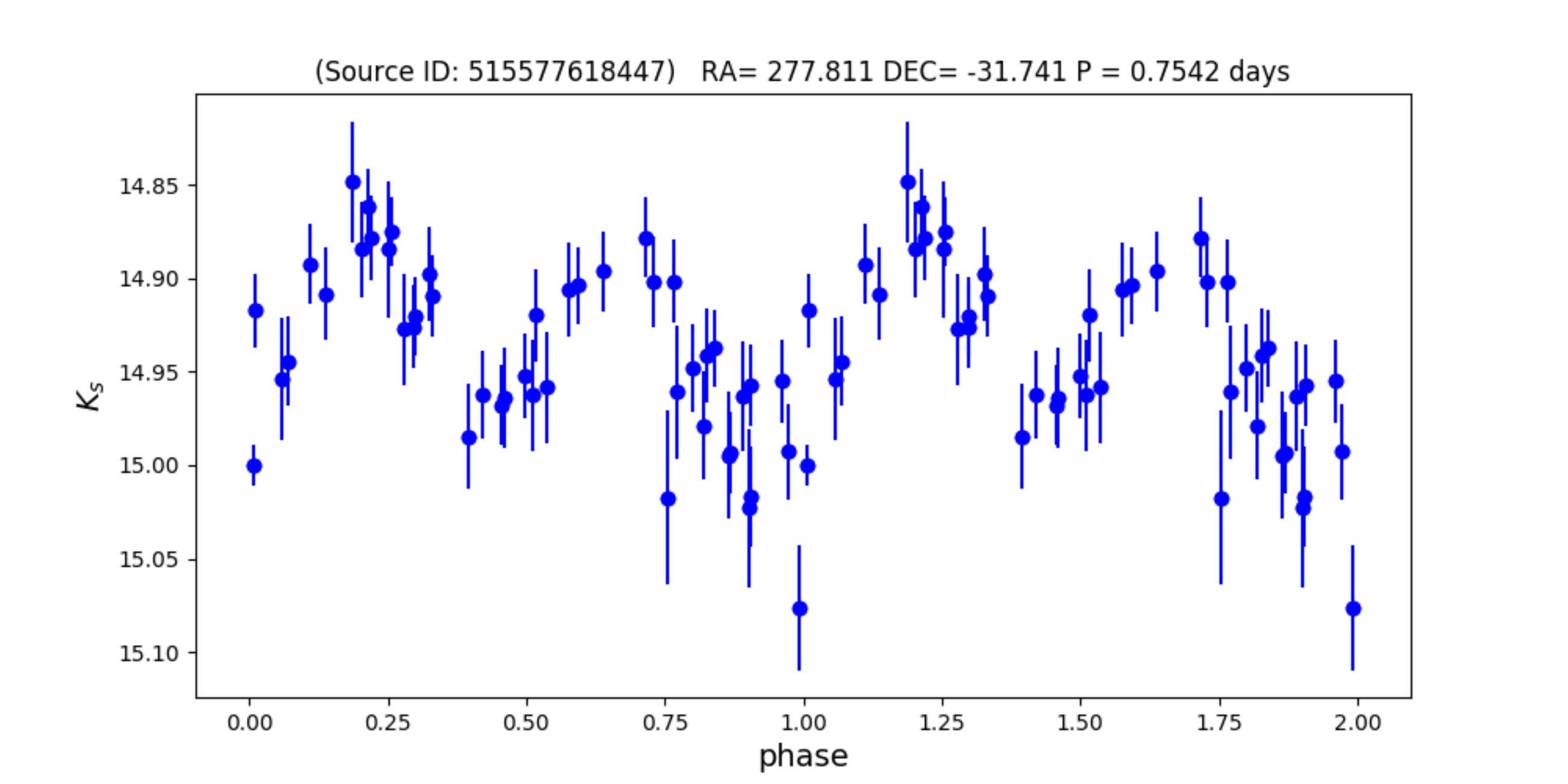} &
    \includegraphics[width=.33\textwidth]{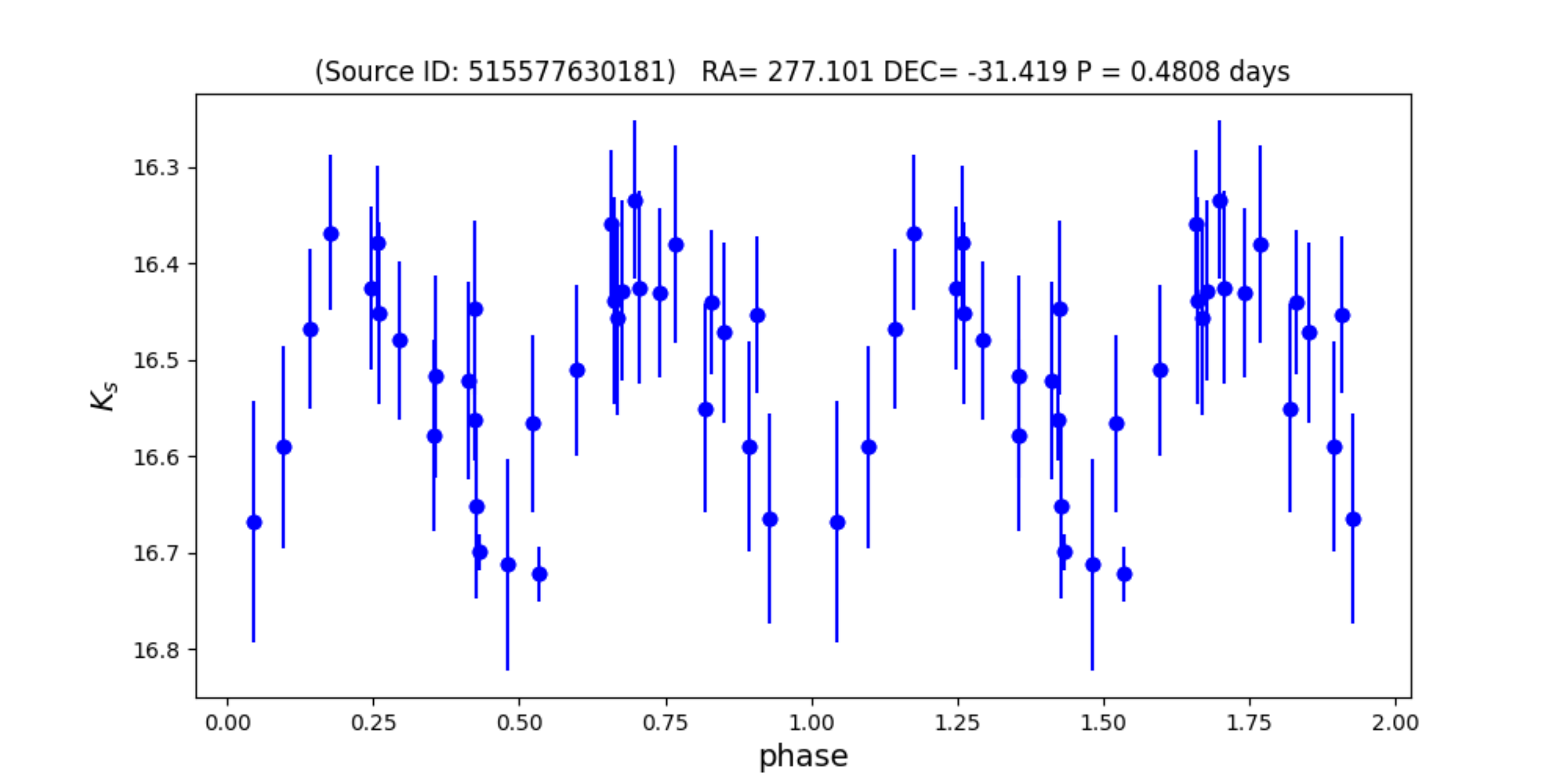} \\ 
  \end{tabular}
  \caption{BHB stars light curves corresponding to $1^{st}$ category eclipsing binaries.}
  \label{fig:bhb4}
\end{figure*}

\begin{figure*}[htb]
\centering
  \begin{tabular}{@{}ccc@{}}    
    \includegraphics[width=.33\textwidth]{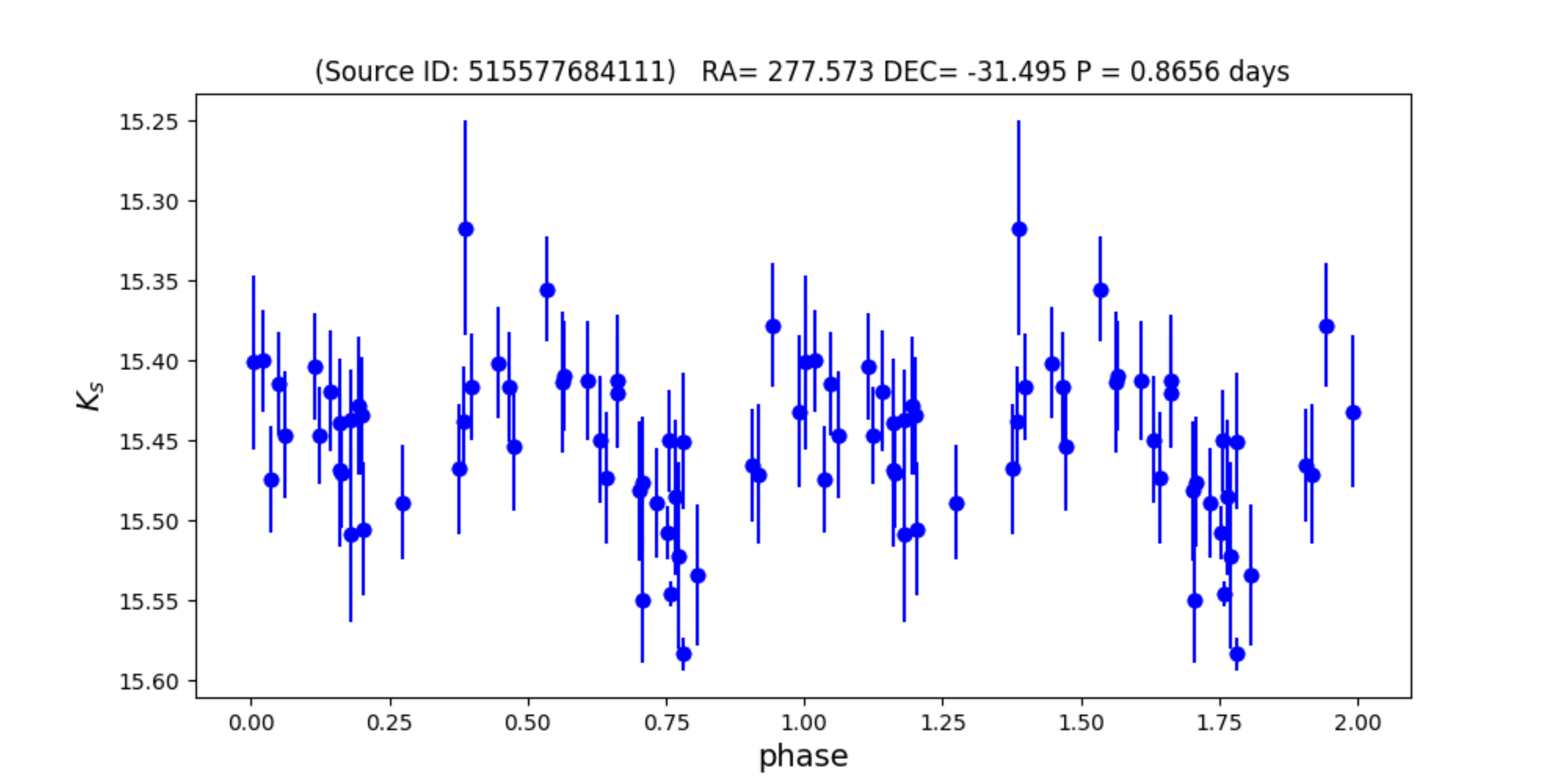} &
    \includegraphics[width=.33\textwidth]{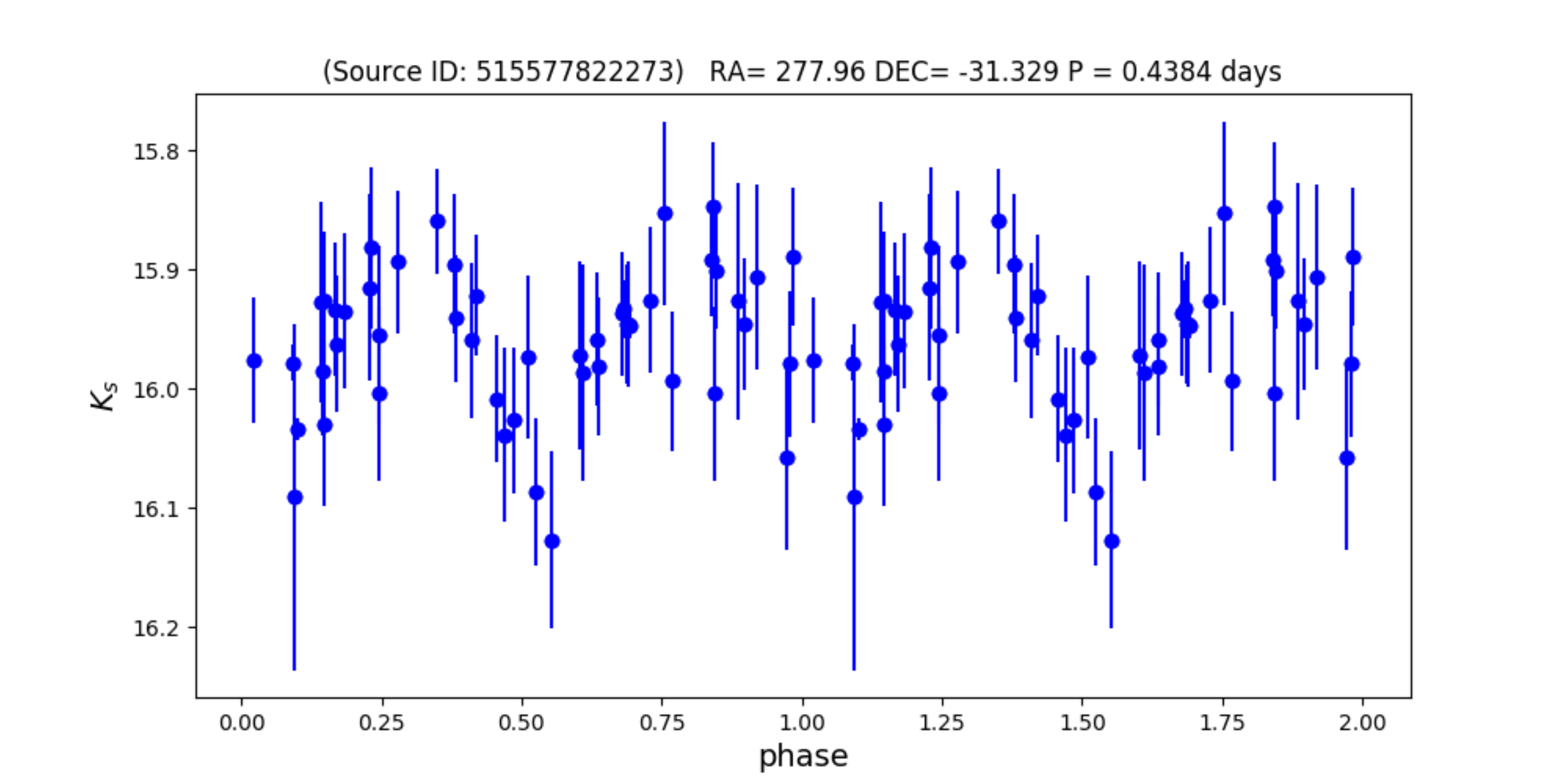} &
    \includegraphics[width=.33\textwidth]{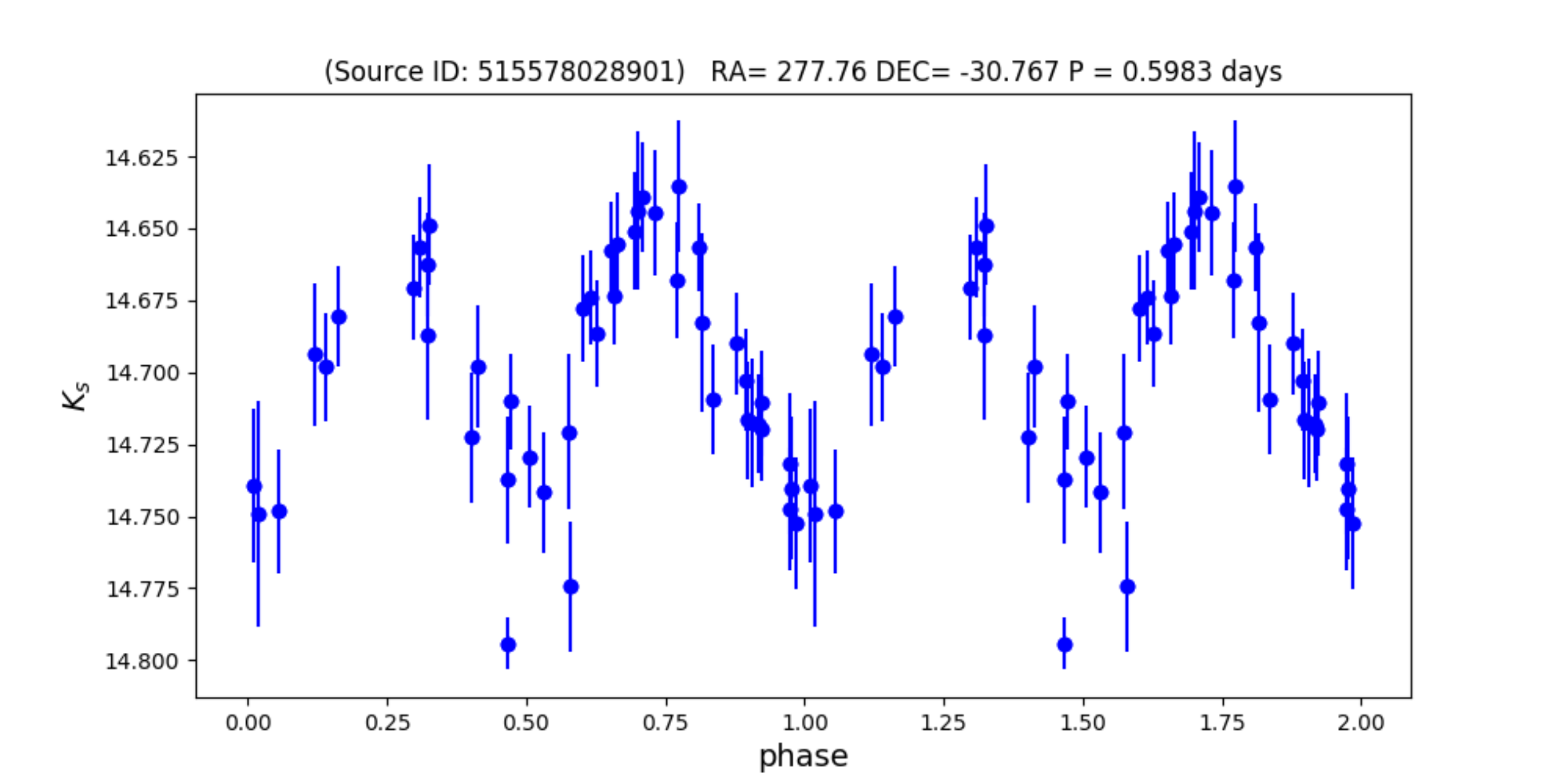} \\
    \includegraphics[width=.33\textwidth]{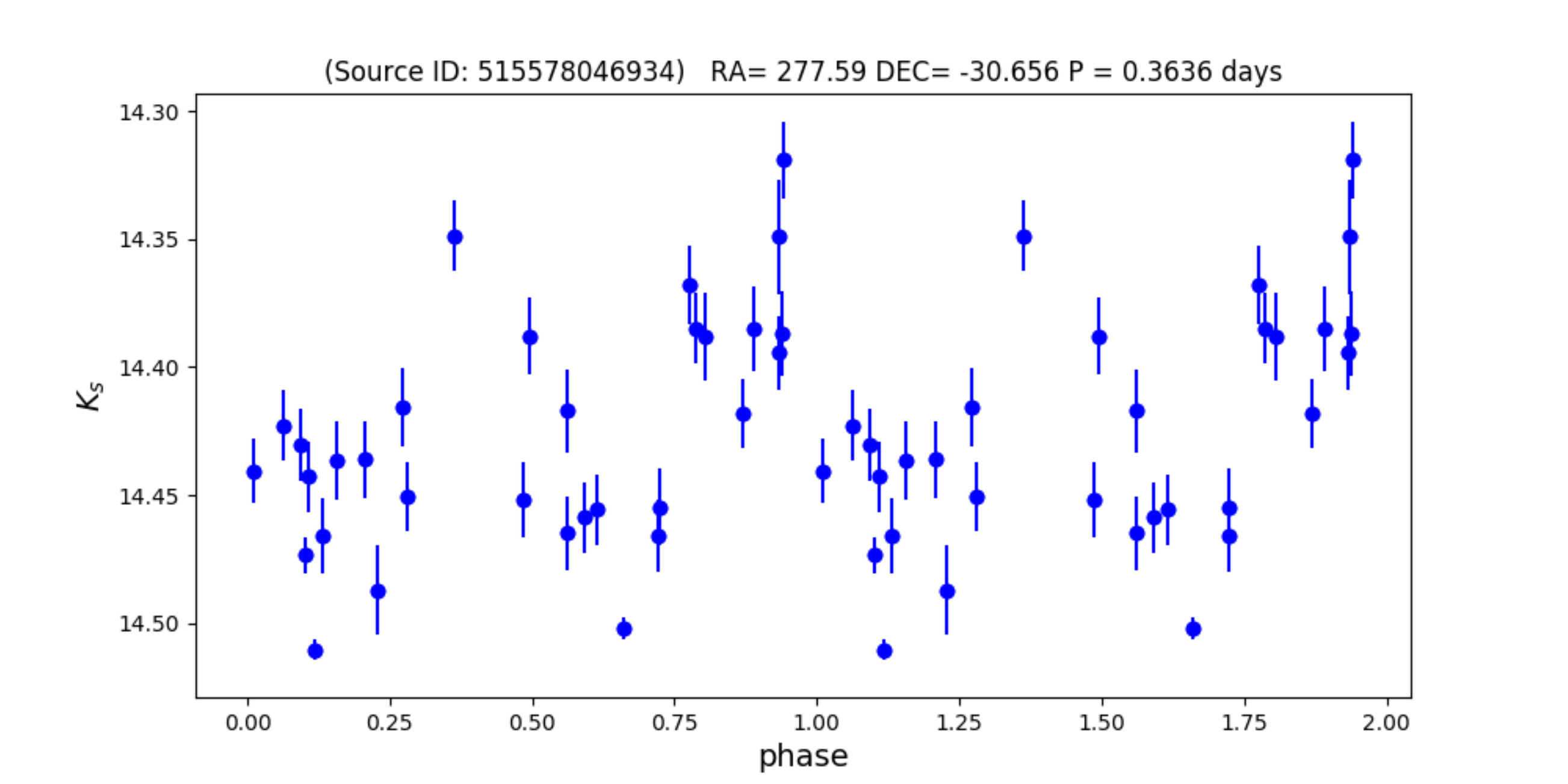} &
    \includegraphics[width=.33\textwidth]{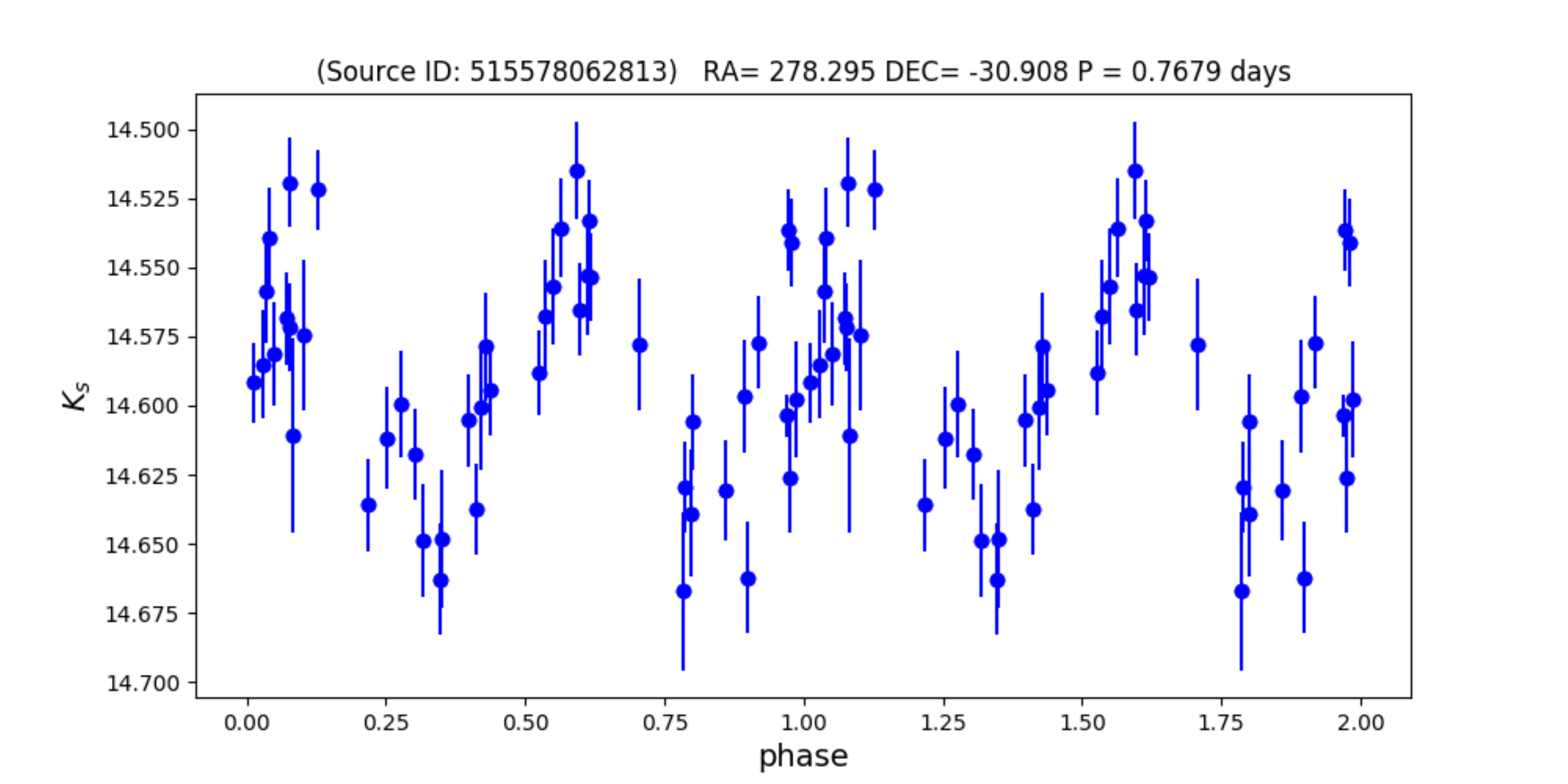} &
    \includegraphics[width=.33\textwidth]{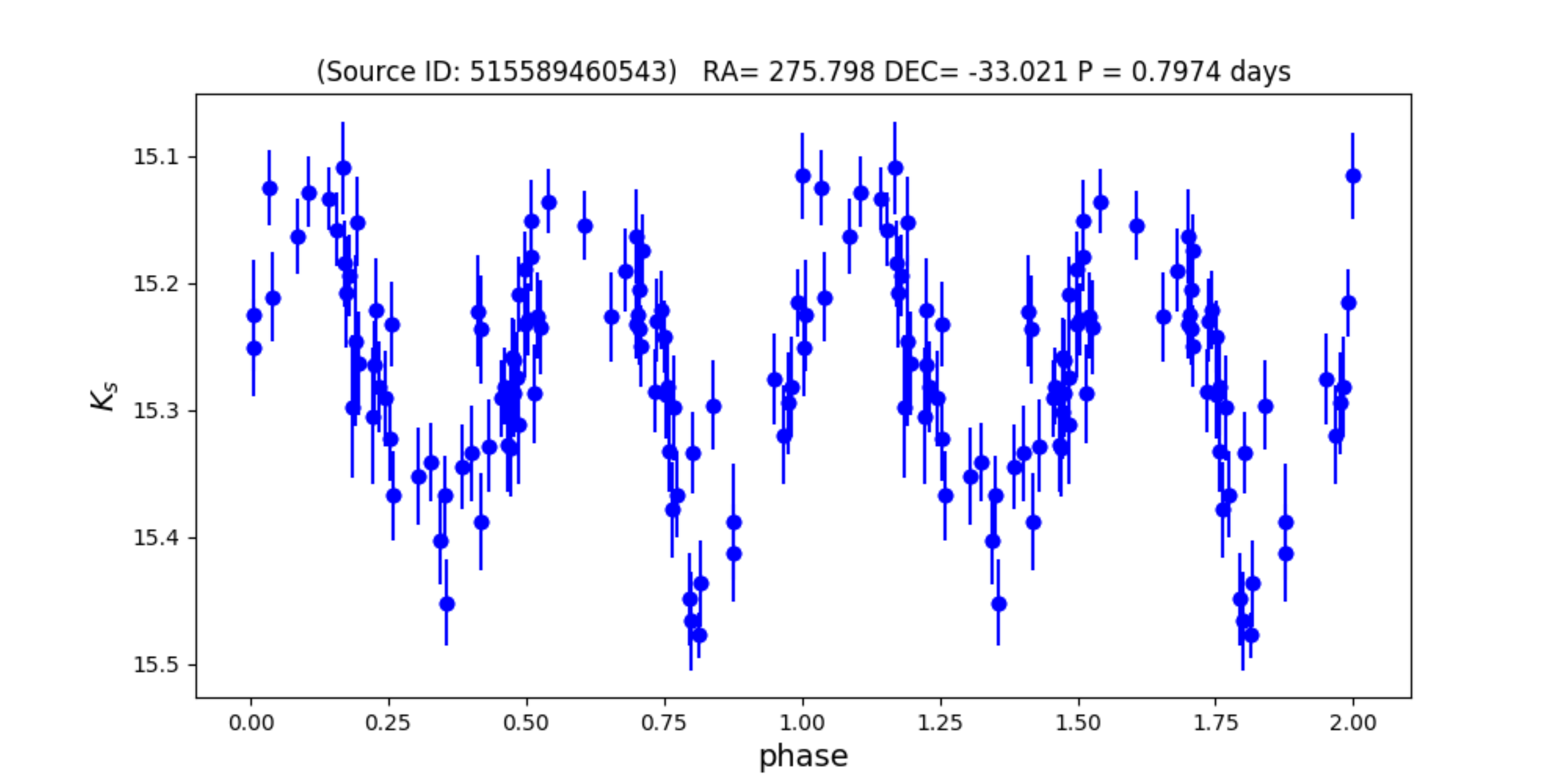} \\
    \includegraphics[width=.33\textwidth]{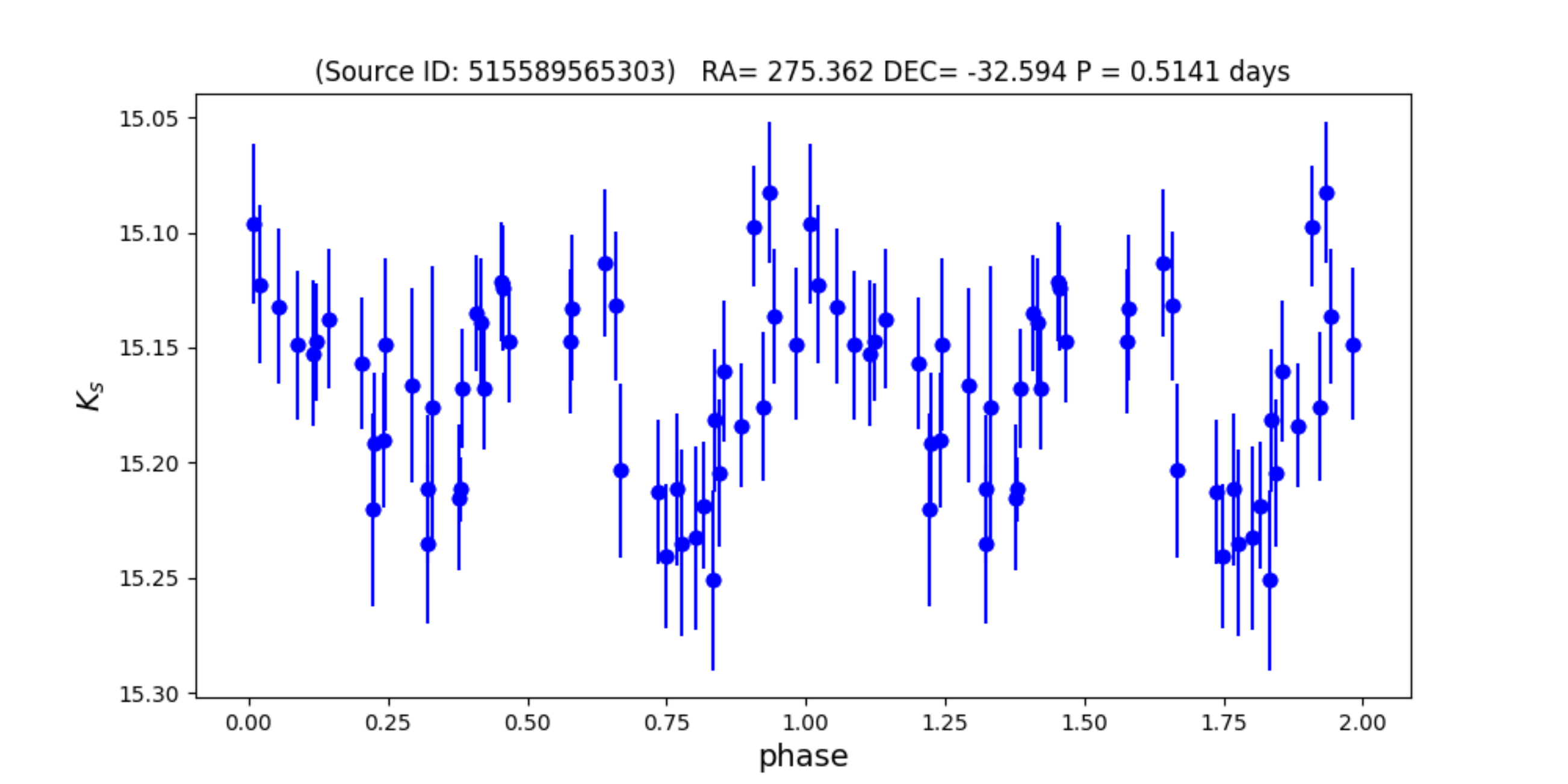} &
    \includegraphics[width=.33\textwidth]{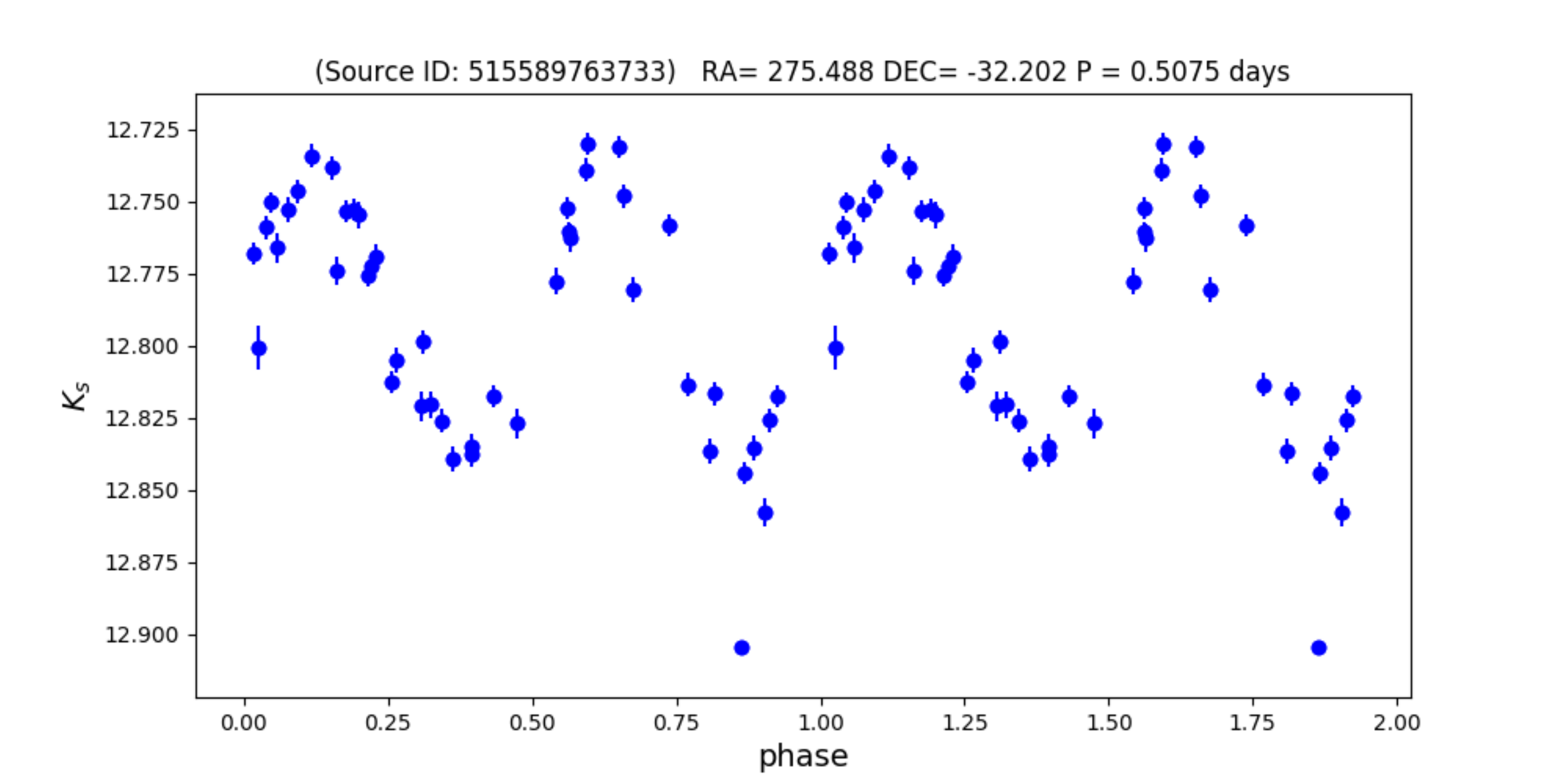} &
    \includegraphics[width=.33\textwidth]{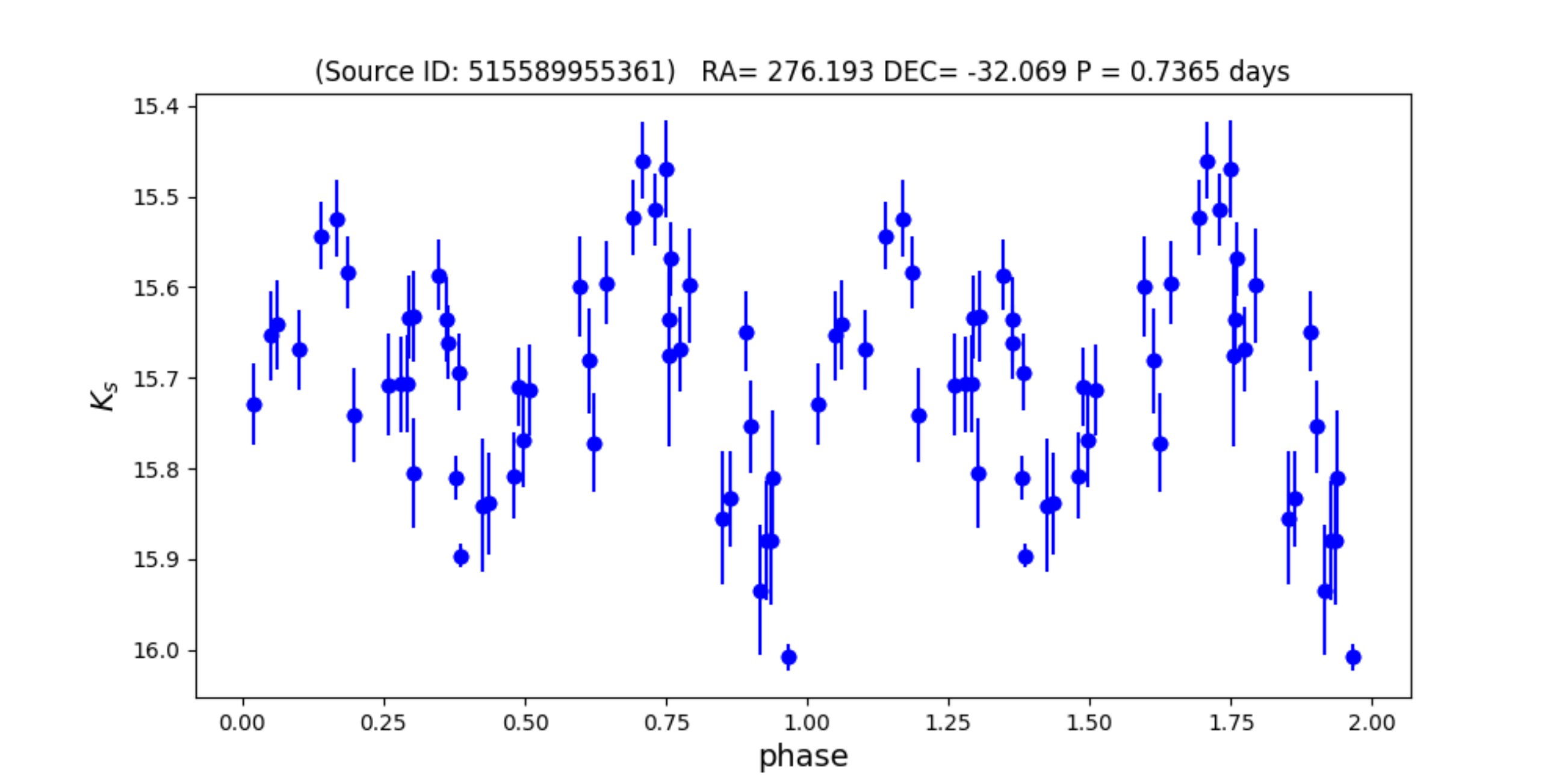} \\
    \includegraphics[width=.33\textwidth]{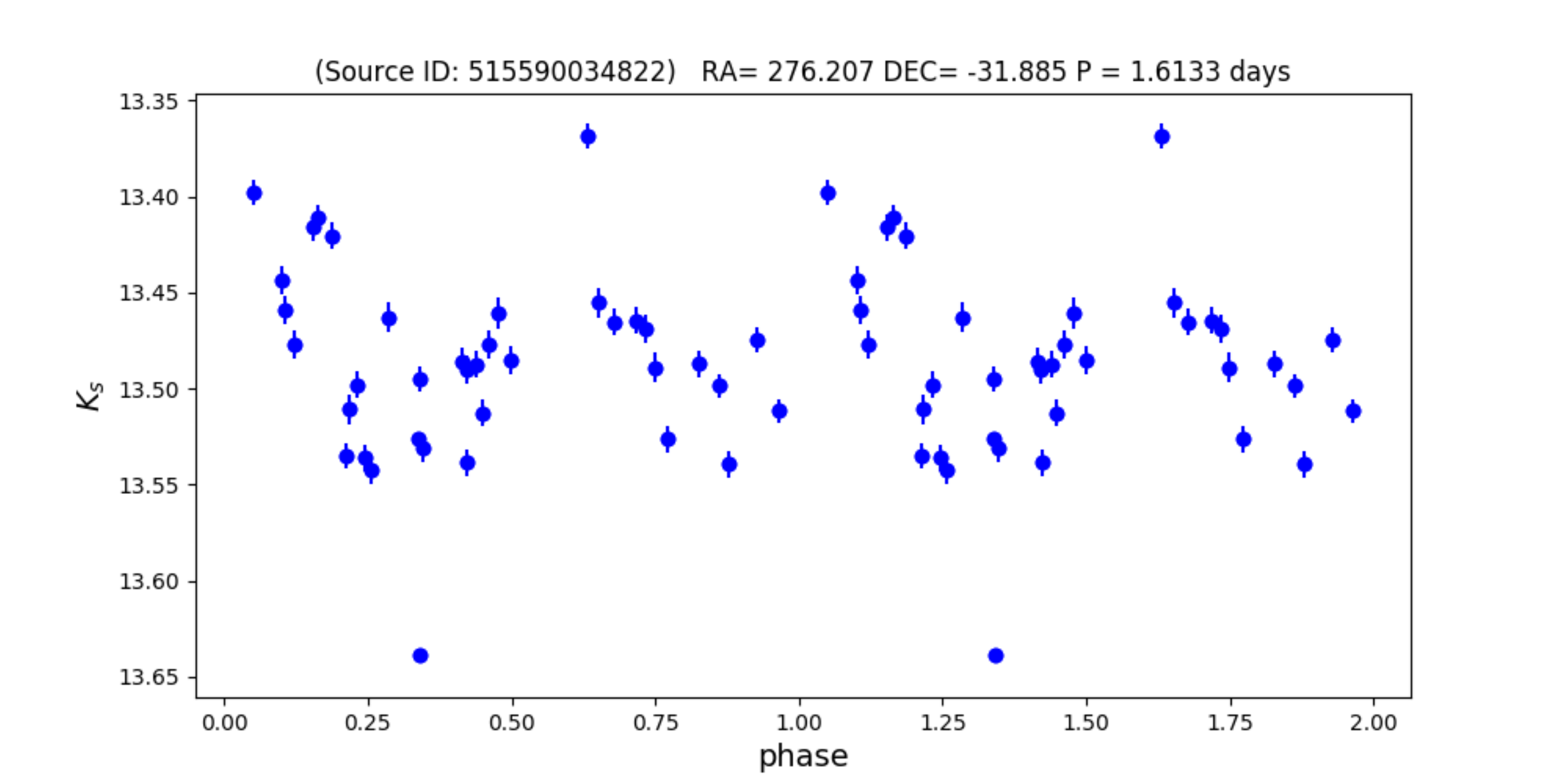} &
    \includegraphics[width=.33\textwidth]{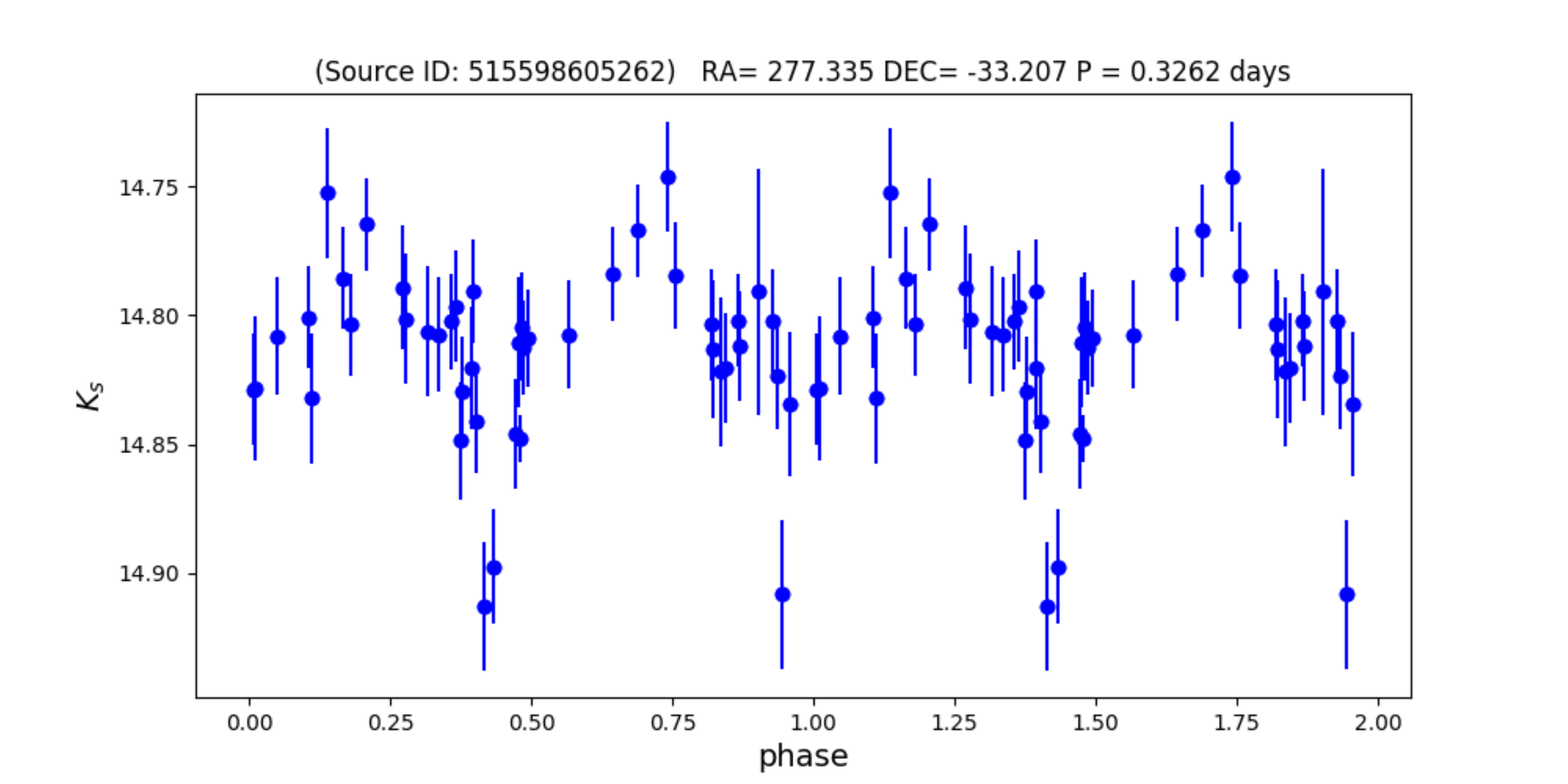} &
    \includegraphics[width=.33\textwidth]{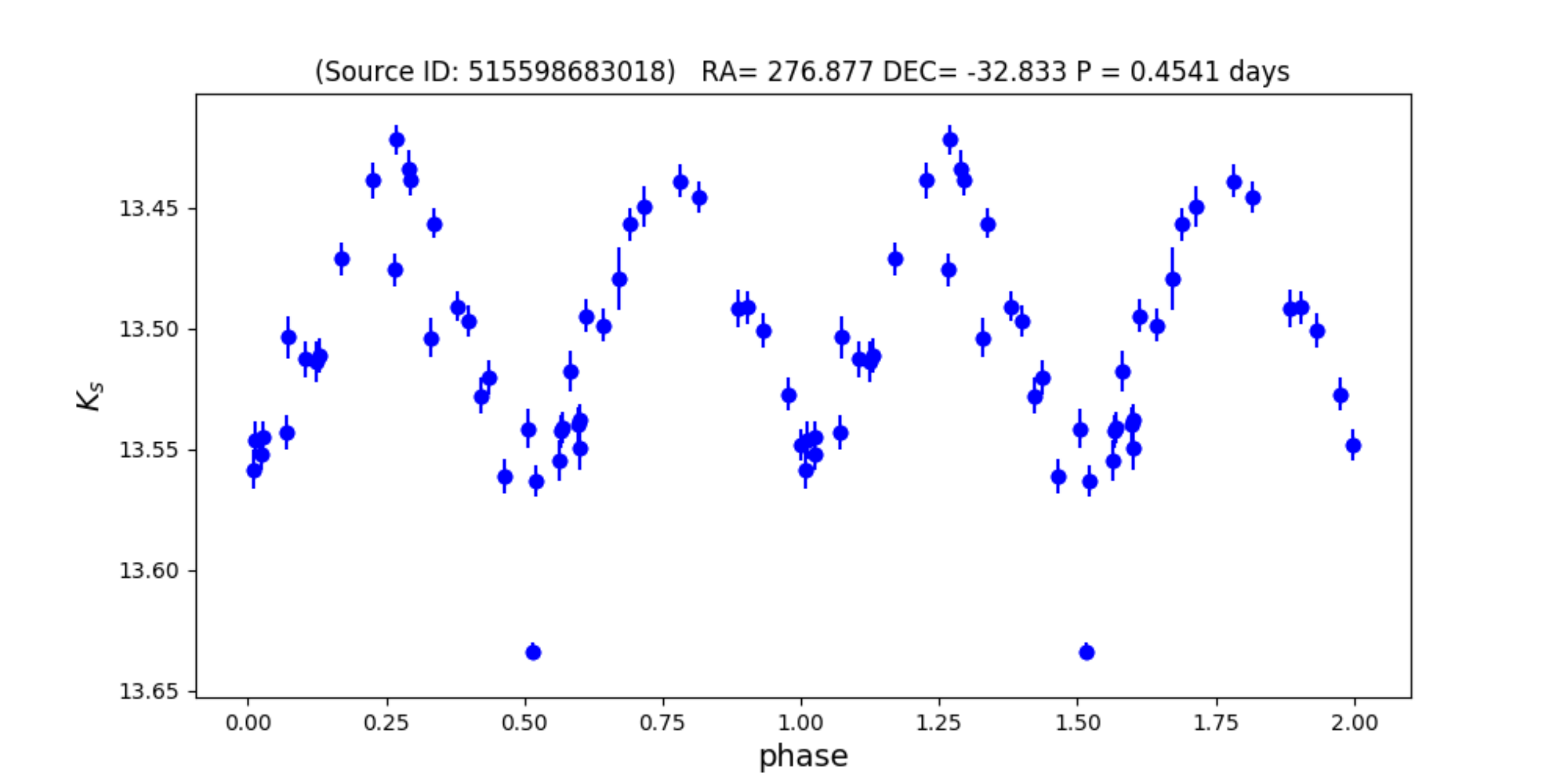} \\
    \includegraphics[width=.33\textwidth]{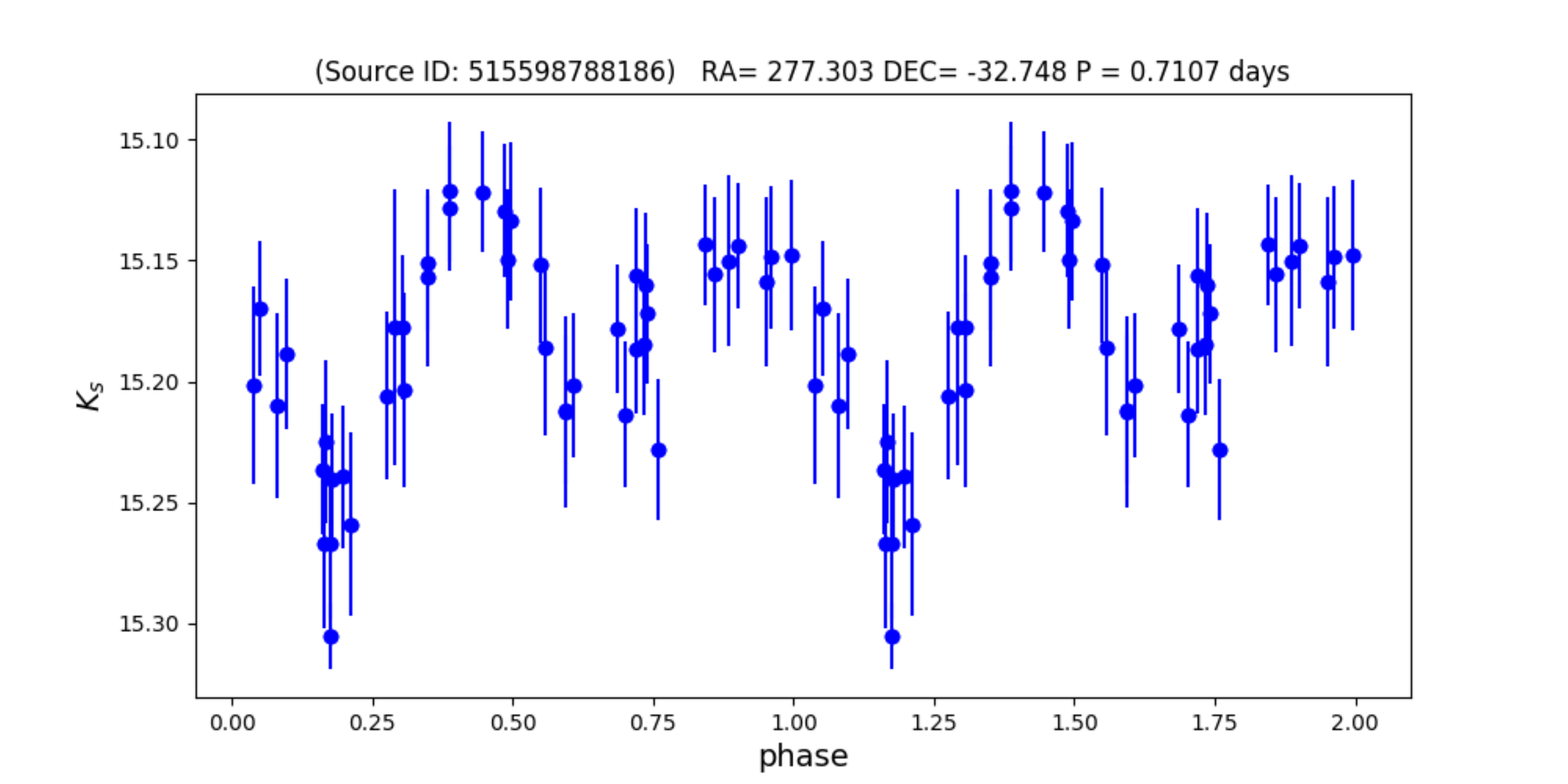} &
    \includegraphics[width=.33\textwidth]{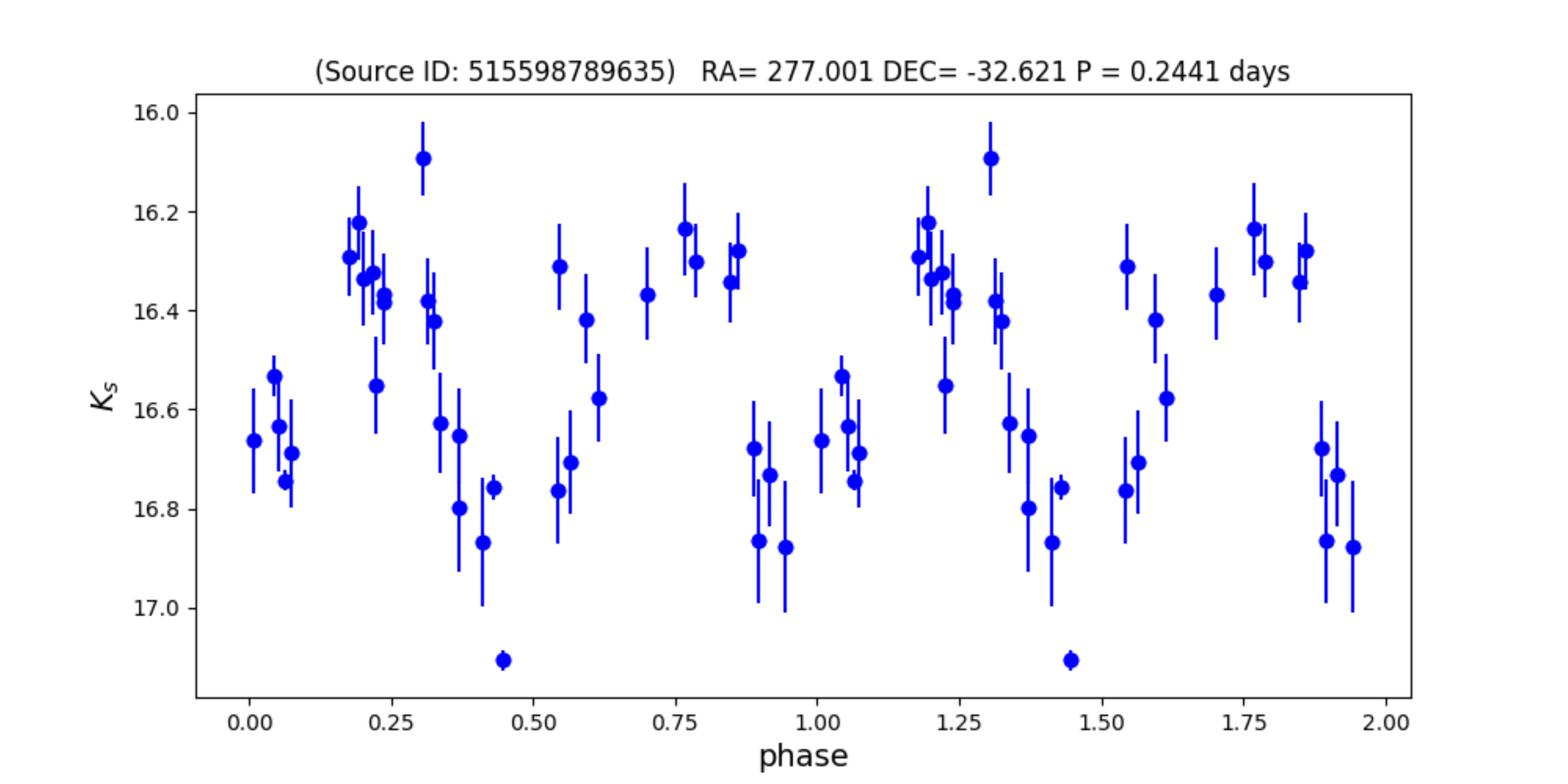} &
    \includegraphics[width=.33\textwidth]{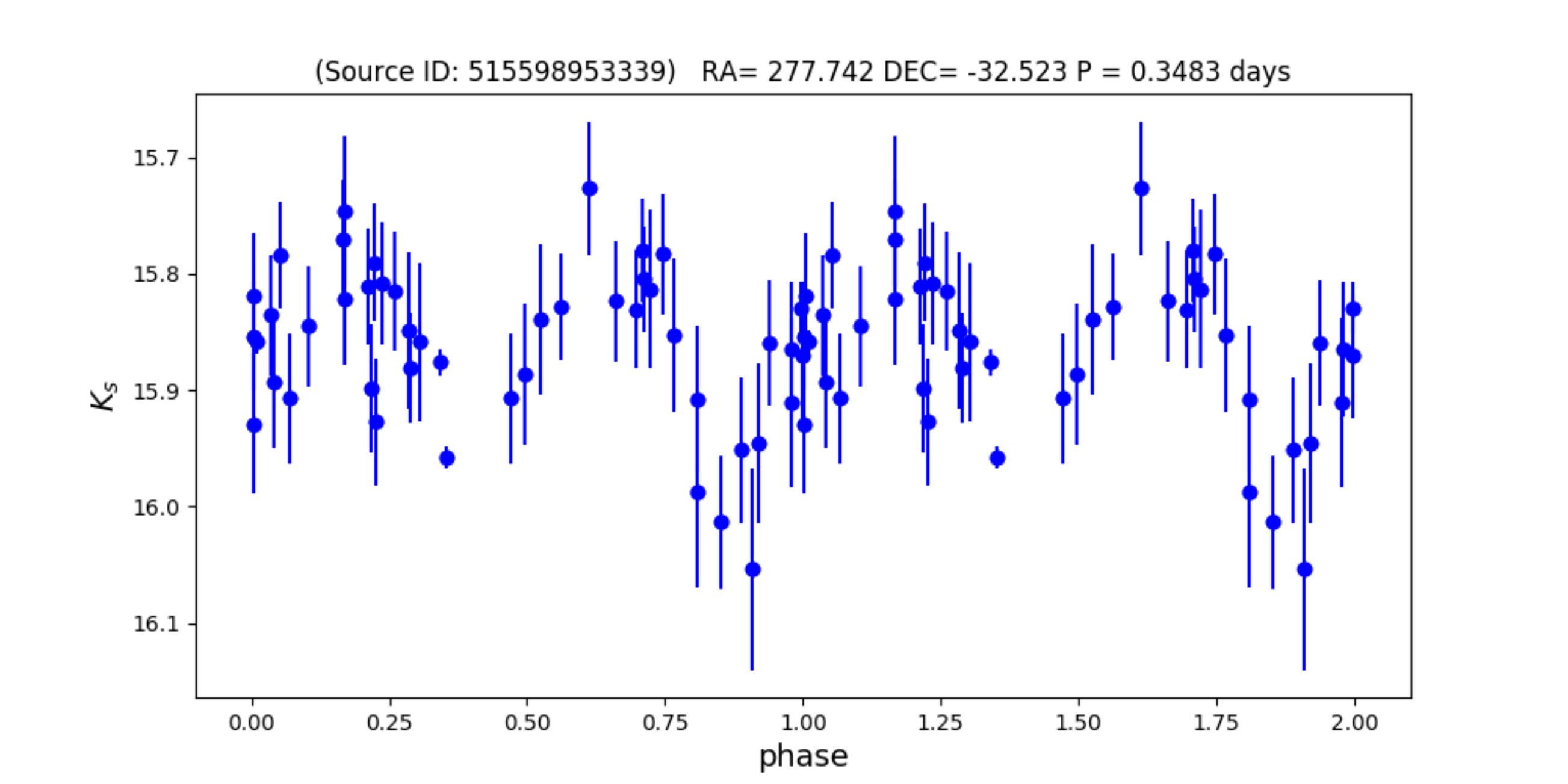} \\
    \includegraphics[width=.33\textwidth]{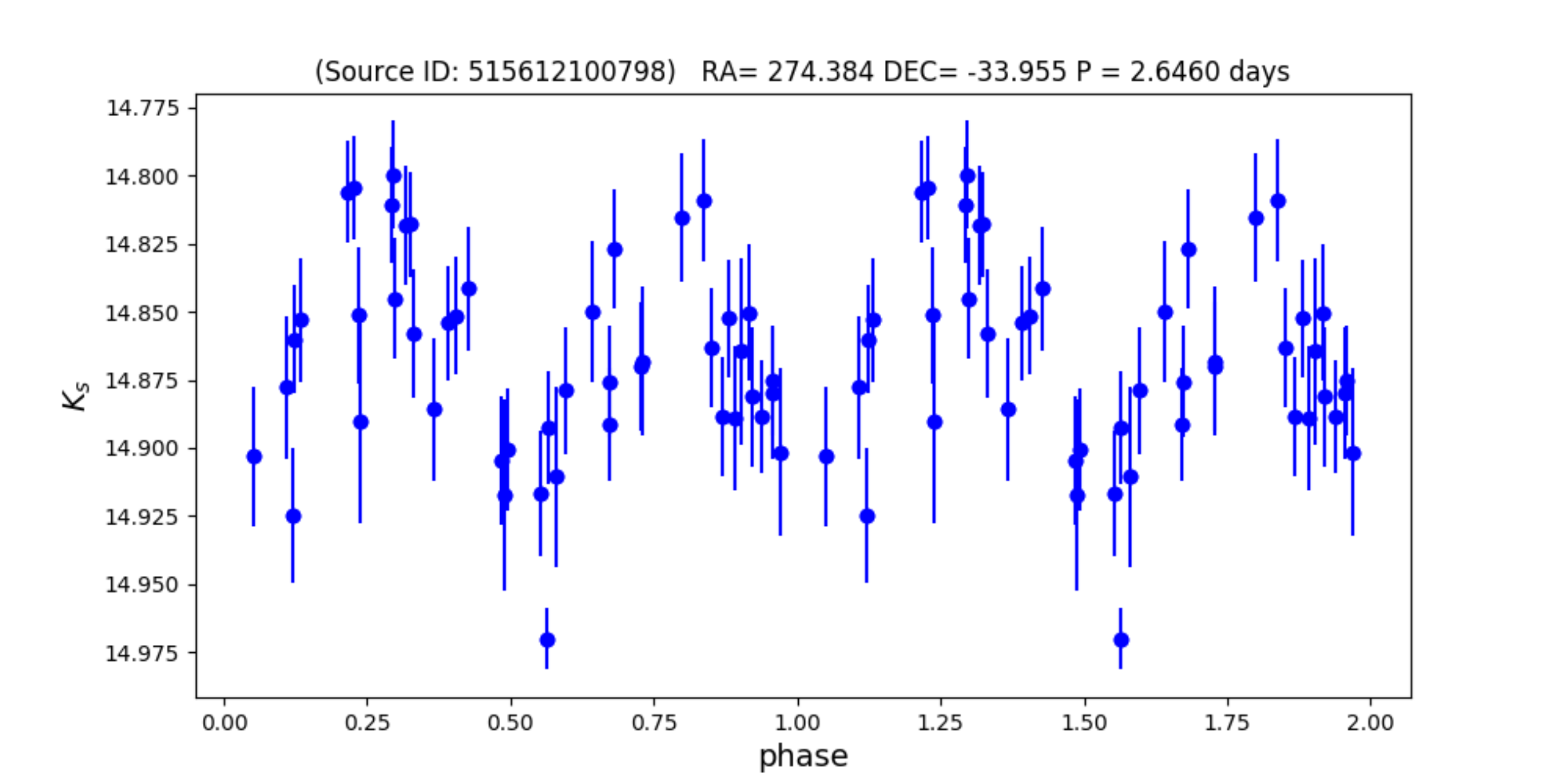} &
    \includegraphics[width=.33\textwidth]{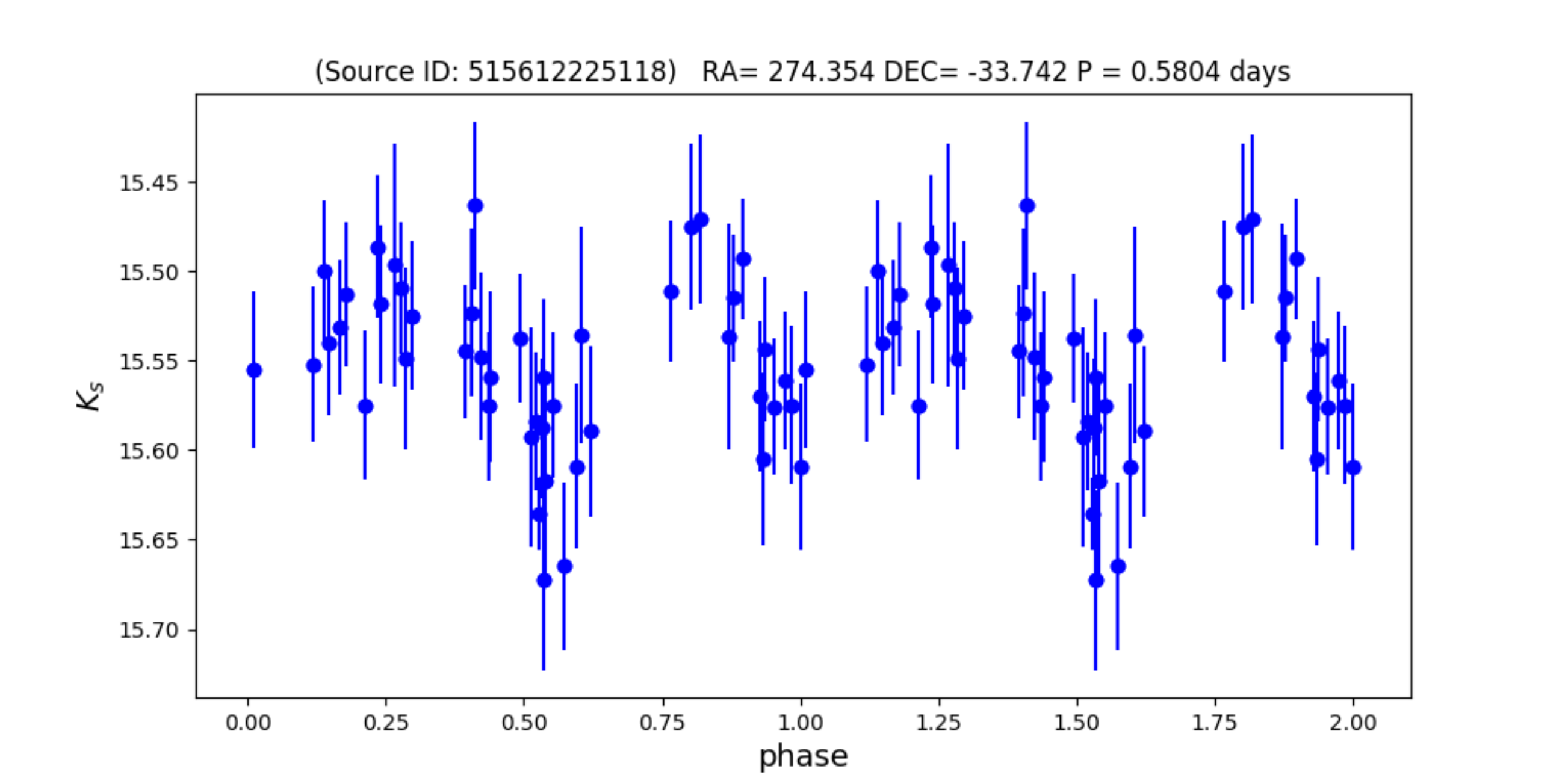} &
    \includegraphics[width=.33\textwidth]{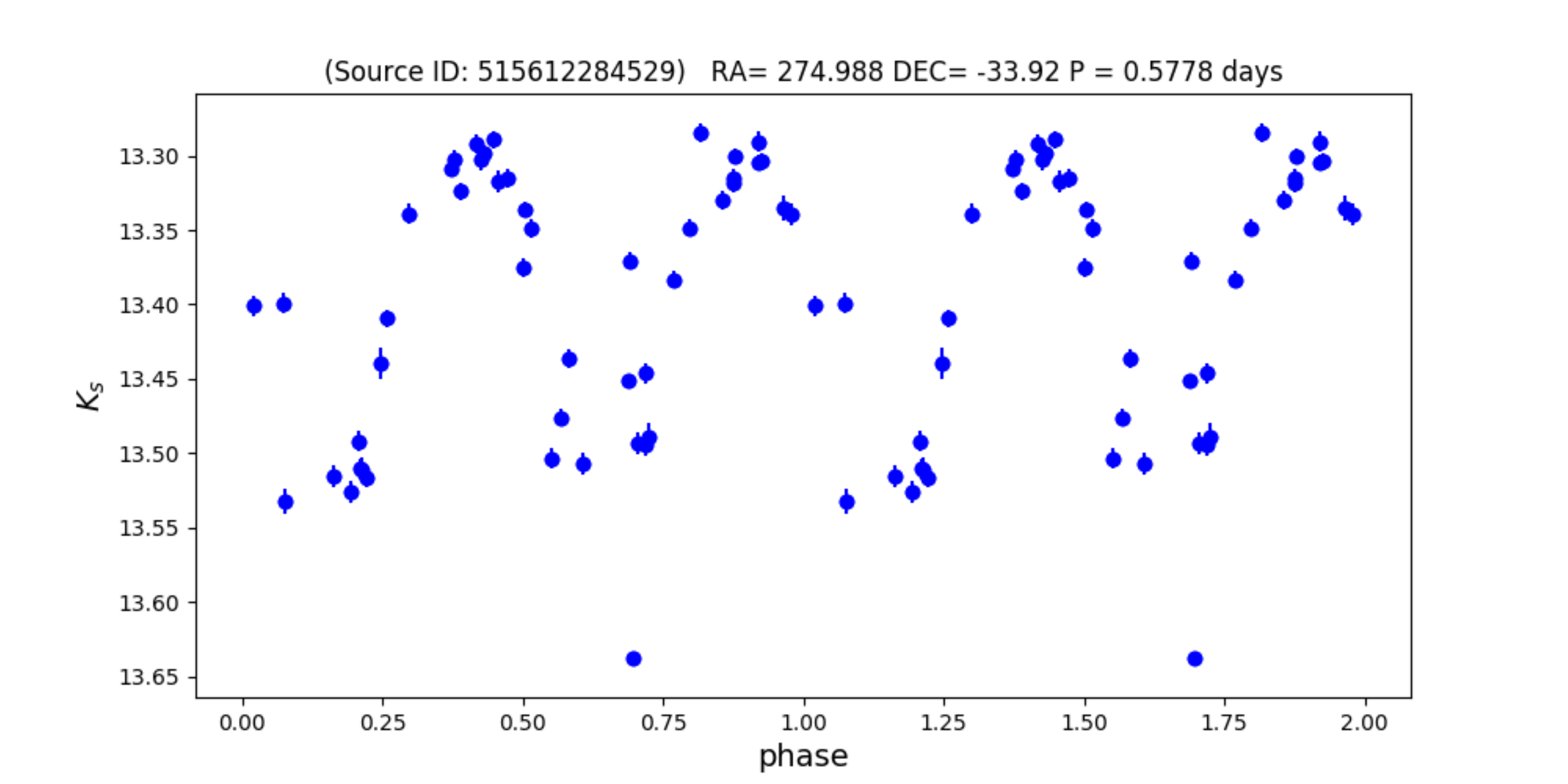} \\
    \includegraphics[width=.33\textwidth]{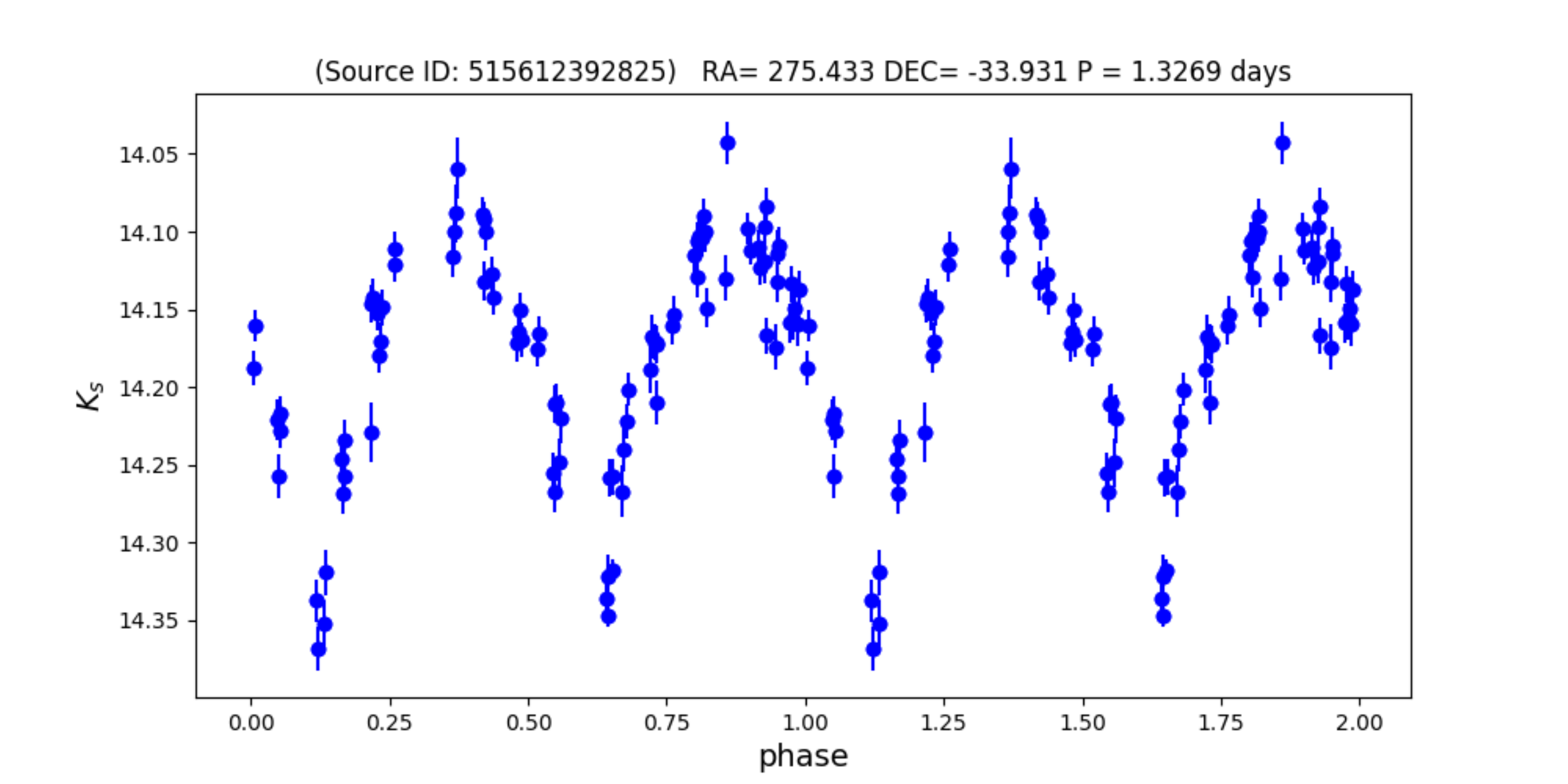} &
    \includegraphics[width=.33\textwidth]{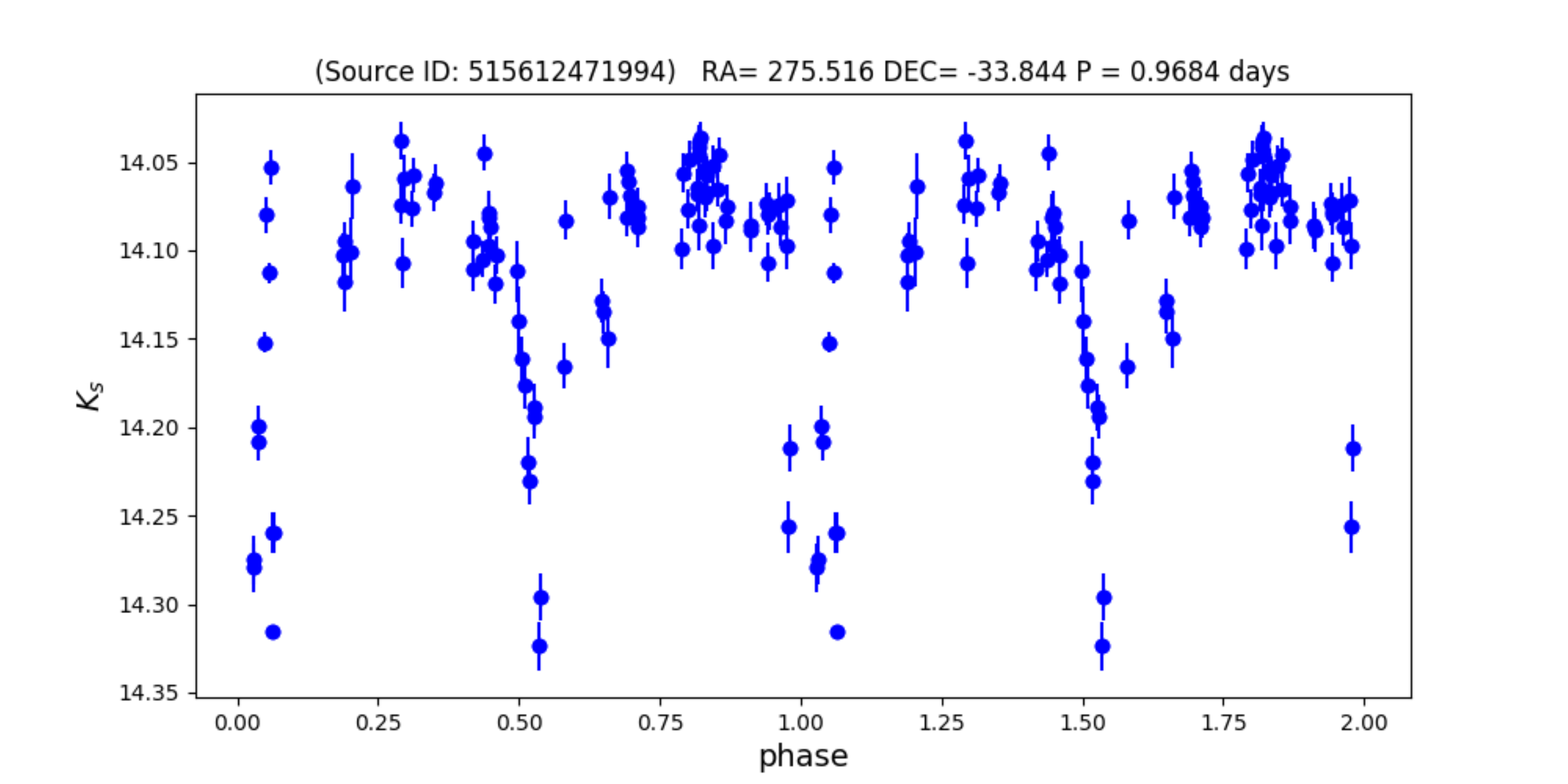} &   
    \includegraphics[width=.33\textwidth]{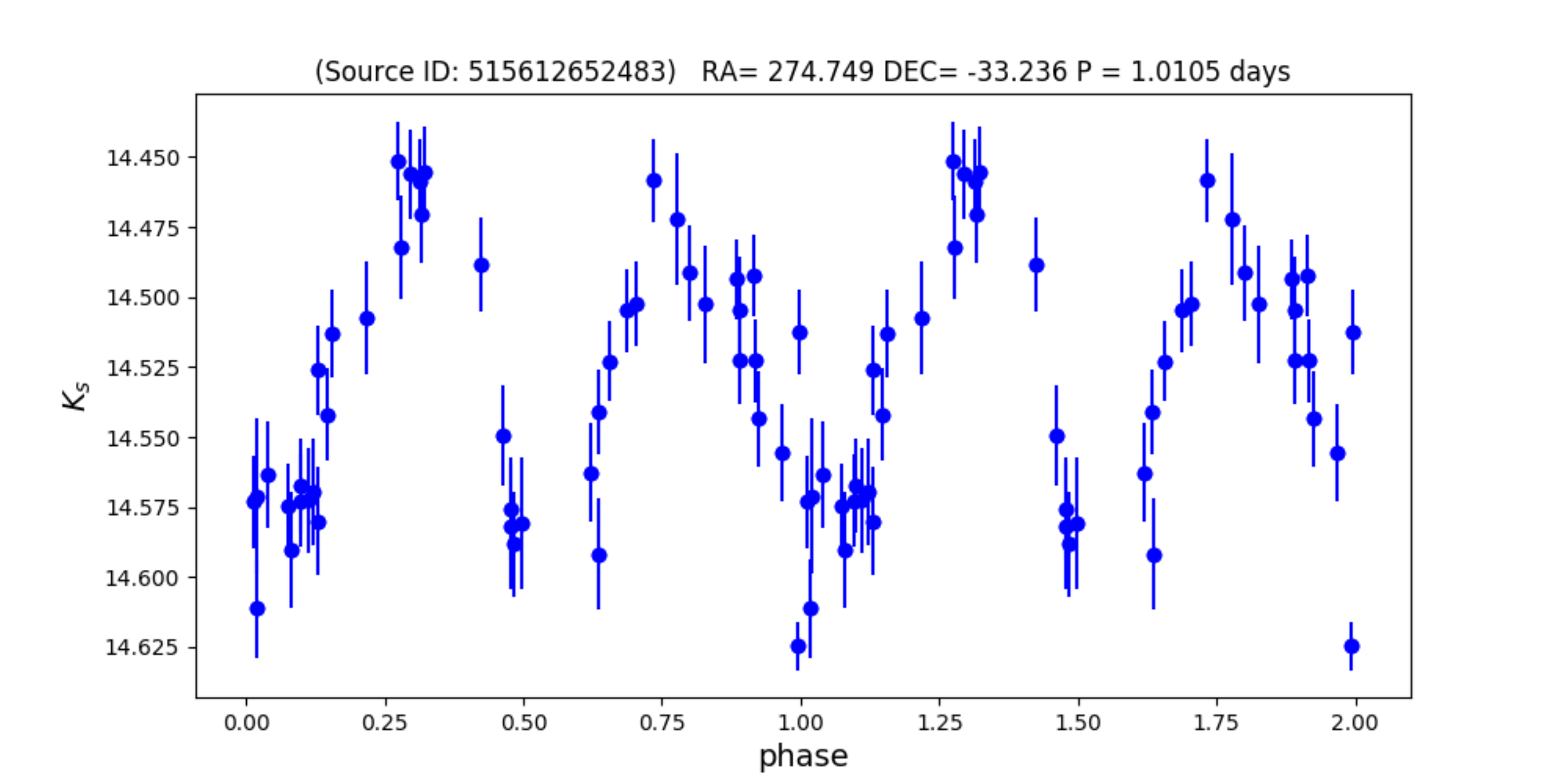} \\
    \includegraphics[width=.33\textwidth]{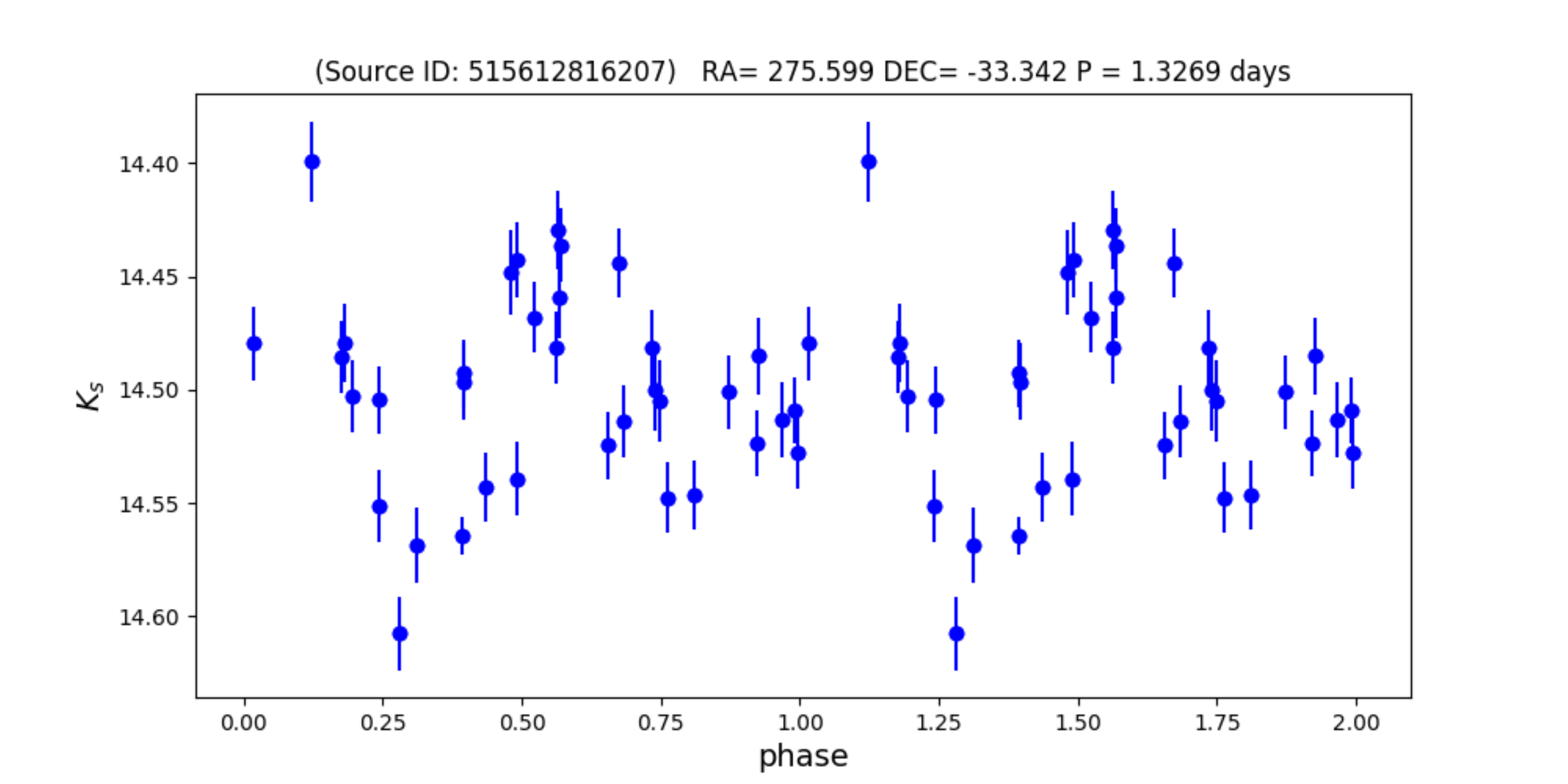} &
    \includegraphics[width=.33\textwidth]{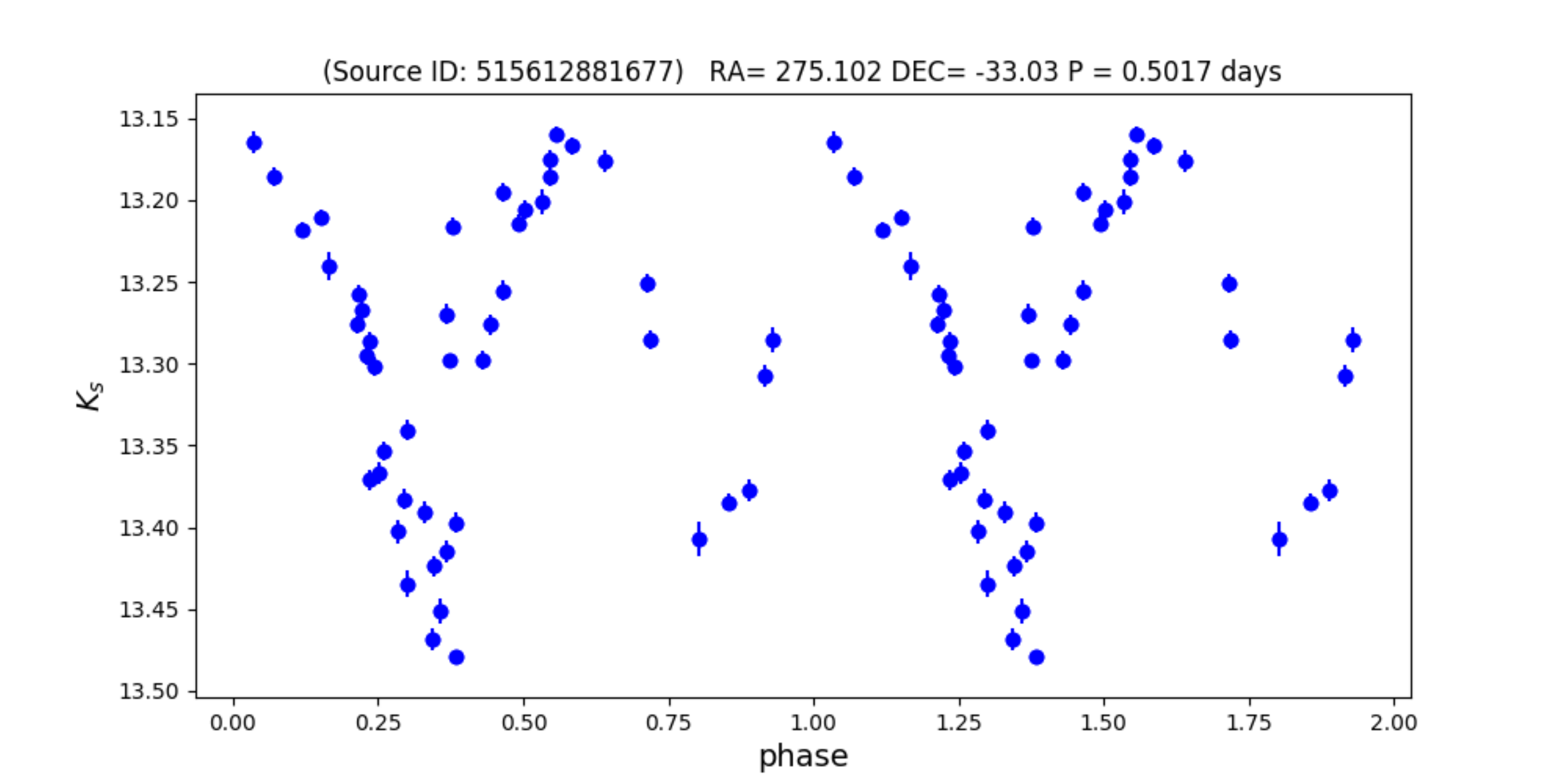} &
    \includegraphics[width=.33\textwidth]{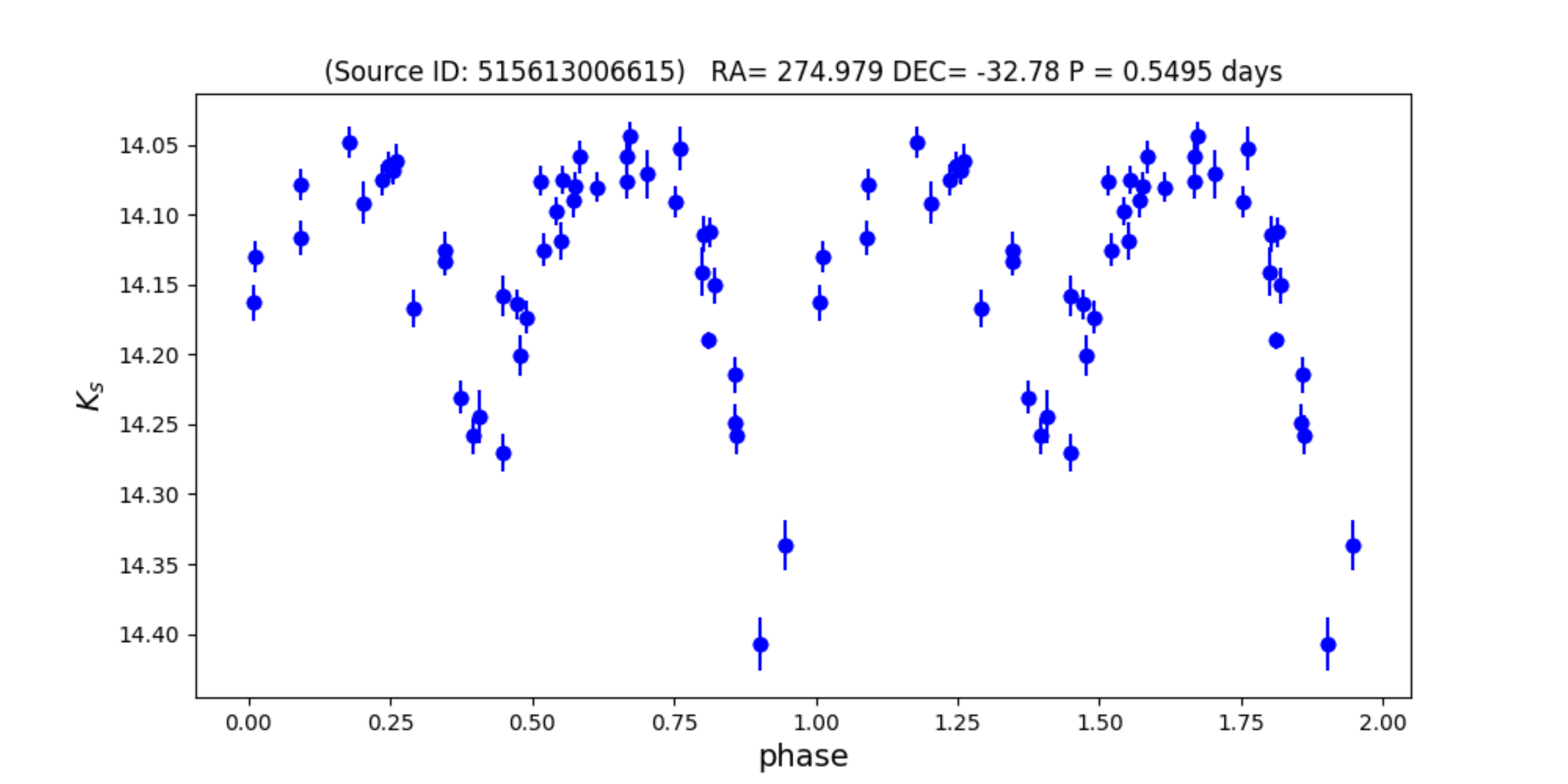} \\ 
  \end{tabular}
  \caption{BHB stars light curves corresponding to $1^{st}$ category eclipsing binaries.}
  \label{fig:bhb5}
\end{figure*}

\begin{figure*}[htb]
\centering
  \begin{tabular}{@{}ccc@{}}    
    \includegraphics[width=.33\textwidth]{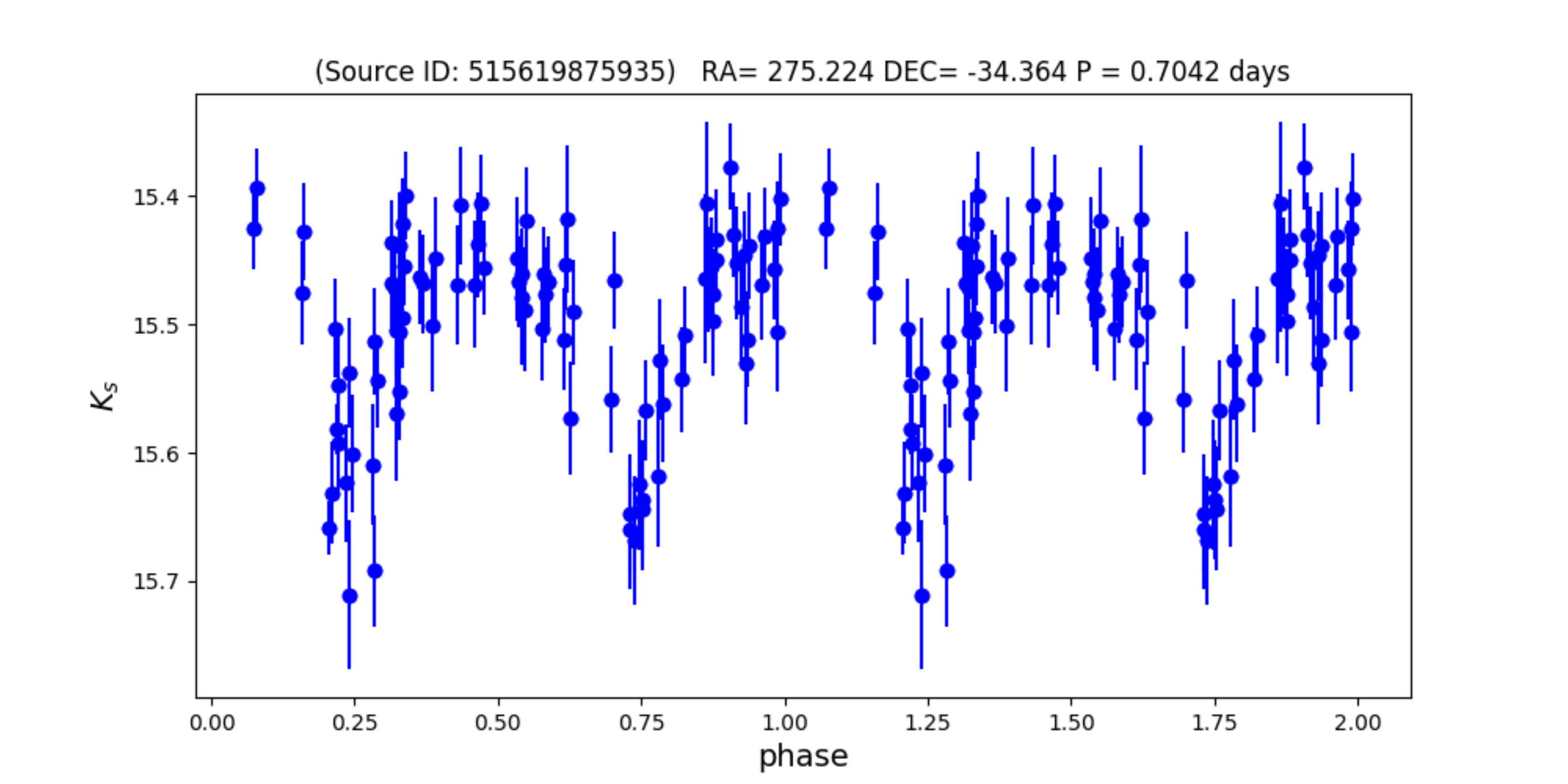} &
    \includegraphics[width=.33\textwidth]{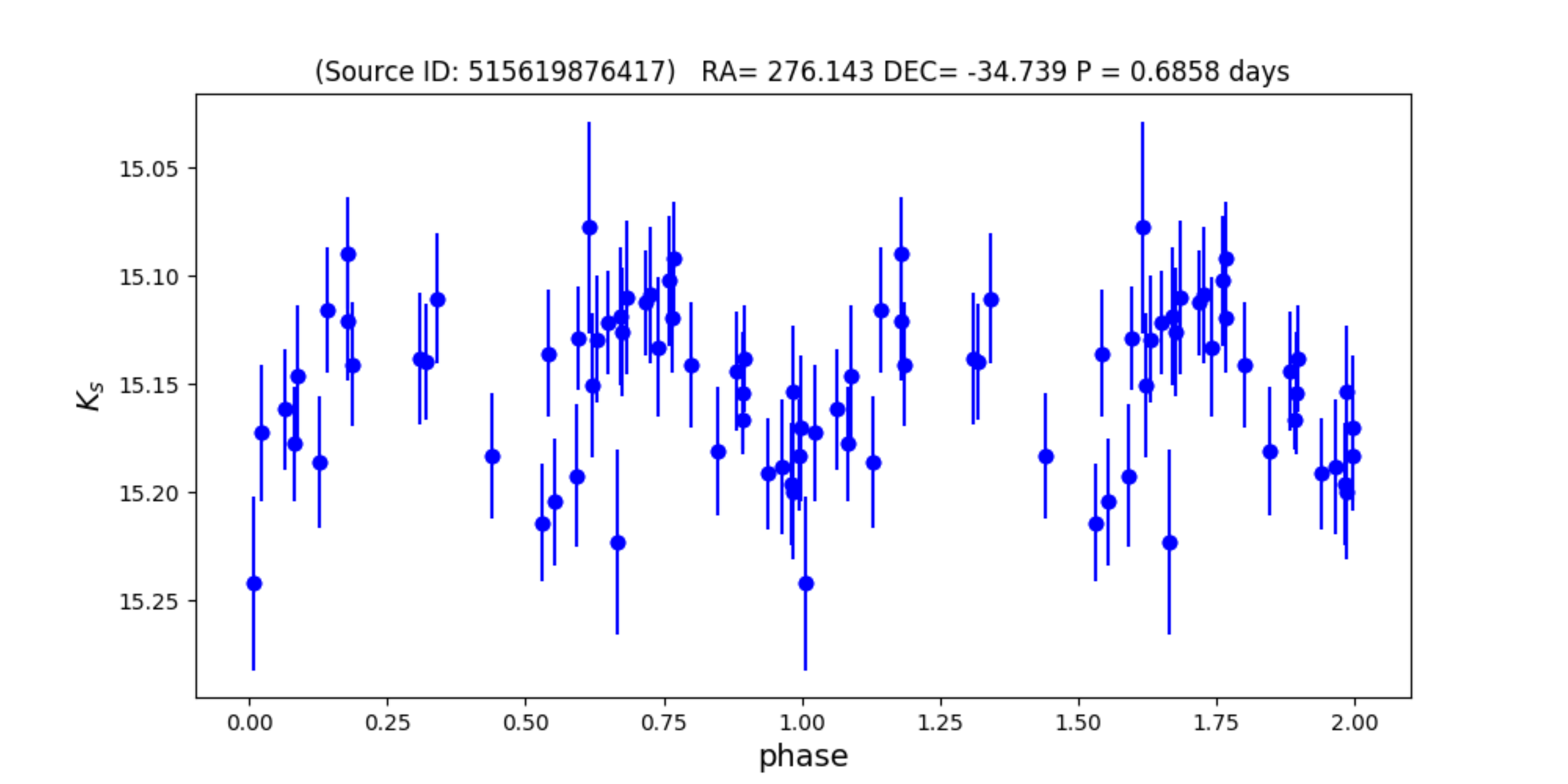} &
    \includegraphics[width=.33\textwidth]{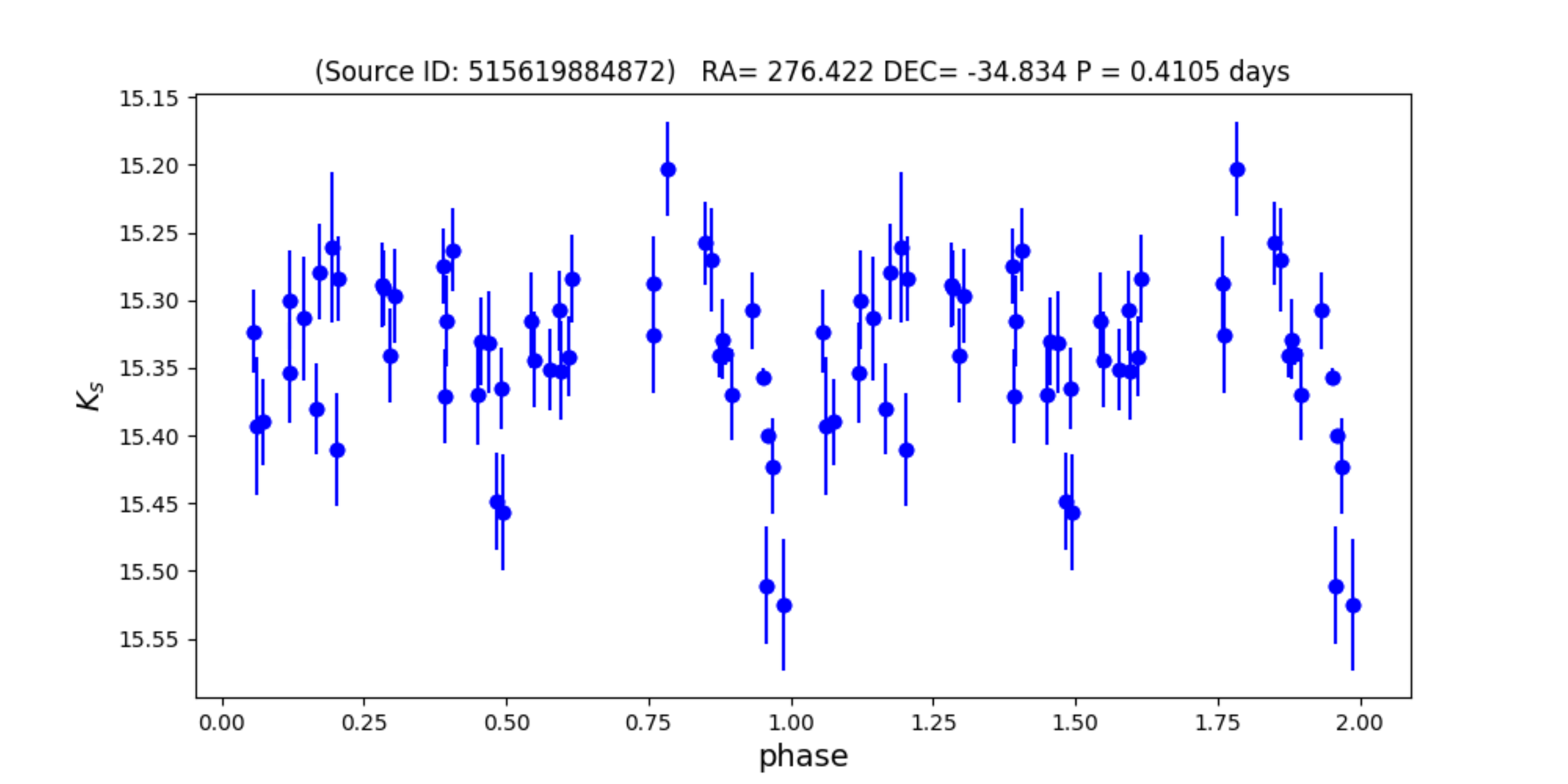} \\
    \includegraphics[width=.33\textwidth]{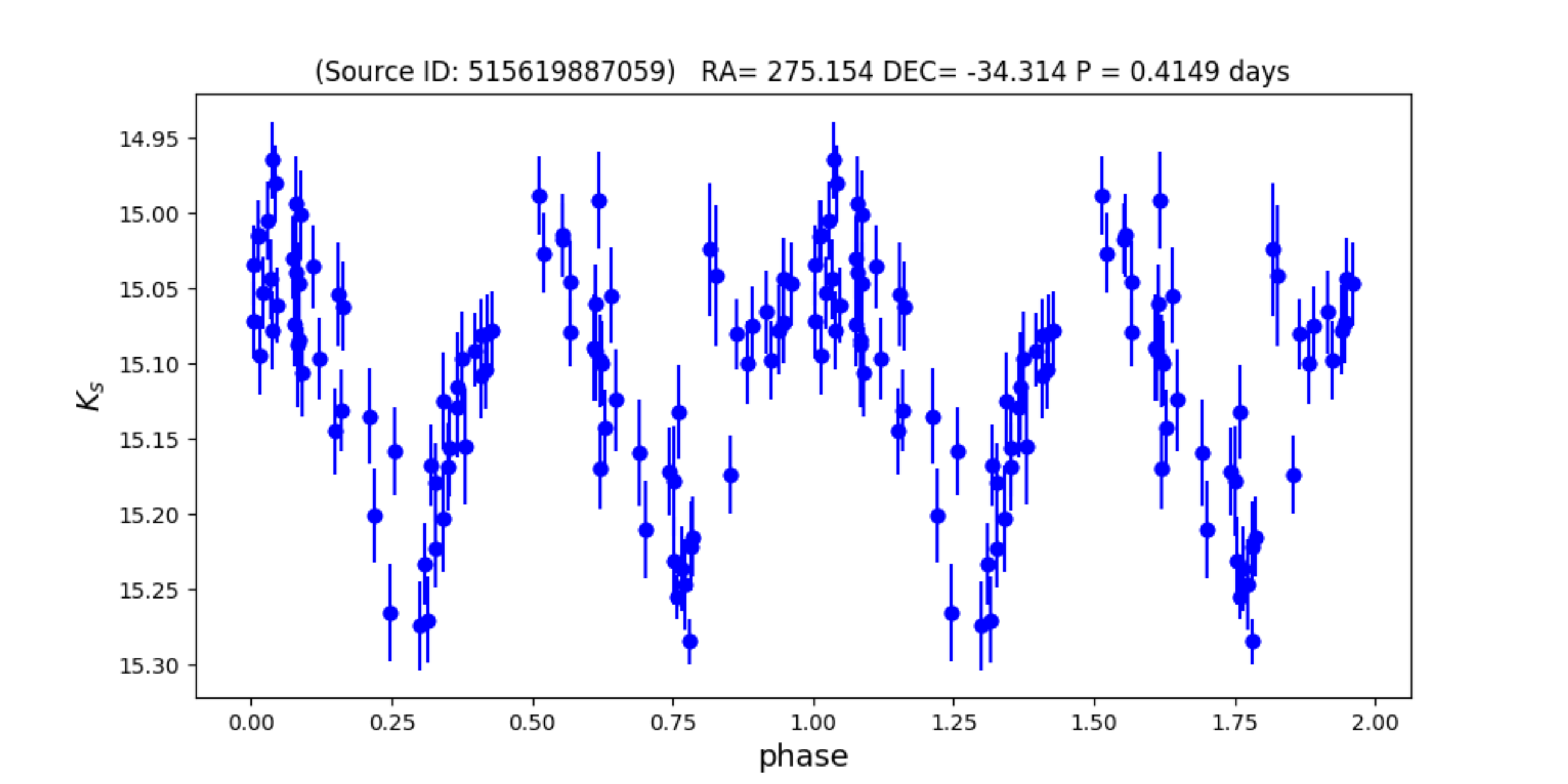} &
    \includegraphics[width=.33\textwidth]{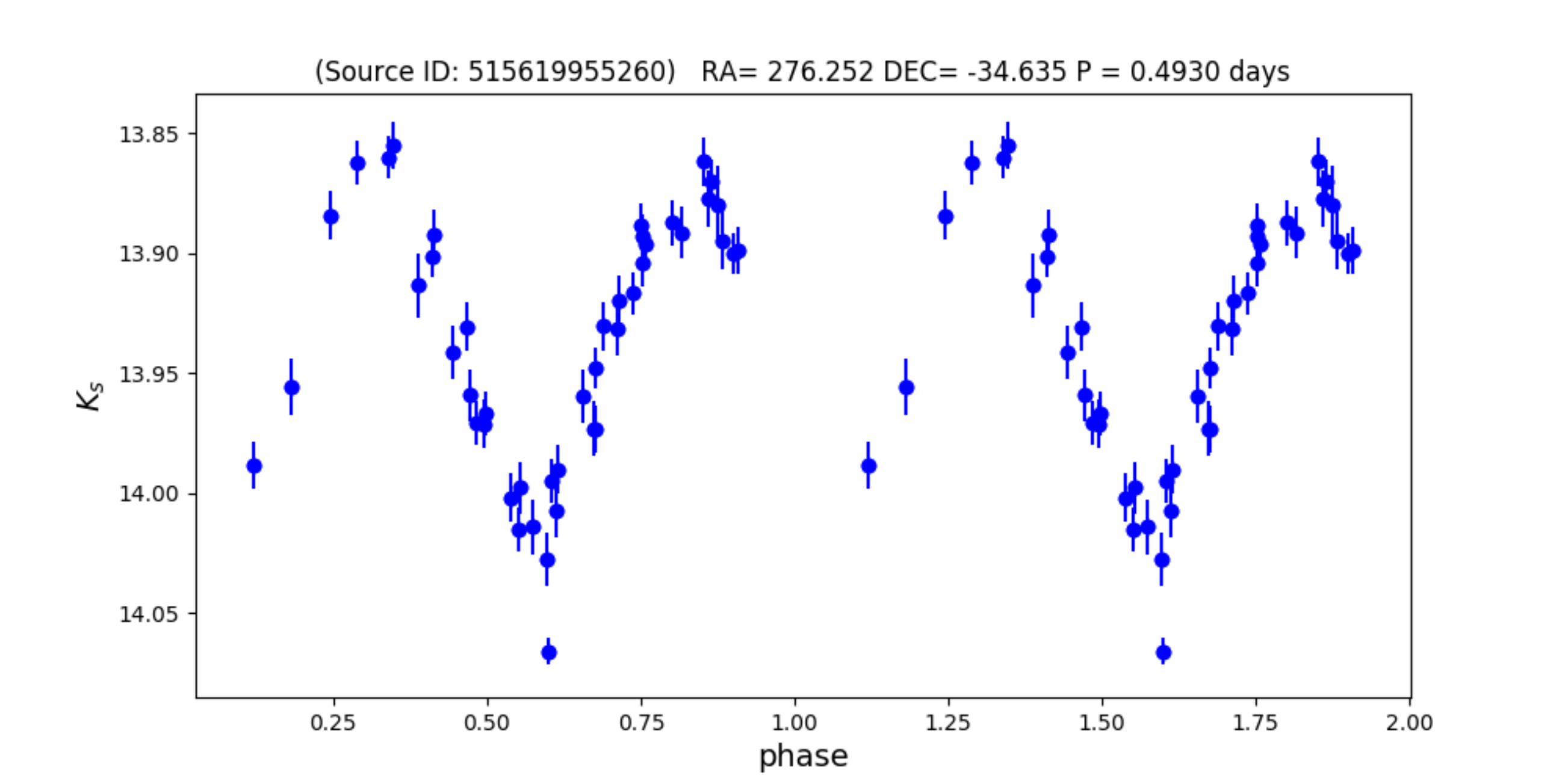} &
    \includegraphics[width=.33\textwidth]{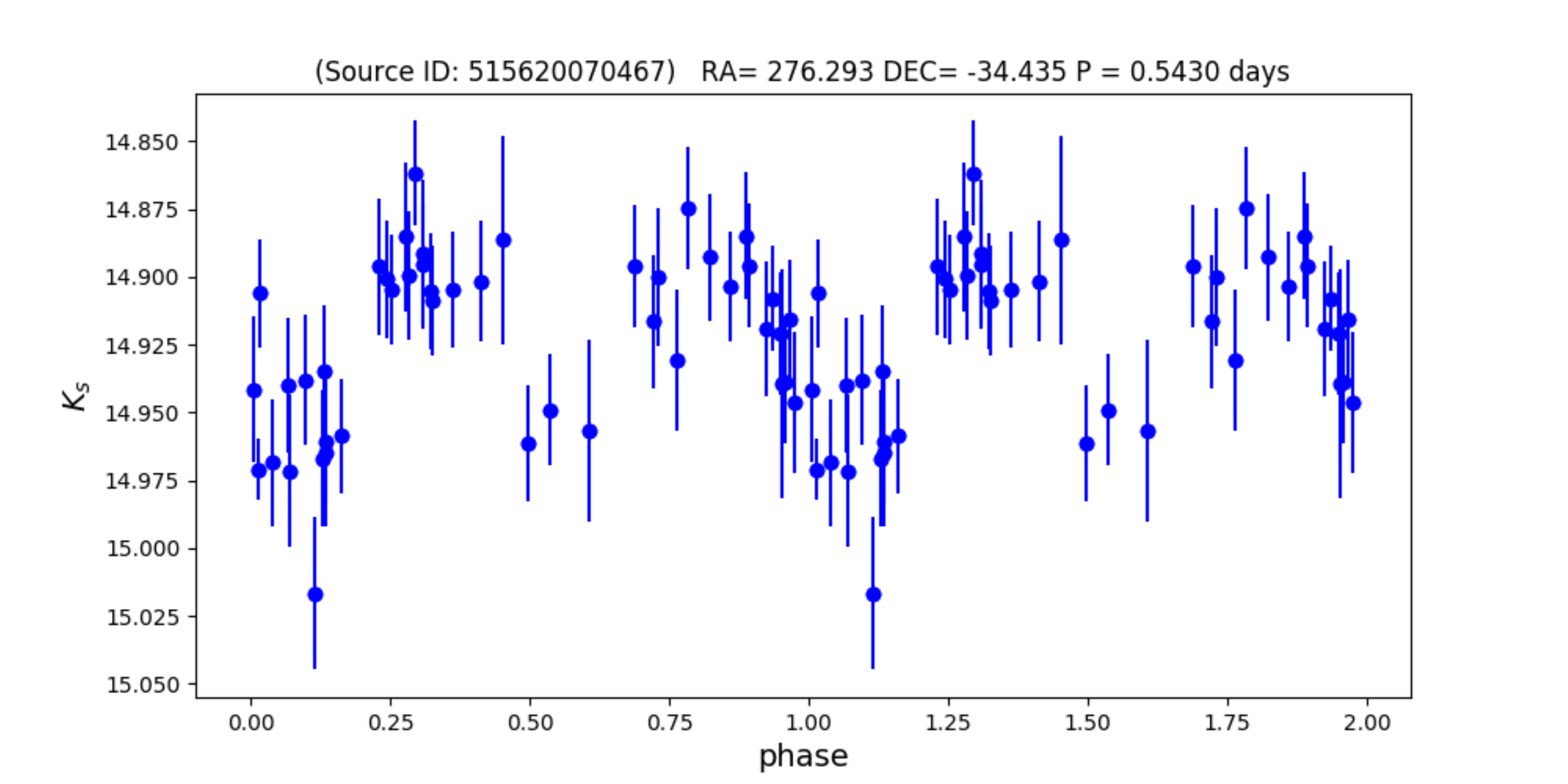} \\
    \includegraphics[width=.33\textwidth]{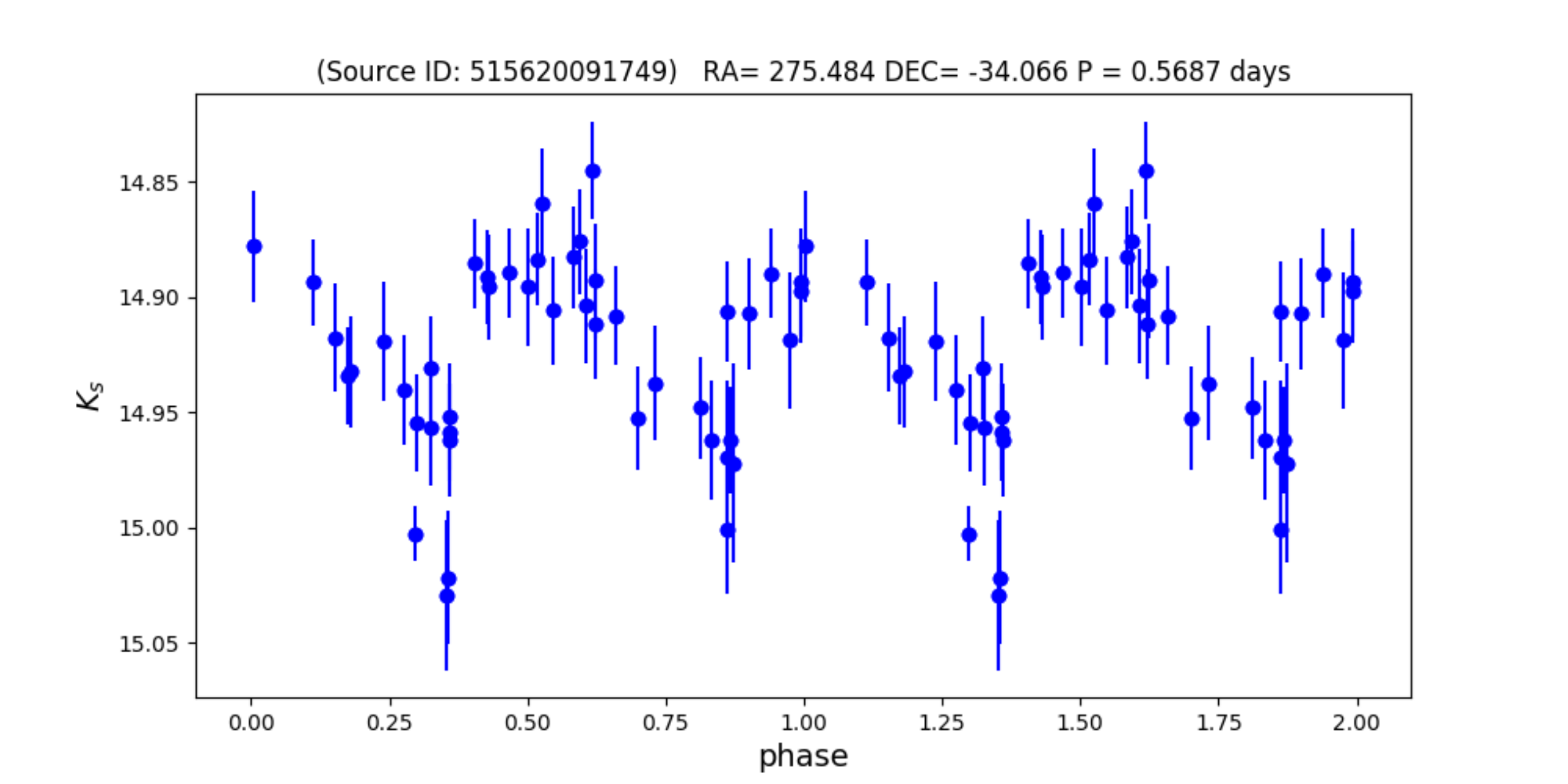} &
    \includegraphics[width=.33\textwidth]{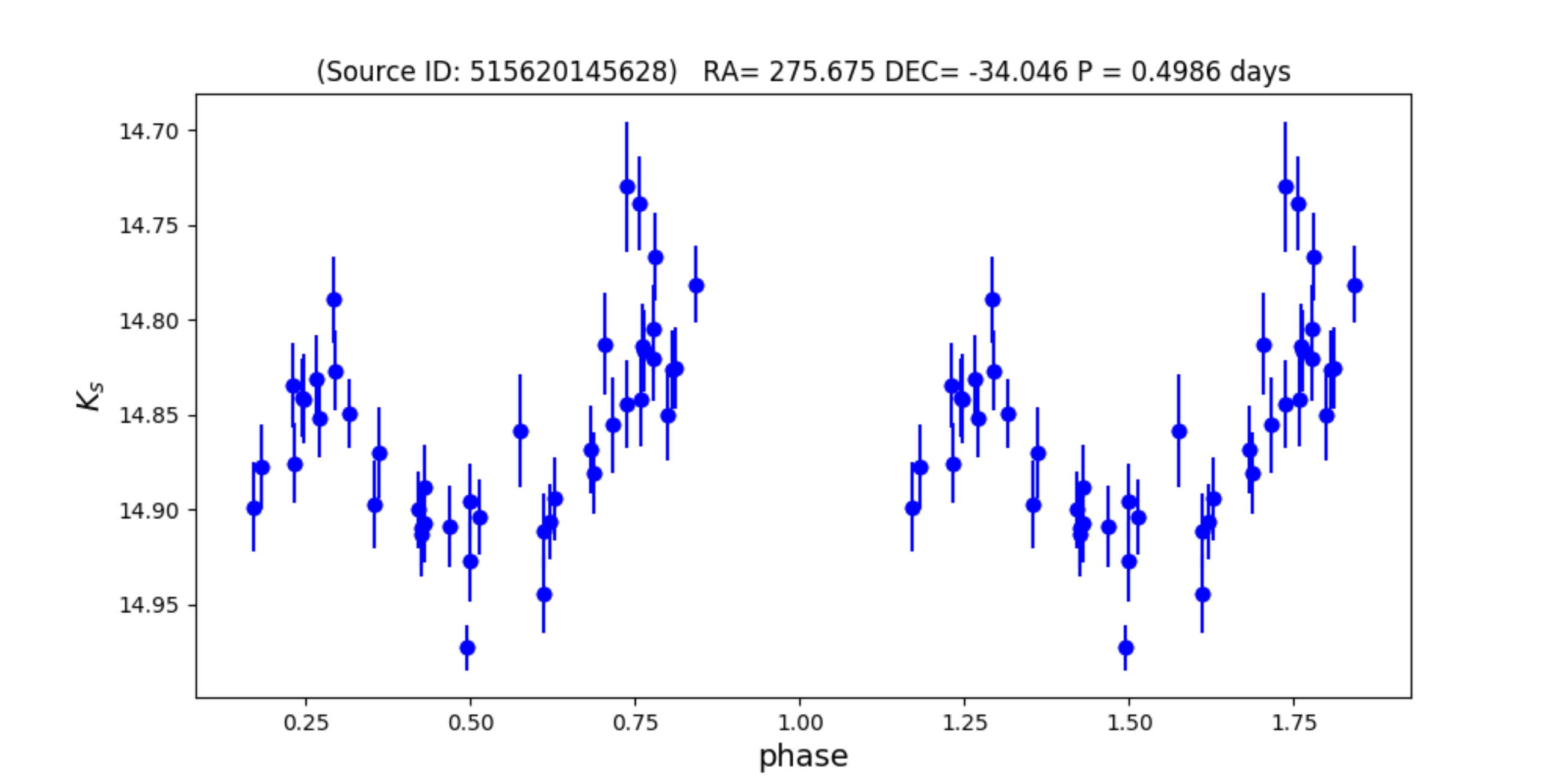} &
    \includegraphics[width=.33\textwidth]{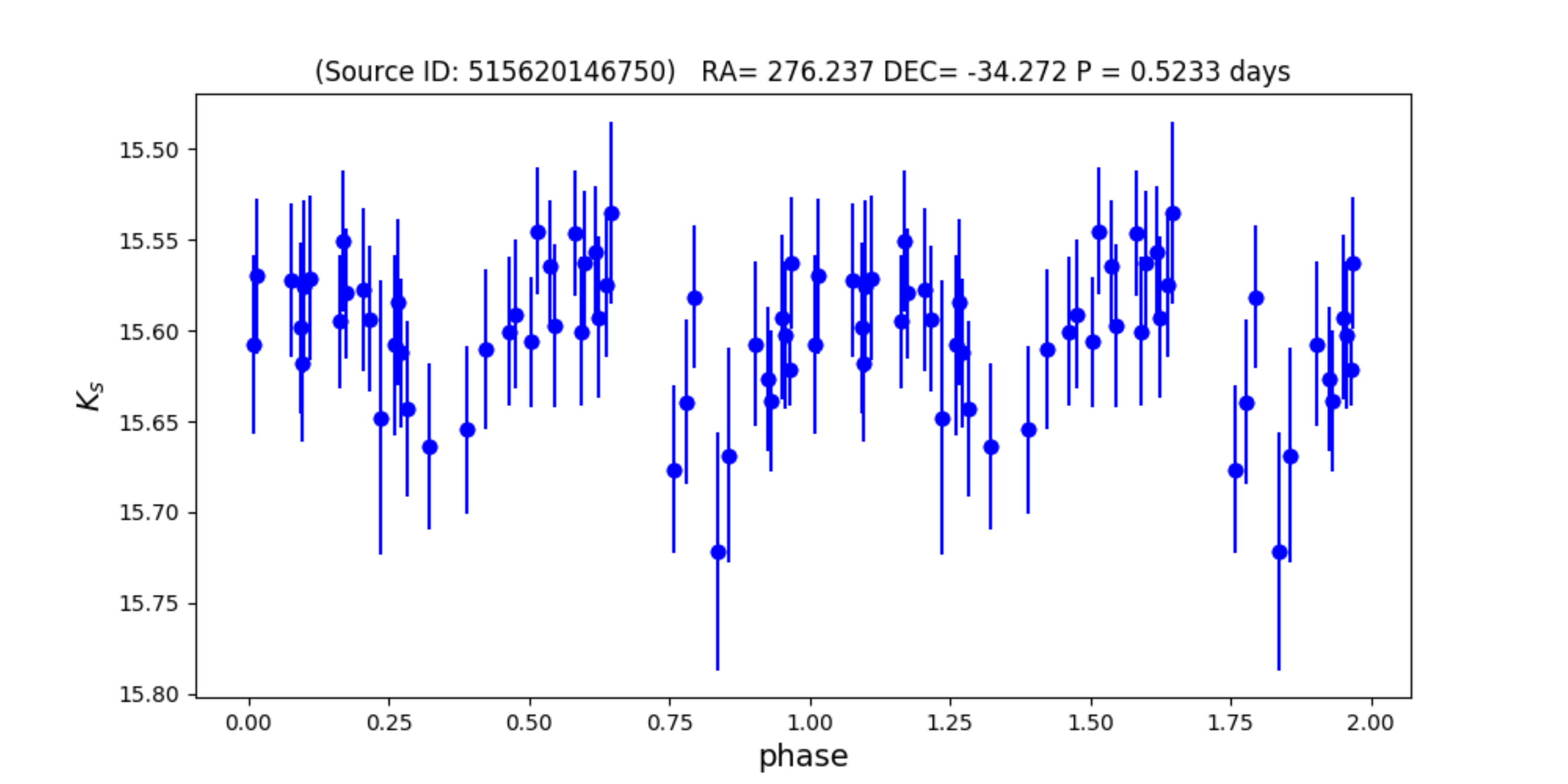} \\
    \includegraphics[width=.33\textwidth]{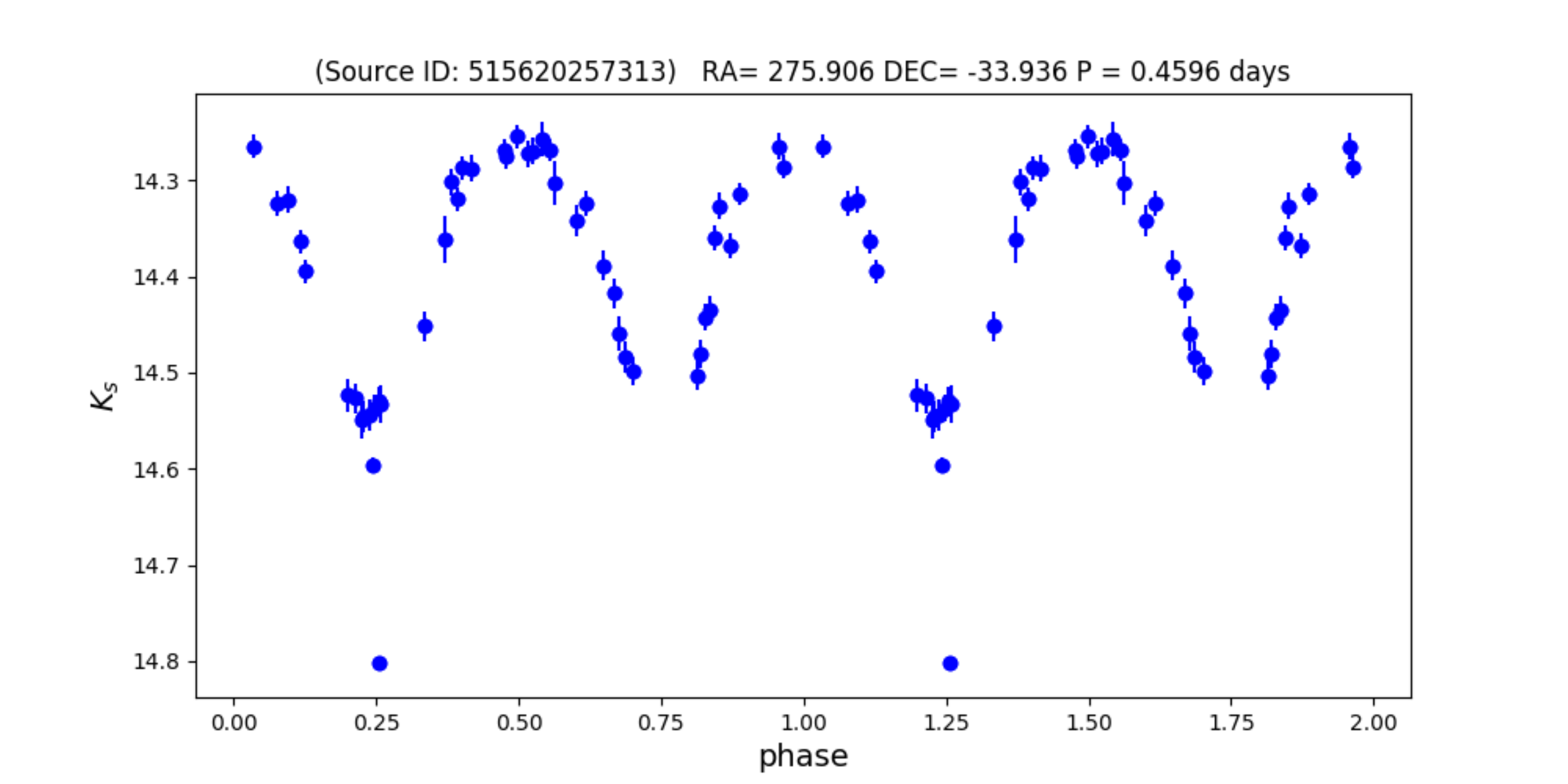} &
    \includegraphics[width=.33\textwidth]{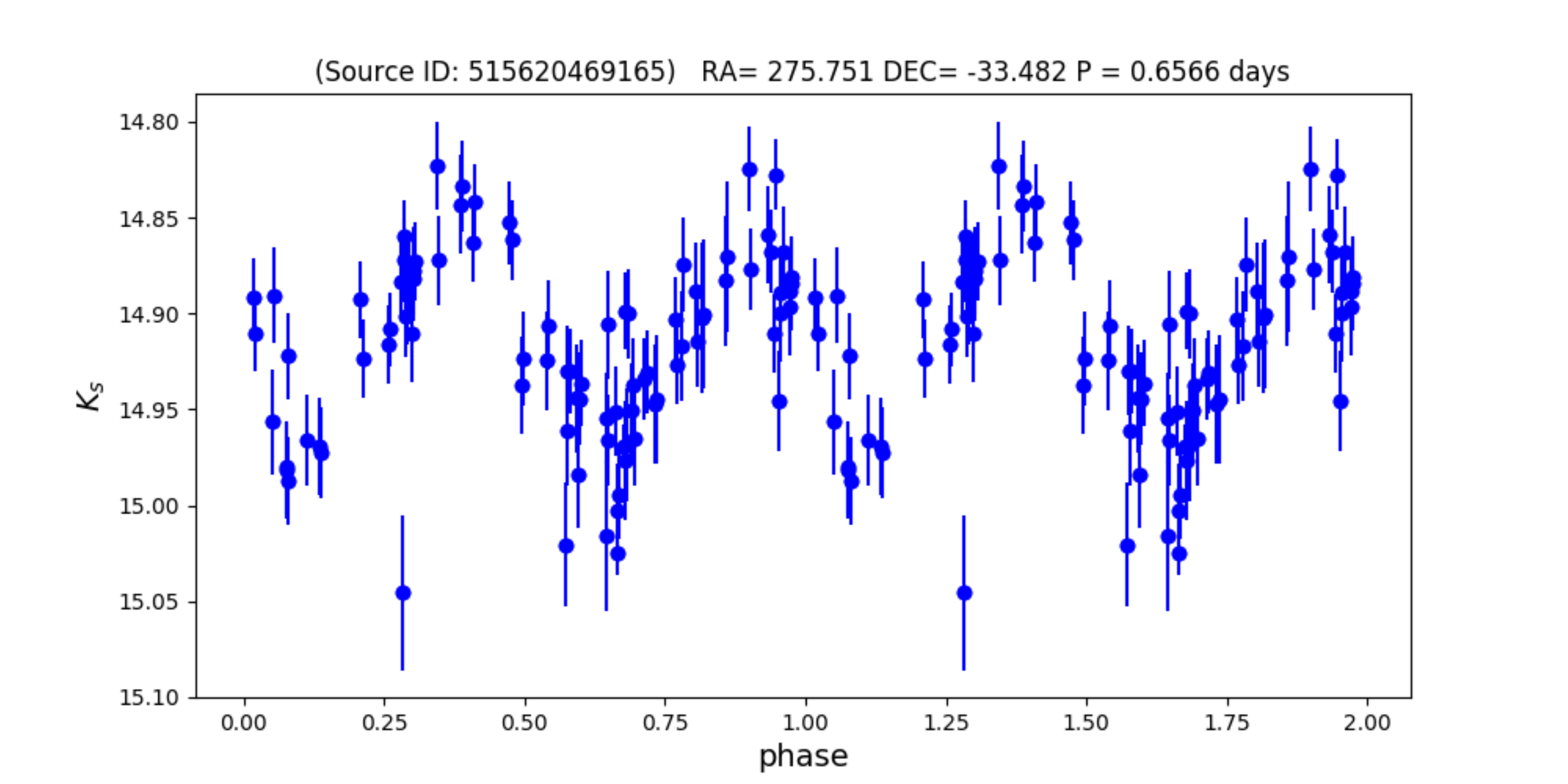} &
    \includegraphics[width=.33\textwidth]{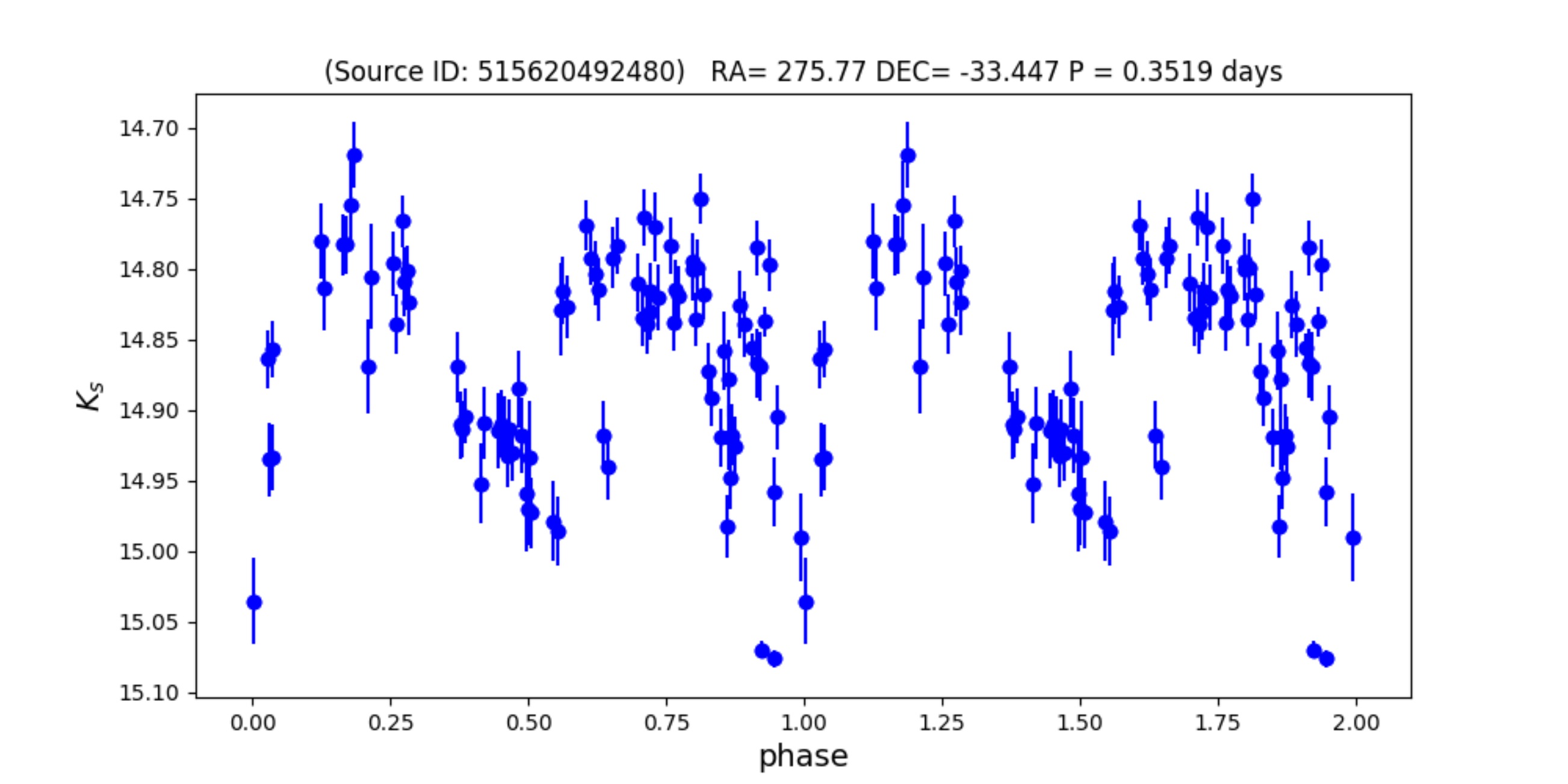} \\
    \includegraphics[width=.33\textwidth]{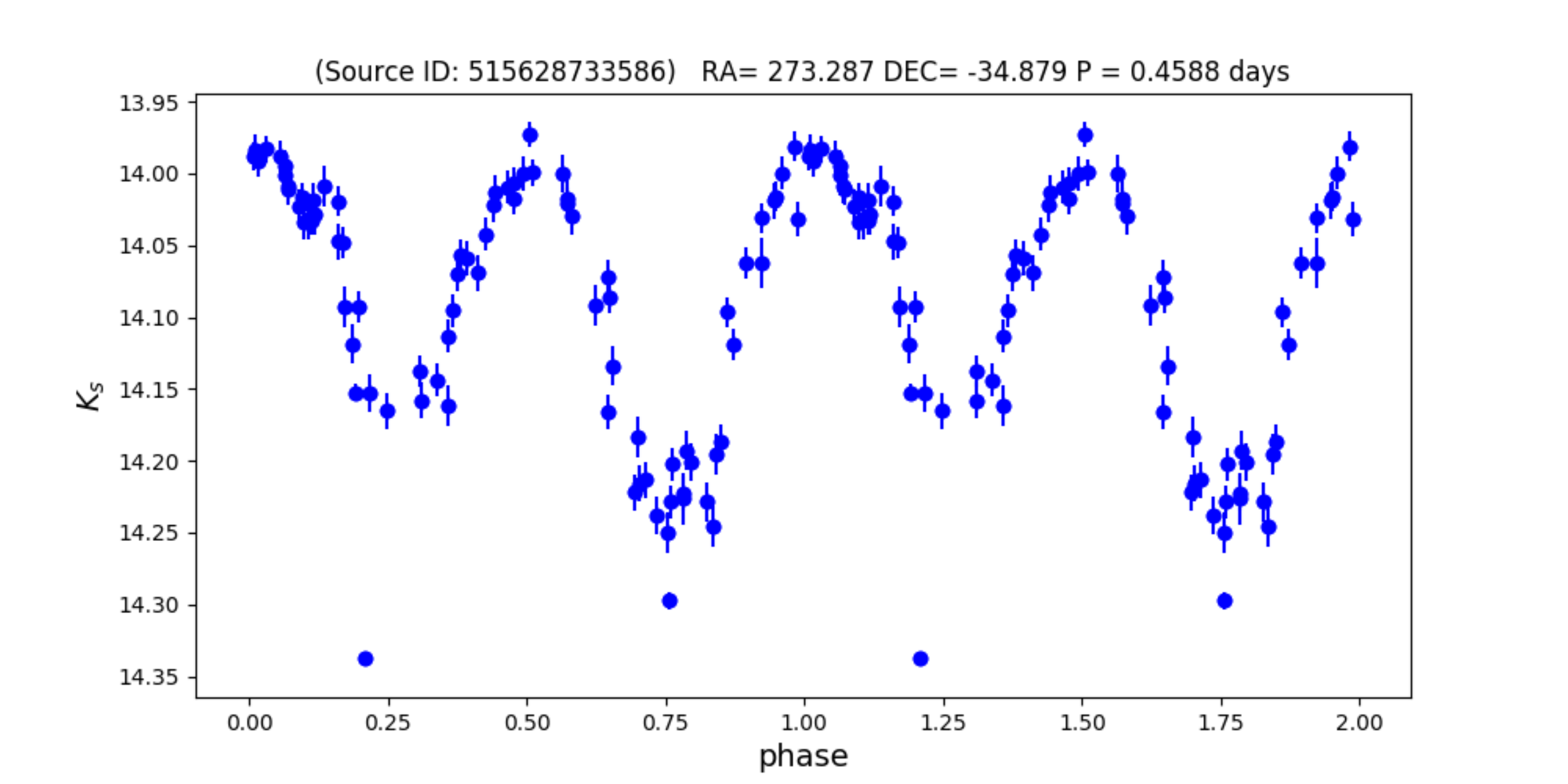} &
    \includegraphics[width=.33\textwidth]{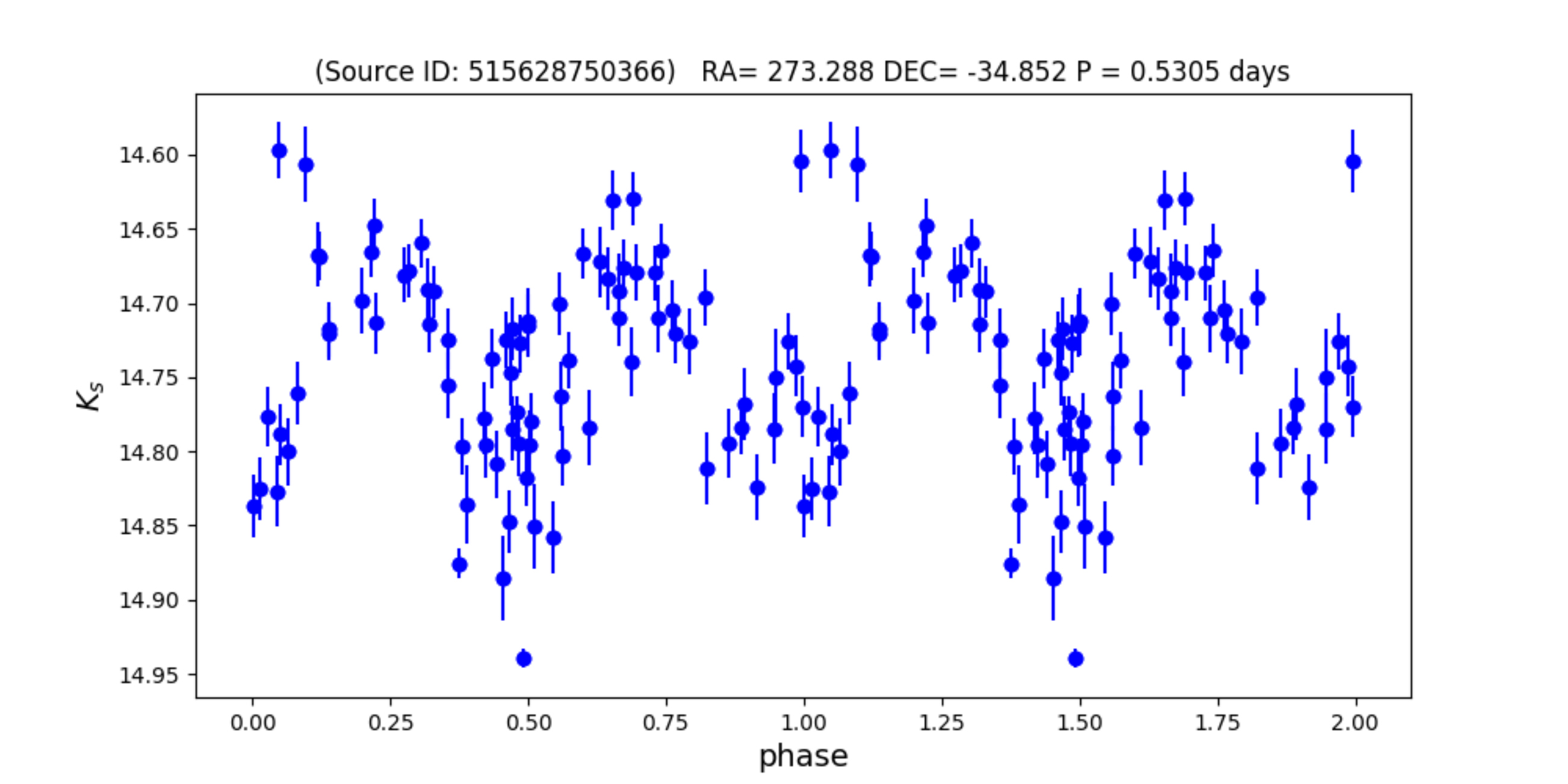} &
    \includegraphics[width=.33\textwidth]{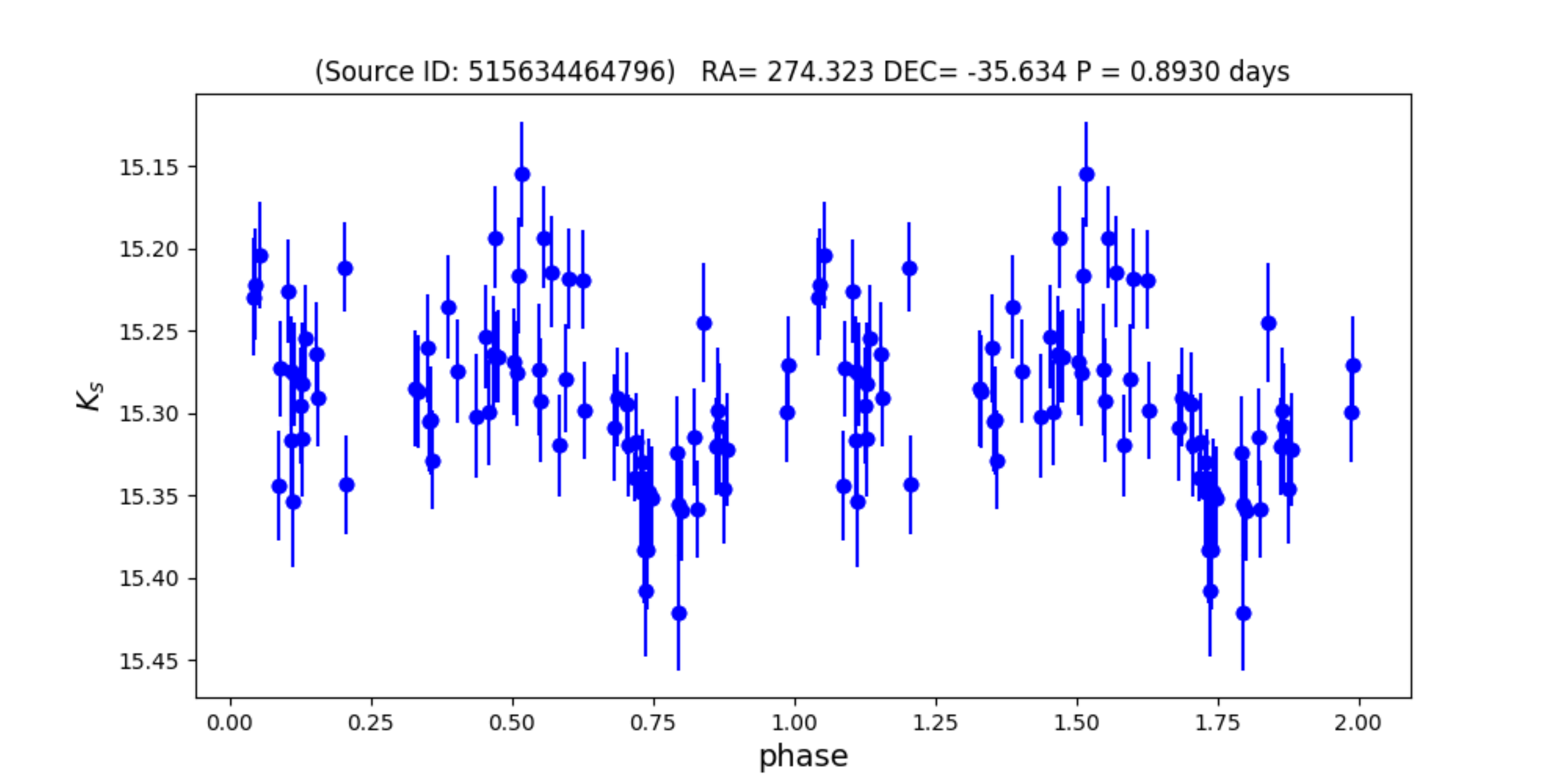} \\
    \includegraphics[width=.33\textwidth]{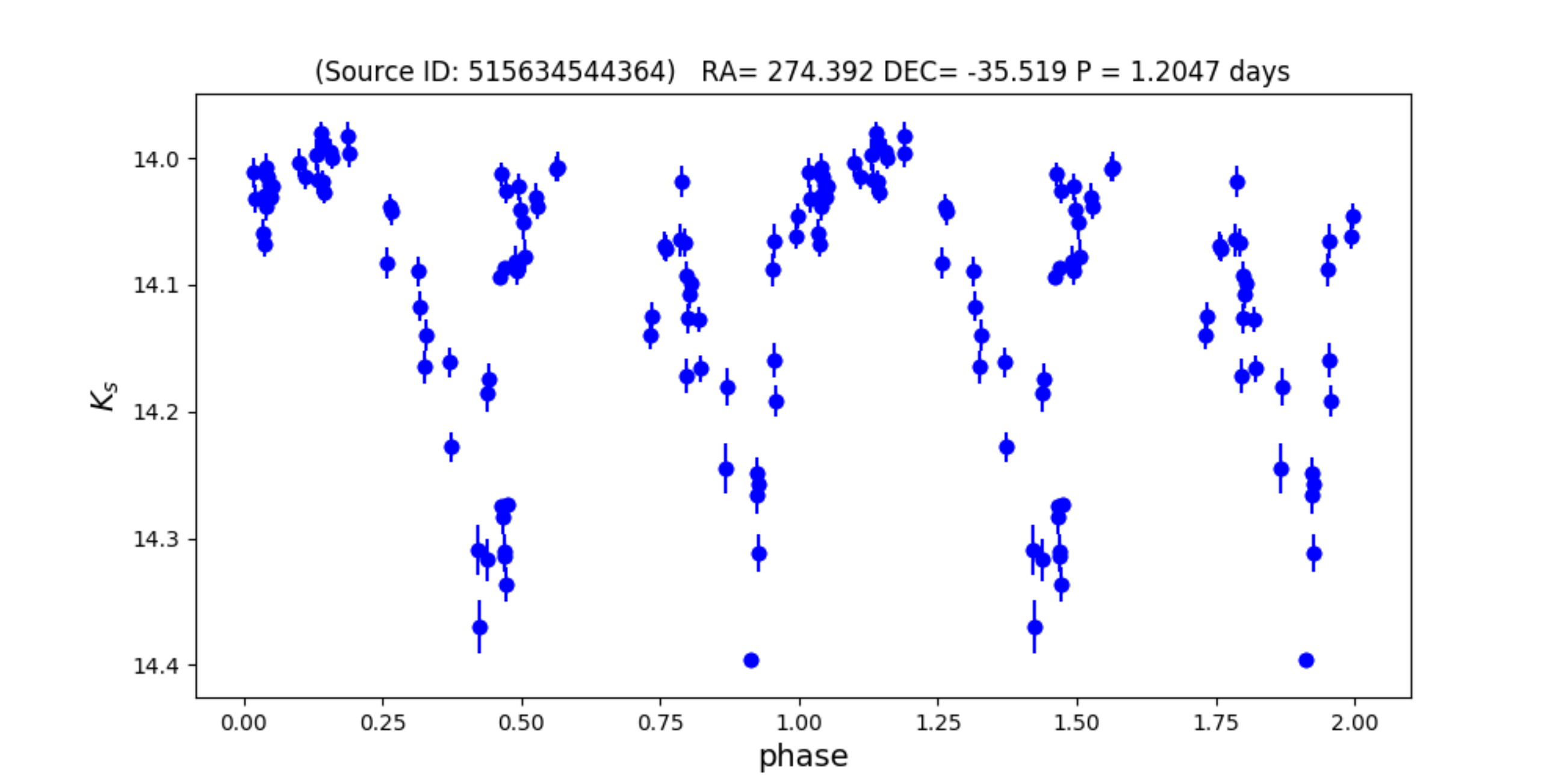} &
    \includegraphics[width=.33\textwidth]{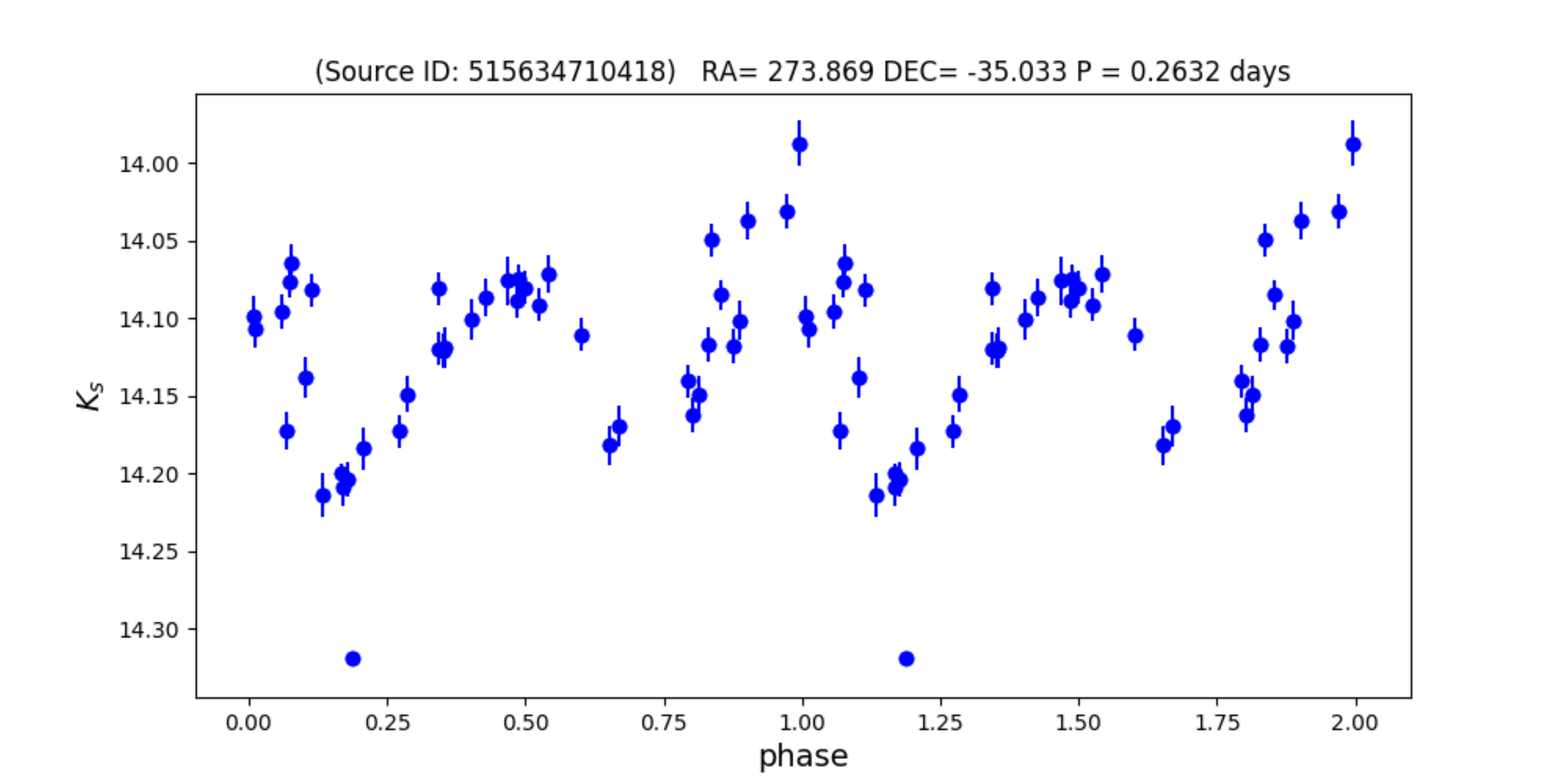} &
    \includegraphics[width=.33\textwidth]{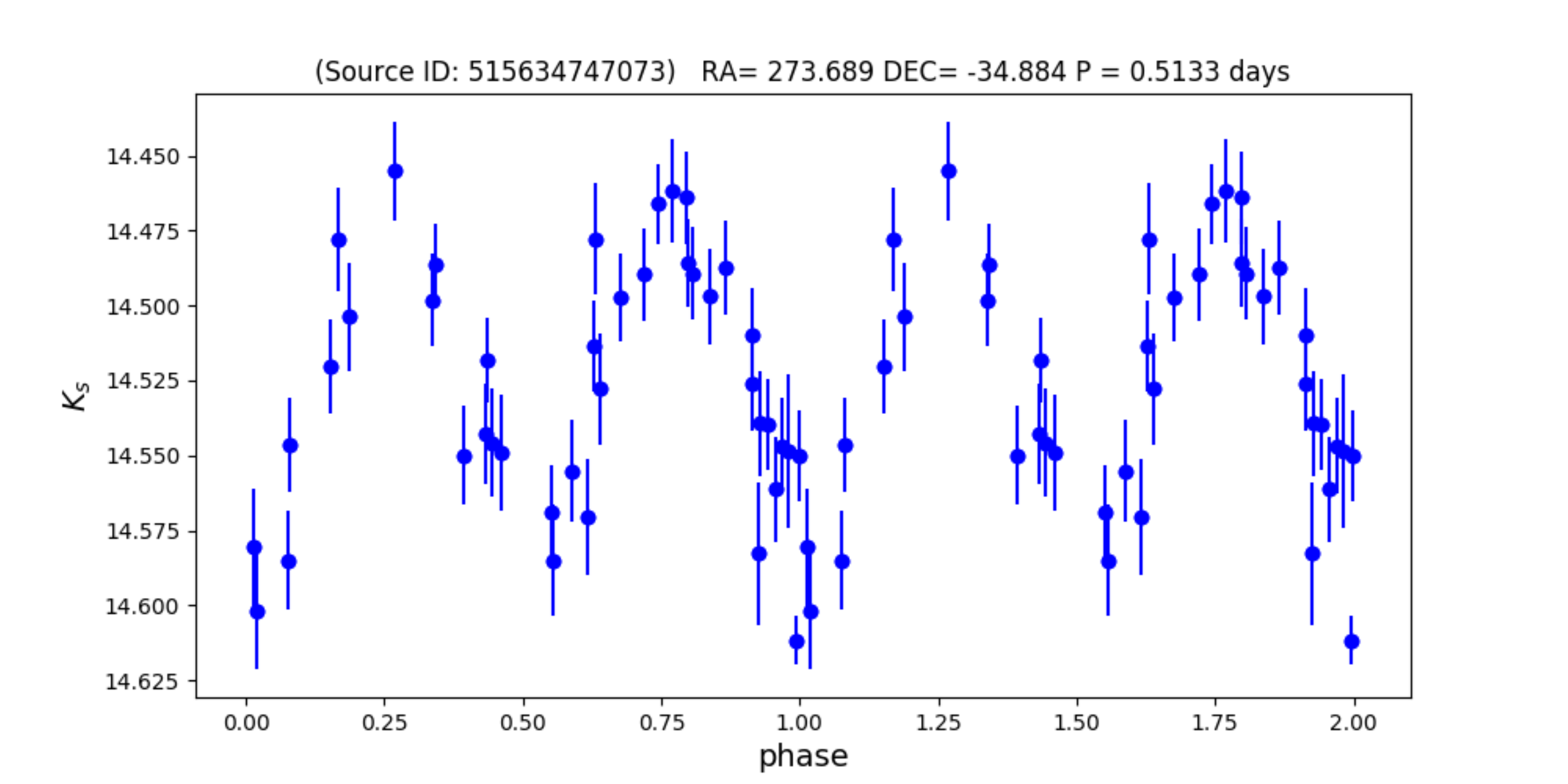} \\
    \includegraphics[width=.33\textwidth]{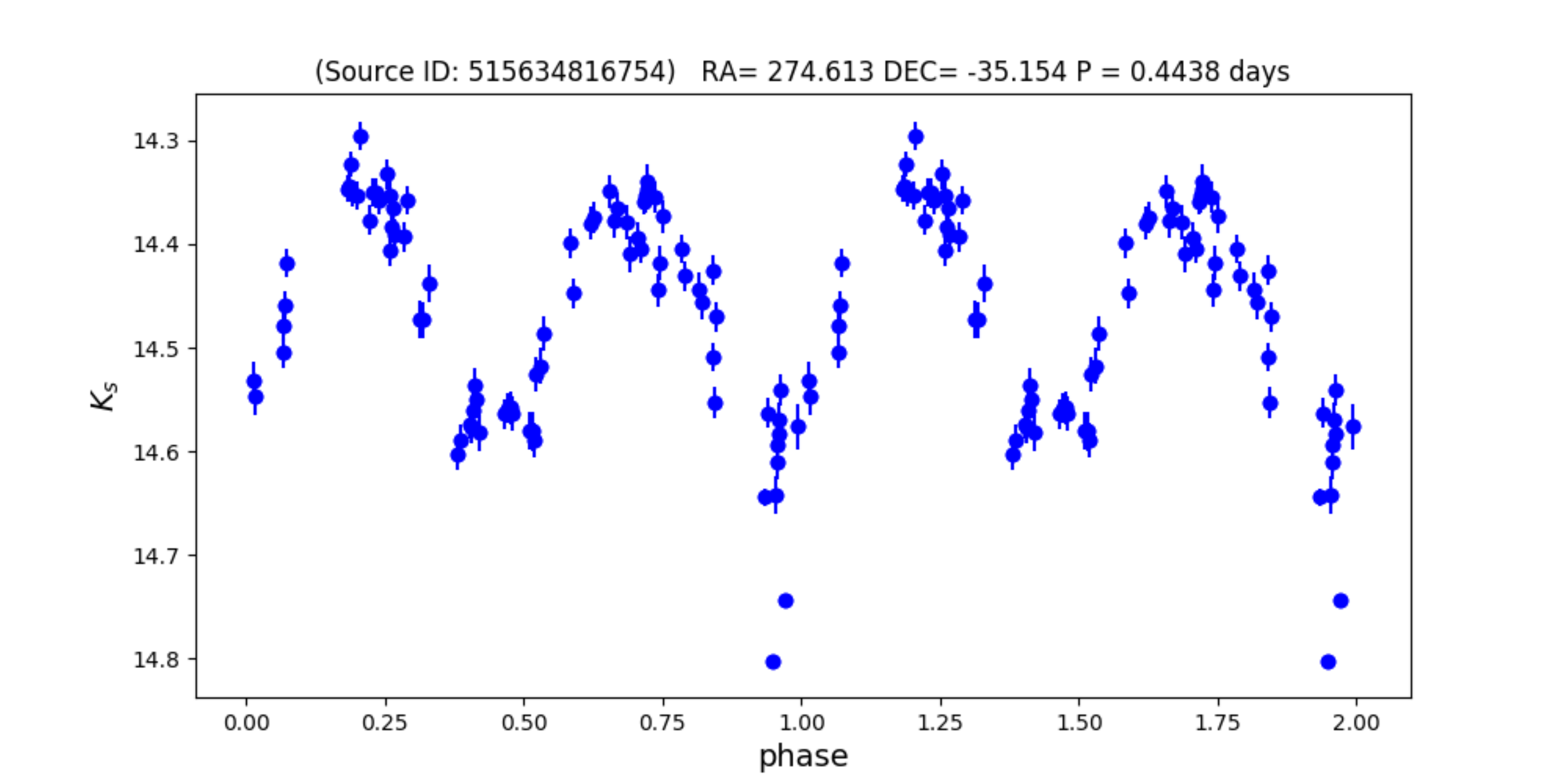} &
    \includegraphics[width=.33\textwidth]{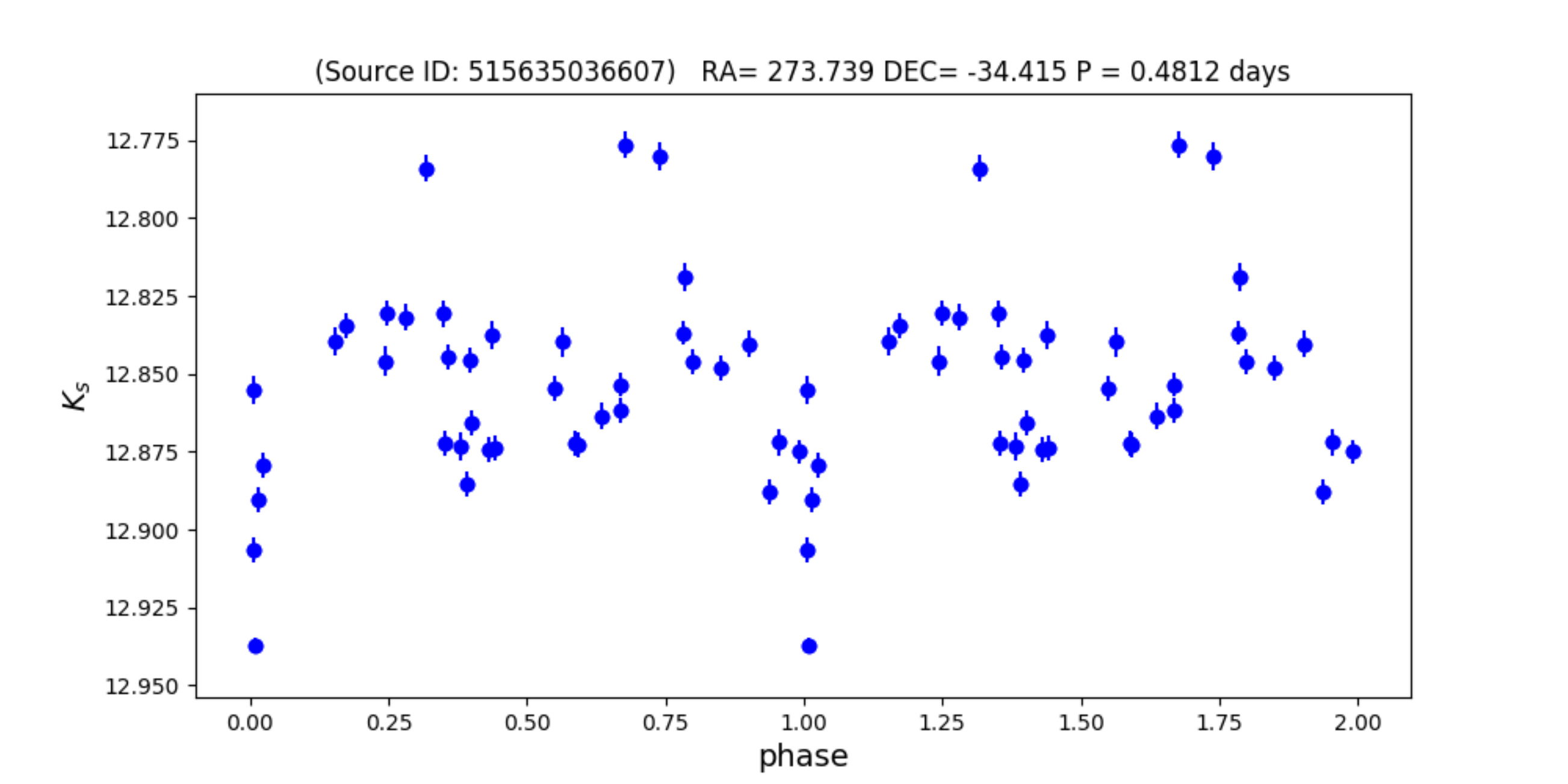} &
    \includegraphics[width=.33\textwidth]{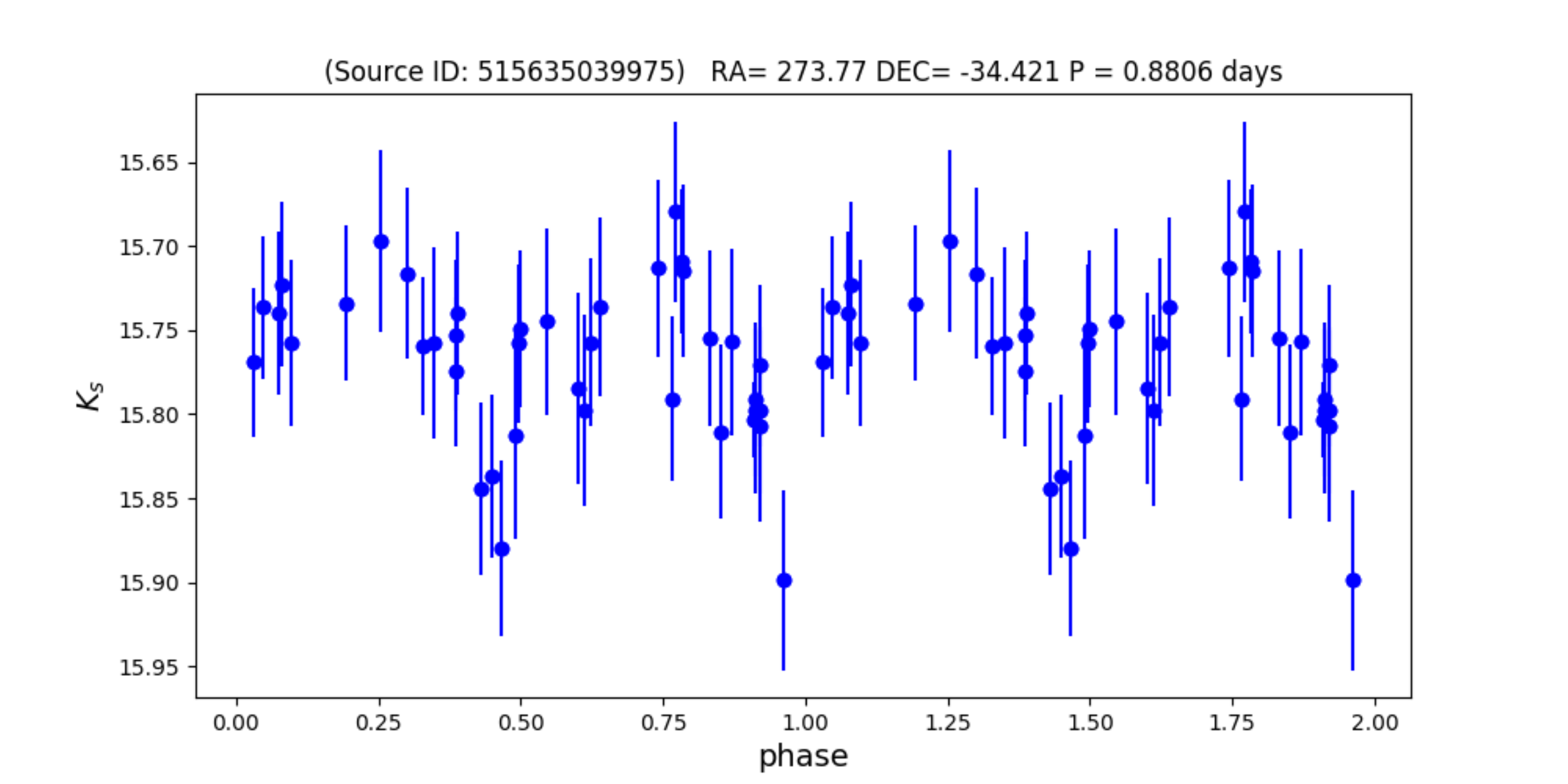} \\
    \includegraphics[width=.33\textwidth]{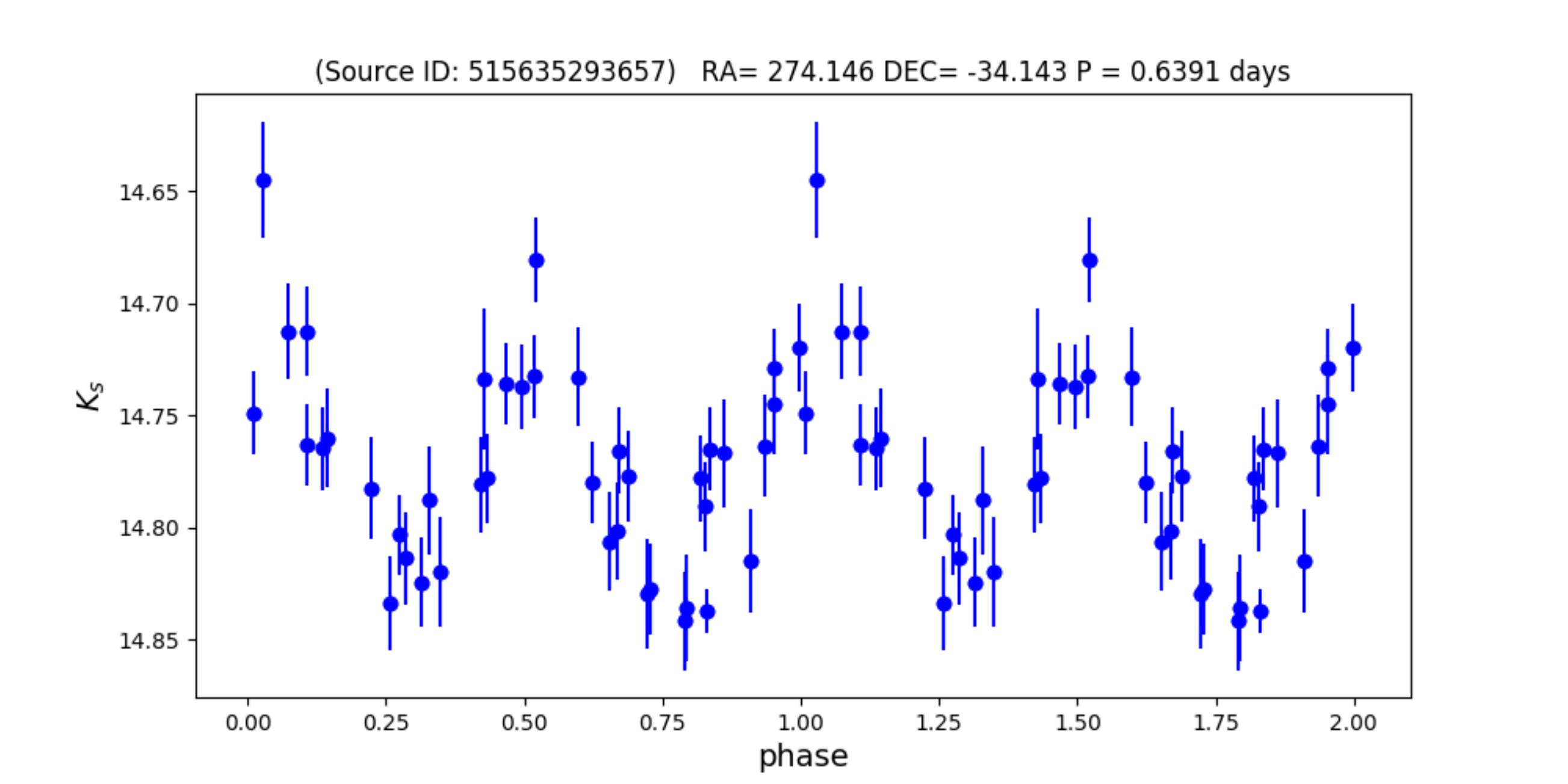} &
    \includegraphics[width=.33\textwidth]{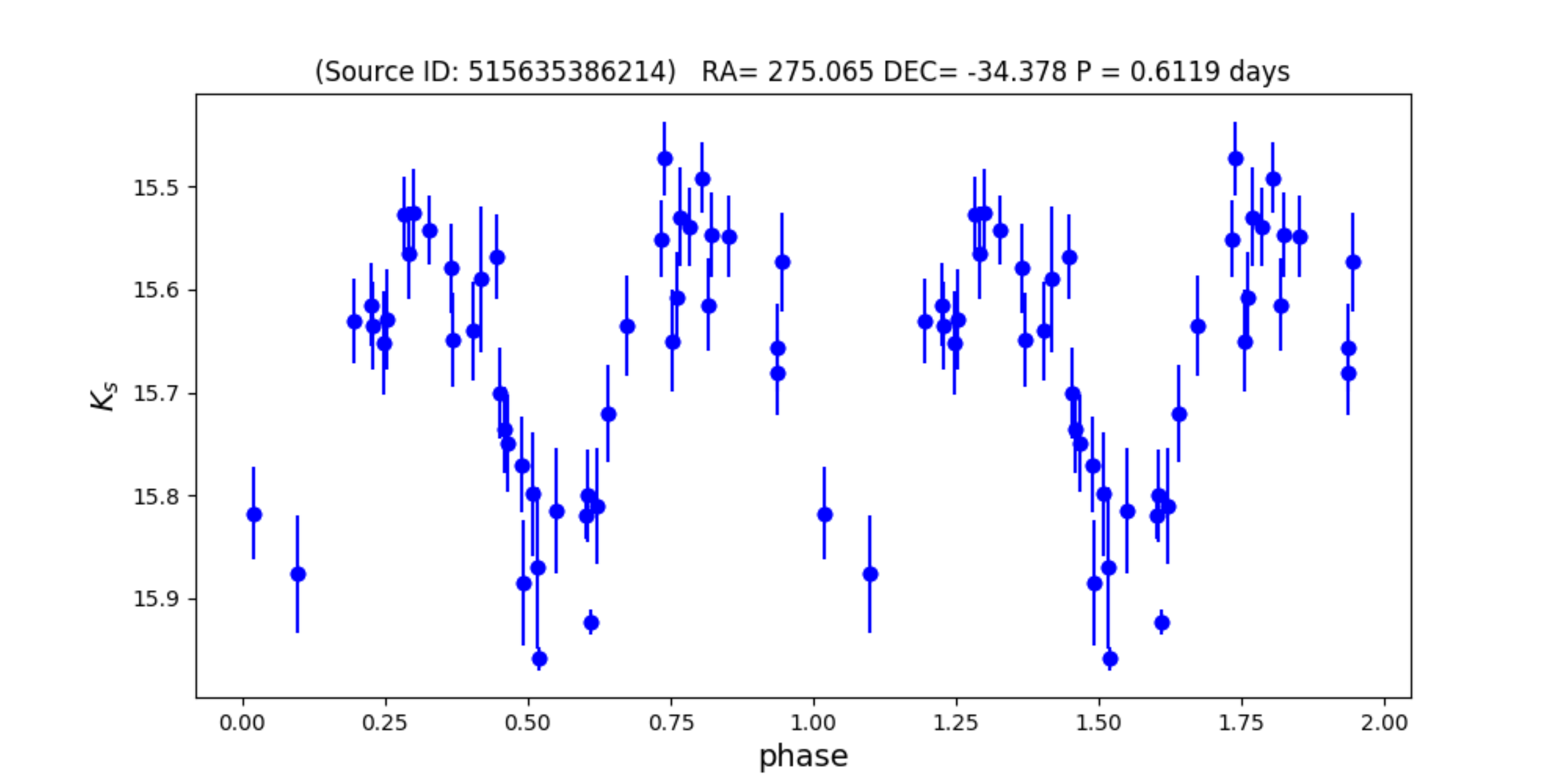} &
    \includegraphics[width=.33\textwidth]{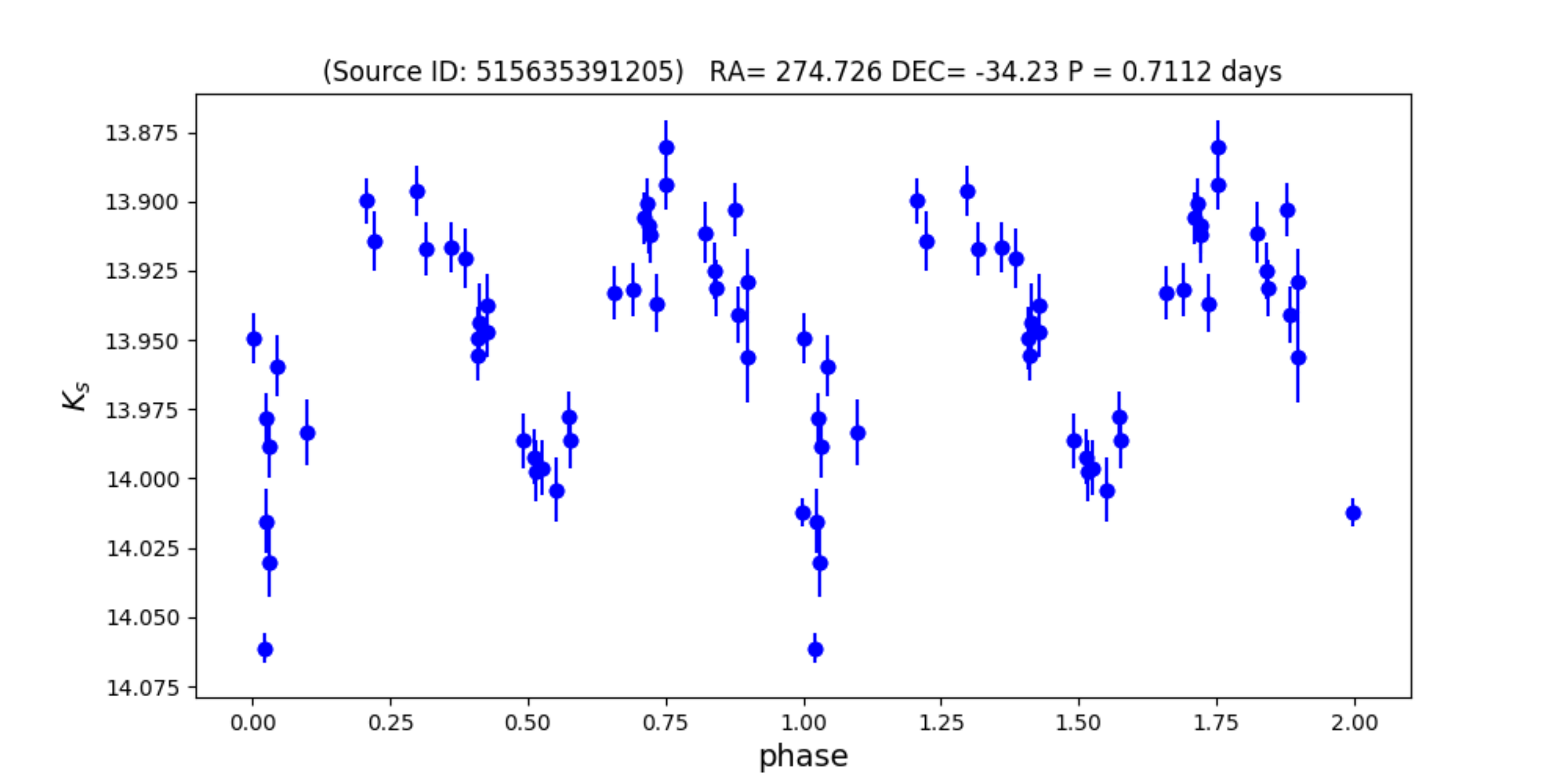} \\  
  \end{tabular}
  \caption{BHB stars light curves corresponding to $1^{st}$ category eclipsing binaries.}
  \label{fig:bhb6}
\end{figure*}

\begin{figure*}[htb]
\centering
  \begin{tabular}{@{}ccc@{}}  
    \includegraphics[width=.33\textwidth]{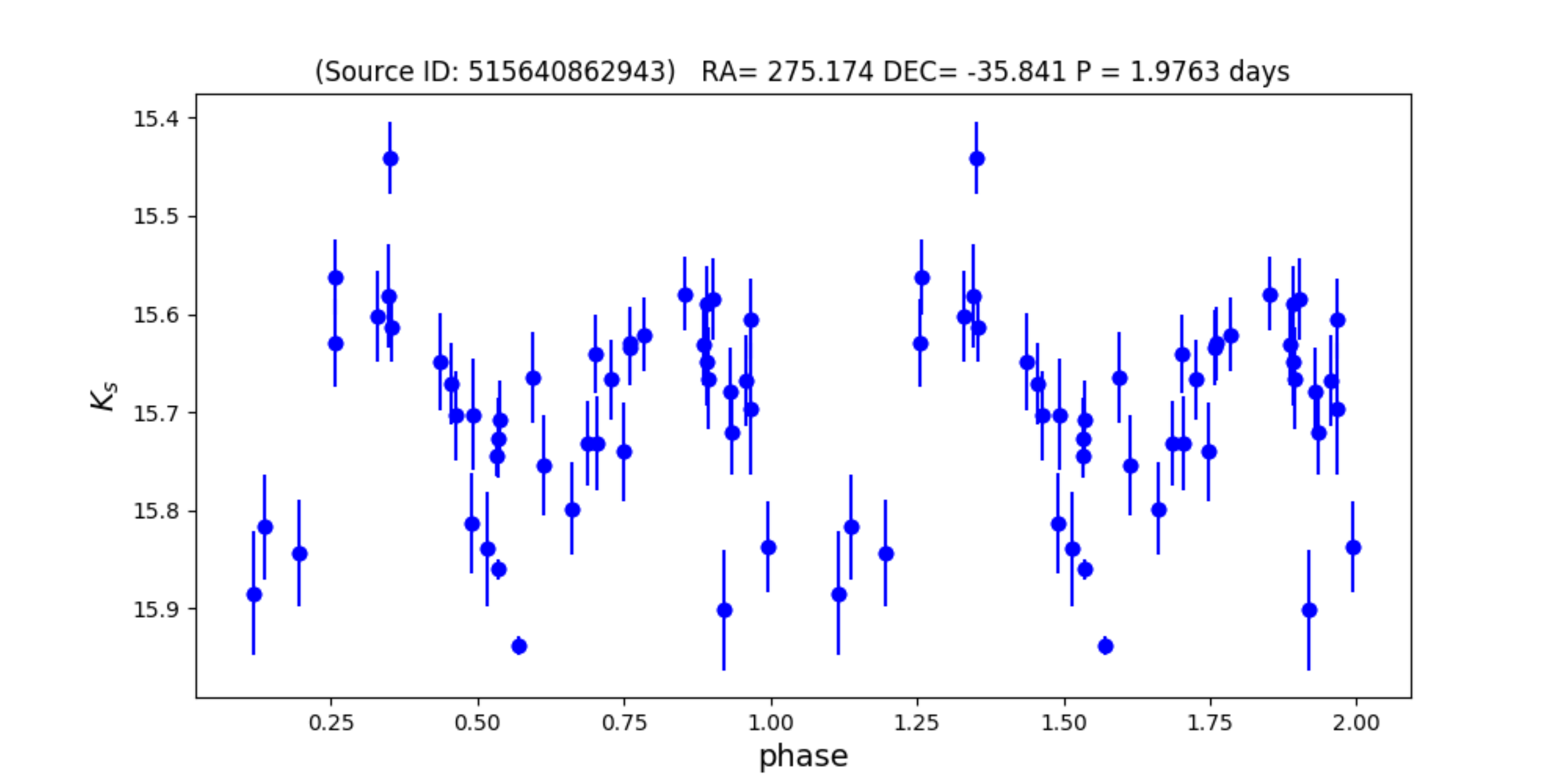} &   
    \includegraphics[width=.33\textwidth]{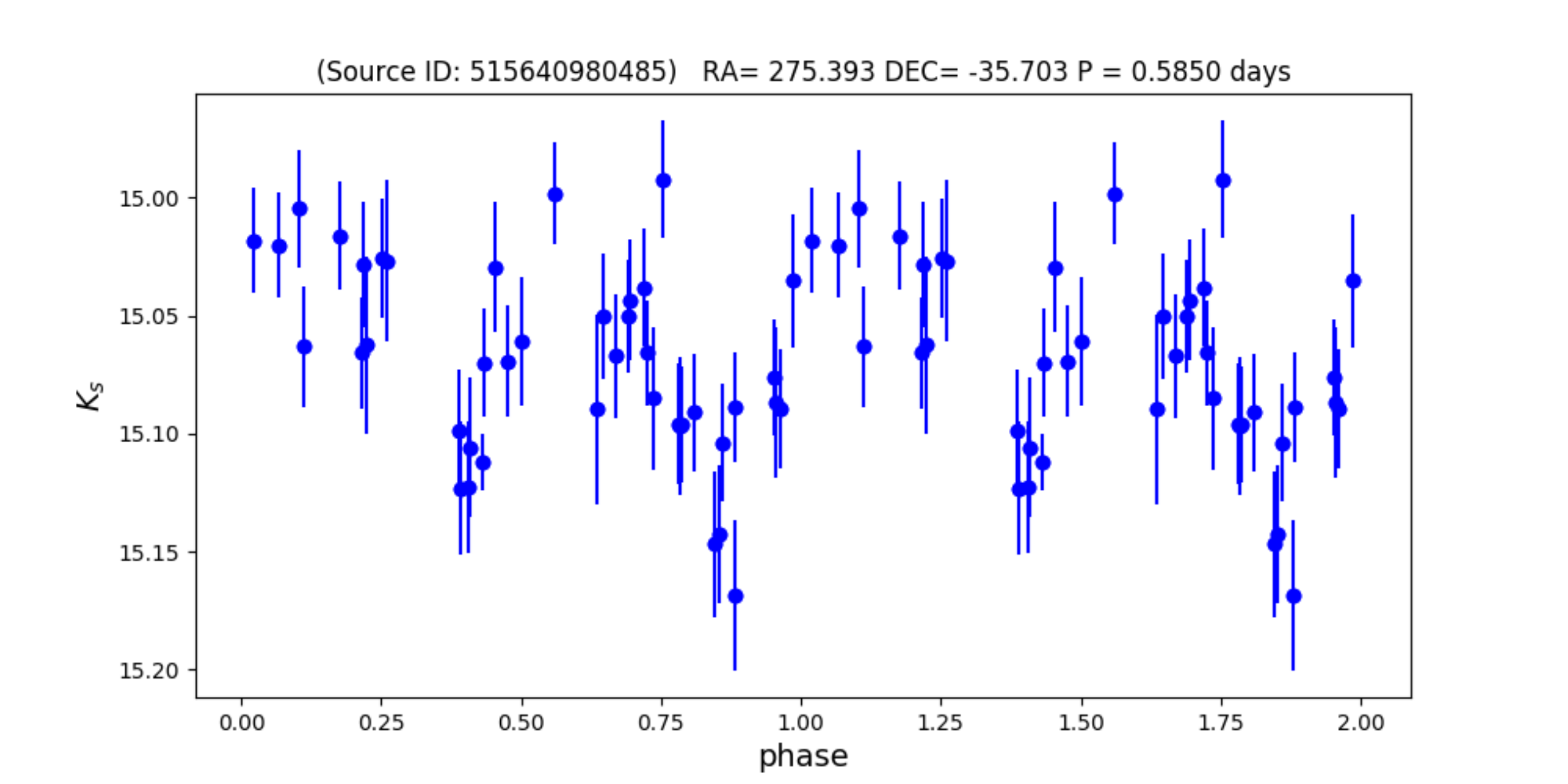} &
    \includegraphics[width=.33\textwidth]{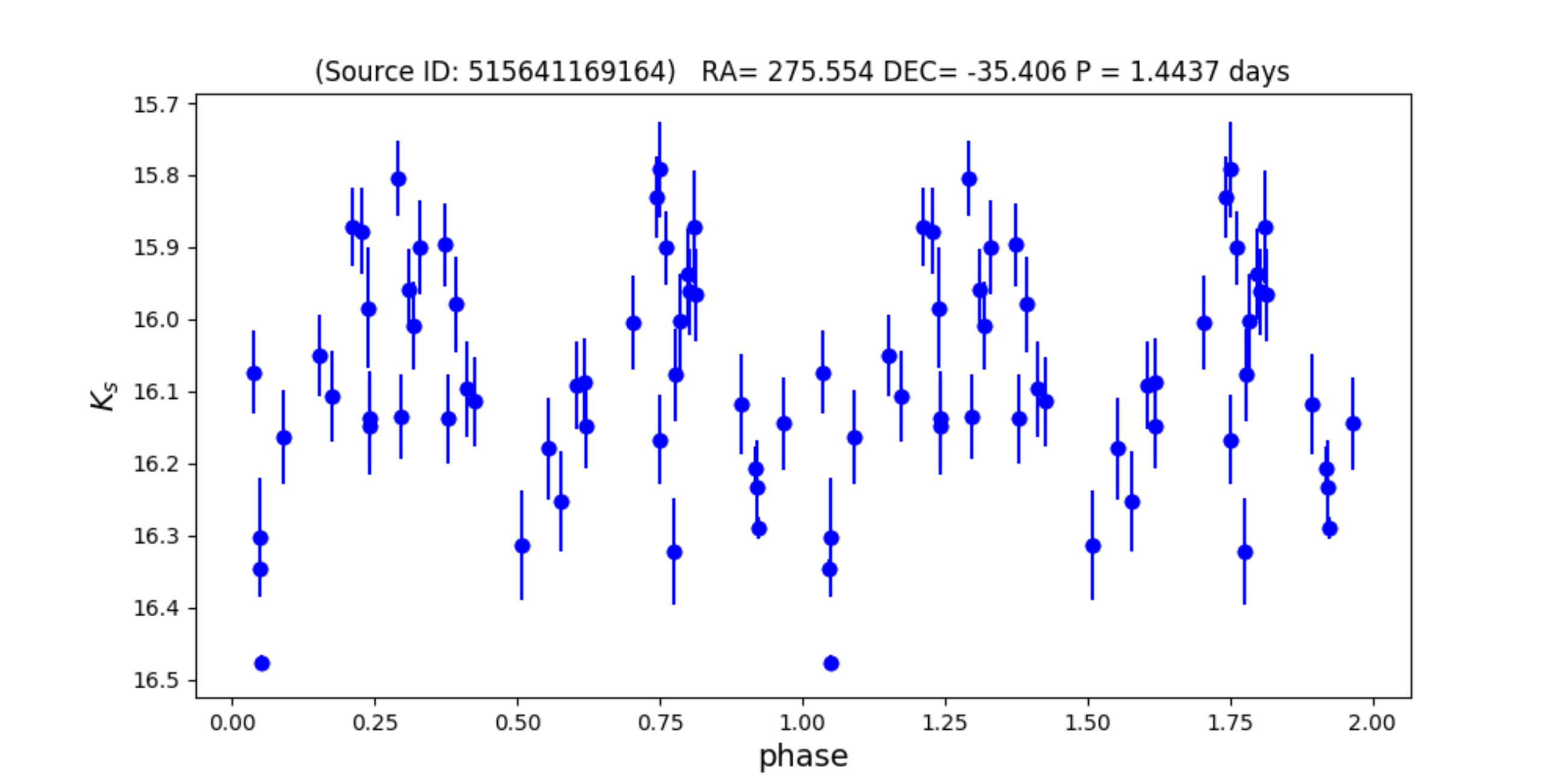} \\
    \includegraphics[width=.33\textwidth]{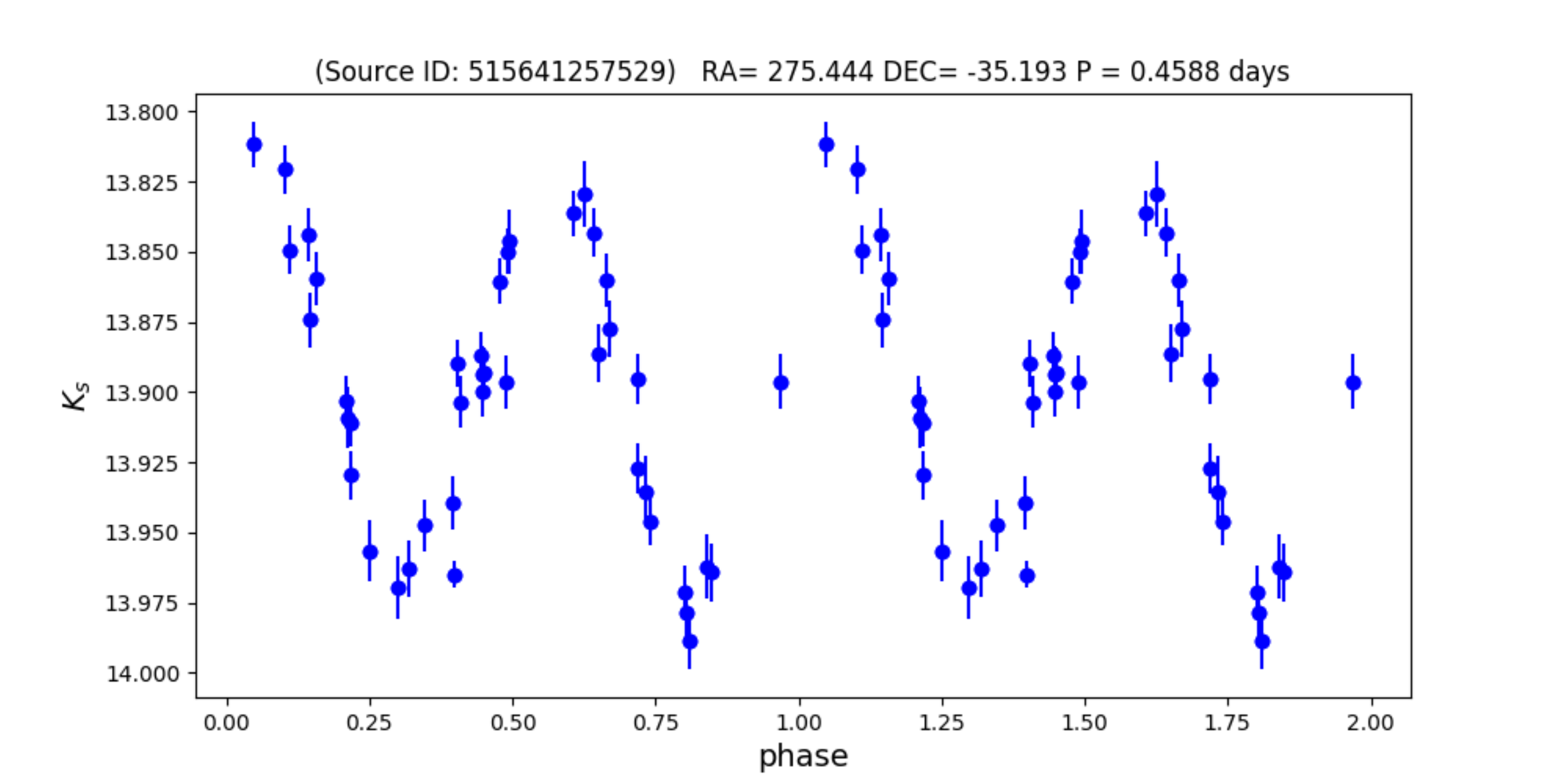} &
    \includegraphics[width=.33\textwidth]{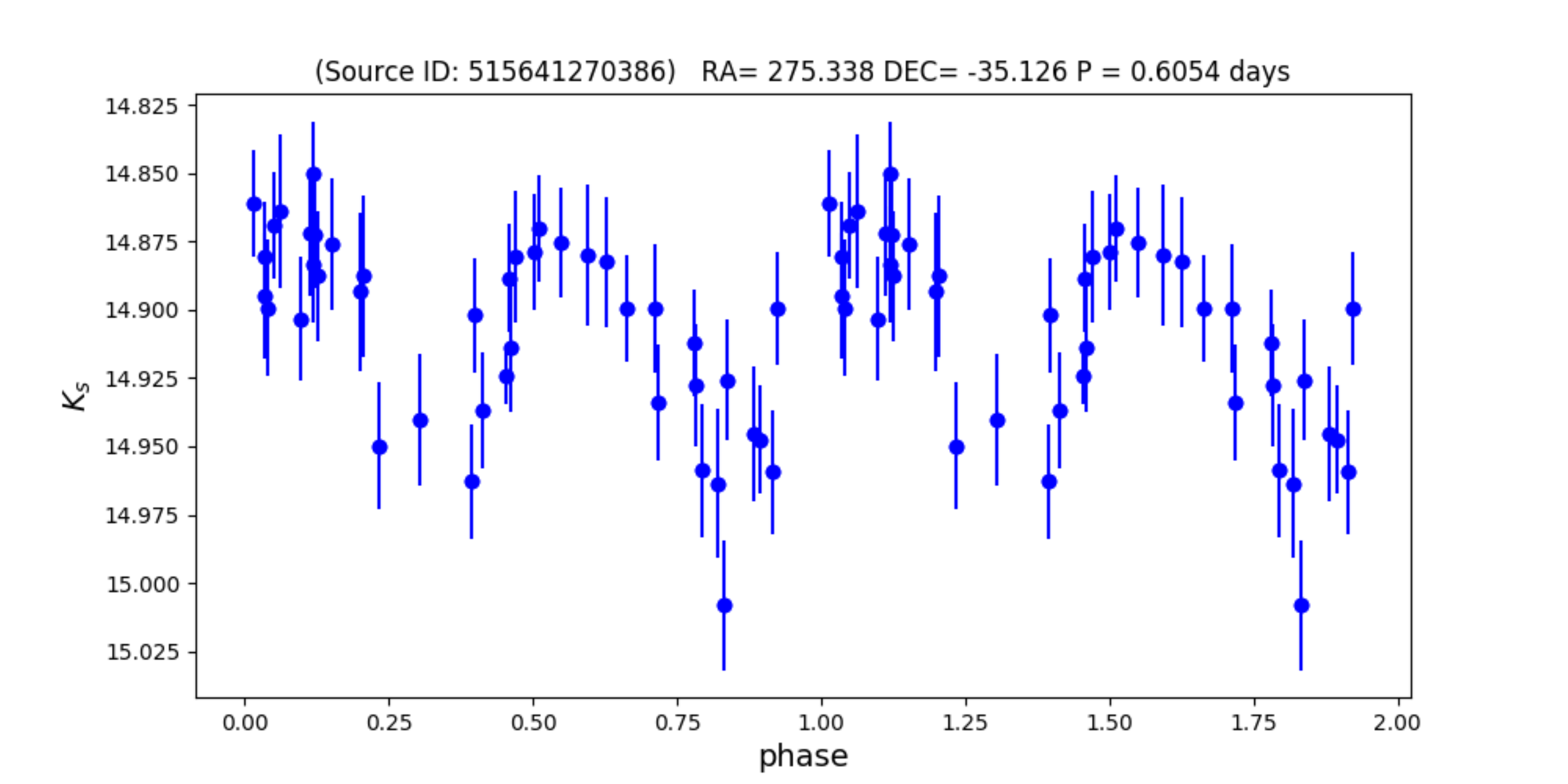} &
    \includegraphics[width=.33\textwidth]{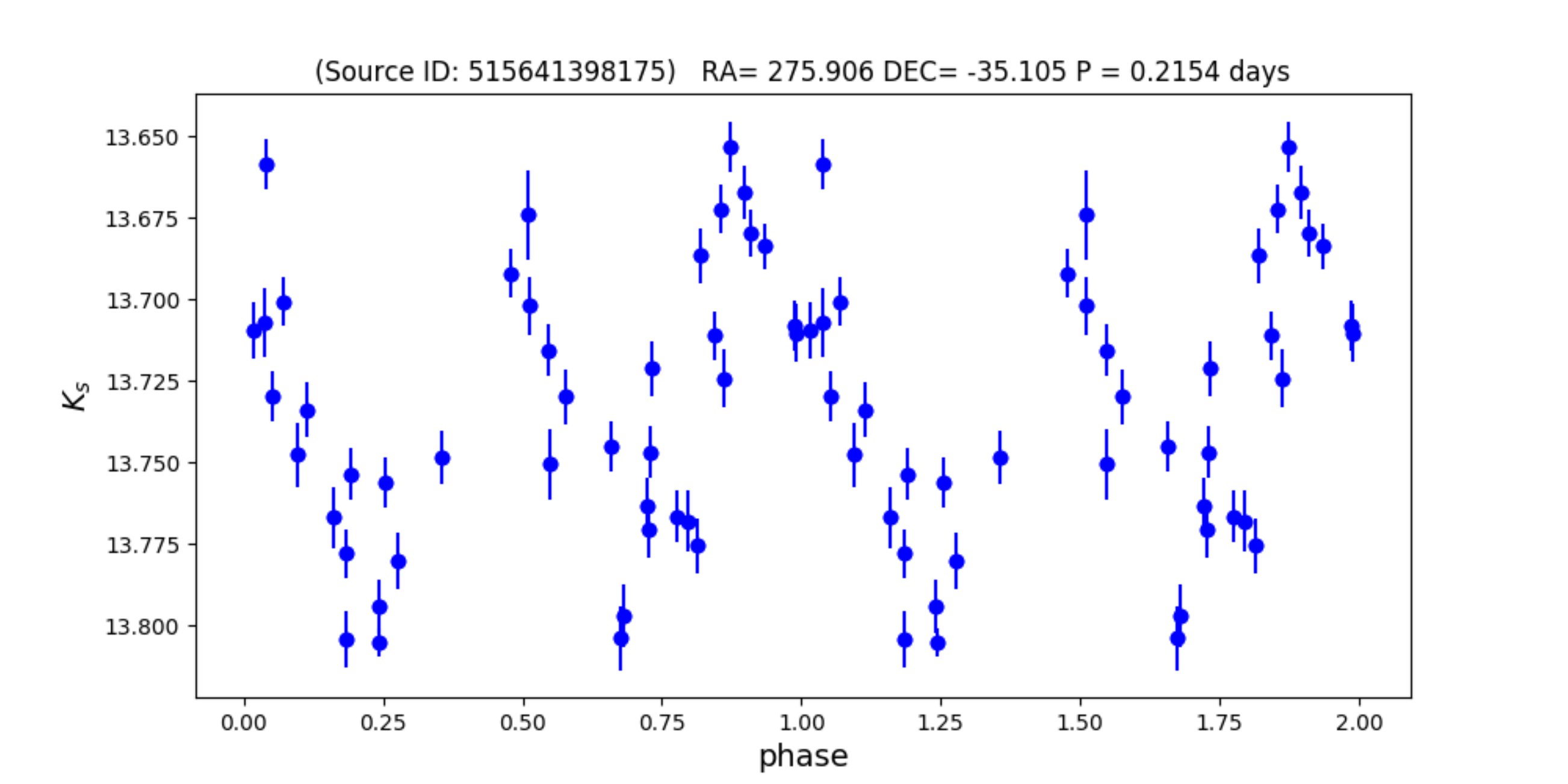} \\
    \includegraphics[width=.33\textwidth]{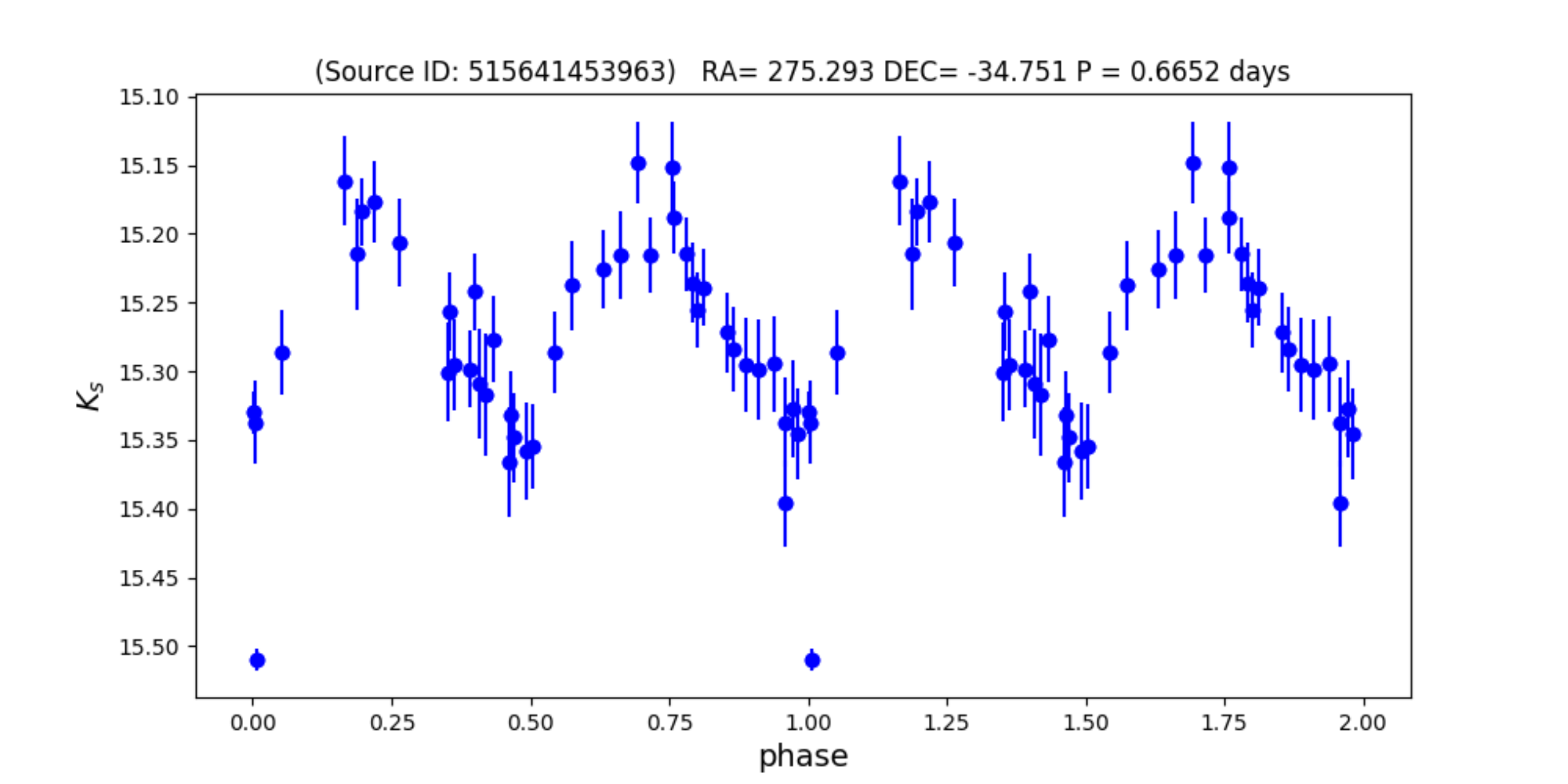} &
    \includegraphics[width=.33\textwidth]{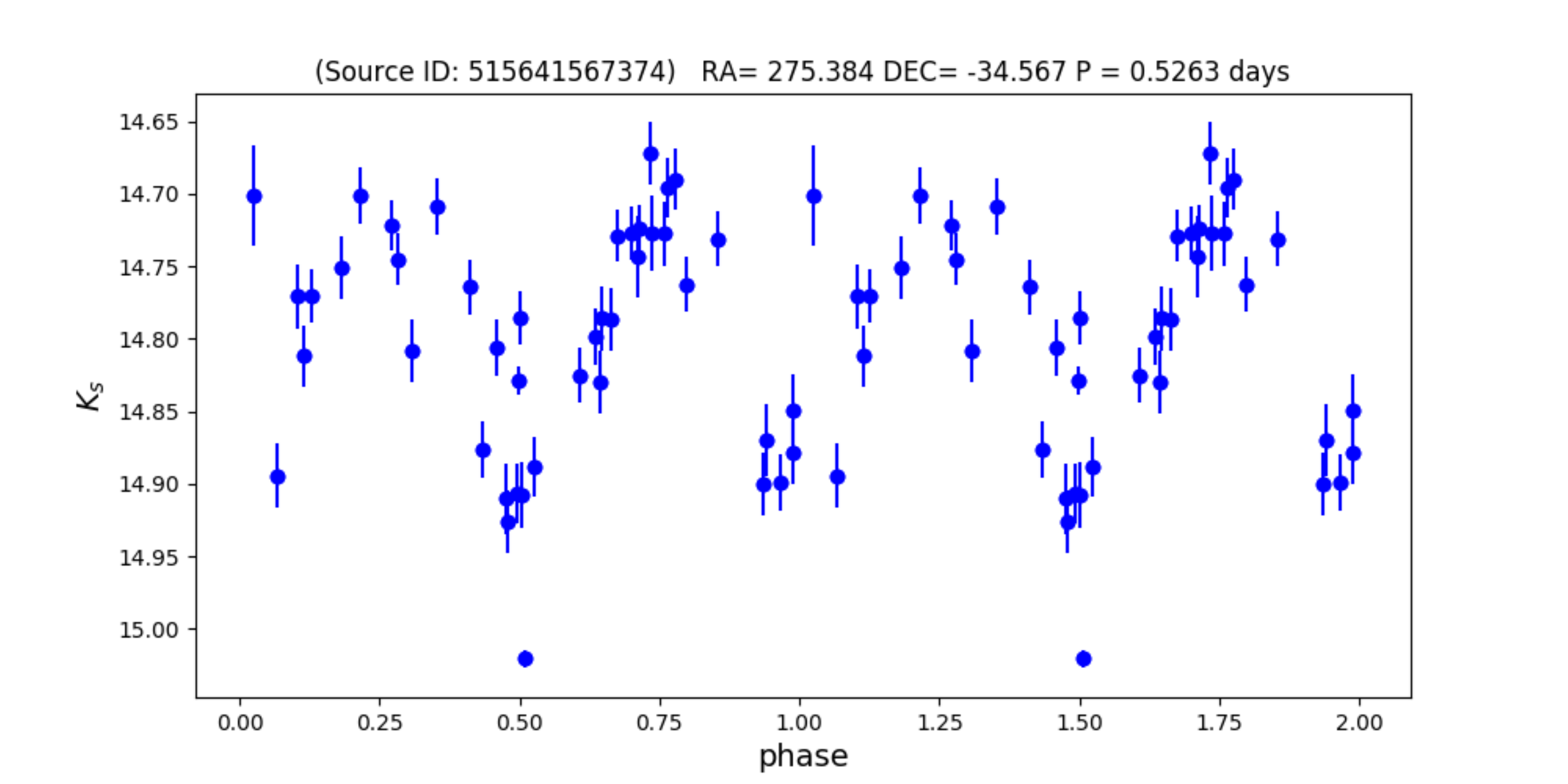} &
    \includegraphics[width=.33\textwidth]{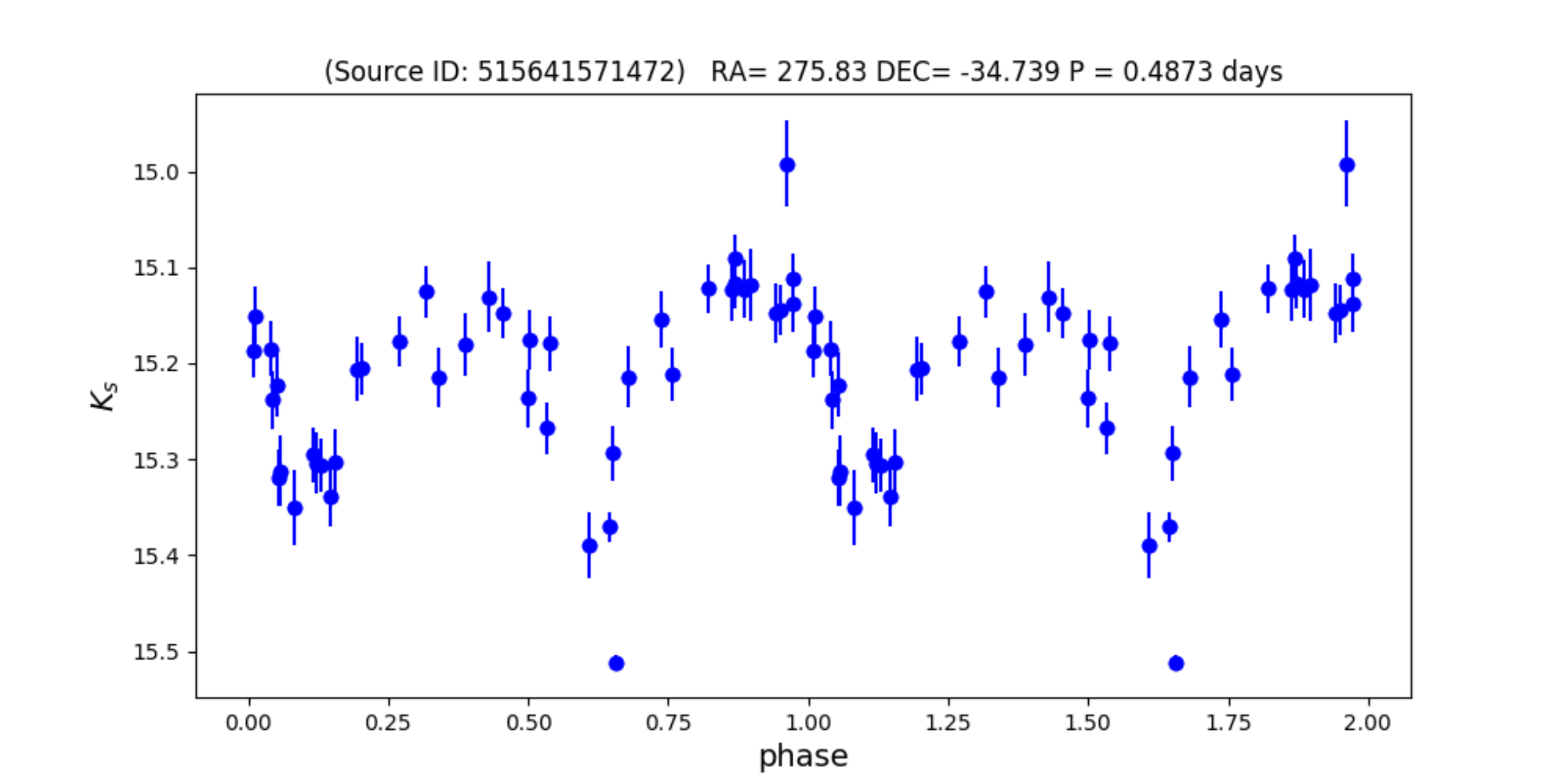} \\
    \includegraphics[width=.33\textwidth]{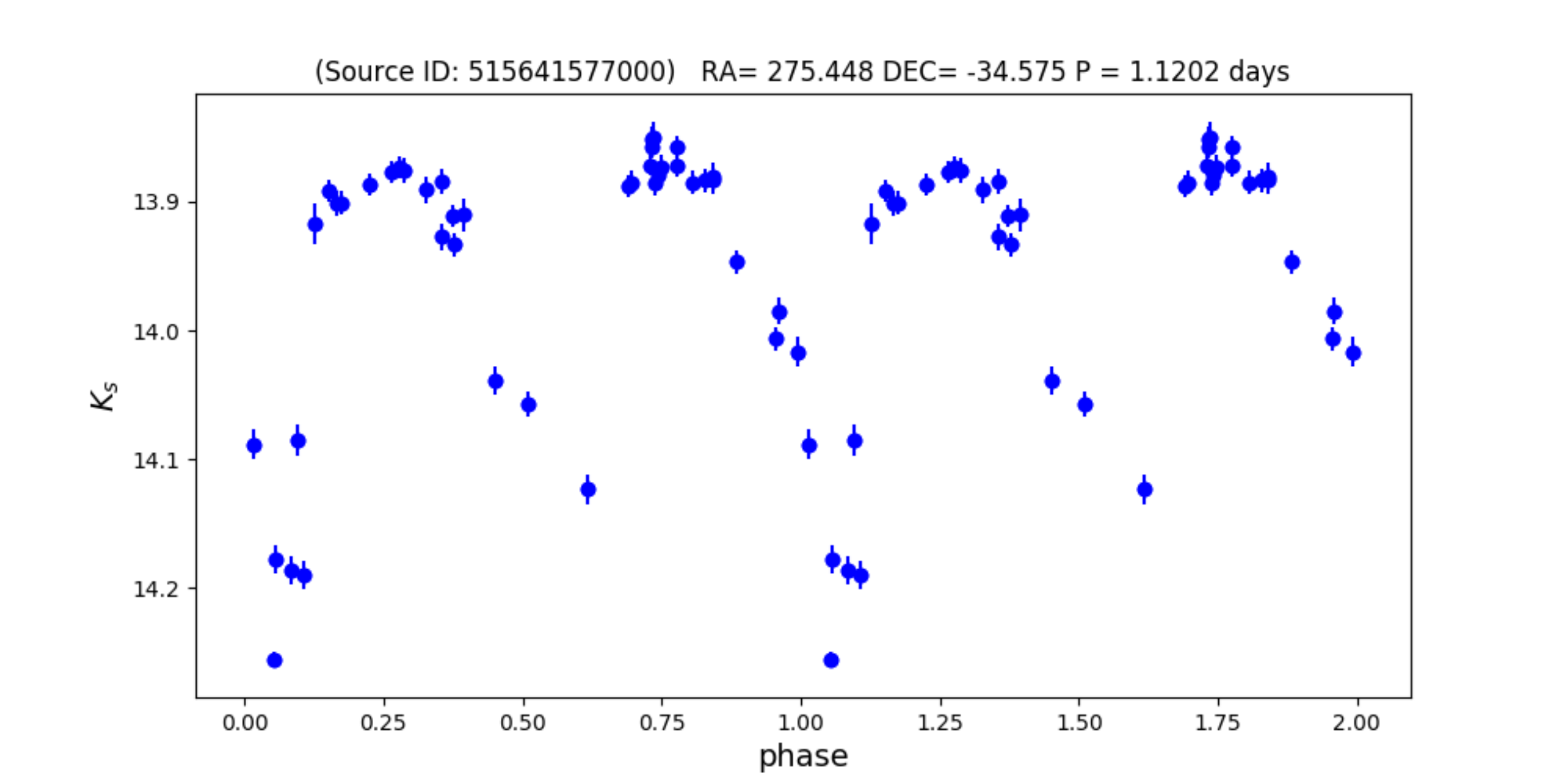} &
    \includegraphics[width=.33\textwidth]{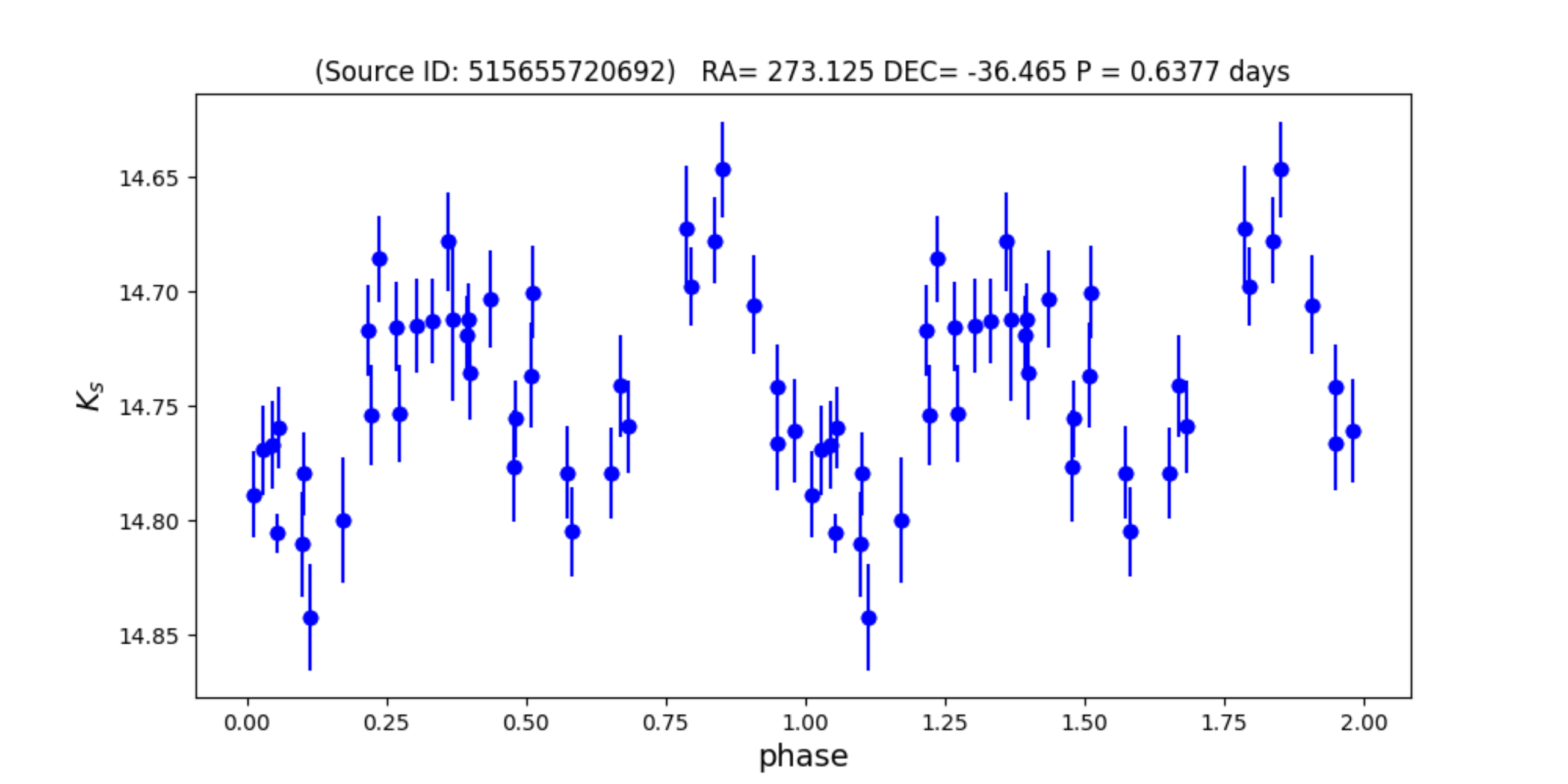} &
    \includegraphics[width=.33\textwidth]{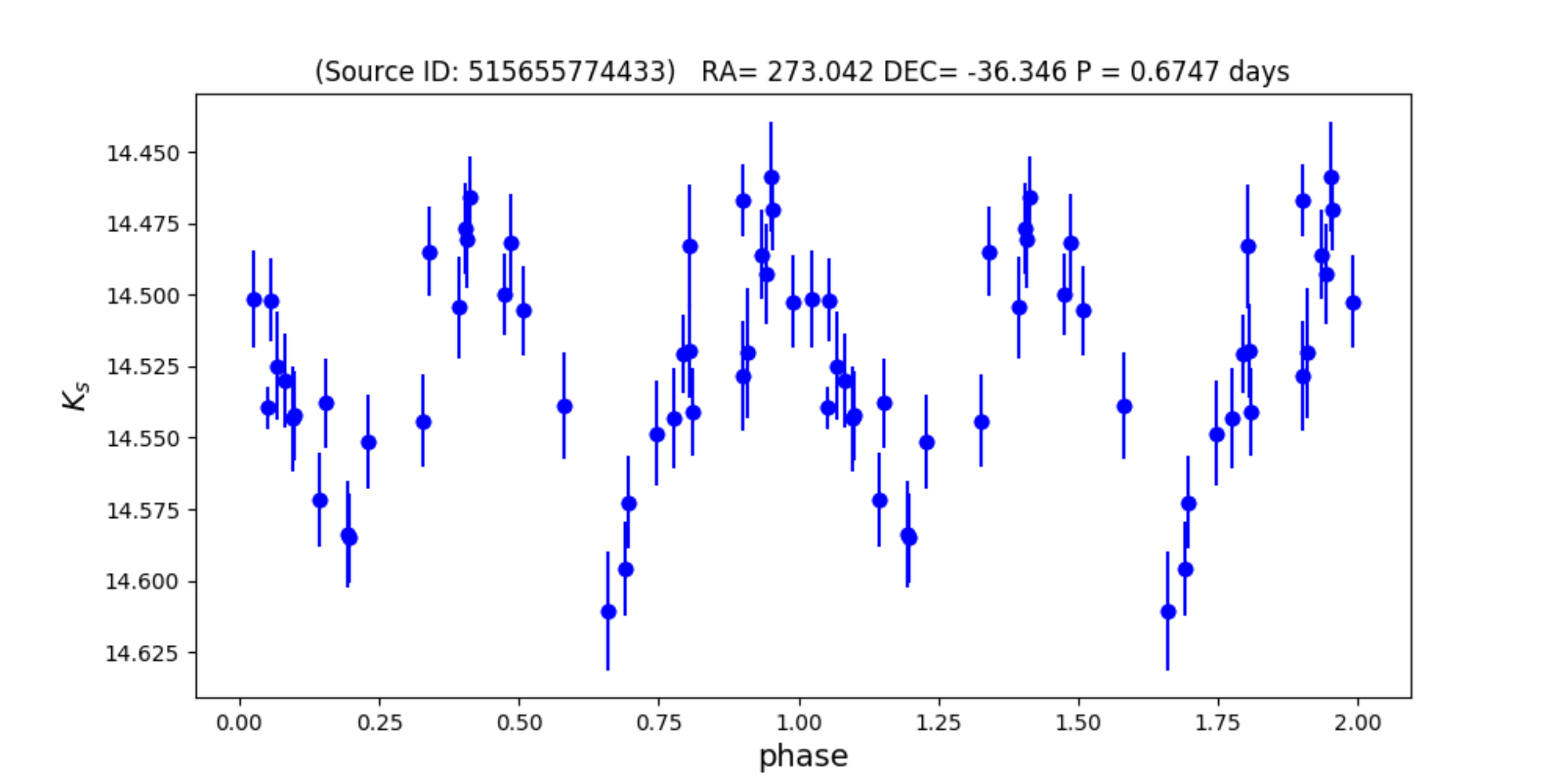} \\
    \includegraphics[width=.33\textwidth]{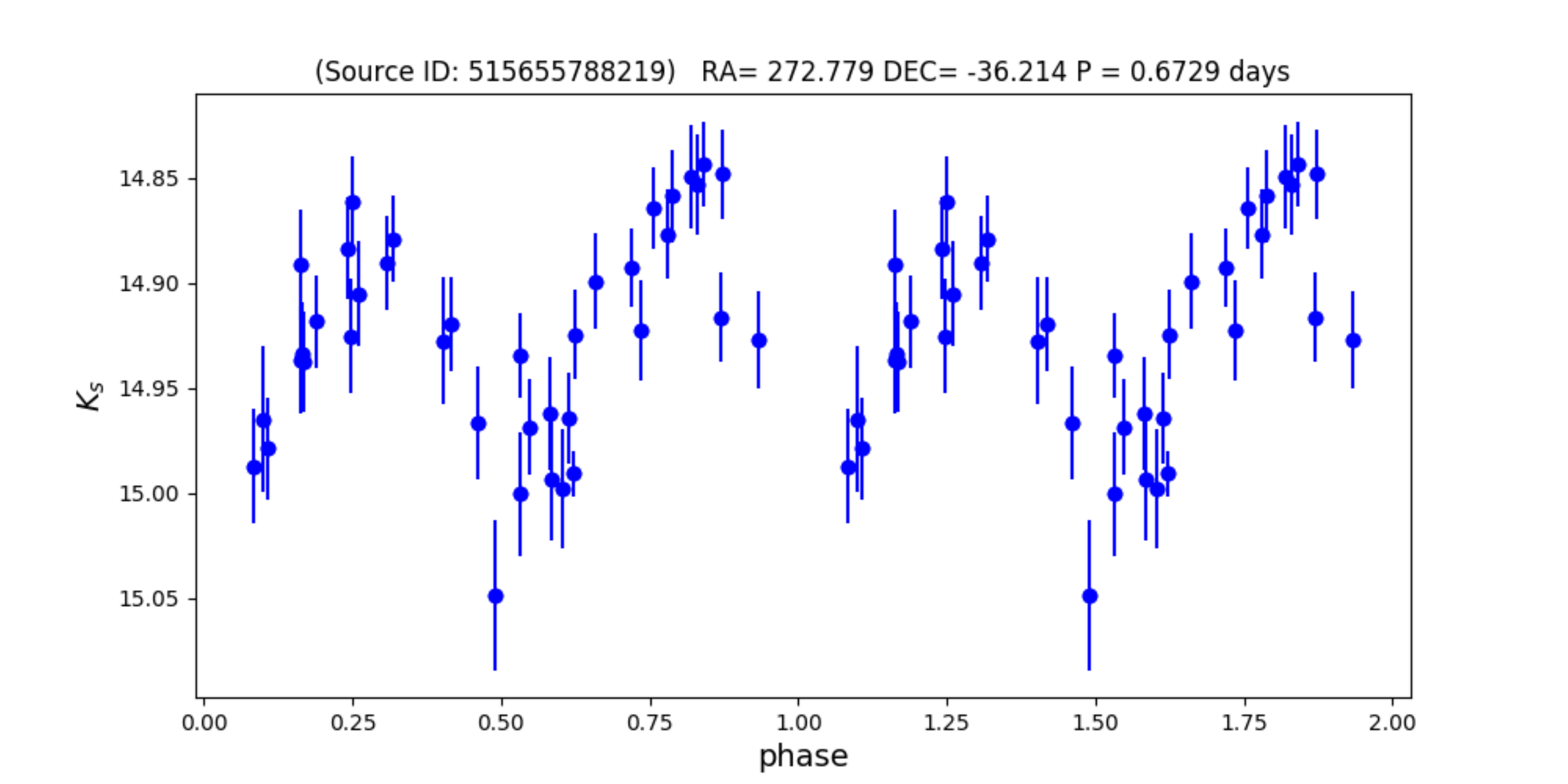} &
    \includegraphics[width=.33\textwidth]{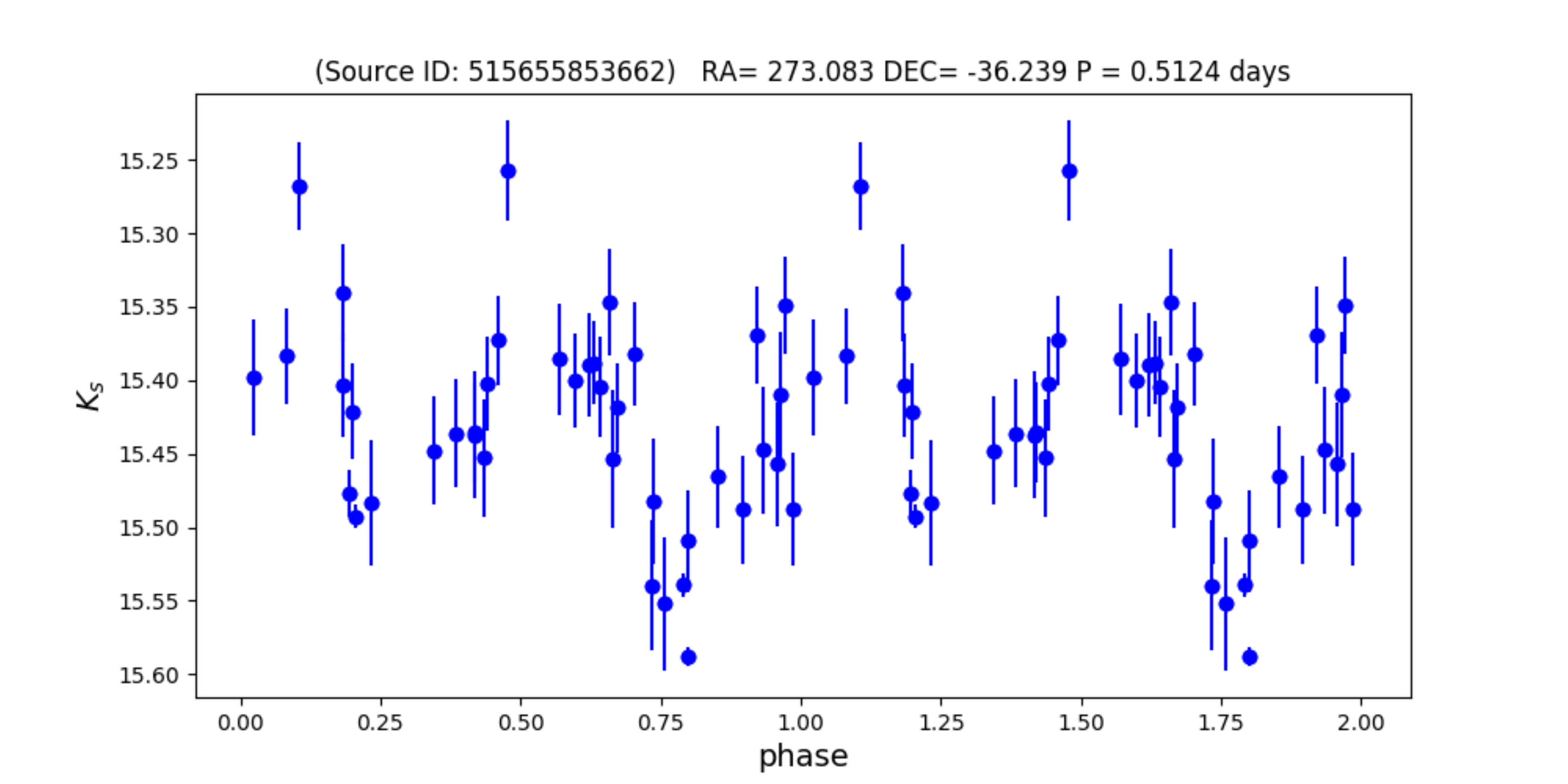} &
    \includegraphics[width=.33\textwidth]{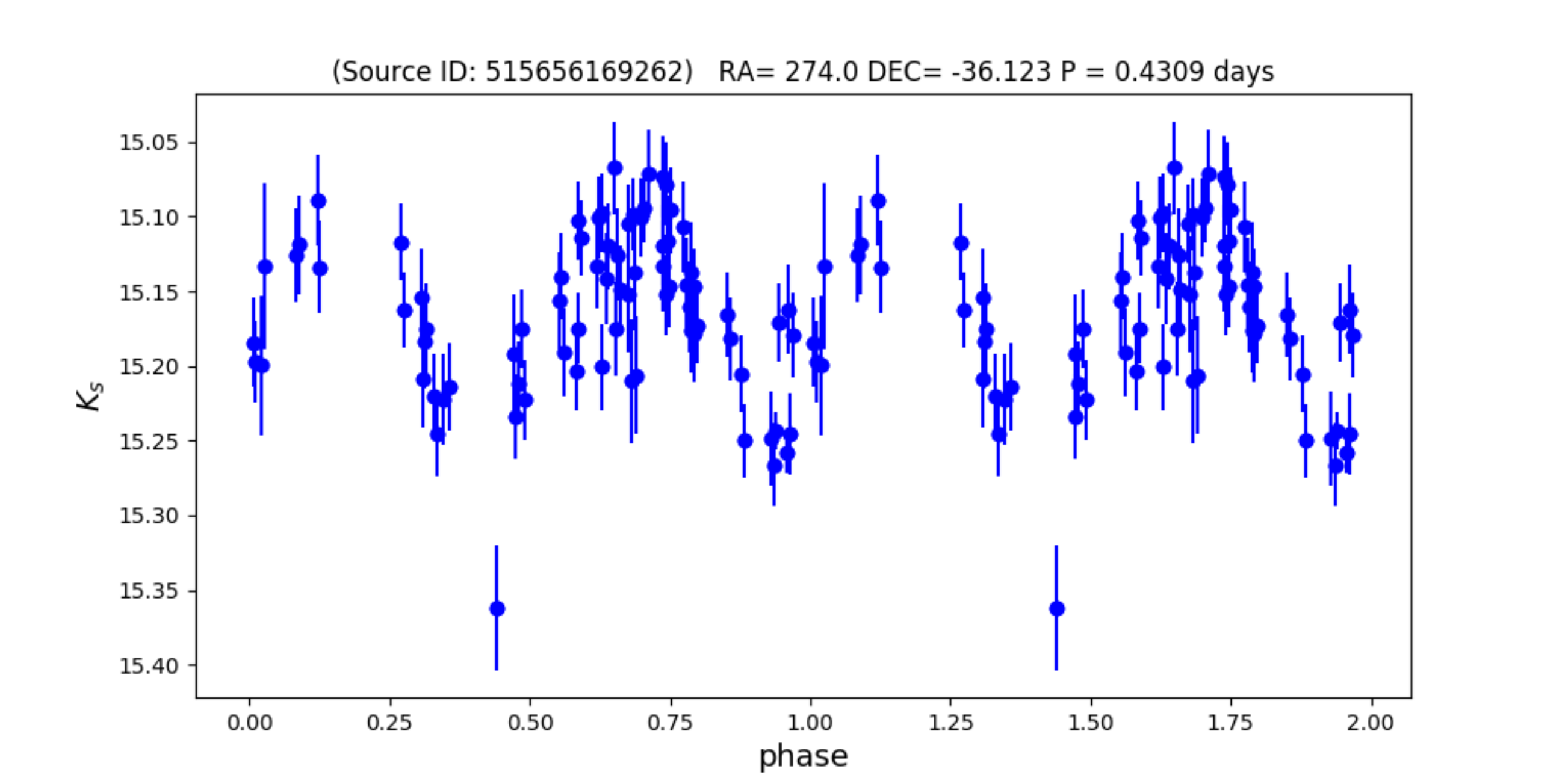} \\
    \includegraphics[width=.33\textwidth]{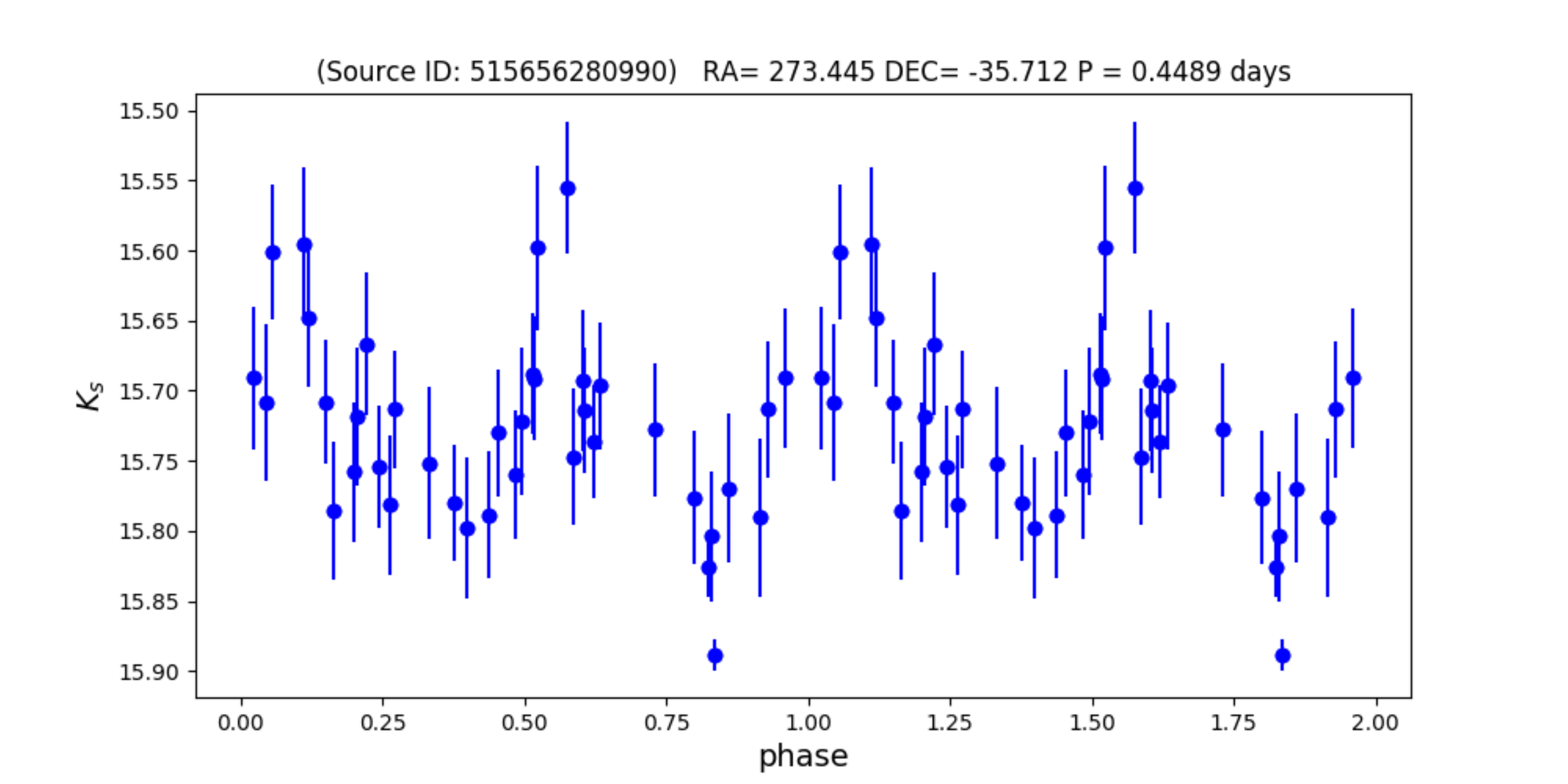} &
    \includegraphics[width=.33\textwidth]{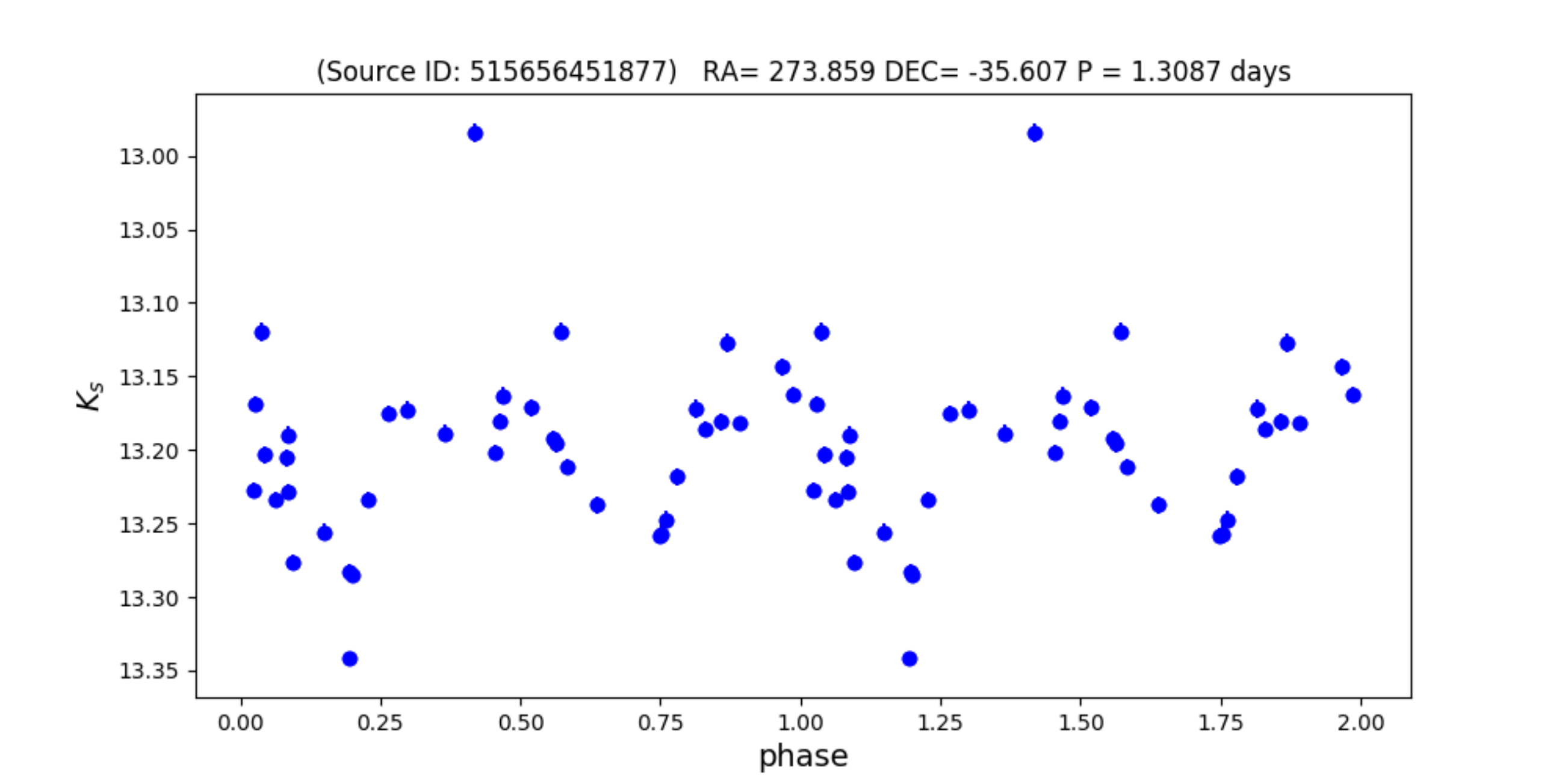} &
    \includegraphics[width=.33\textwidth]{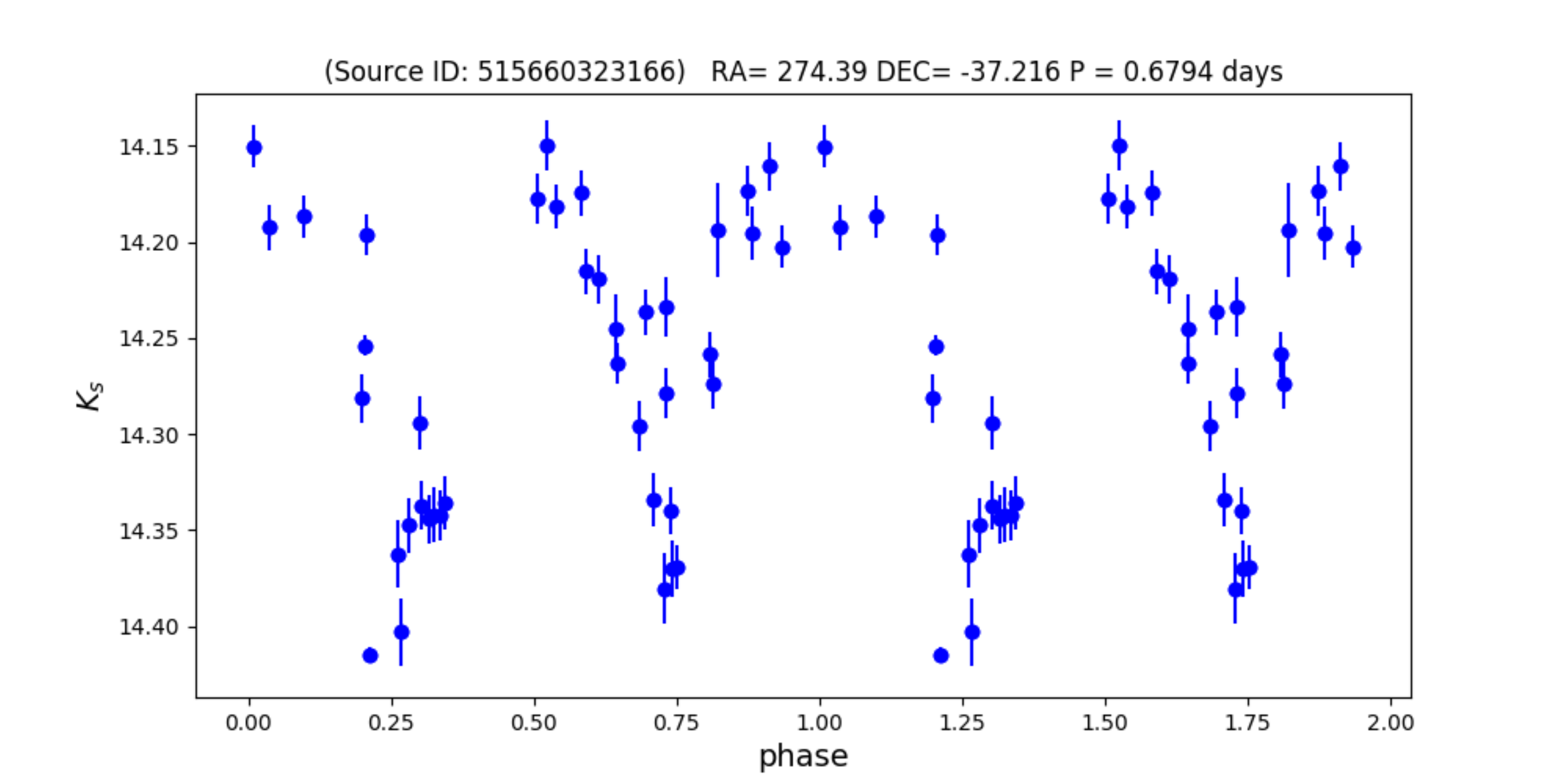} \\
    \includegraphics[width=.33\textwidth]{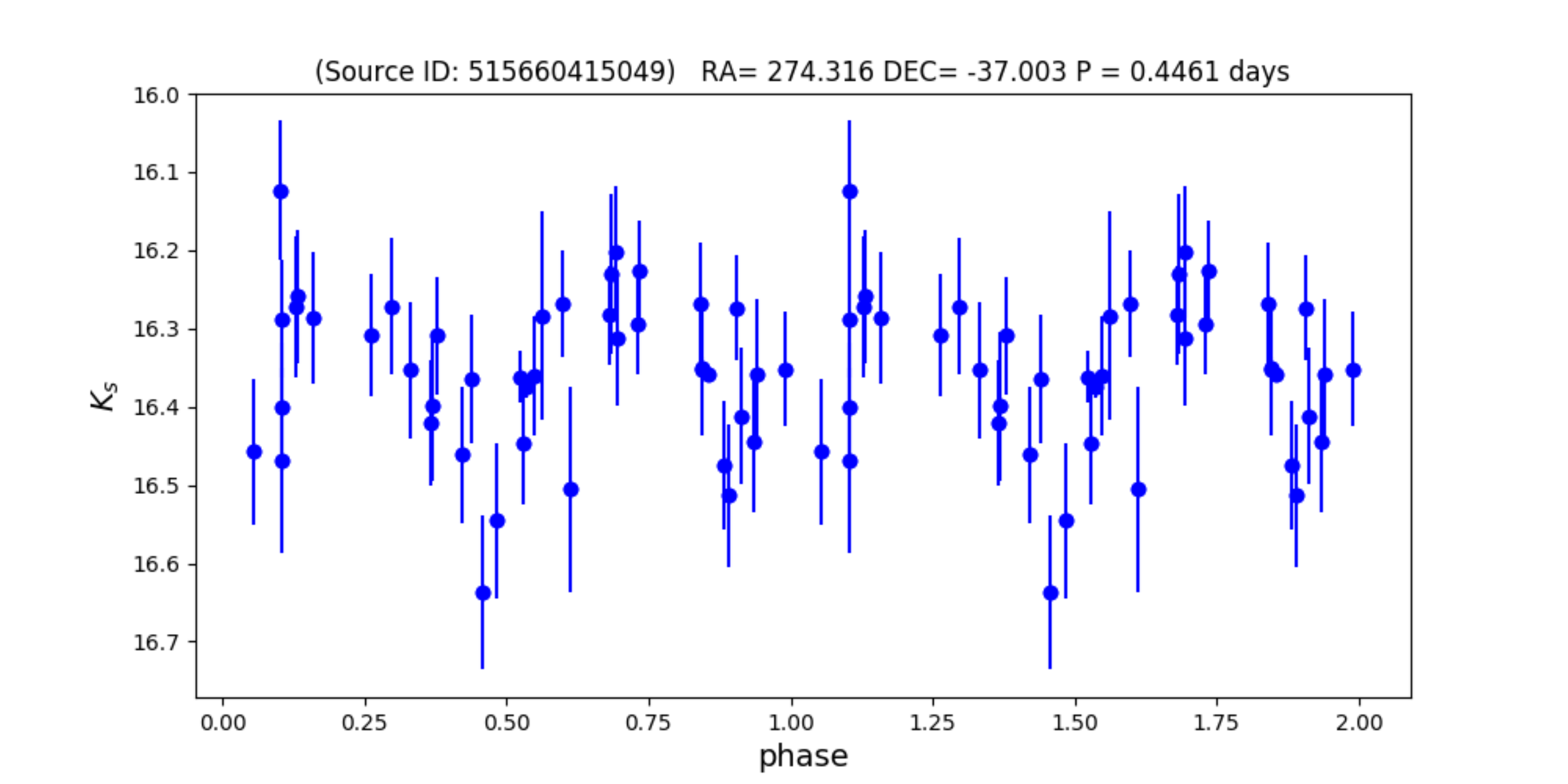} &
    \includegraphics[width=.33\textwidth]{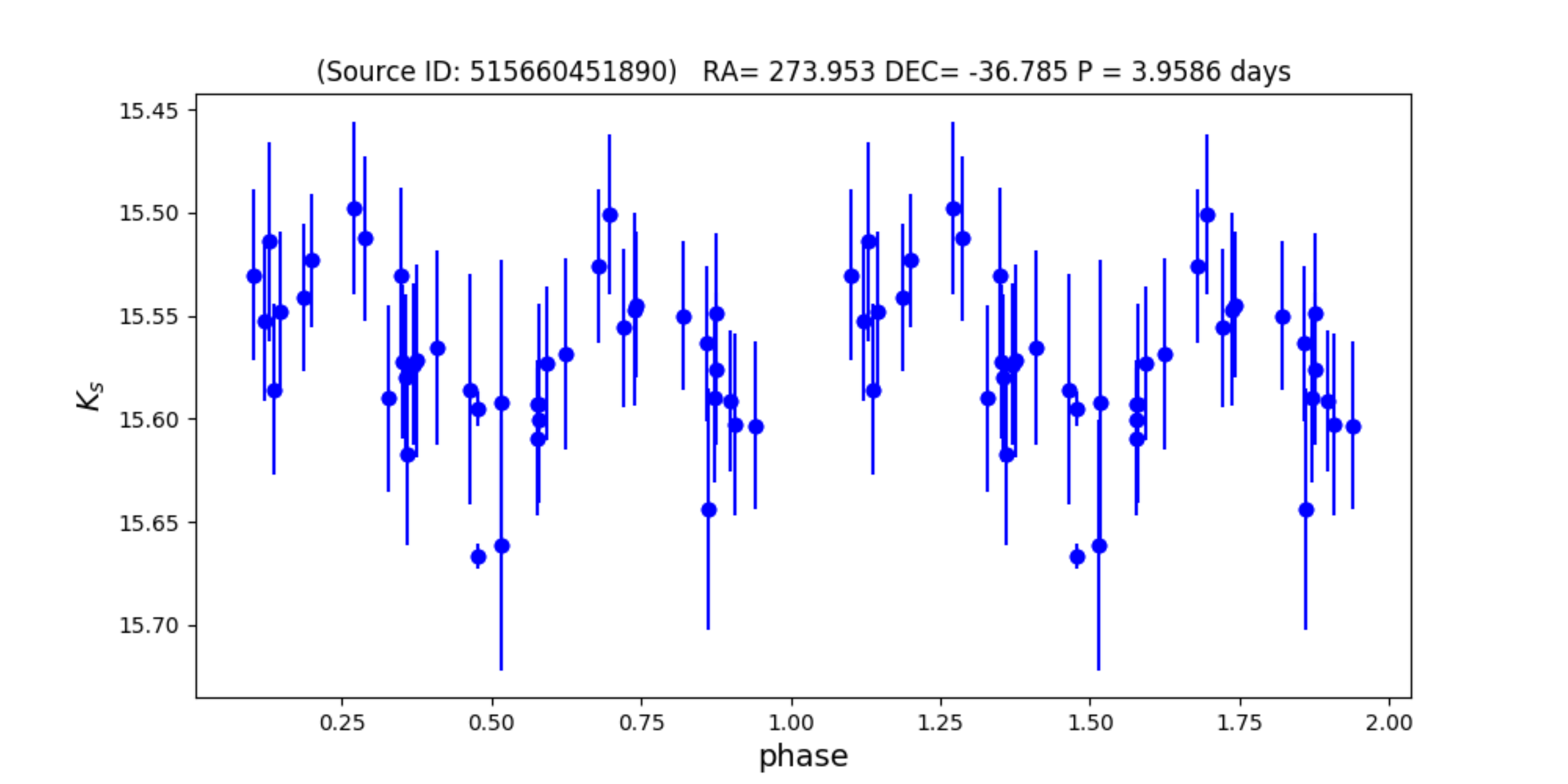} &
    \includegraphics[width=.33\textwidth]{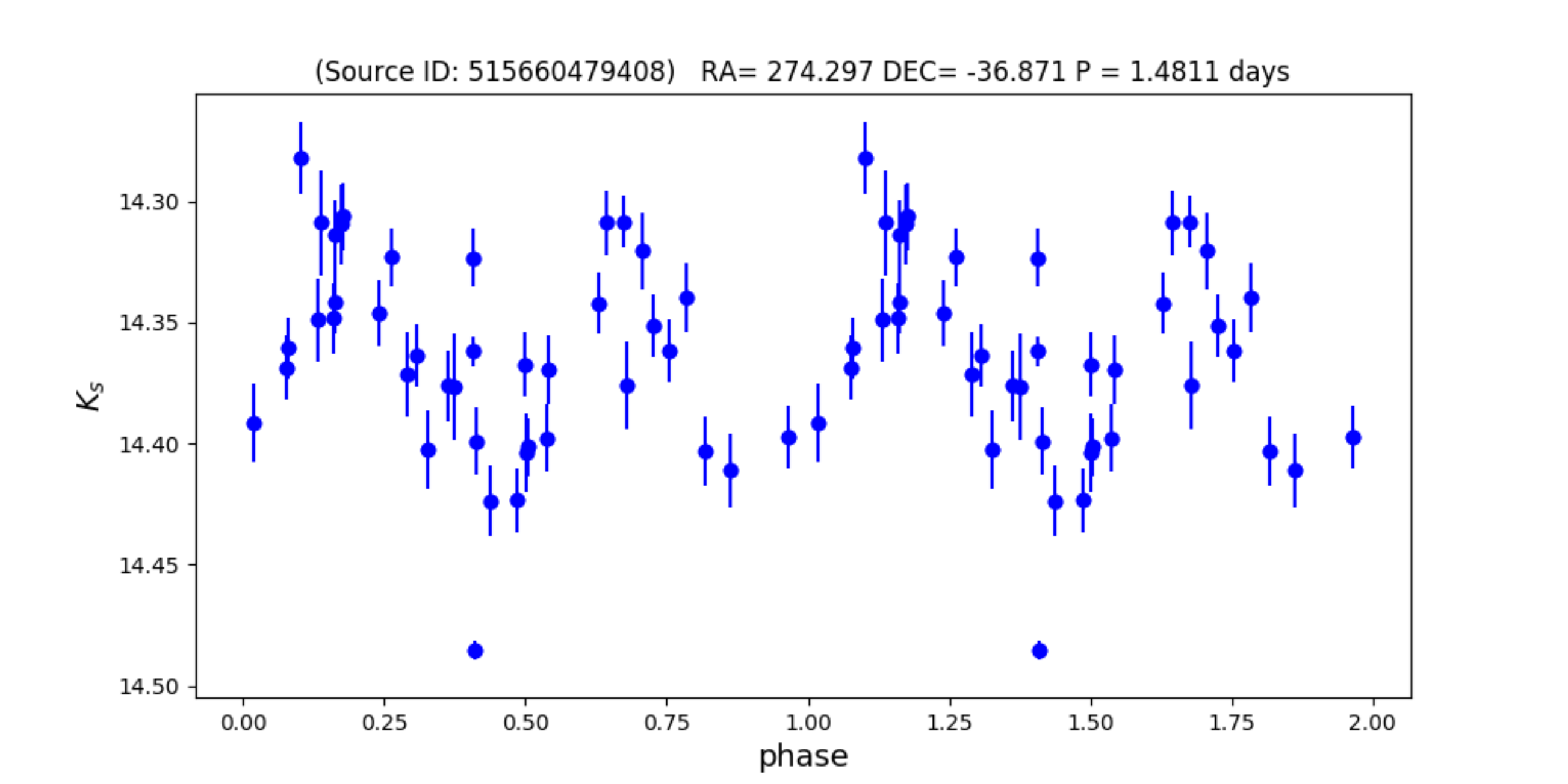} \\
    \includegraphics[width=.33\textwidth]{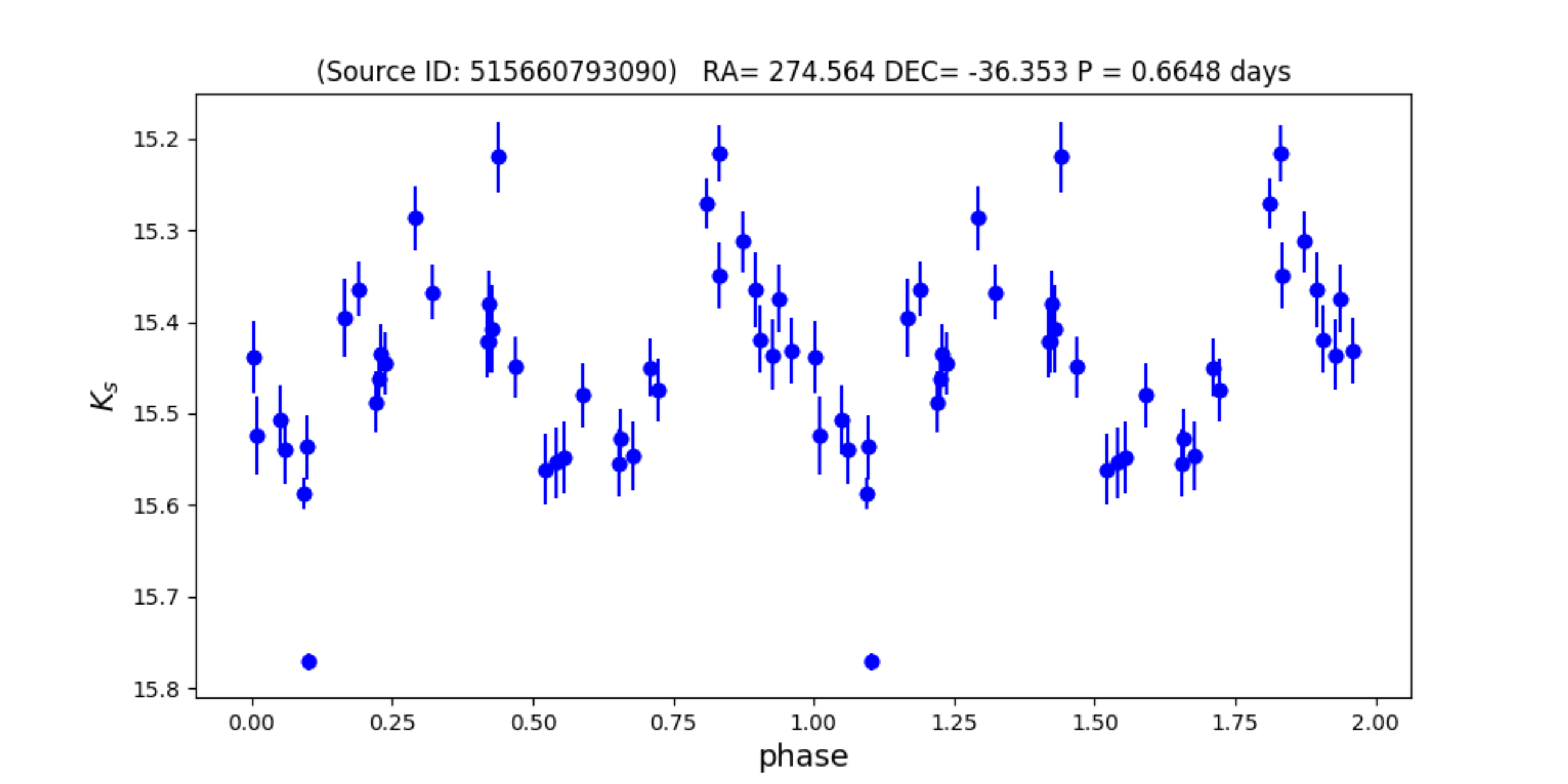} &
    \includegraphics[width=.33\textwidth]{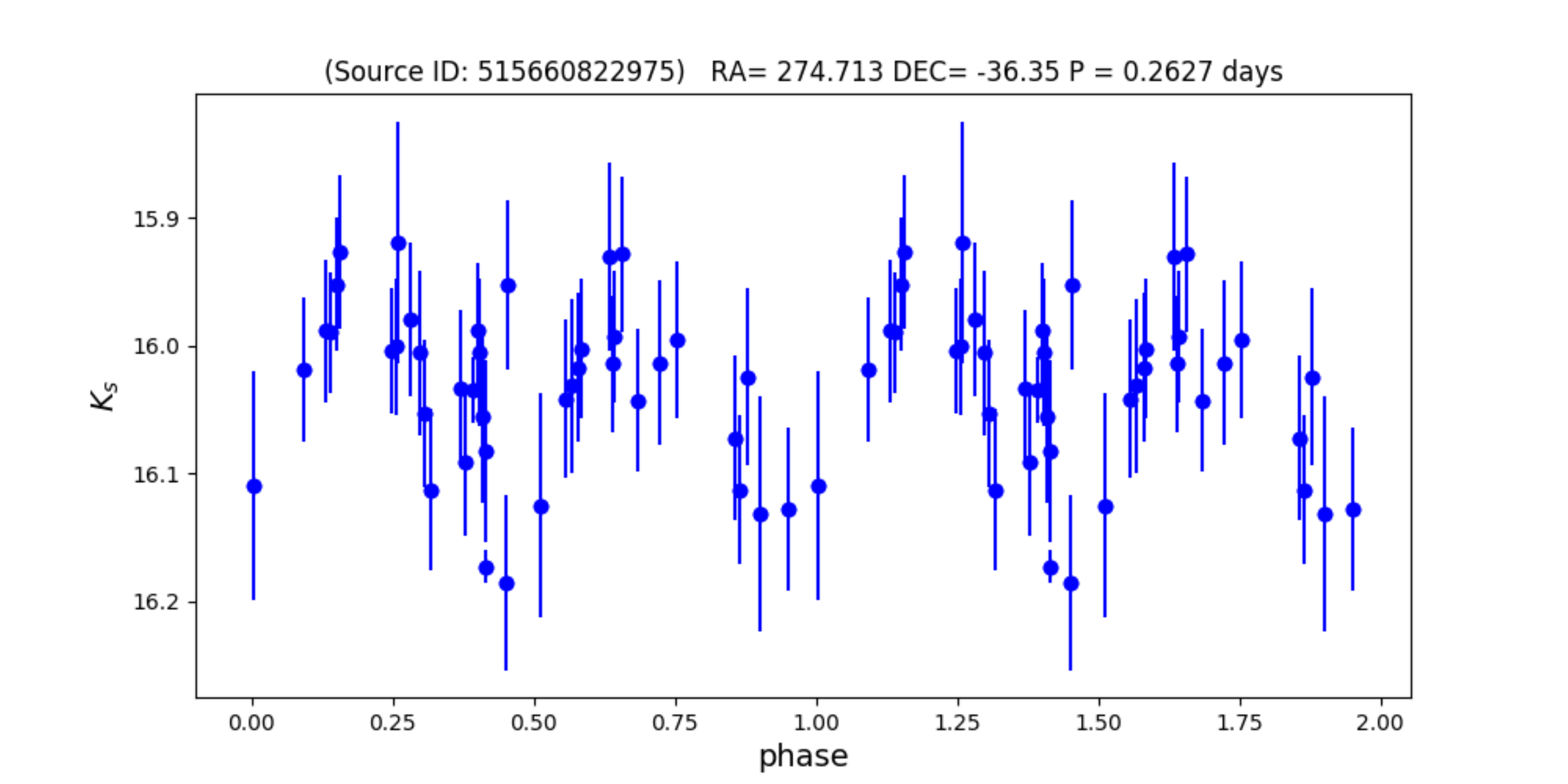} &
    \includegraphics[width=.33\textwidth]{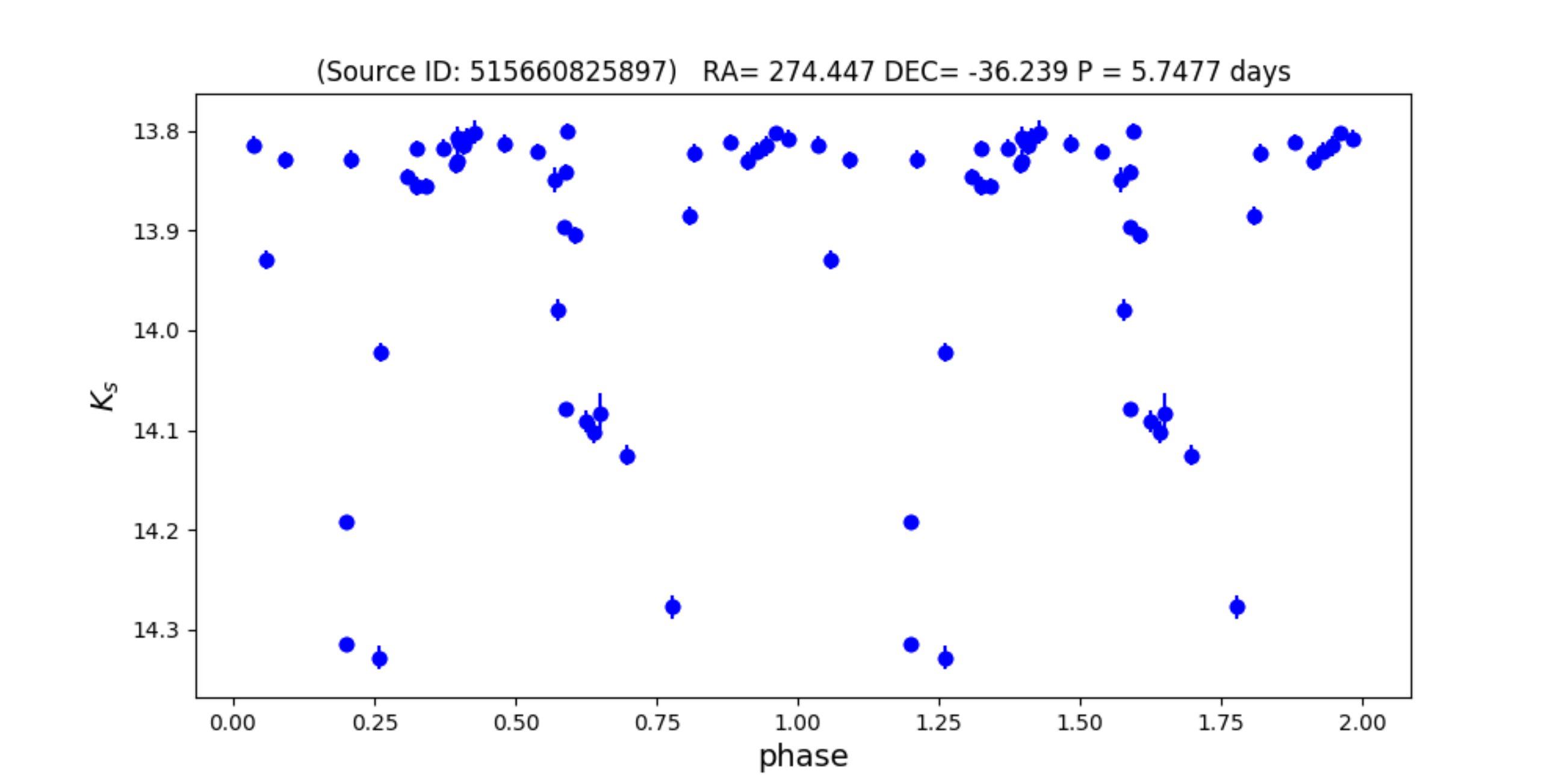} \\ 
  \end{tabular}
  \caption{BHB stars light curves corresponding to $1^{st}$ category eclipsing binaries.}
  \label{fig:bhb7}
\end{figure*}

\begin{figure*}[htb]
\centering
  \begin{tabular}{@{}ccc@{}}     
    \includegraphics[width=.33\textwidth]{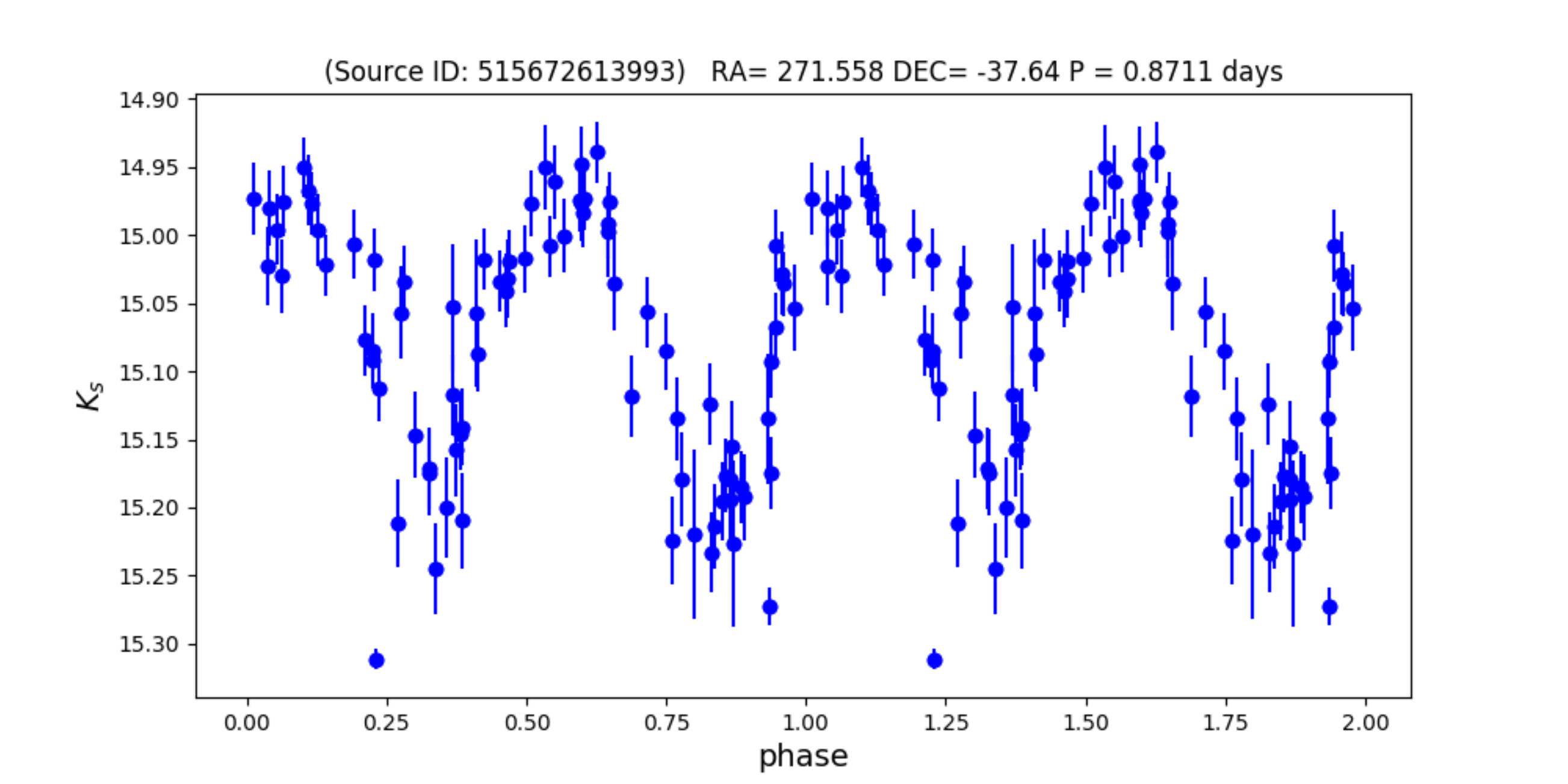} &   
    \includegraphics[width=.33\textwidth]{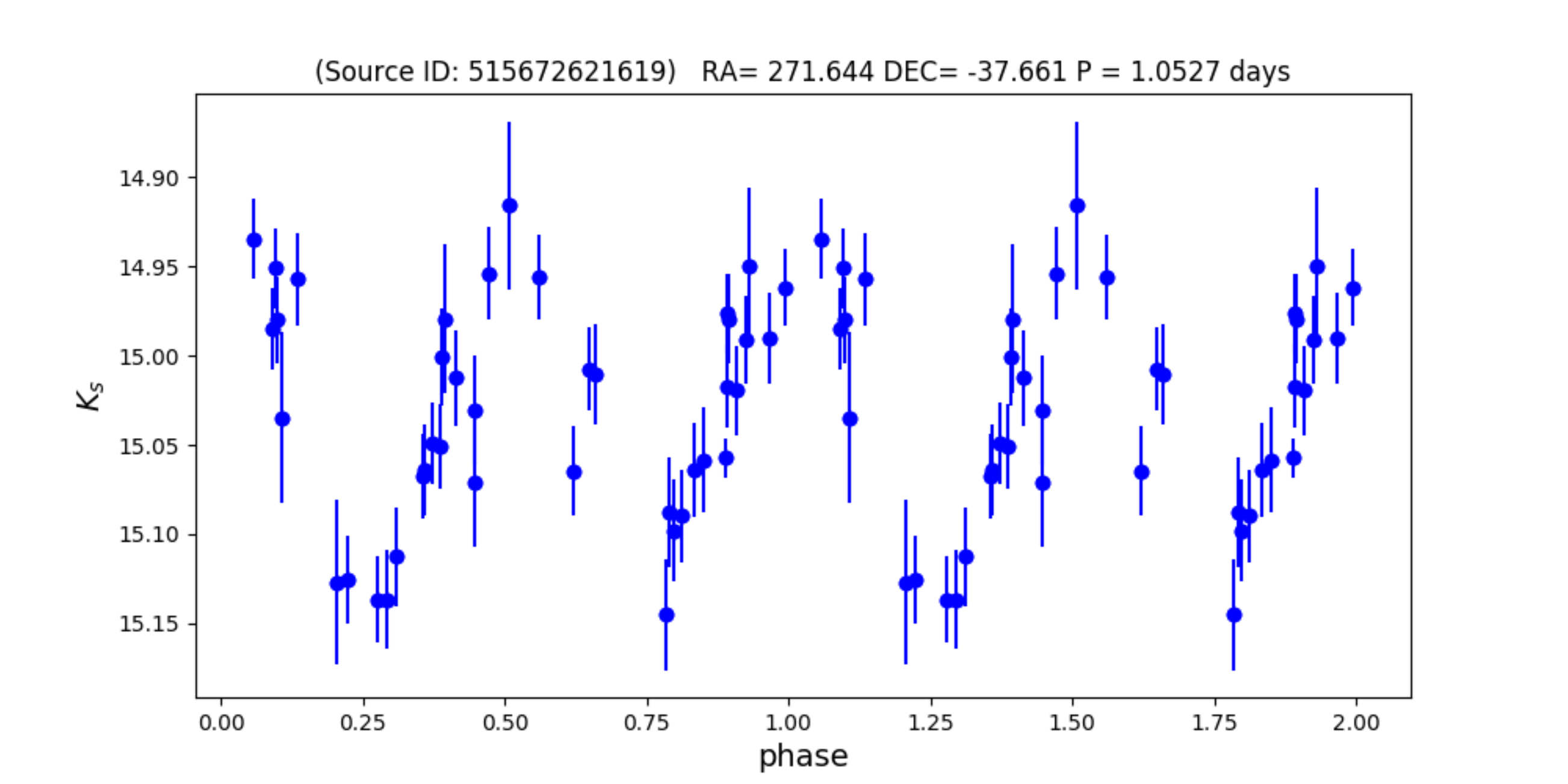} &
    \includegraphics[width=.33\textwidth]{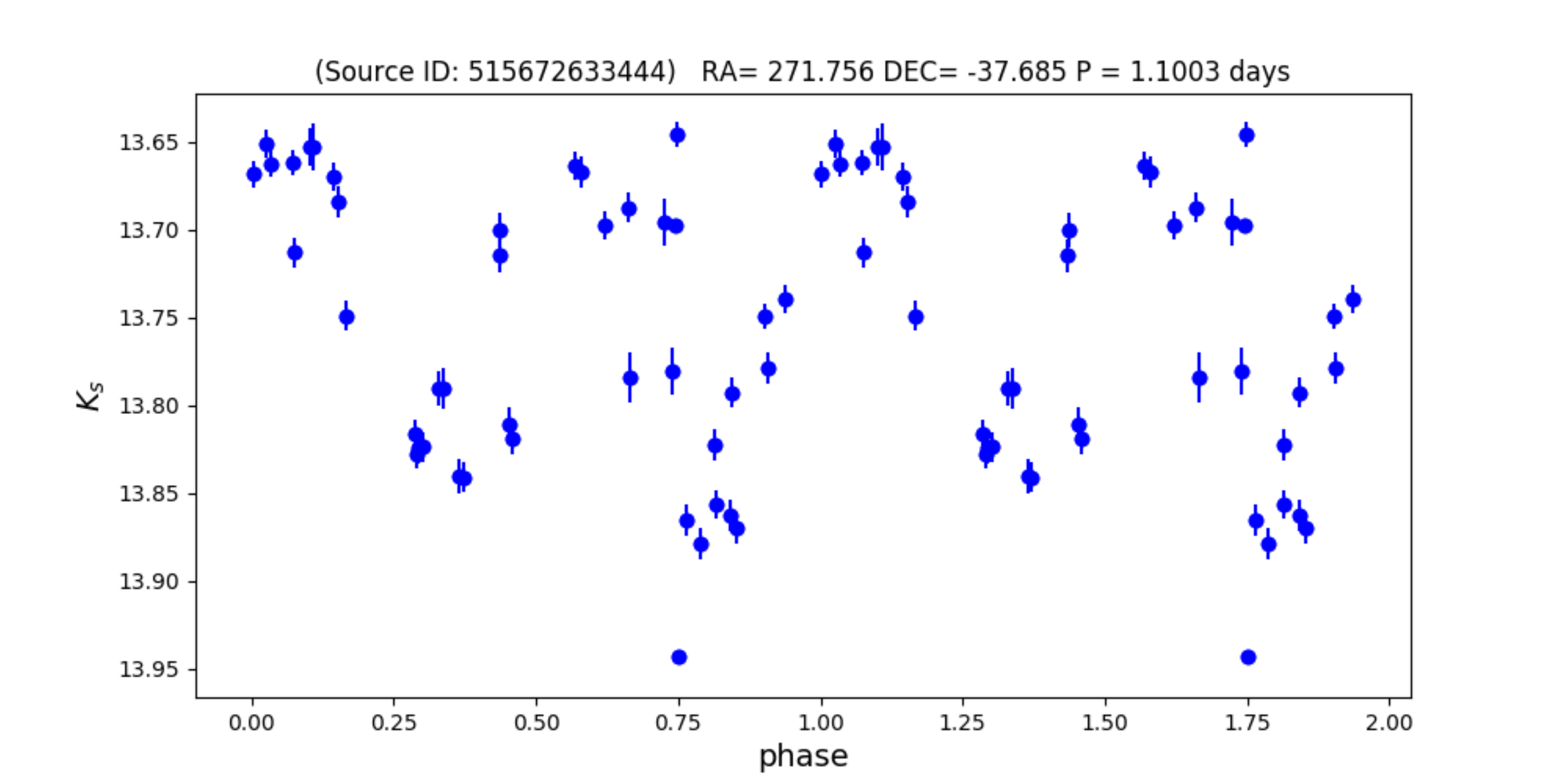} \\
    \includegraphics[width=.33\textwidth]{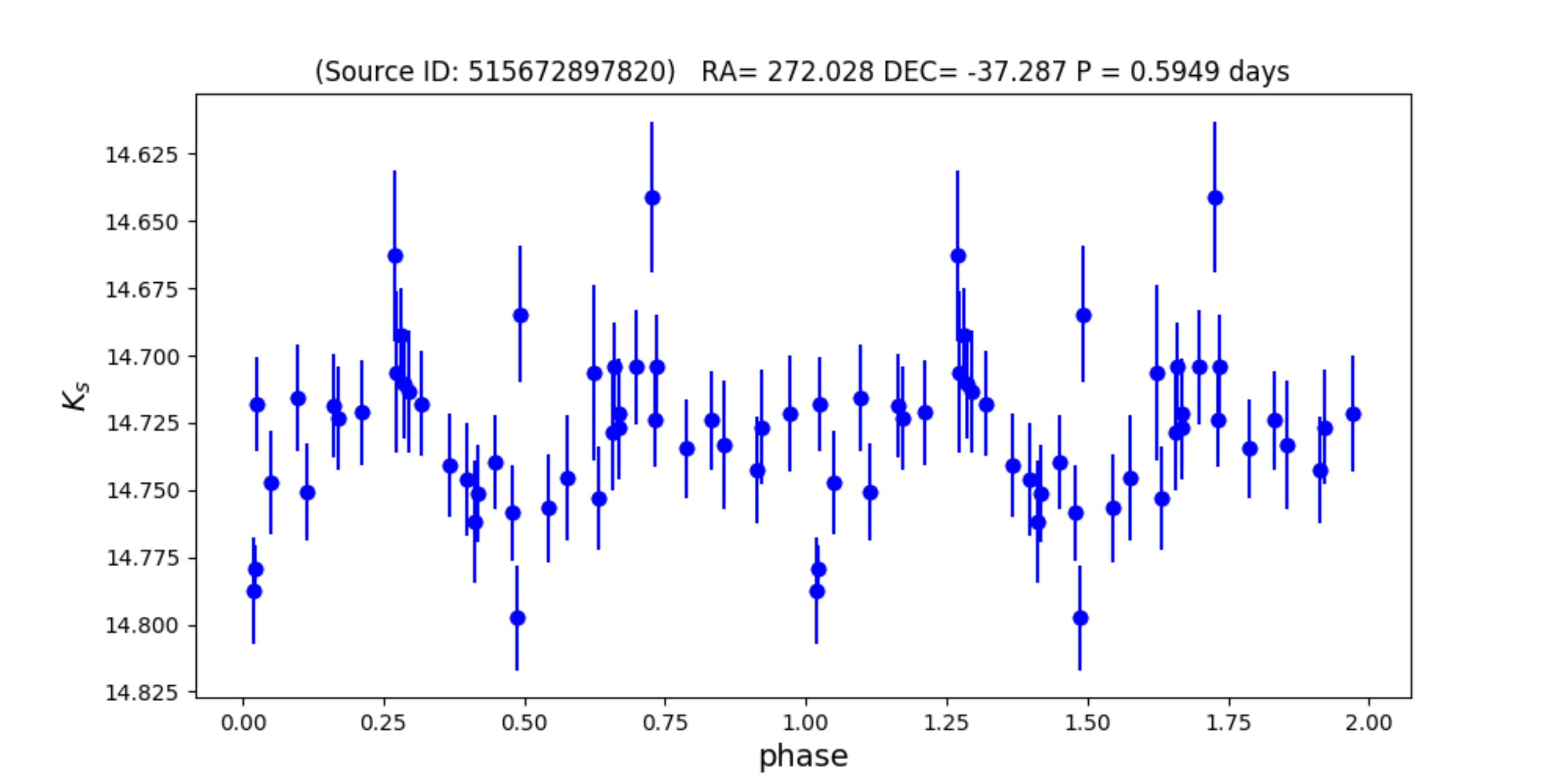} &
    \includegraphics[width=.33\textwidth]{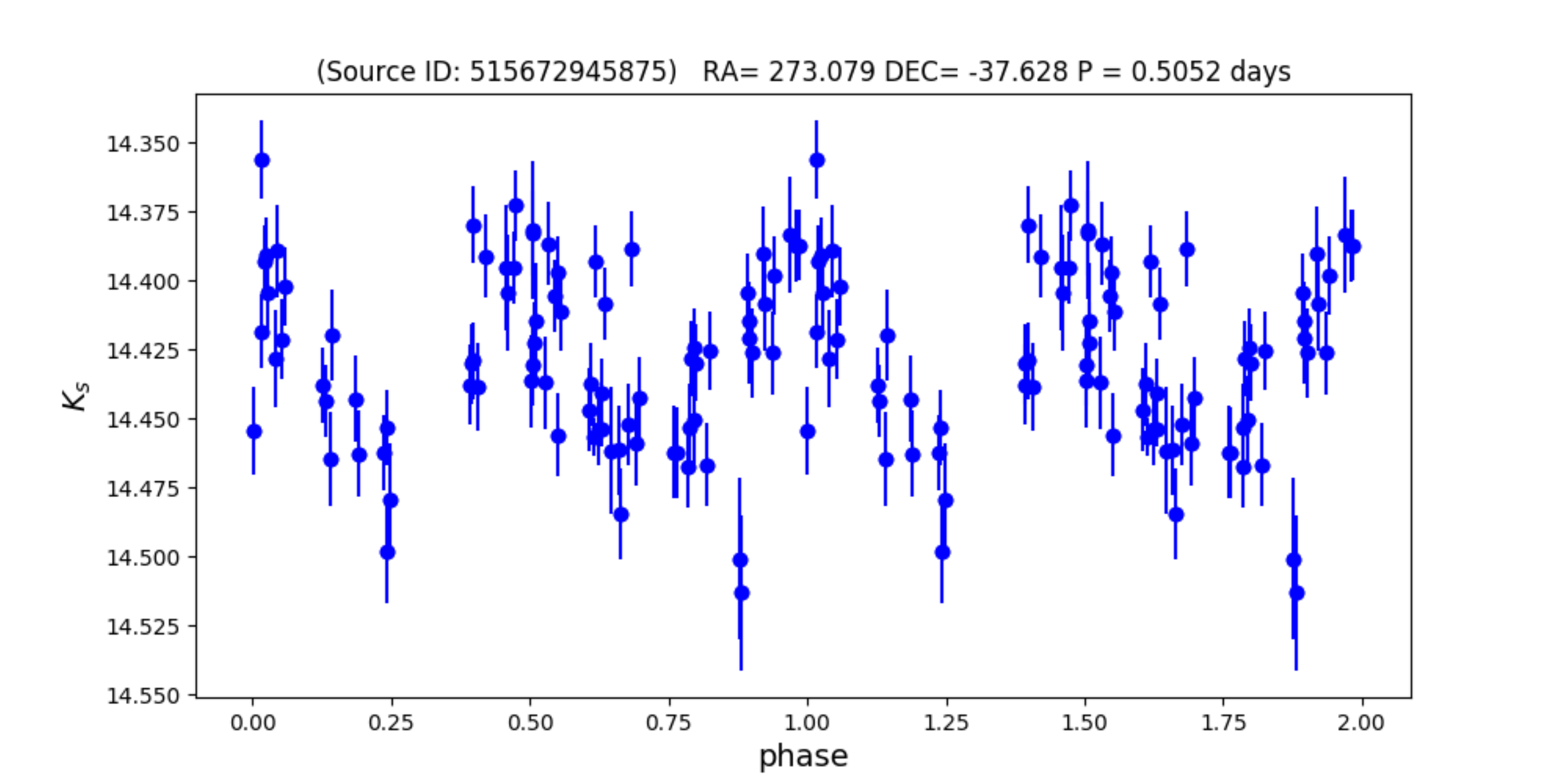} &
    \includegraphics[width=.33\textwidth]{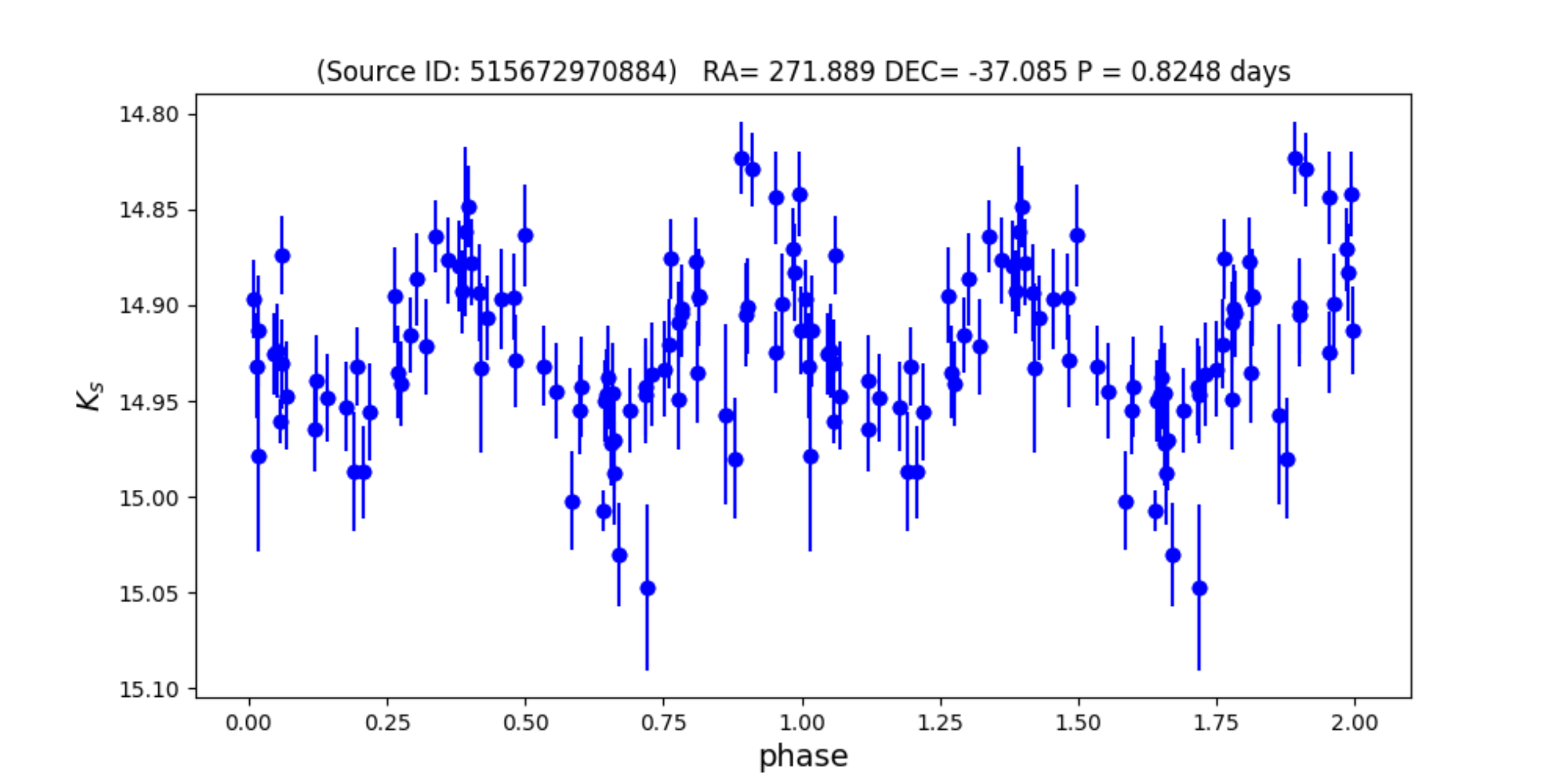} \\
    \includegraphics[width=.33\textwidth]{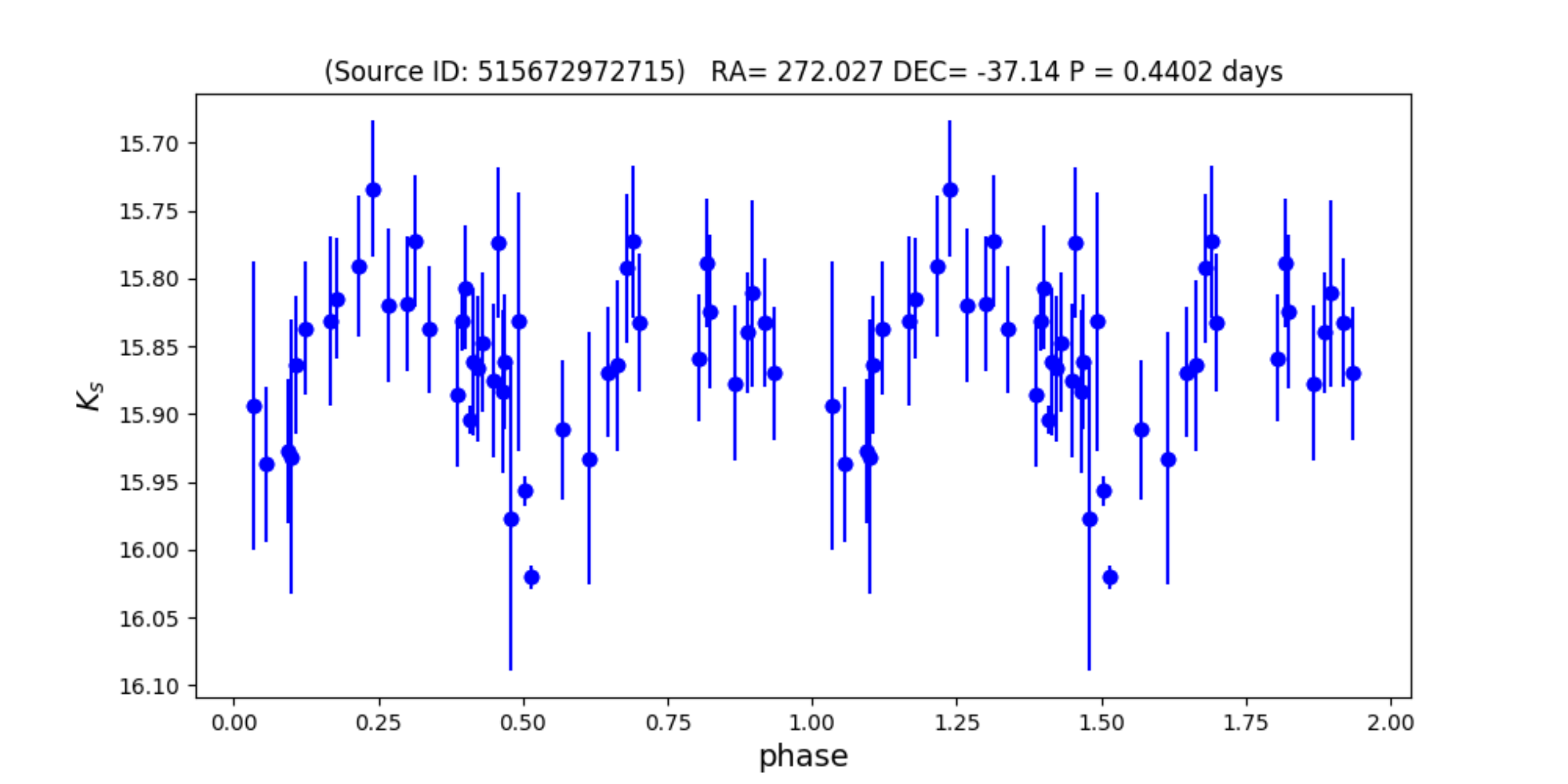} &
    \includegraphics[width=.33\textwidth]{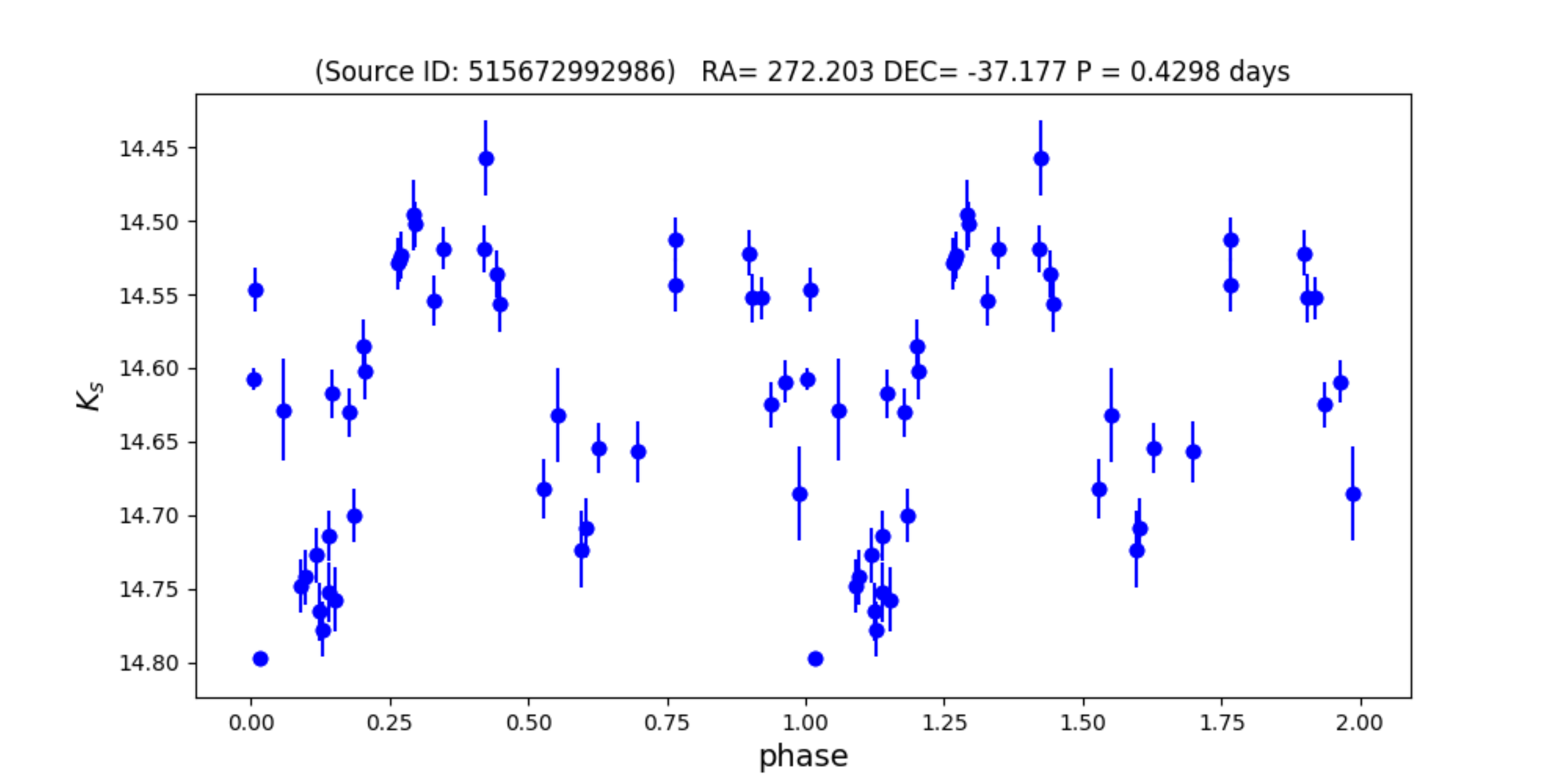} &
    \includegraphics[width=.33\textwidth]{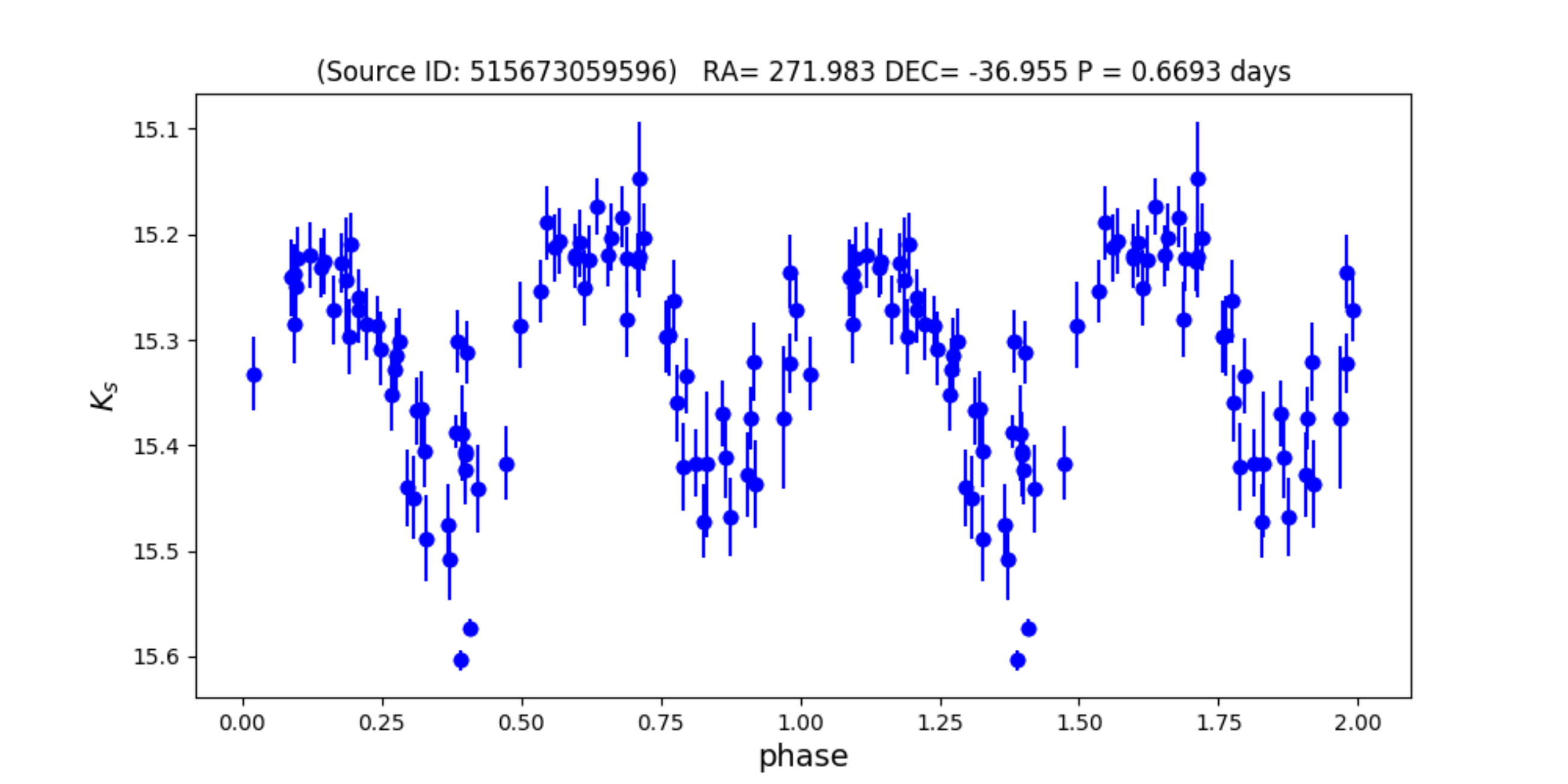} \\
    \includegraphics[width=.33\textwidth]{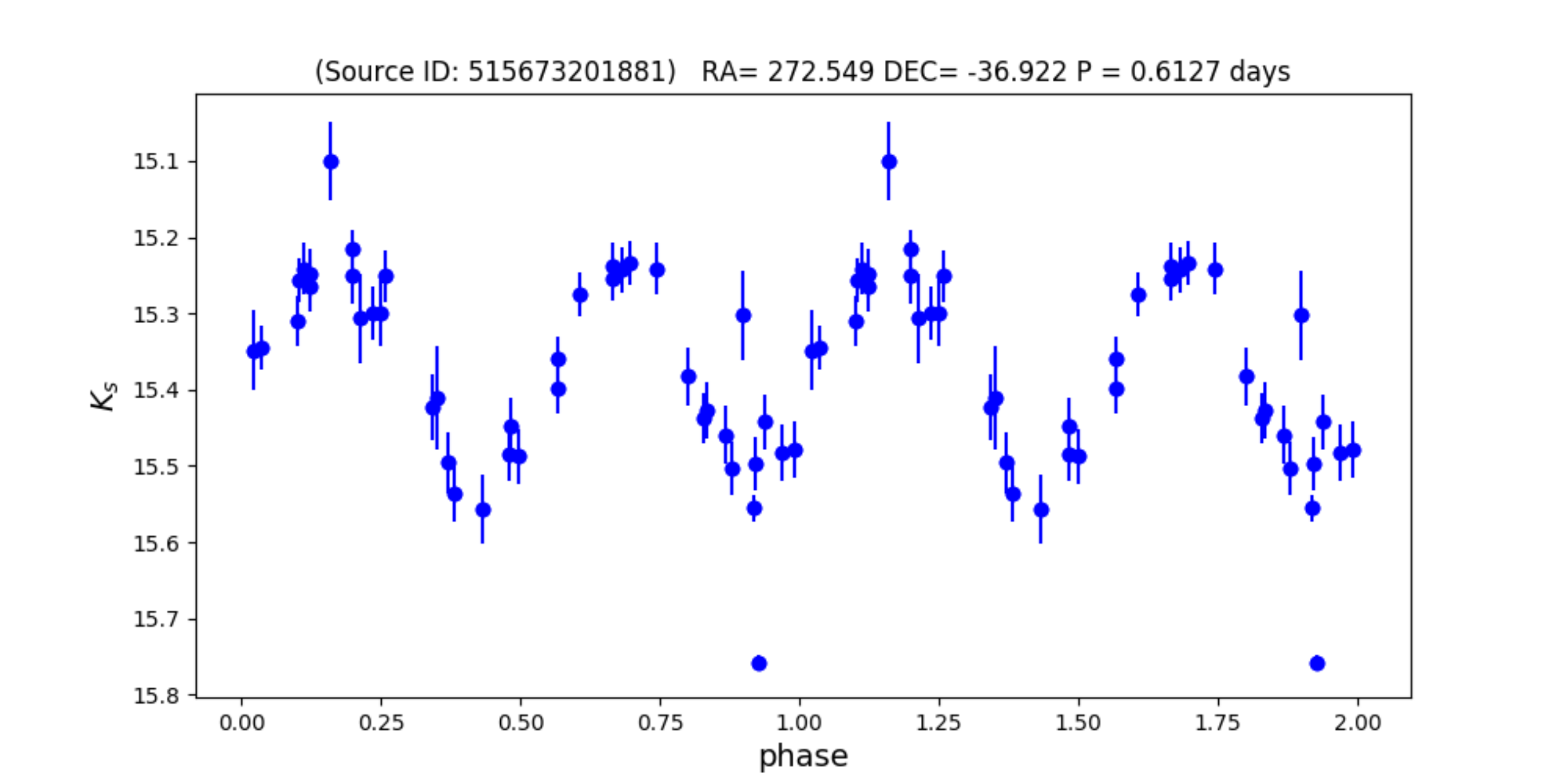} &
    \includegraphics[width=.33\textwidth]{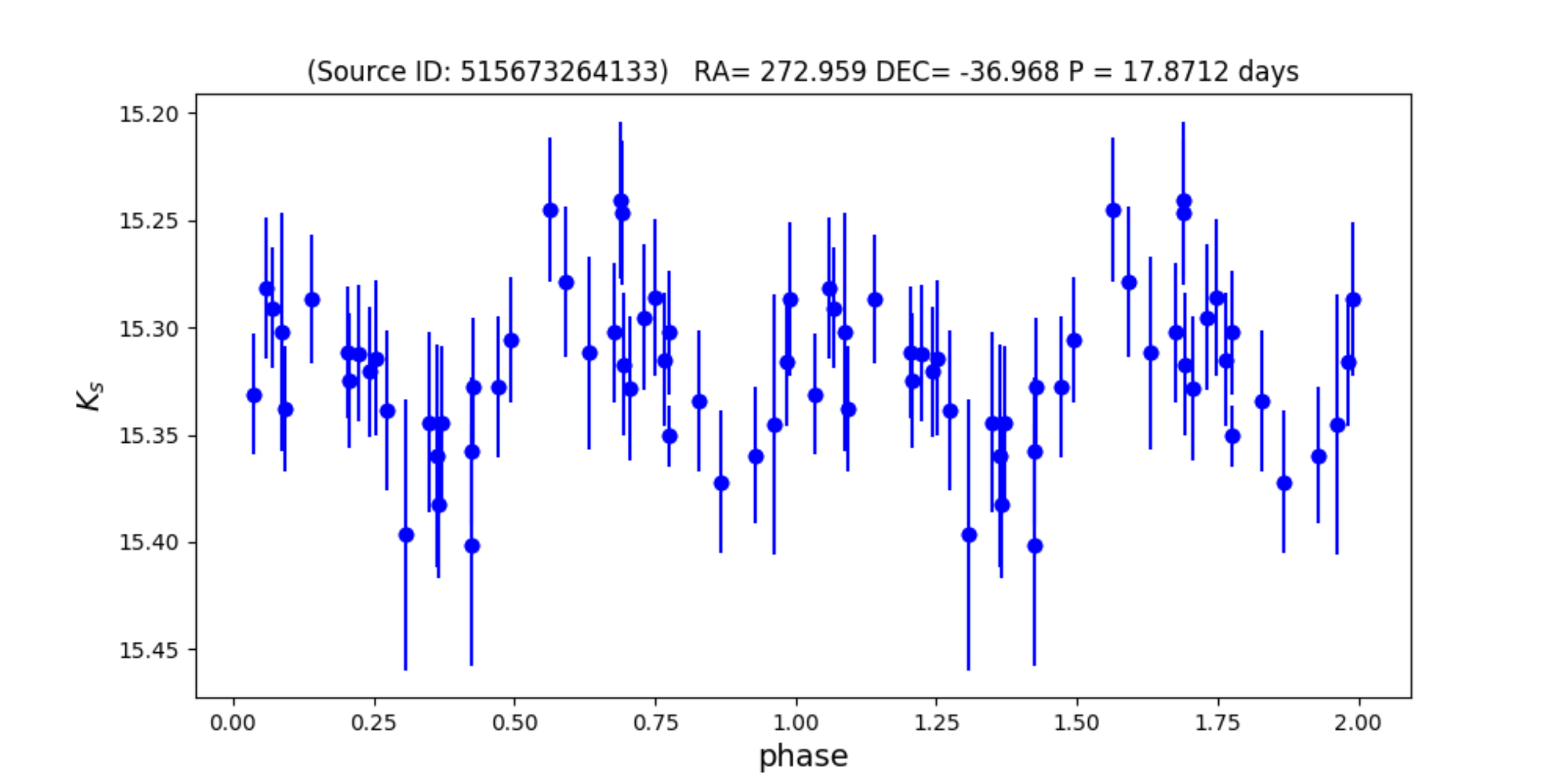} &
    \includegraphics[width=.33\textwidth]{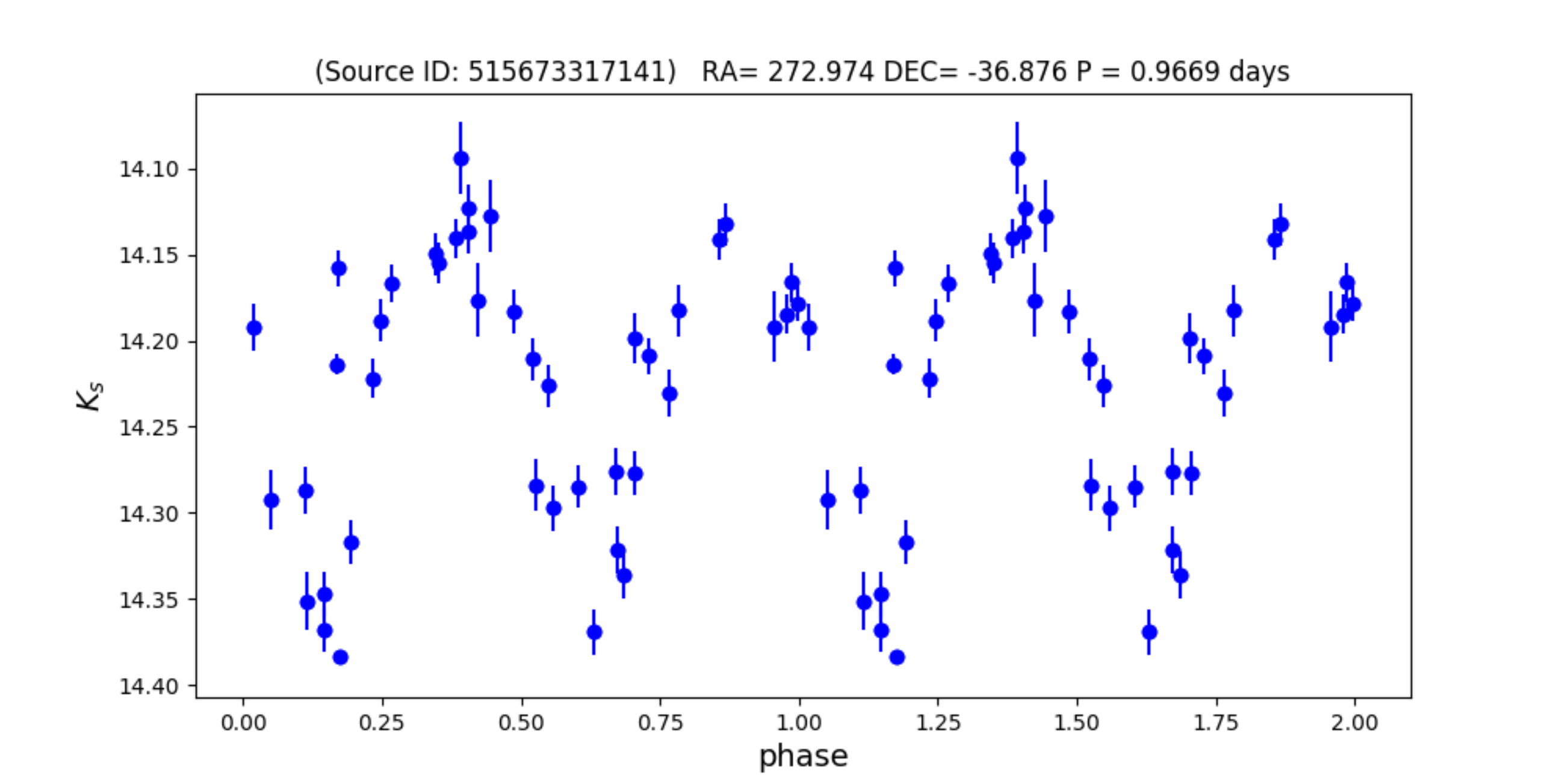} \\
    \includegraphics[width=.33\textwidth]{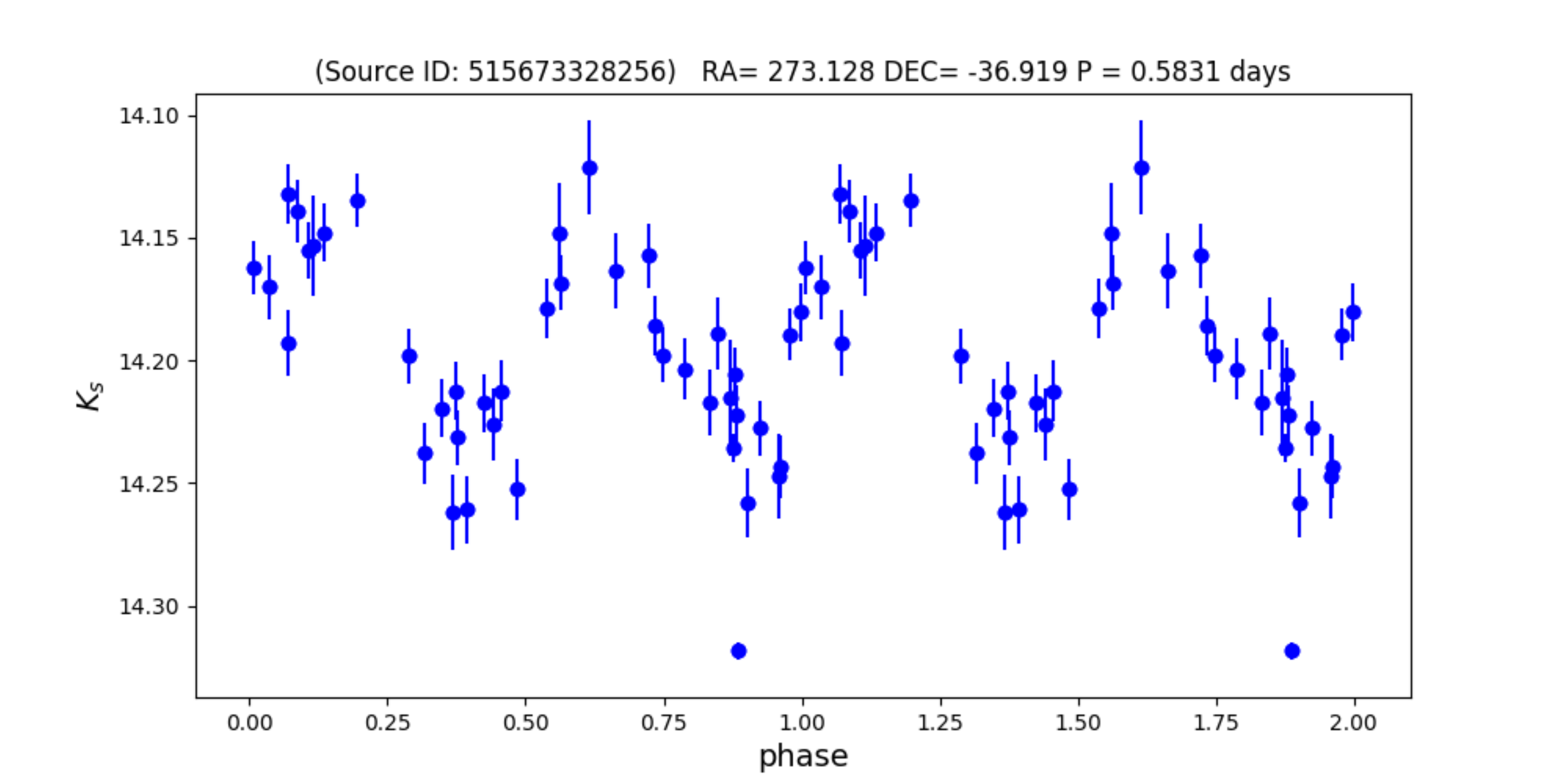} &
    \includegraphics[width=.33\textwidth]{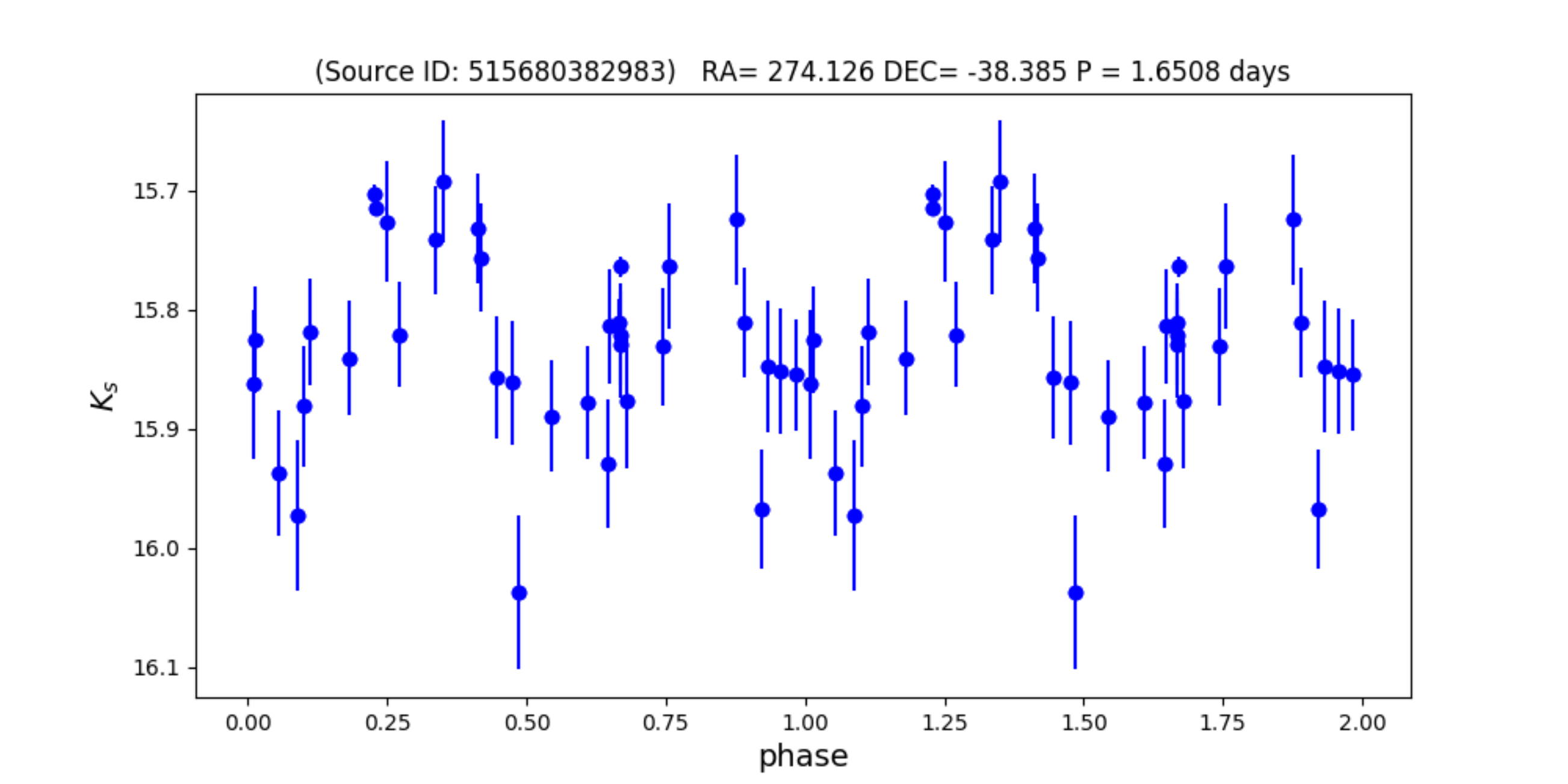} &
    \includegraphics[width=.33\textwidth]{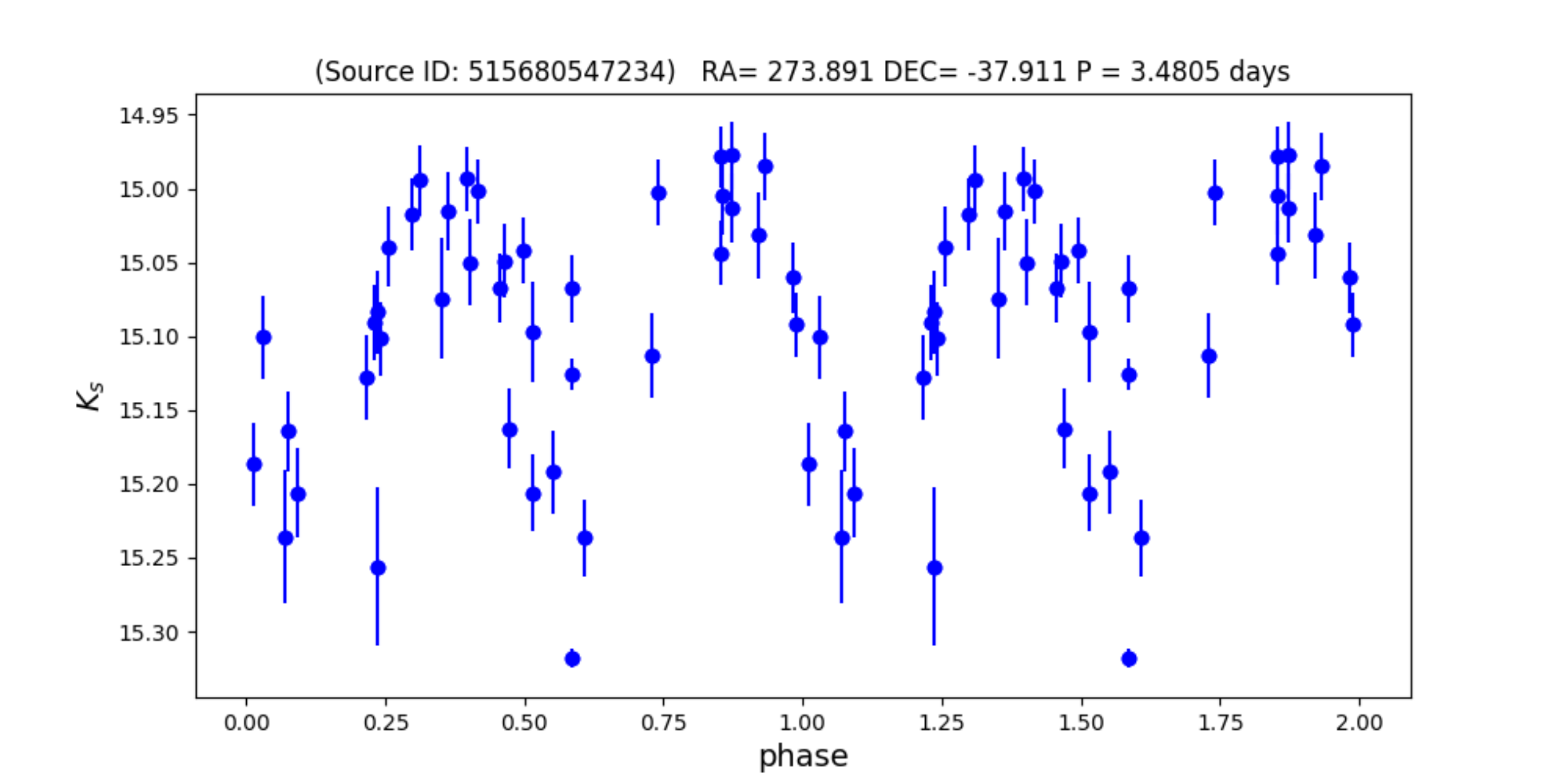} \\
    \includegraphics[width=.33\textwidth]{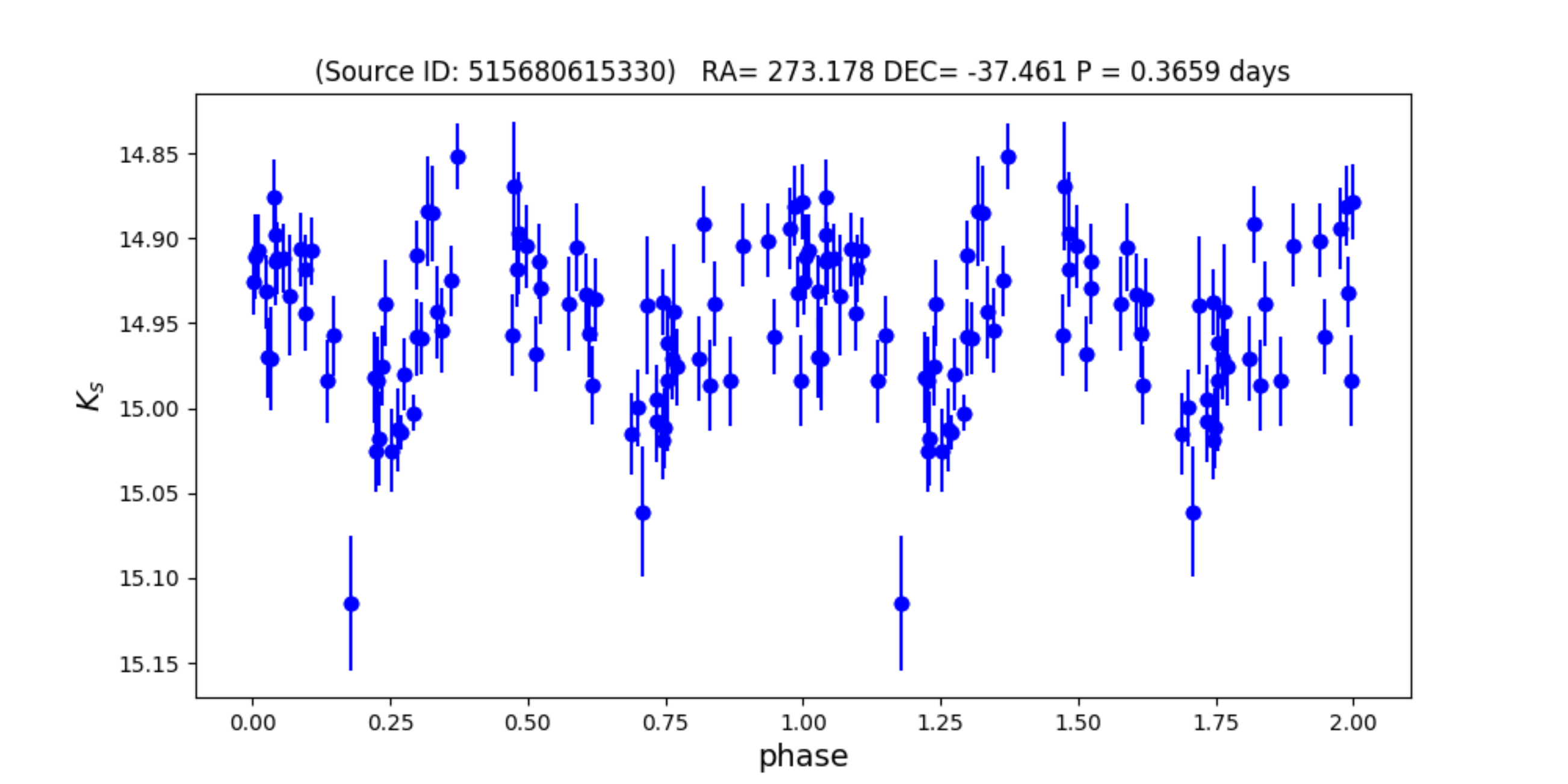} &
    \includegraphics[width=.33\textwidth]{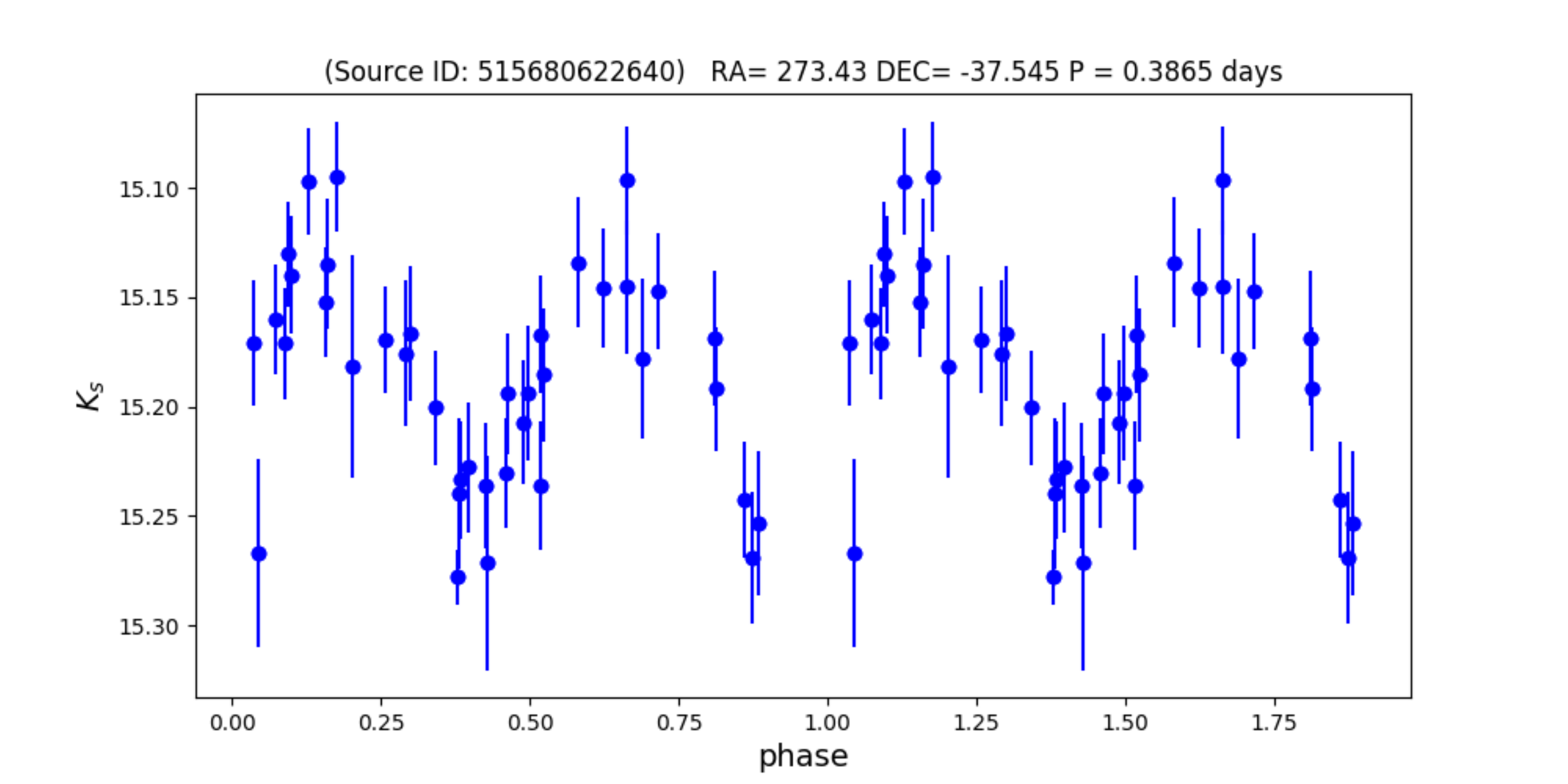} &
    \includegraphics[width=.33\textwidth]{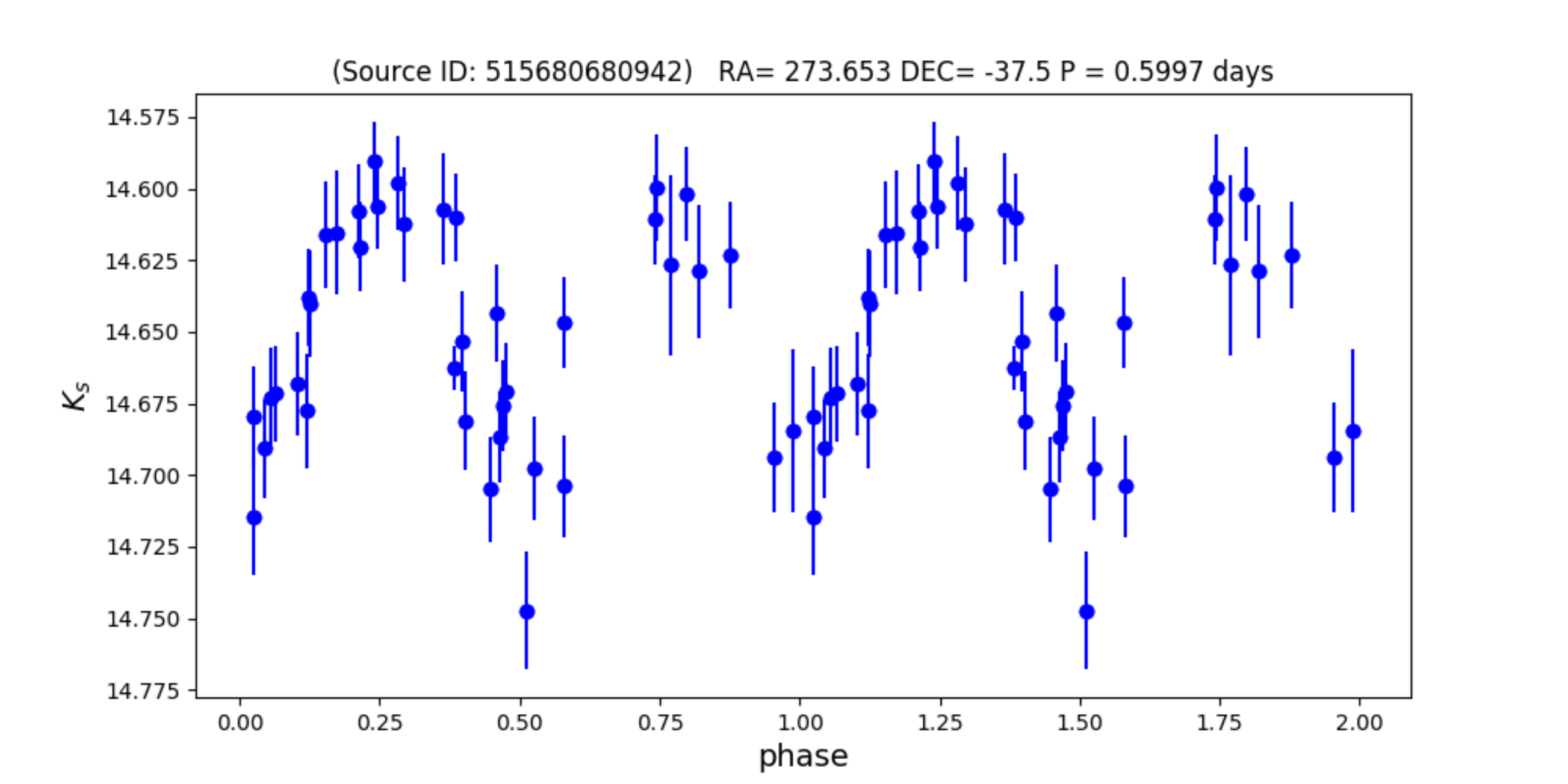} \\
    \includegraphics[width=.33\textwidth]{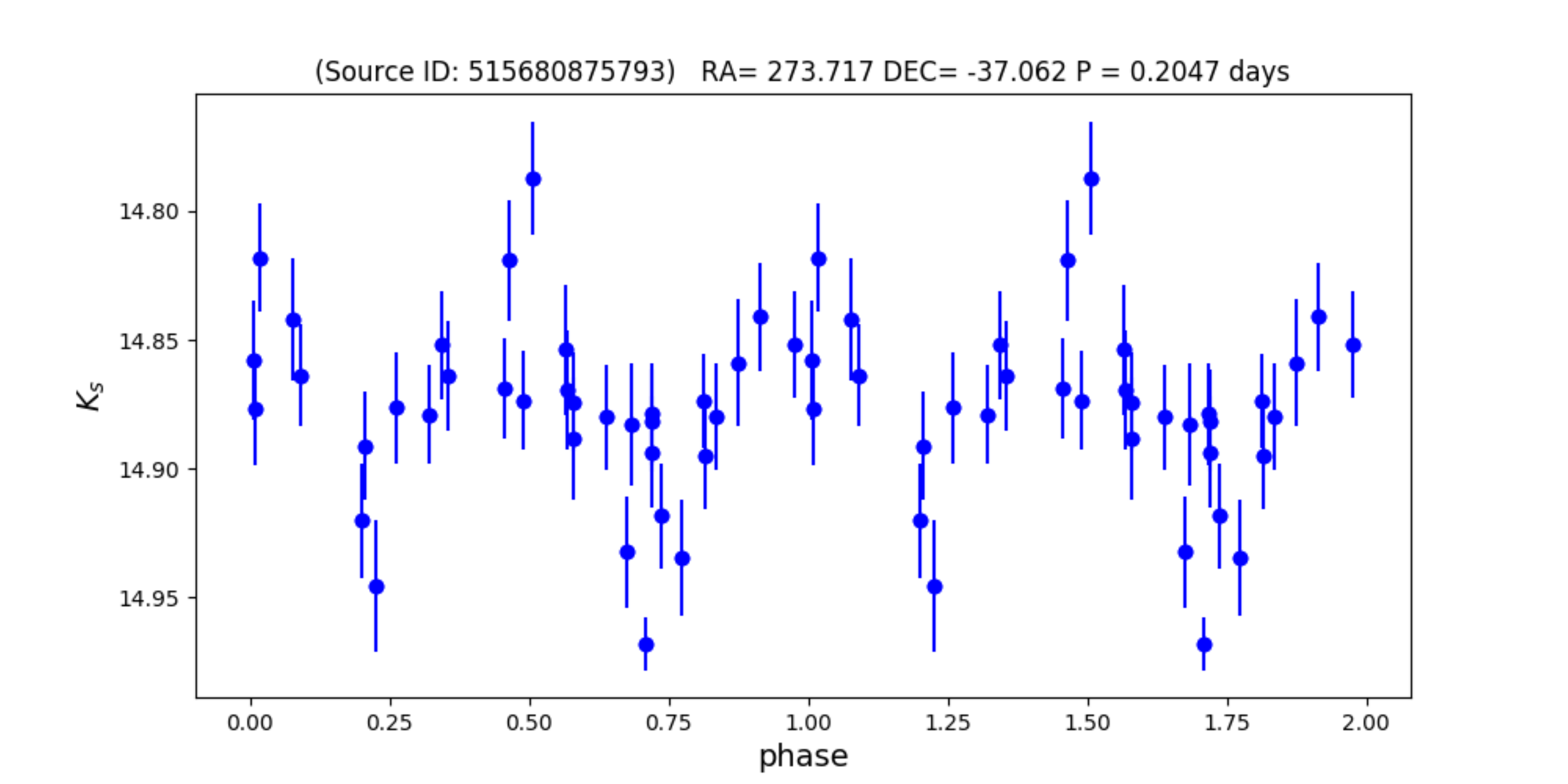} &
    \includegraphics[width=.33\textwidth]{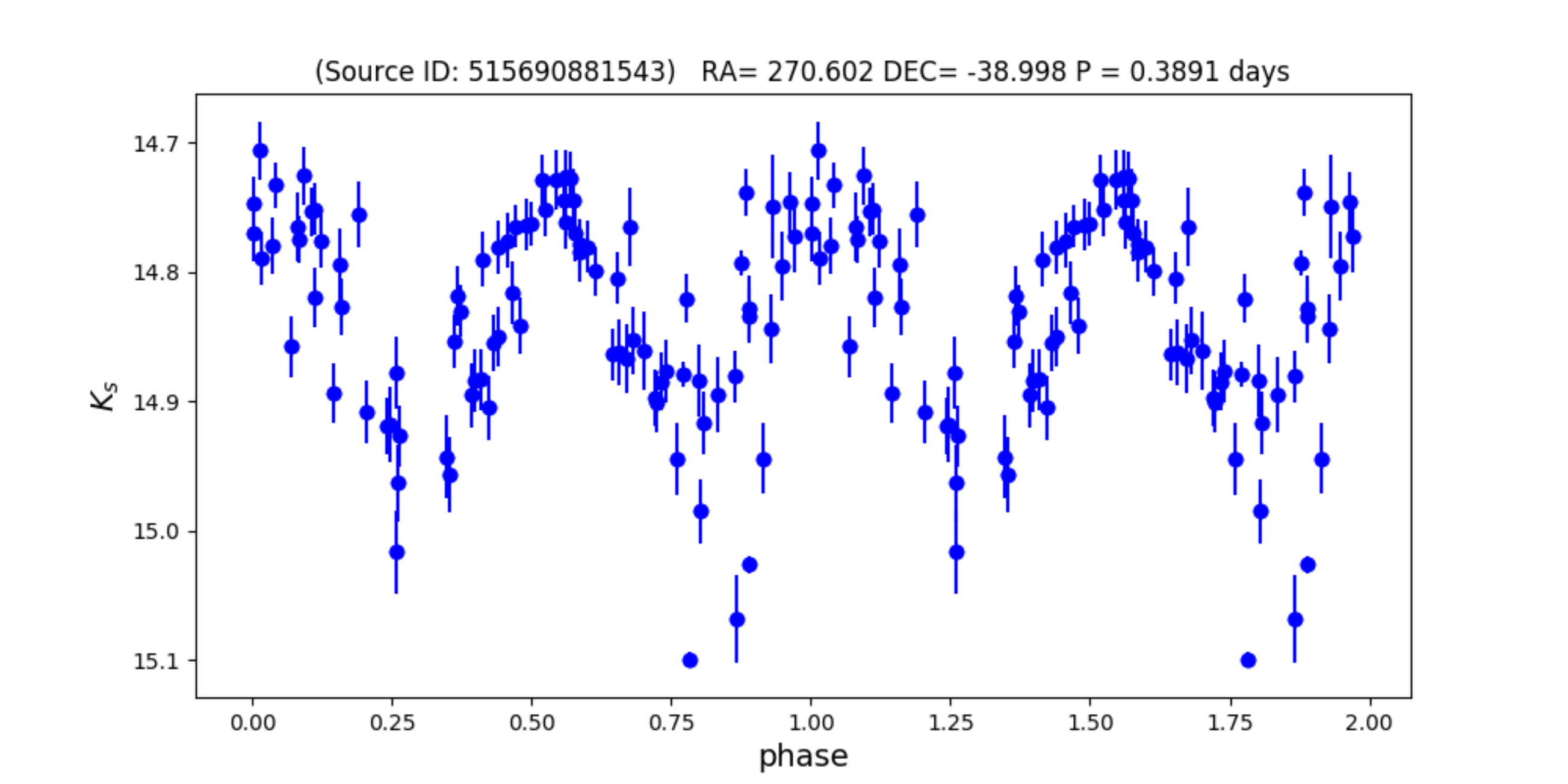} &
    \includegraphics[width=.33\textwidth]{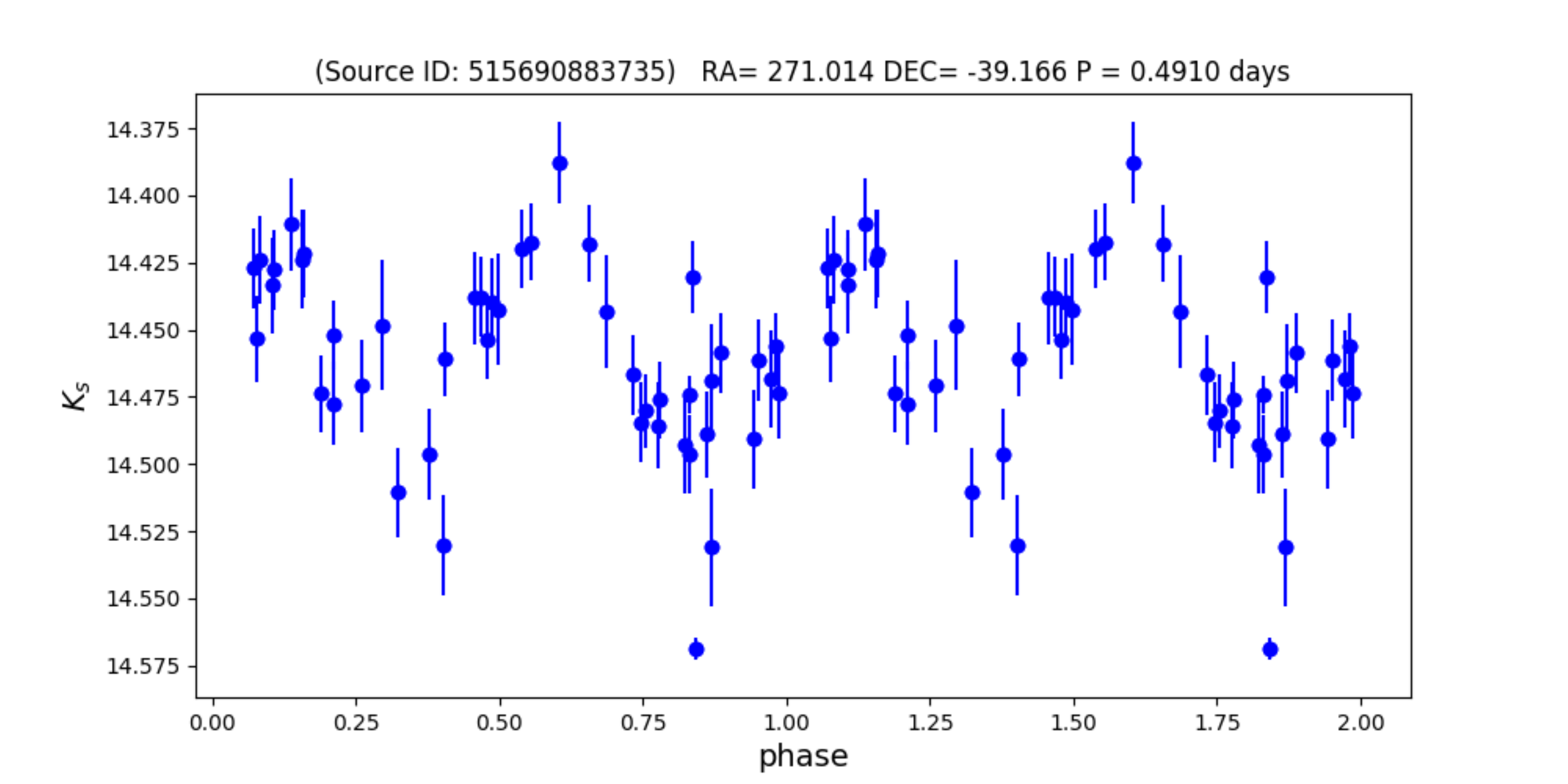} \\
    \includegraphics[width=.33\textwidth]{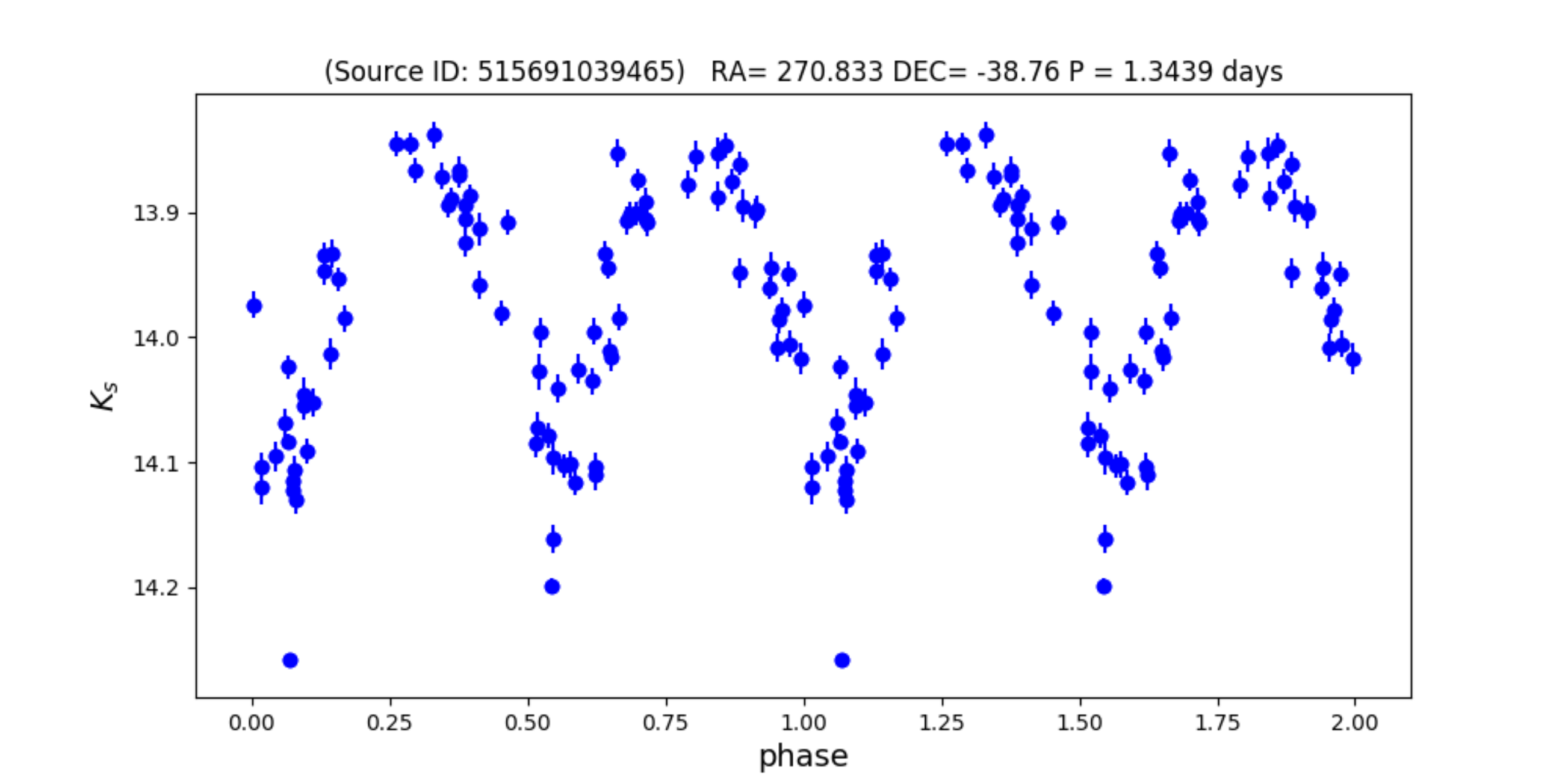} &
    \includegraphics[width=.33\textwidth]{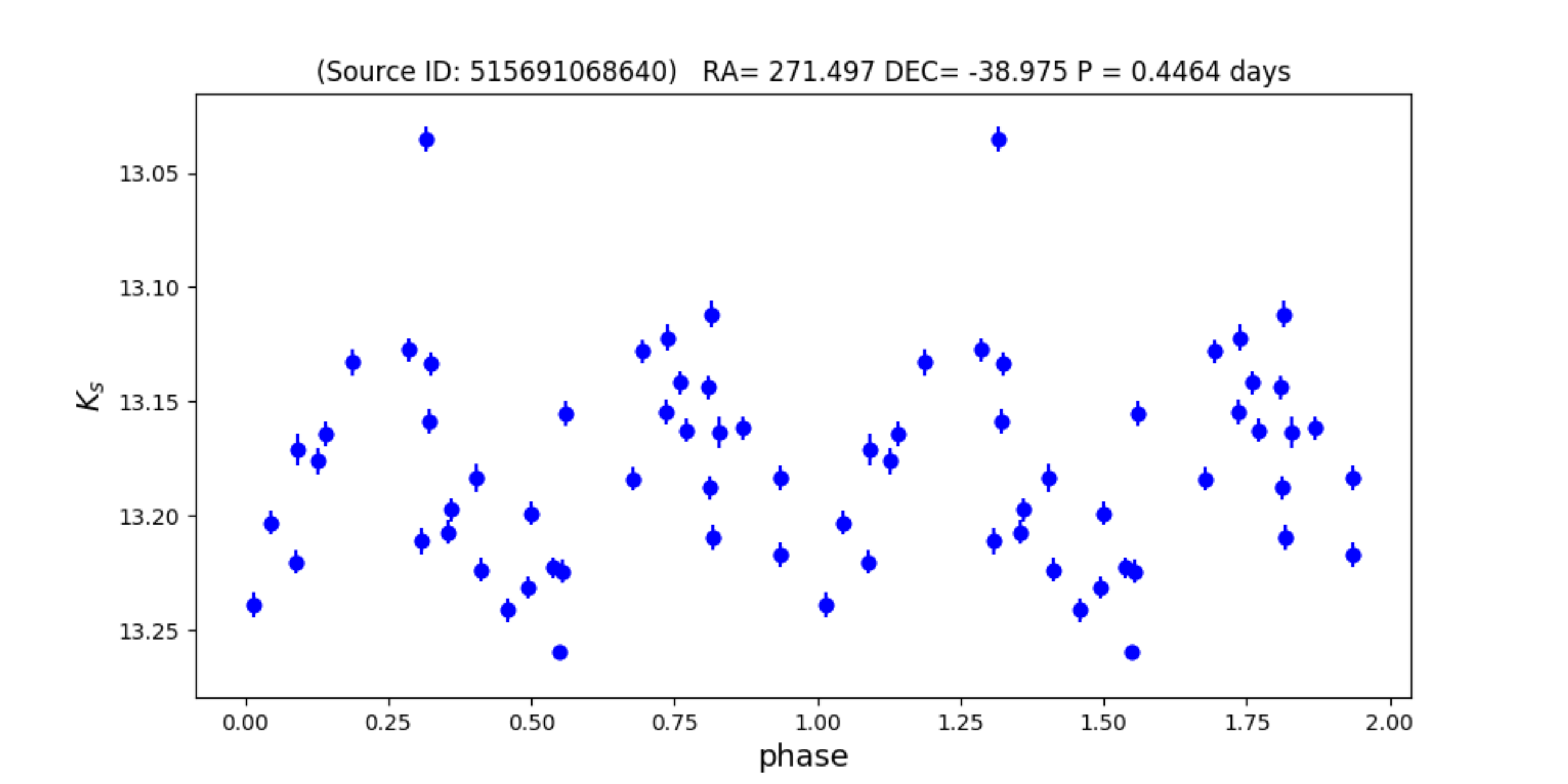} &
    \includegraphics[width=.33\textwidth]{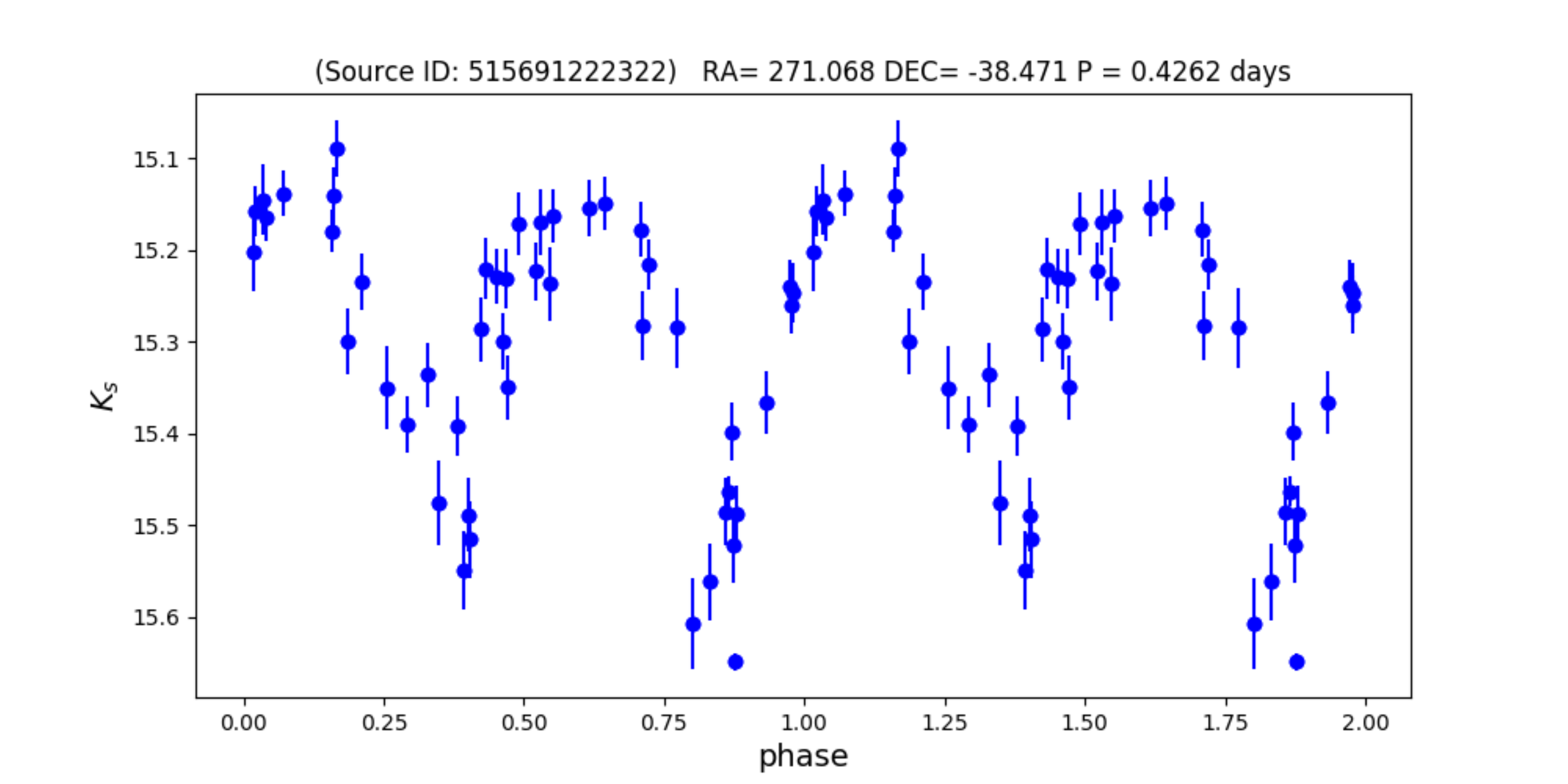} \\ 
  \end{tabular}
  \caption{BHB stars light curves corresponding to $1^{st}$ category eclipsing binaries.}
  \label{fig:bhb8}
\end{figure*}

\begin{figure*}[htb]
\centering
  \begin{tabular}{@{}ccc@{}}     
    \includegraphics[width=.33\textwidth]{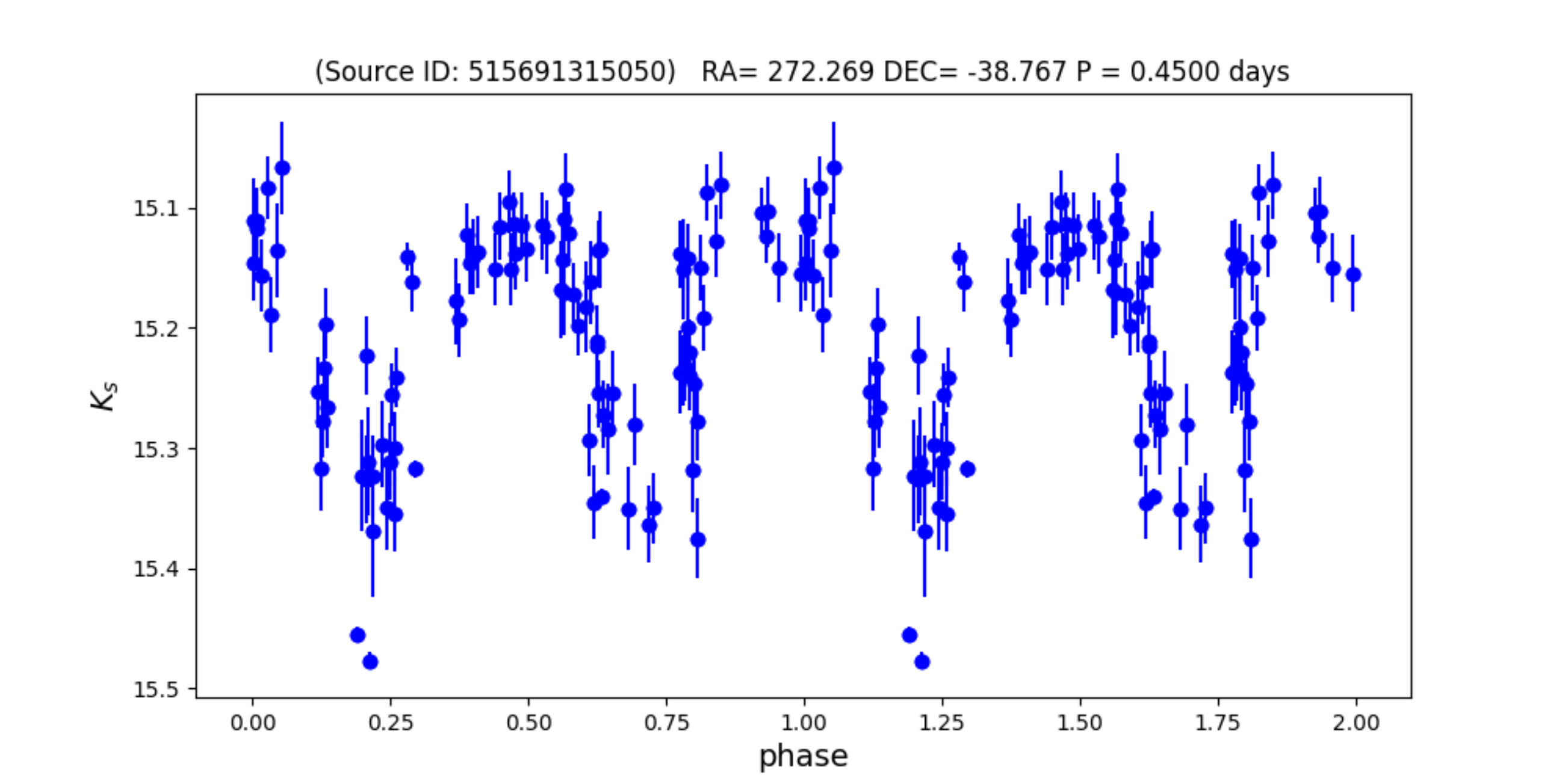} &
    \includegraphics[width=.33\textwidth]{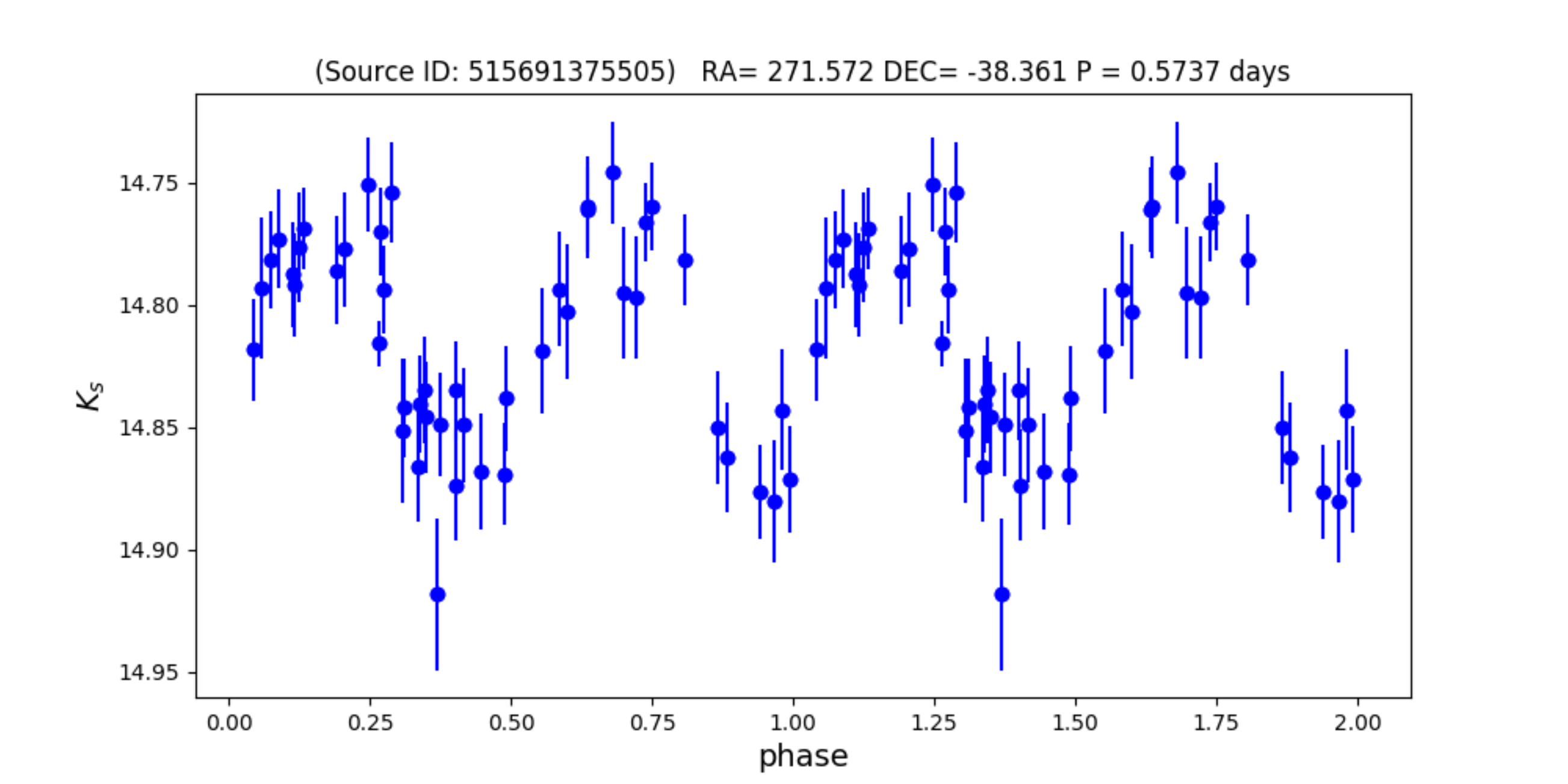} &
    \includegraphics[width=.33\textwidth]{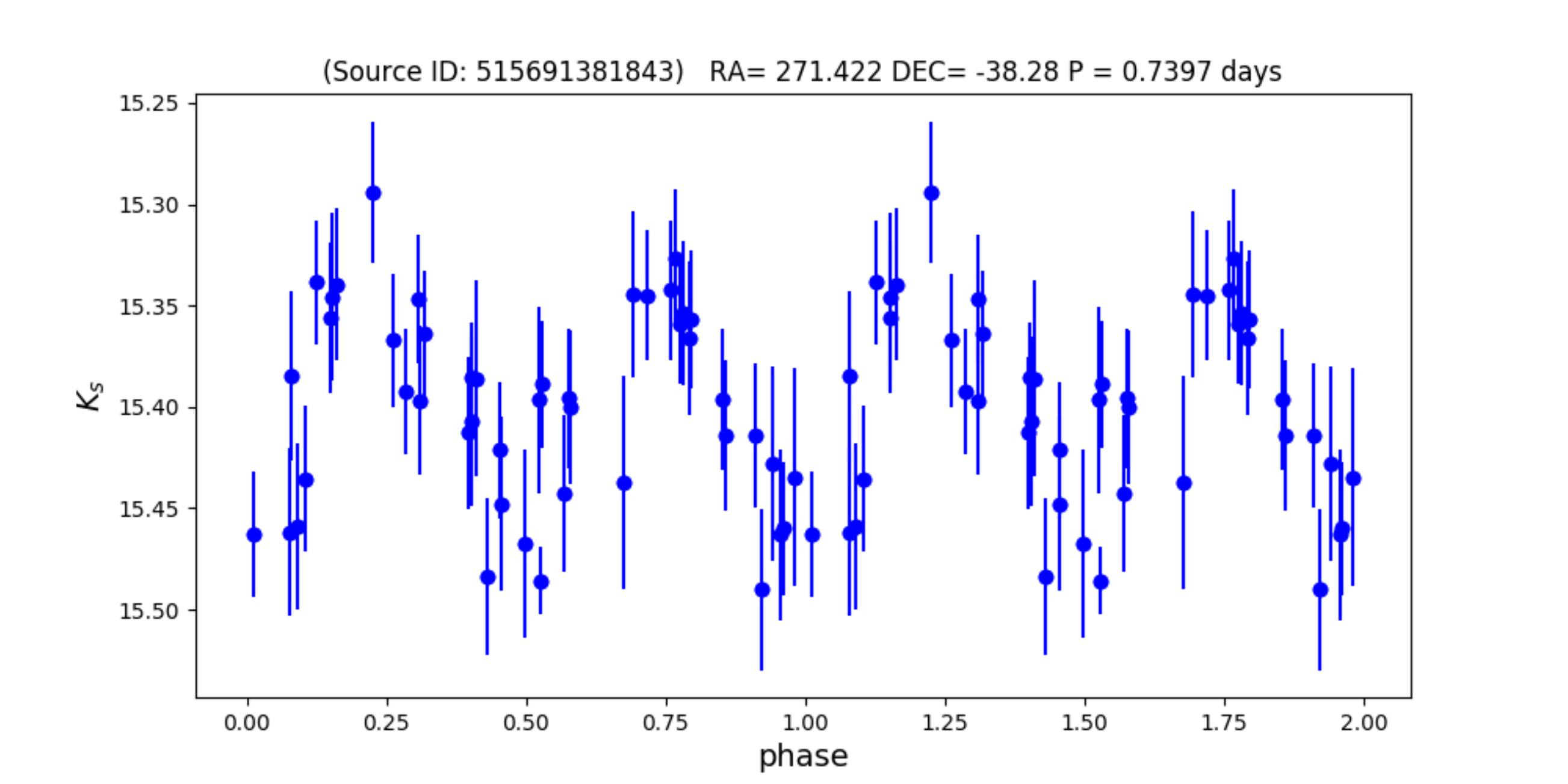} \\
    \includegraphics[width=.33\textwidth]{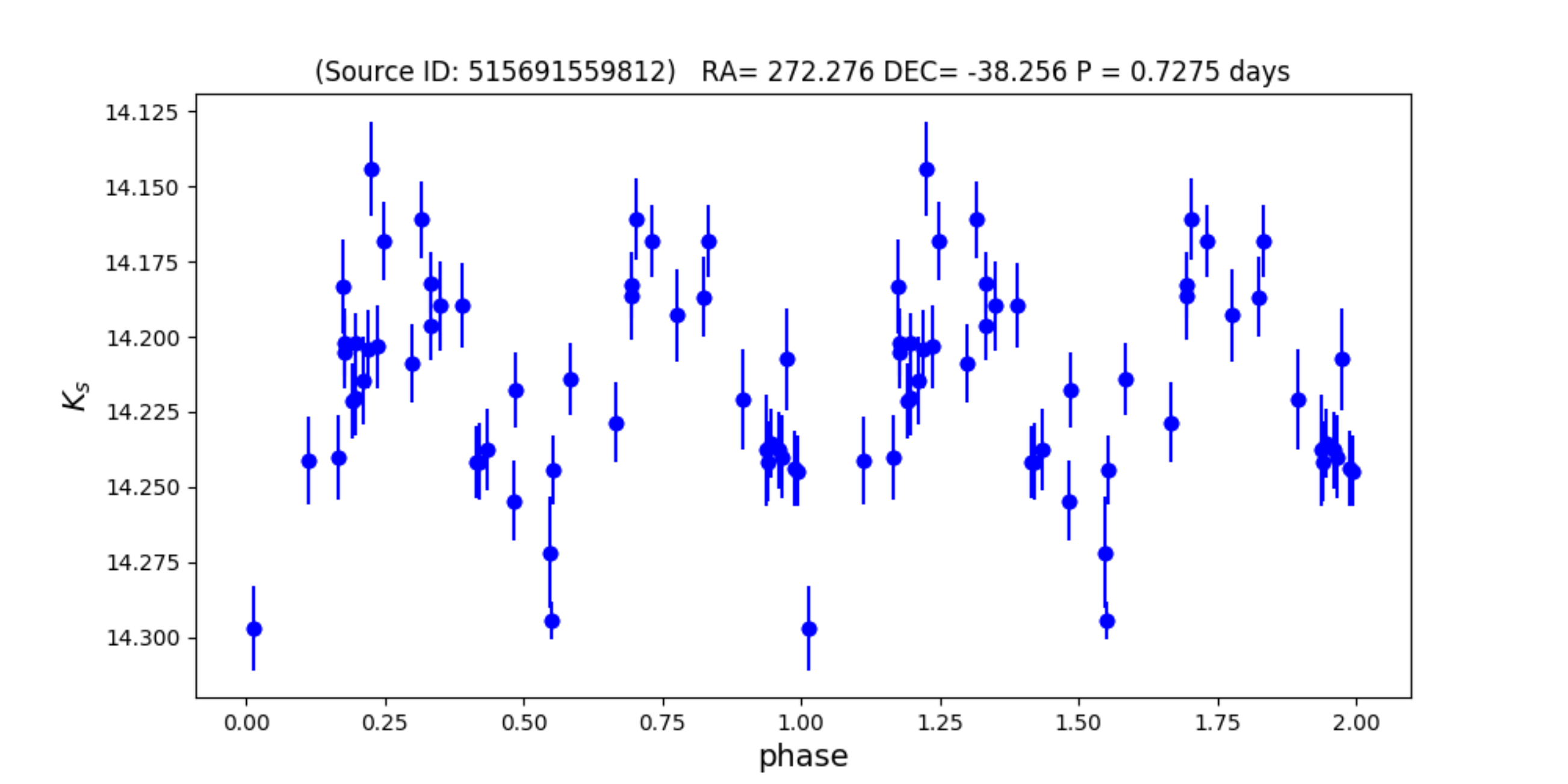} &
    \includegraphics[width=.33\textwidth]{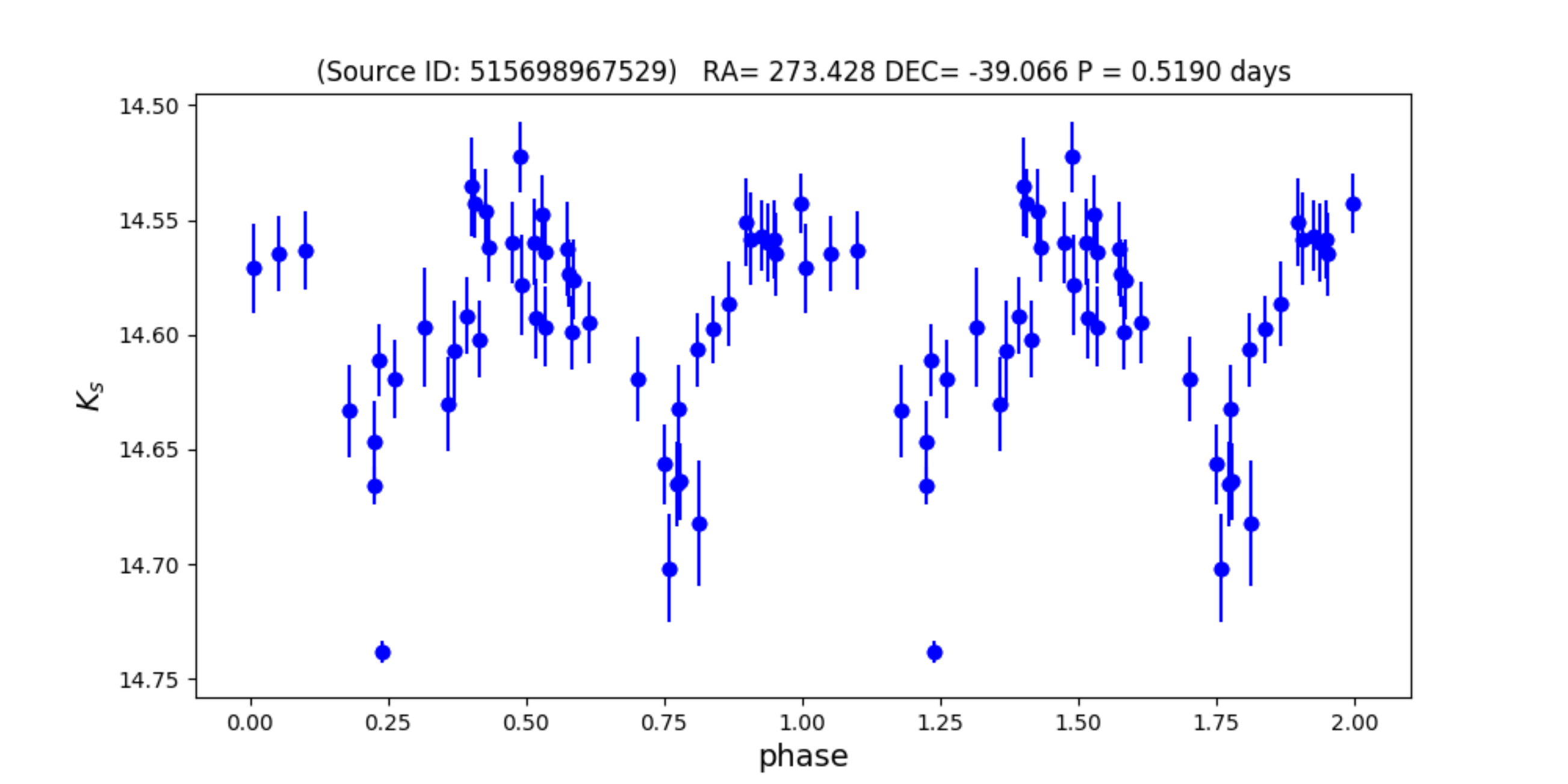} &
    \includegraphics[width=.33\textwidth]{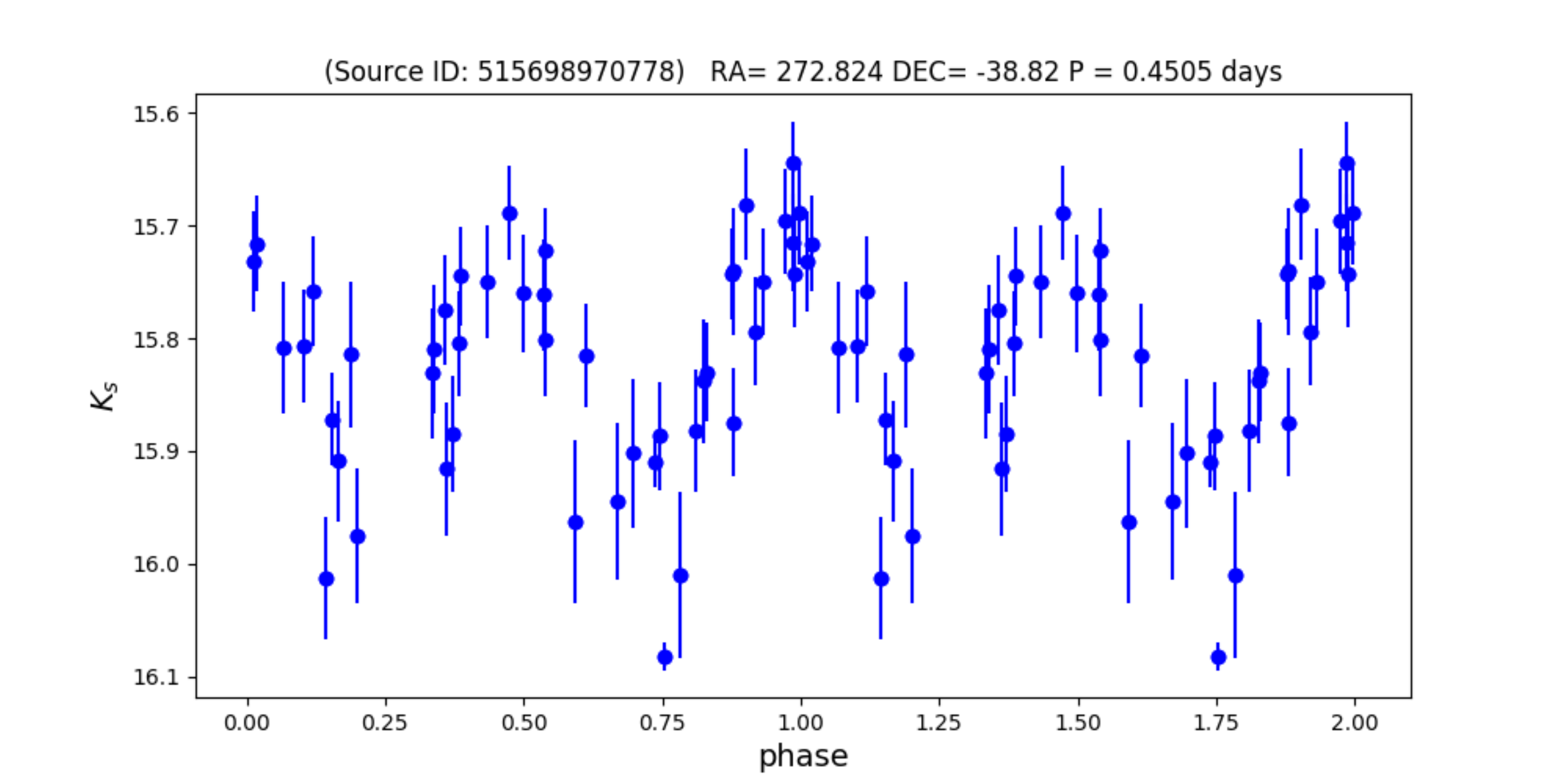} \\
    \includegraphics[width=.33\textwidth]{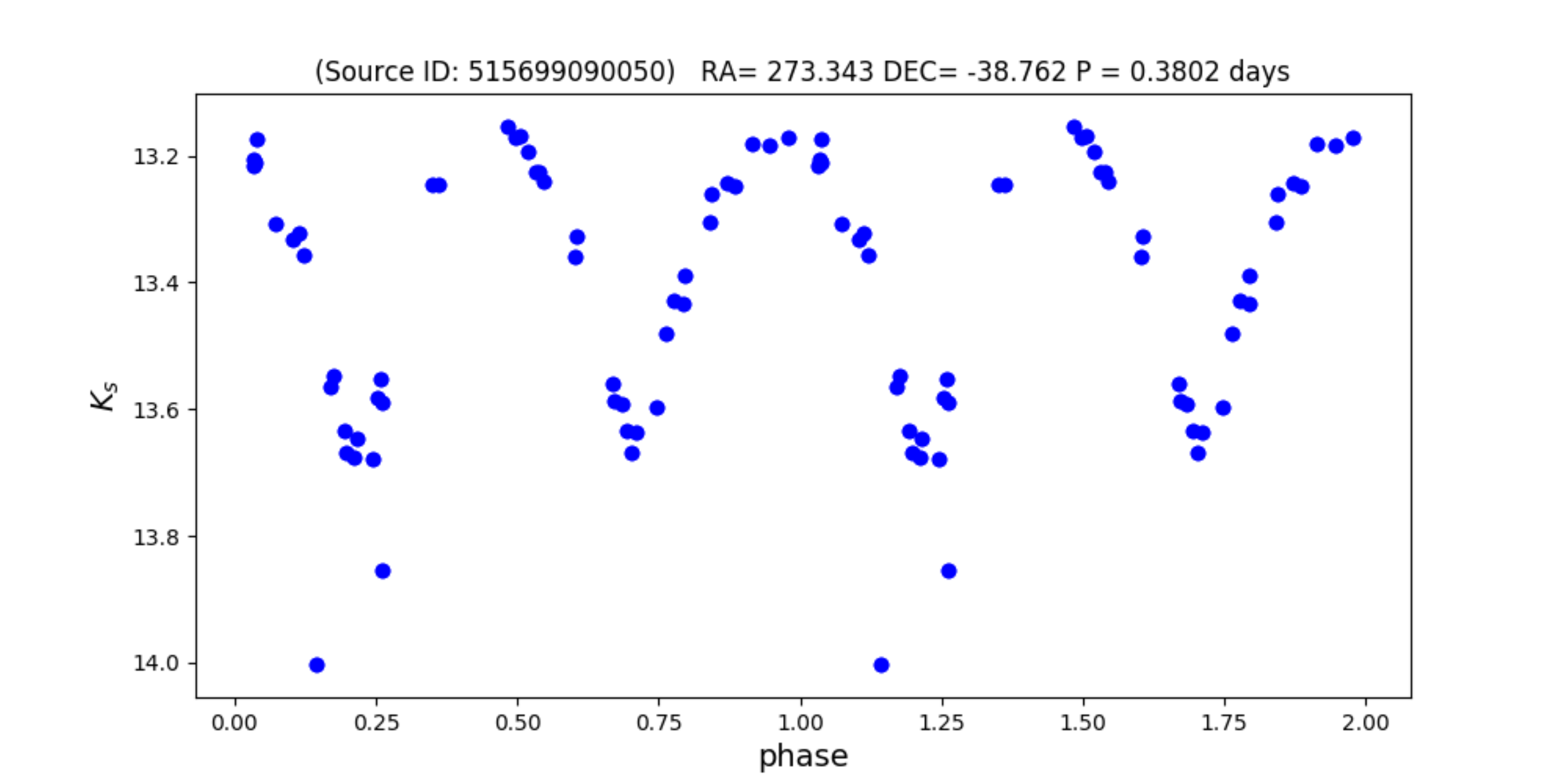} &
    \includegraphics[width=.33\textwidth]{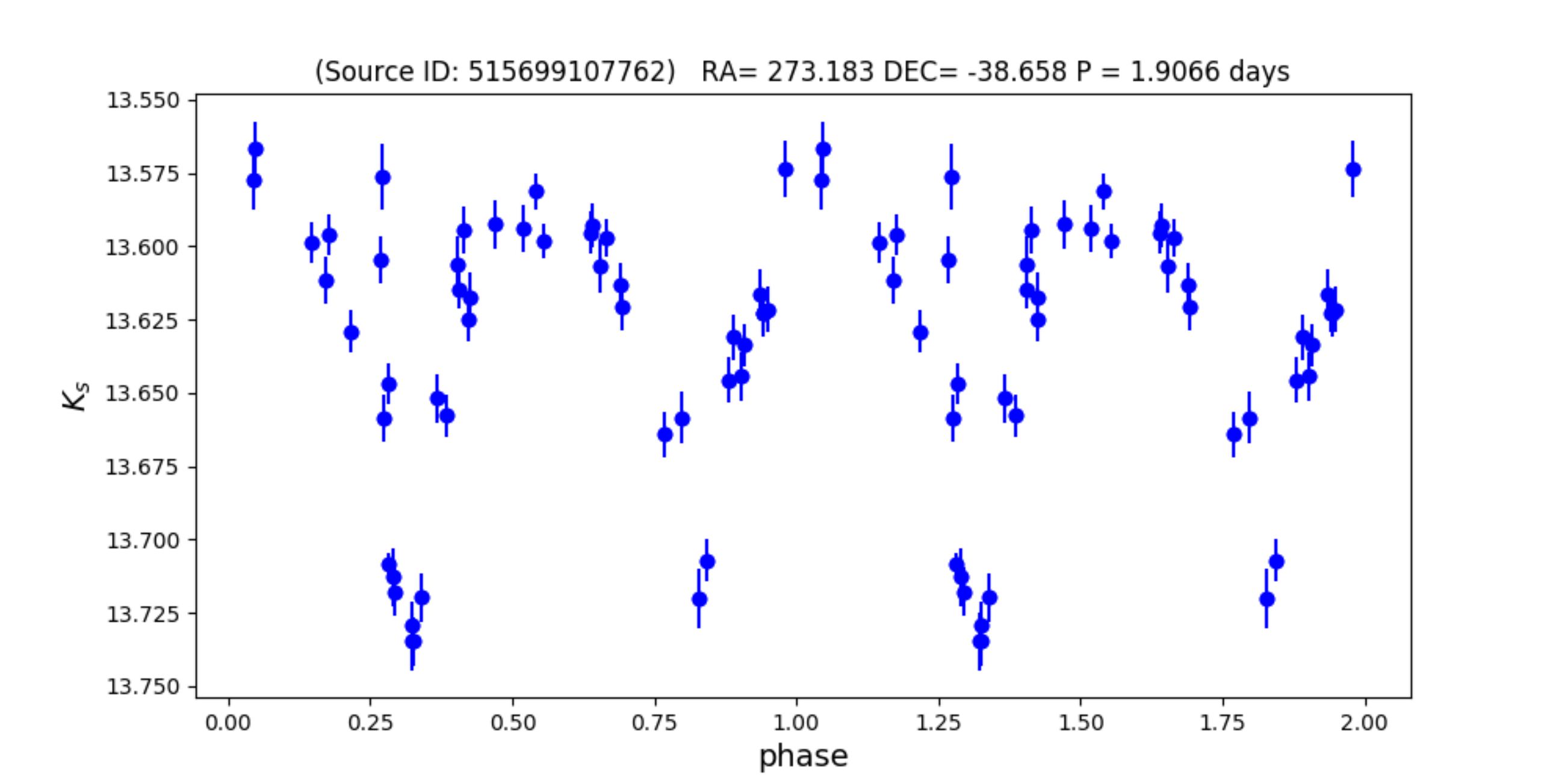} &
    \includegraphics[width=.33\textwidth]{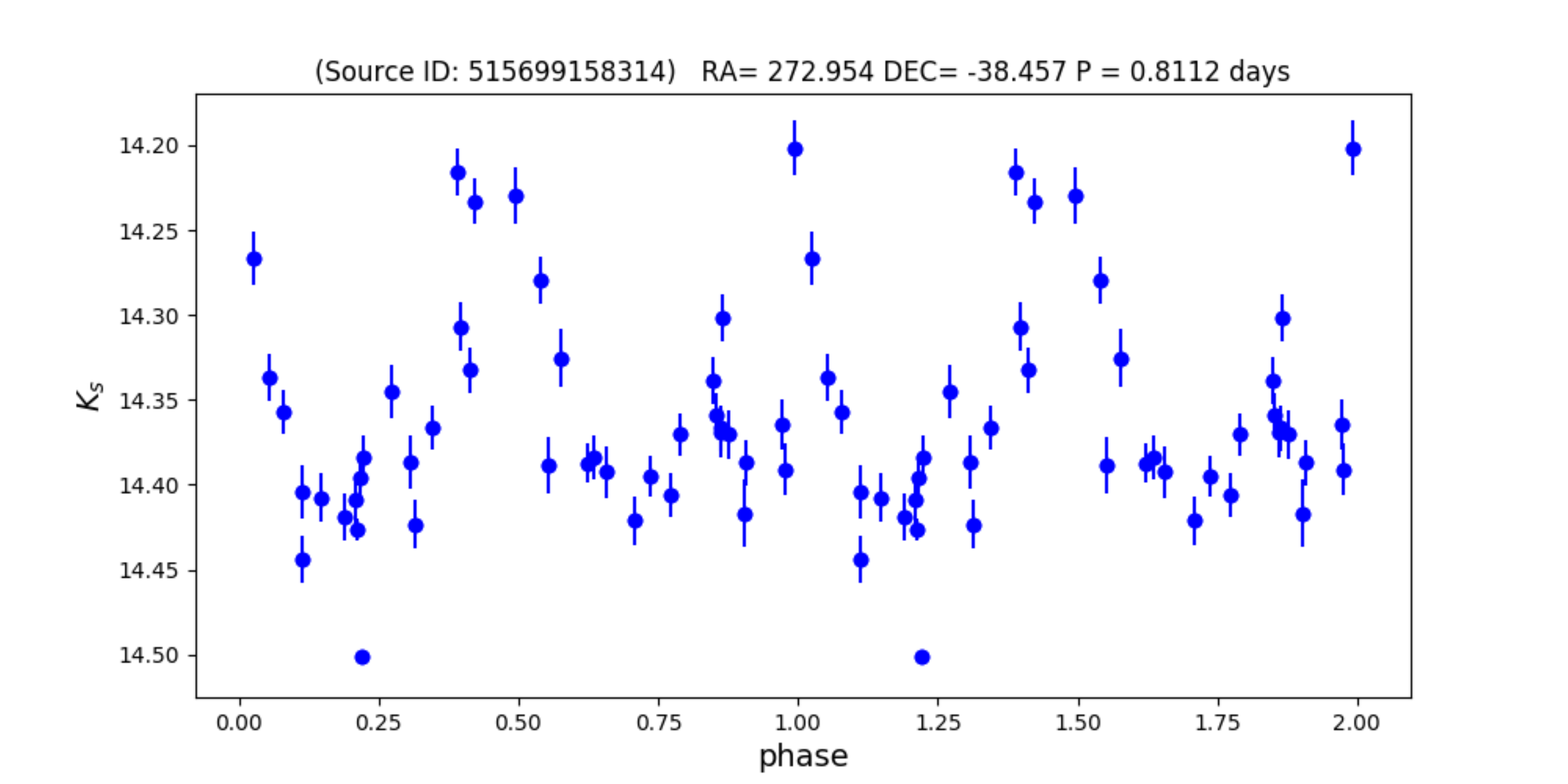} \\
    \includegraphics[width=.33\textwidth]{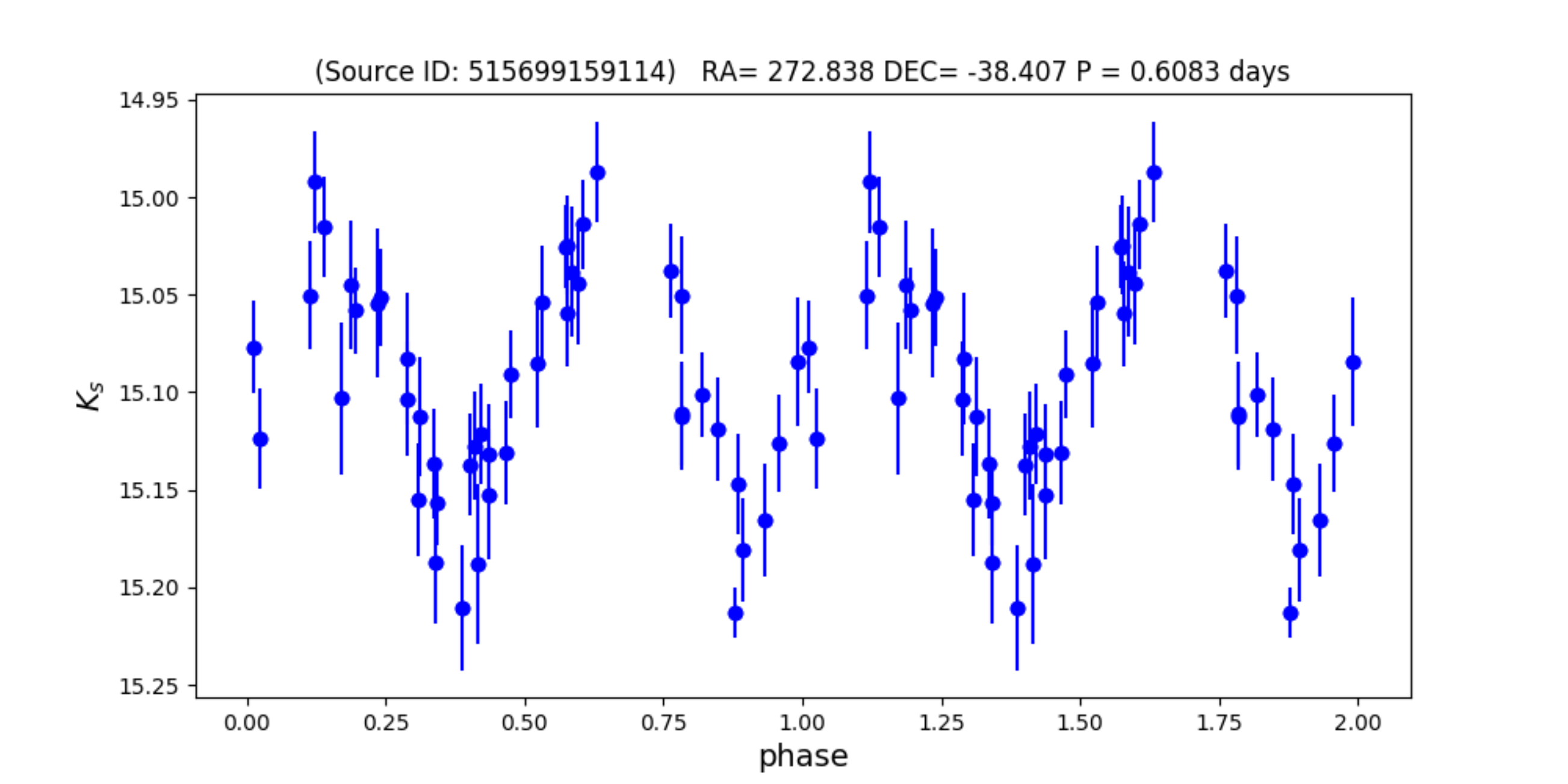} &
    \includegraphics[width=.33\textwidth]{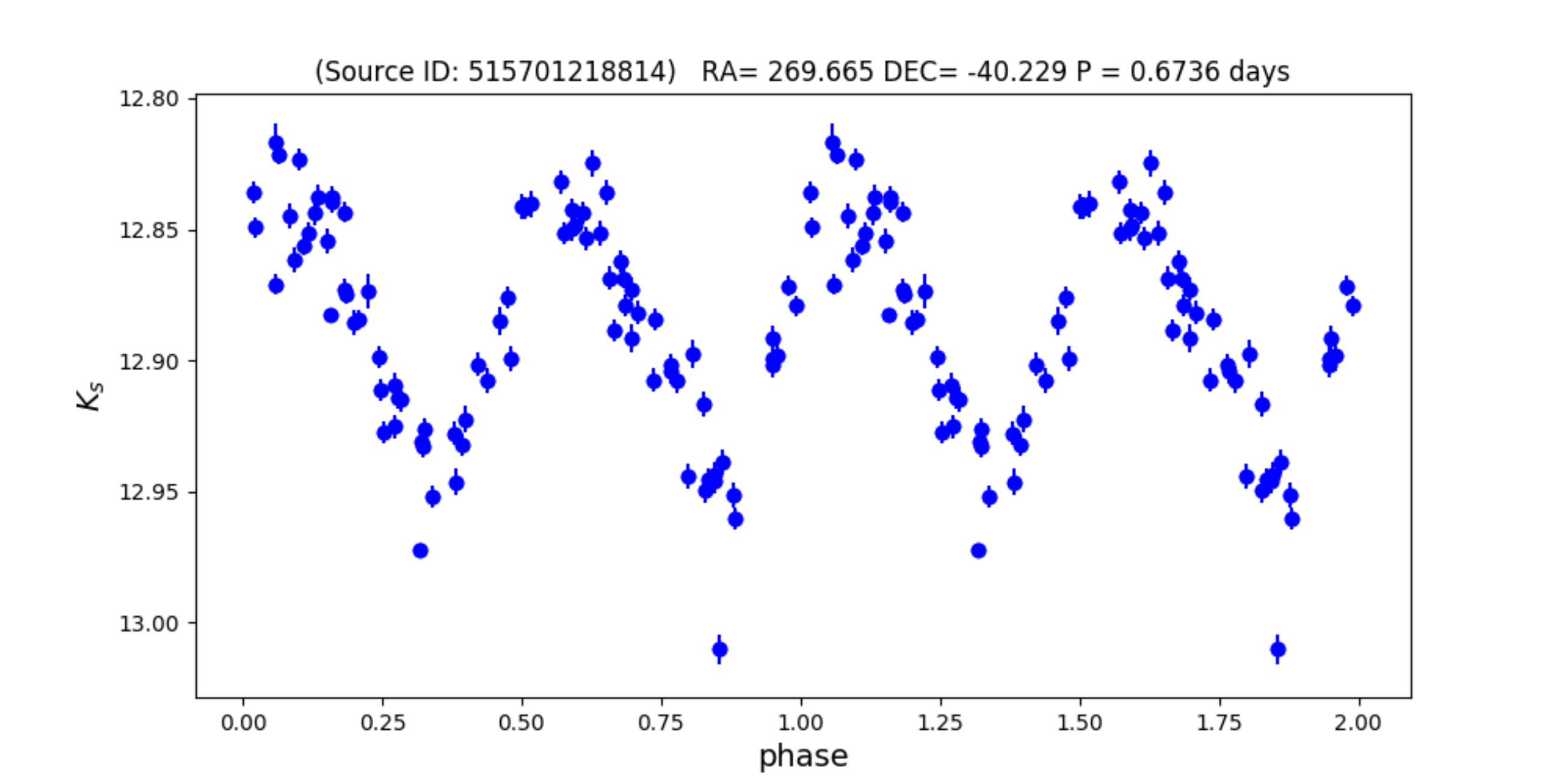} &
    \includegraphics[width=.33\textwidth]{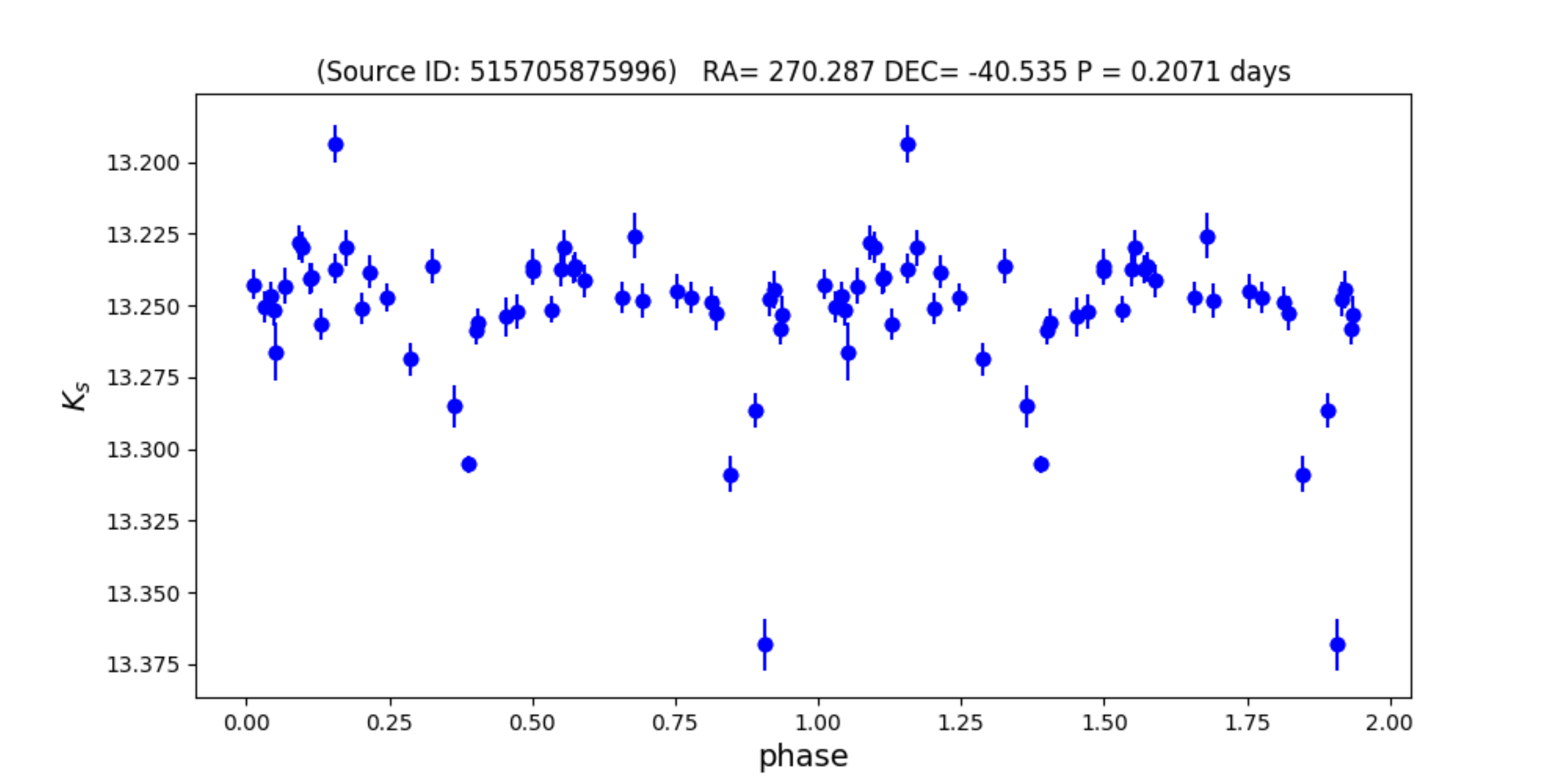} \\
    \includegraphics[width=.33\textwidth]{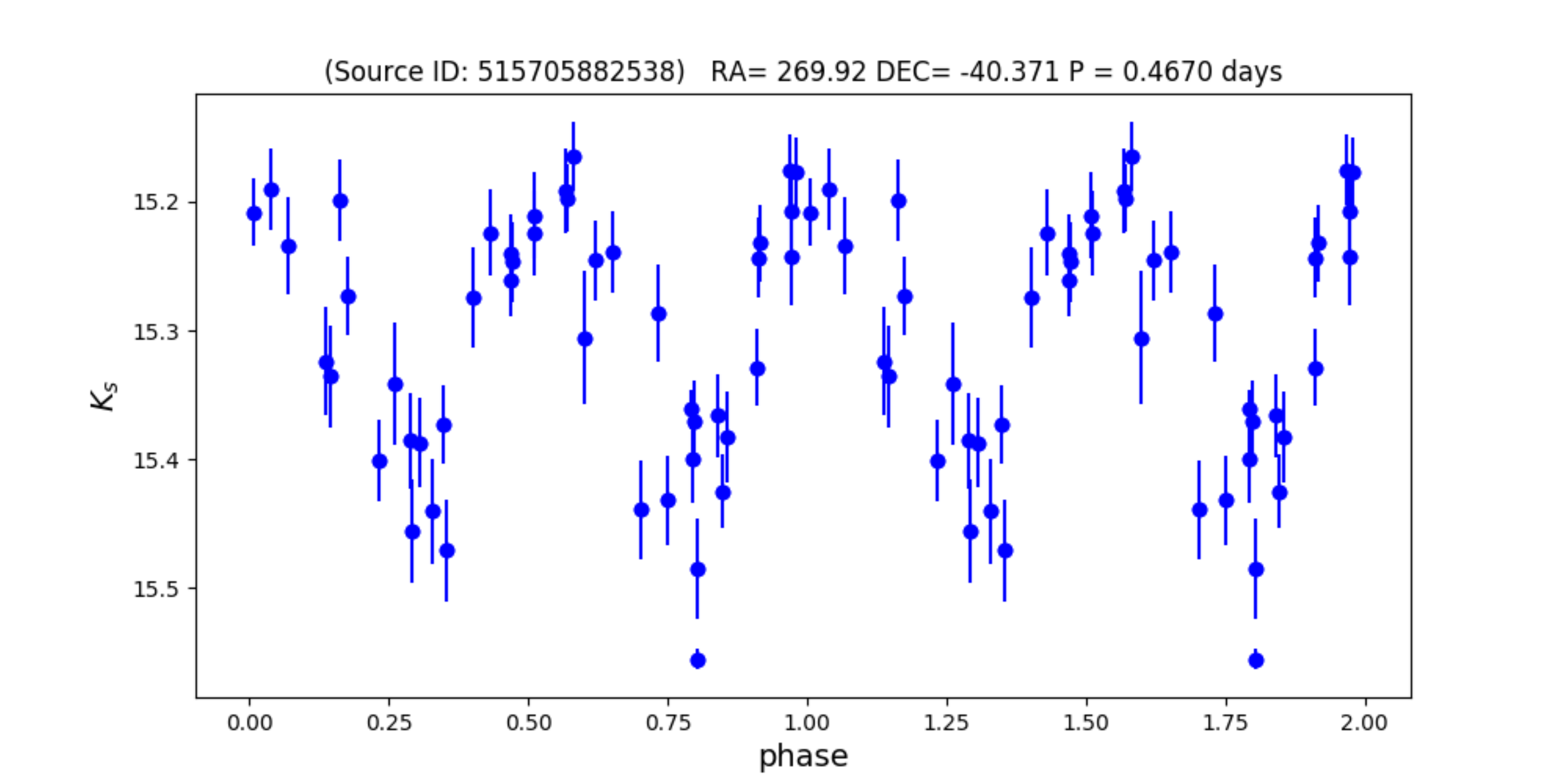} &
    \includegraphics[width=.33\textwidth]{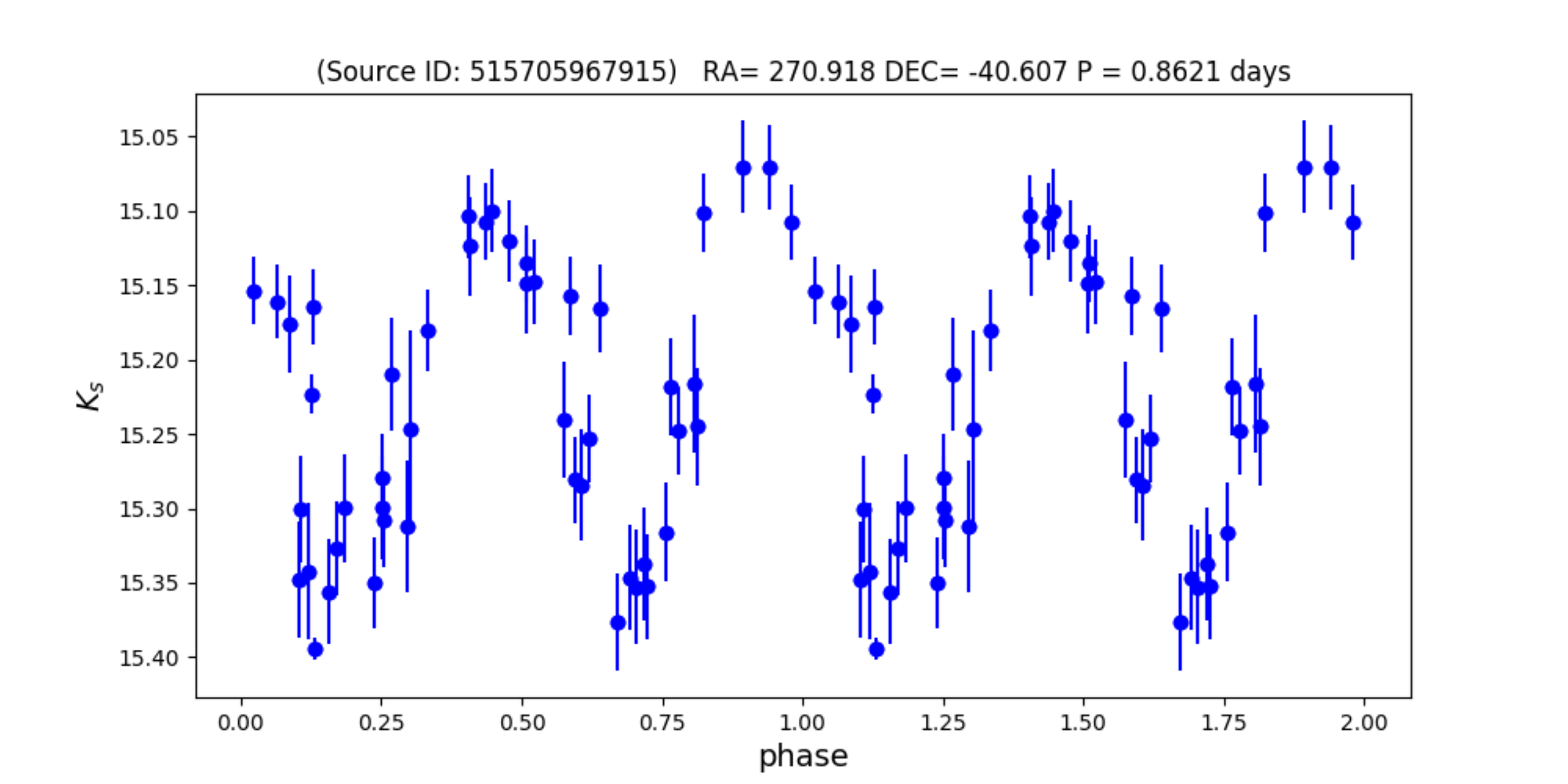} &
    \includegraphics[width=.33\textwidth]{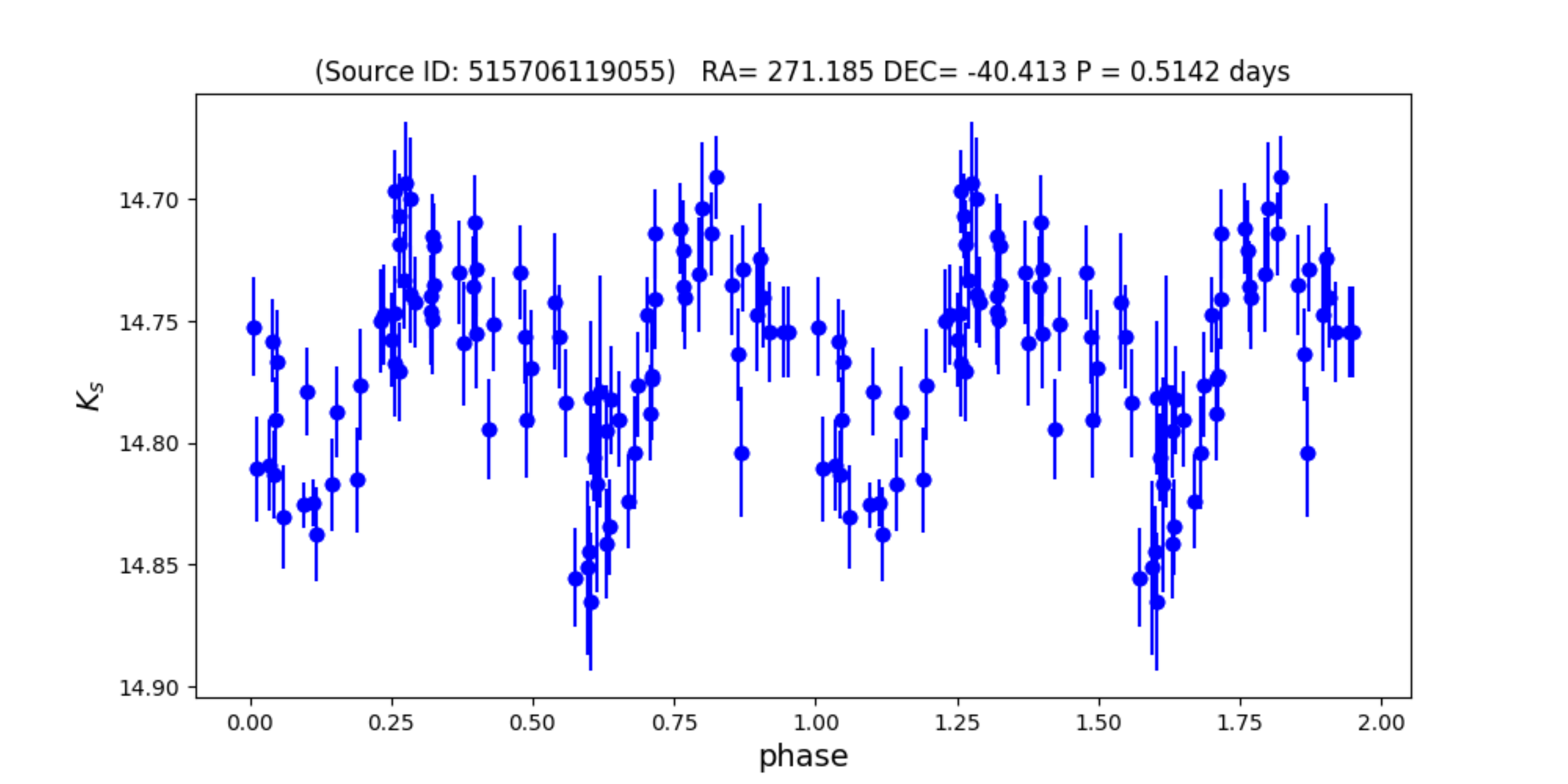} \\
    \includegraphics[width=.33\textwidth]{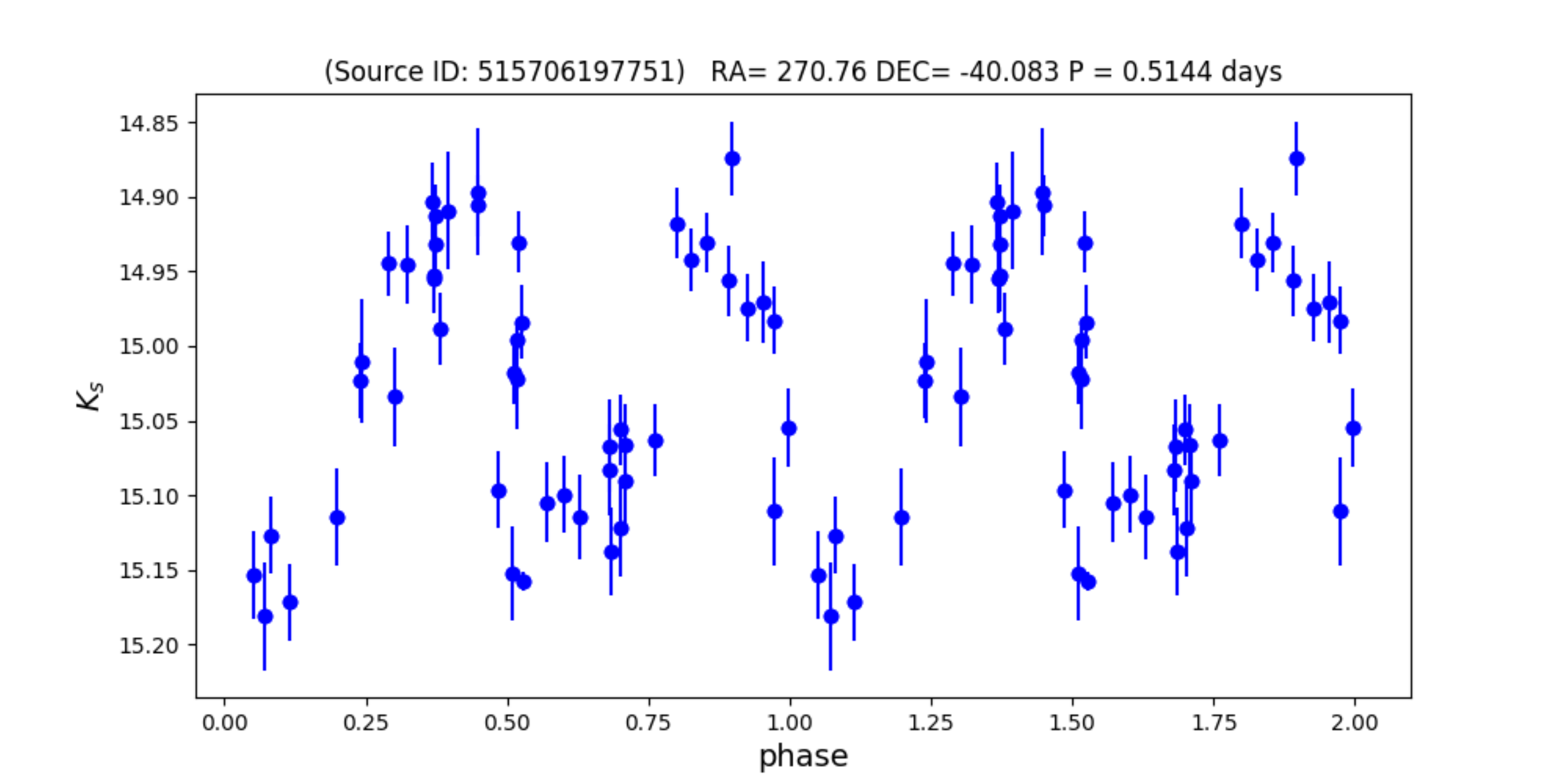} &
    \includegraphics[width=.33\textwidth]{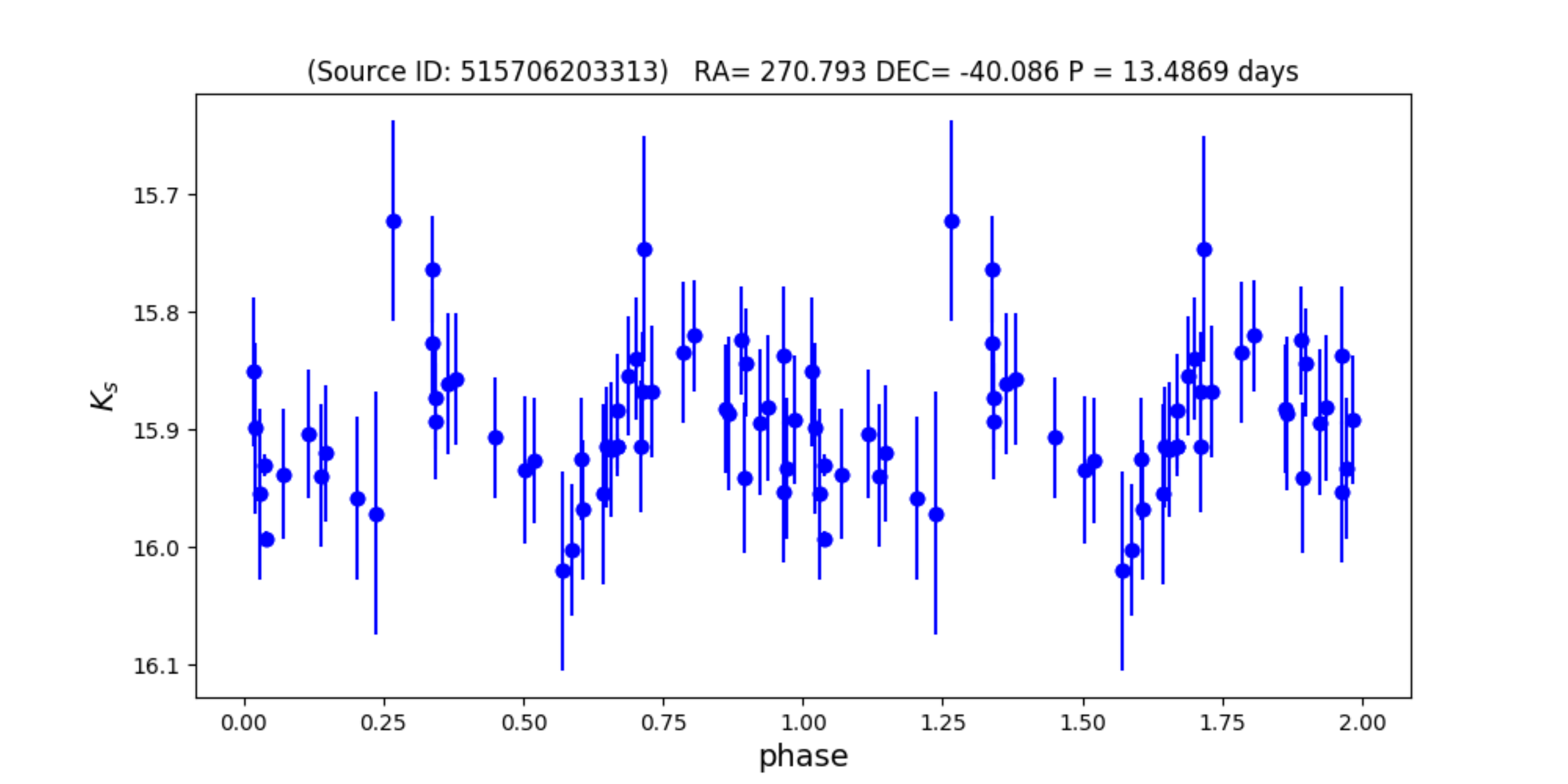} &
    \includegraphics[width=.33\textwidth]{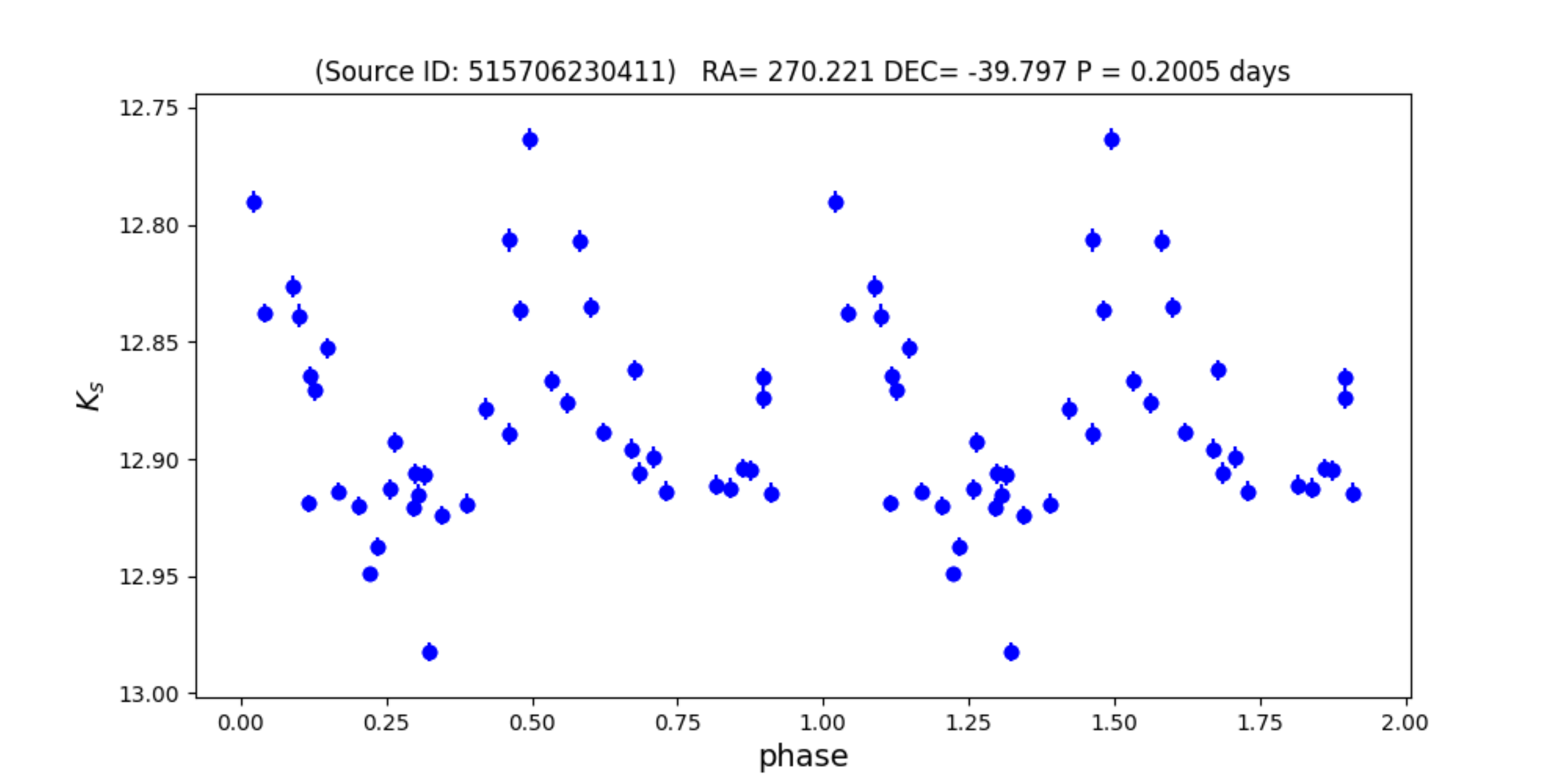} \\
    \includegraphics[width=.33\textwidth]{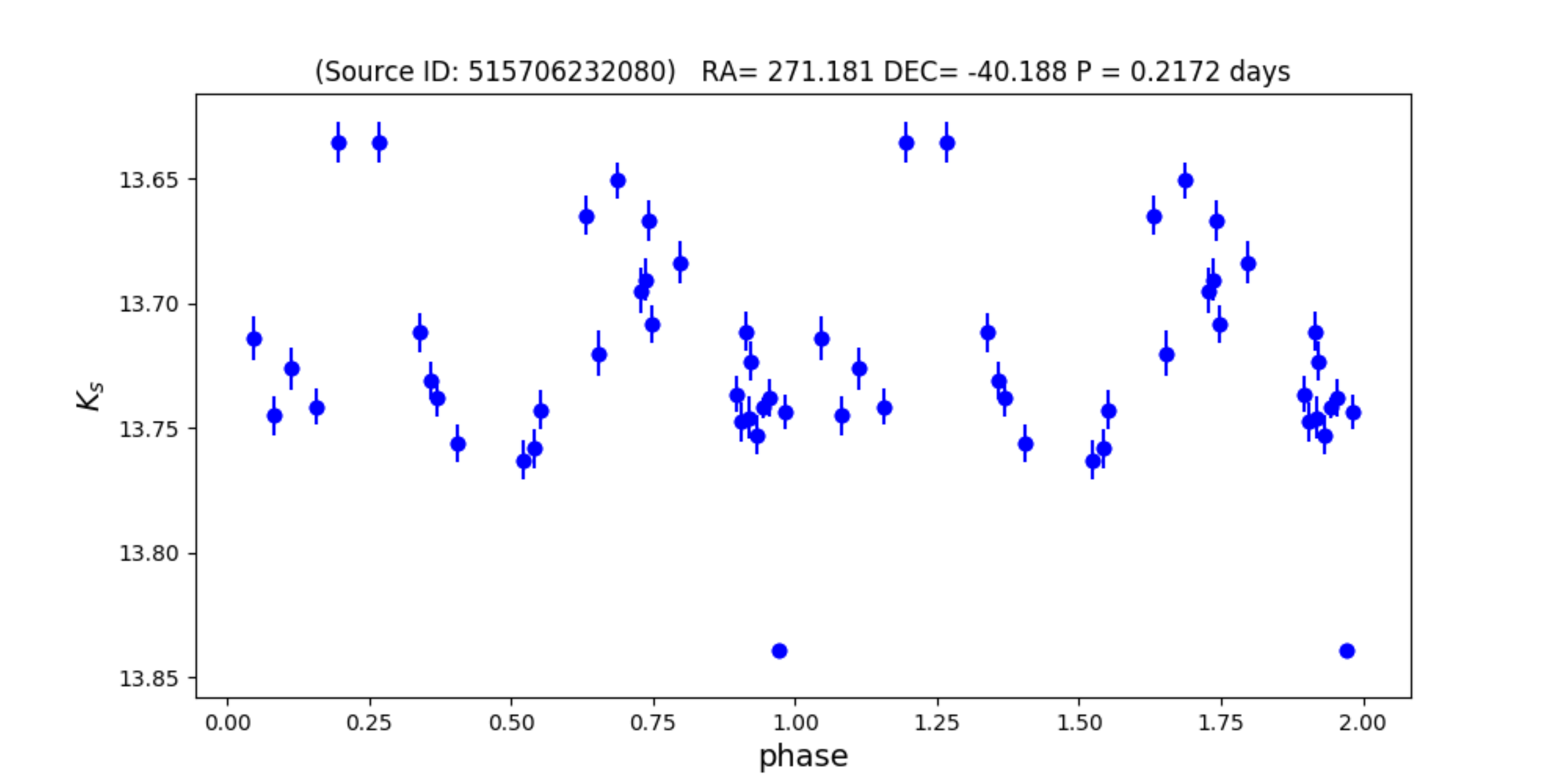} &
    \includegraphics[width=.33\textwidth]{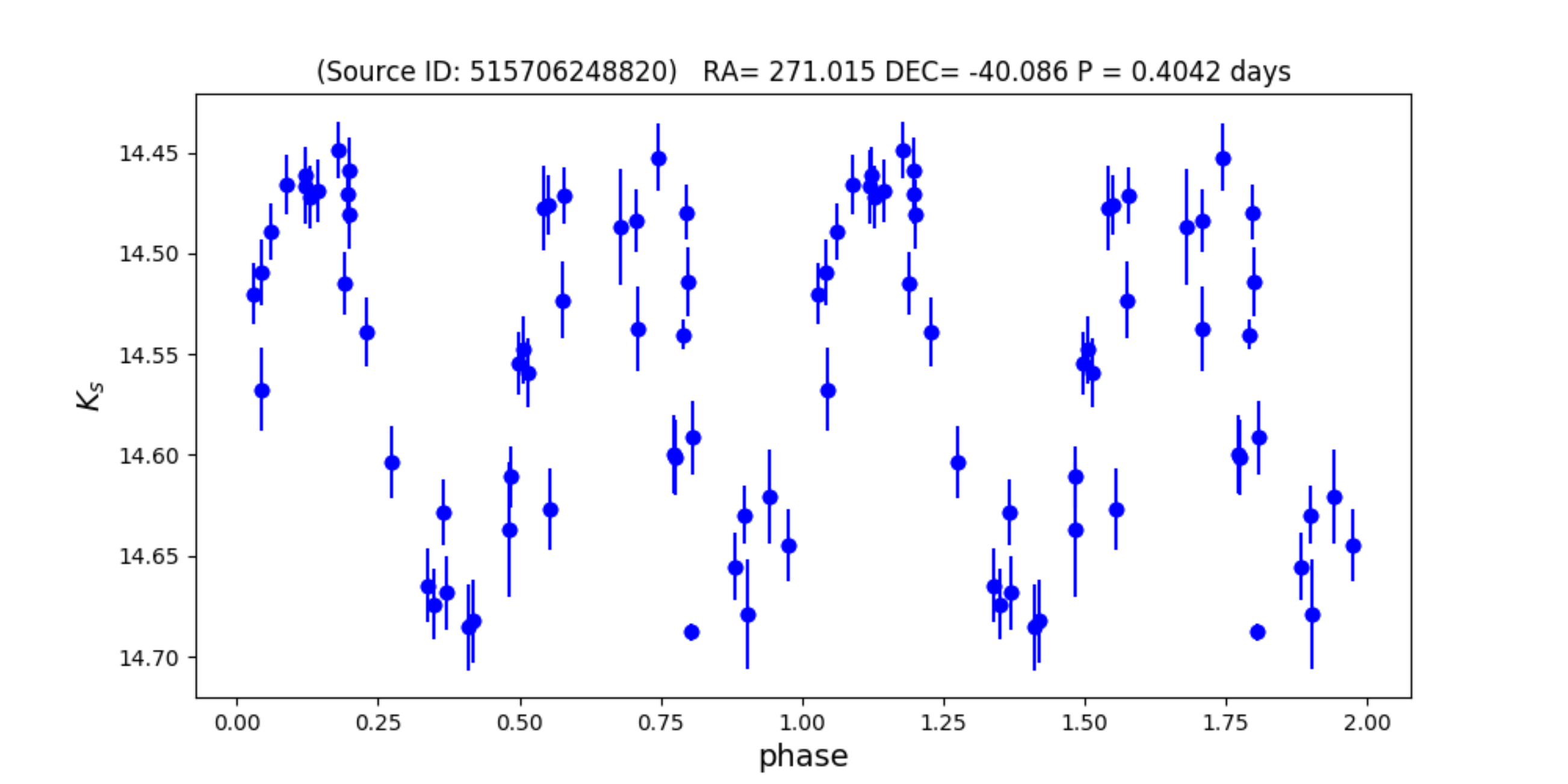} &
    \includegraphics[width=.33\textwidth]{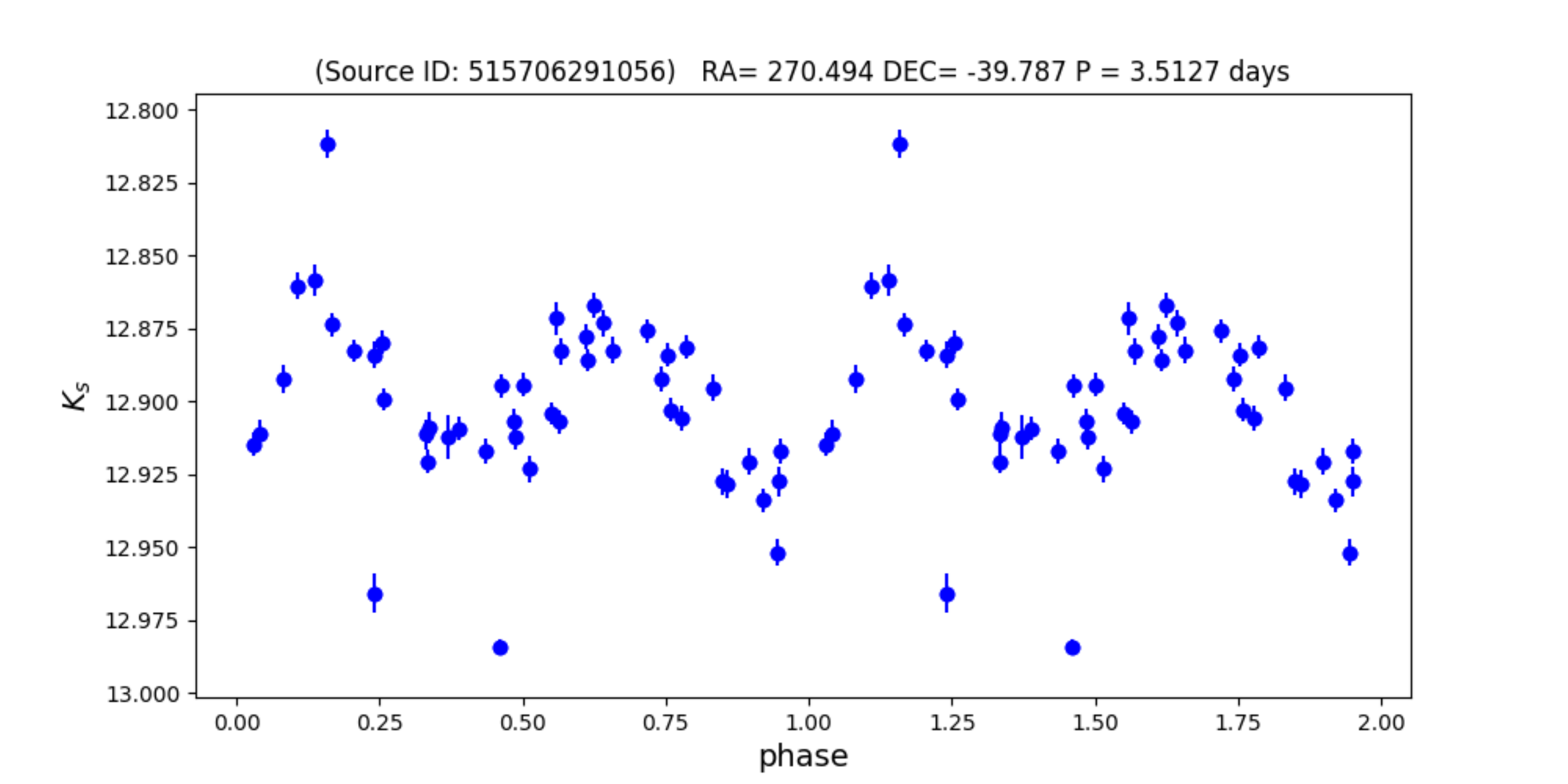} \\
    \includegraphics[width=.33\textwidth]{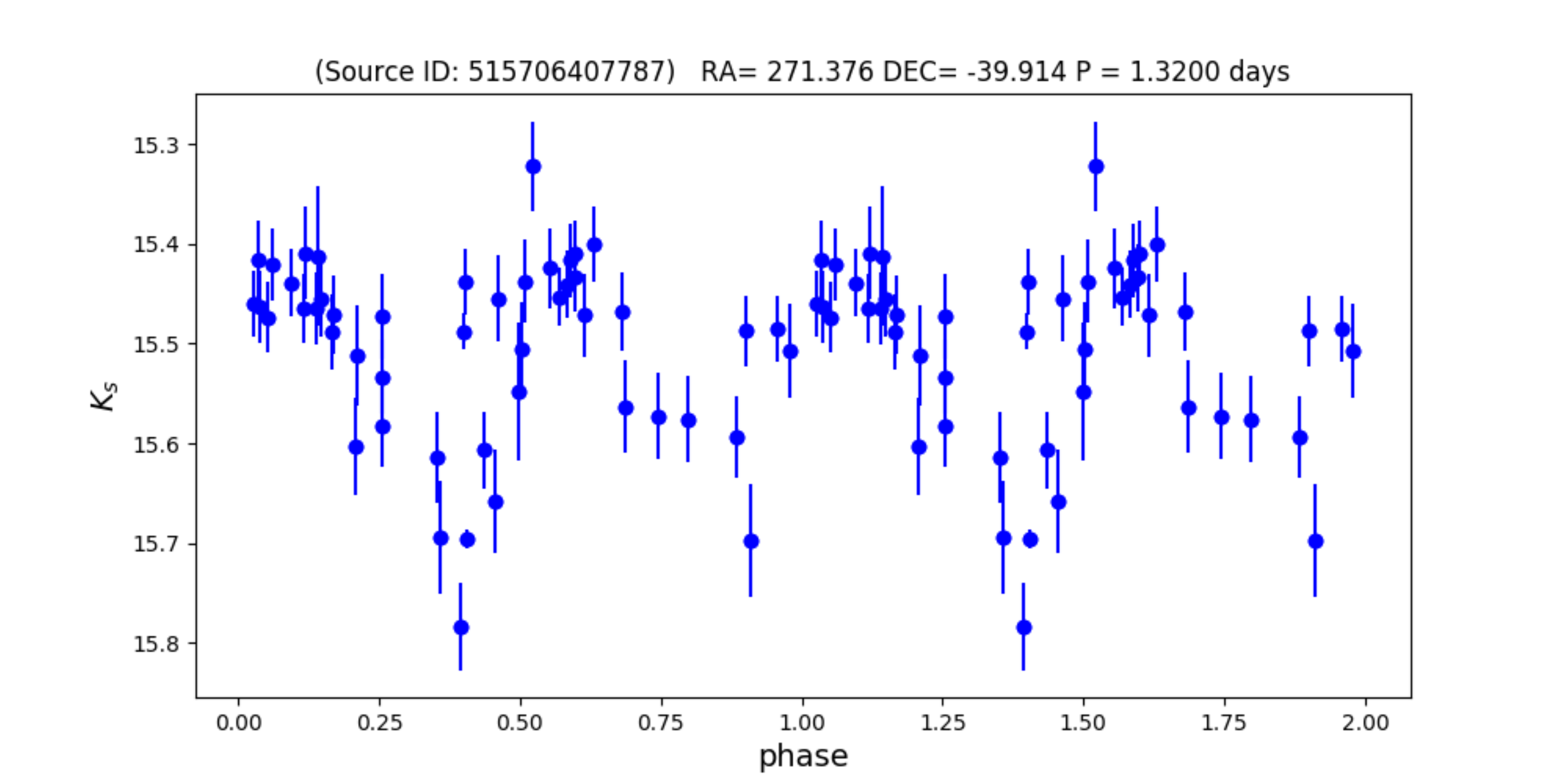} &
    \includegraphics[width=.33\textwidth]{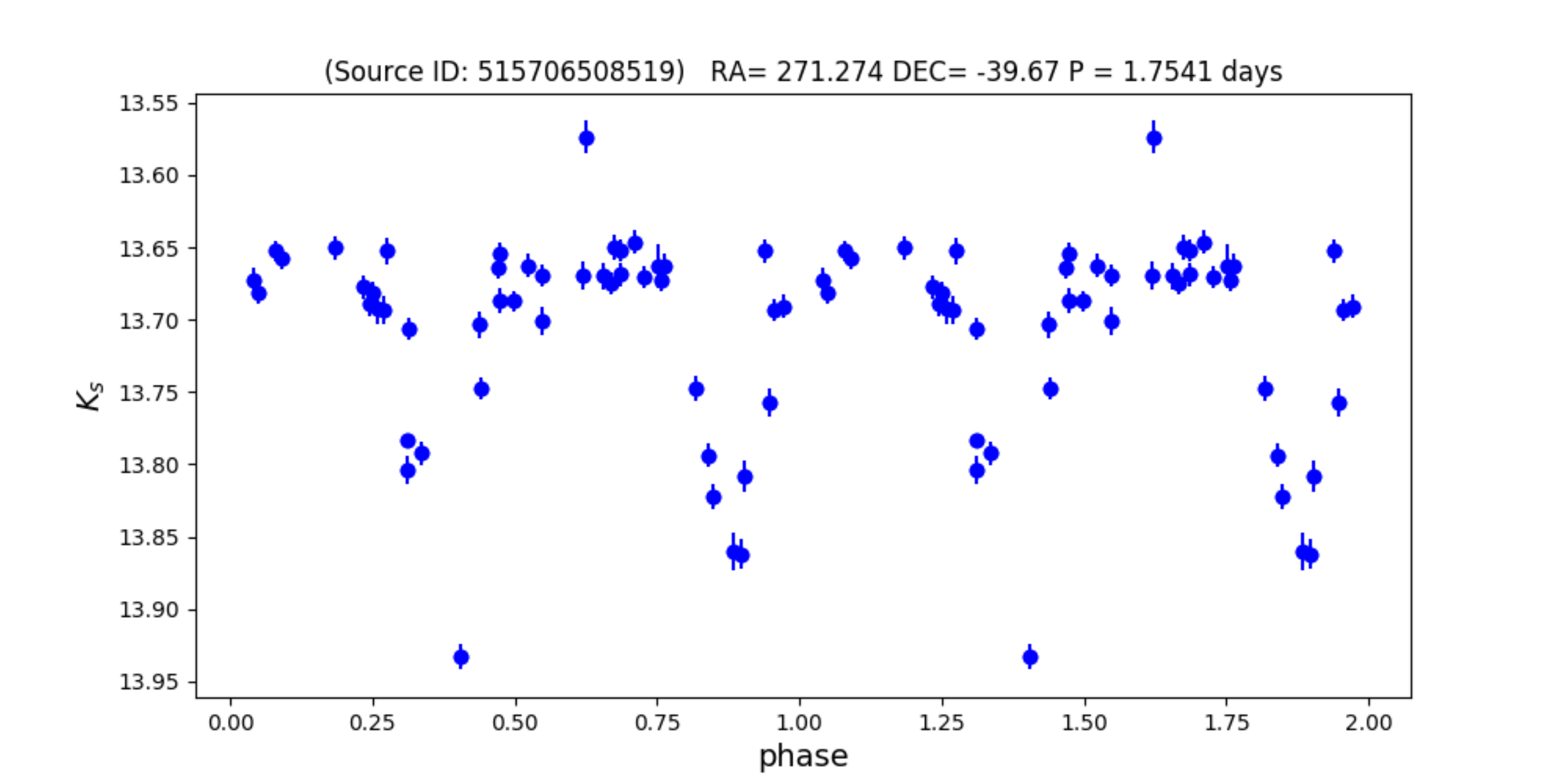} &
    \includegraphics[width=.33\textwidth]{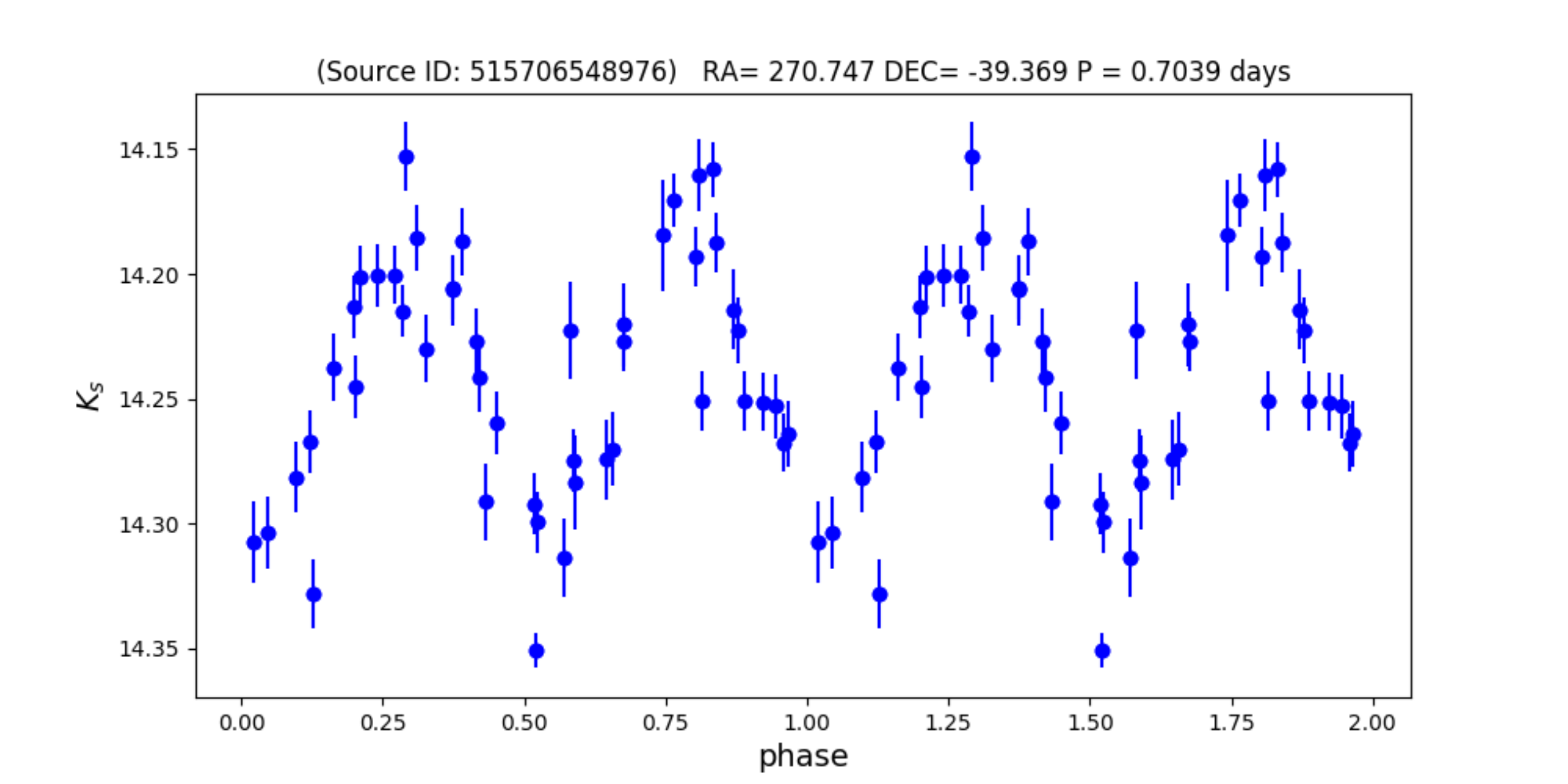} \\
  \end{tabular}  
  \caption{BHB stars light curves corresponding to $1^{st}$ category eclipsing binaries.}
  \label{fig:bhb9}
\end{figure*}

\begin{figure*}[htb]
\centering
  \begin{tabular}{@{}ccc@{}} 
    \includegraphics[width=.33\textwidth]{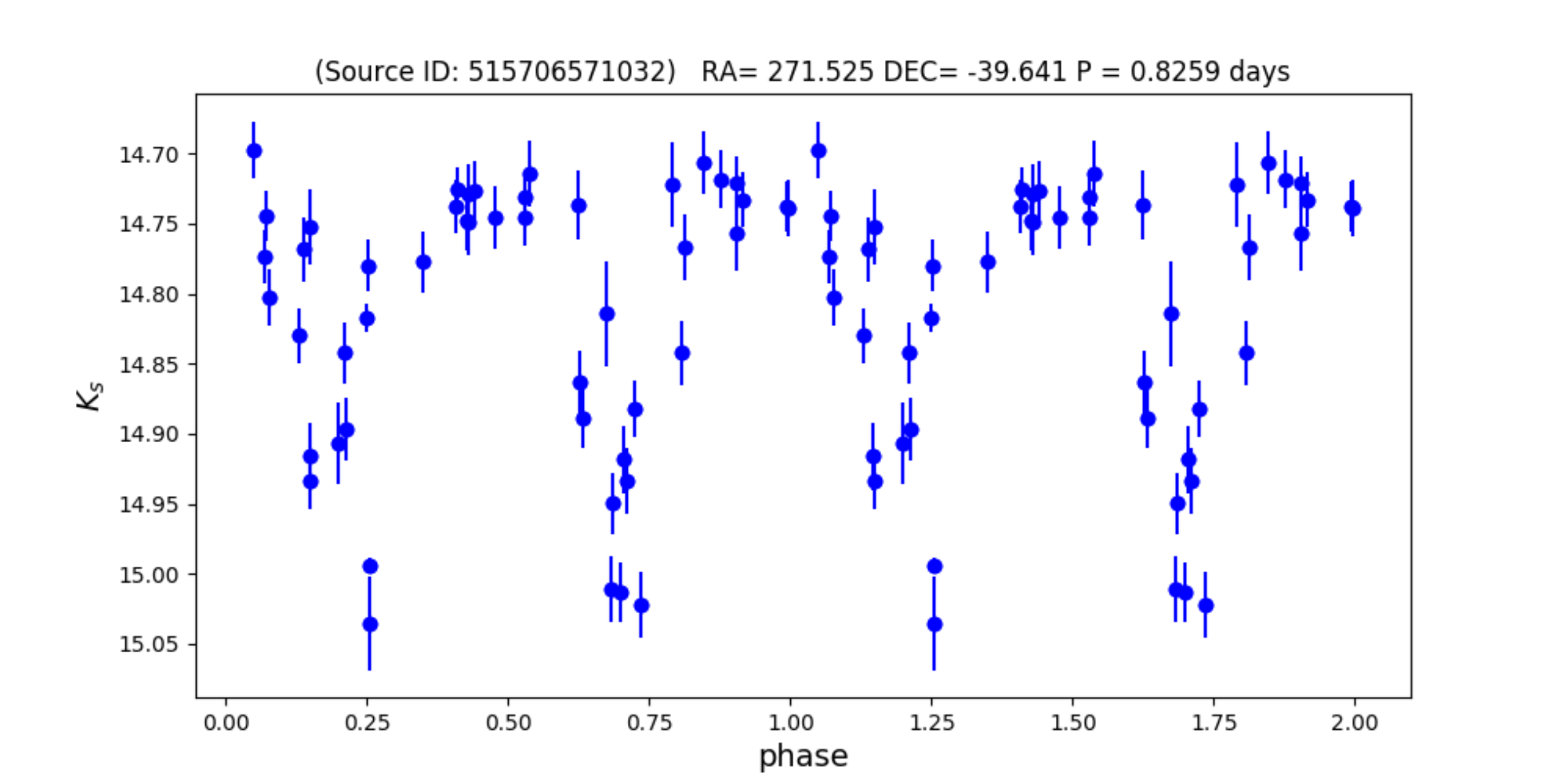} &
    \includegraphics[width=.33\textwidth]{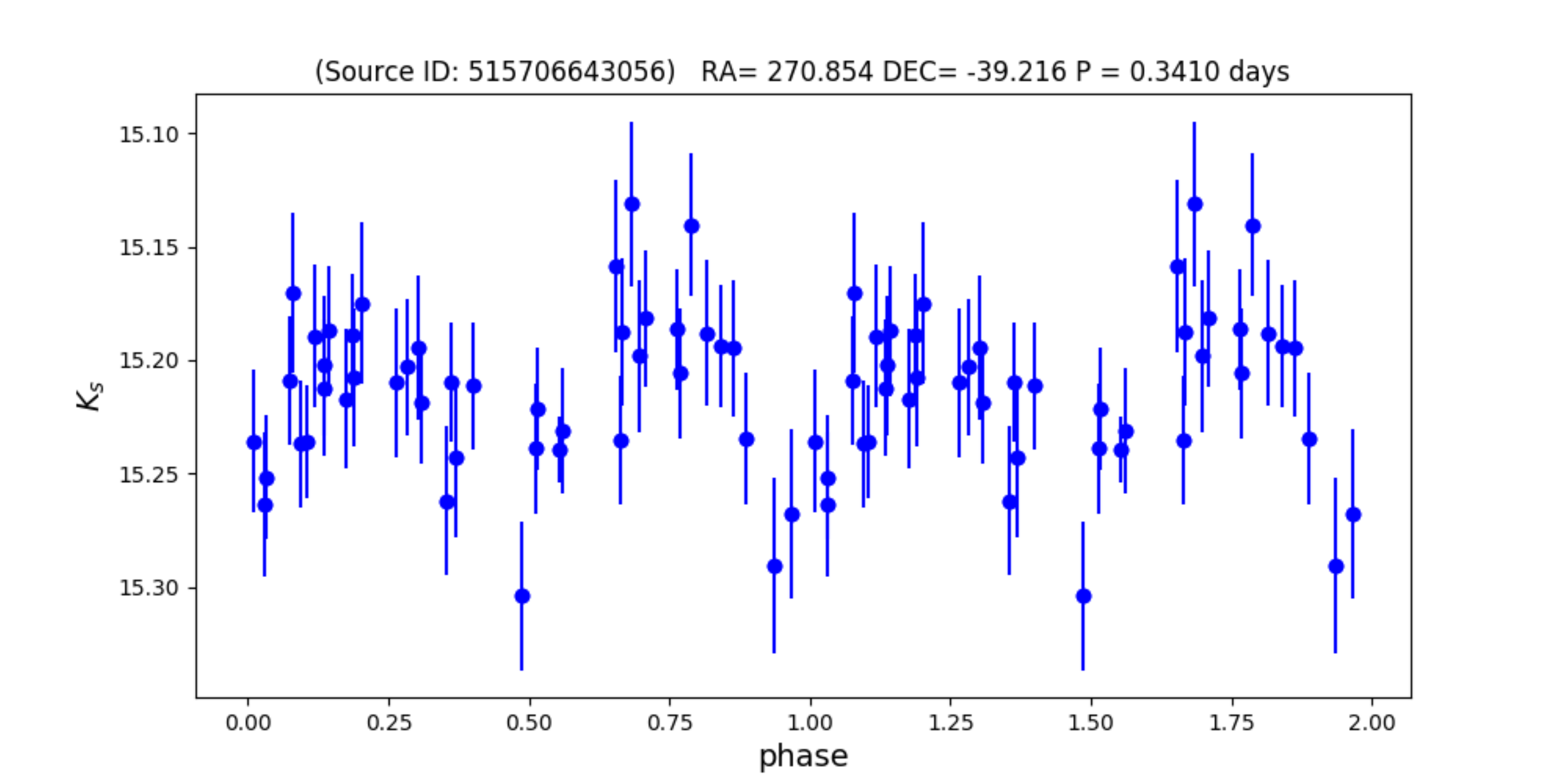} &
    \includegraphics[width=.33\textwidth]{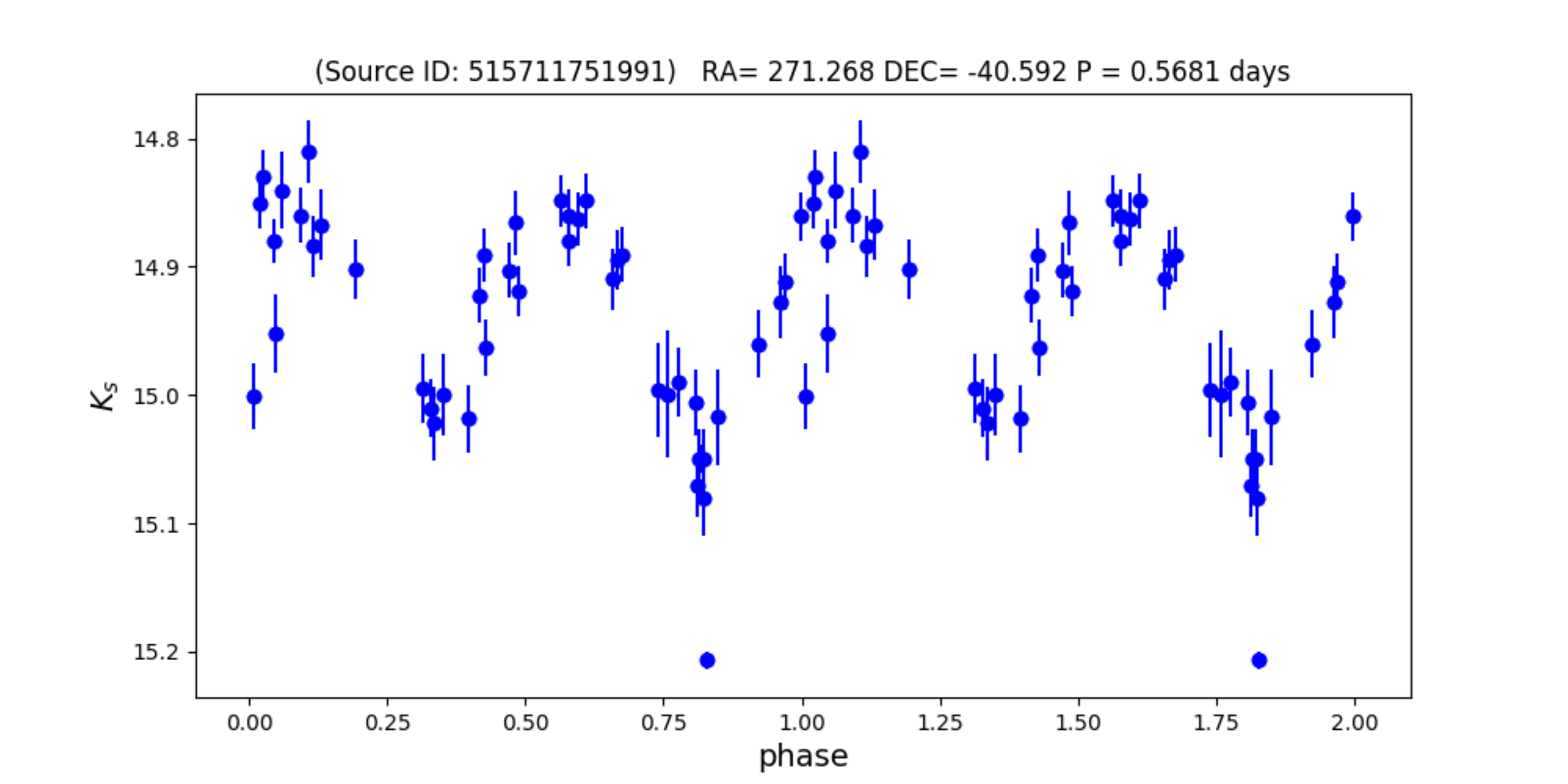} \\
    \includegraphics[width=.33\textwidth]{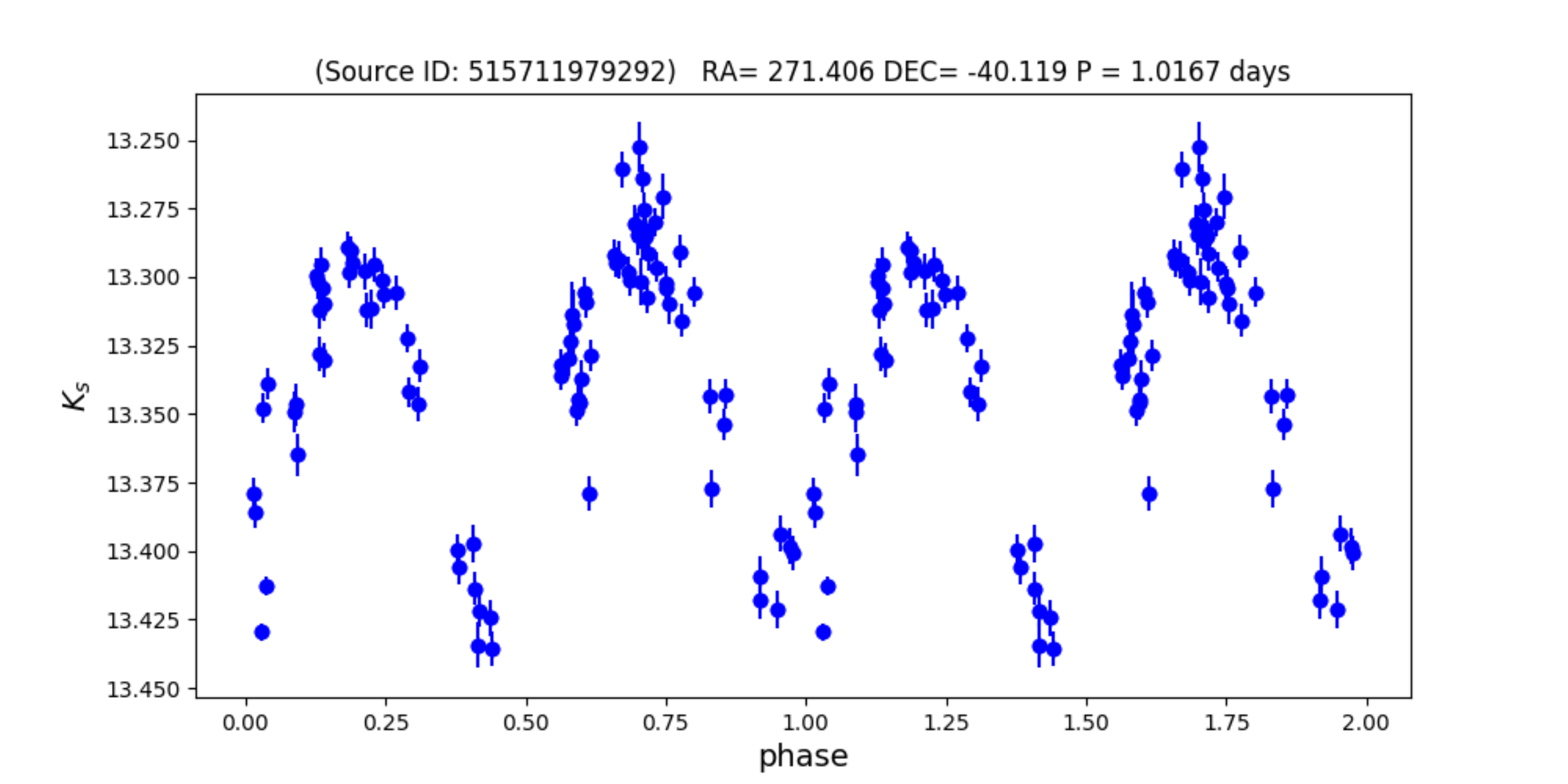} &
    \includegraphics[width=.33\textwidth]{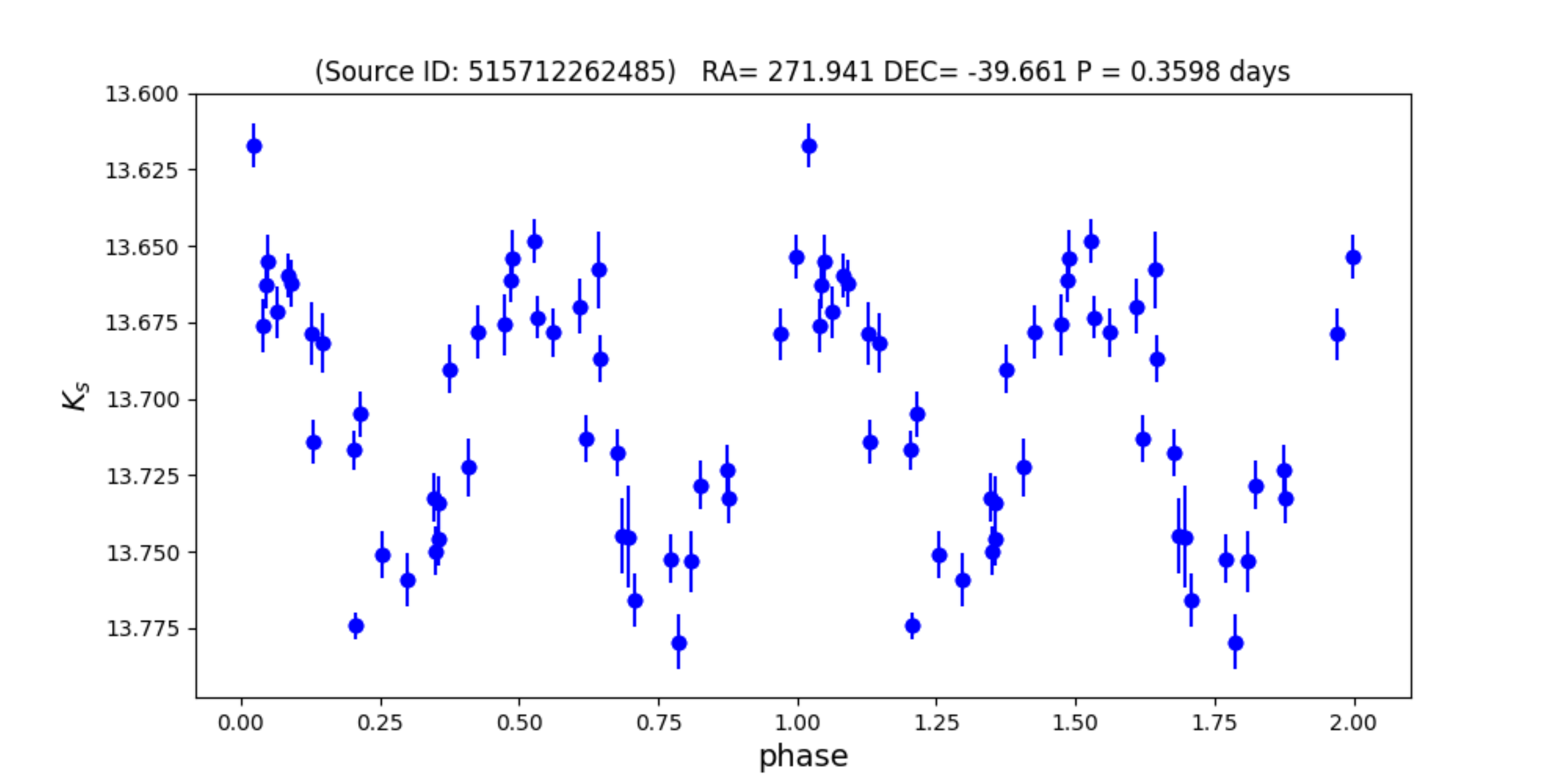} &
    \includegraphics[width=.33\textwidth]{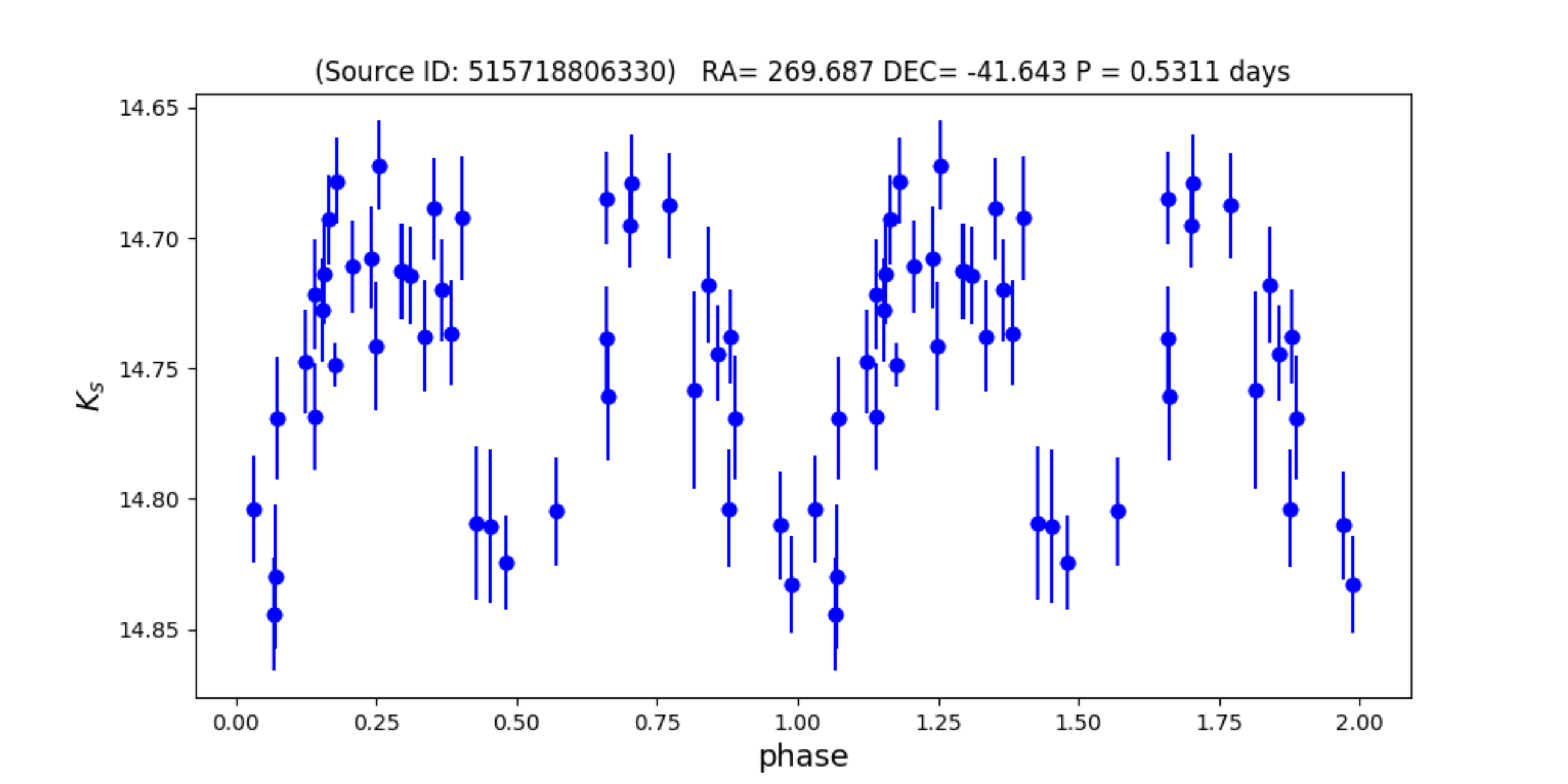} \\ 
    \includegraphics[width=.33\textwidth]{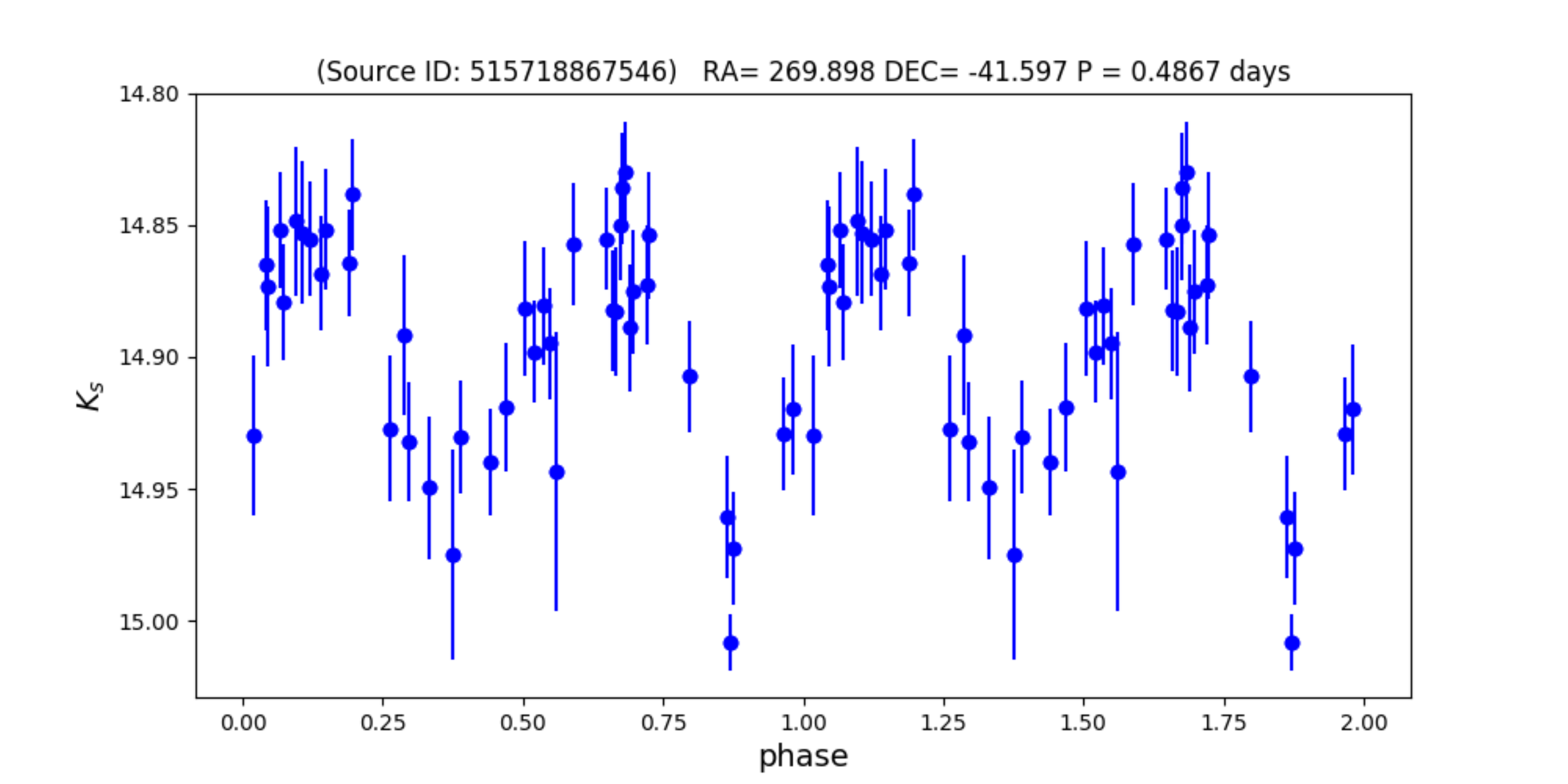} & 
    \includegraphics[width=.33\textwidth]{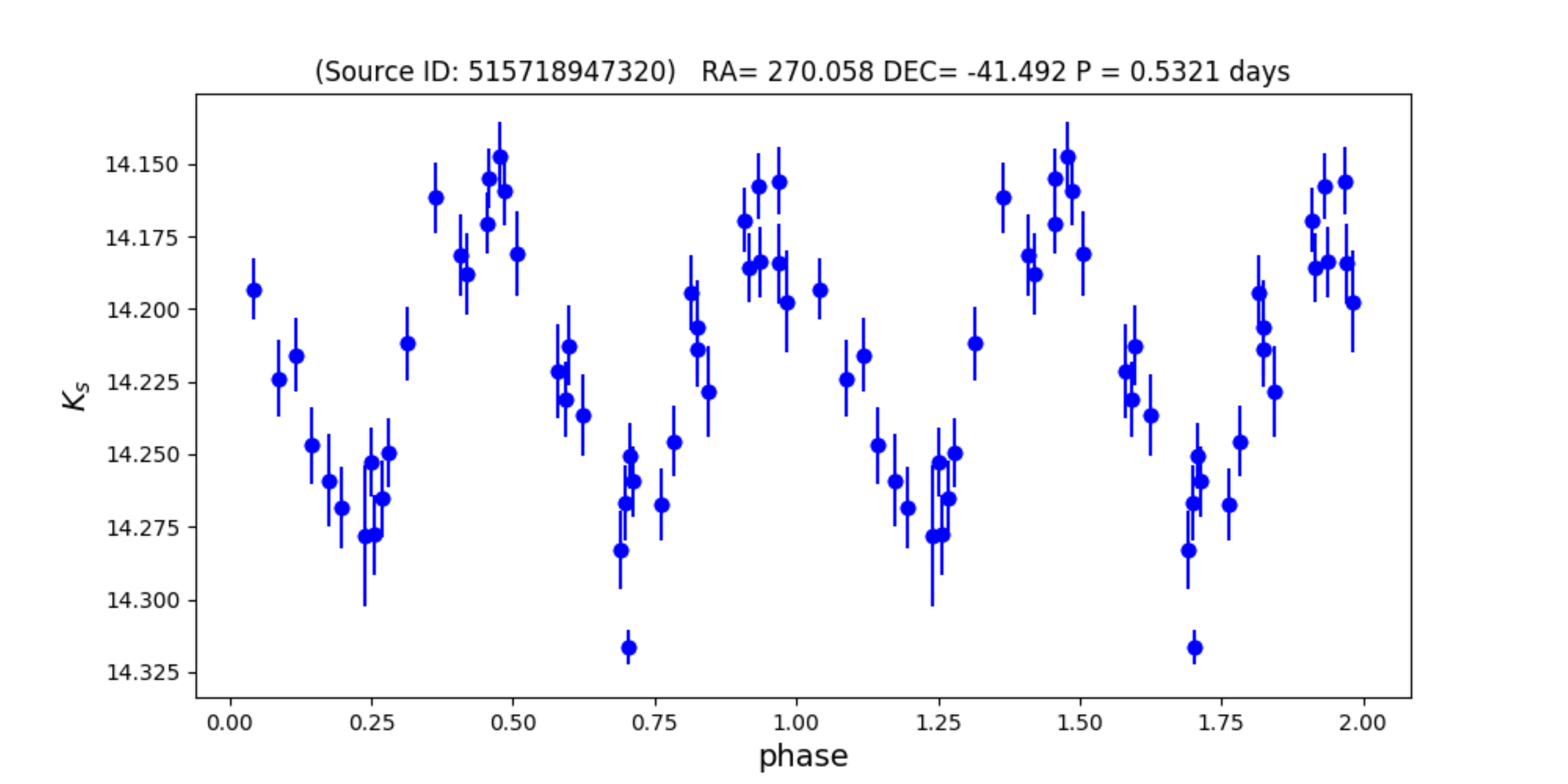} &
    \includegraphics[width=.33\textwidth]{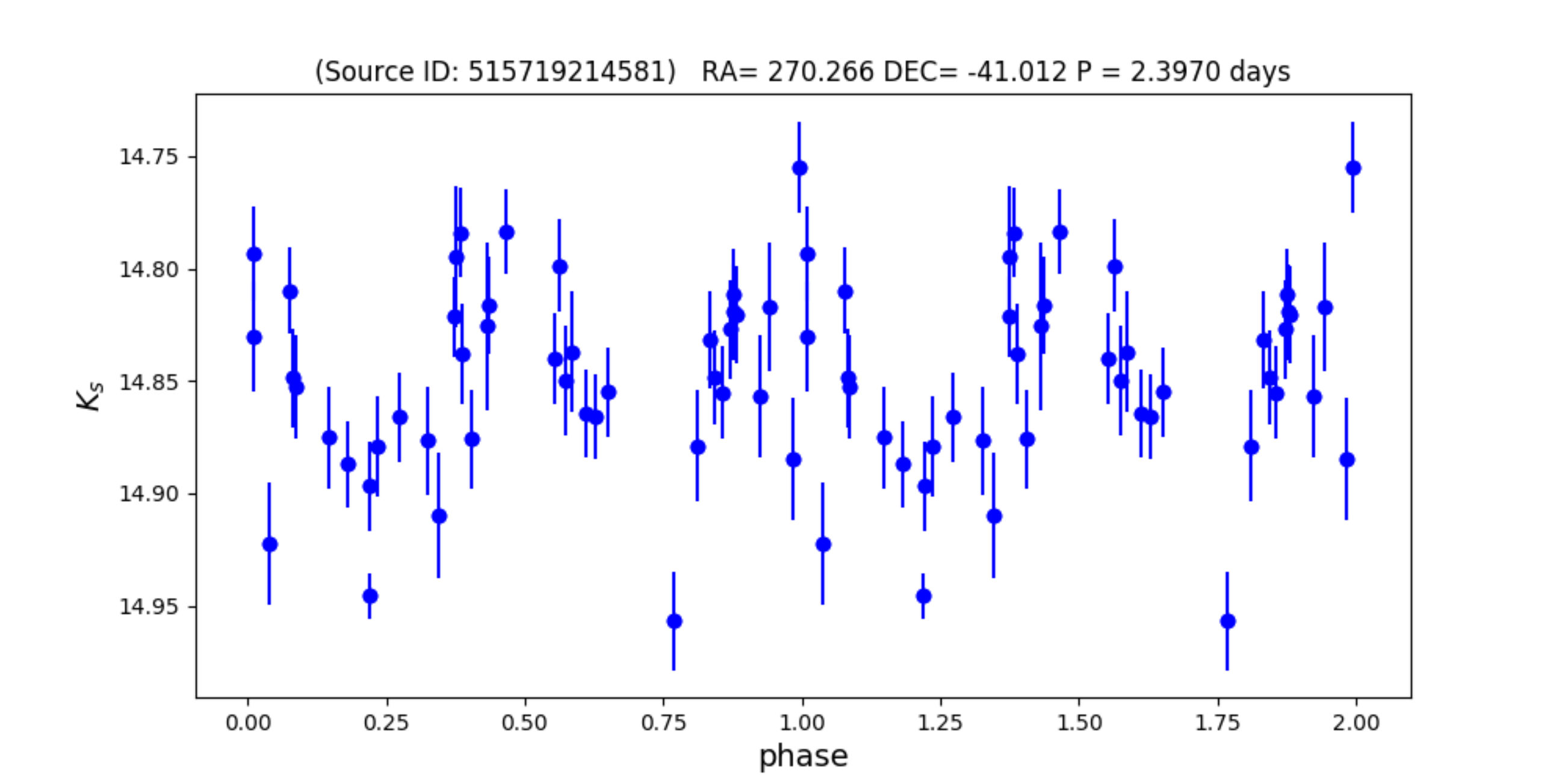} \\
    \includegraphics[width=.33\textwidth]{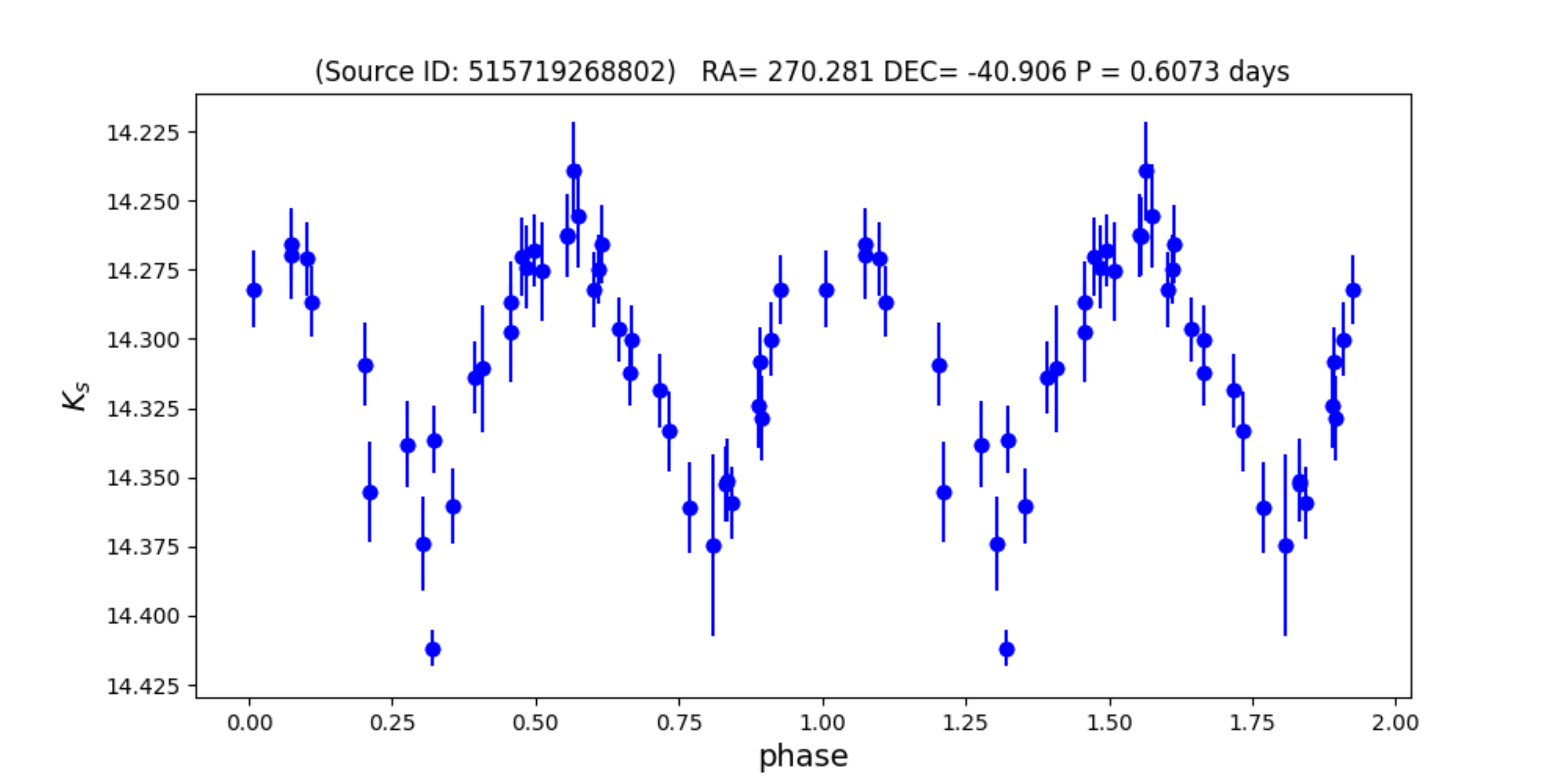} &
    \includegraphics[width=.33\textwidth]{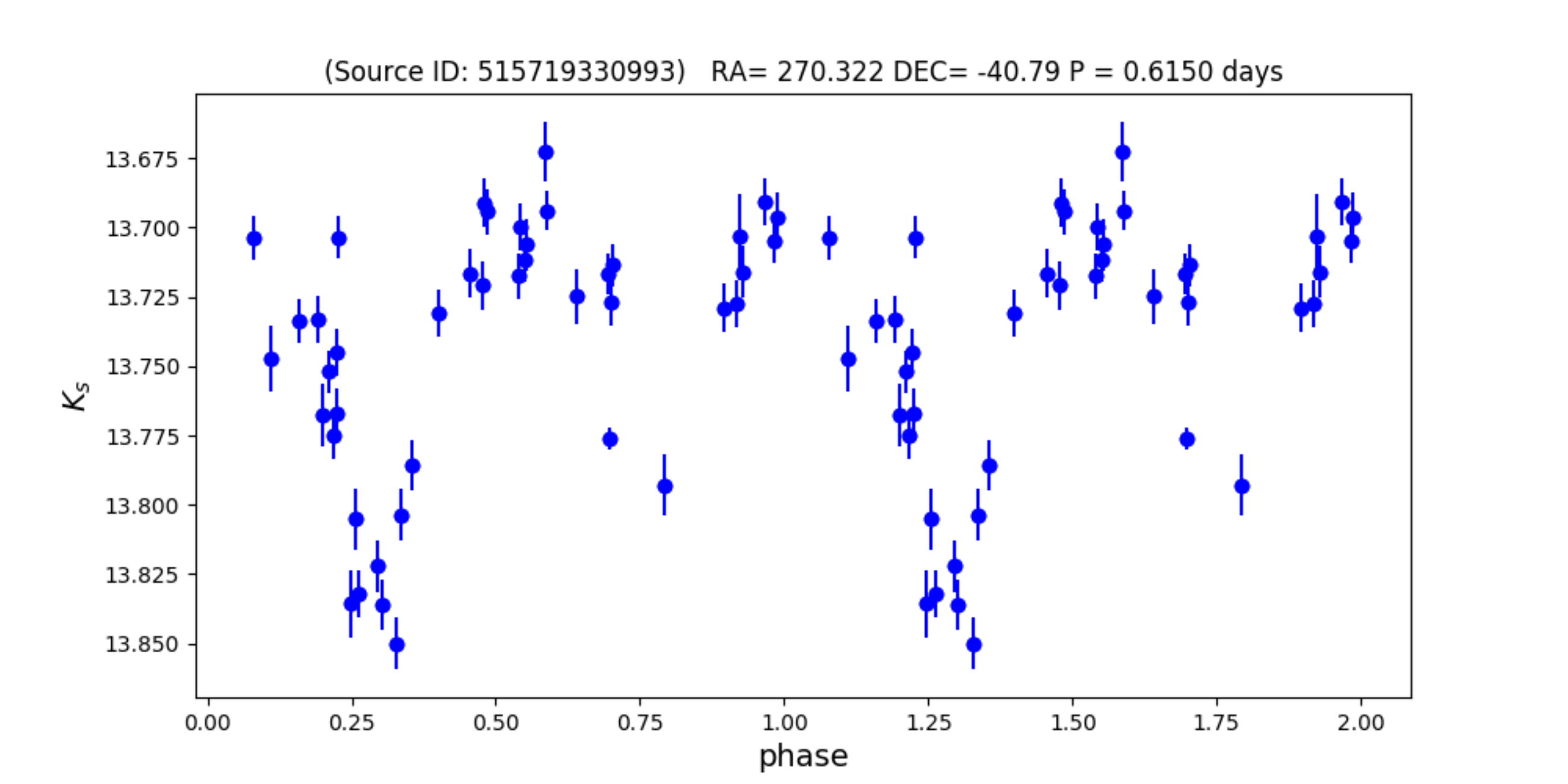} &
    \includegraphics[width=.33\textwidth]{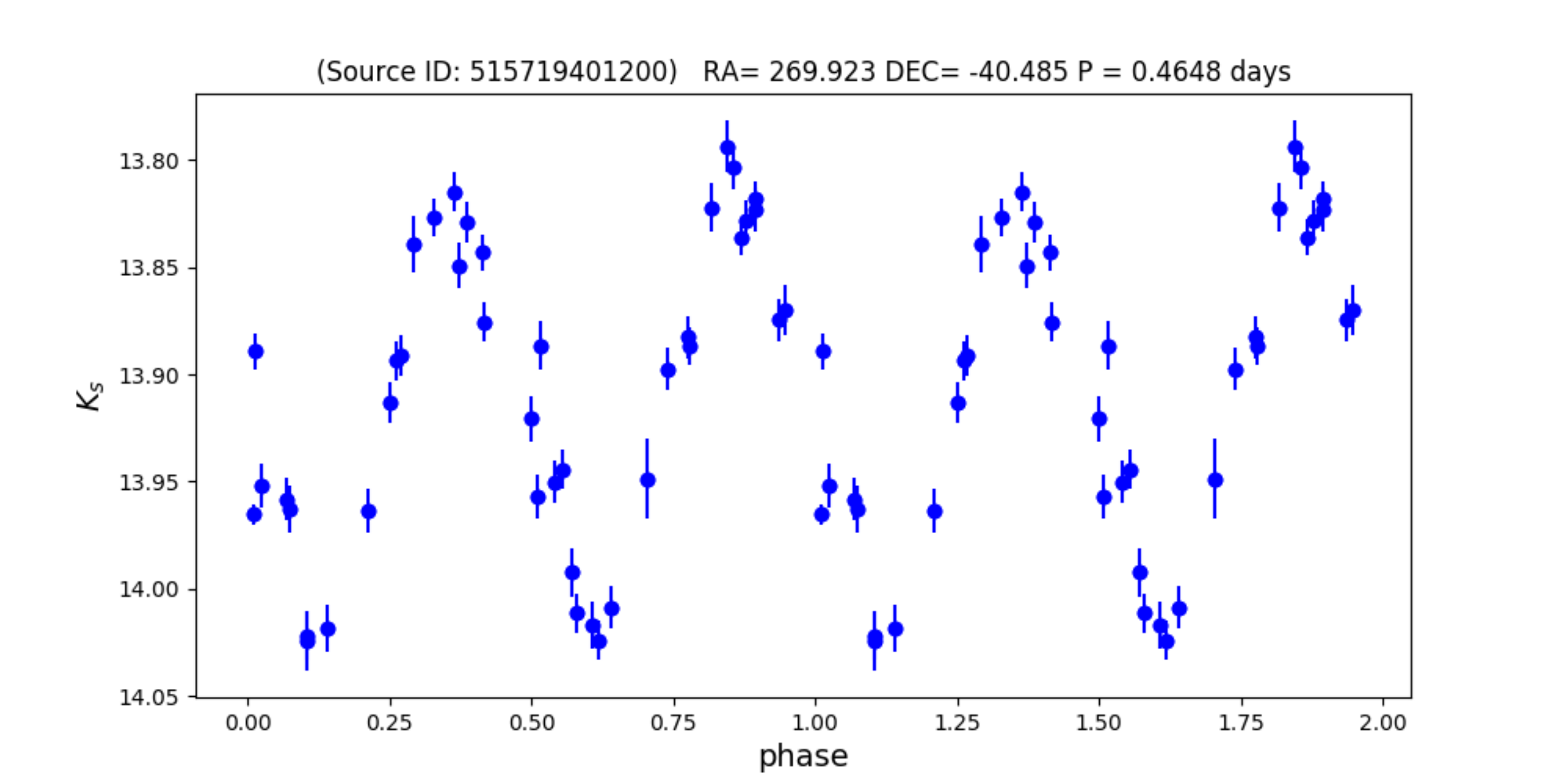} \\  
    \includegraphics[width=.33\textwidth]{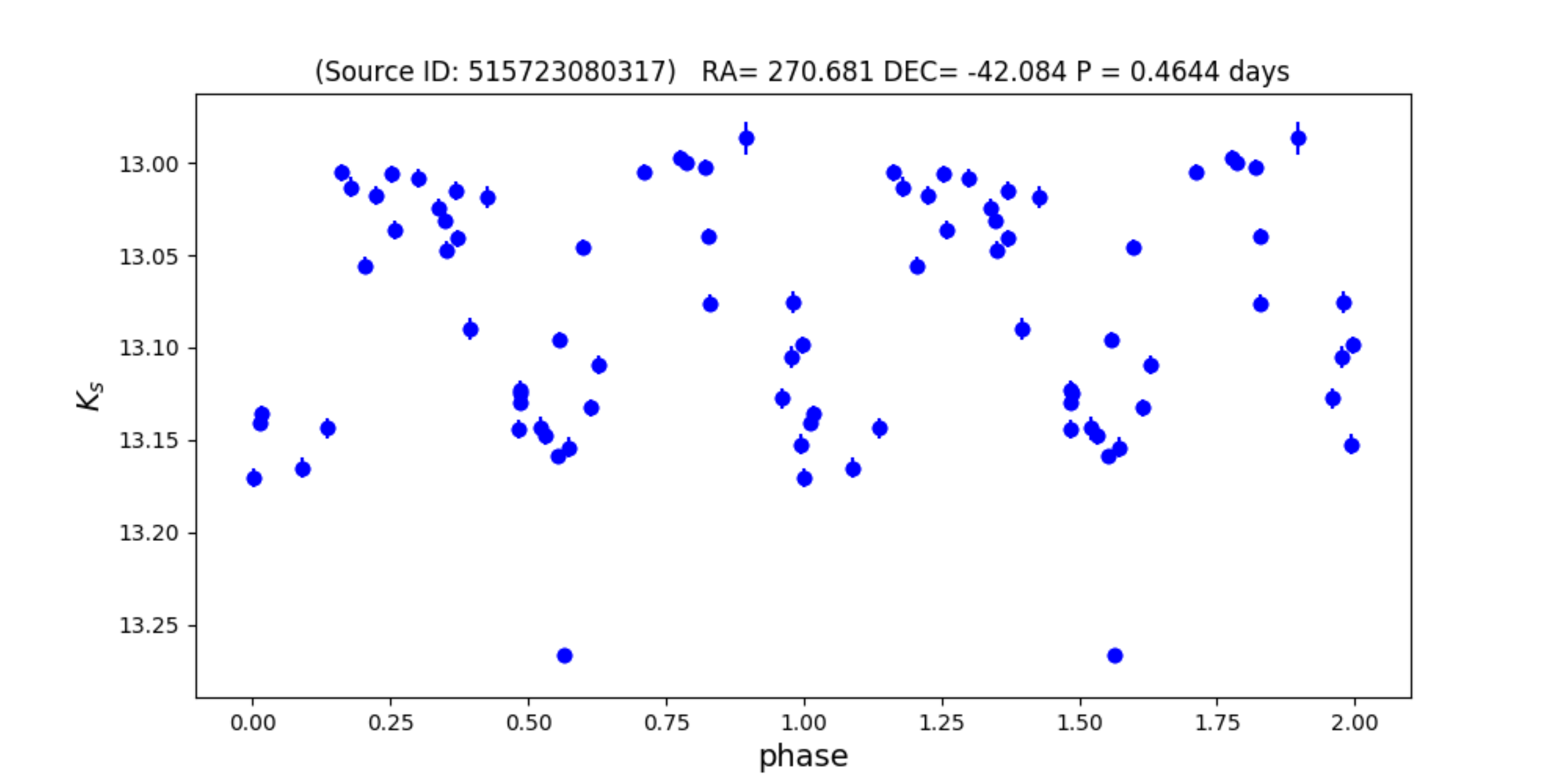} &
    \includegraphics[width=.33\textwidth]{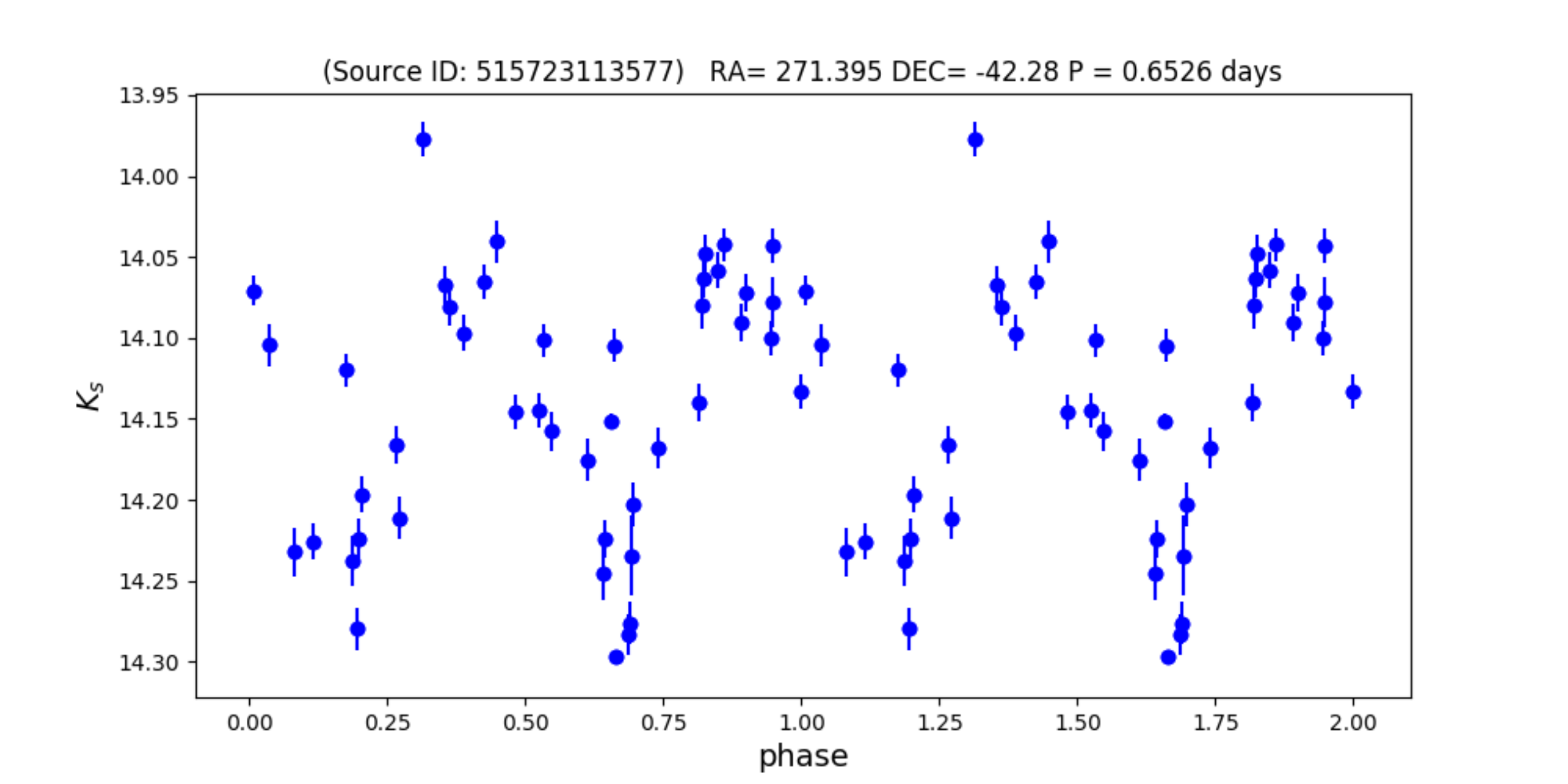} &
    \includegraphics[width=.33\textwidth]{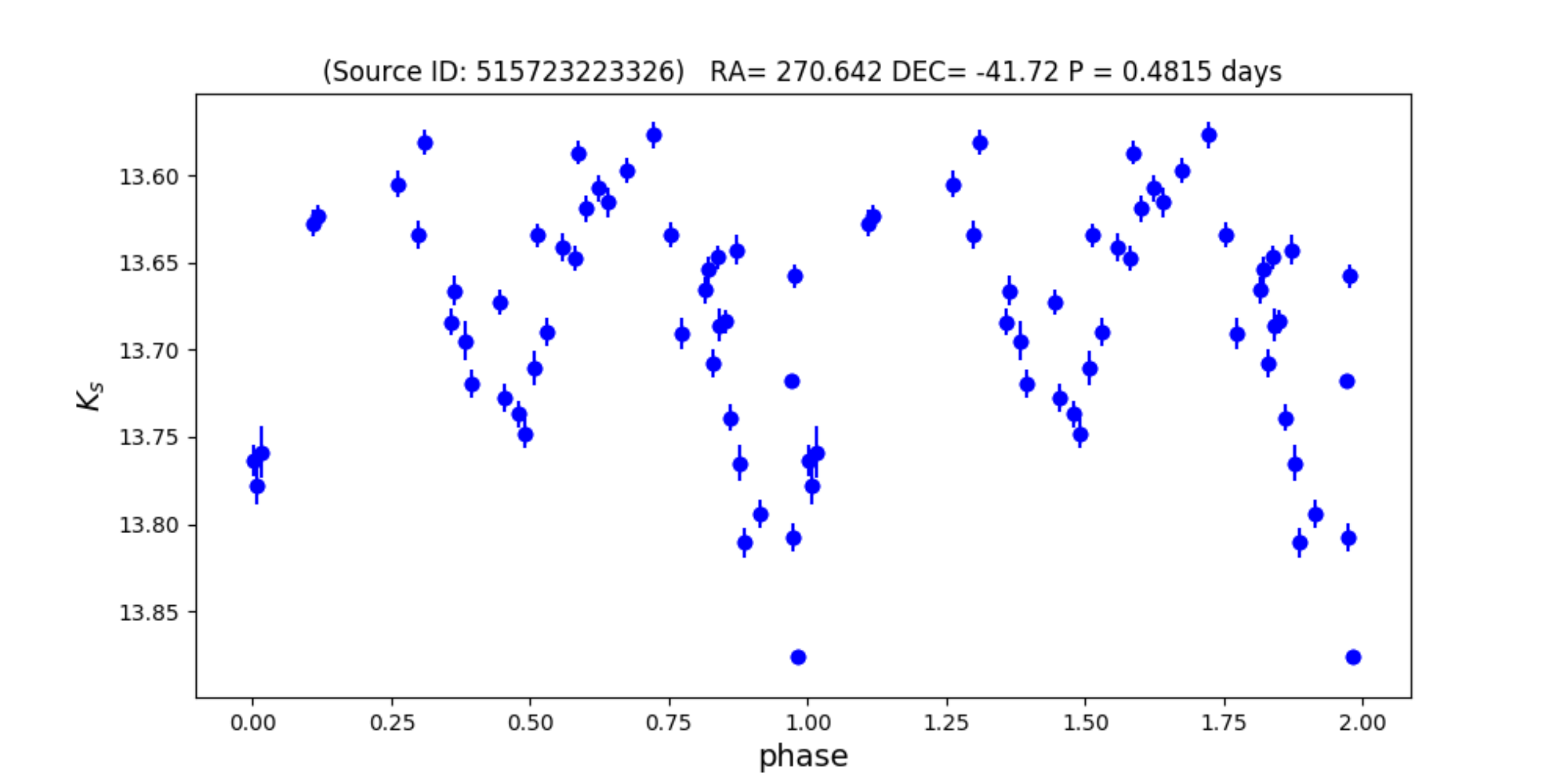} \\ 
    &
    \includegraphics[width=.33\textwidth]{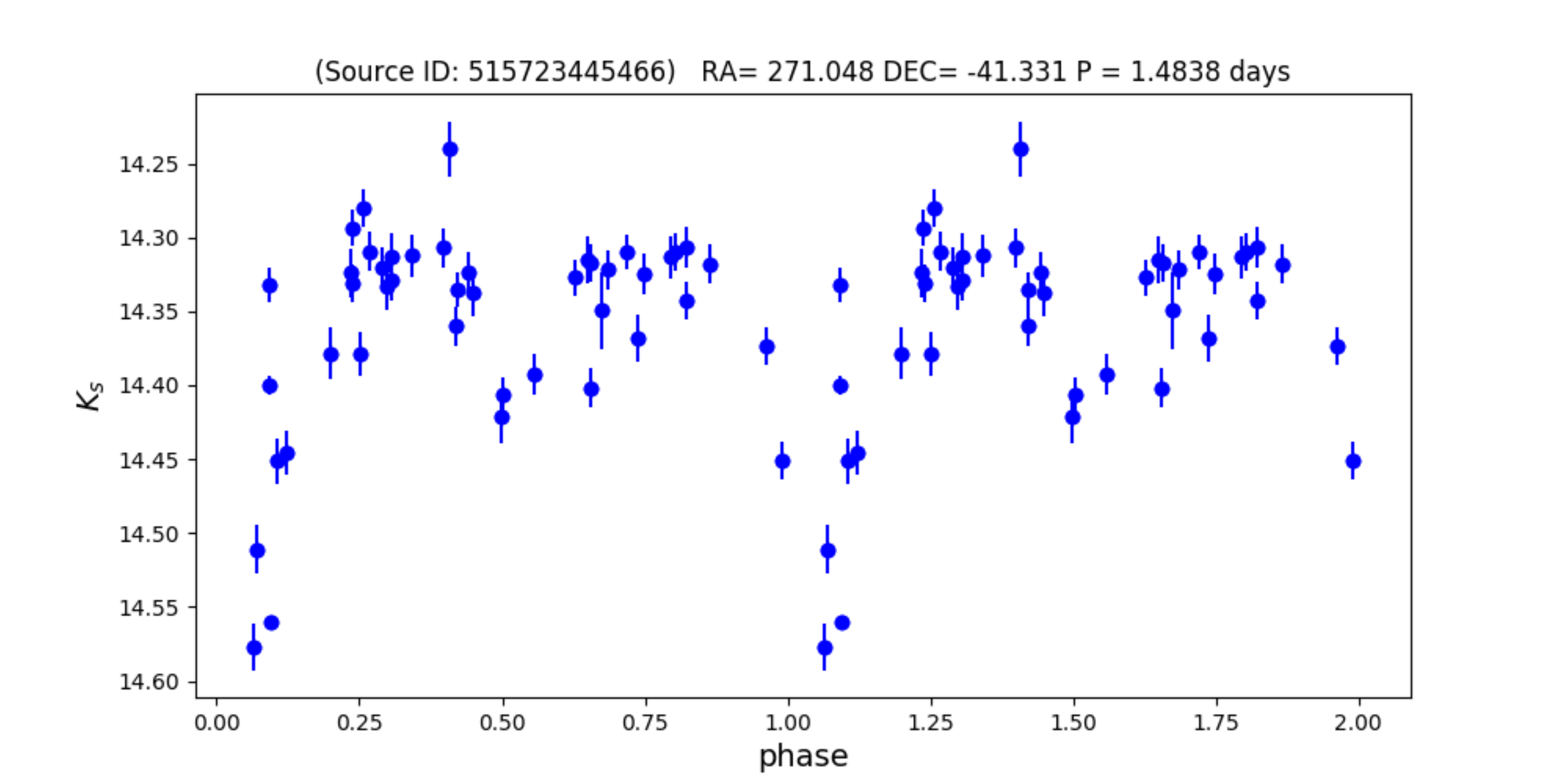} \\ 
    &
  \end{tabular}  
  \caption{BHB stars light curves corresponding to $1^{st}$ category eclipsing binaries.}
  \label{fig:bhb10}
\end{figure*}

\begin{figure*}[htb]
\centering
  \begin{tabular}{@{}ccc@{}}    
    \includegraphics[width=.33\textwidth]{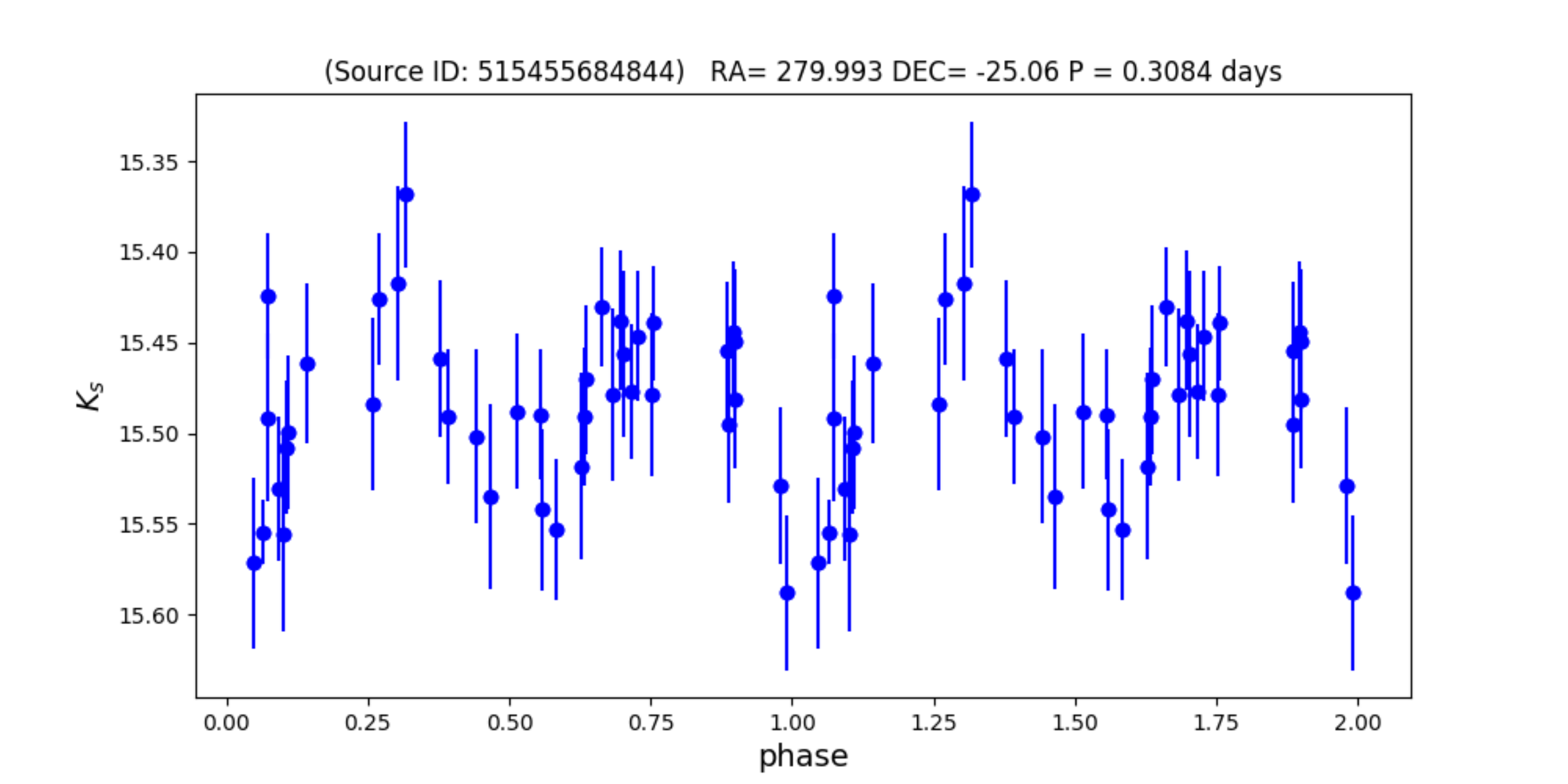} &    
    \includegraphics[width=.33\textwidth]{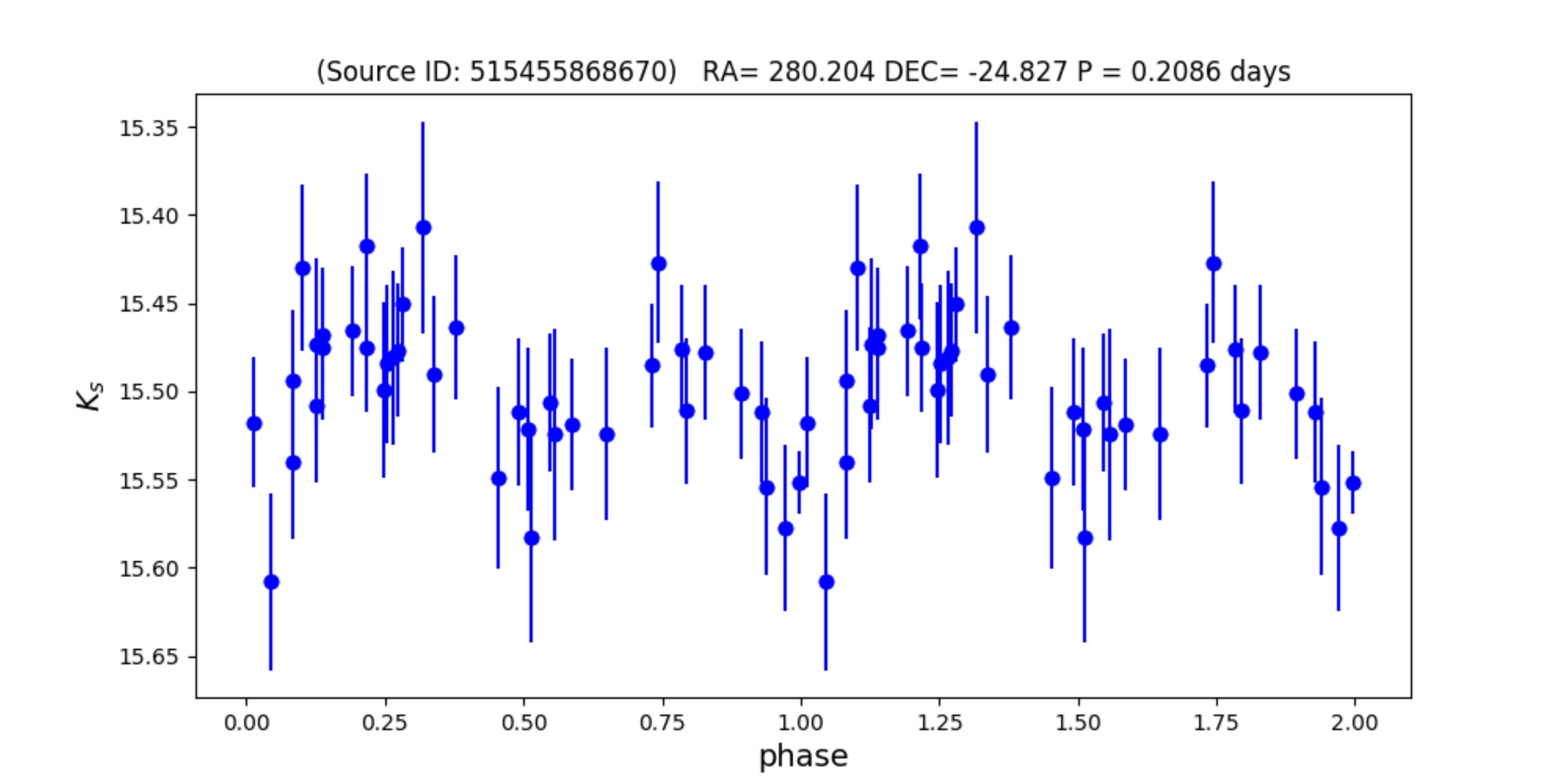} &
    \includegraphics[width=.33\textwidth]{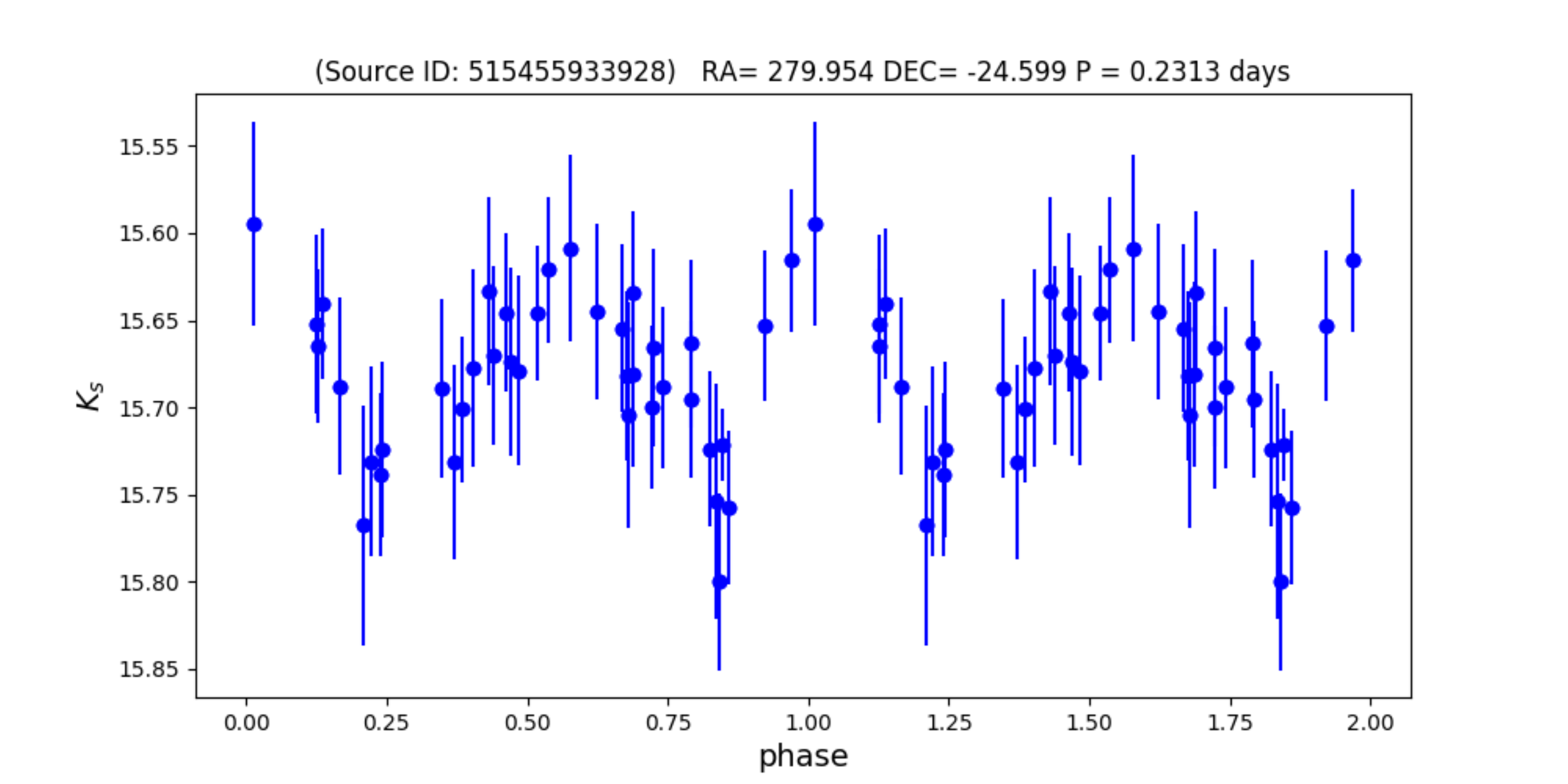} \\
    \includegraphics[width=.33\textwidth]{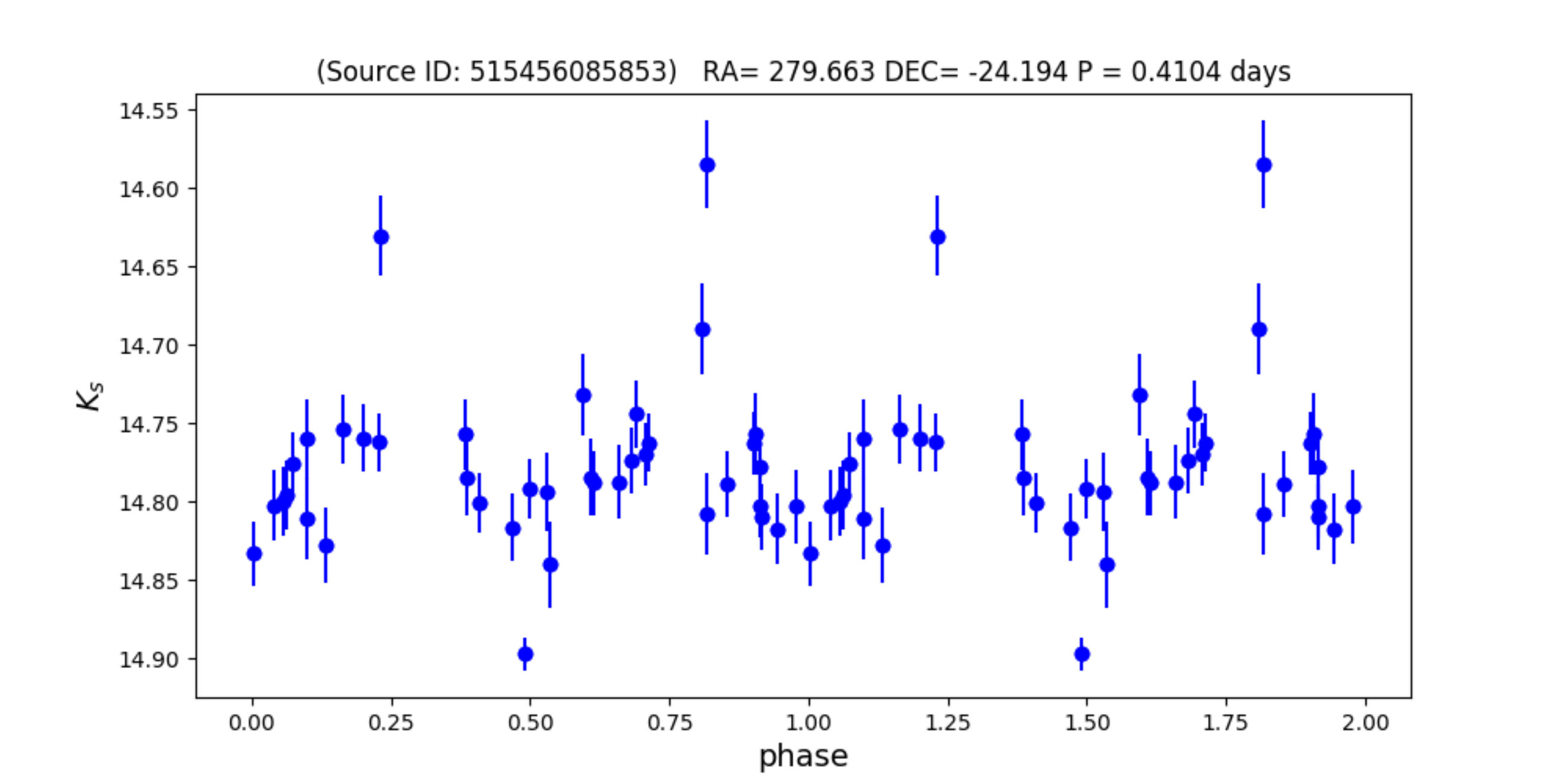} &    
    \includegraphics[width=.33\textwidth]{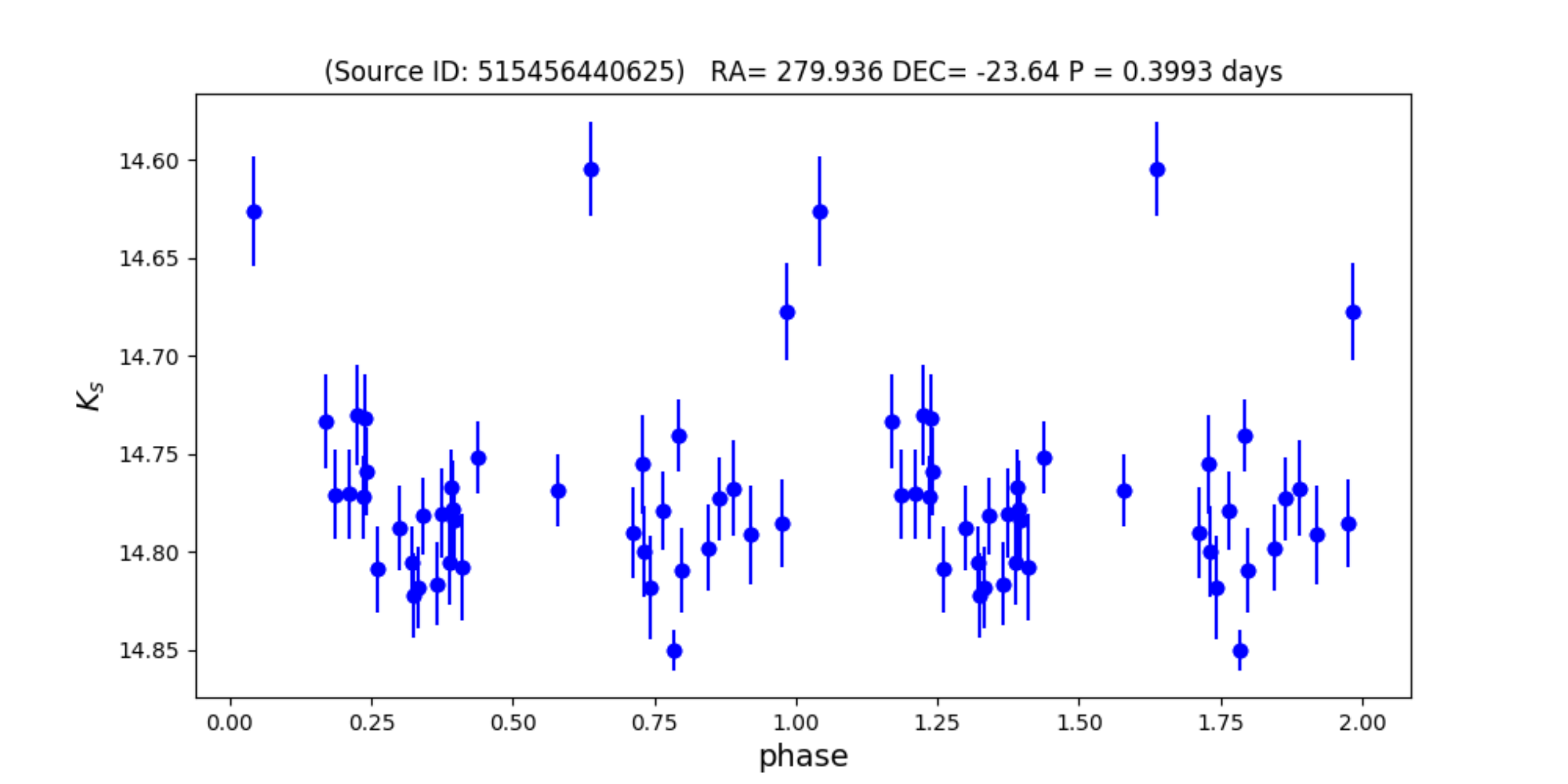} &    
    \includegraphics[width=.33\textwidth]{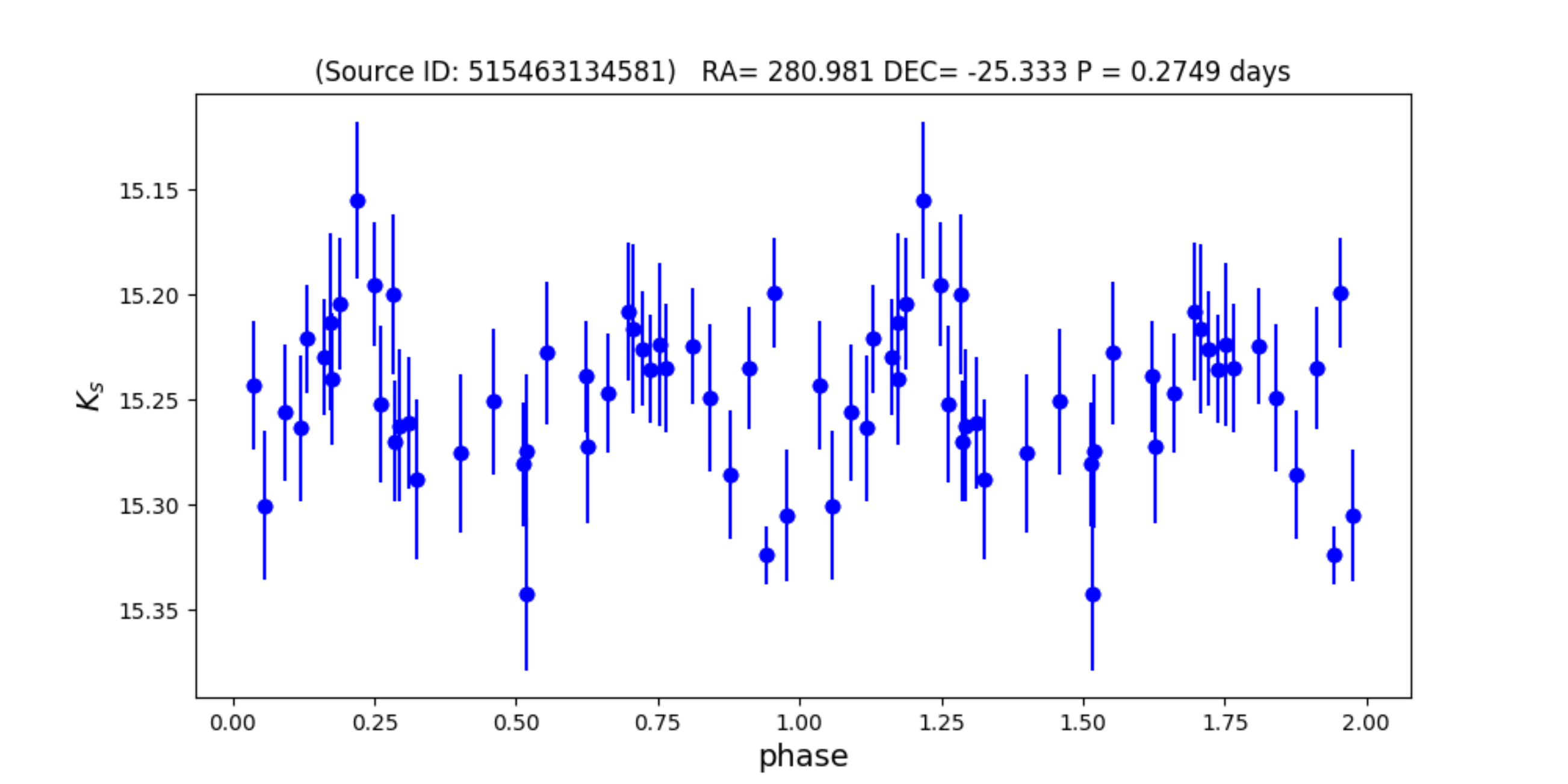} \\
    \includegraphics[width=.33\textwidth]{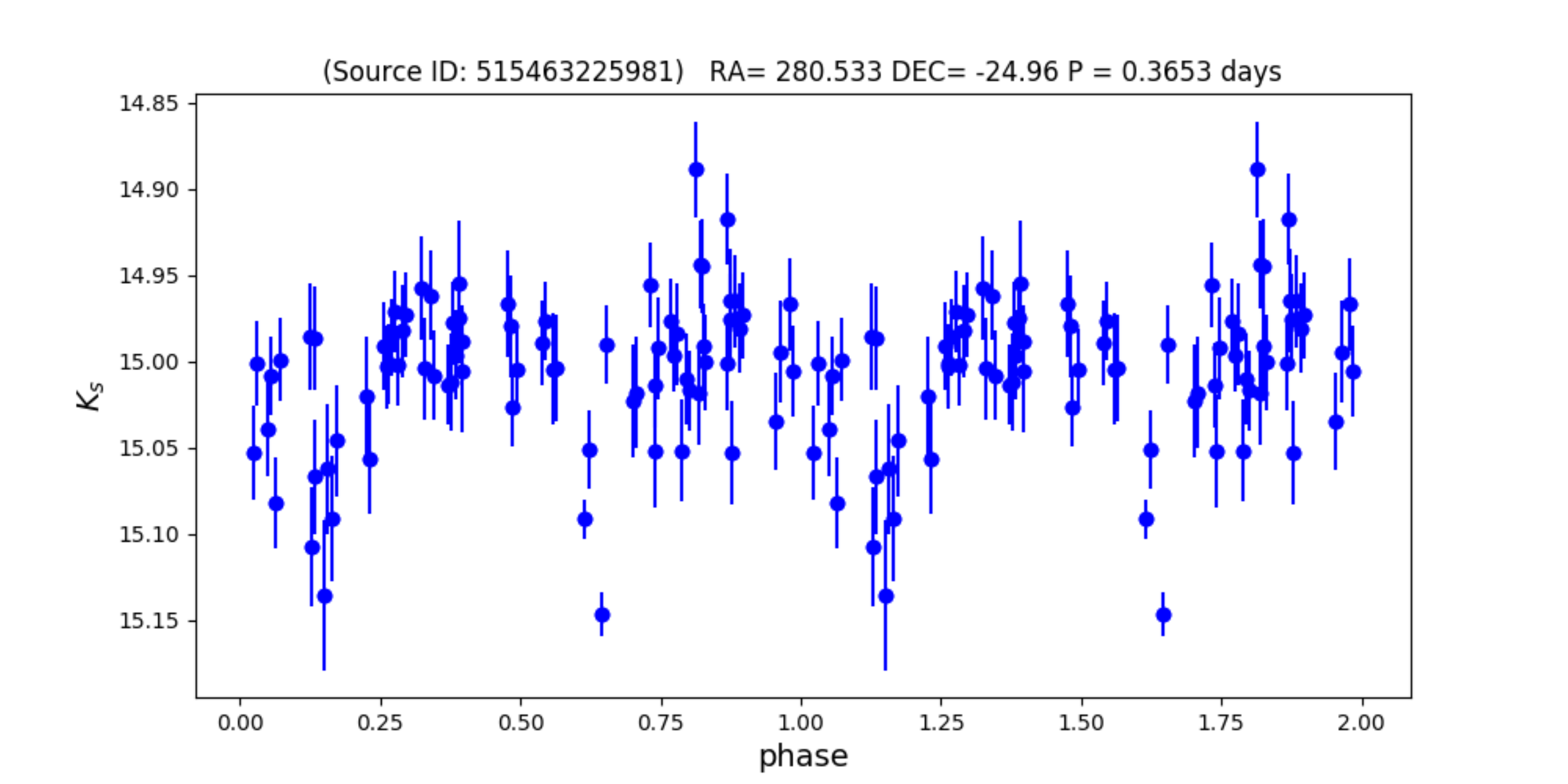} &    
    \includegraphics[width=.33\textwidth]{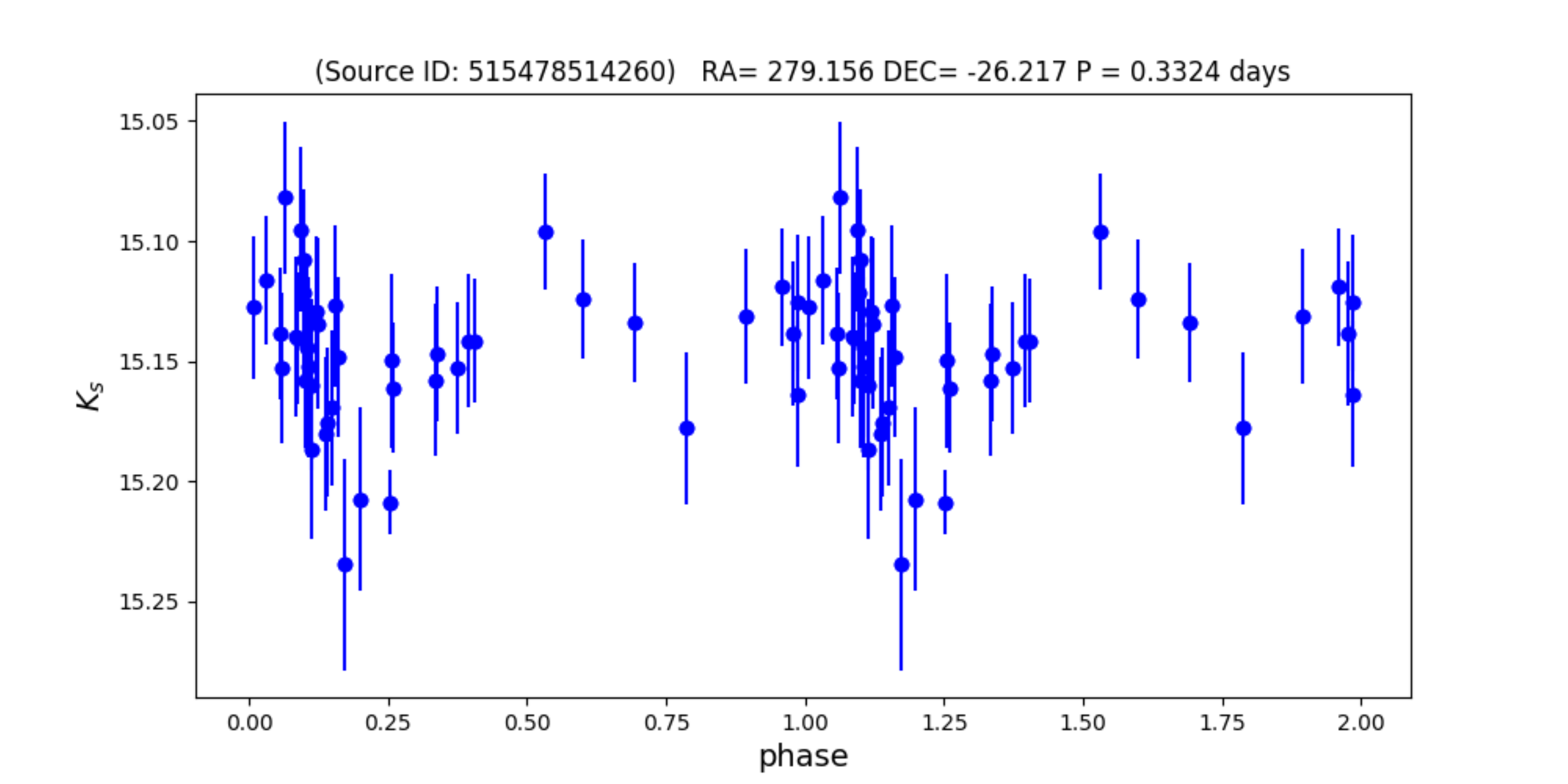} &
    \includegraphics[width=.33\textwidth]{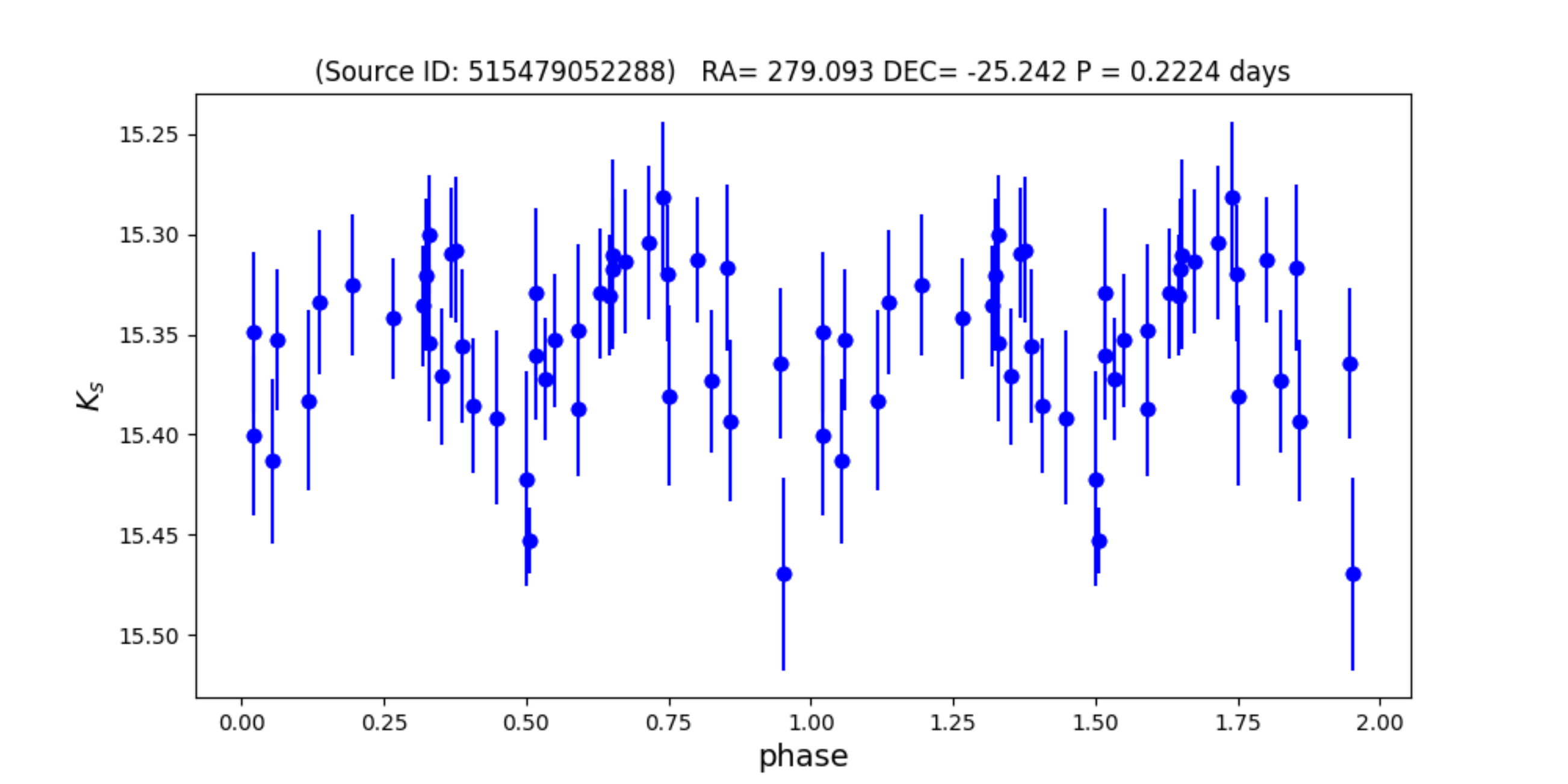} \\
    \includegraphics[width=.33\textwidth]{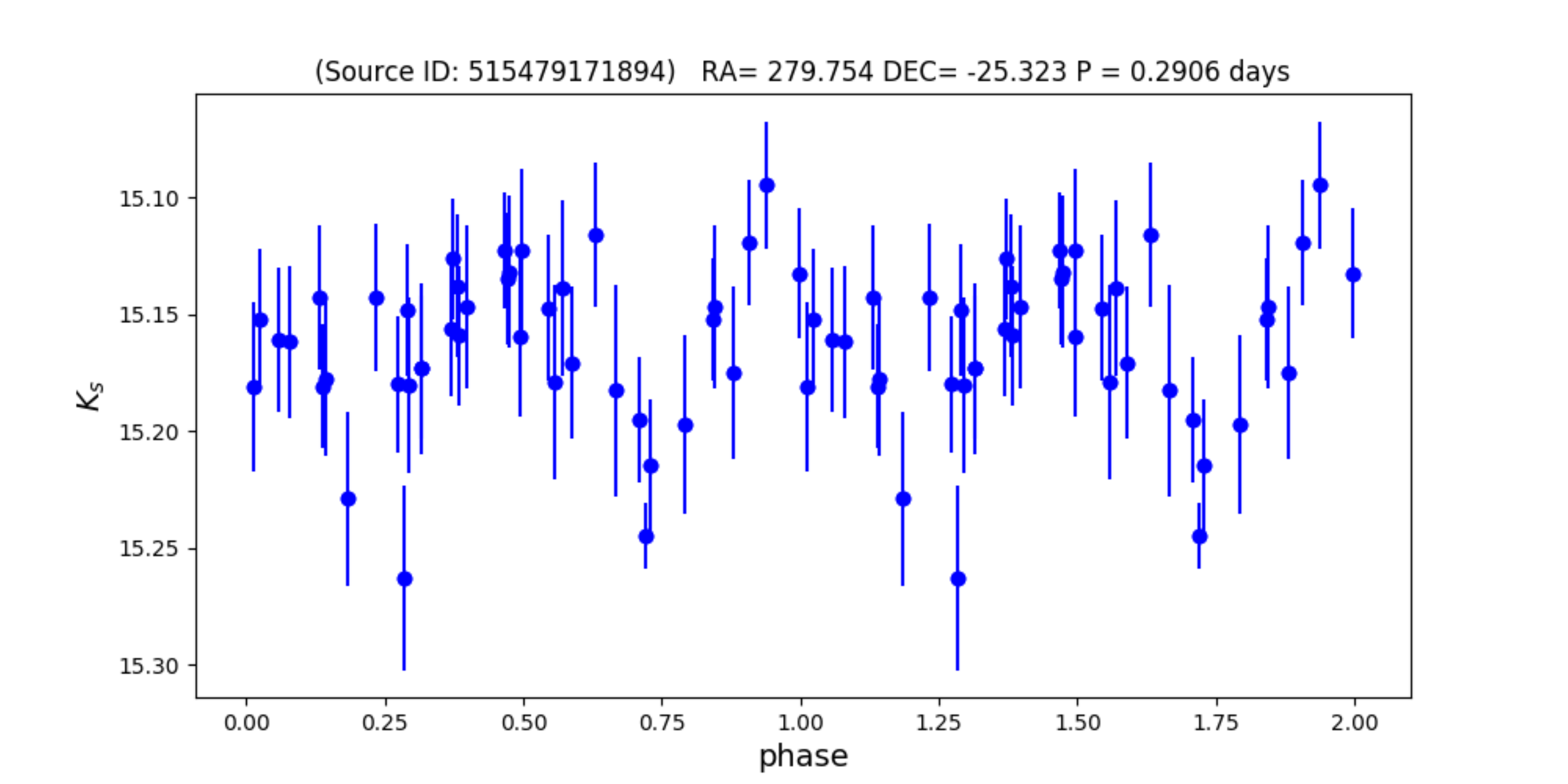} &
    \includegraphics[width=.33\textwidth]{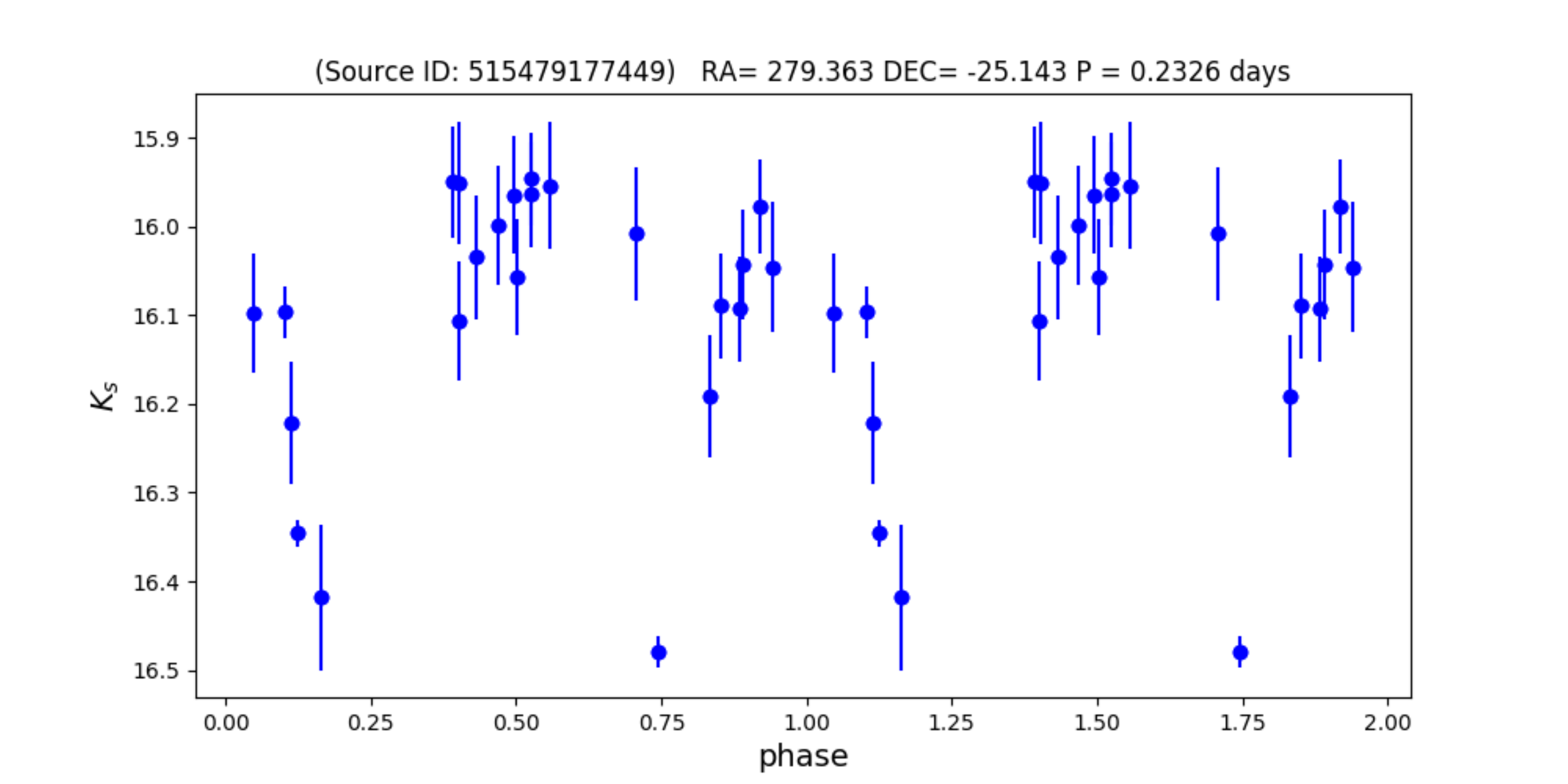} &
    \includegraphics[width=.33\textwidth]{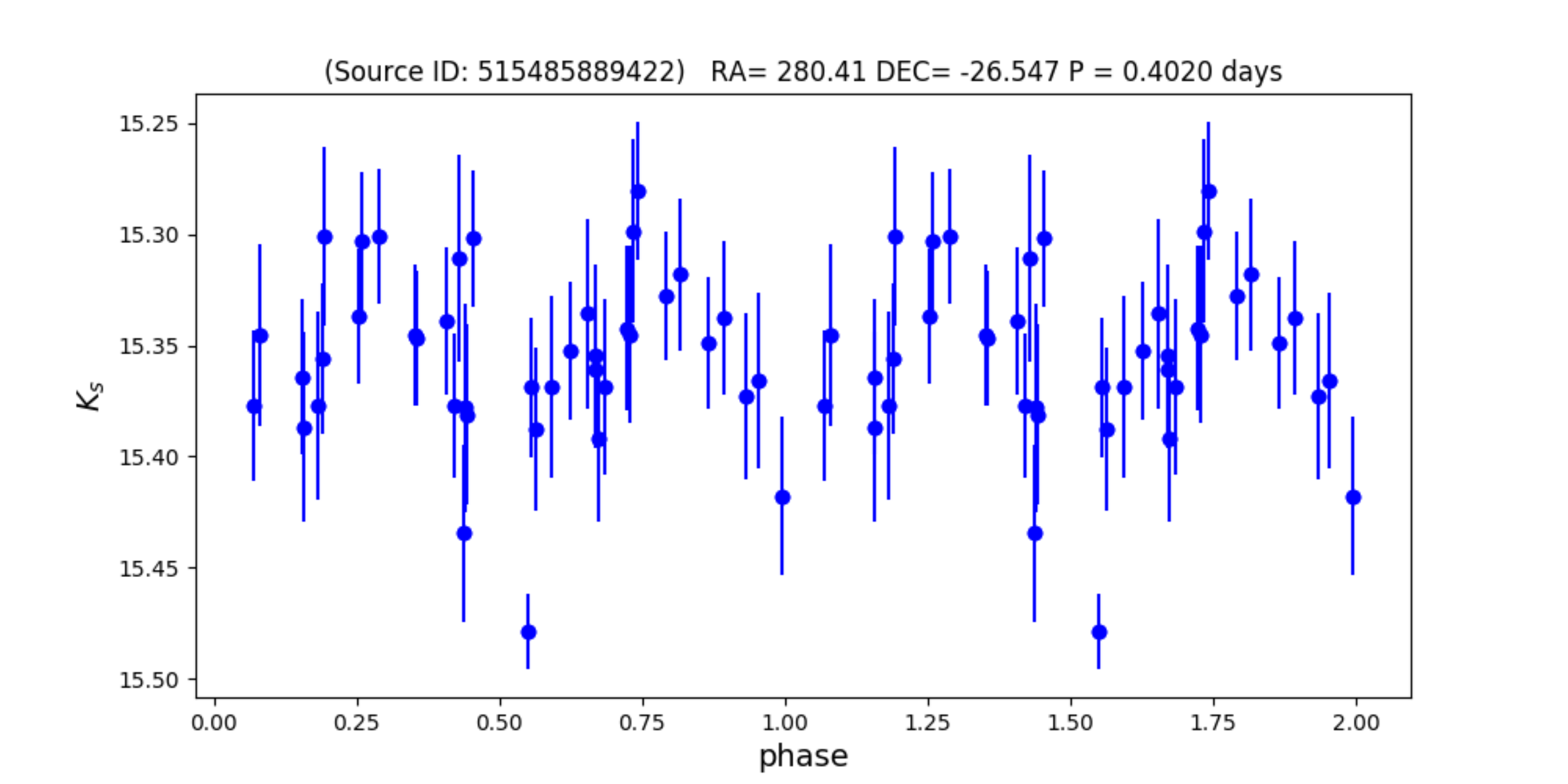} \\    
    \includegraphics[width=.33\textwidth]{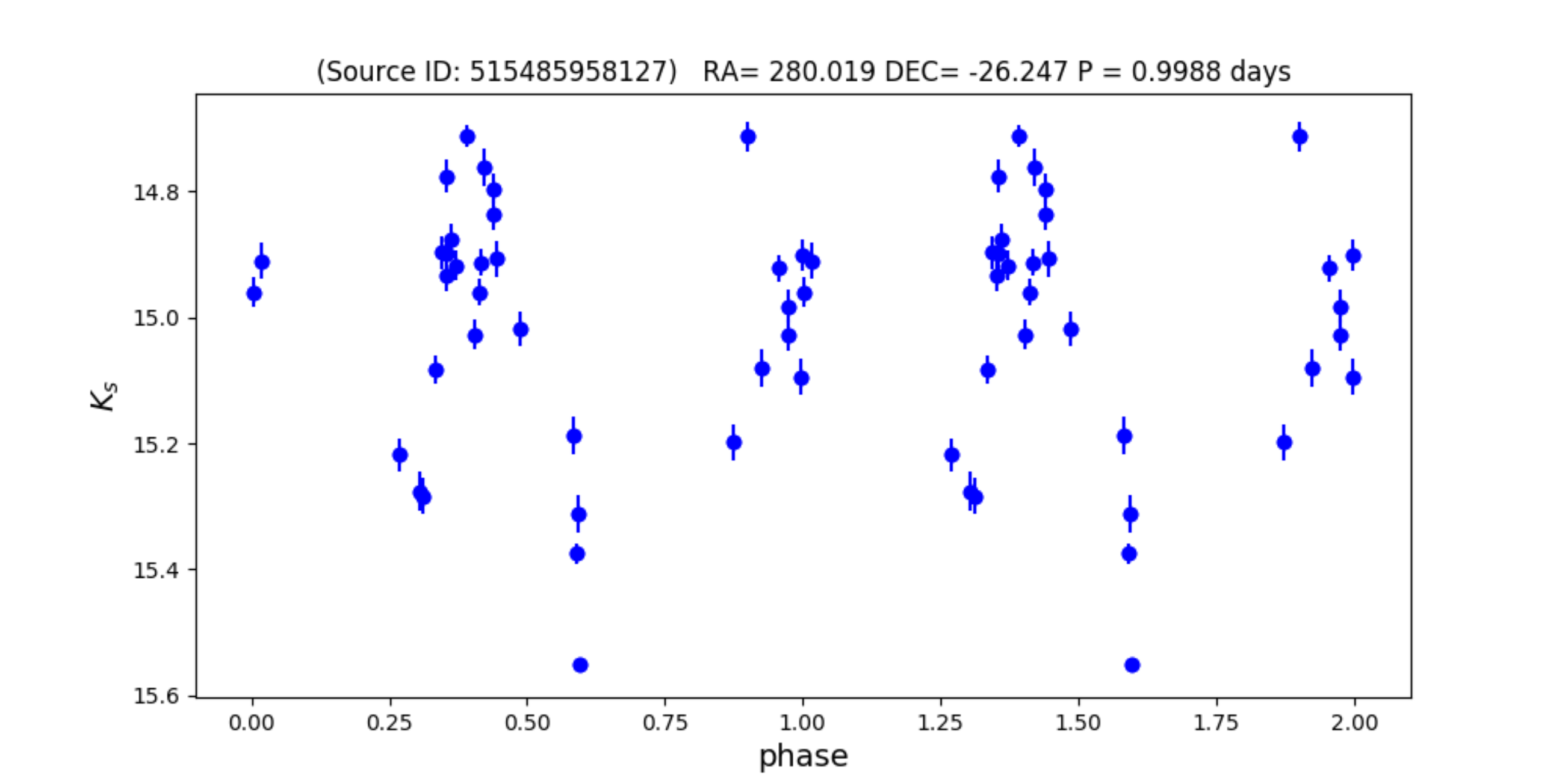} &   
    \includegraphics[width=.33\textwidth]{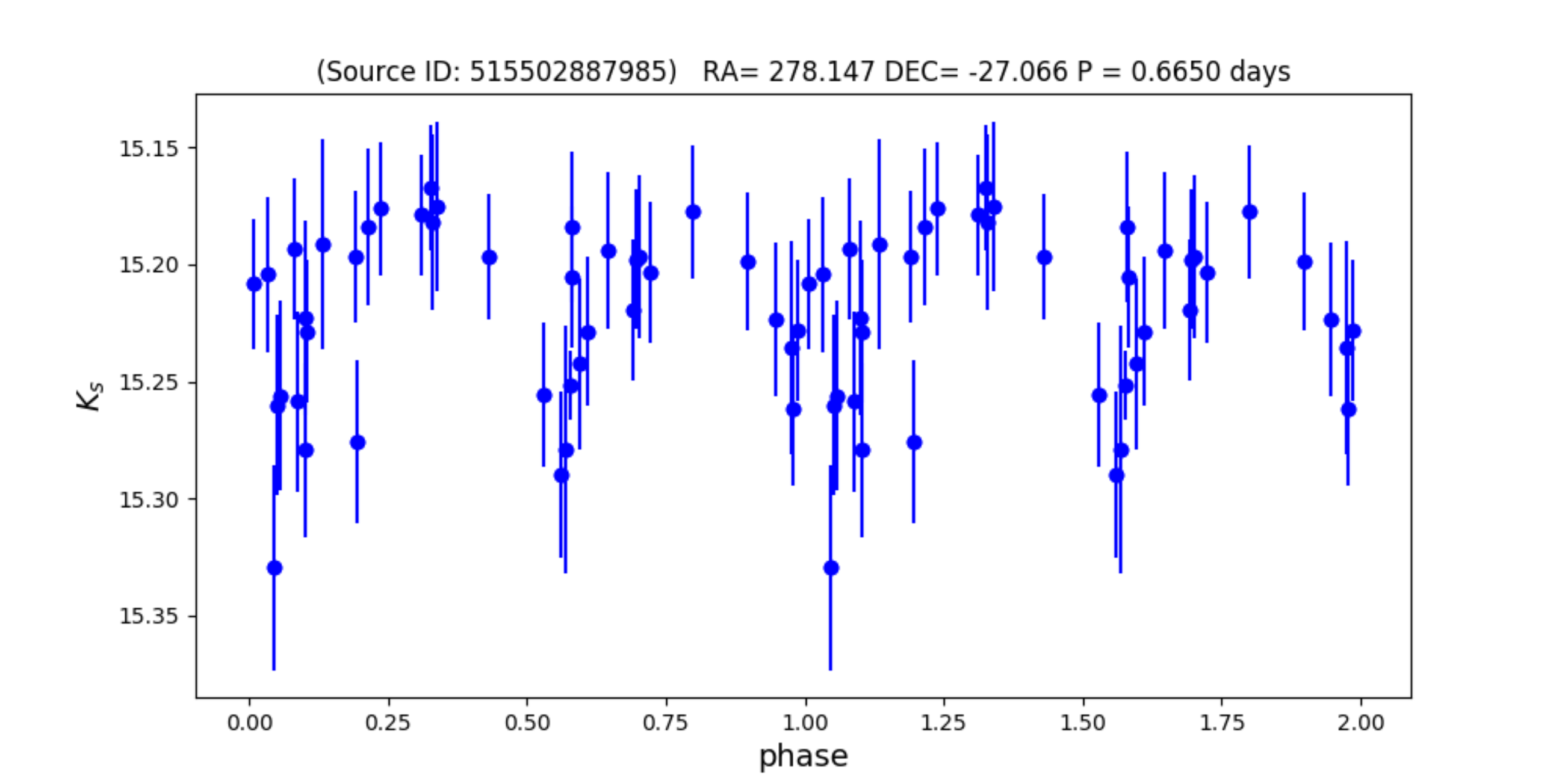} &    
    \includegraphics[width=.33\textwidth]{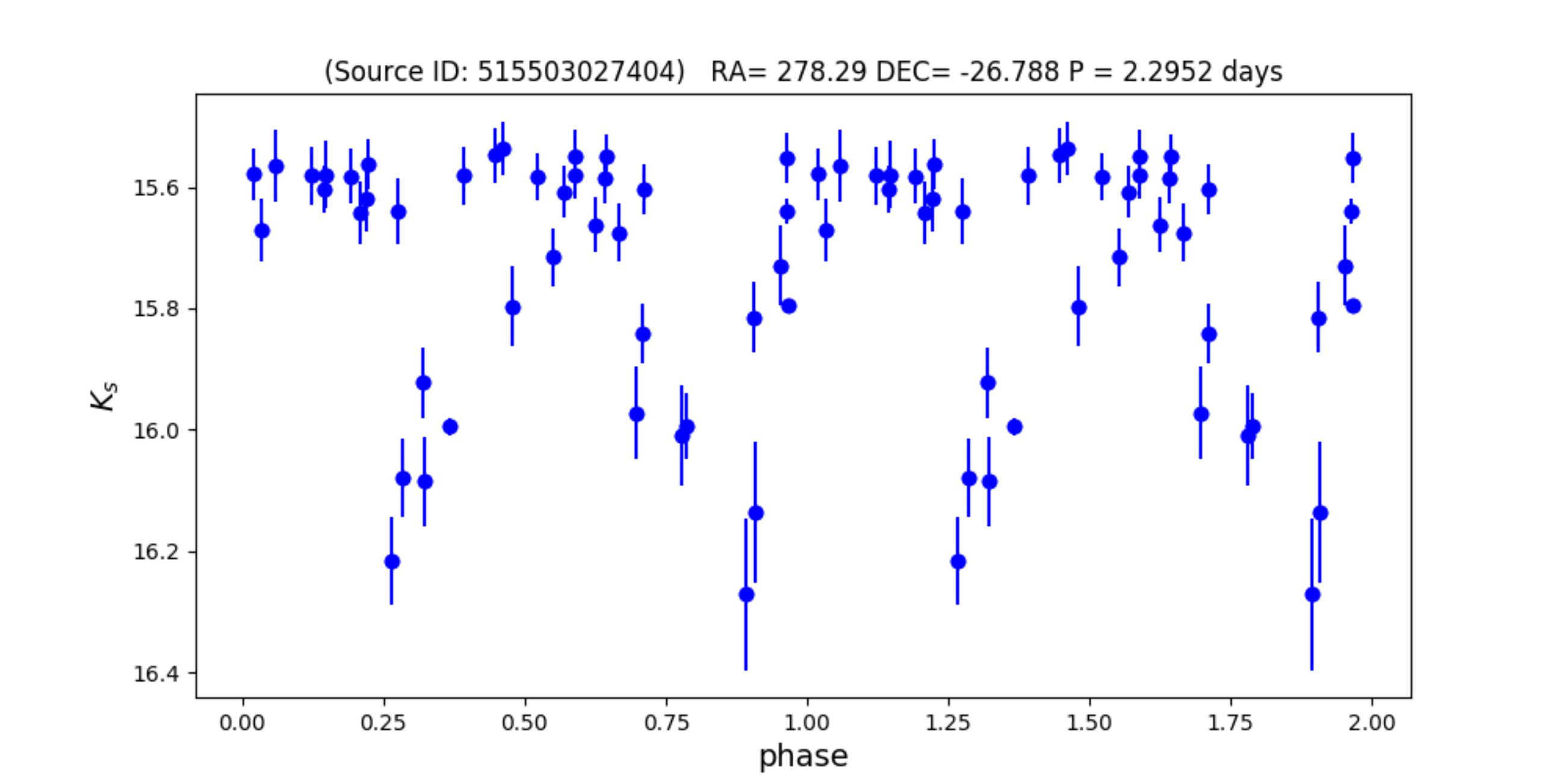} \\
    \includegraphics[width=.33\textwidth]{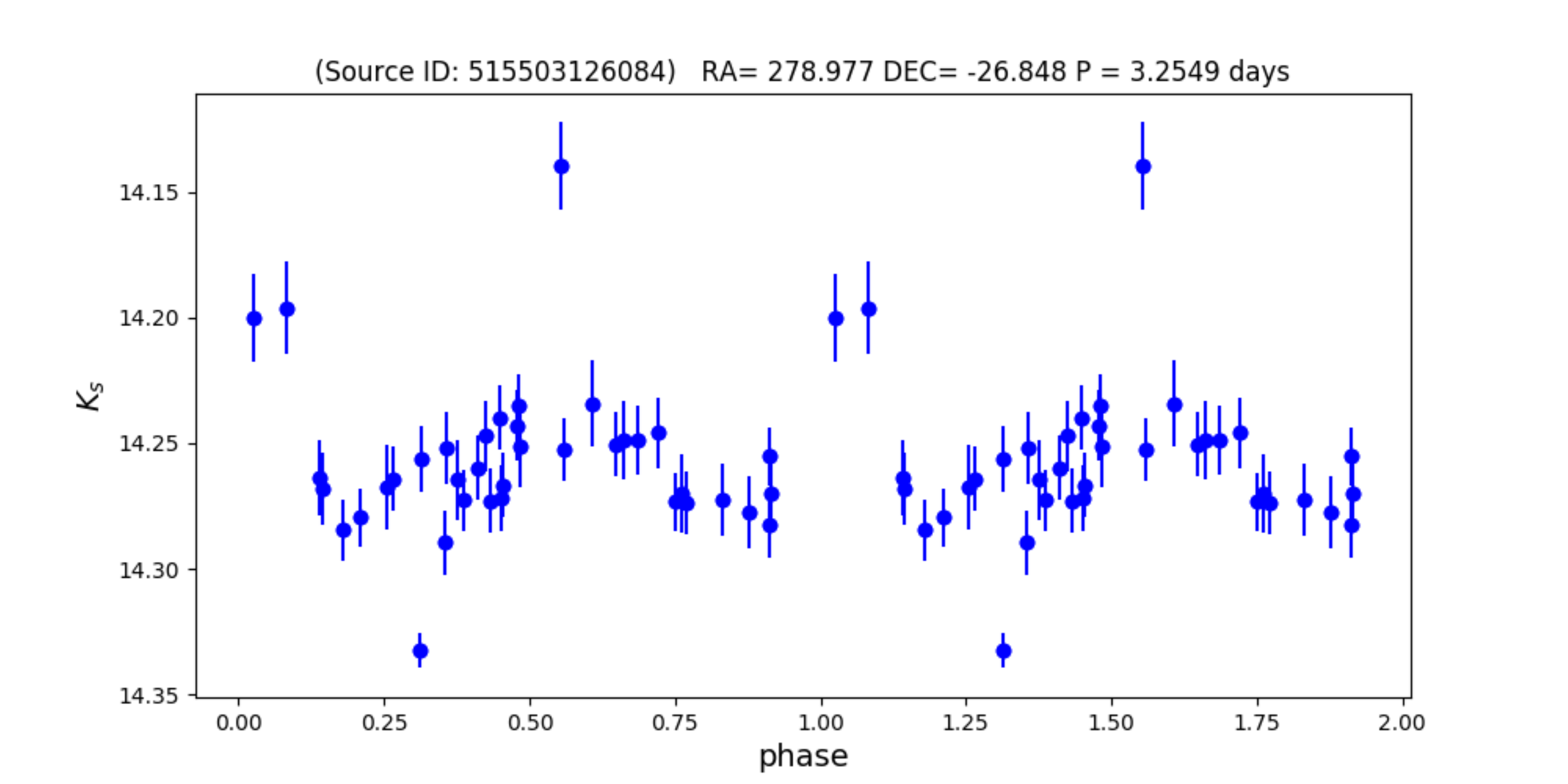} &
    \includegraphics[width=.33\textwidth]{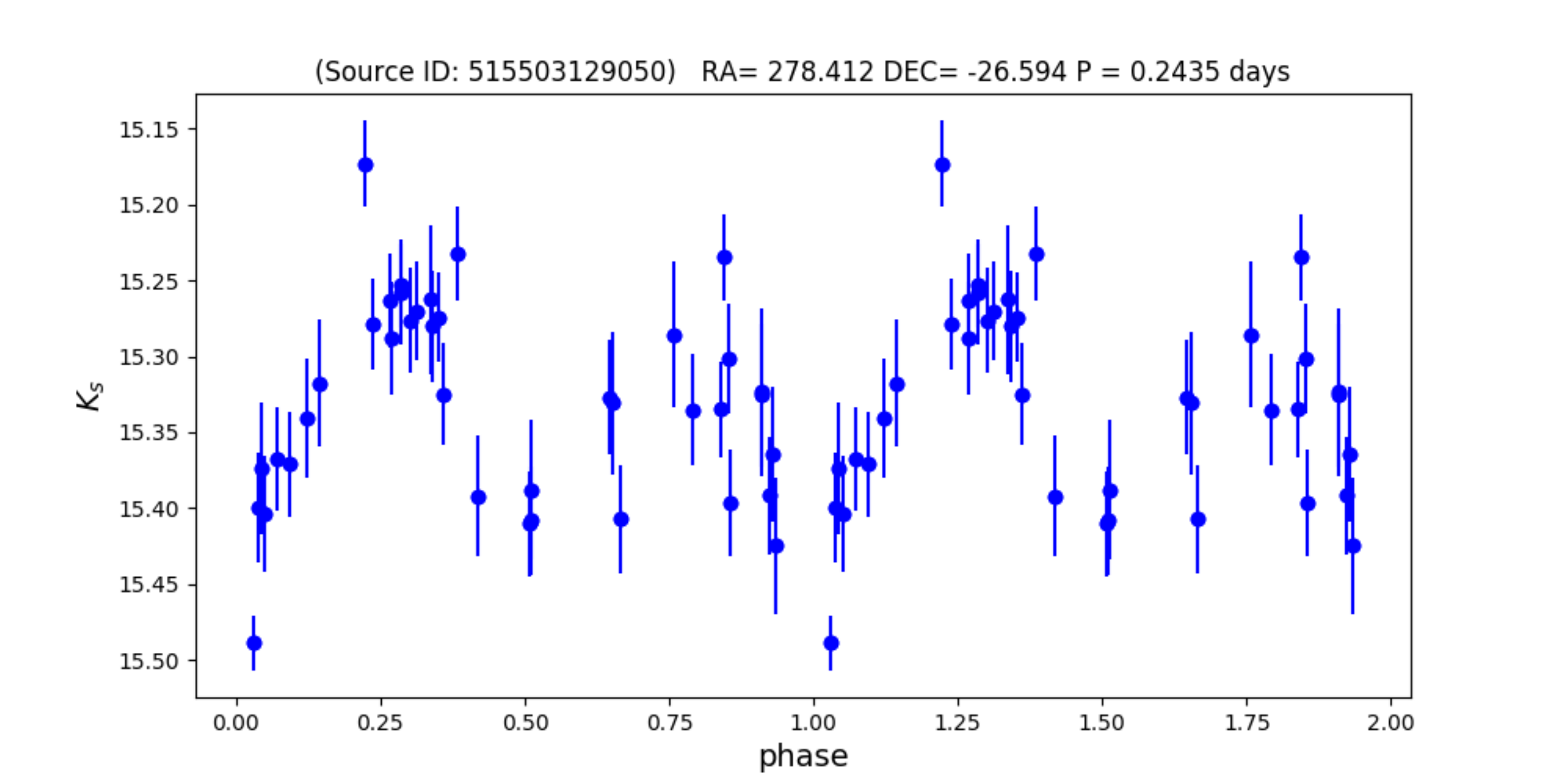} &
    \includegraphics[width=.33\textwidth]{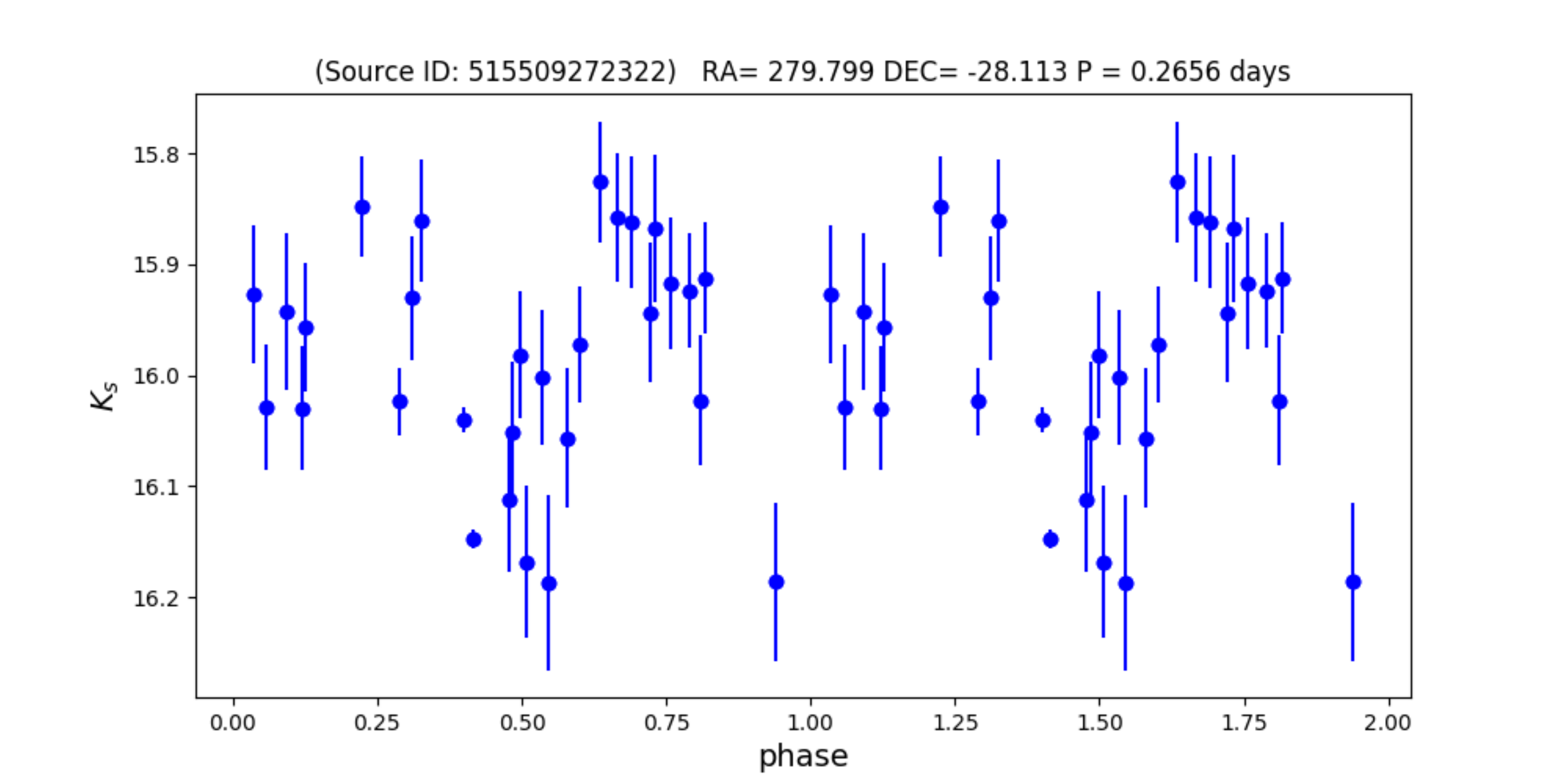} \\
    \includegraphics[width=.33\textwidth]{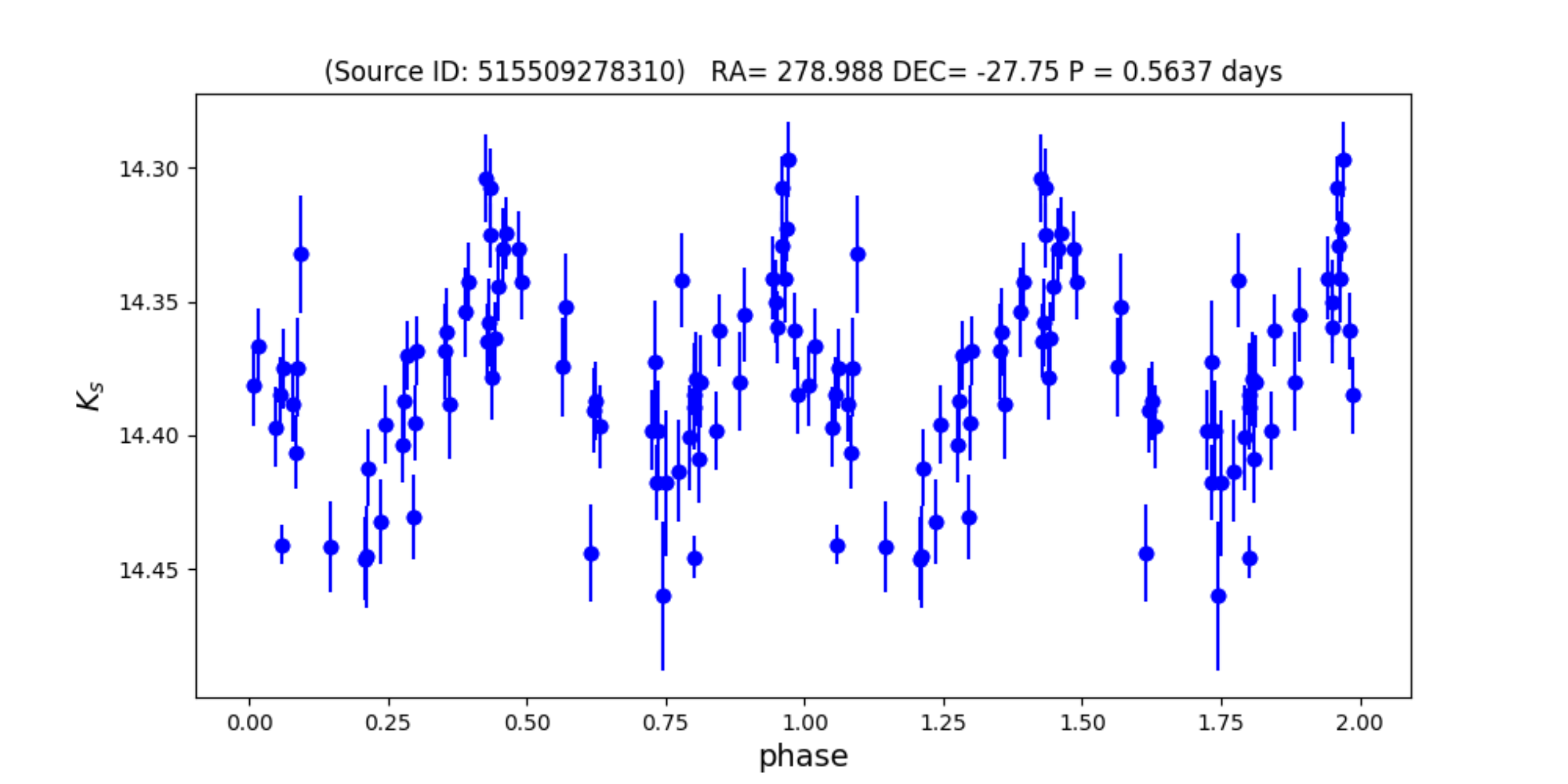} &
    \includegraphics[width=.33\textwidth]{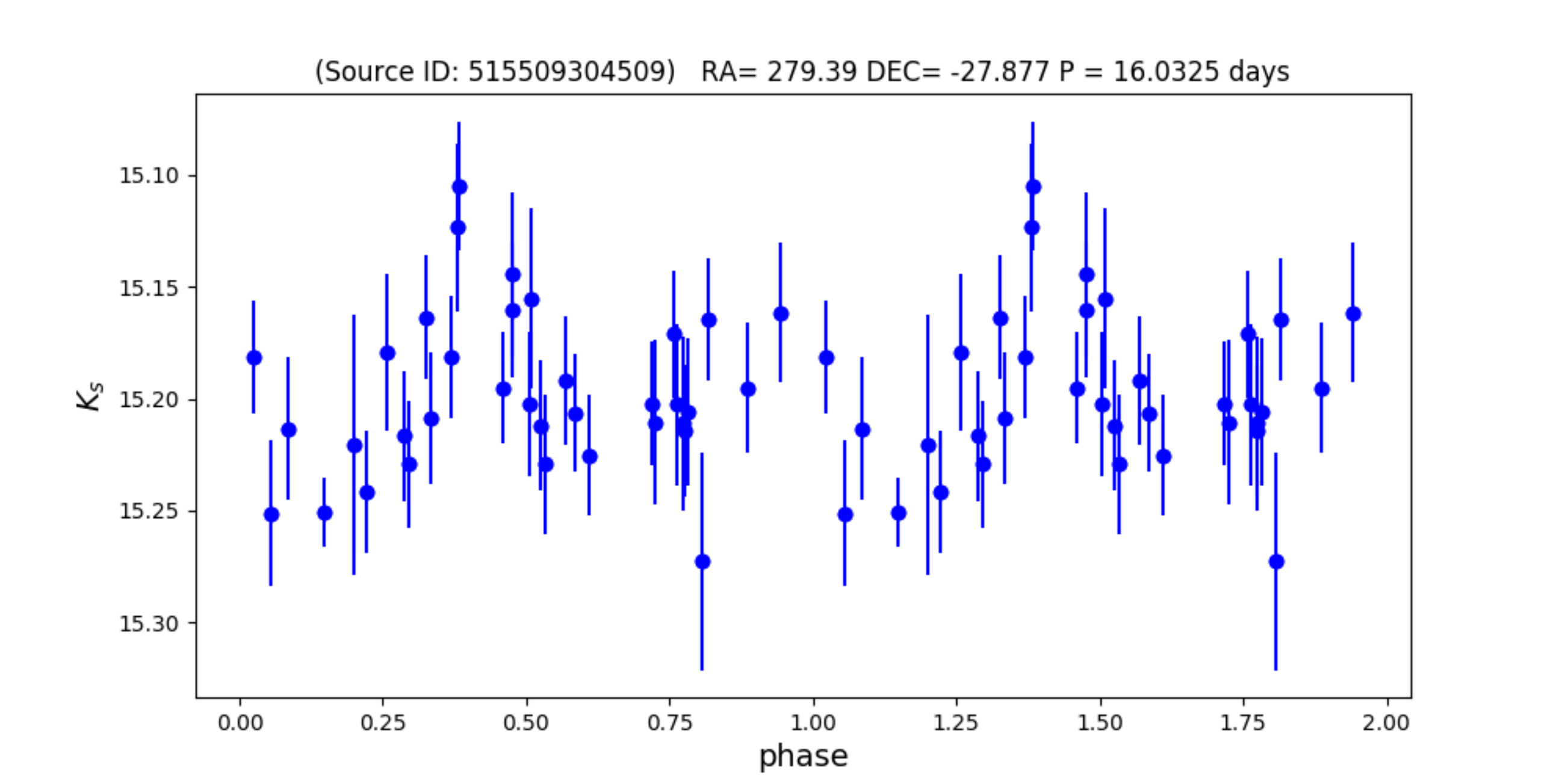} &
    \includegraphics[width=.33\textwidth]{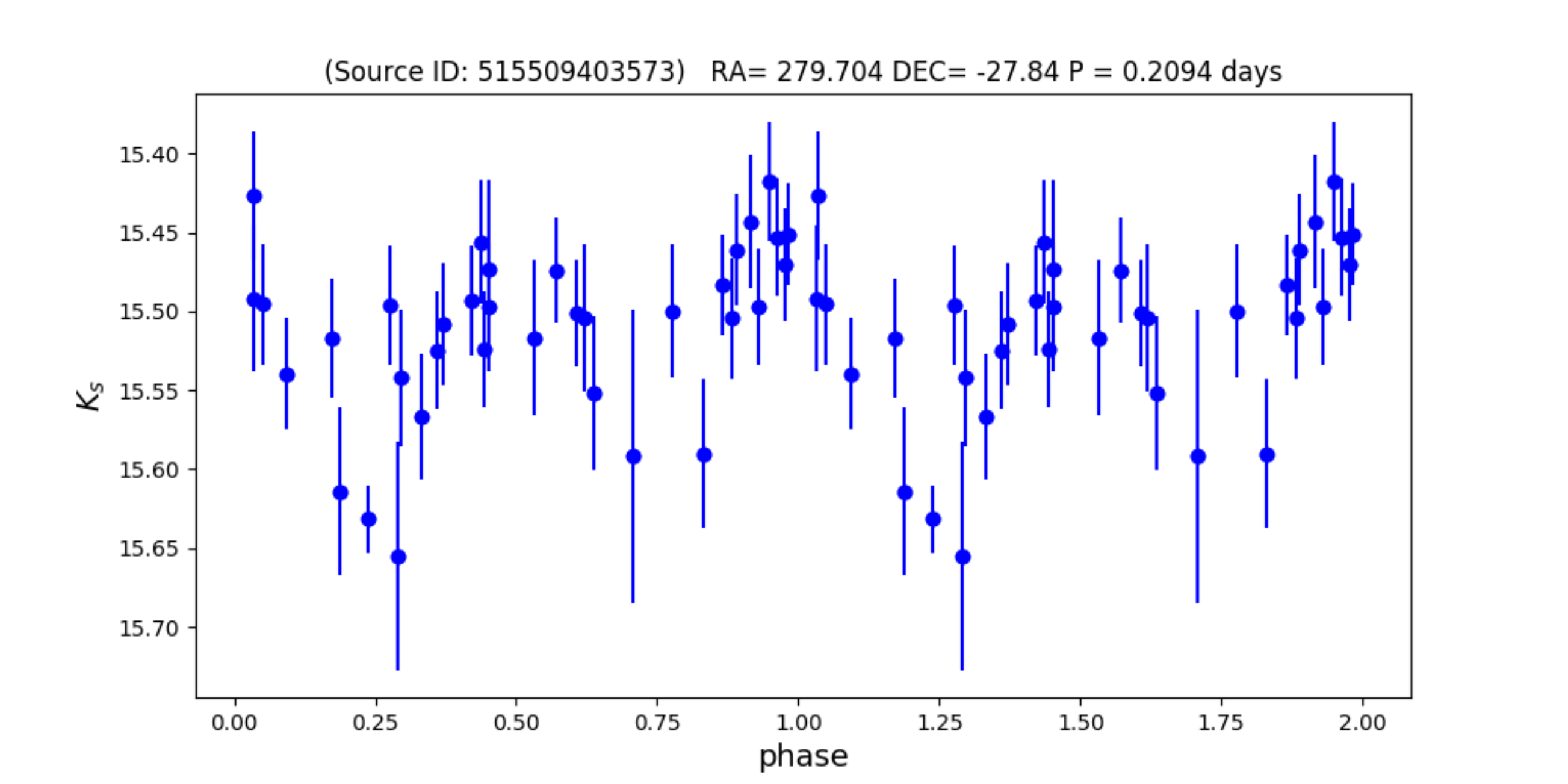} \\
    \includegraphics[width=.33\textwidth]{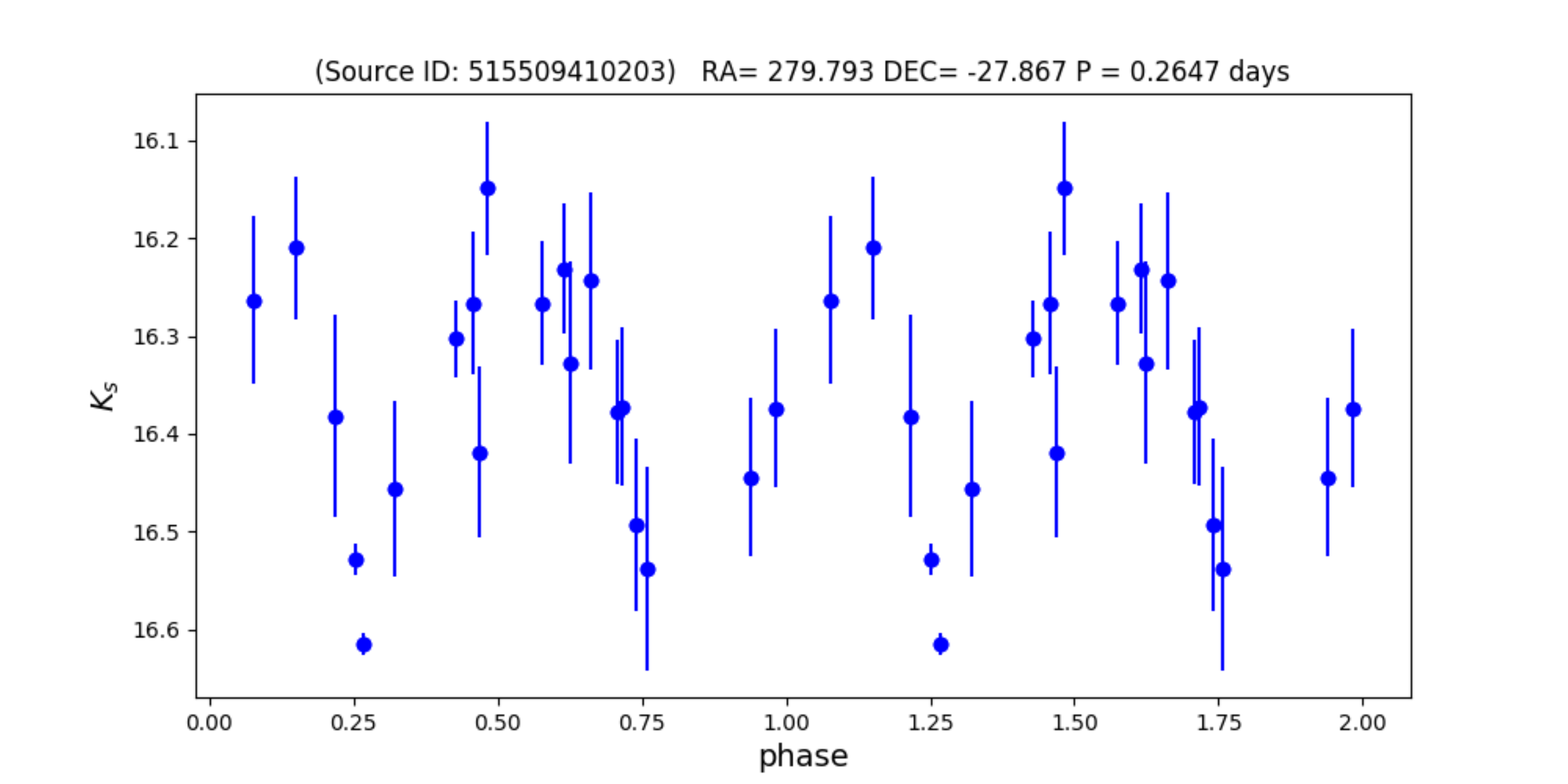} &
    \includegraphics[width=.33\textwidth]{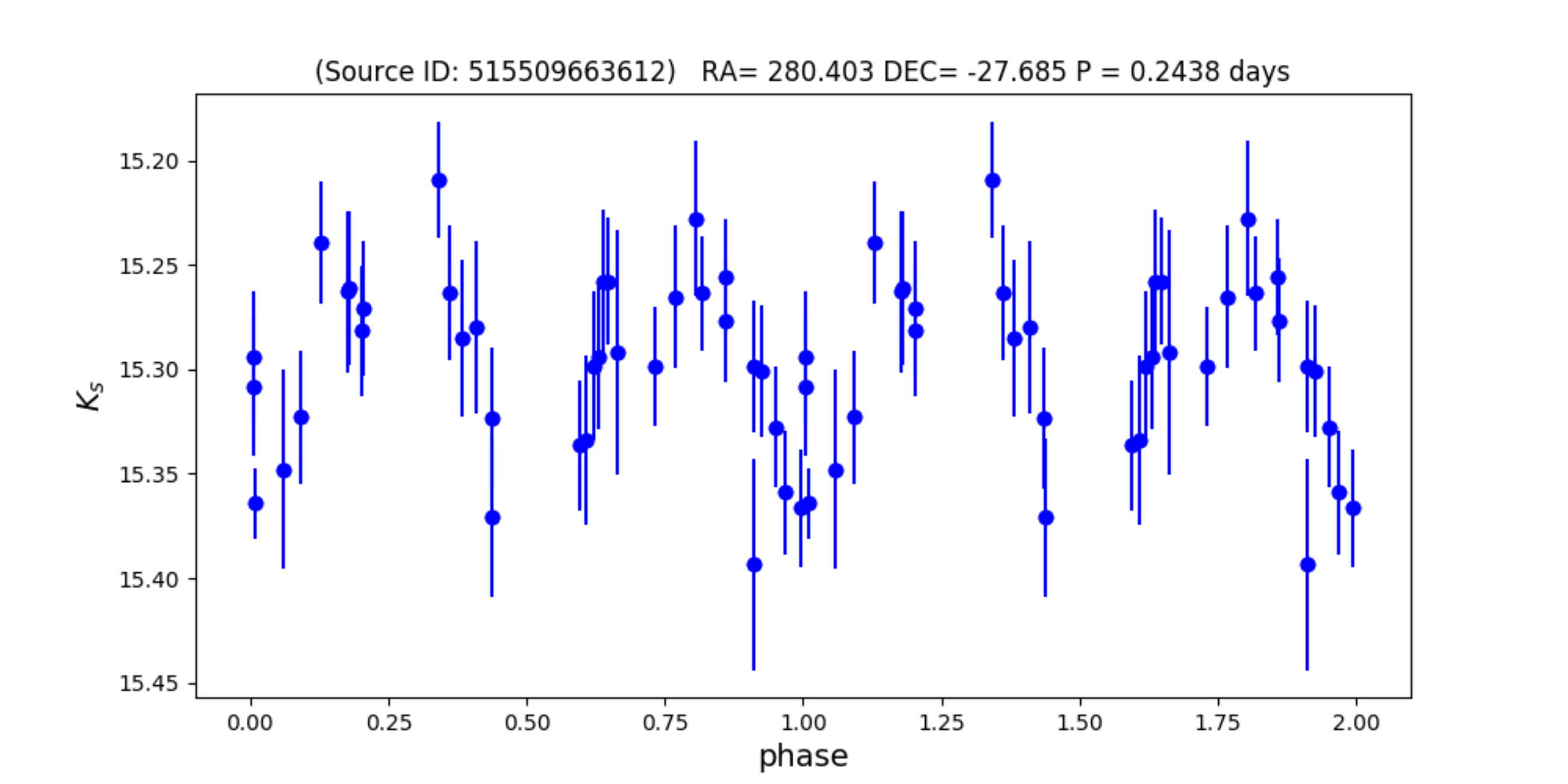} &
    \includegraphics[width=.33\textwidth]{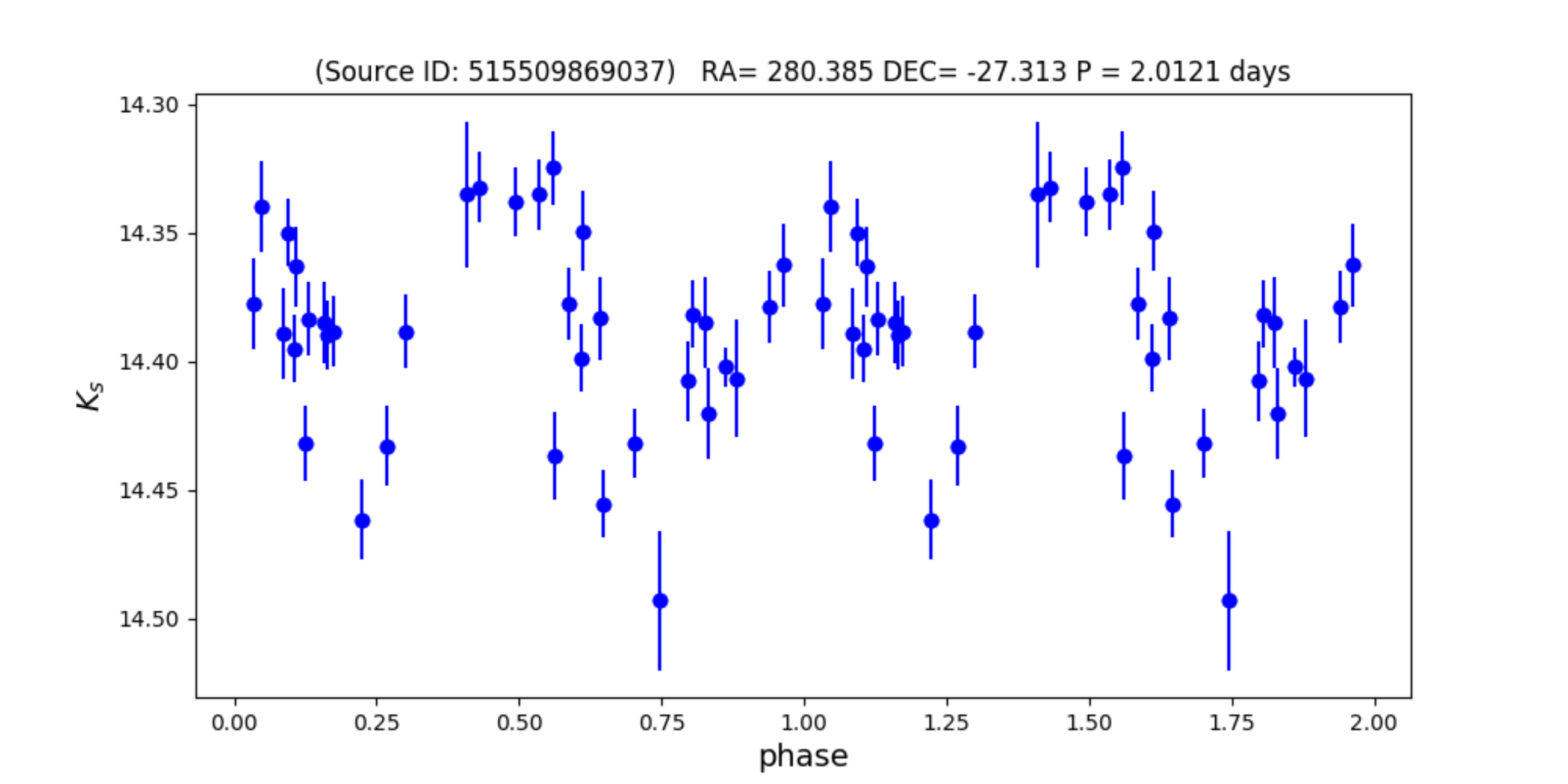} \\
  \end{tabular}
  \caption{BHB stars light curves corresponding to $2^{nd}$ category eclipsing binaries.}
  \label{fig:bhbB1}
\end{figure*}

\begin{figure*}[htb]
\centering
  \begin{tabular}{@{}ccc@{}}            
    \includegraphics[width=.33\textwidth]{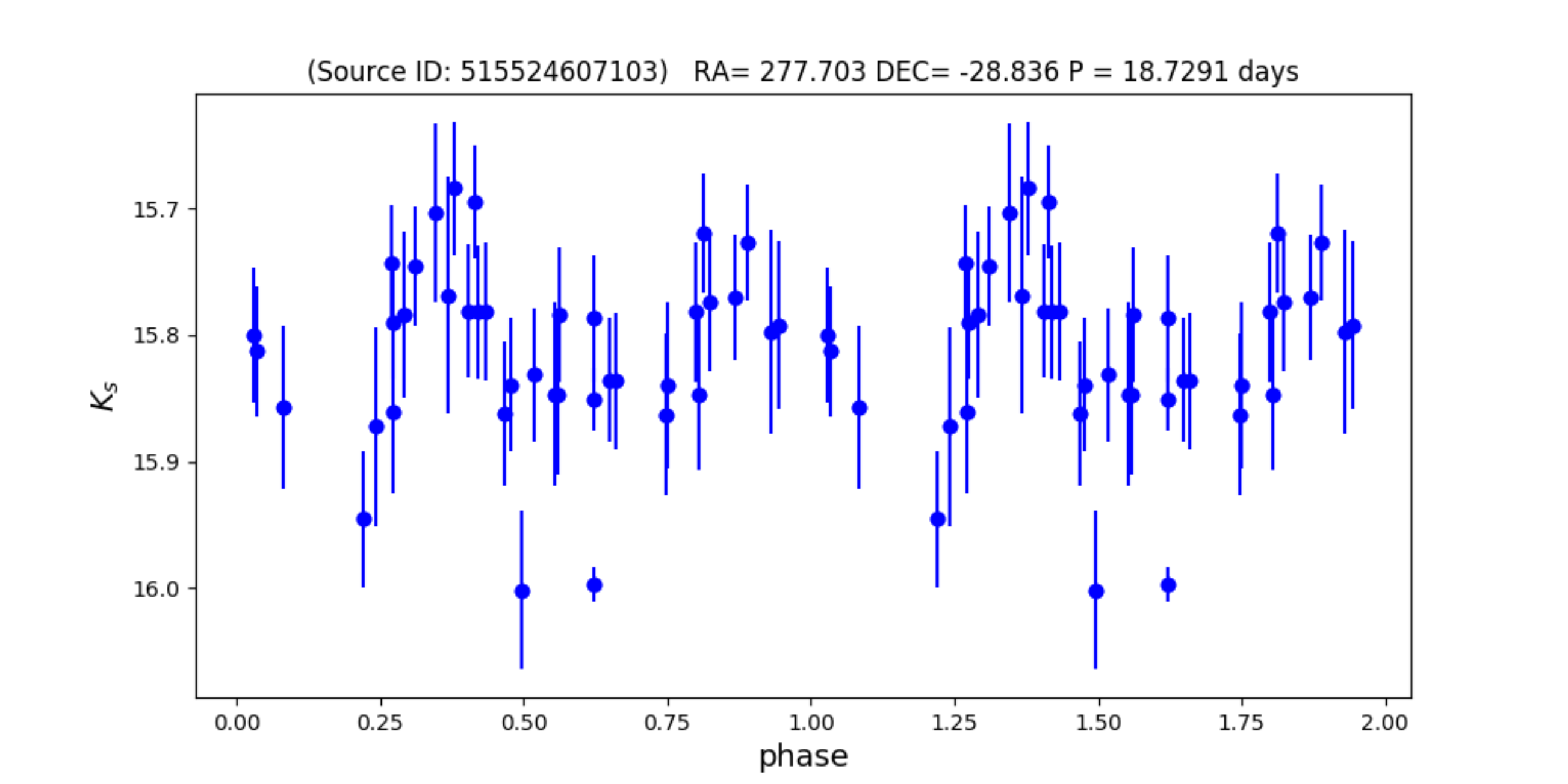} &
    \includegraphics[width=.33\textwidth]{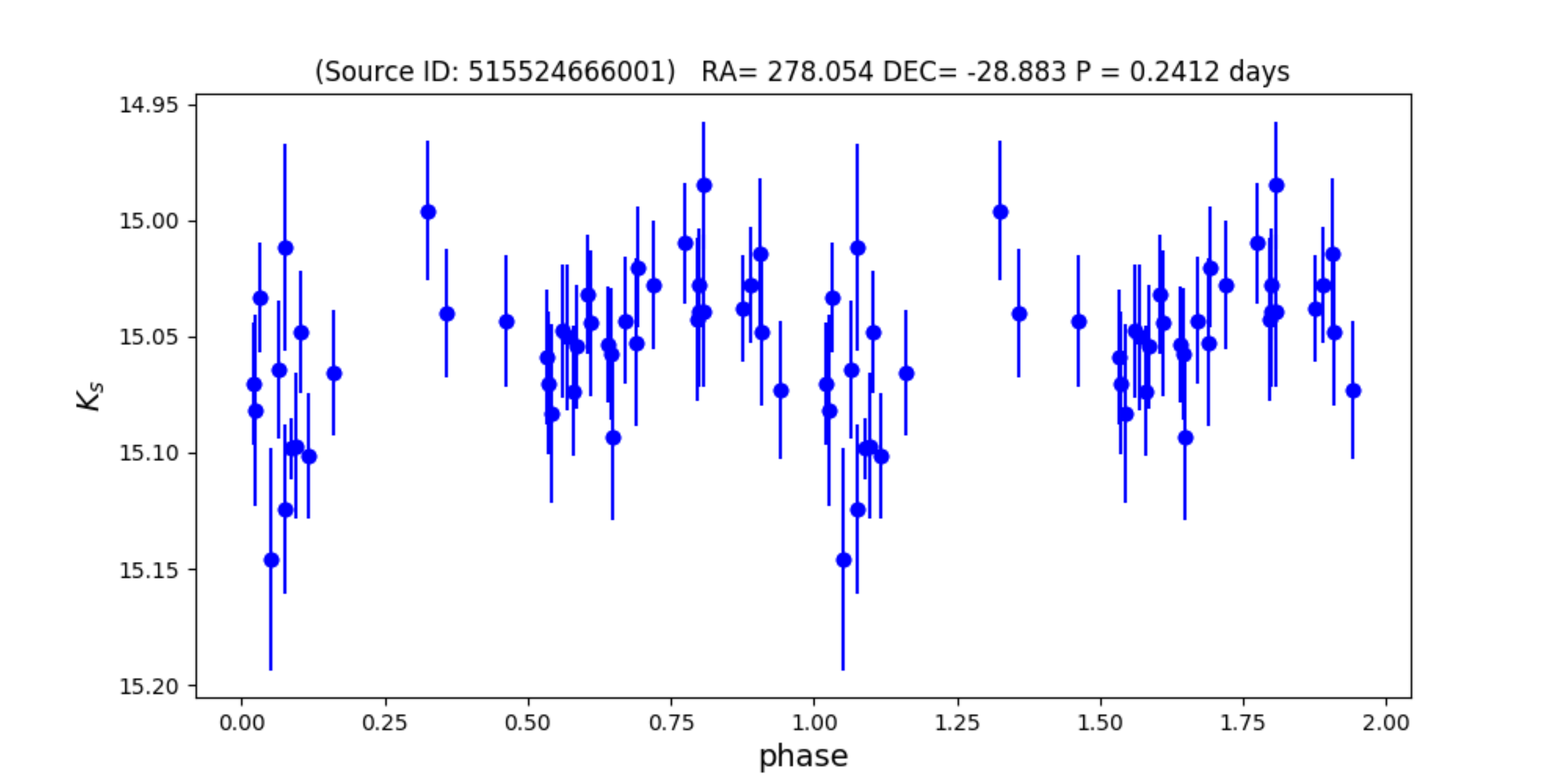} &
    \includegraphics[width=.33\textwidth]{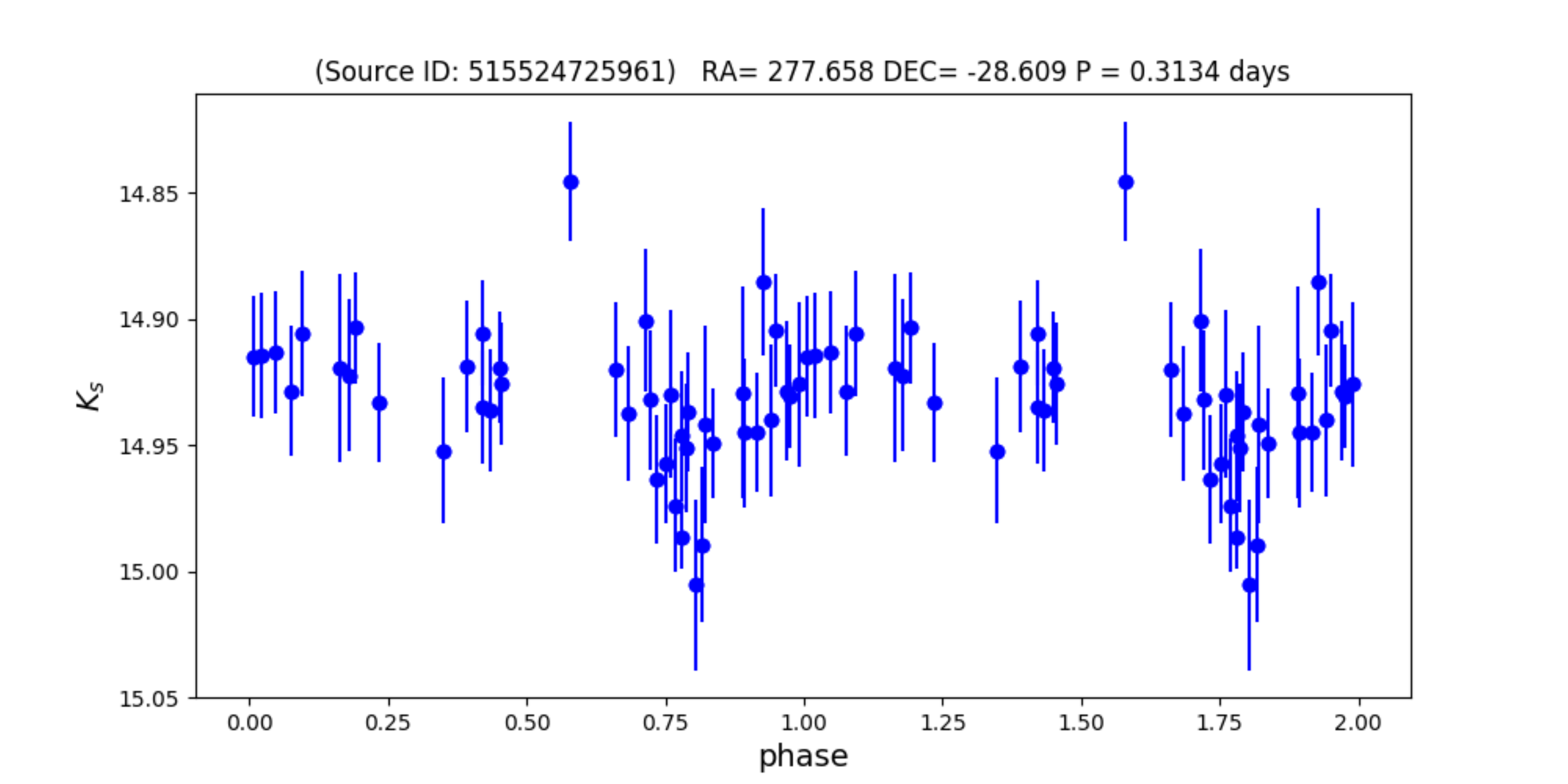} \\
    \includegraphics[width=.33\textwidth]{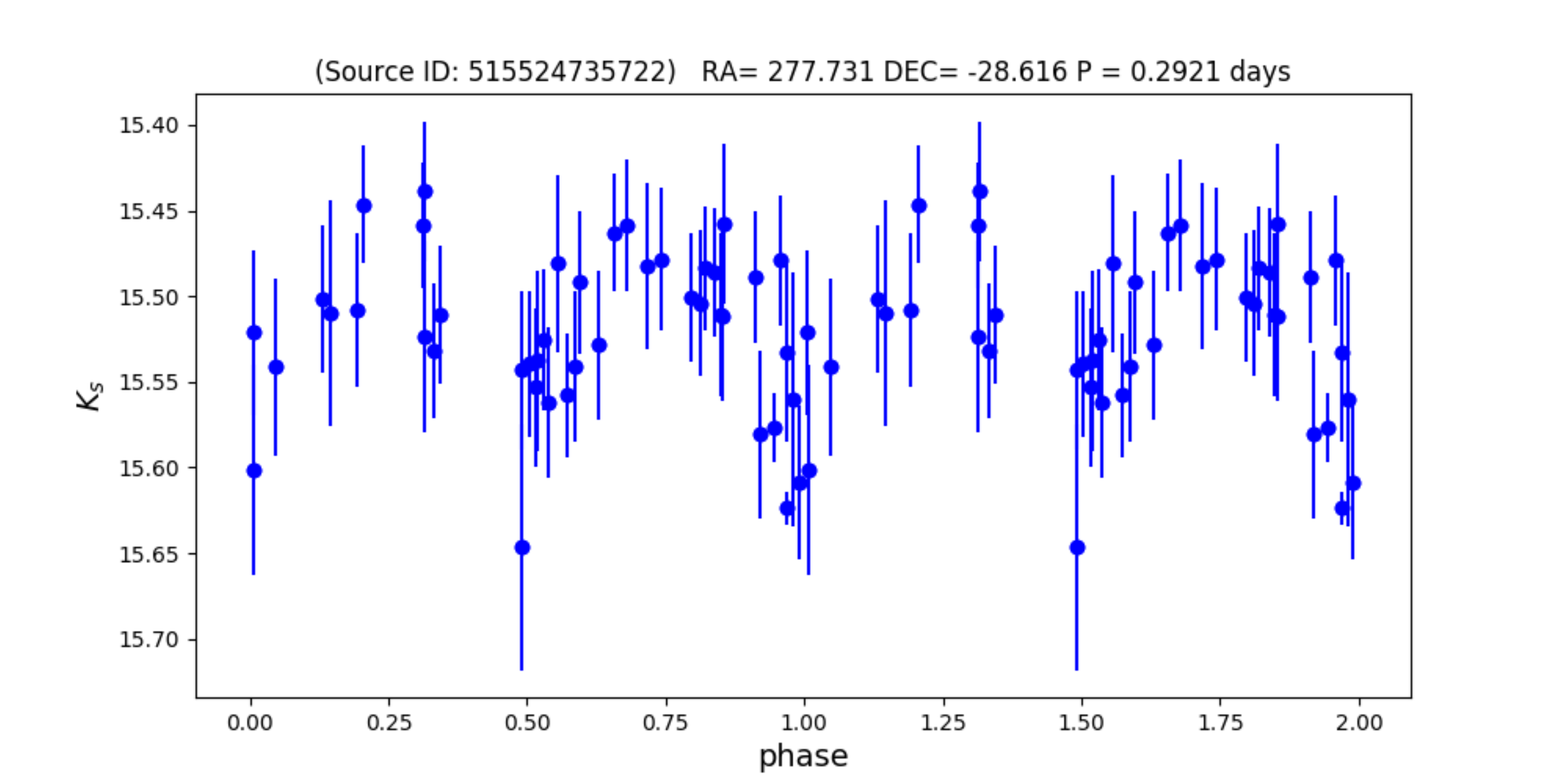} &
    \includegraphics[width=.33\textwidth]{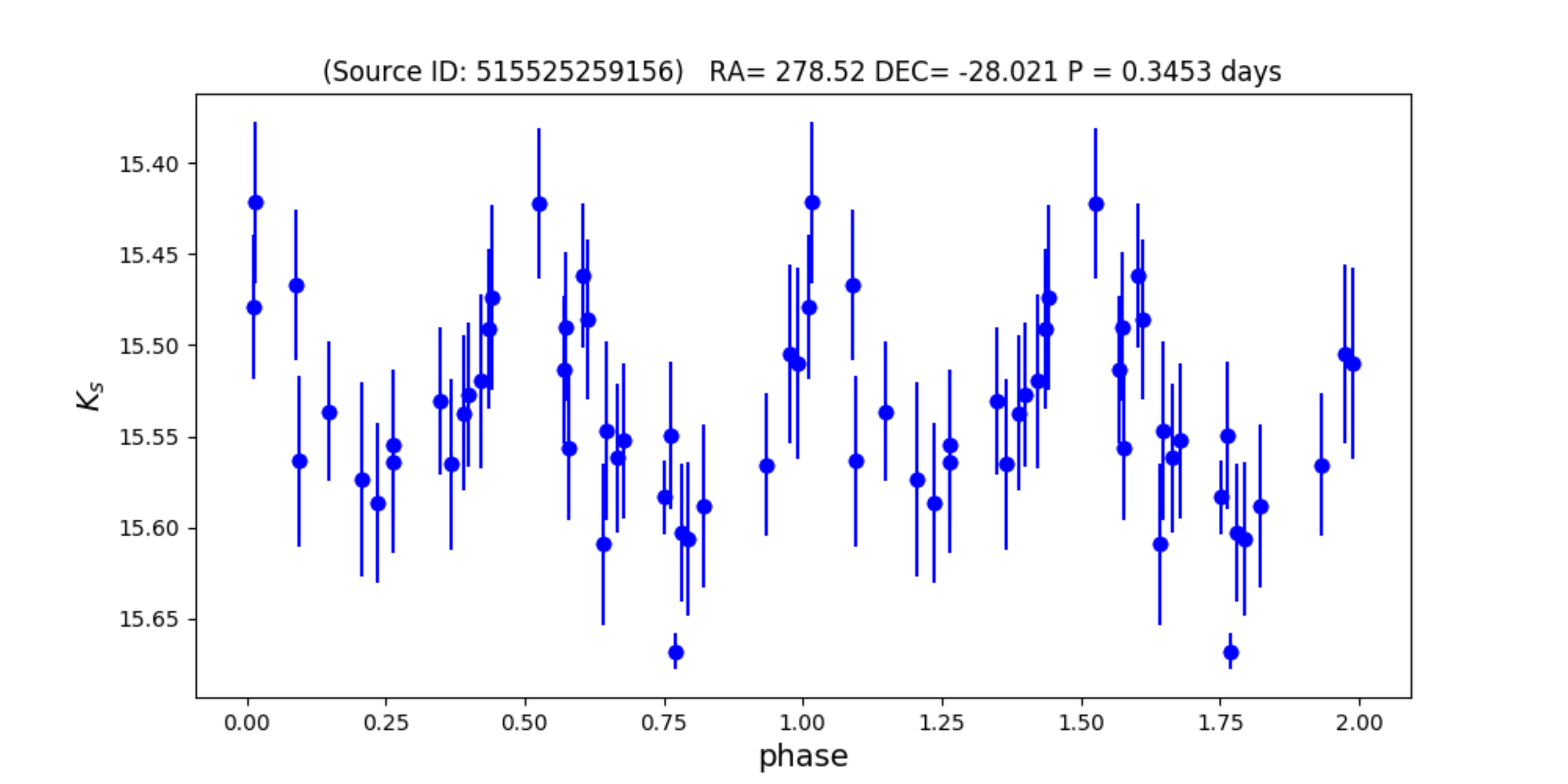} &
    \includegraphics[width=.33\textwidth]{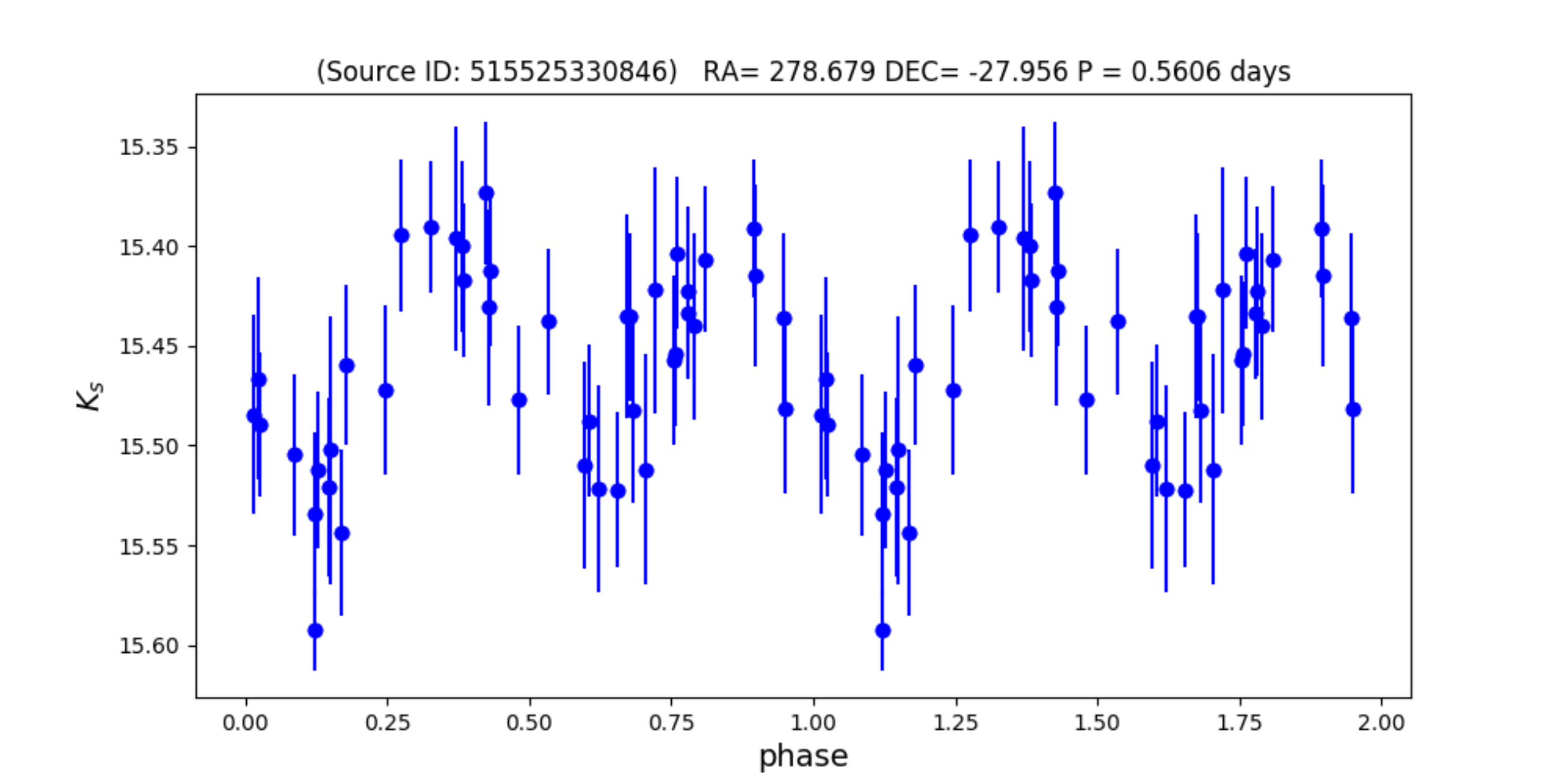} \\
    \includegraphics[width=.33\textwidth]{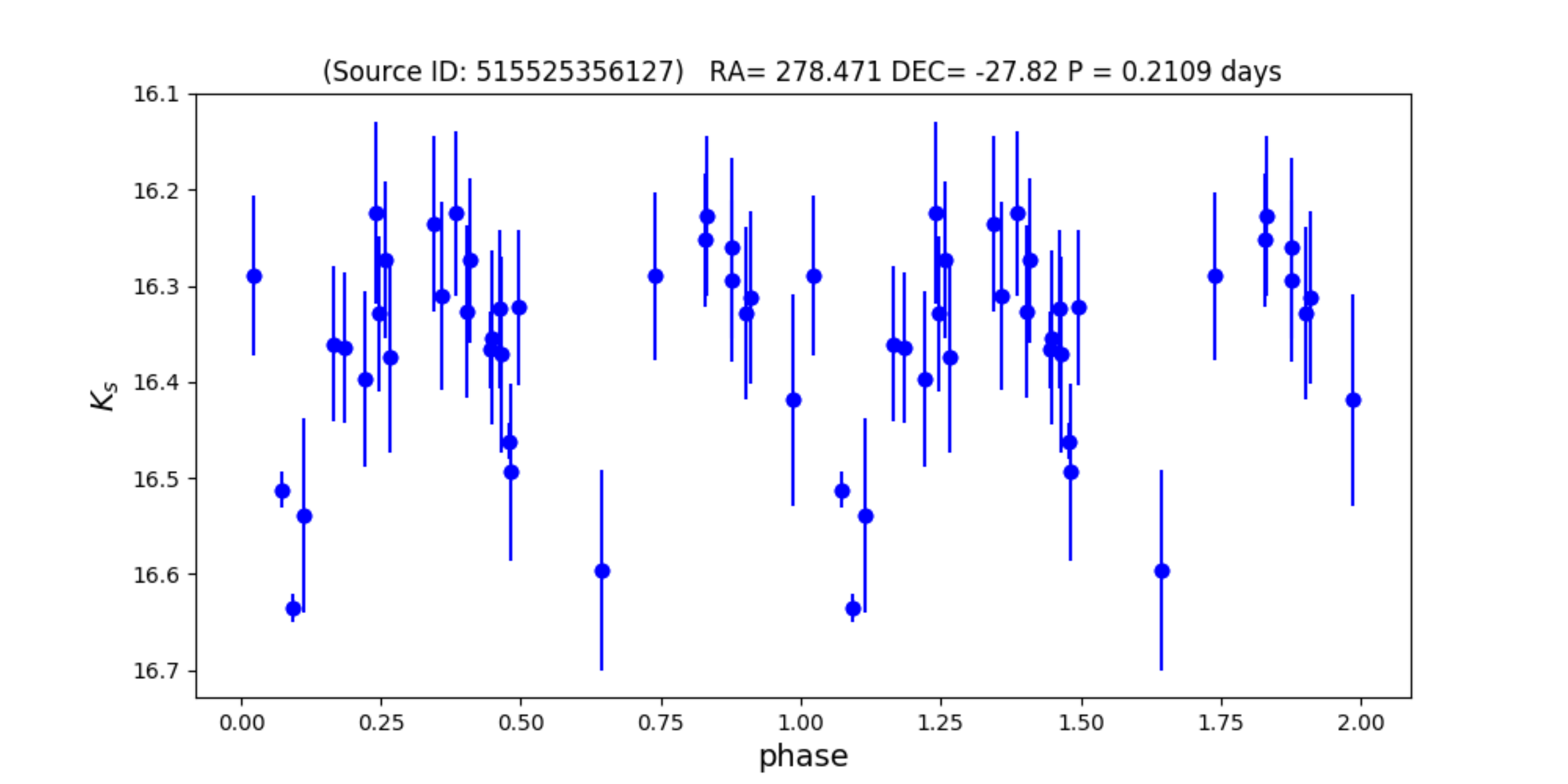} &
    \includegraphics[width=.33\textwidth]{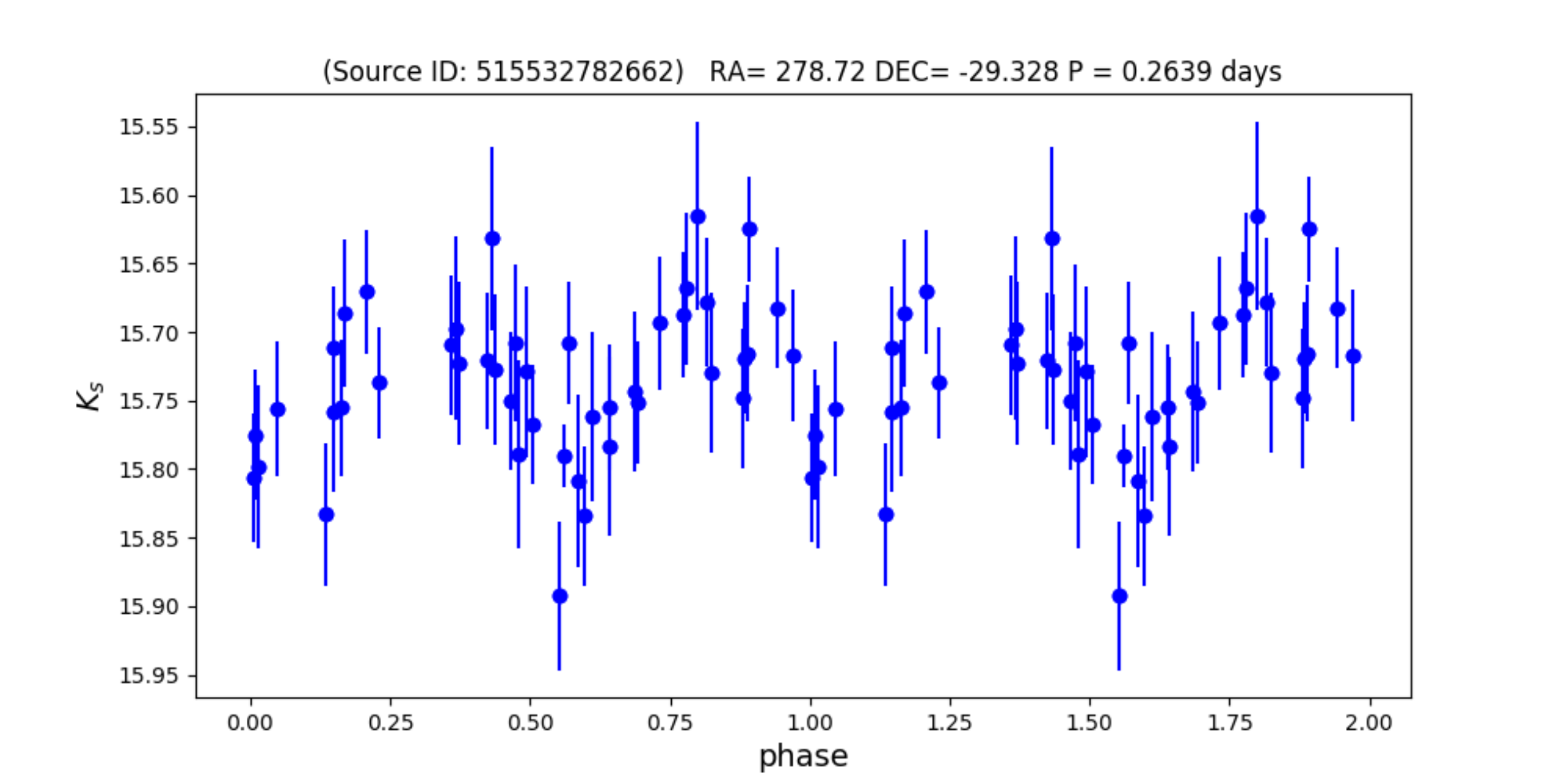} &
    \includegraphics[width=.33\textwidth]{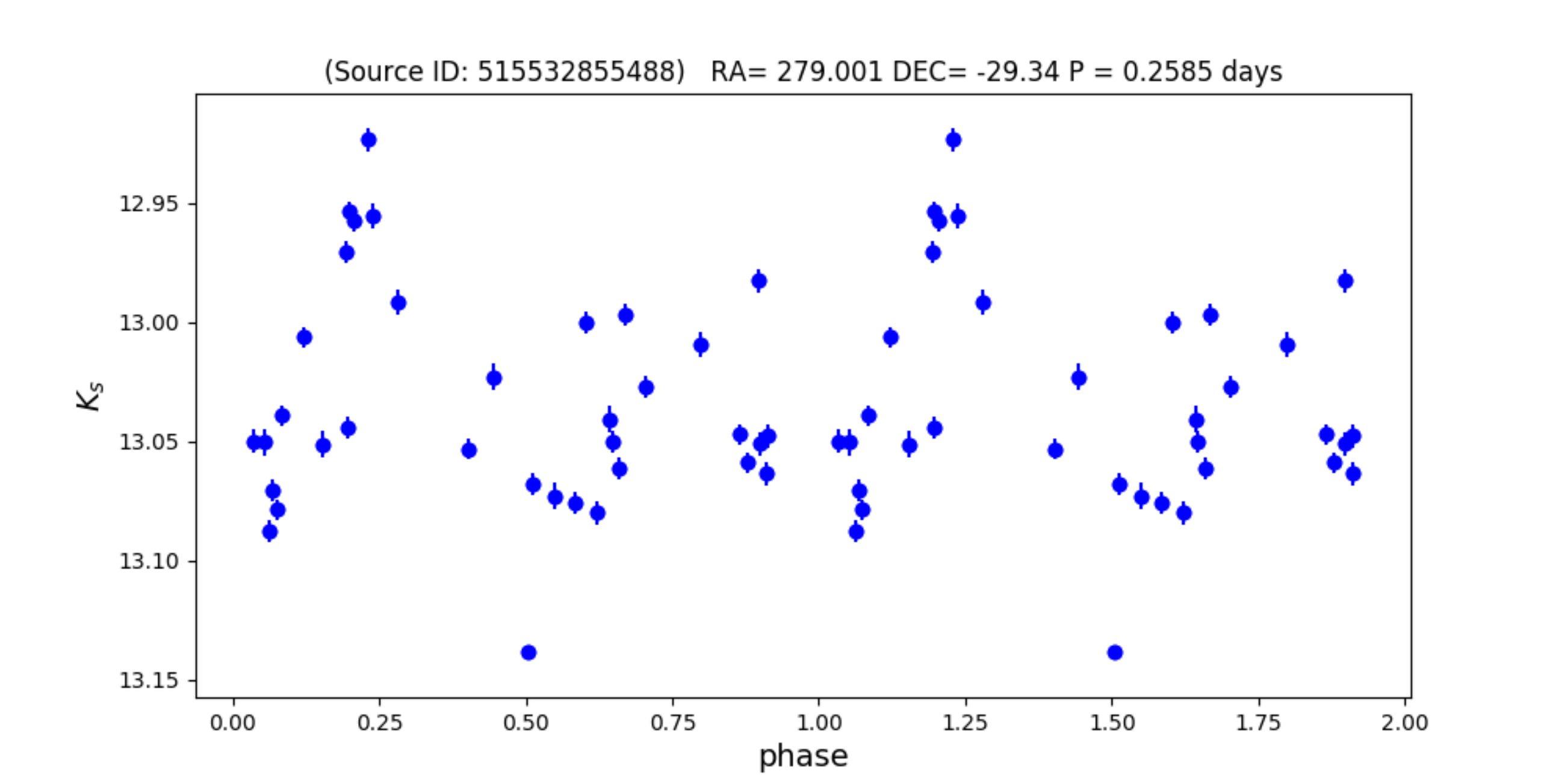} \\
    \includegraphics[width=.33\textwidth]{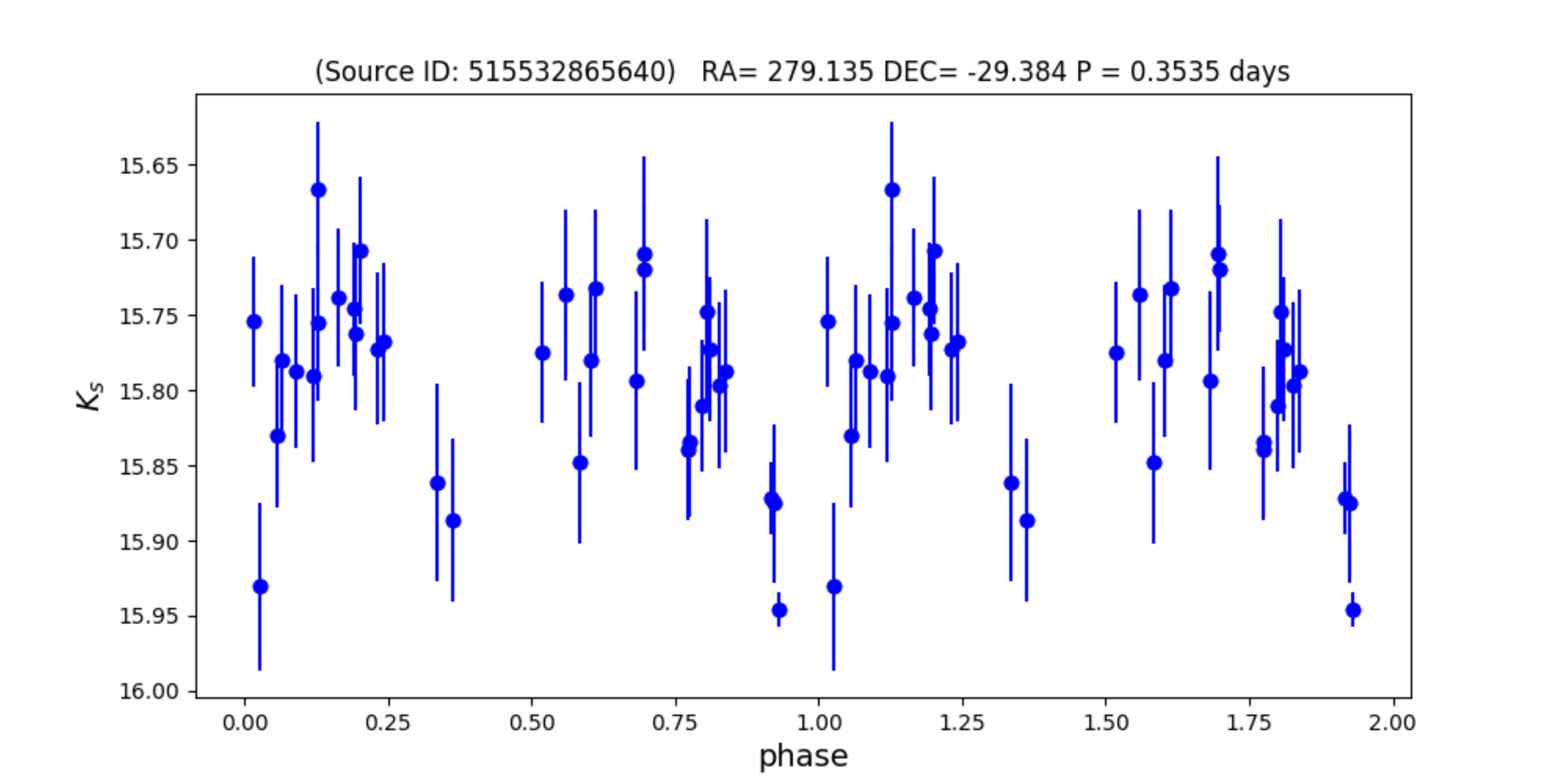} &
    \includegraphics[width=.33\textwidth]{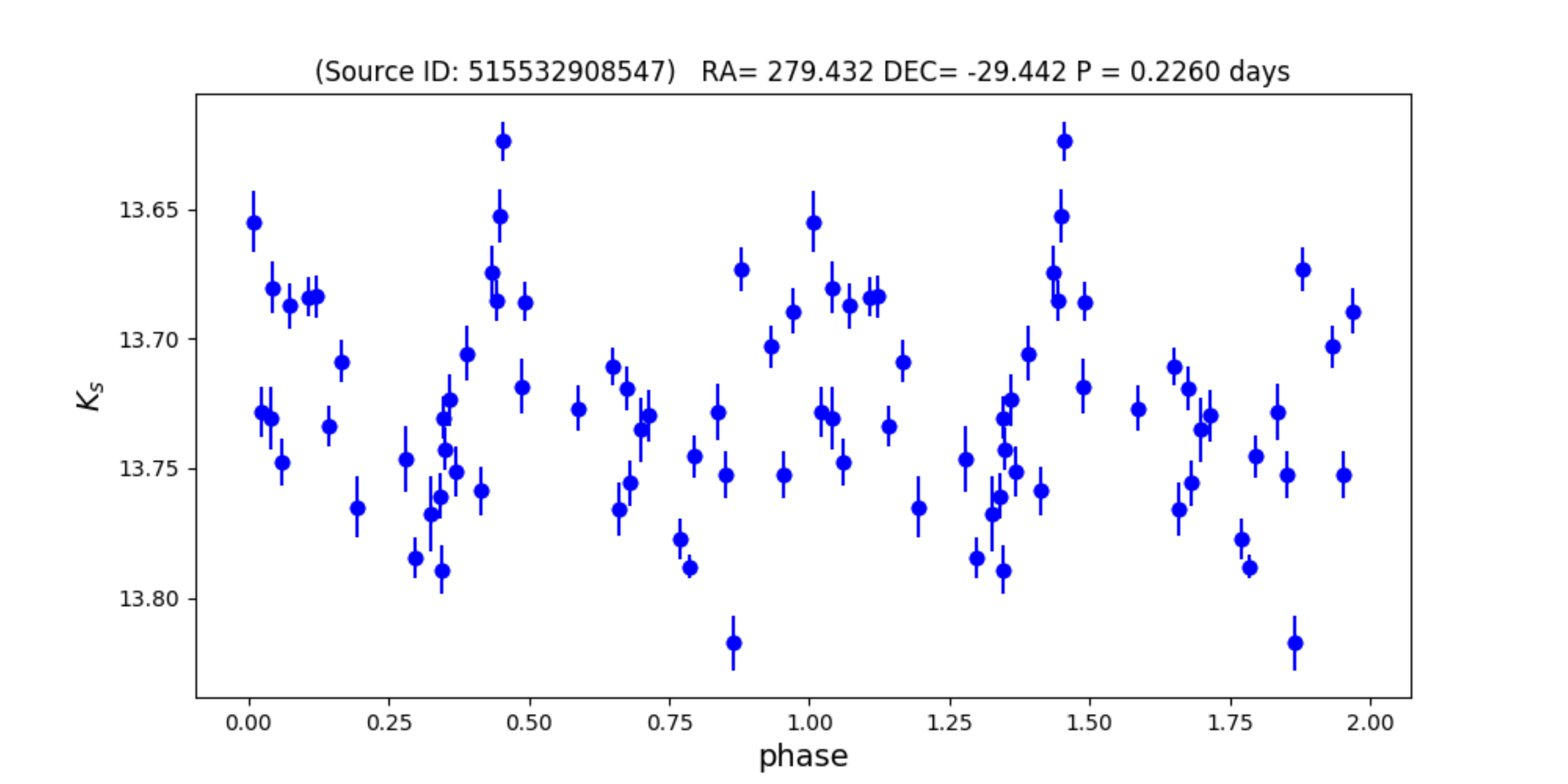} &
    \includegraphics[width=.33\textwidth]{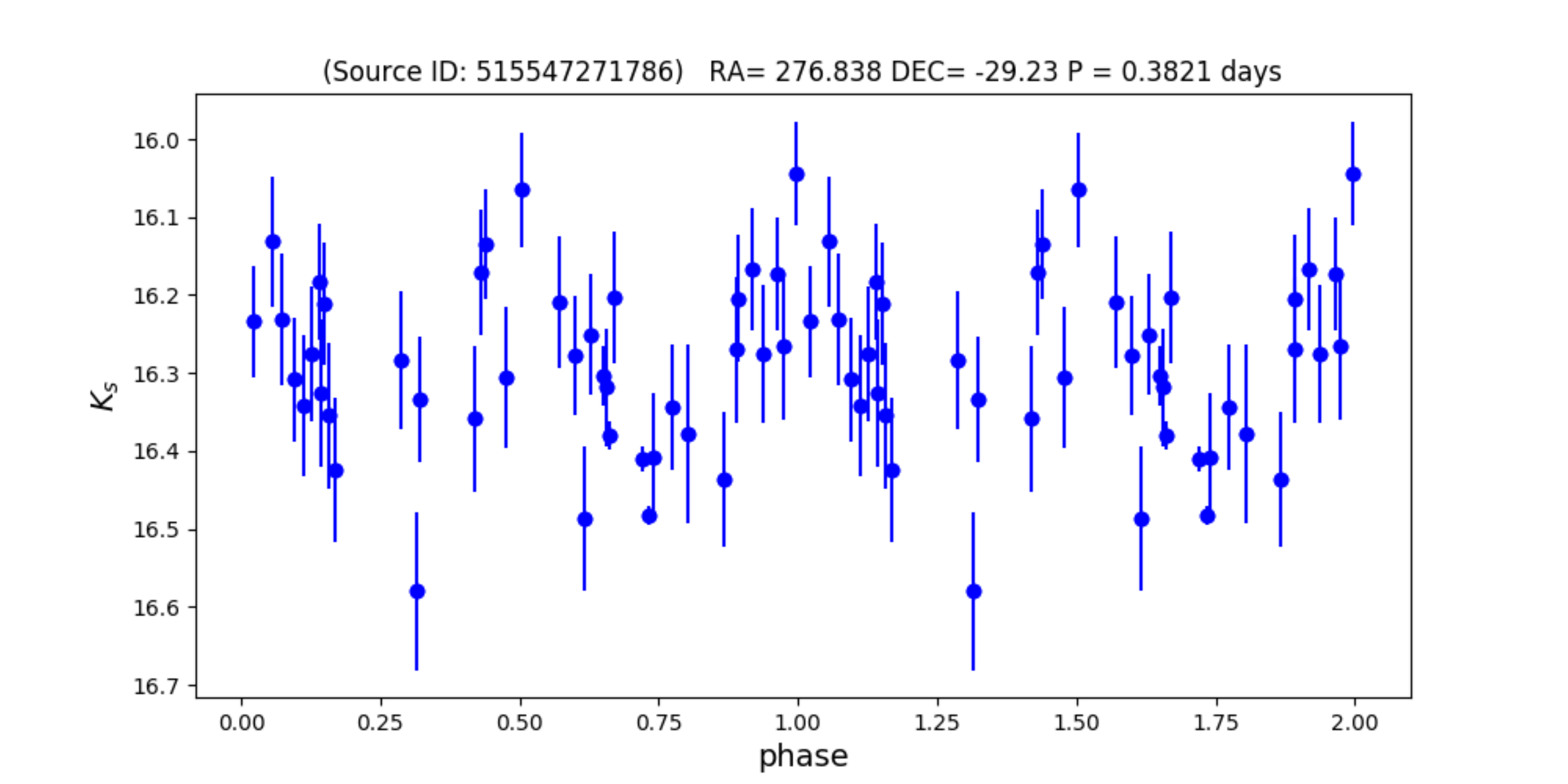} \\    
    \includegraphics[width=.33\textwidth]{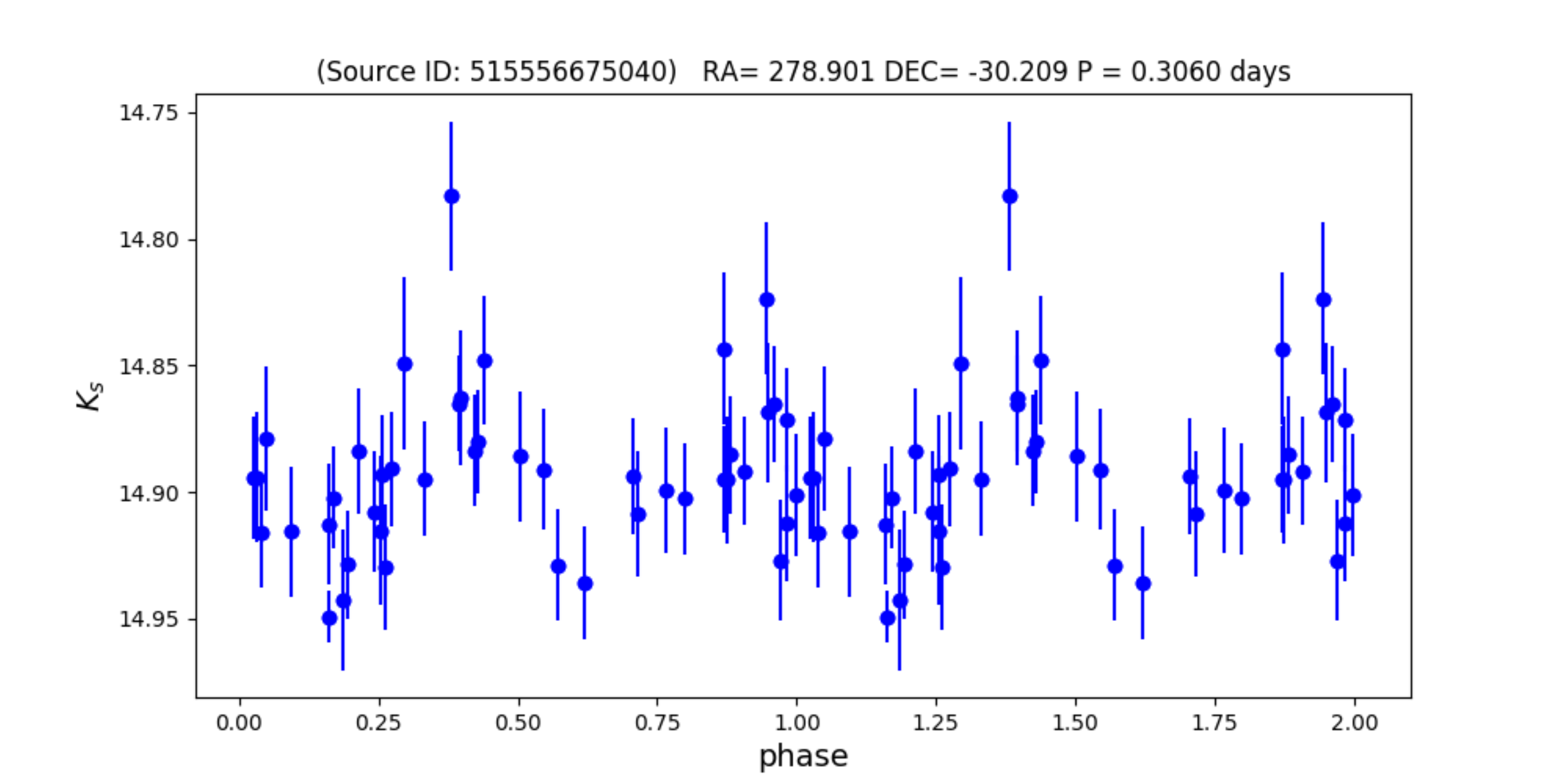} &
    \includegraphics[width=.33\textwidth]{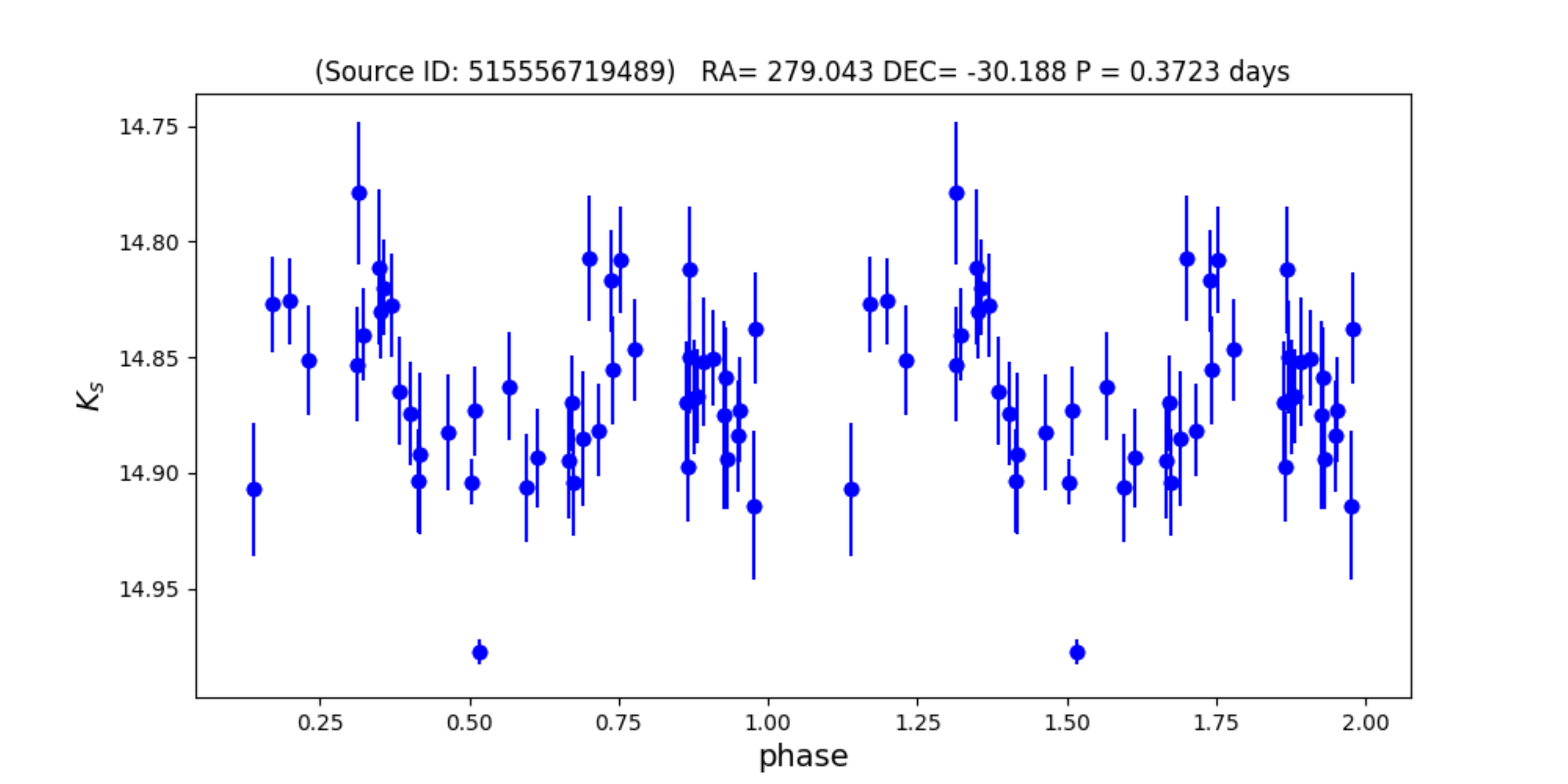} &    
    \includegraphics[width=.33\textwidth]{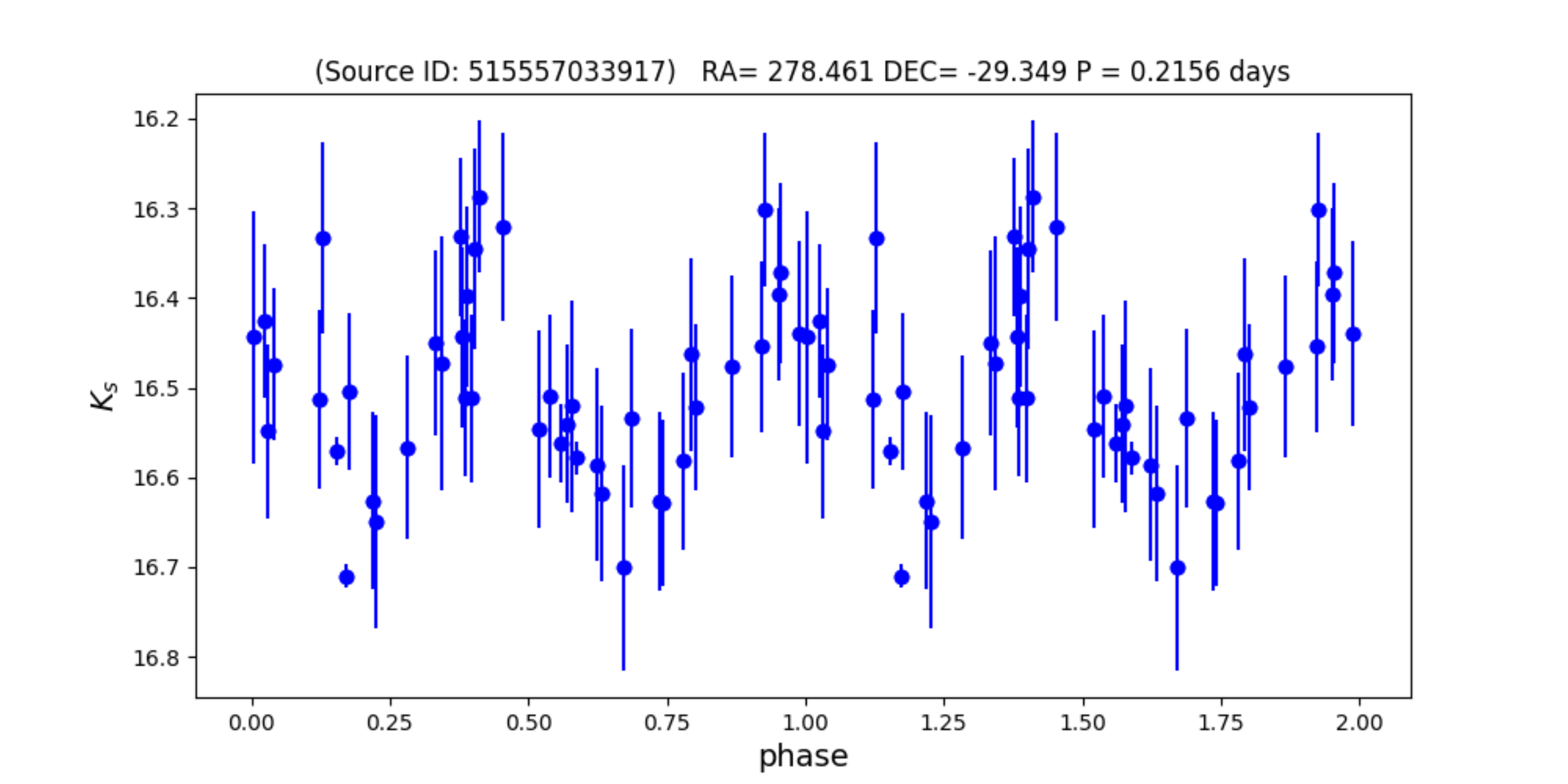} \\
    \includegraphics[width=.33\textwidth]{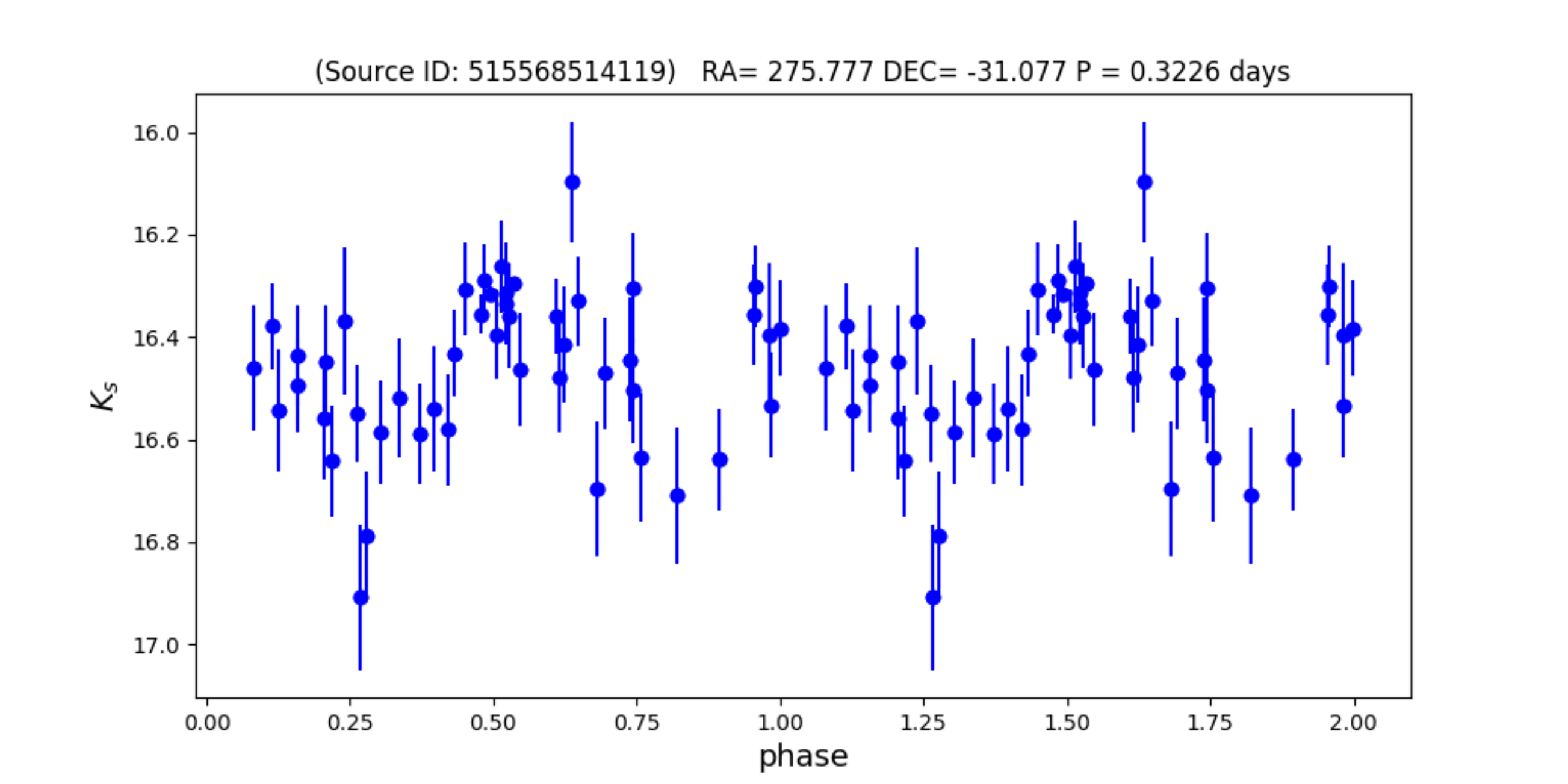} &
    \includegraphics[width=.33\textwidth]{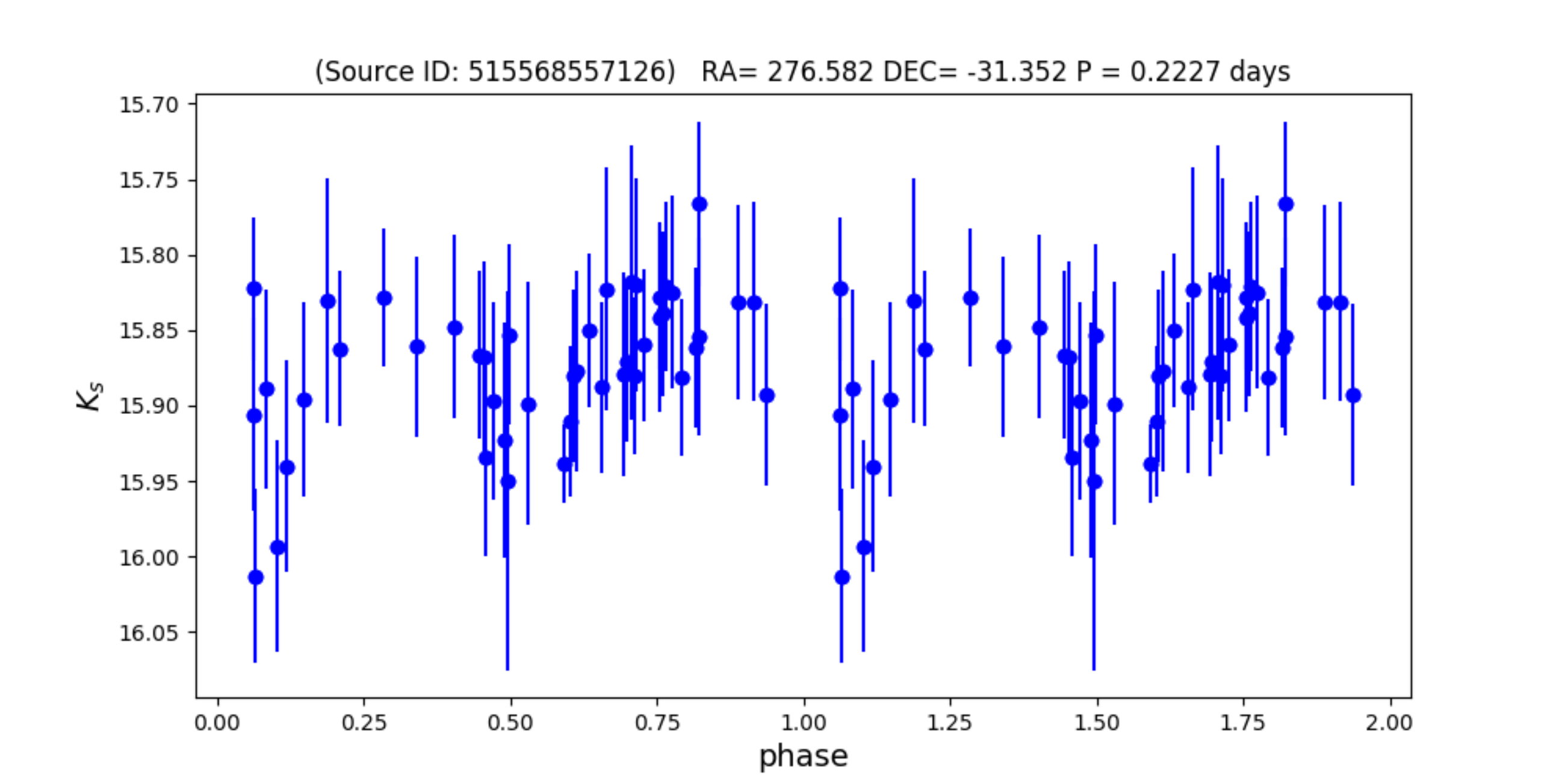} &
    \includegraphics[width=.33\textwidth]{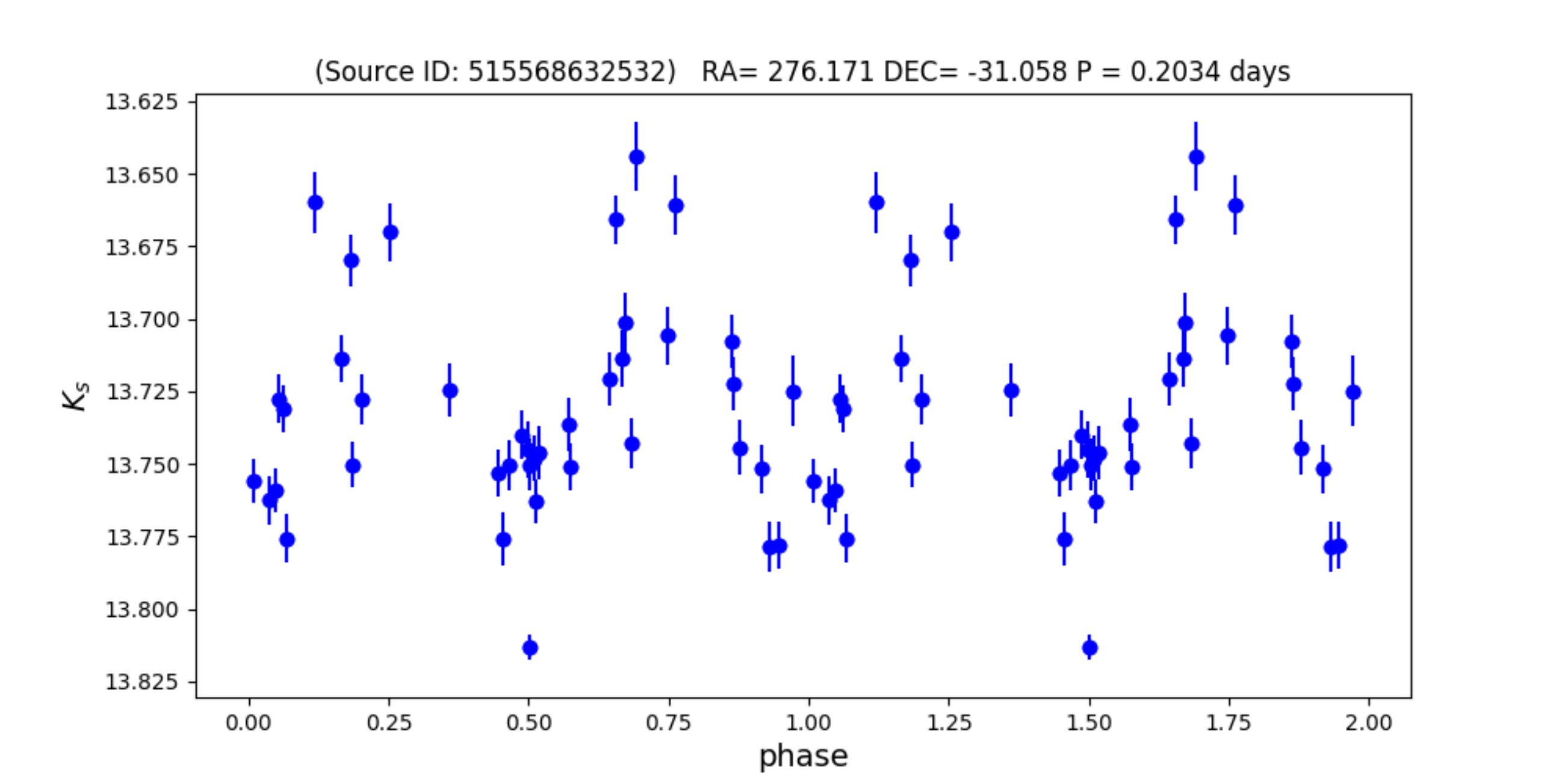} \\
    \includegraphics[width=.33\textwidth]{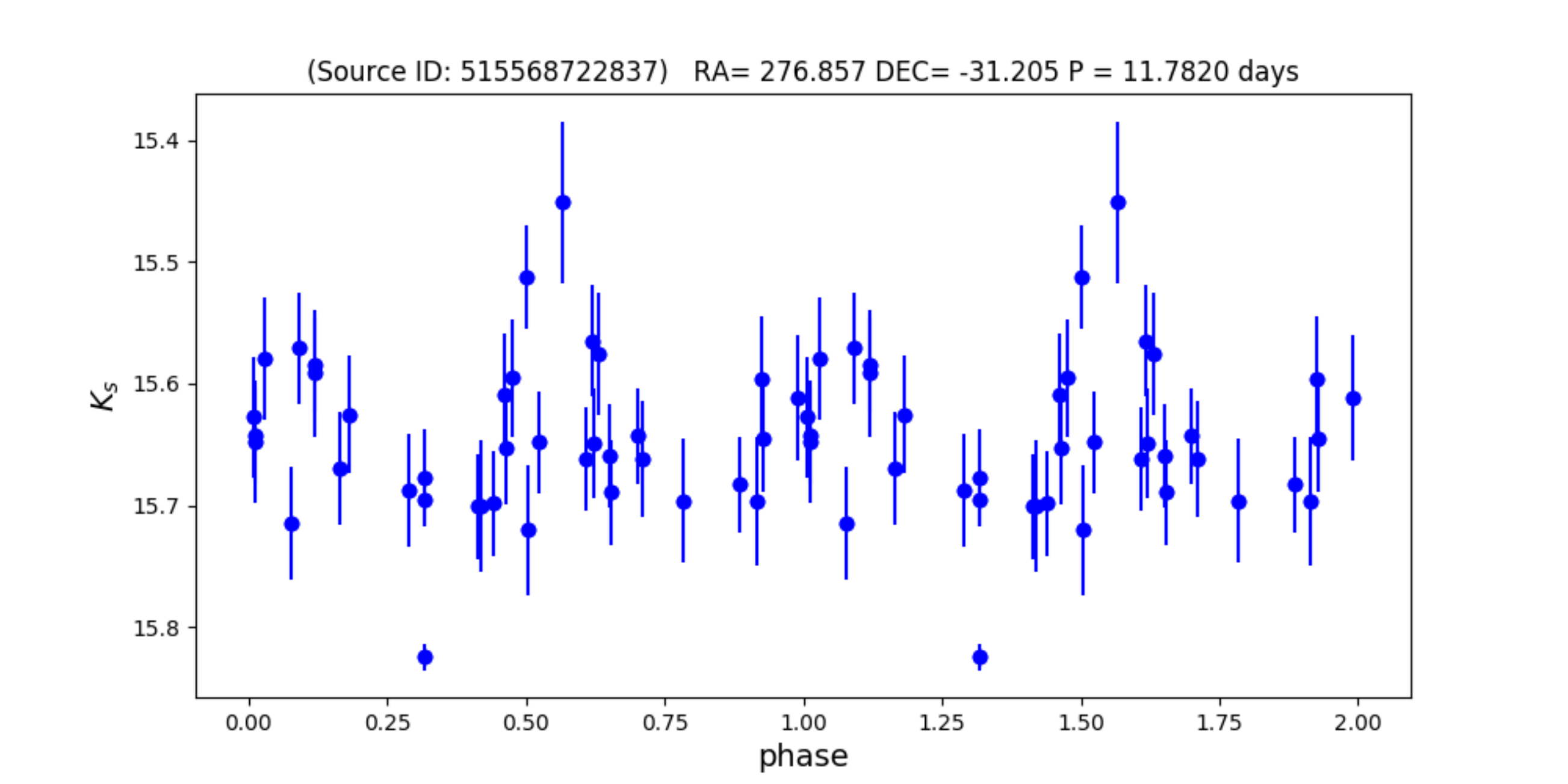} &
    \includegraphics[width=.33\textwidth]{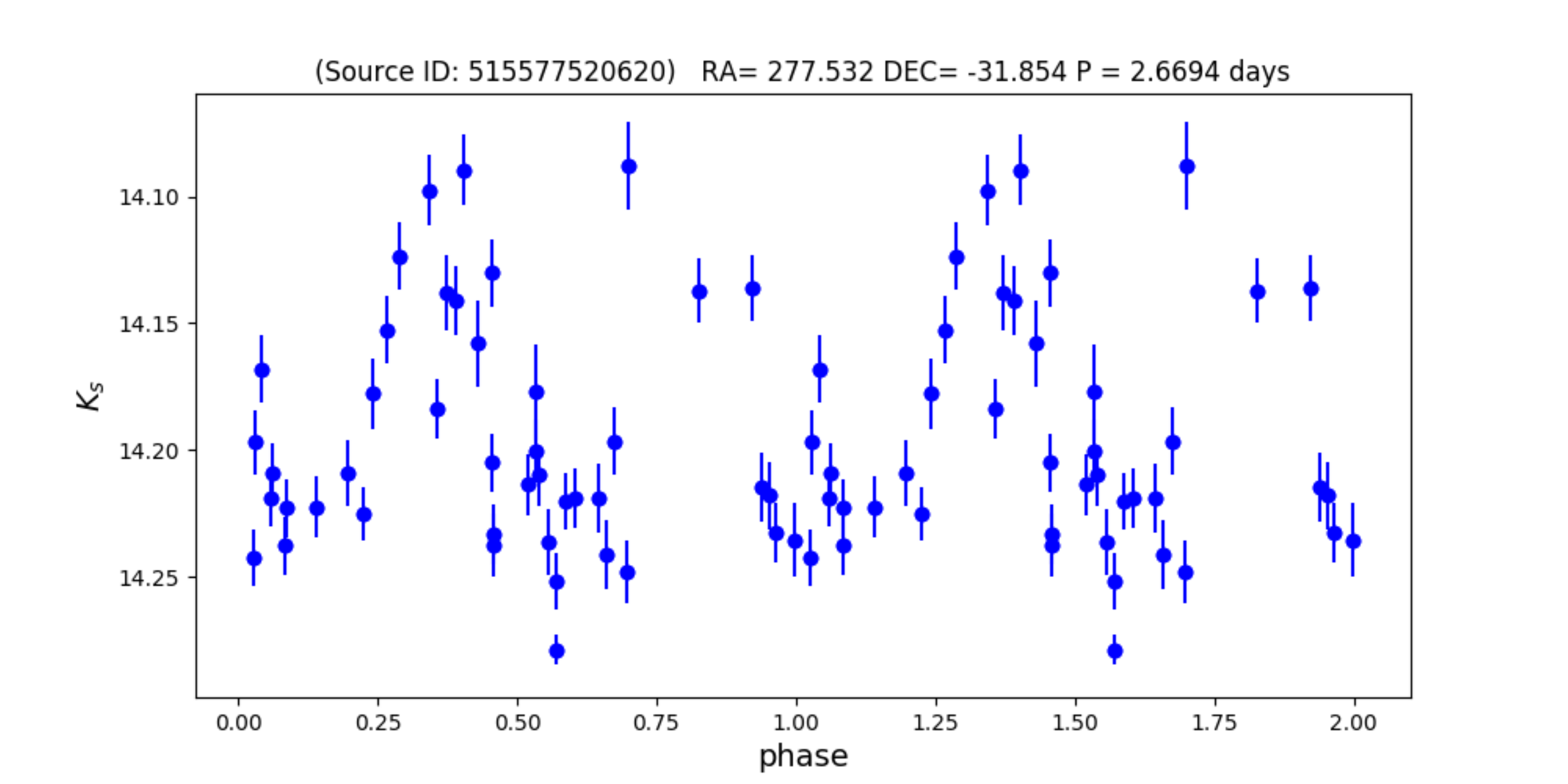} &
    \includegraphics[width=.33\textwidth]{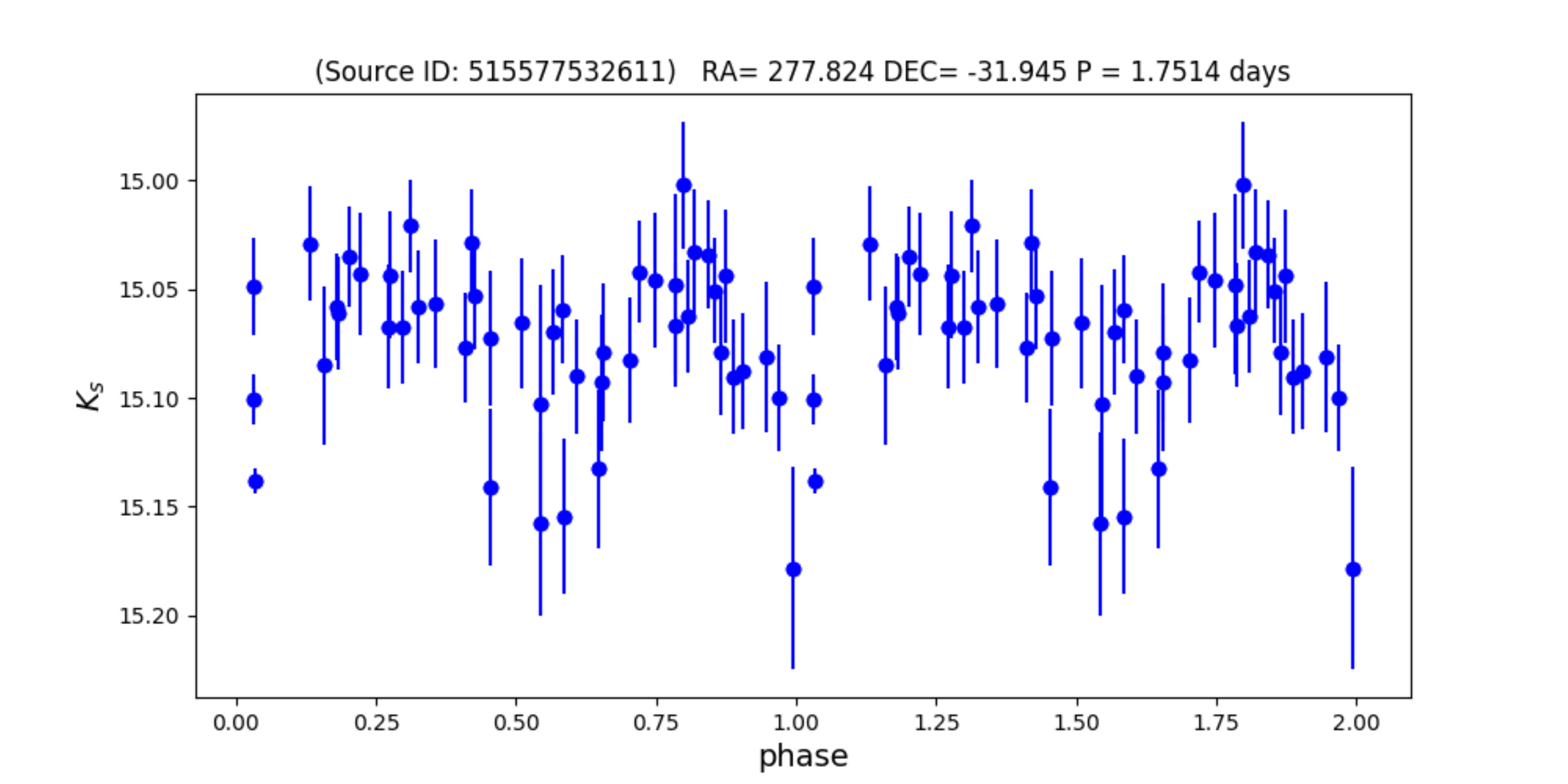} \\    
    \includegraphics[width=.33\textwidth]{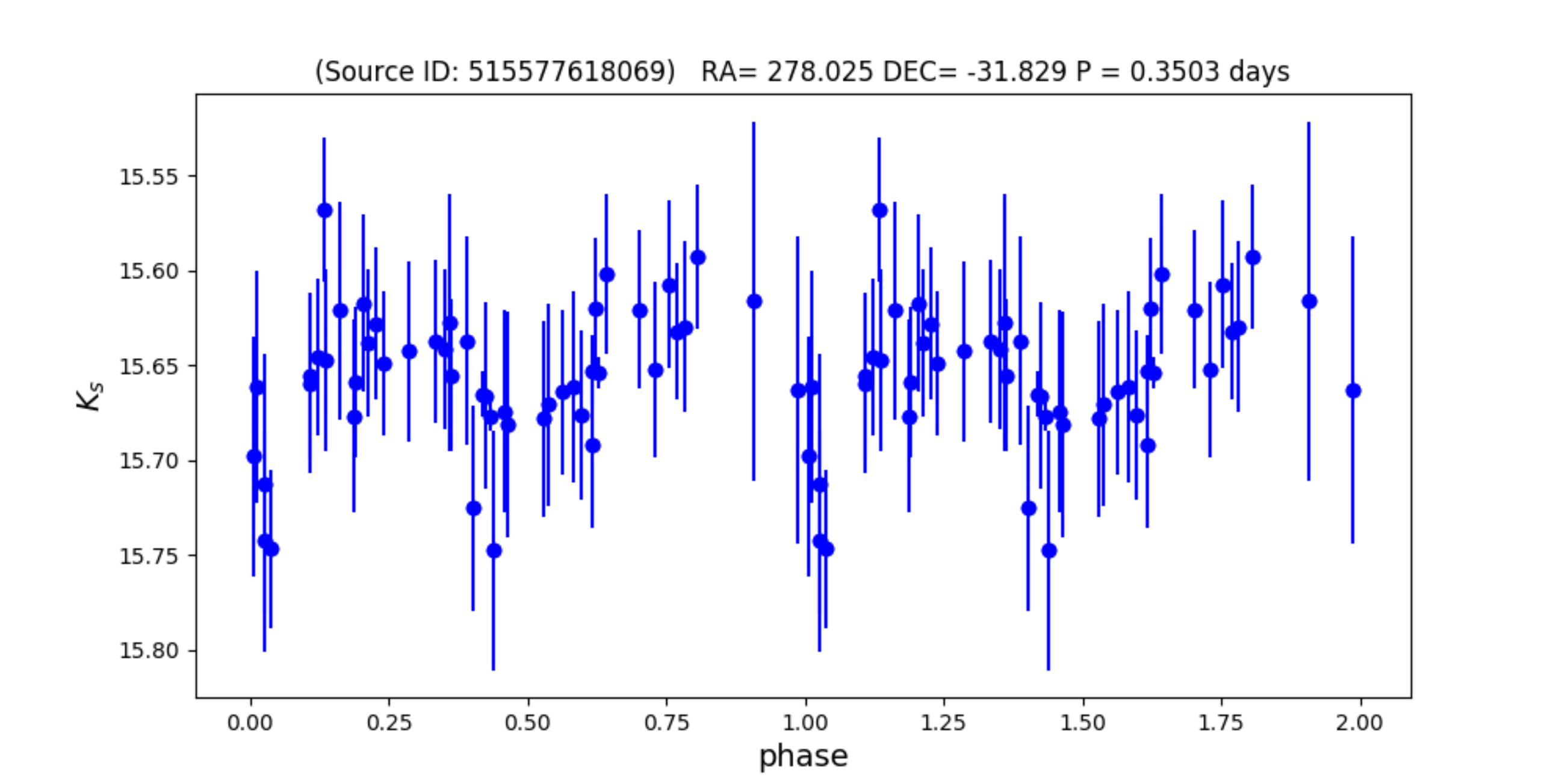} &    
    \includegraphics[width=.33\textwidth]{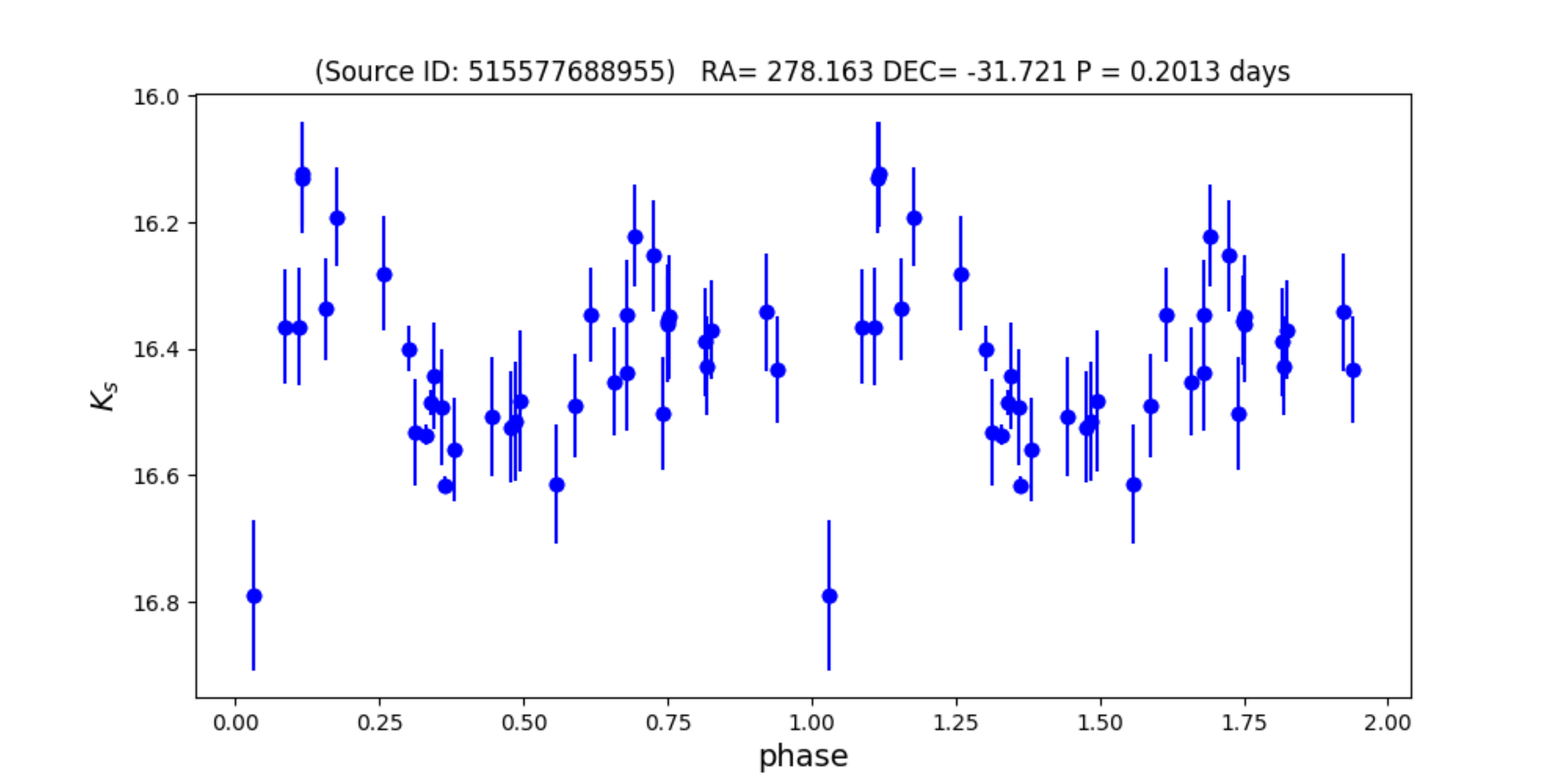} &
    \includegraphics[width=.33\textwidth]{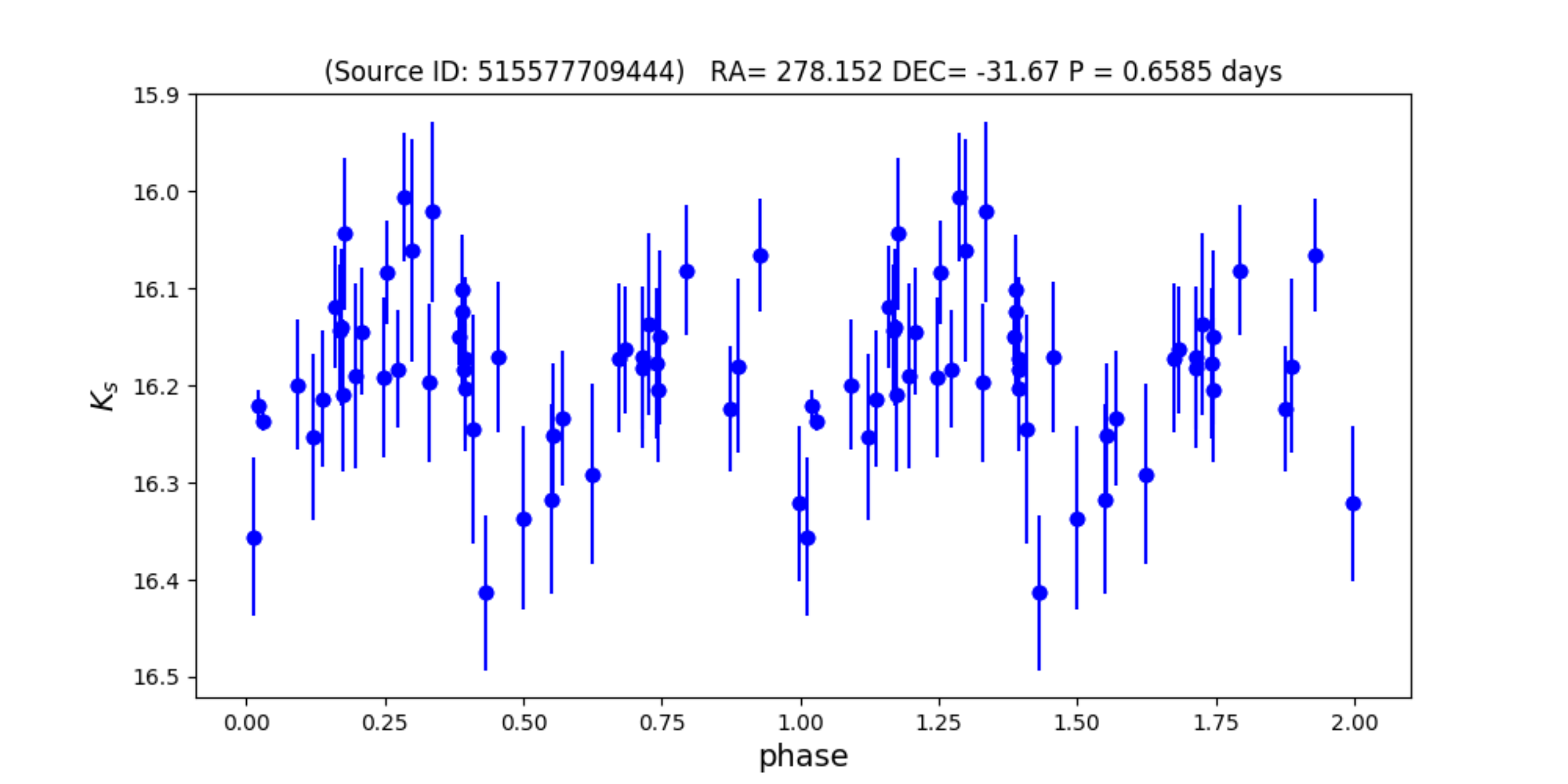} \\    
  \end{tabular}
  \caption{BHB stars light curves corresponding to $2^{nd}$ category eclipsing binaries.}
  \label{fig:bhbB2}
\end{figure*}

\begin{figure*}[htb]
\centering
  \begin{tabular}{@{}ccc@{}}           
    \includegraphics[width=.33\textwidth]{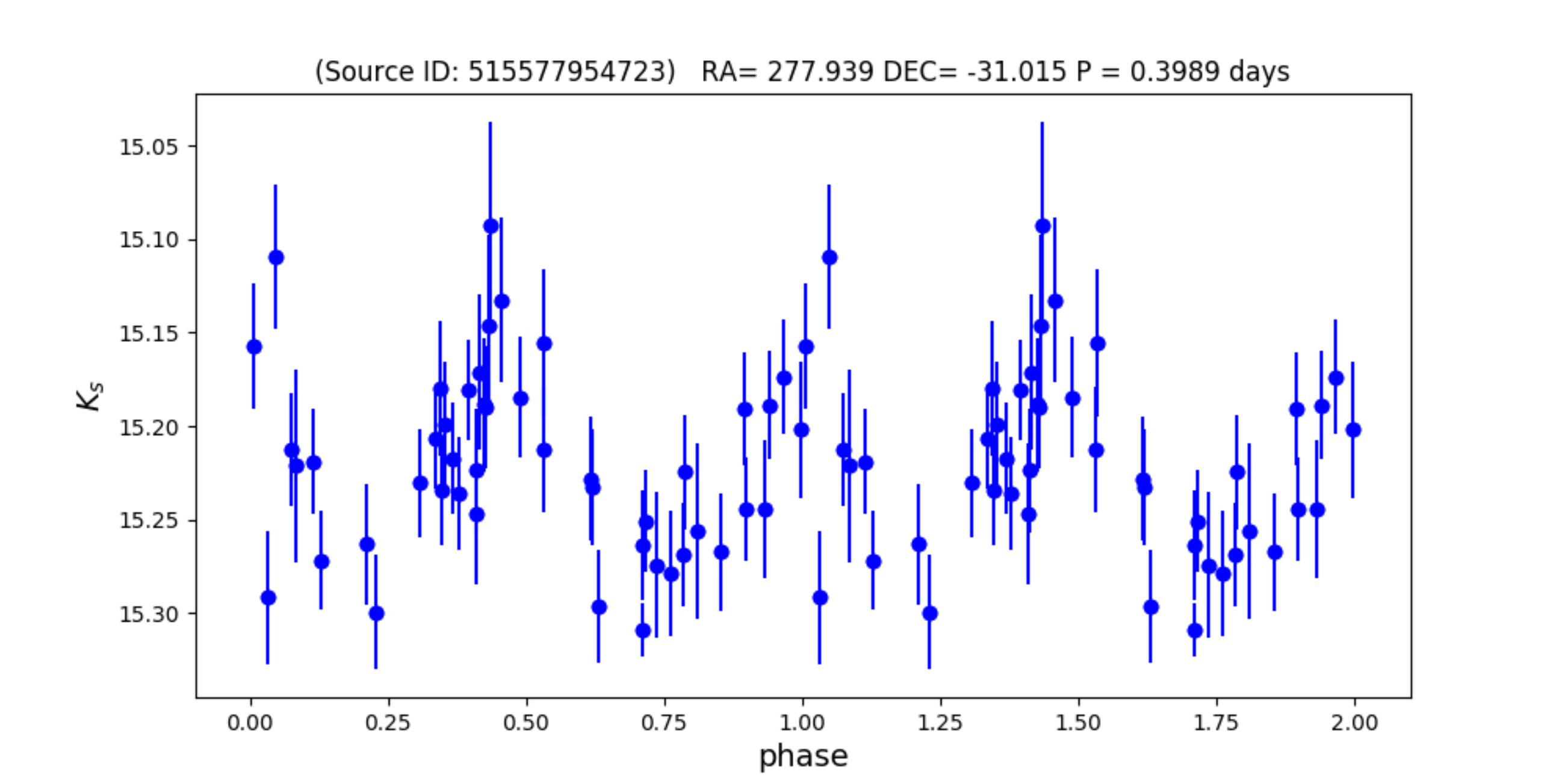} &
    \includegraphics[width=.33\textwidth]{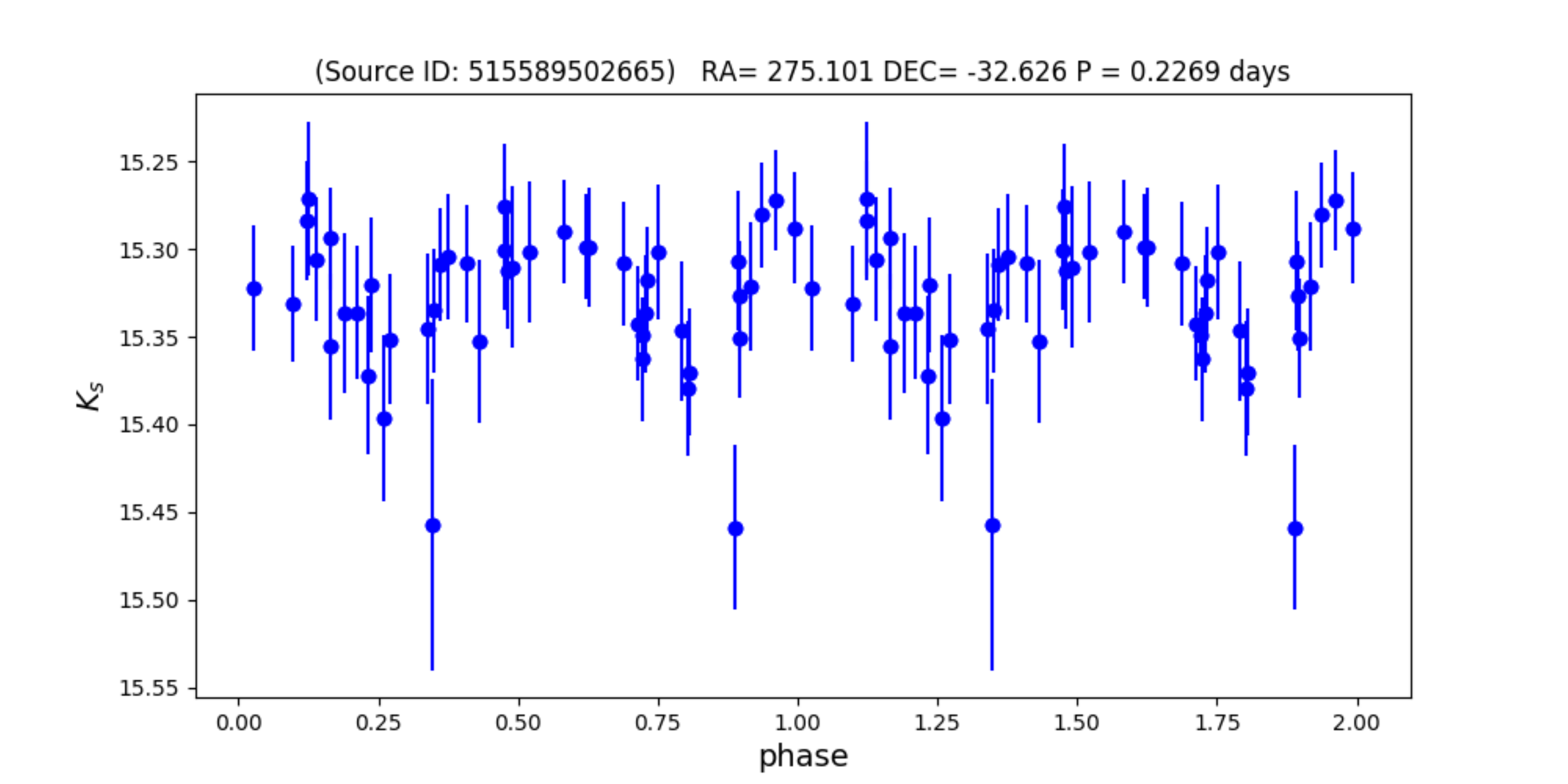} &
    \includegraphics[width=.33\textwidth]{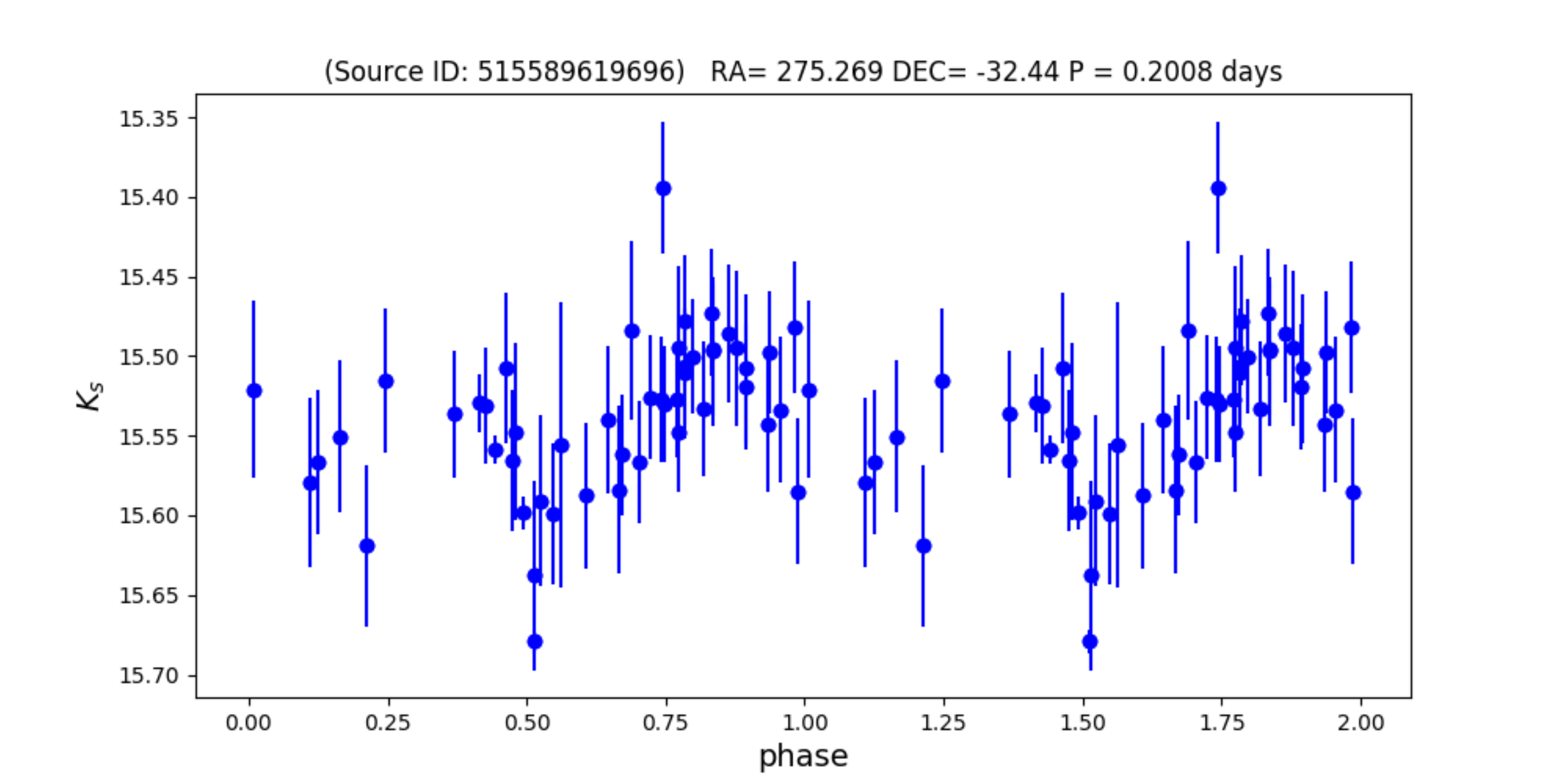} \\
    \includegraphics[width=.33\textwidth]{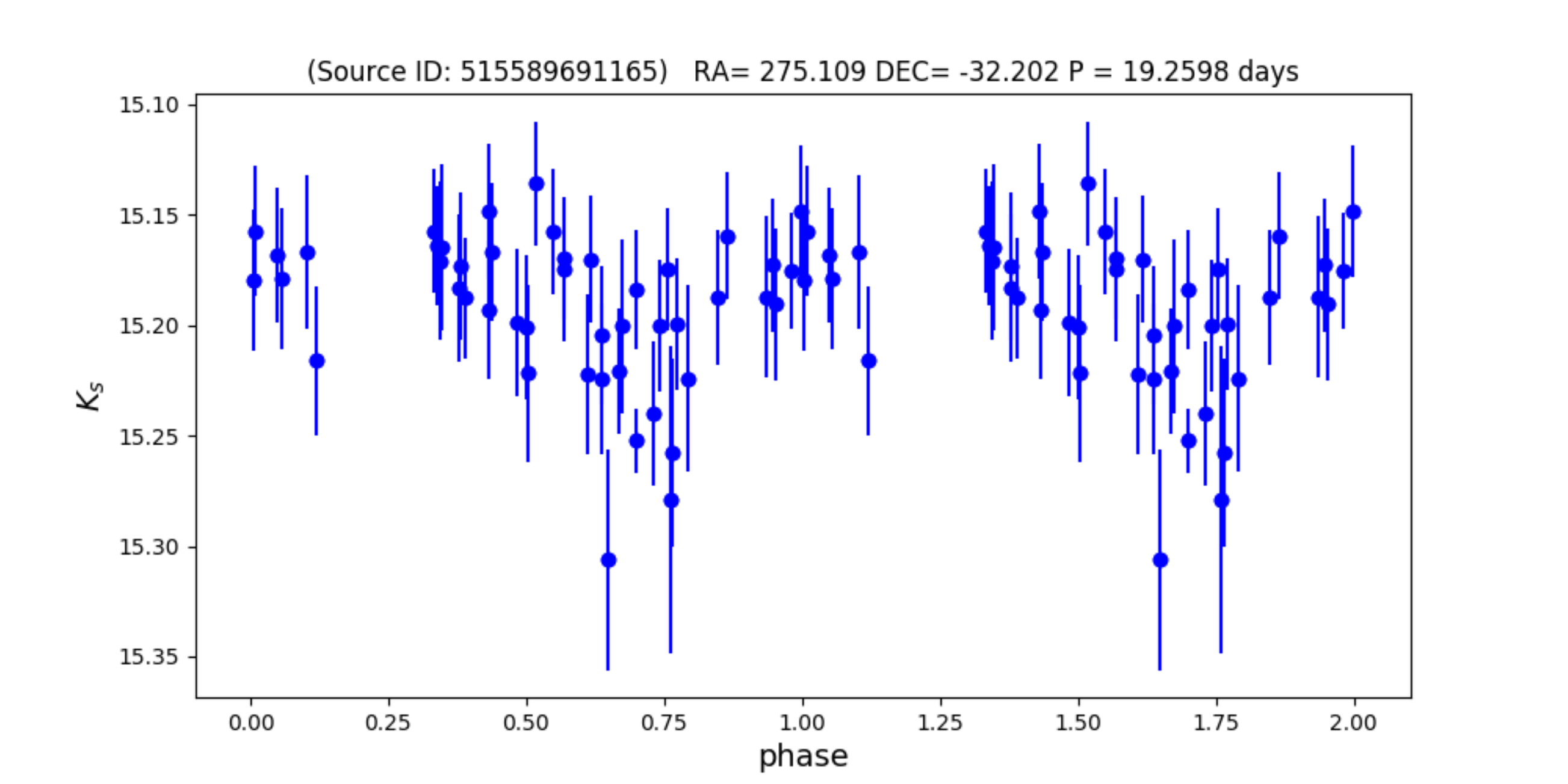} &
    \includegraphics[width=.33\textwidth]{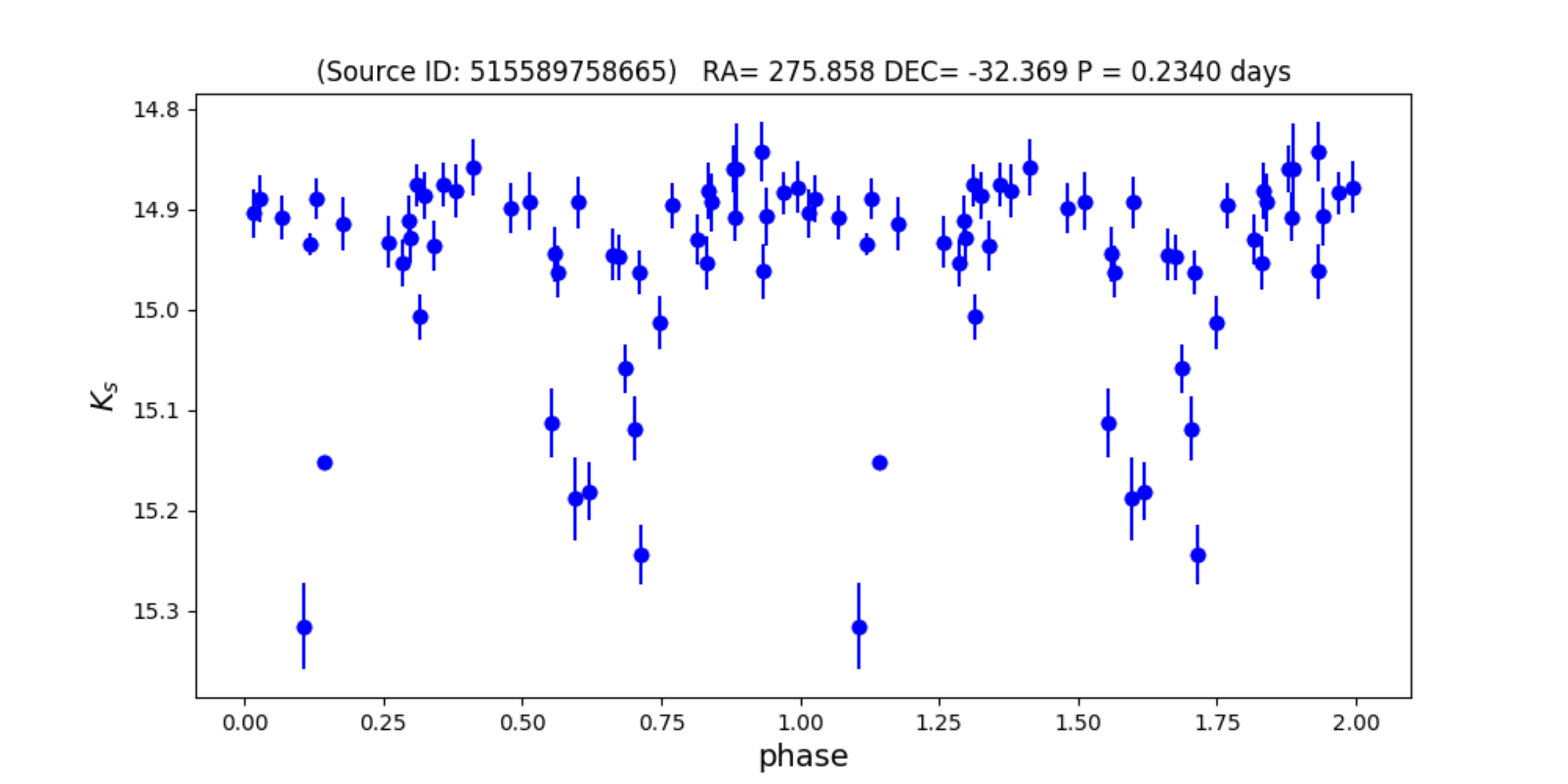} &
    \includegraphics[width=.33\textwidth]{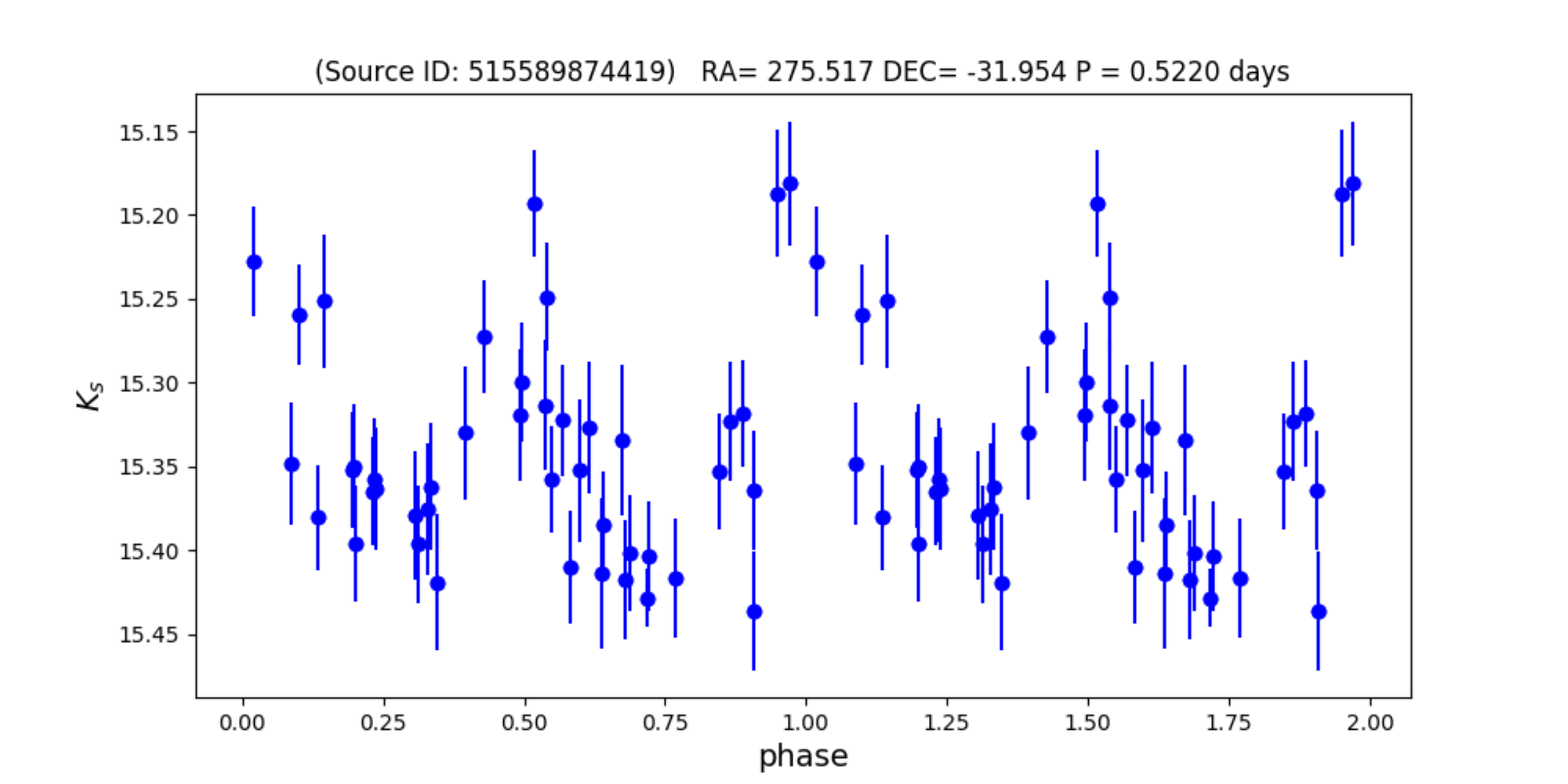} \\
    \includegraphics[width=.33\textwidth]{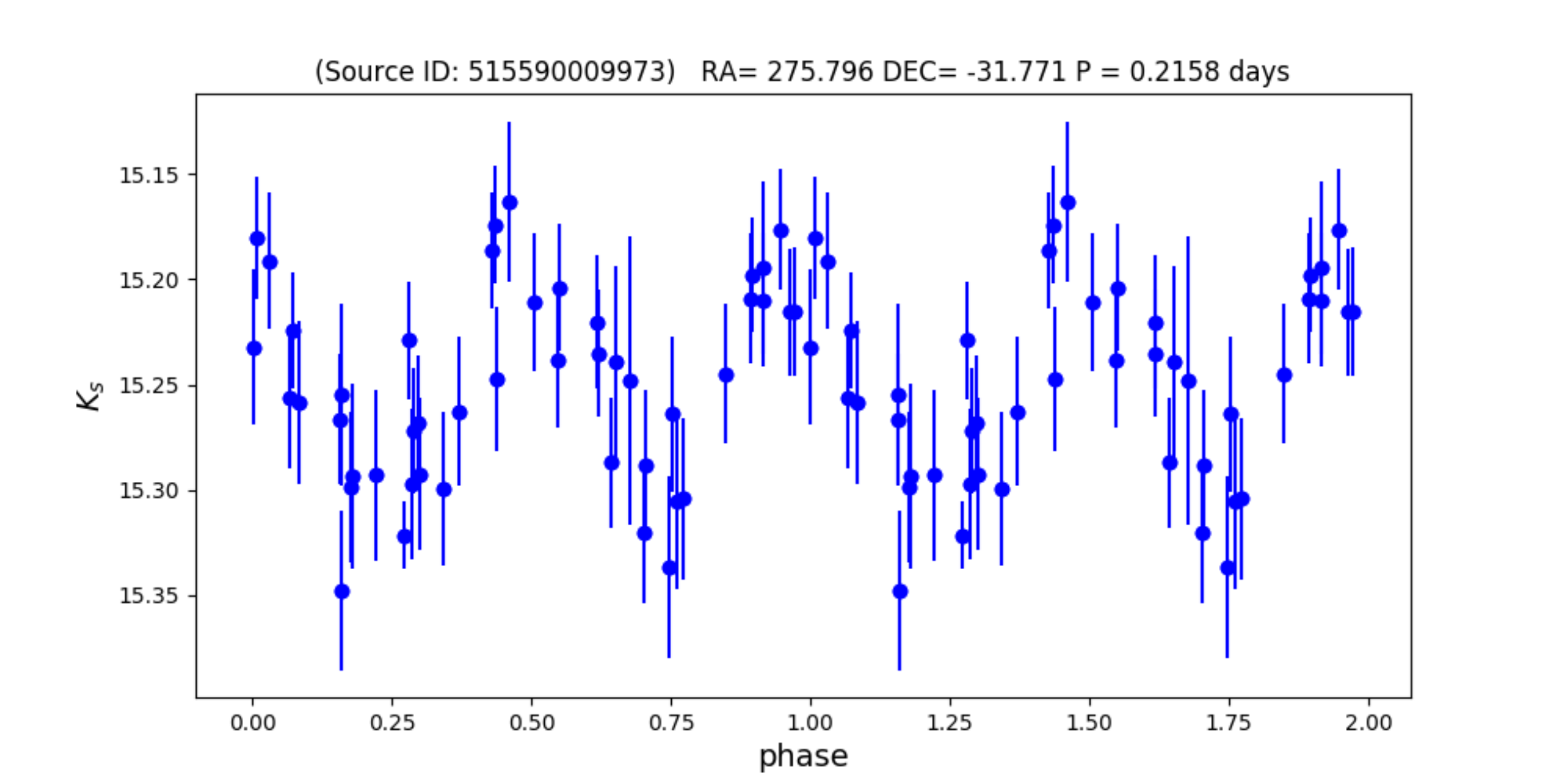} &    
    \includegraphics[width=.33\textwidth]{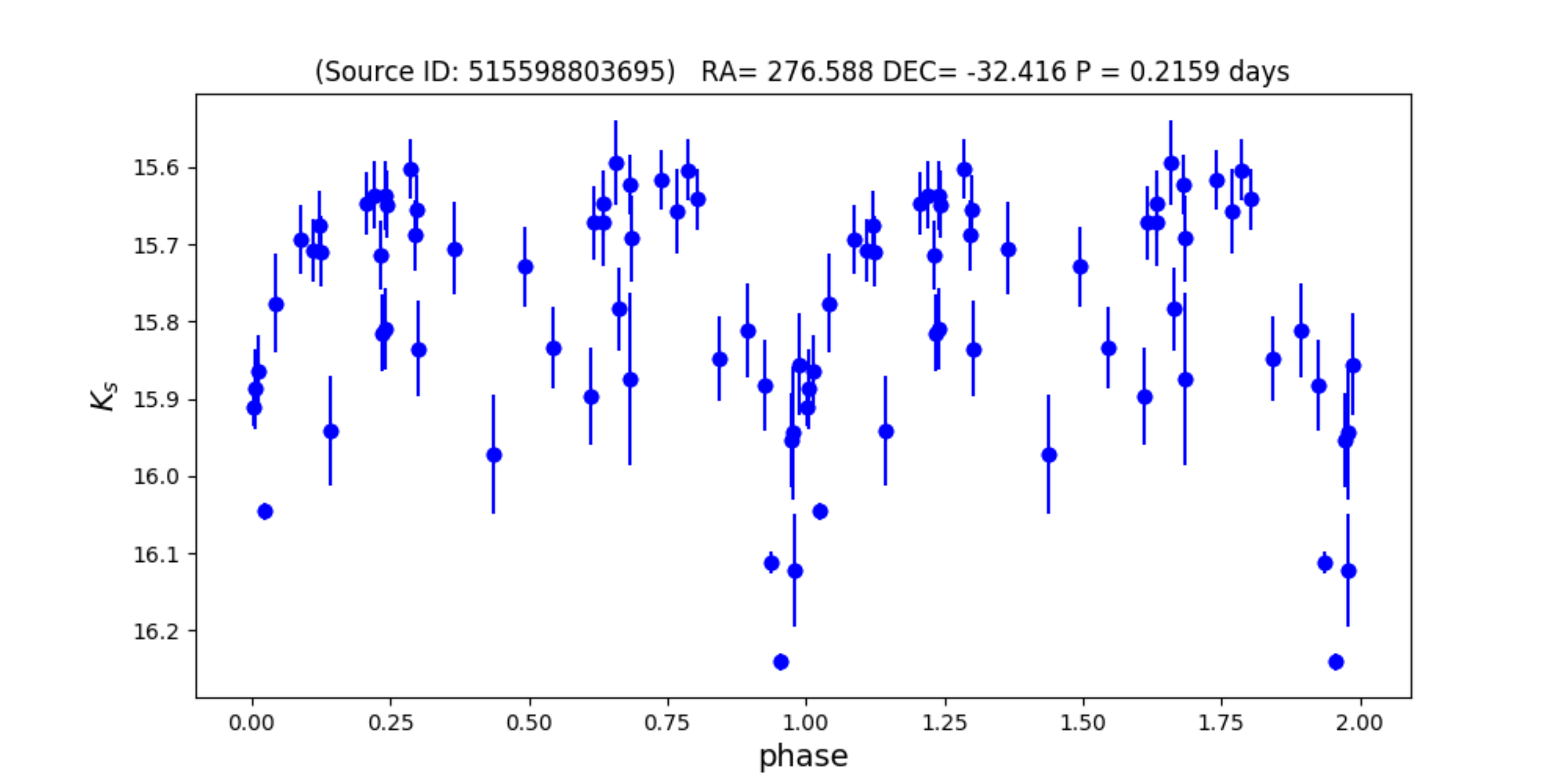} &    
    \includegraphics[width=.33\textwidth]{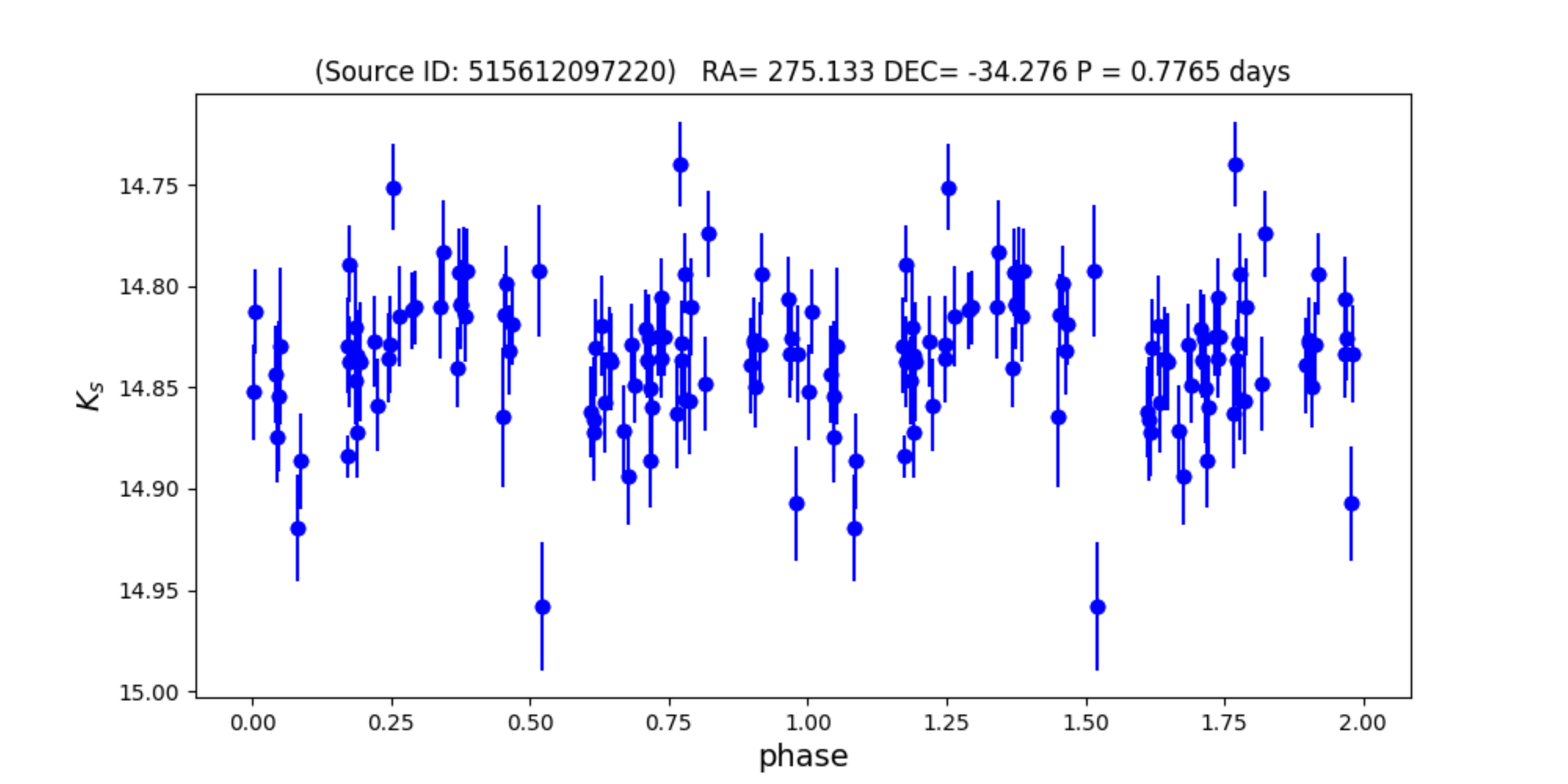} \\    
    \includegraphics[width=.33\textwidth]{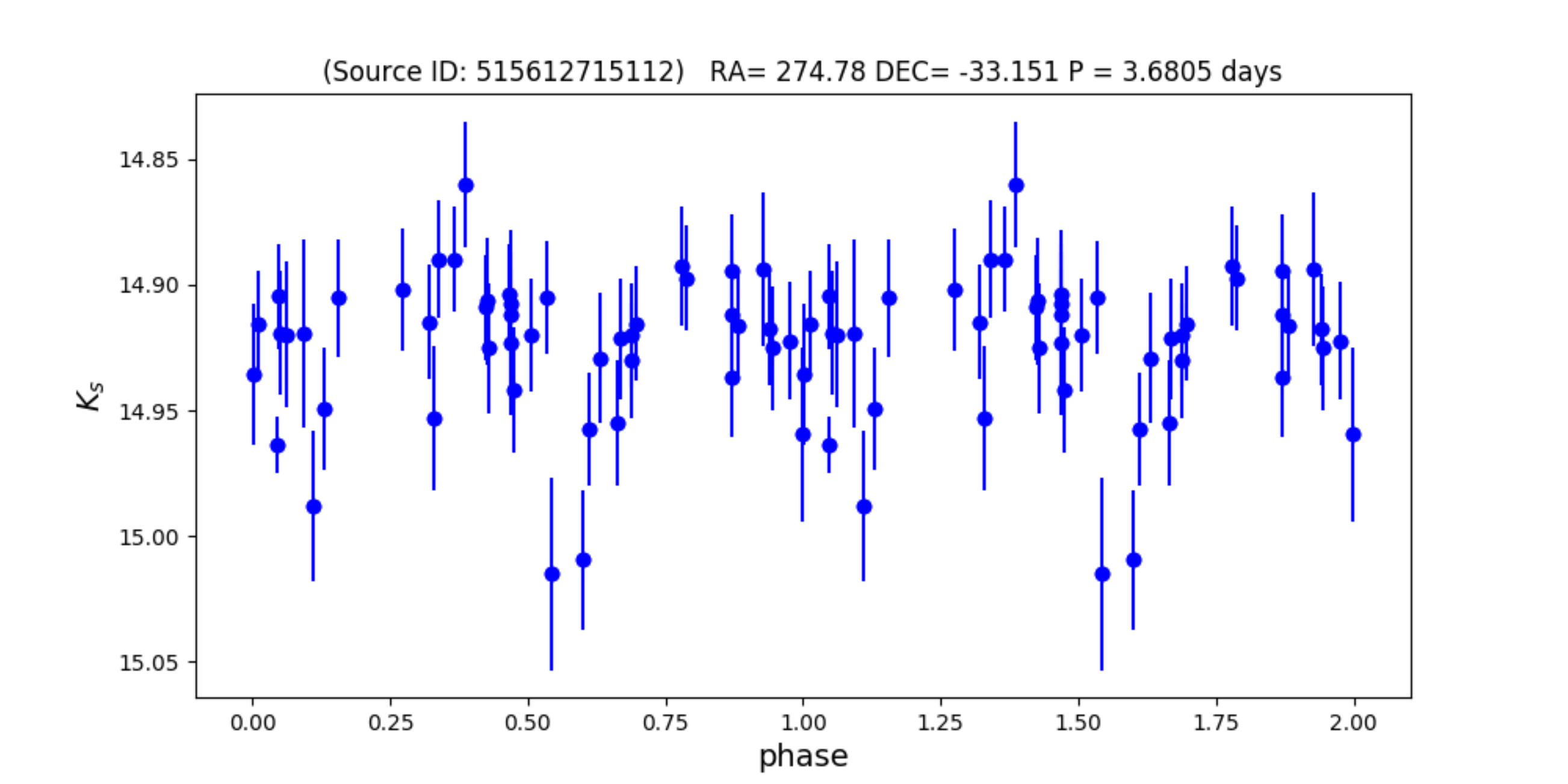} &        
    \includegraphics[width=.33\textwidth]{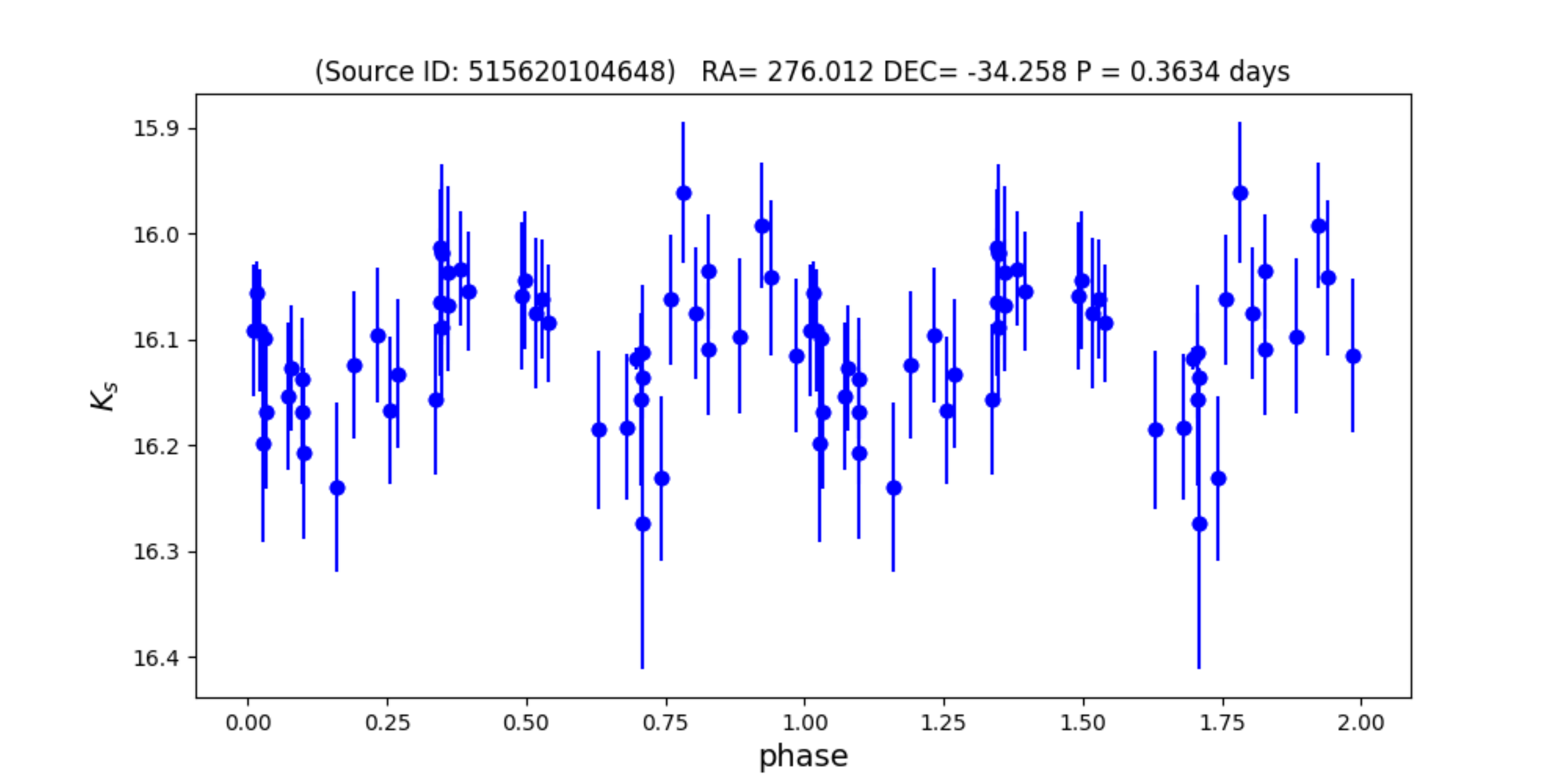} &    
    \includegraphics[width=.33\textwidth]{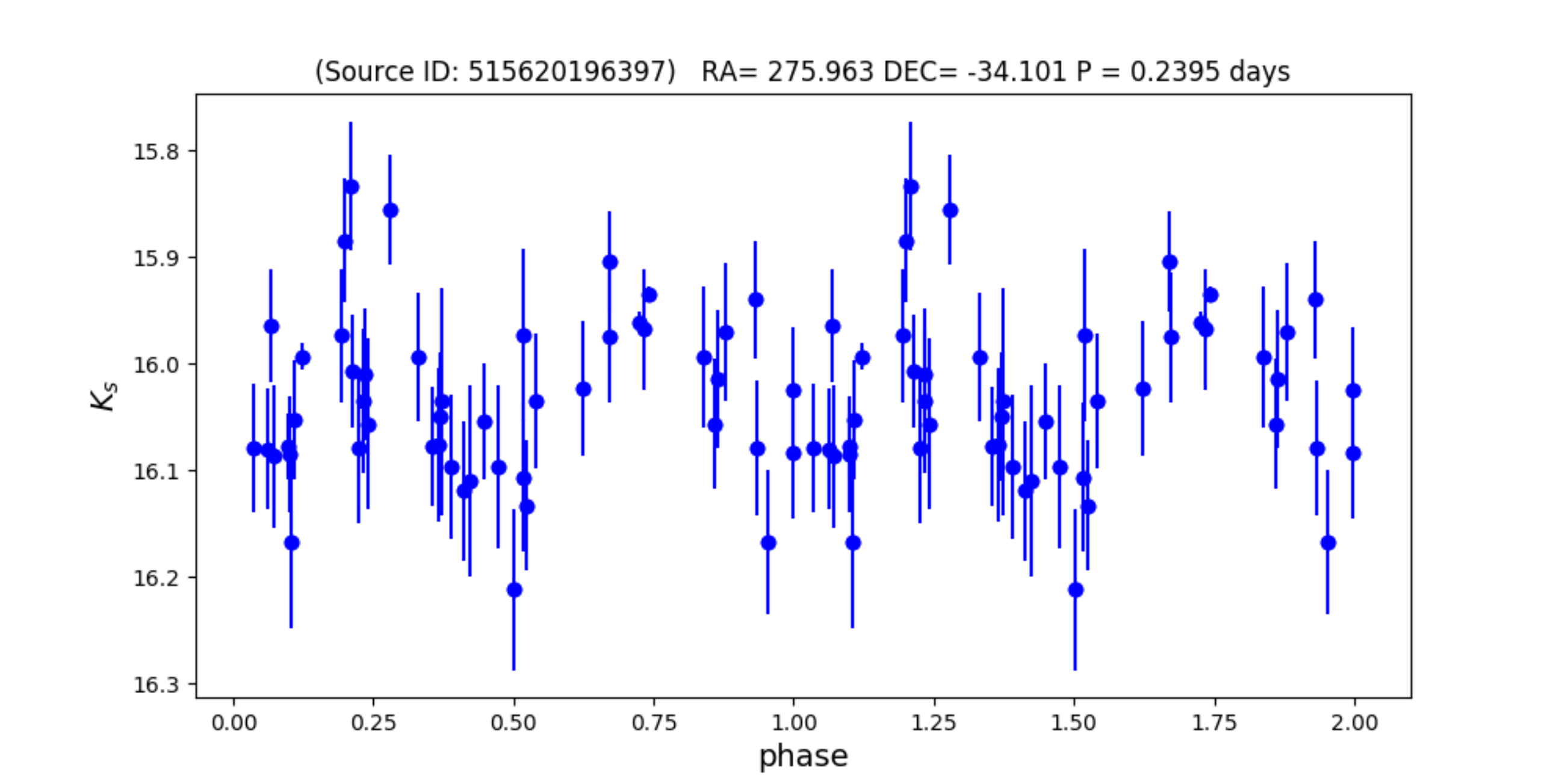} \\
    \includegraphics[width=.33\textwidth]{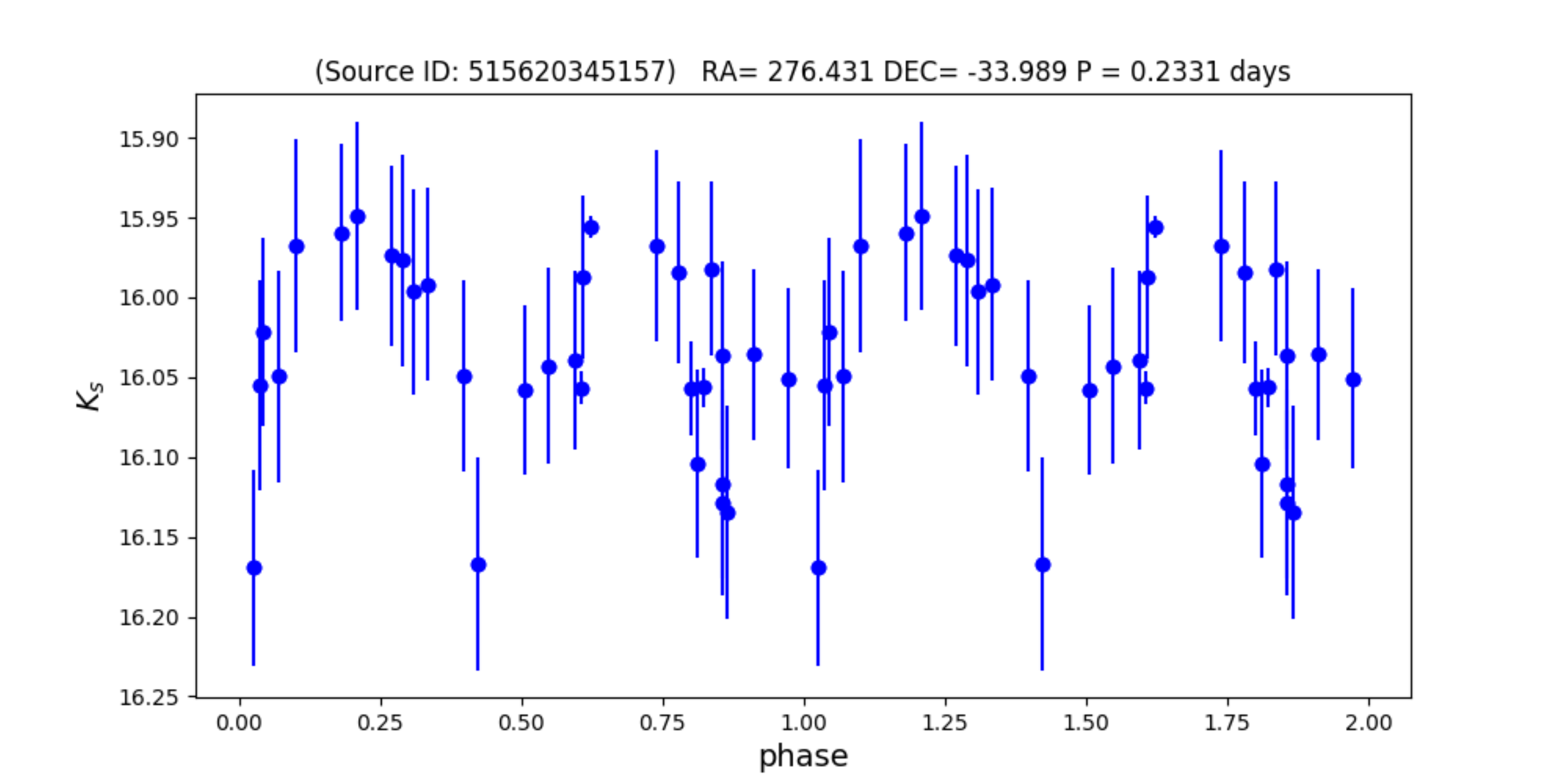} &
    \includegraphics[width=.33\textwidth]{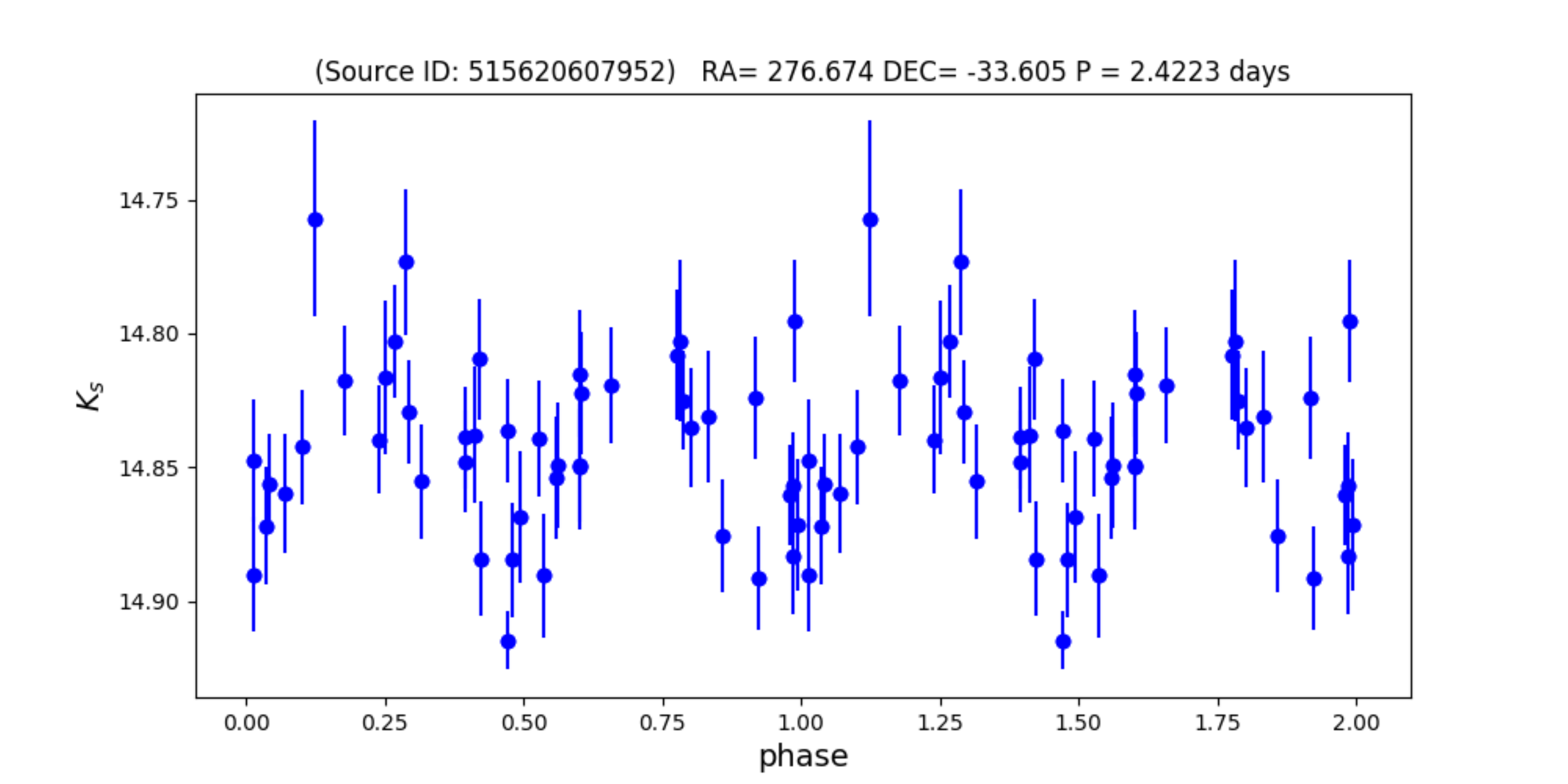} &    
    \includegraphics[width=.33\textwidth]{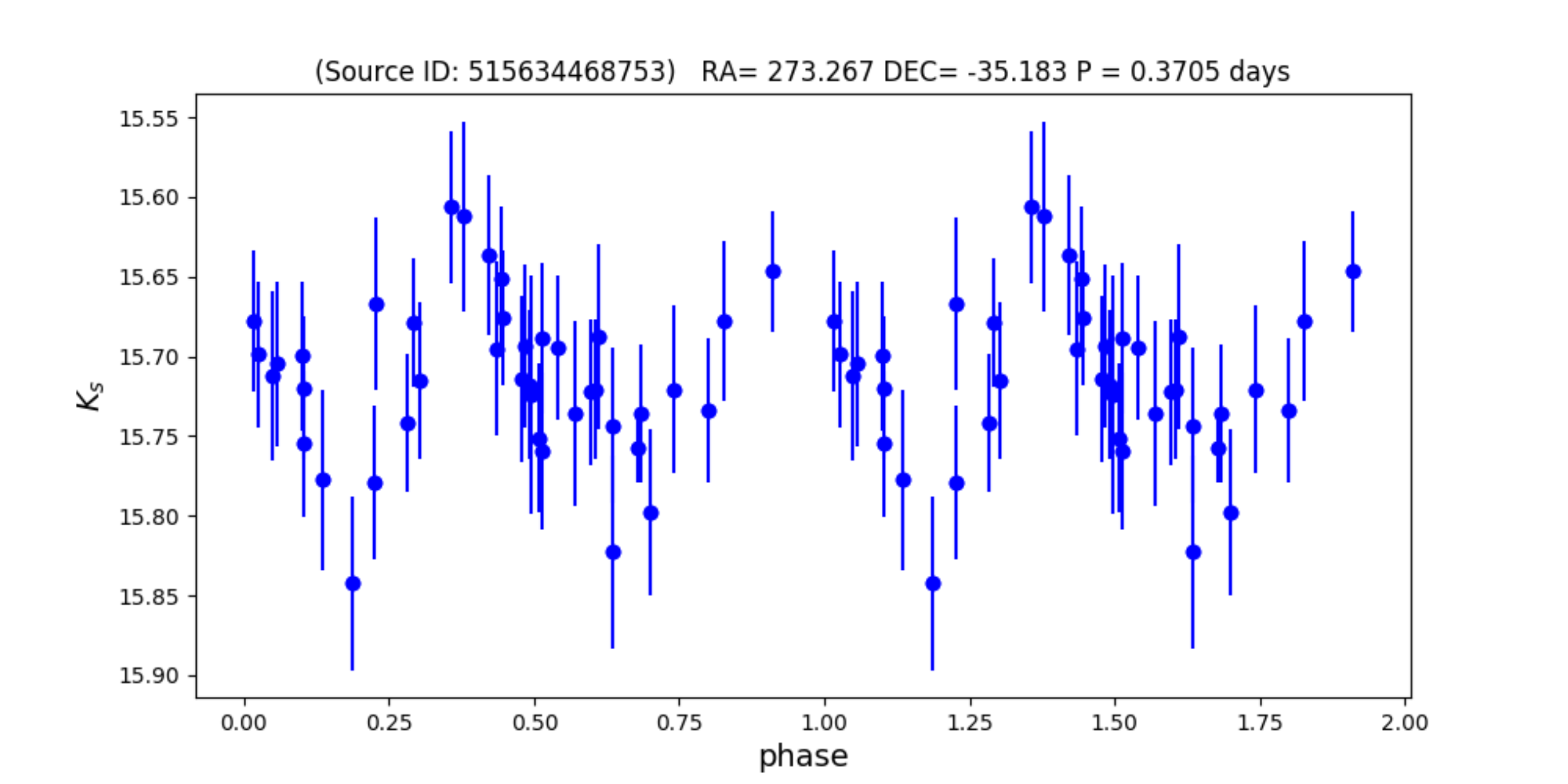} \\
    \includegraphics[width=.33\textwidth]{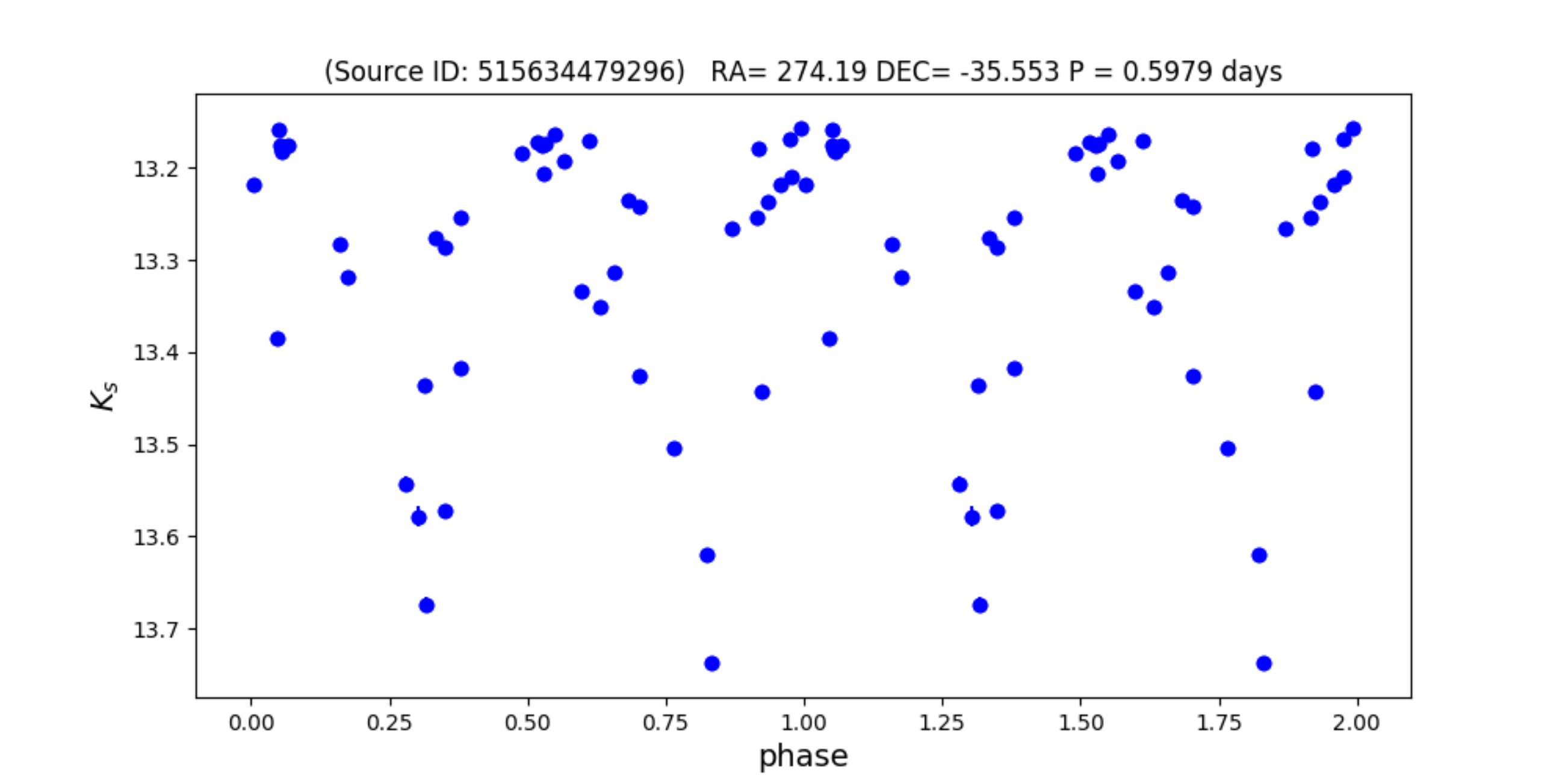} &
    \includegraphics[width=.33\textwidth]{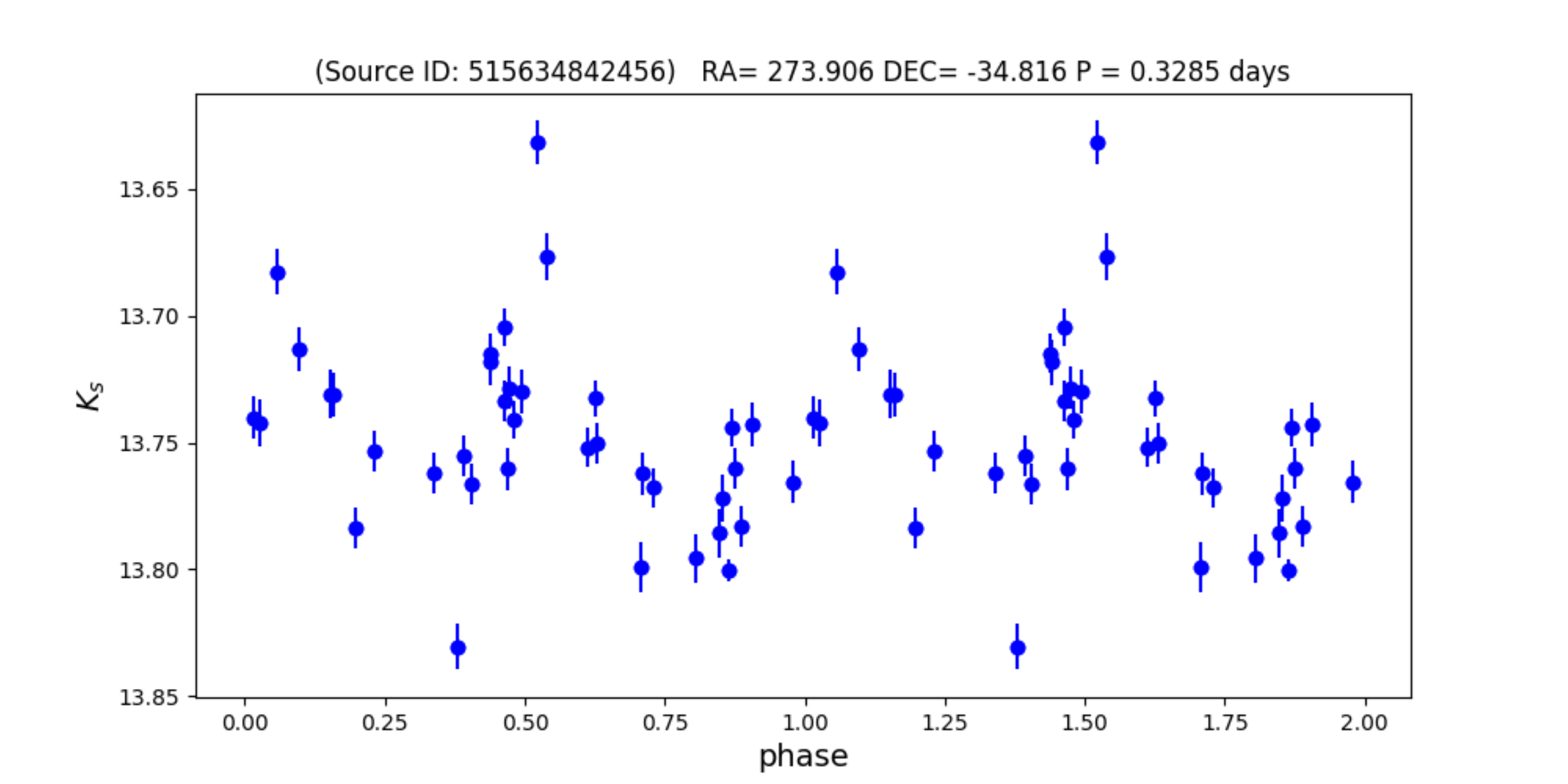} &    
    \includegraphics[width=.33\textwidth]{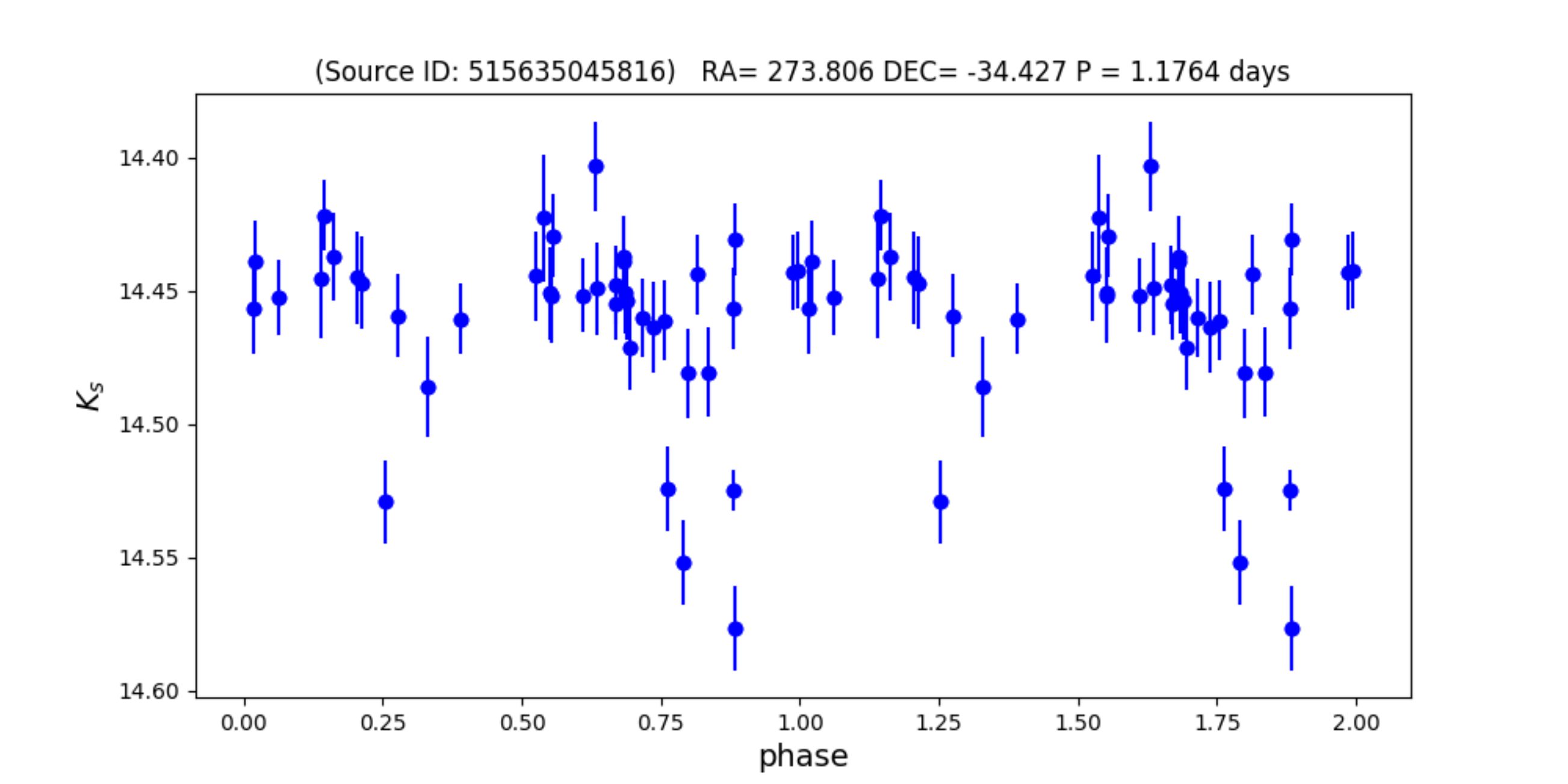} \\ 
    \includegraphics[width=.33\textwidth]{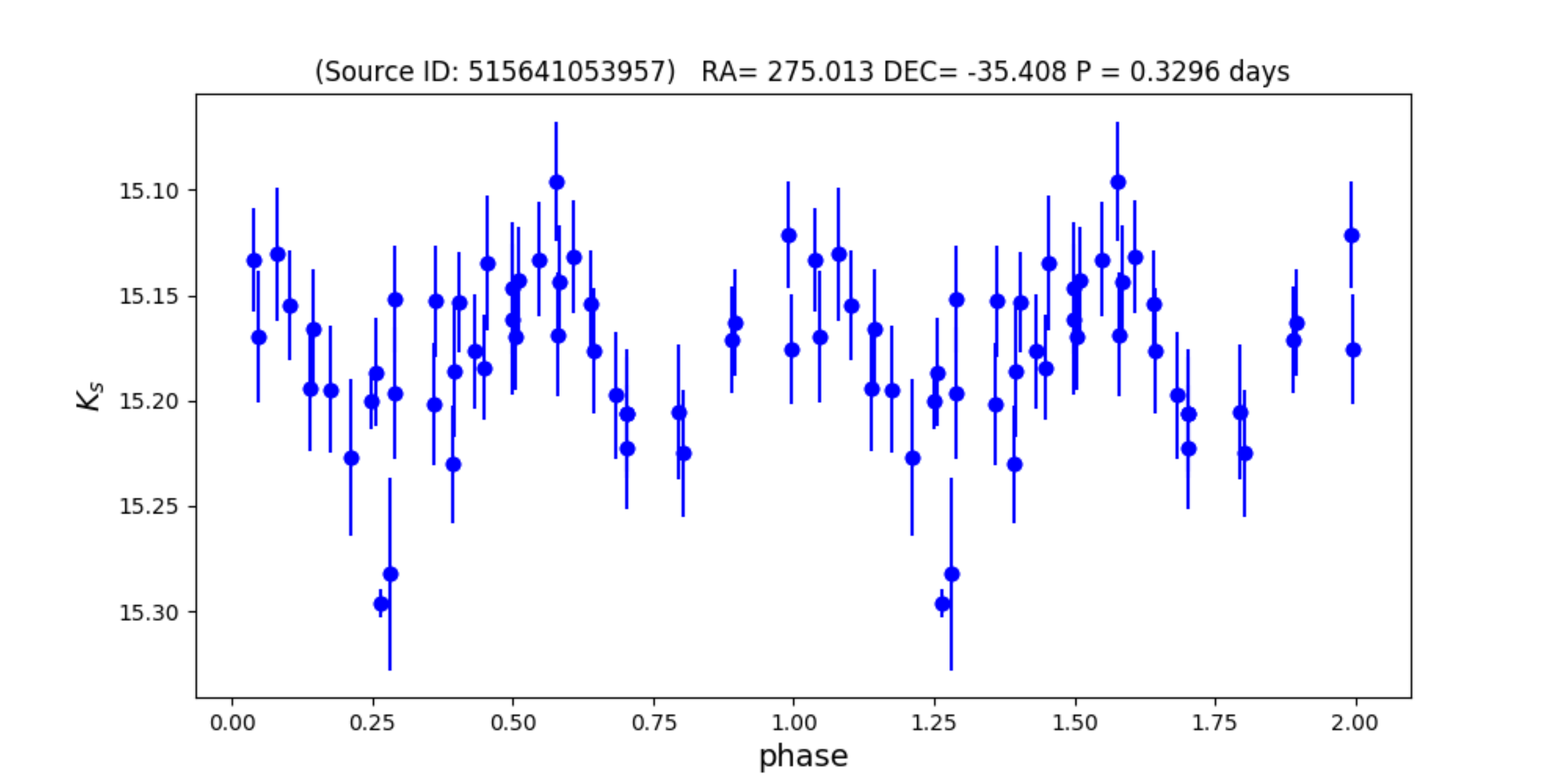} &
    \includegraphics[width=.33\textwidth]{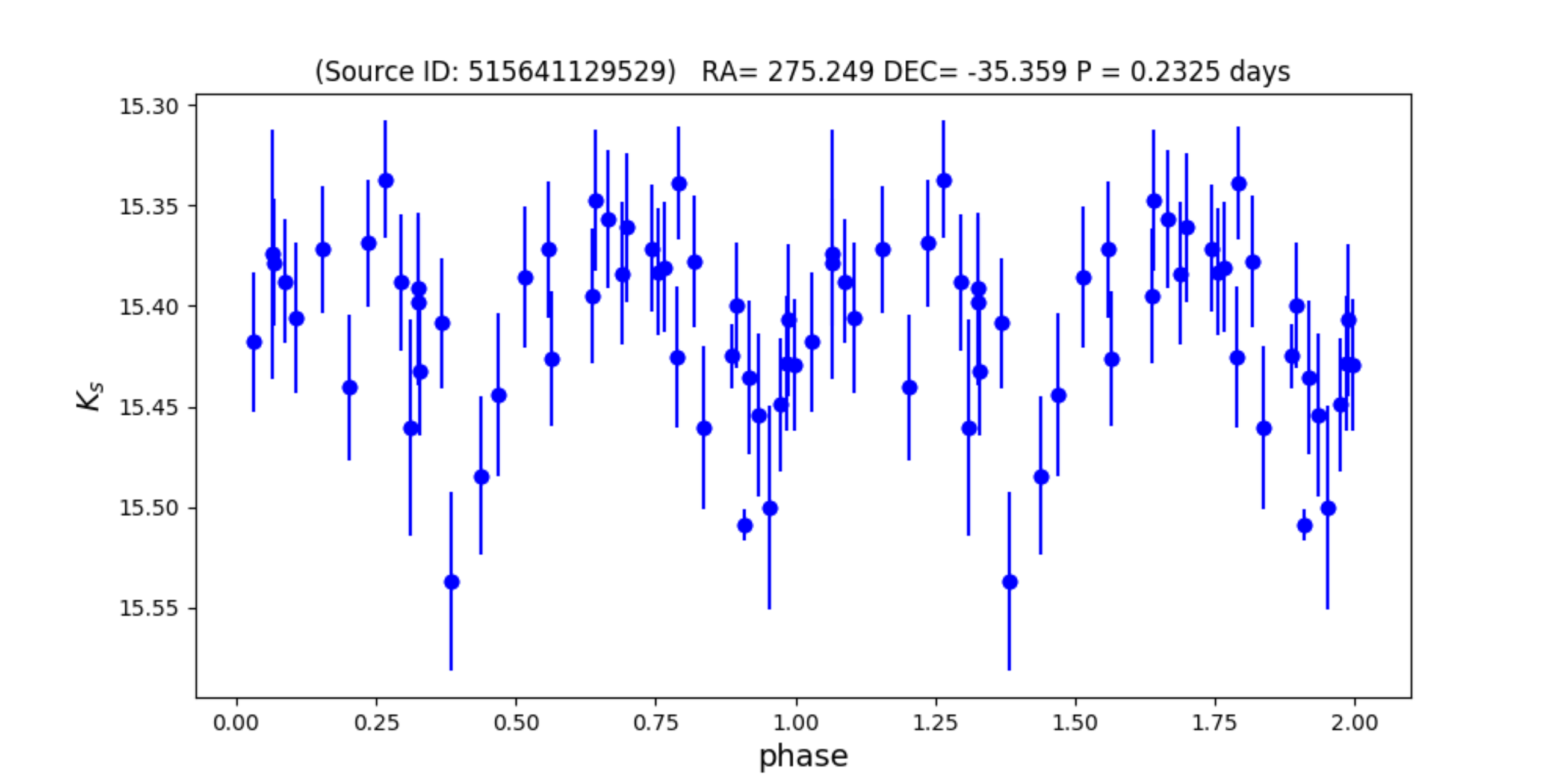} &    
    \includegraphics[width=.33\textwidth]{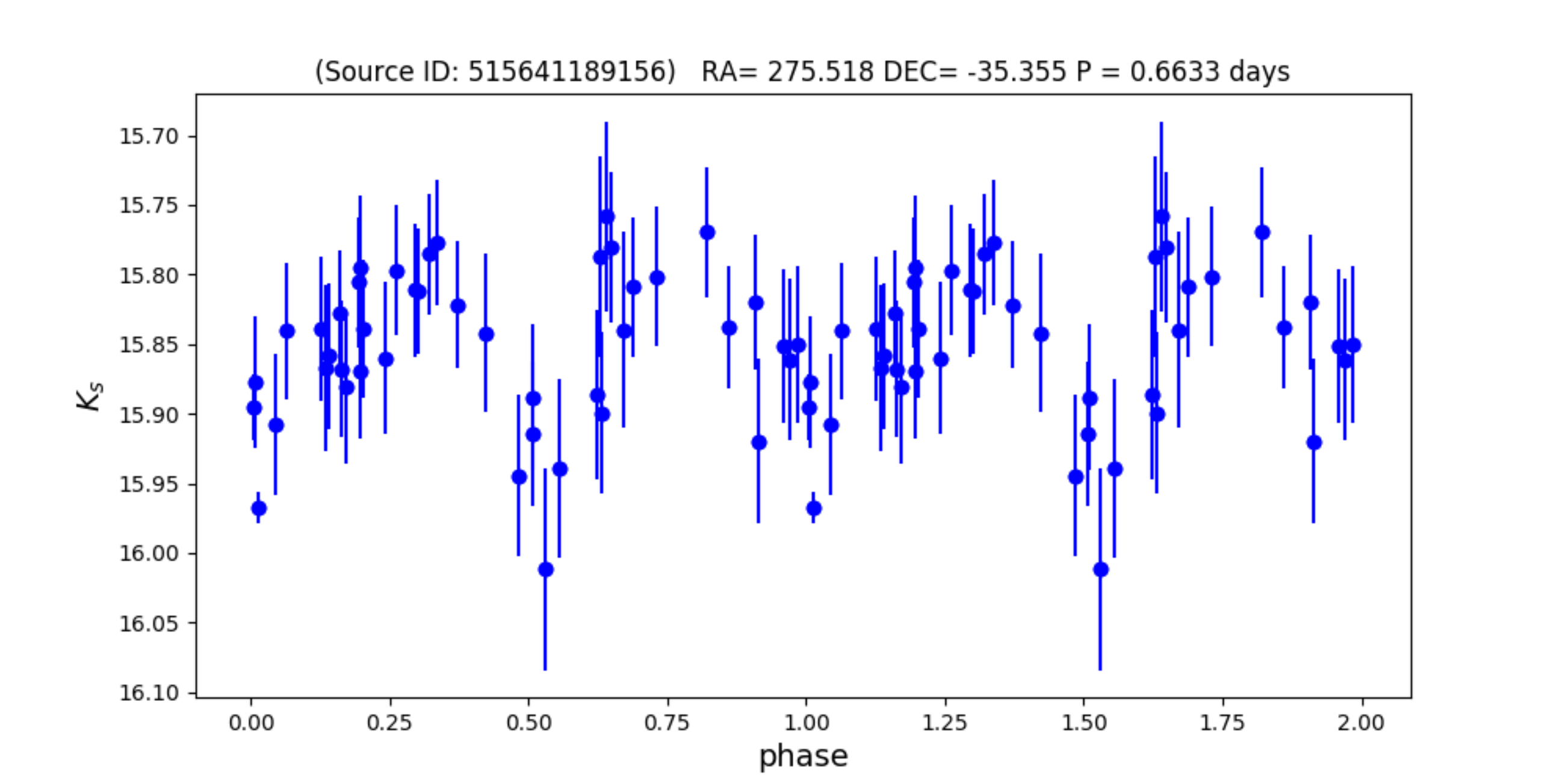} \\
    \includegraphics[width=.33\textwidth]{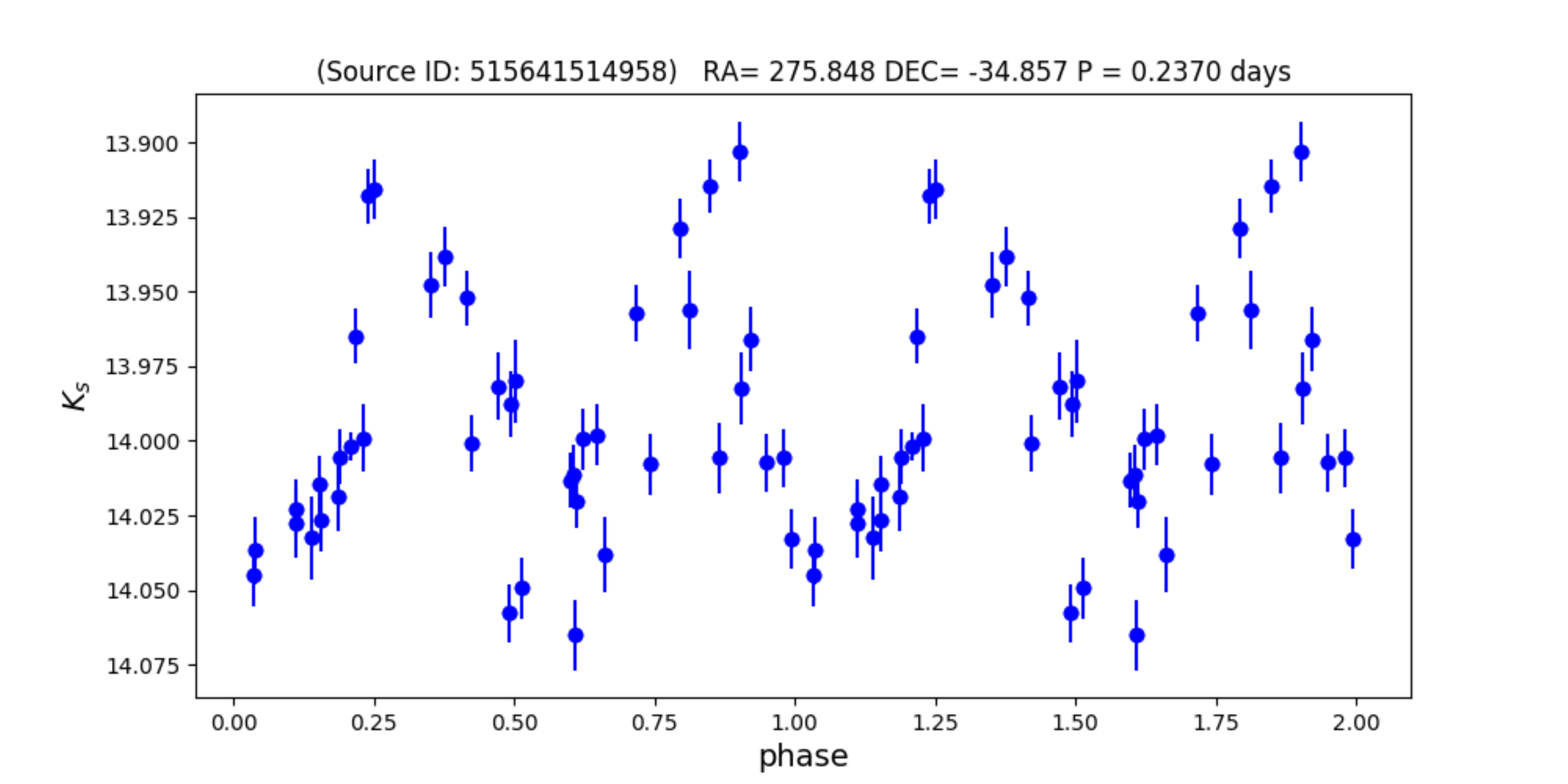} &
    \includegraphics[width=.33\textwidth]{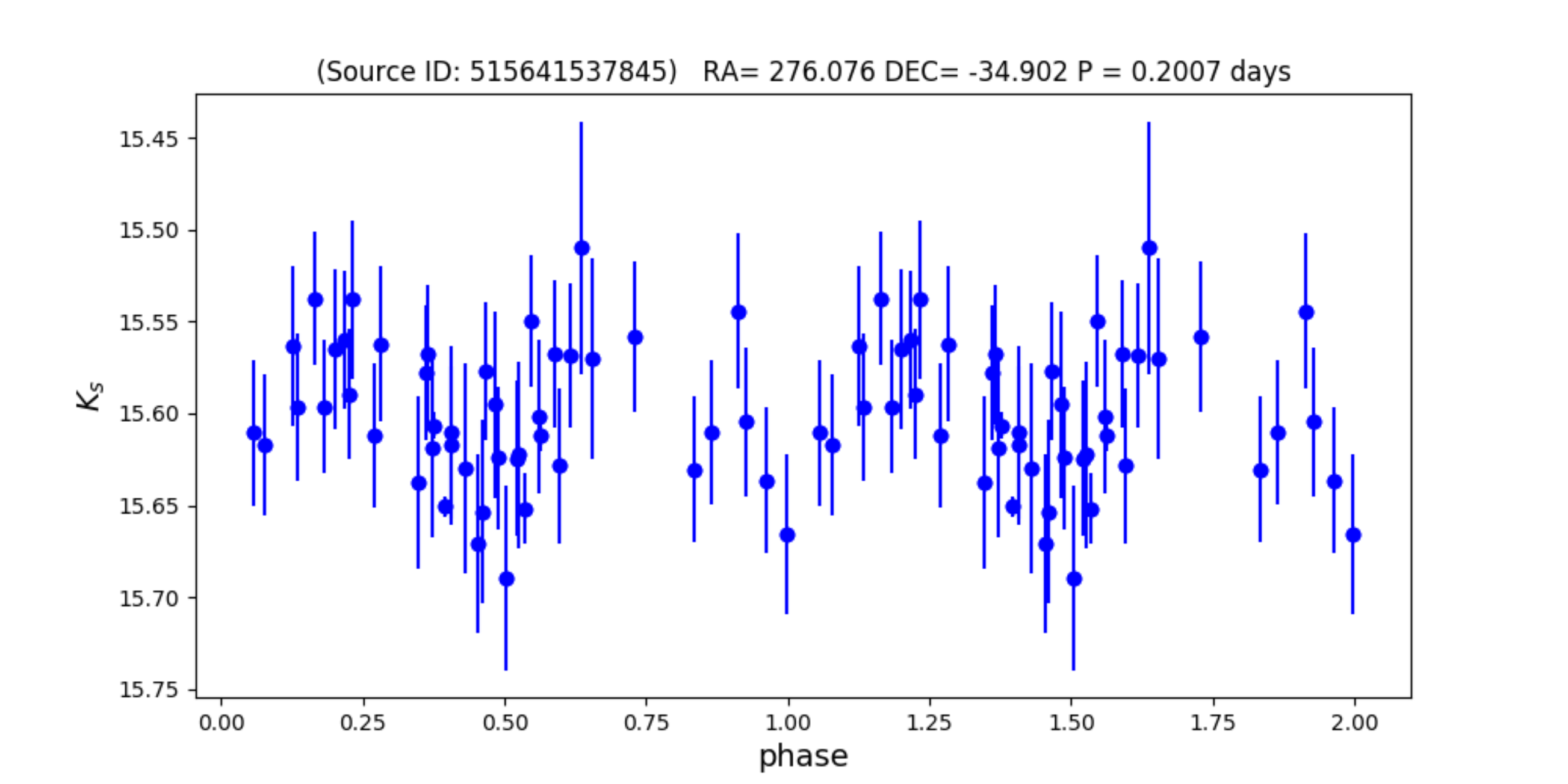} &
    \includegraphics[width=.33\textwidth]{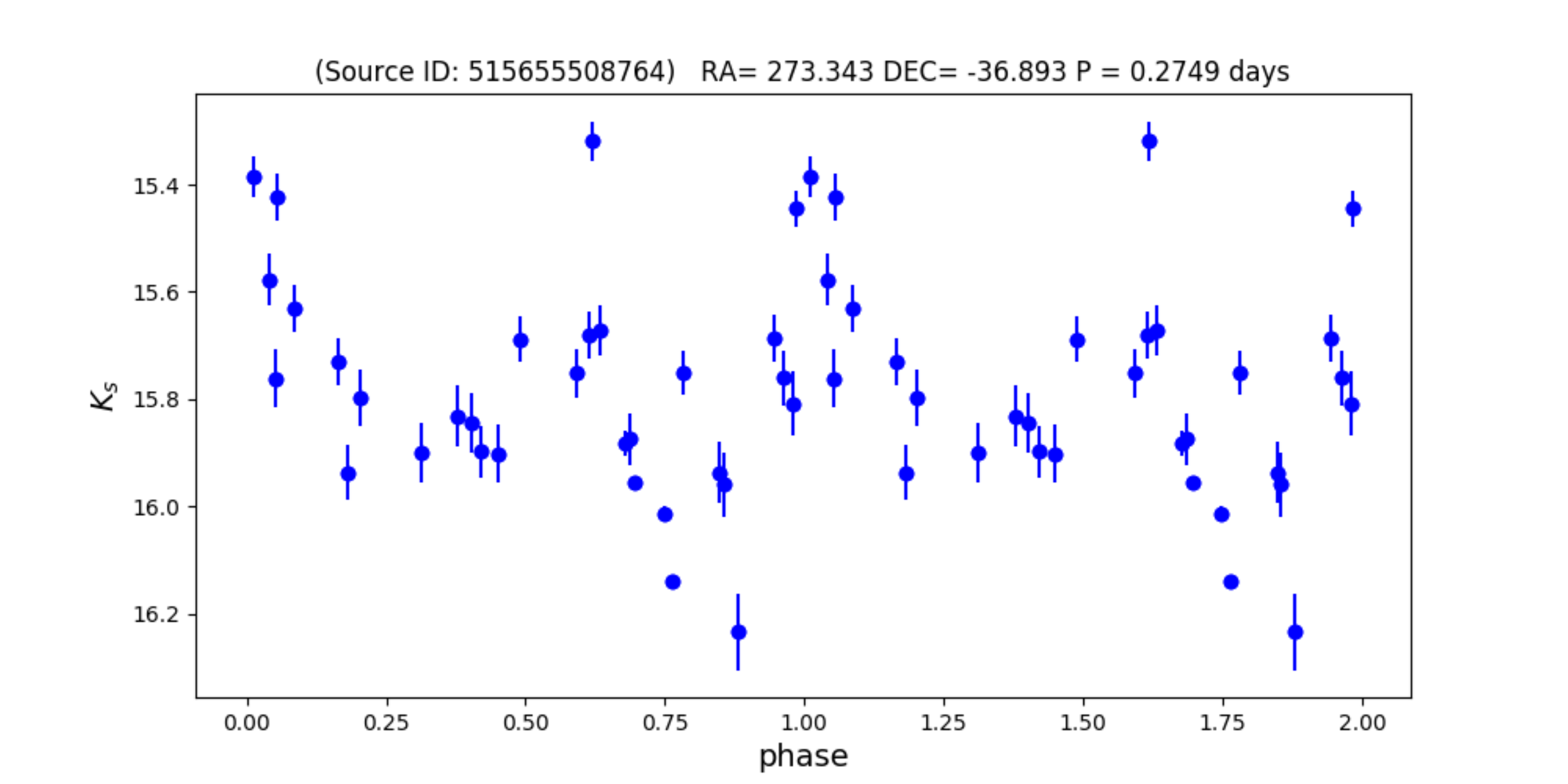} \\    
  \end{tabular}
  \caption{BHB stars light curves corresponding to $2^{nd}$ category eclipsing binaries.}
  \label{fig:bhbB3}
\end{figure*}

\begin{figure*}[htb]
\centering
  \begin{tabular}{@{}ccc@{}}    
    \includegraphics[width=.33\textwidth]{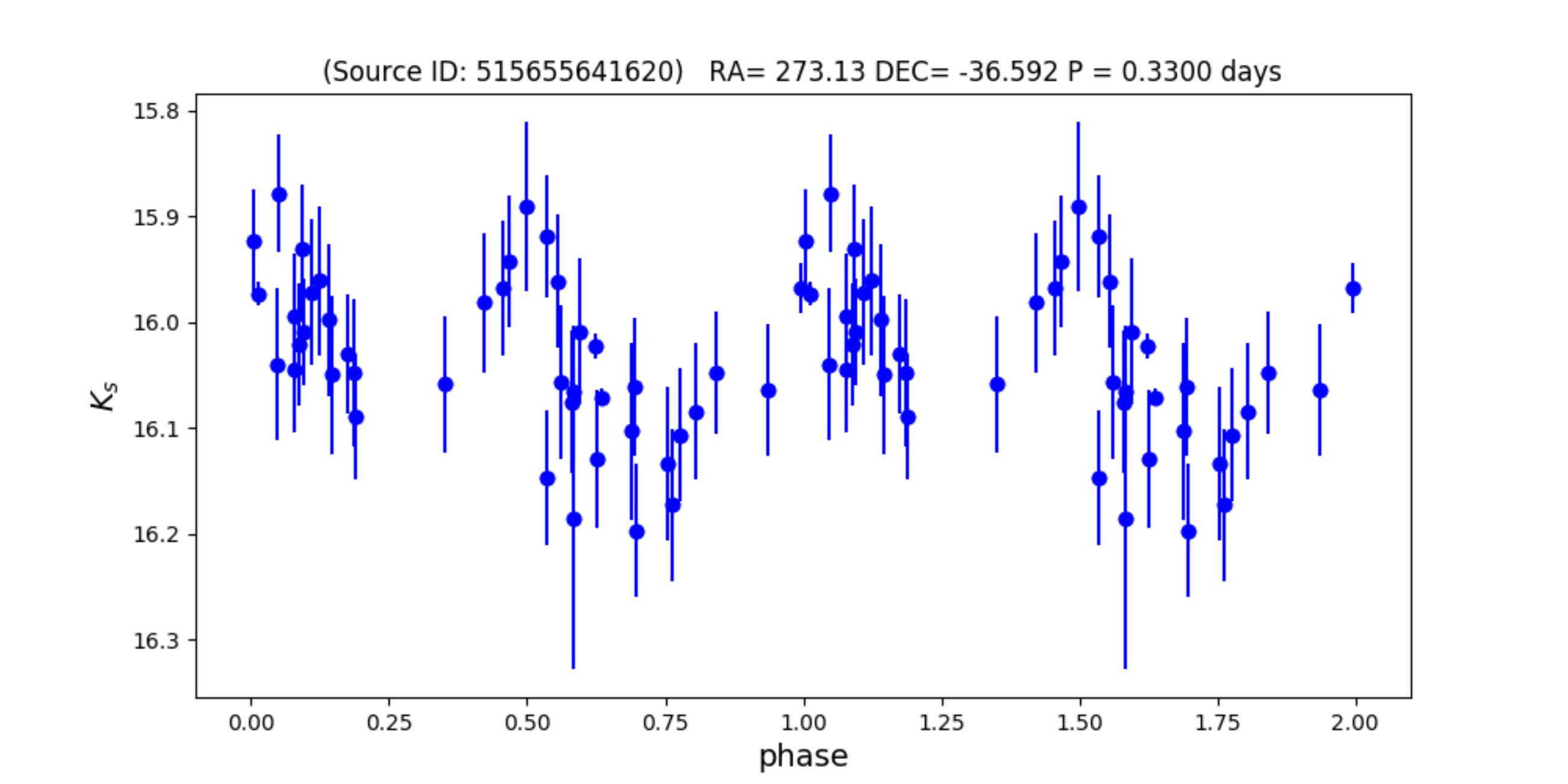} &
    \includegraphics[width=.33\textwidth]{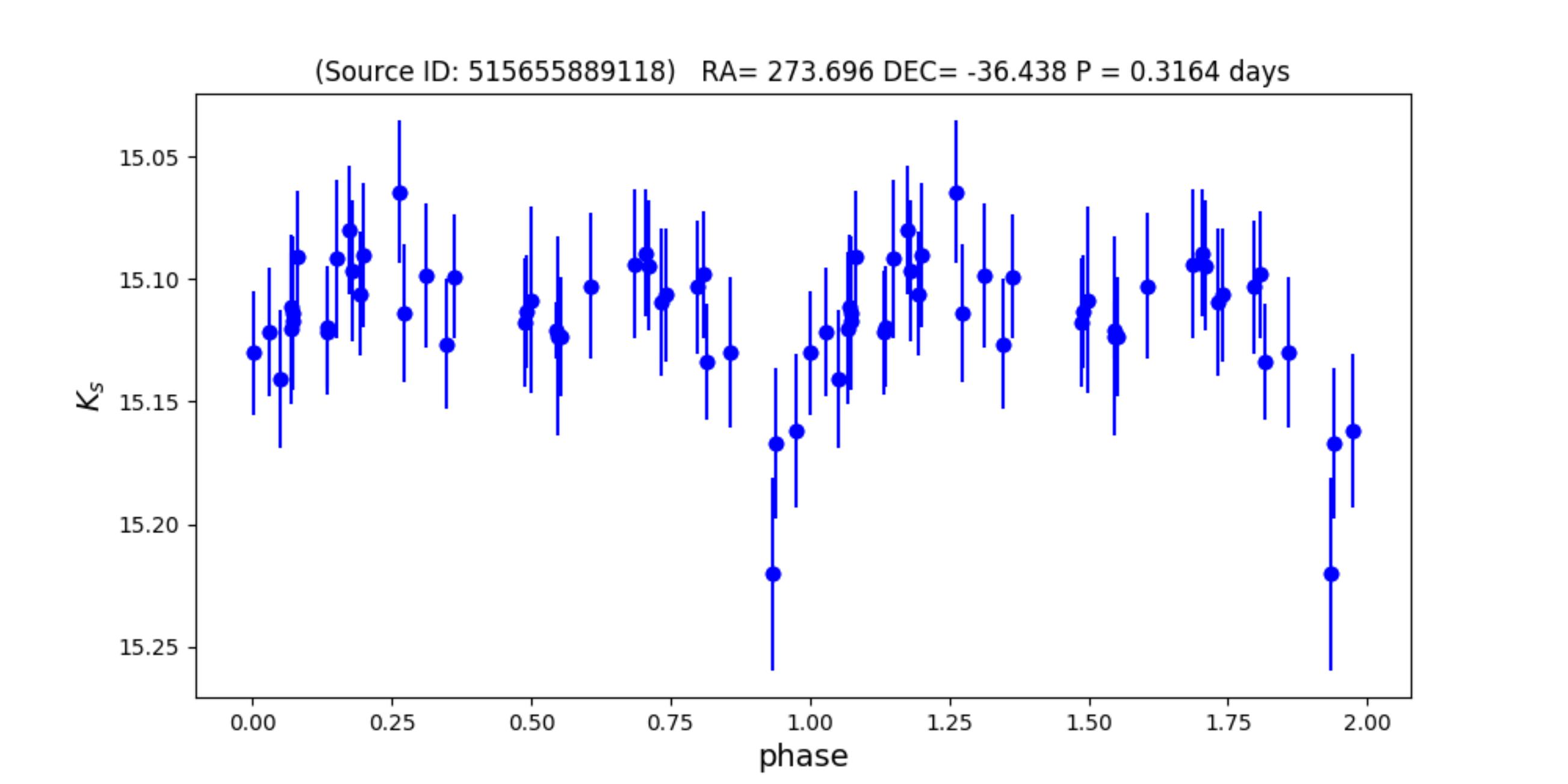} &
    \includegraphics[width=.33\textwidth]{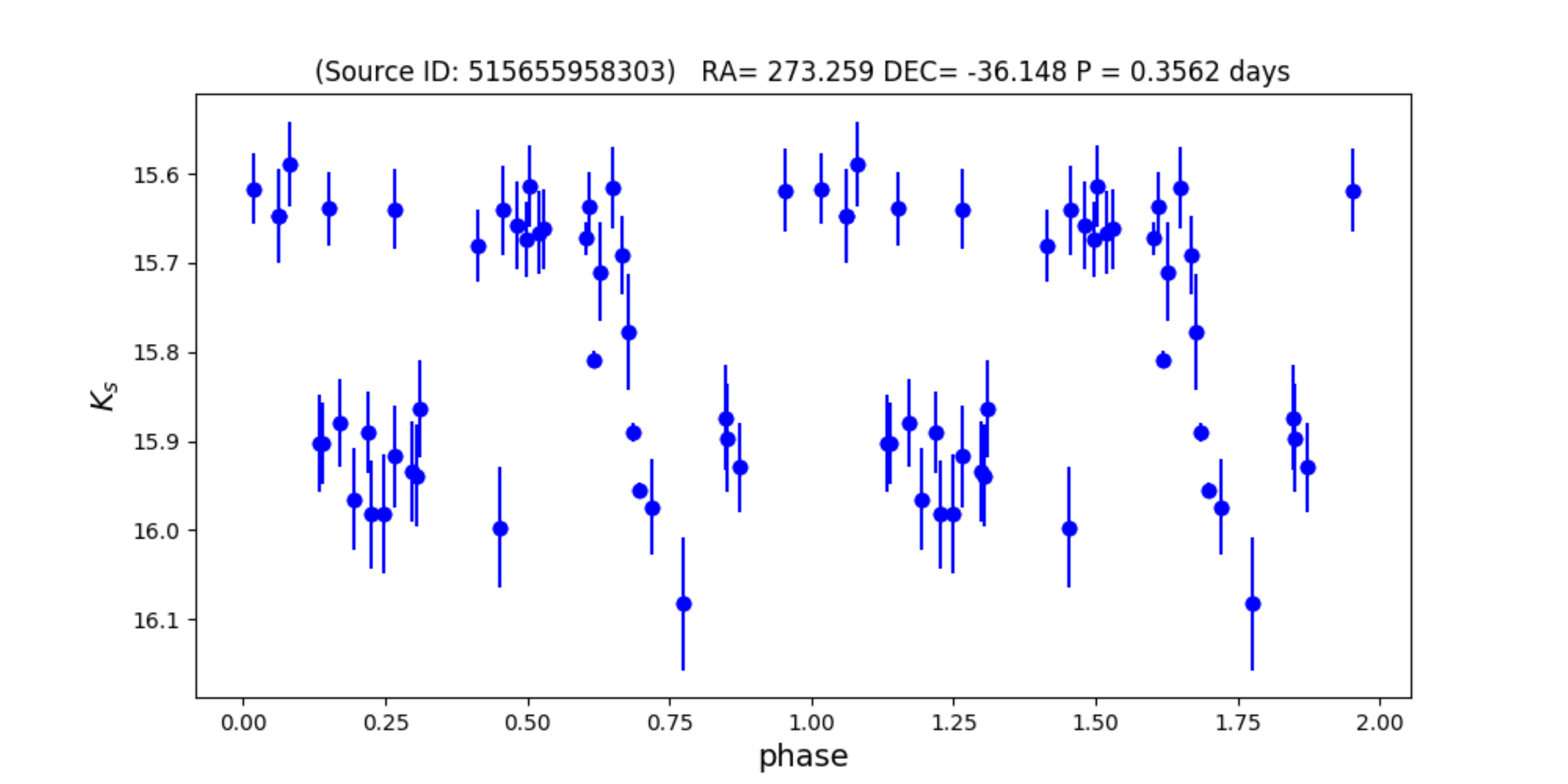} \\   
    \includegraphics[width=.33\textwidth]{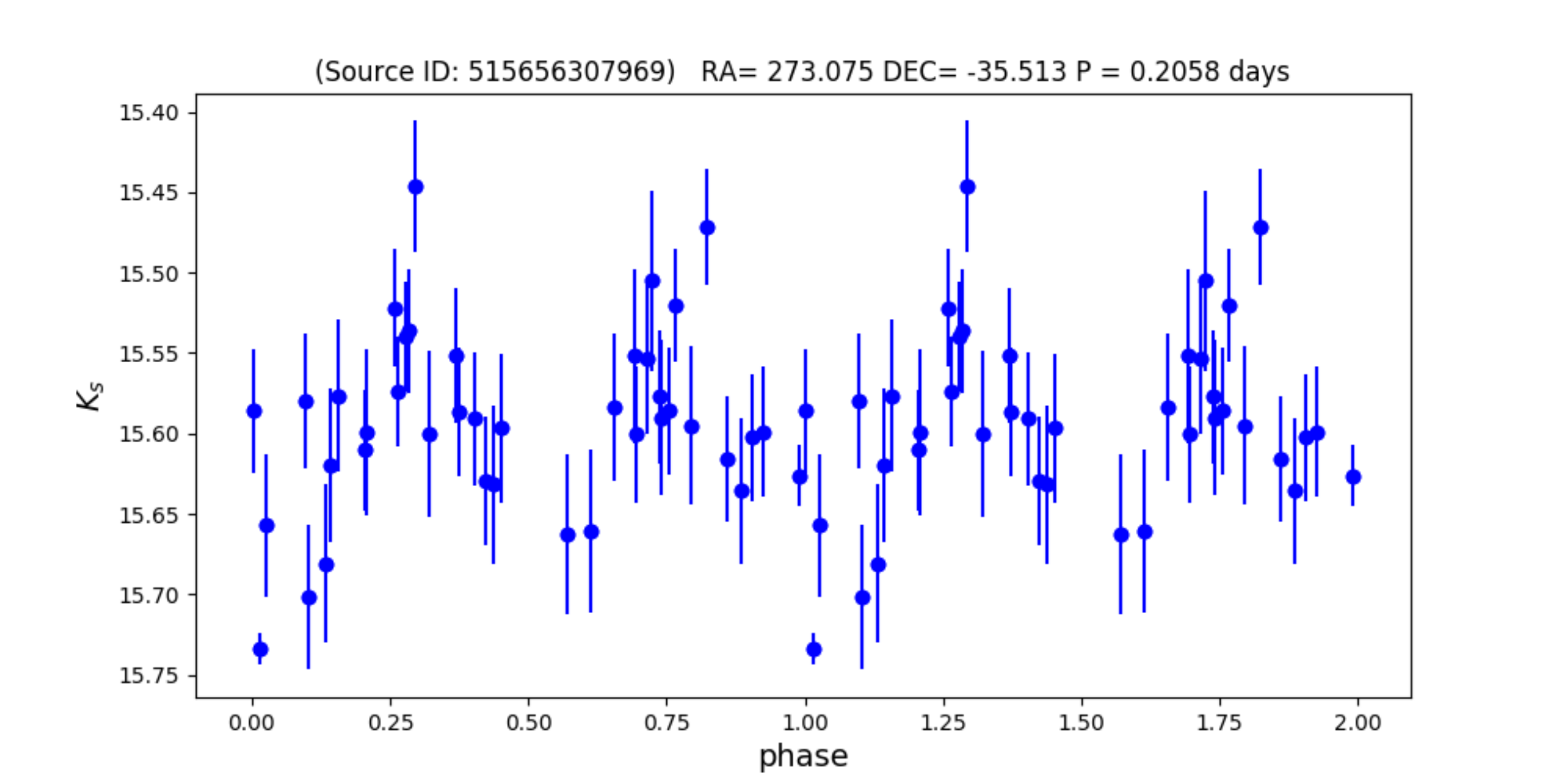} &
    \includegraphics[width=.33\textwidth]{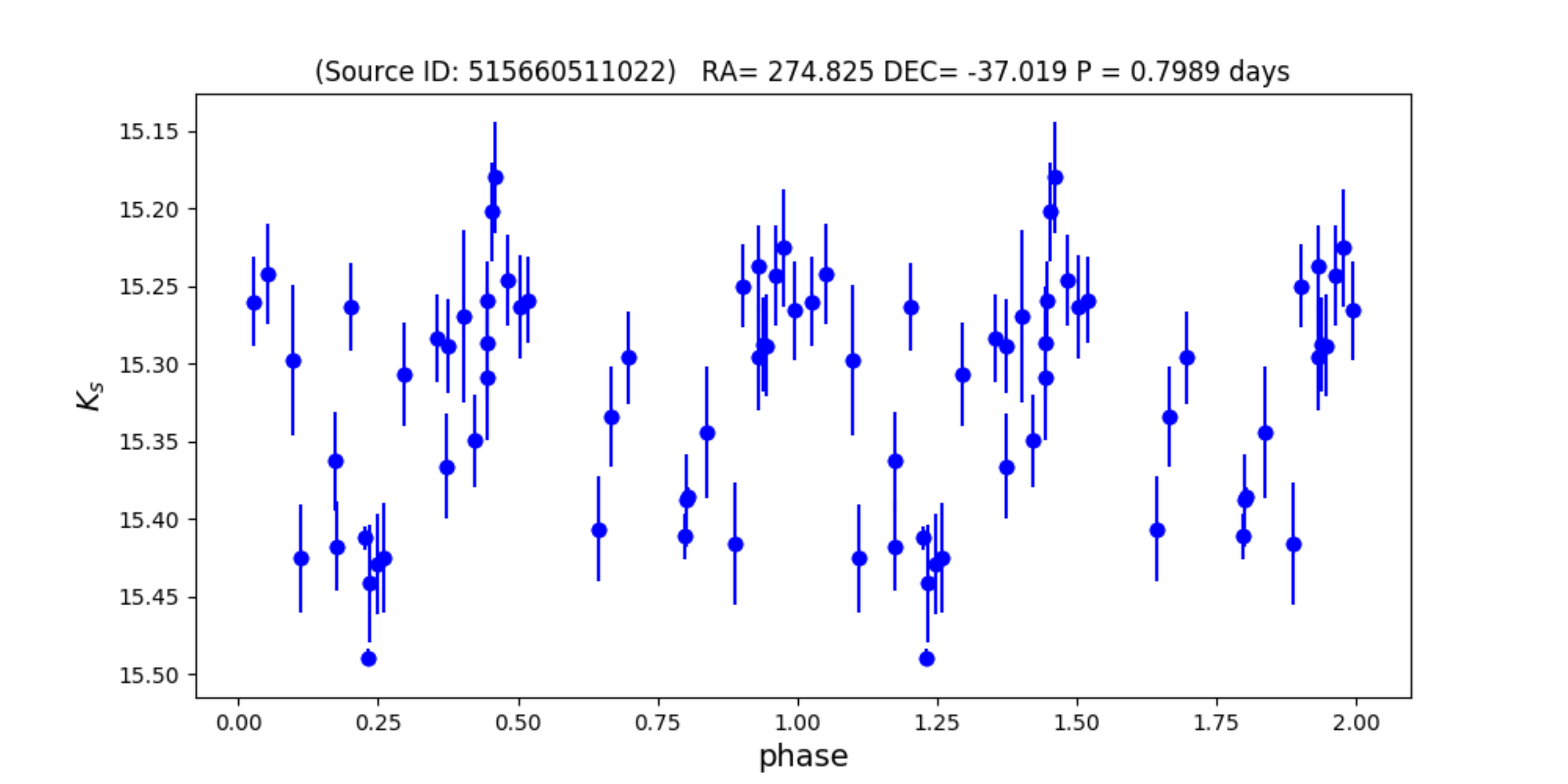} &
    \includegraphics[width=.33\textwidth]{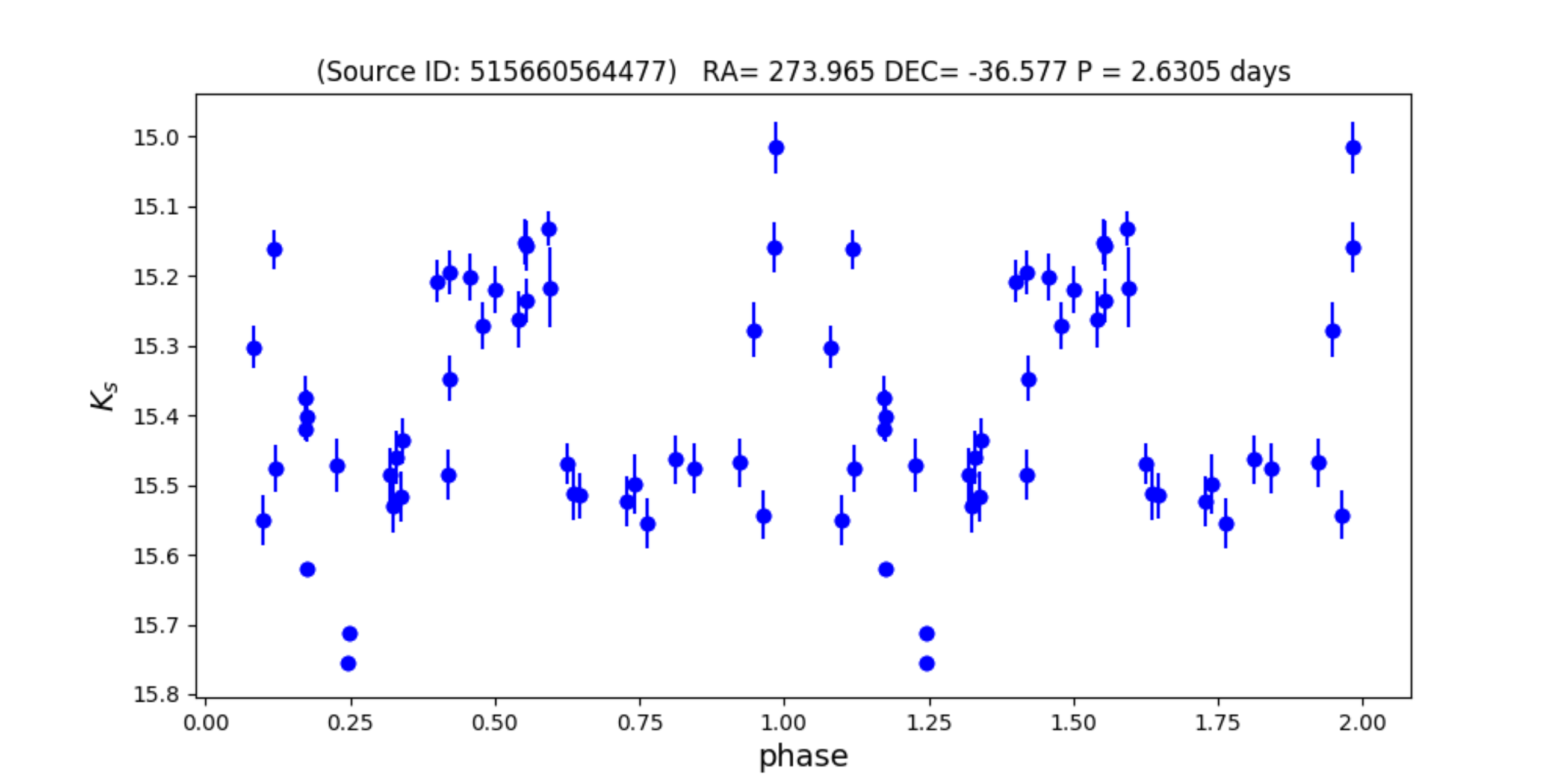} \\
    \includegraphics[width=.33\textwidth]{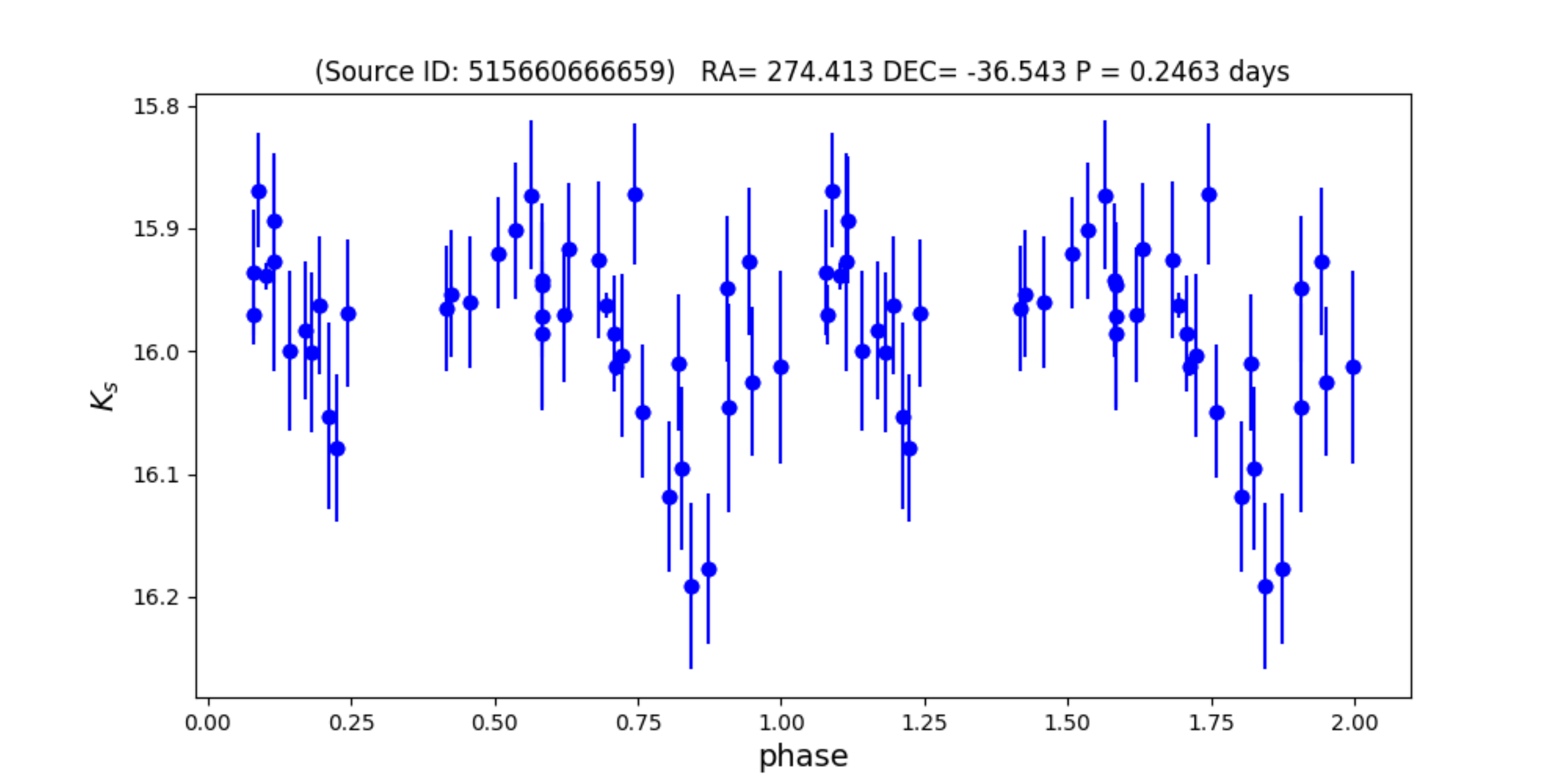} &
    \includegraphics[width=.33\textwidth]{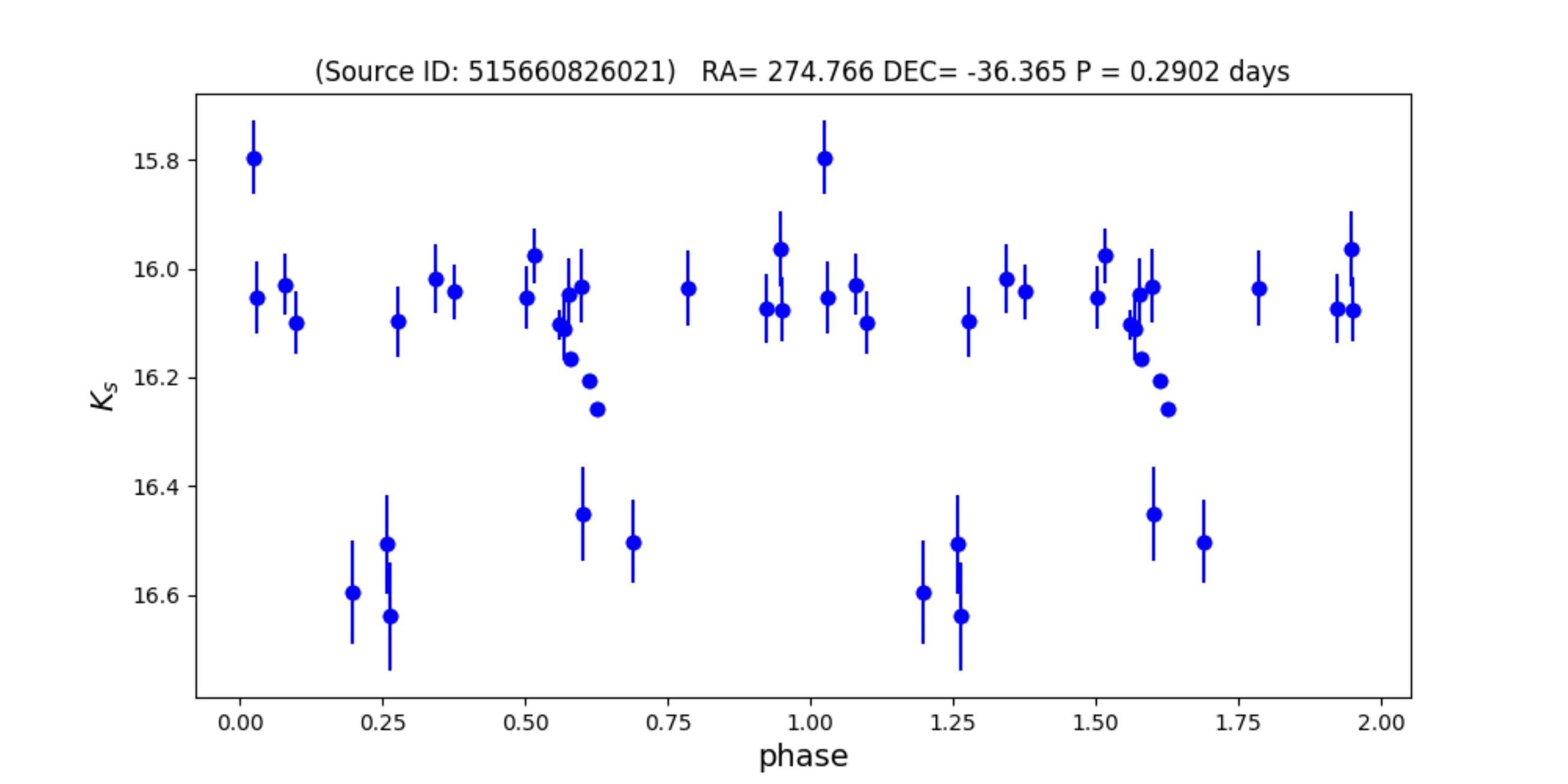} &
    \includegraphics[width=.33\textwidth]{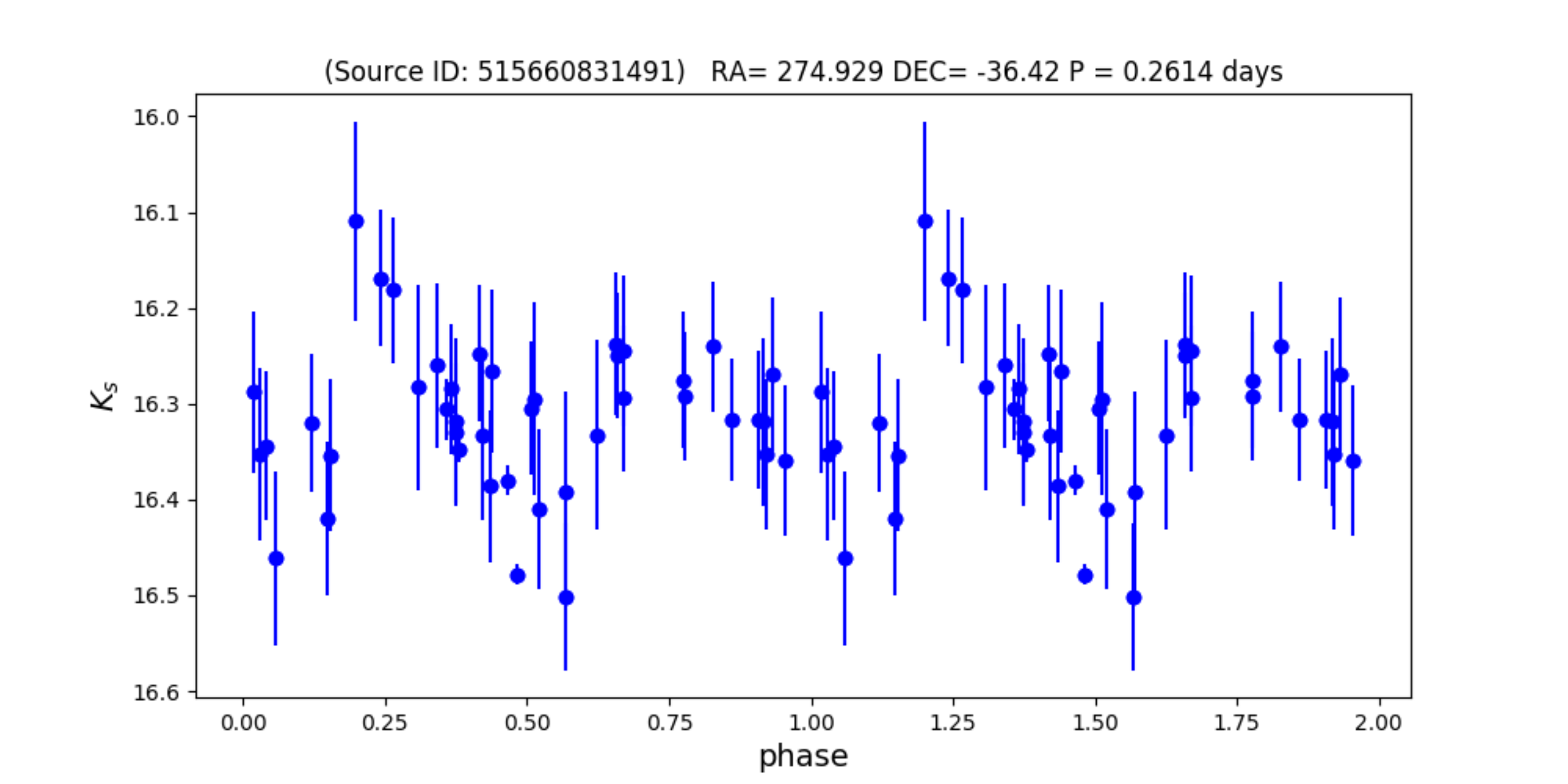} \\    
    \includegraphics[width=.33\textwidth]{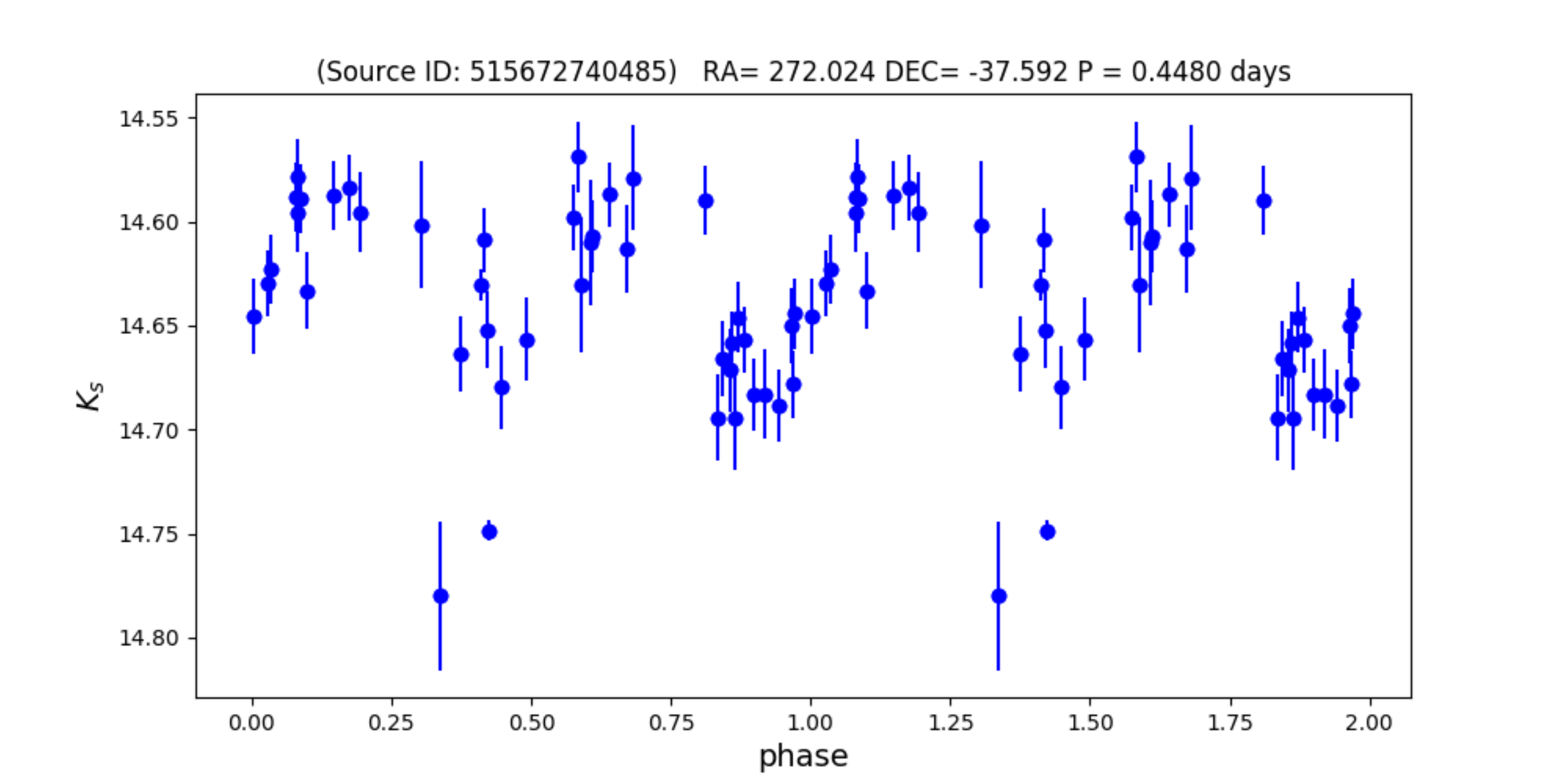} &        
    \includegraphics[width=.33\textwidth]{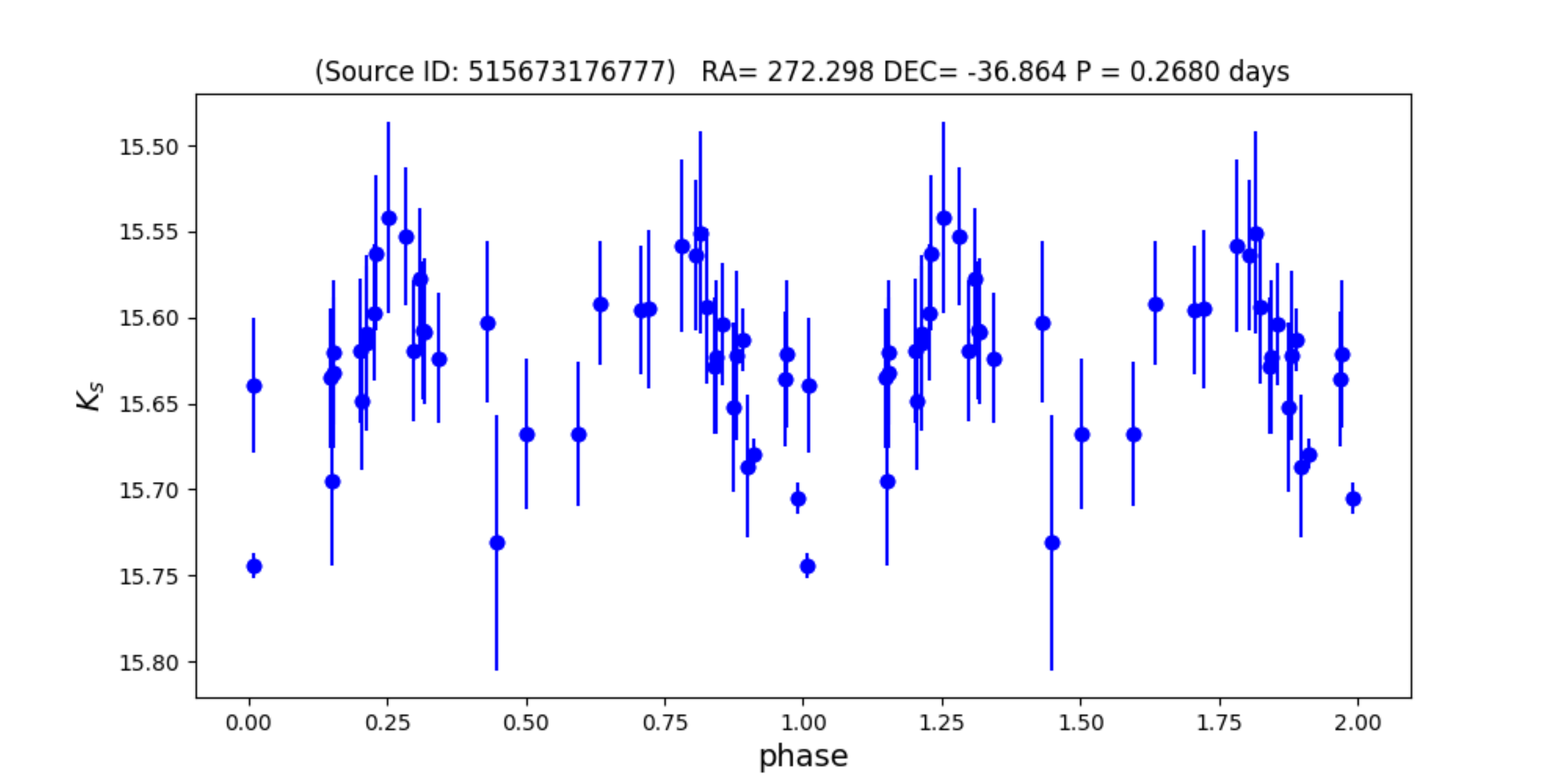} &
    \includegraphics[width=.33\textwidth]{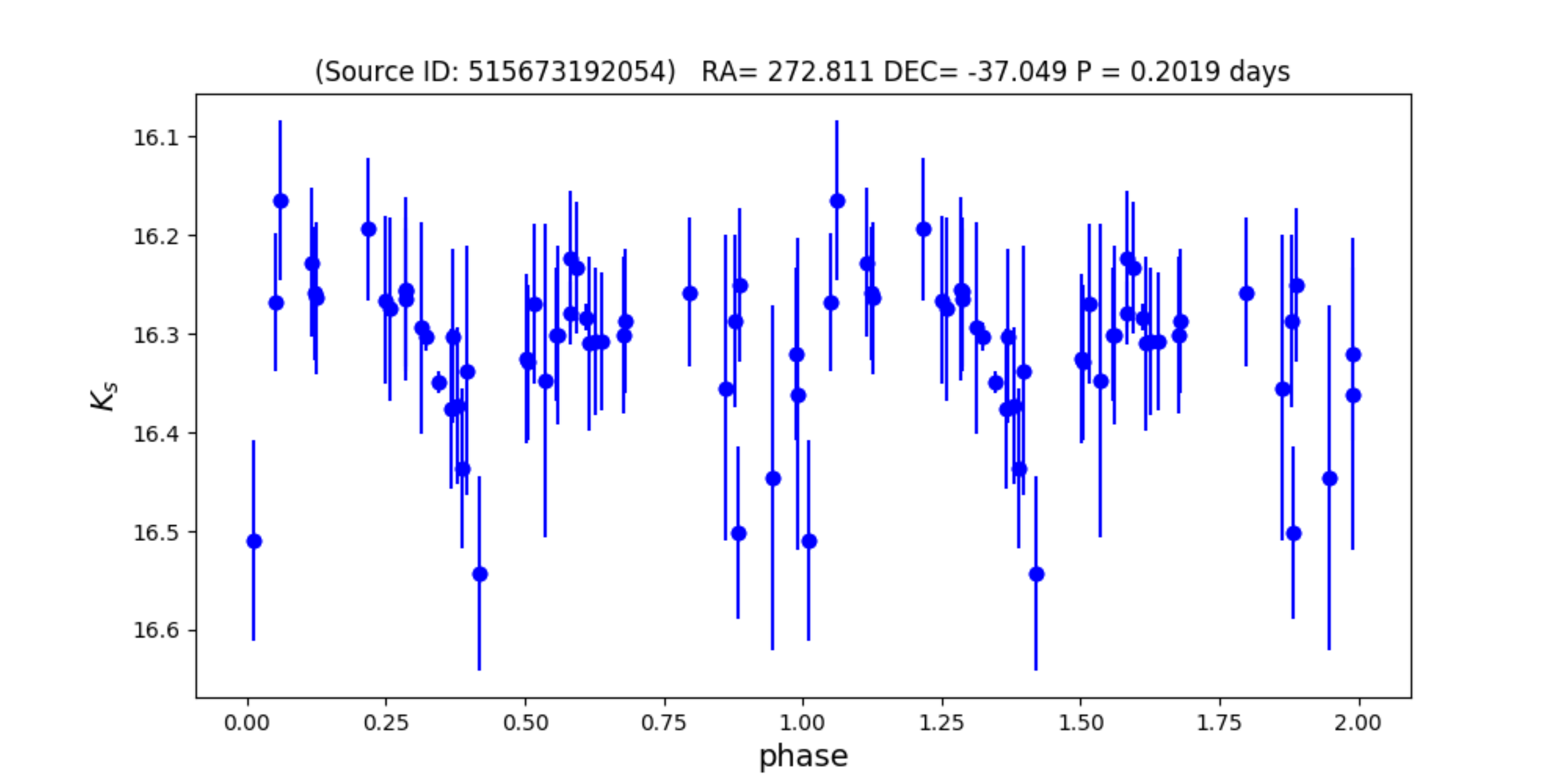} \\    
    \includegraphics[width=.33\textwidth]{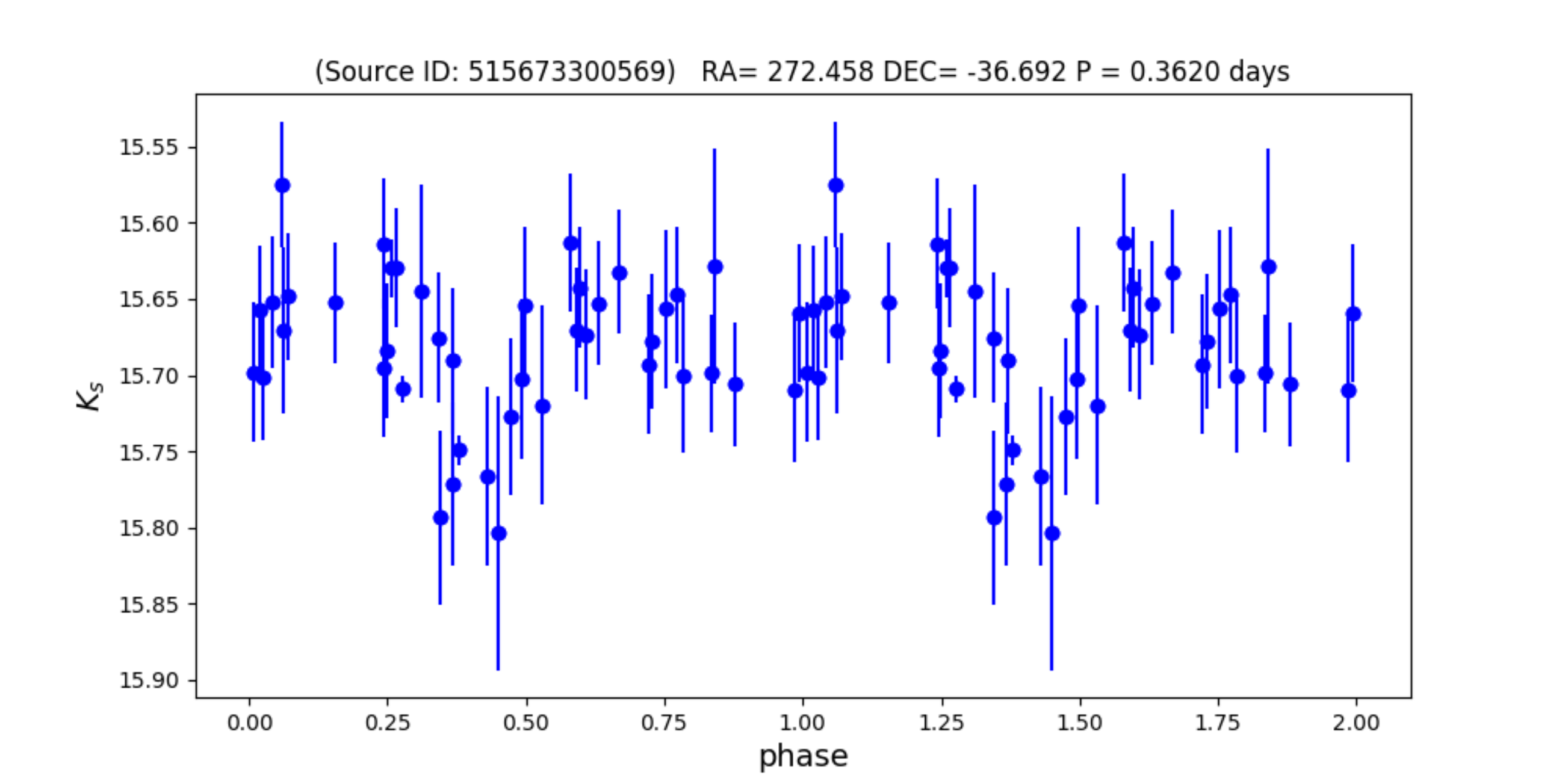} &
    \includegraphics[width=.33\textwidth]{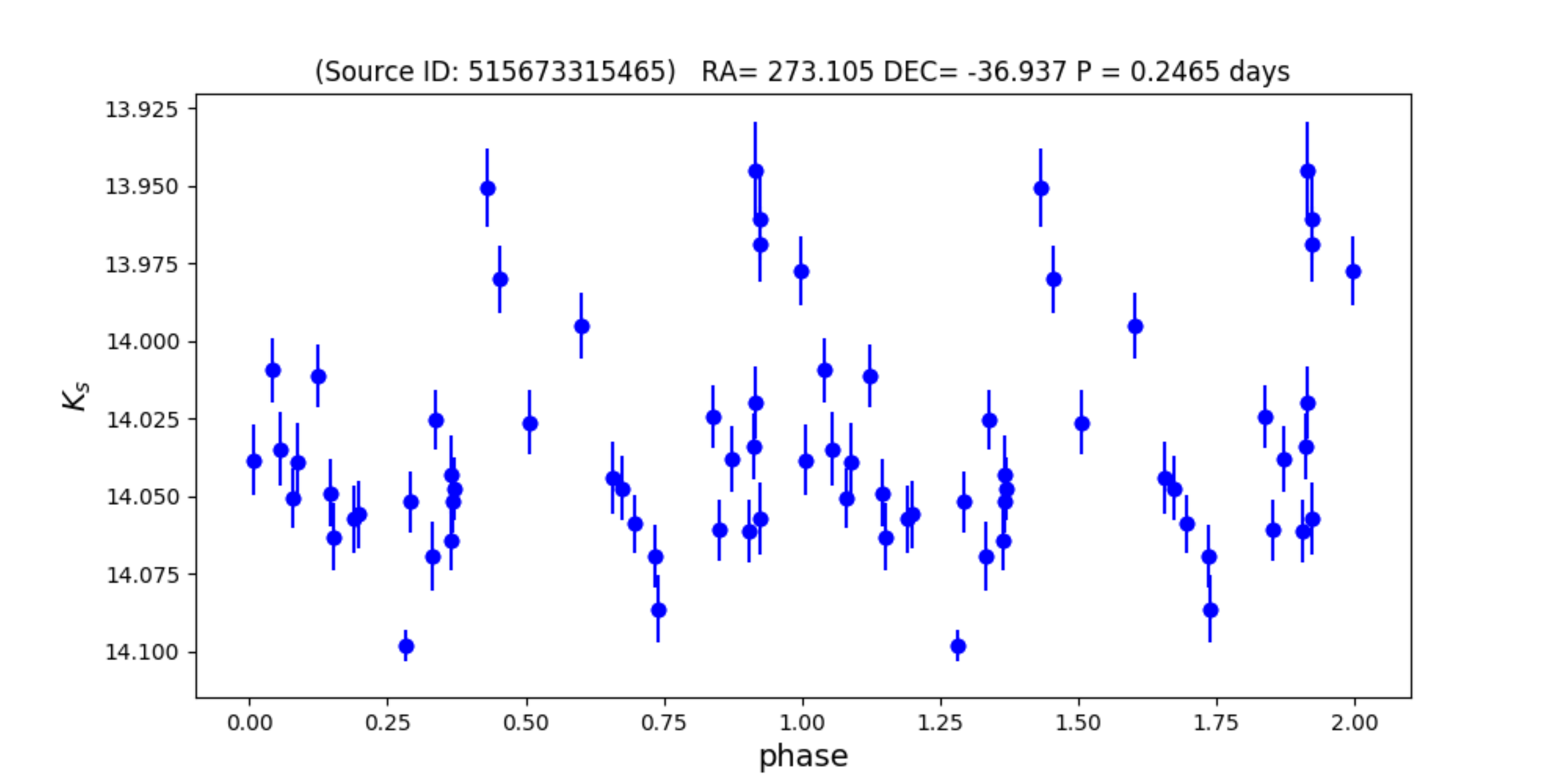} &
    \includegraphics[width=.33\textwidth]{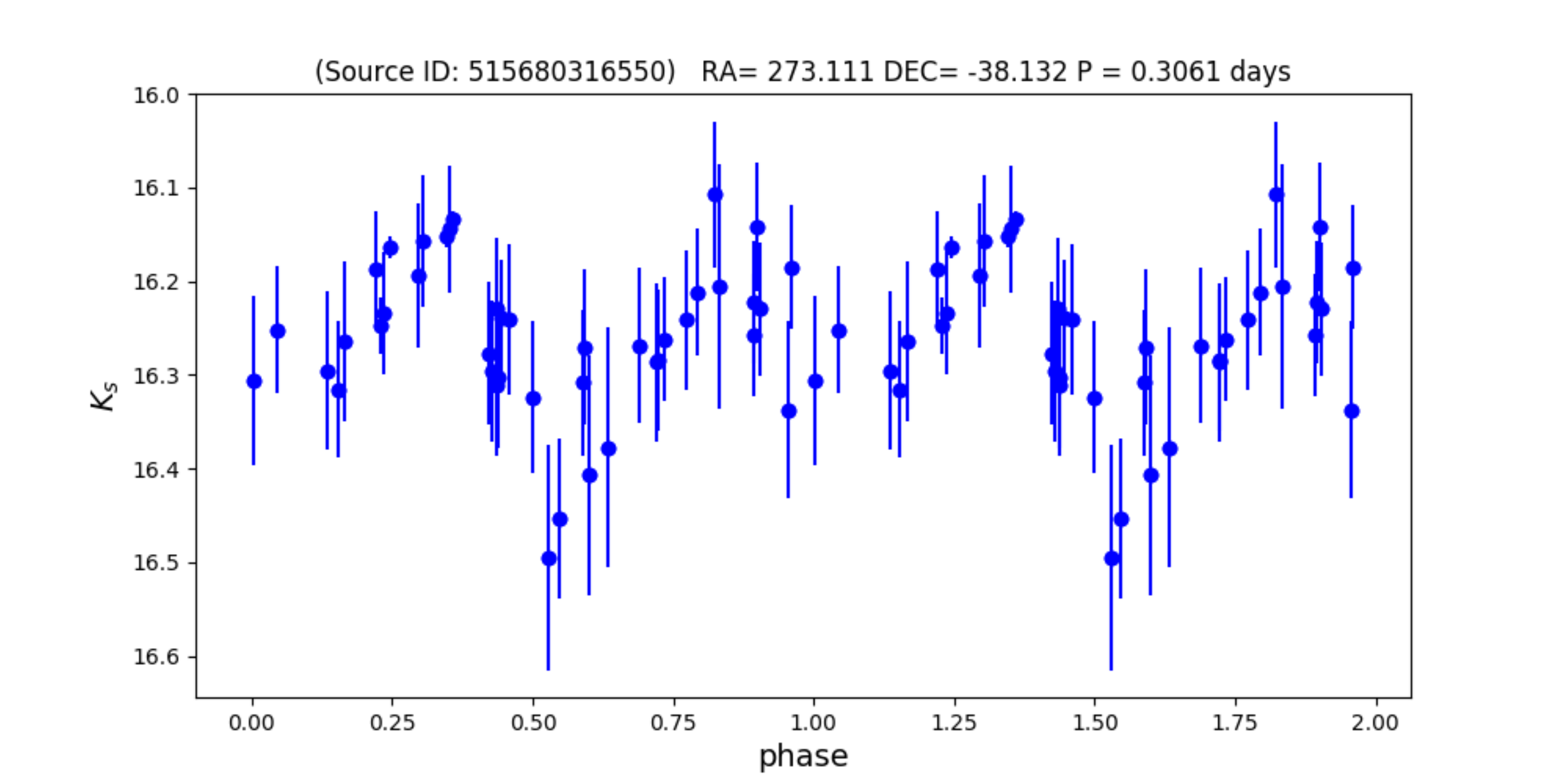} \\
    \includegraphics[width=.33\textwidth]{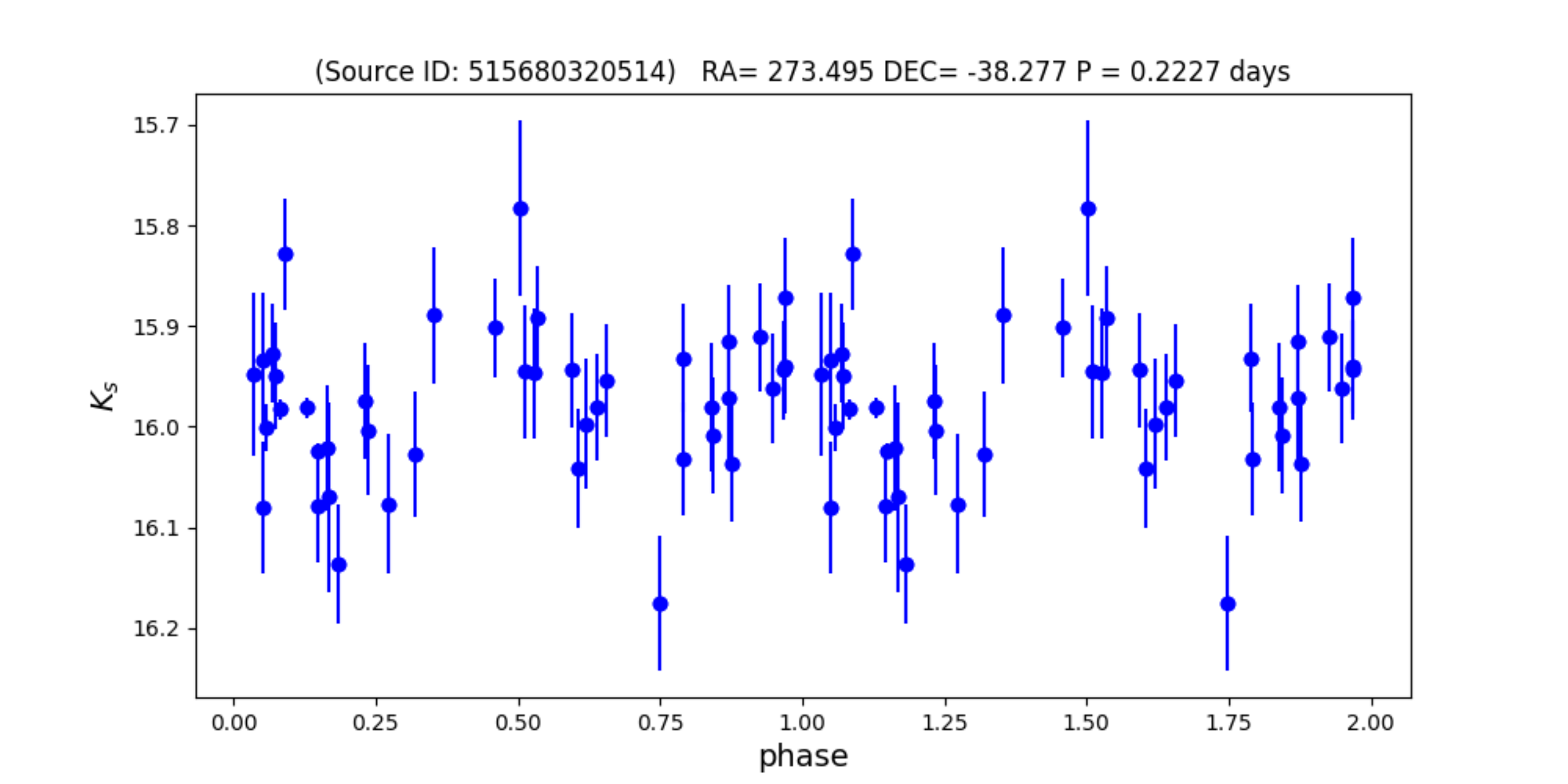} &    
    \includegraphics[width=.33\textwidth]{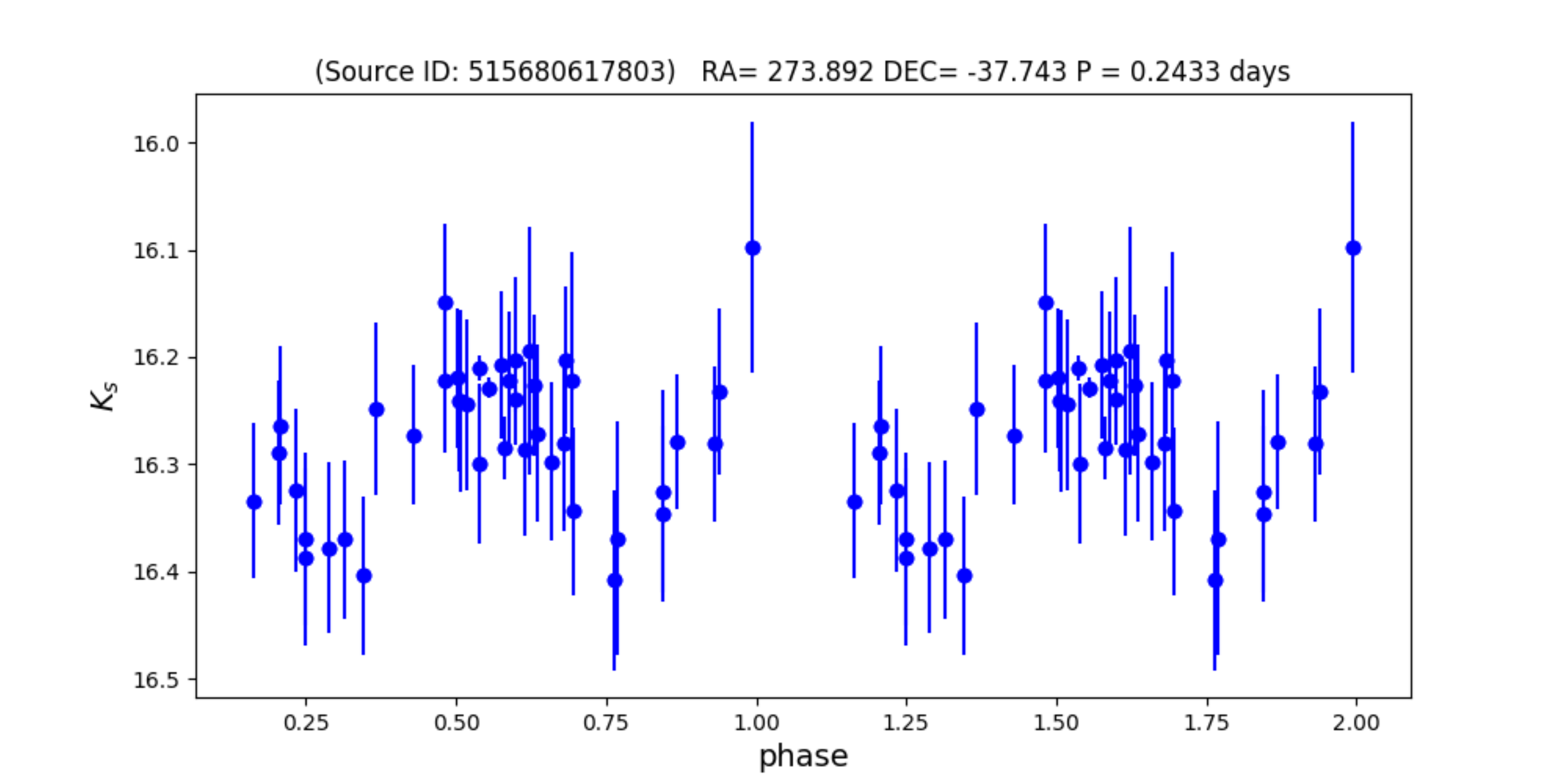} &
    \includegraphics[width=.33\textwidth]{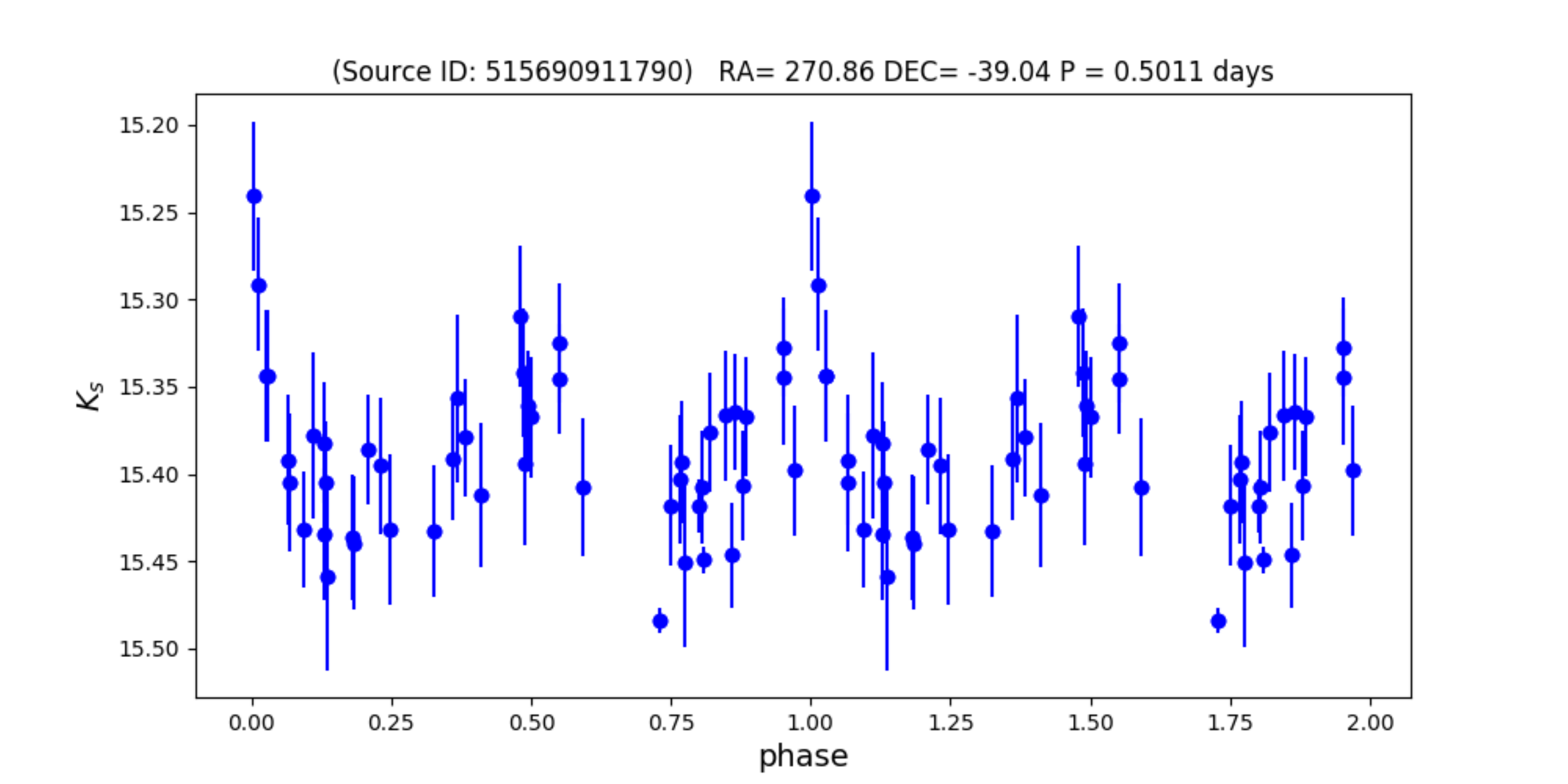} \\
    \includegraphics[width=.33\textwidth]{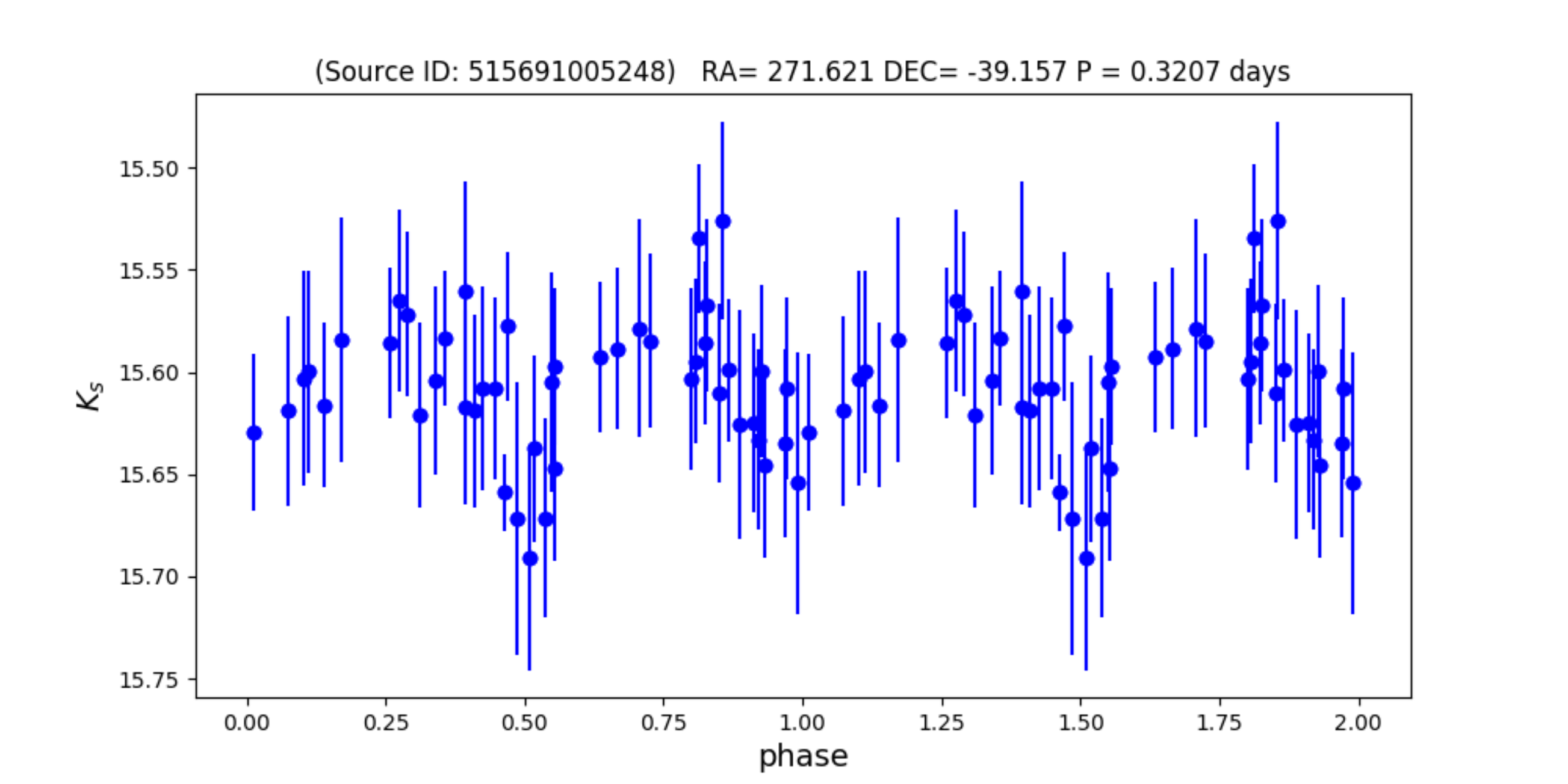} &    
    \includegraphics[width=.33\textwidth]{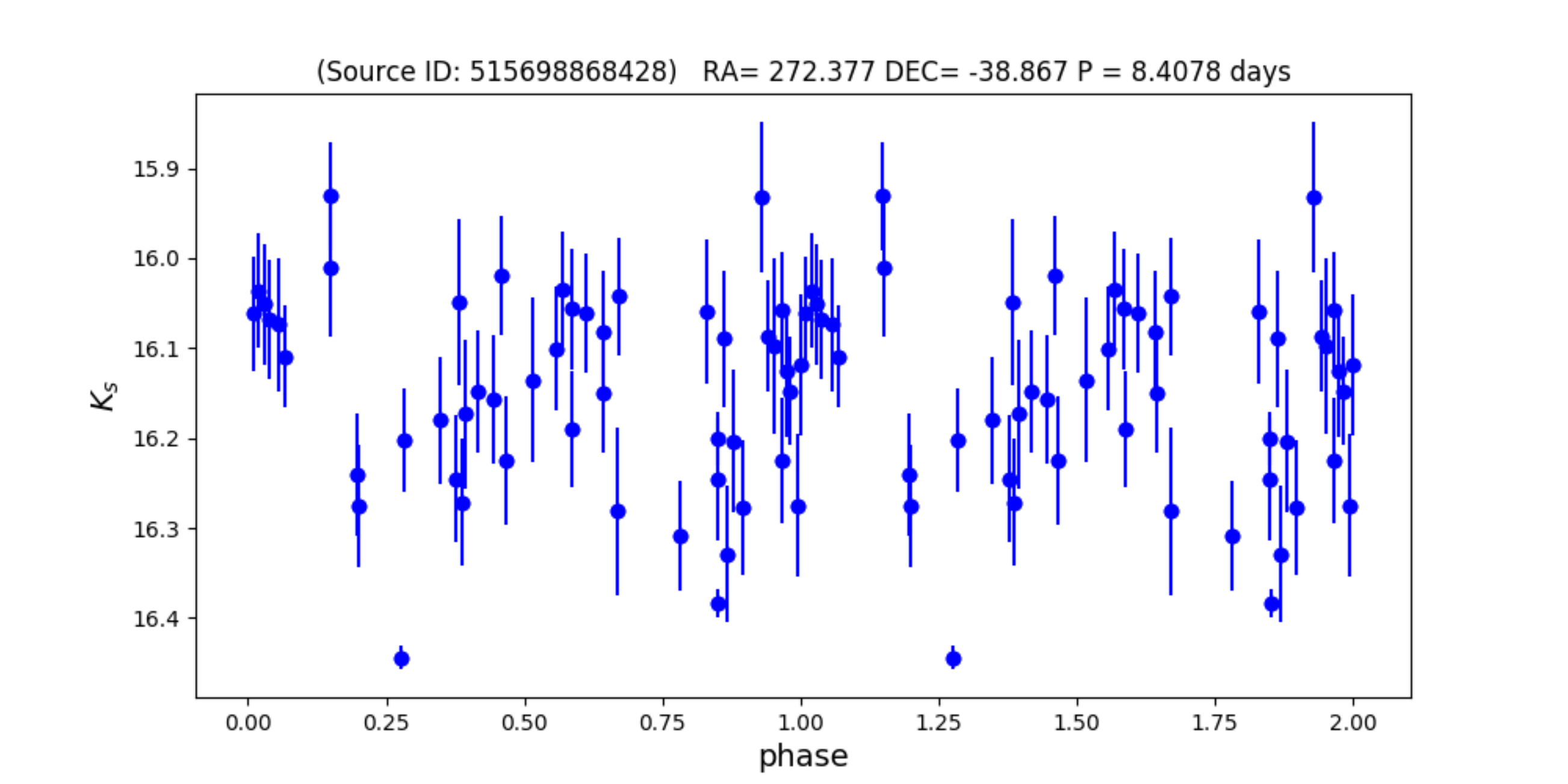} &    
    \includegraphics[width=.33\textwidth]{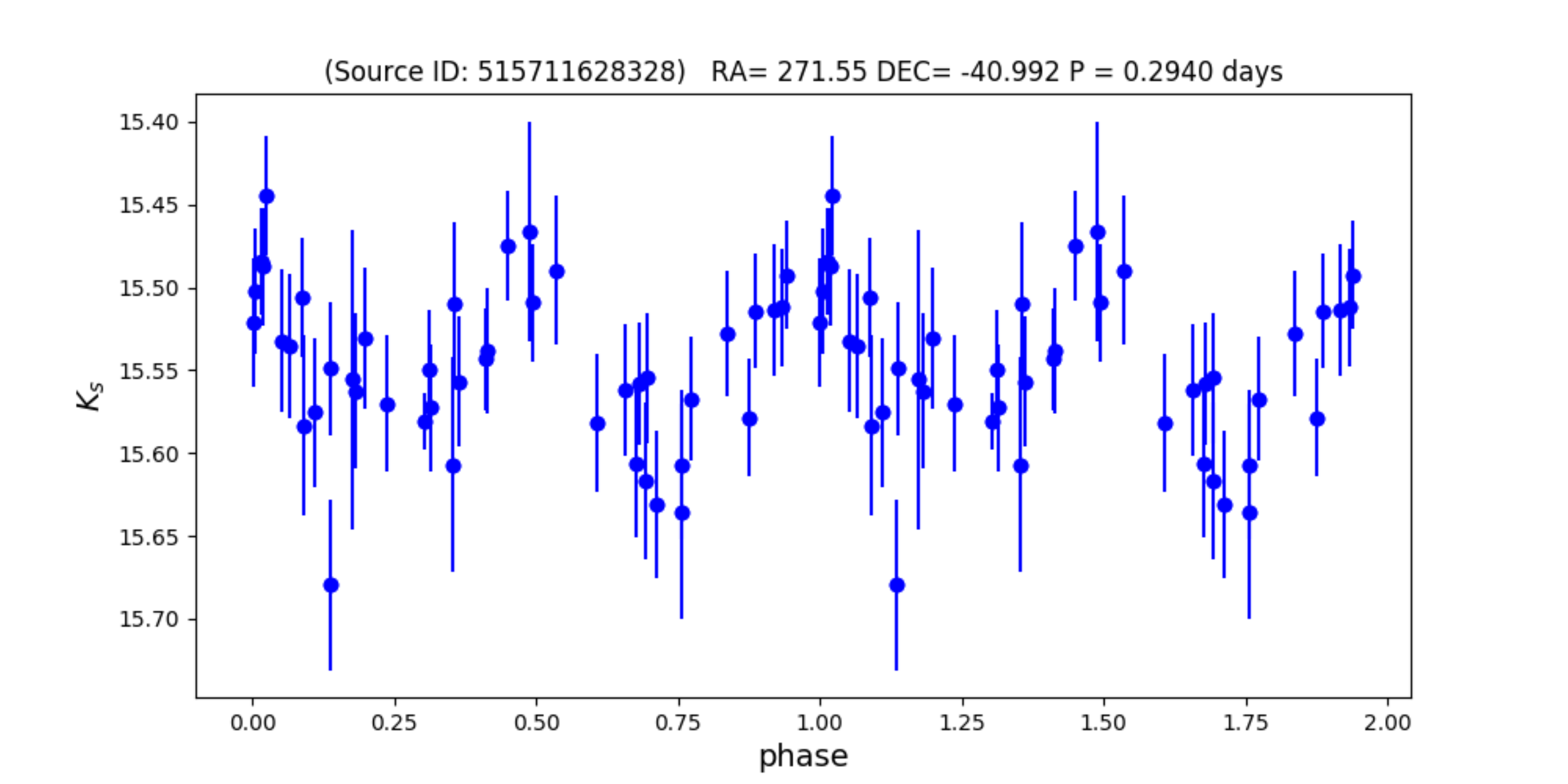} \\
    \includegraphics[width=.33\textwidth]{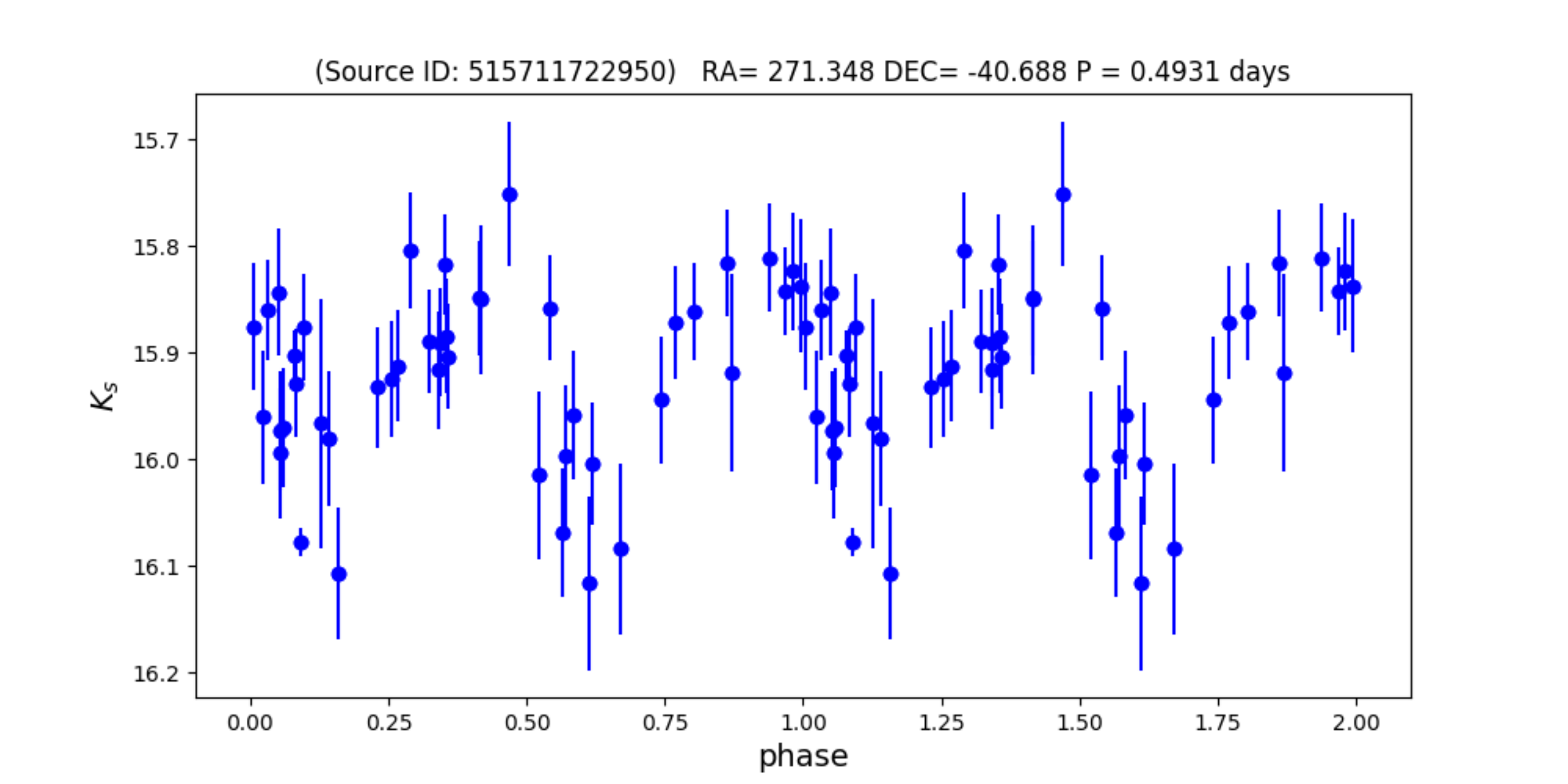} &
    \includegraphics[width=.33\textwidth]{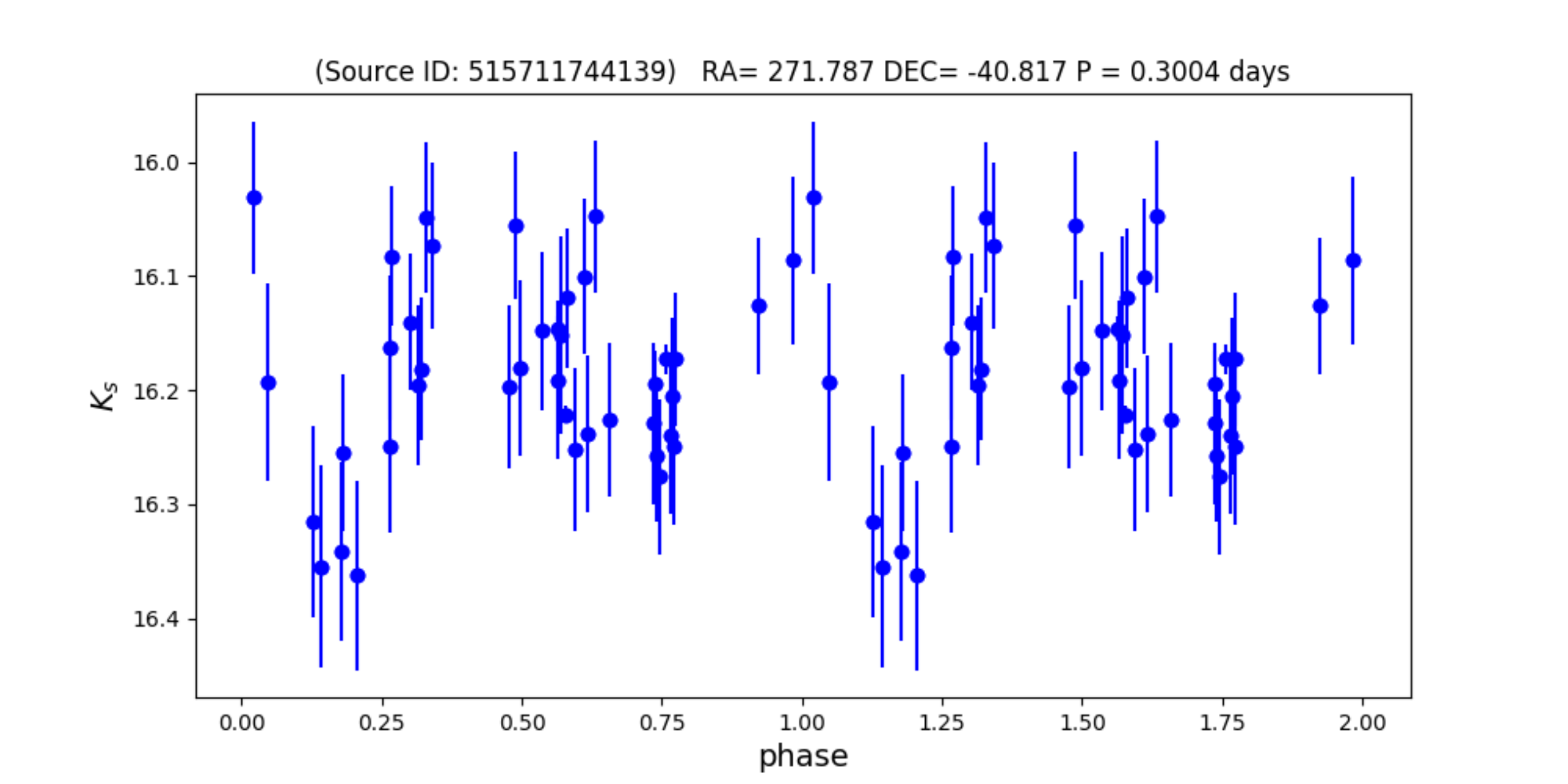} &
    \includegraphics[width=.33\textwidth]{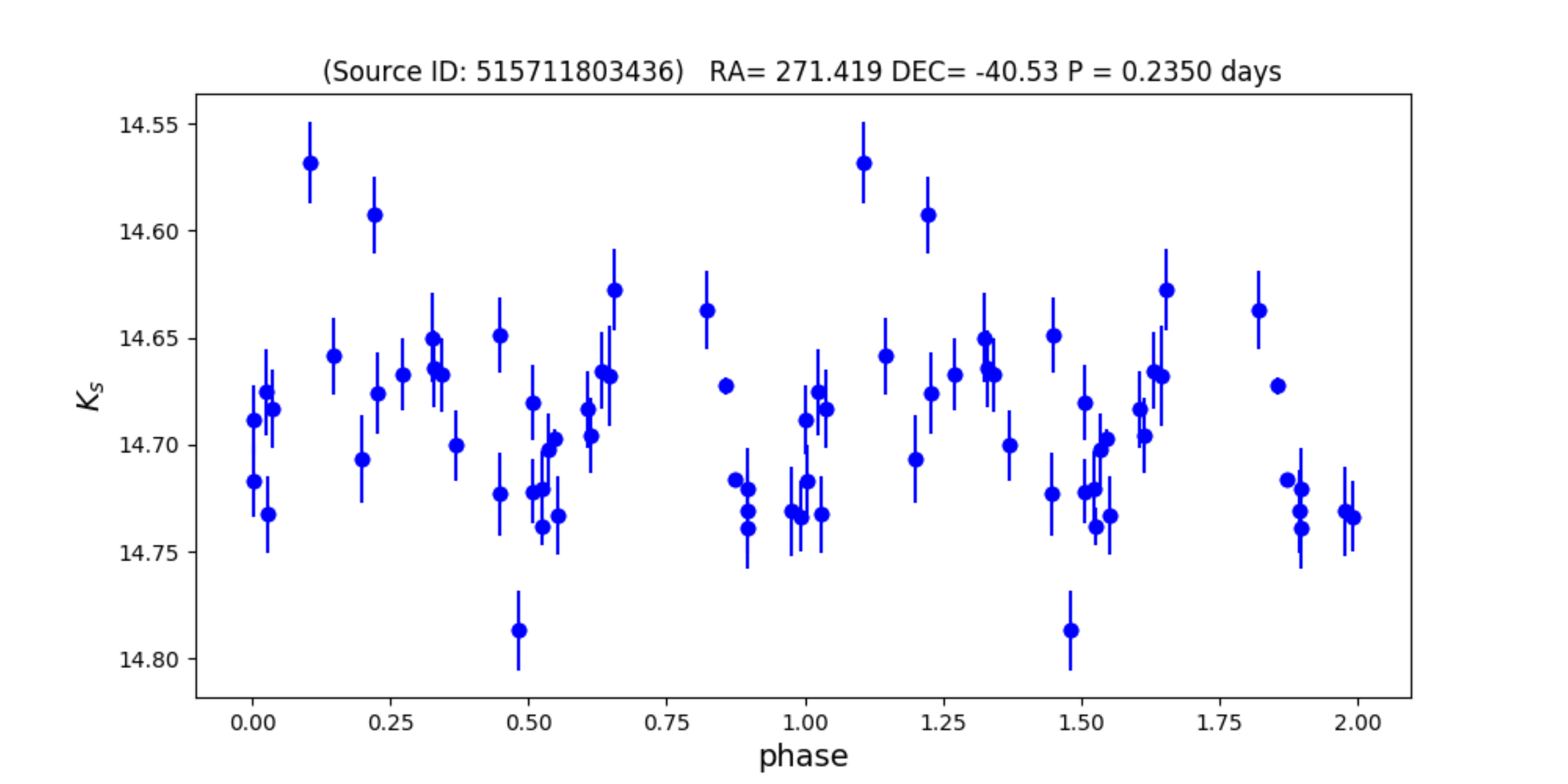} \\
  \end{tabular}
  \caption{BHB stars light curves corresponding to $2^{nd}$ category eclipsing binaries.}
  \label{fig:bhbB4}
\end{figure*}

\begin{figure*}[htb]
\centering
  \begin{tabular}{@{}ccc@{}}    
    \includegraphics[width=.33\textwidth]{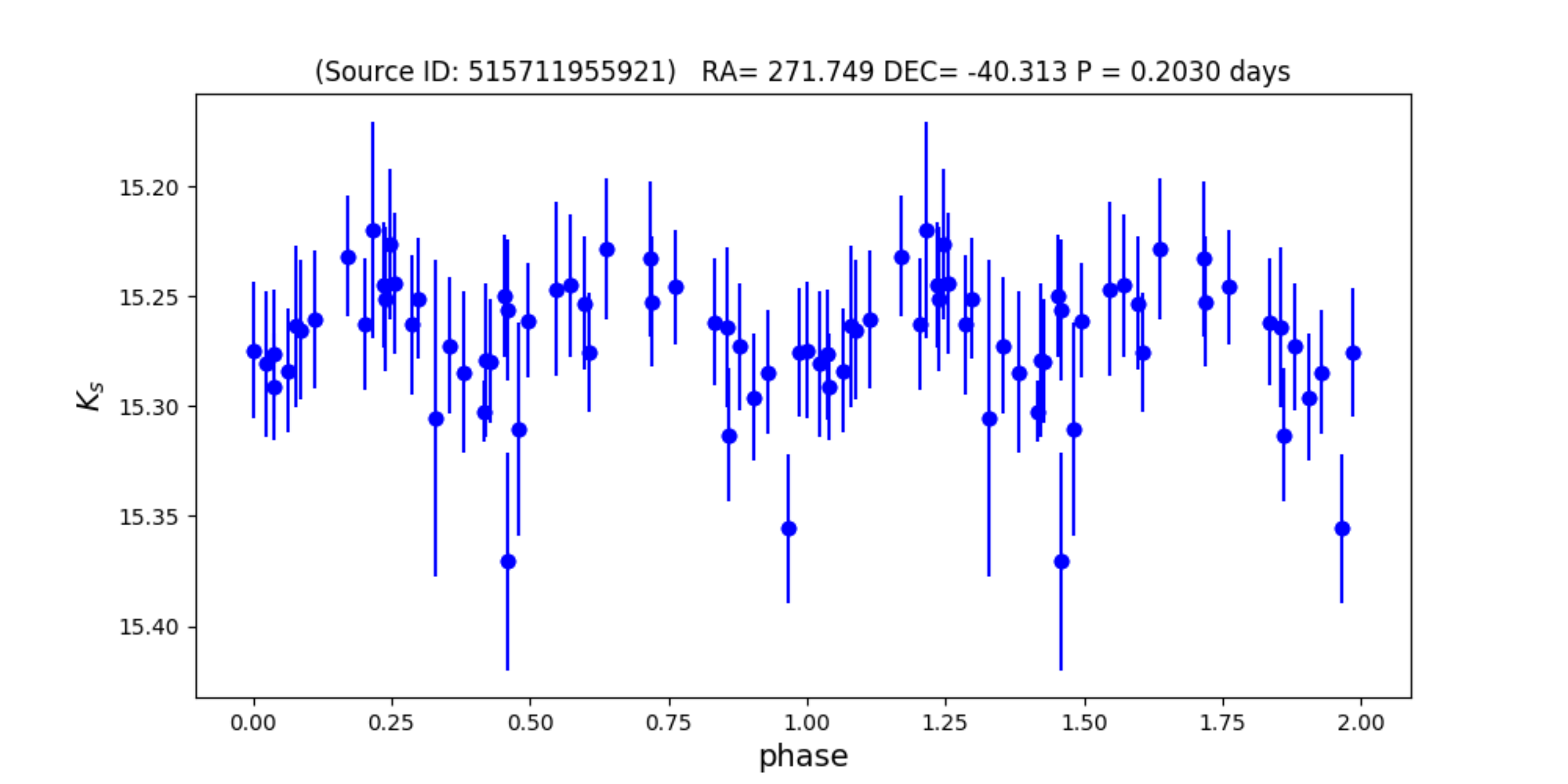} &
    \includegraphics[width=.33\textwidth]{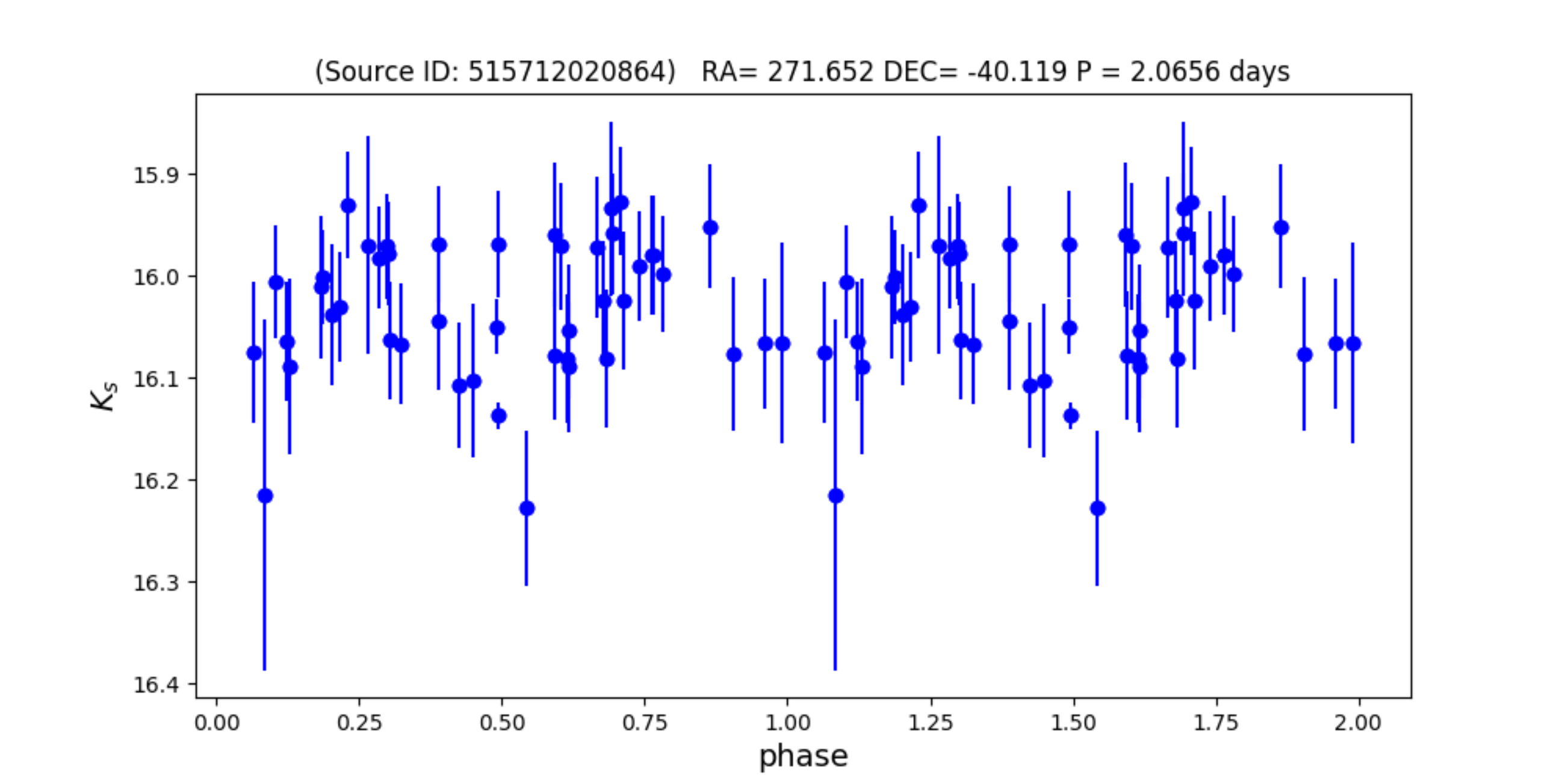} &
    \includegraphics[width=.33\textwidth]{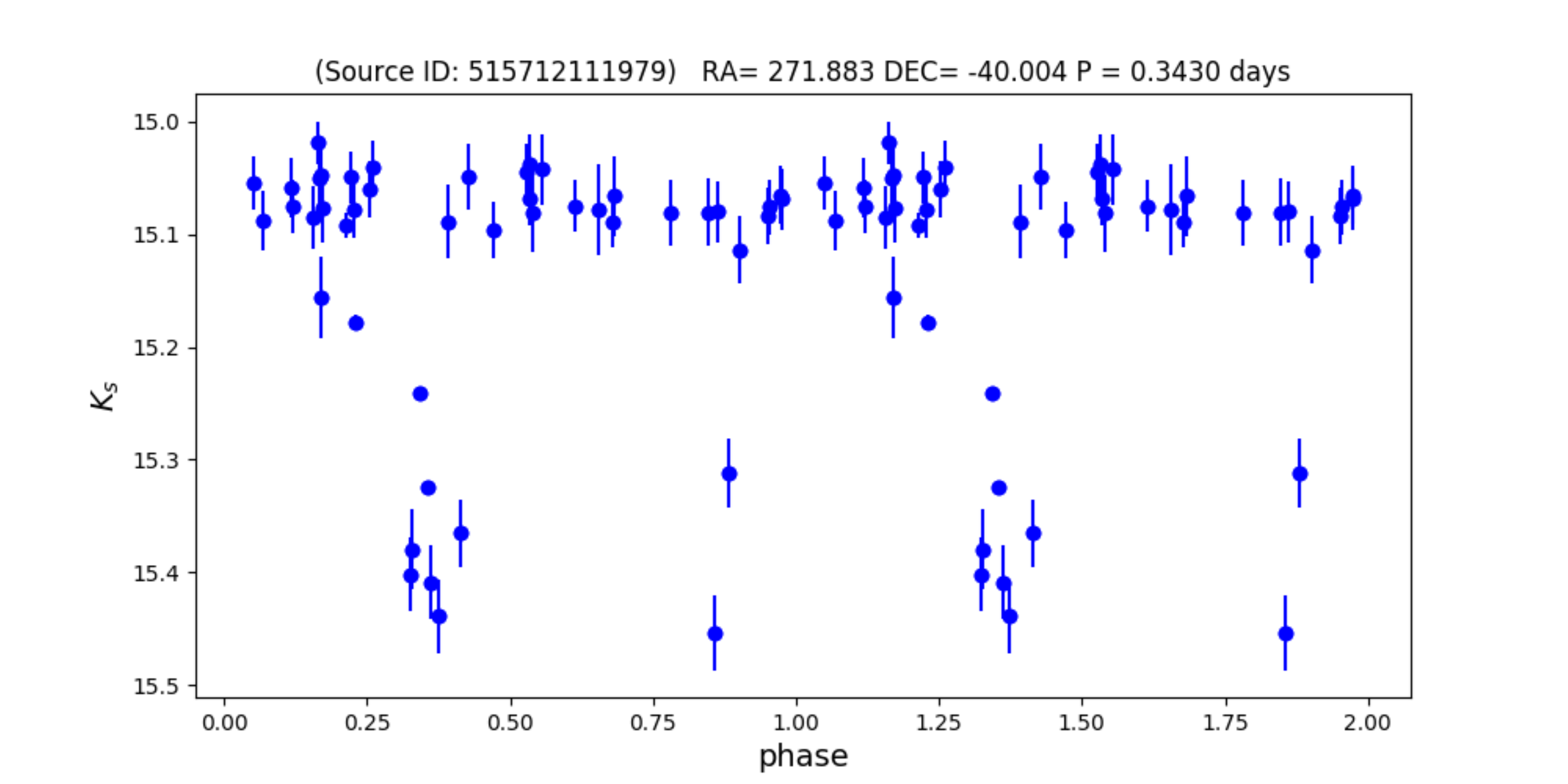} \\
    \includegraphics[width=.33\textwidth]{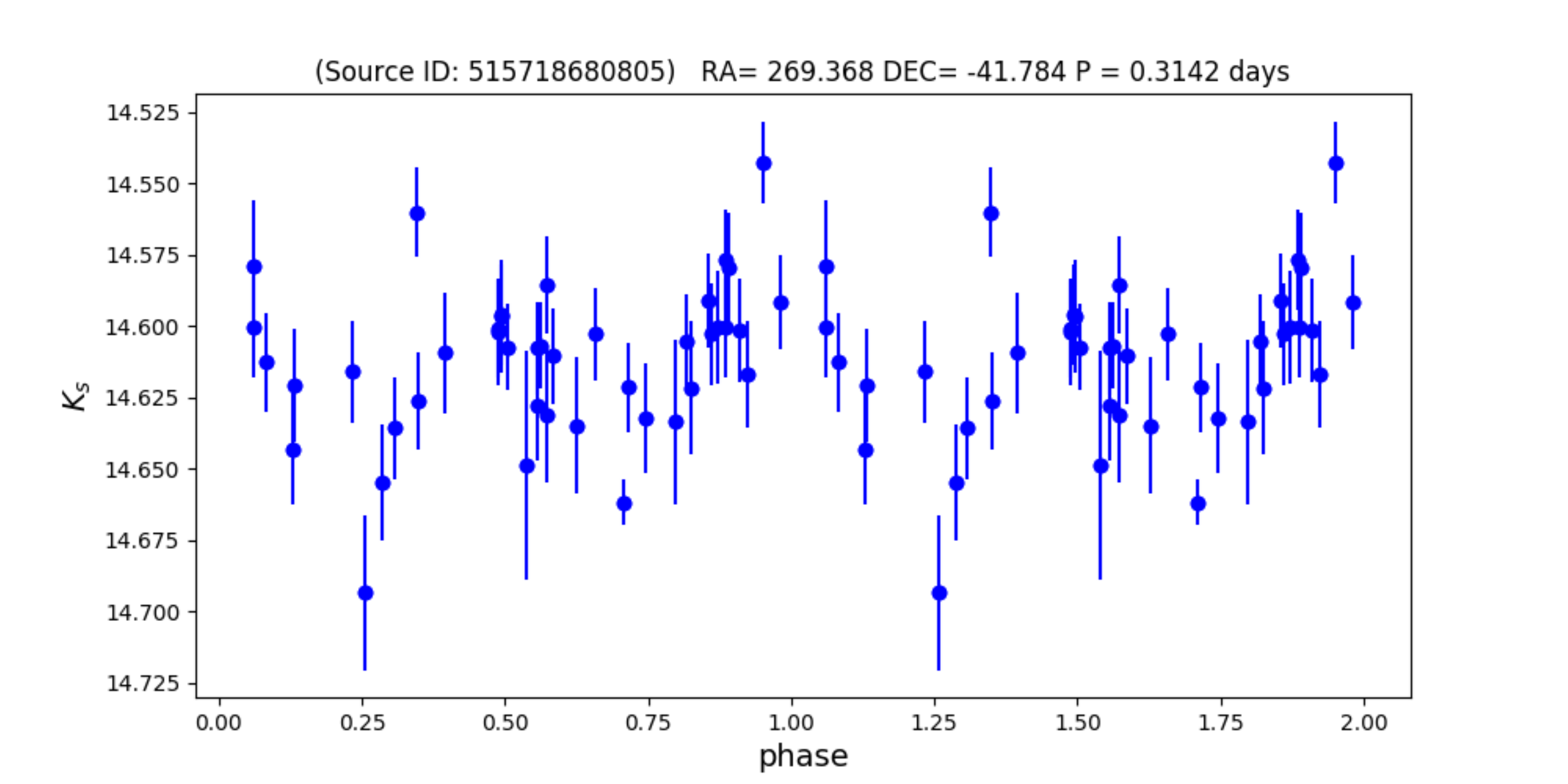} &
    \includegraphics[width=.33\textwidth]{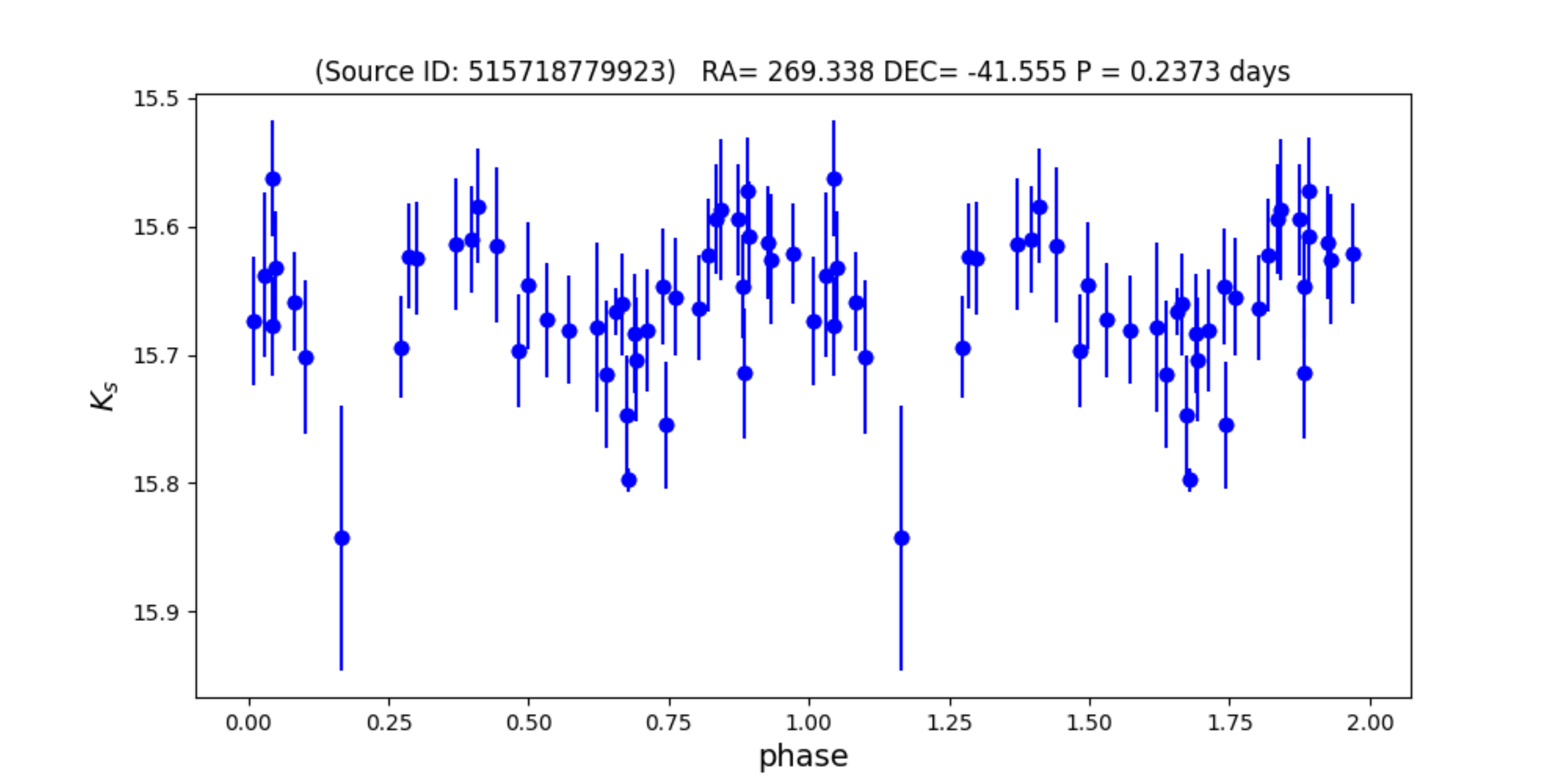} &
    \includegraphics[width=.33\textwidth]{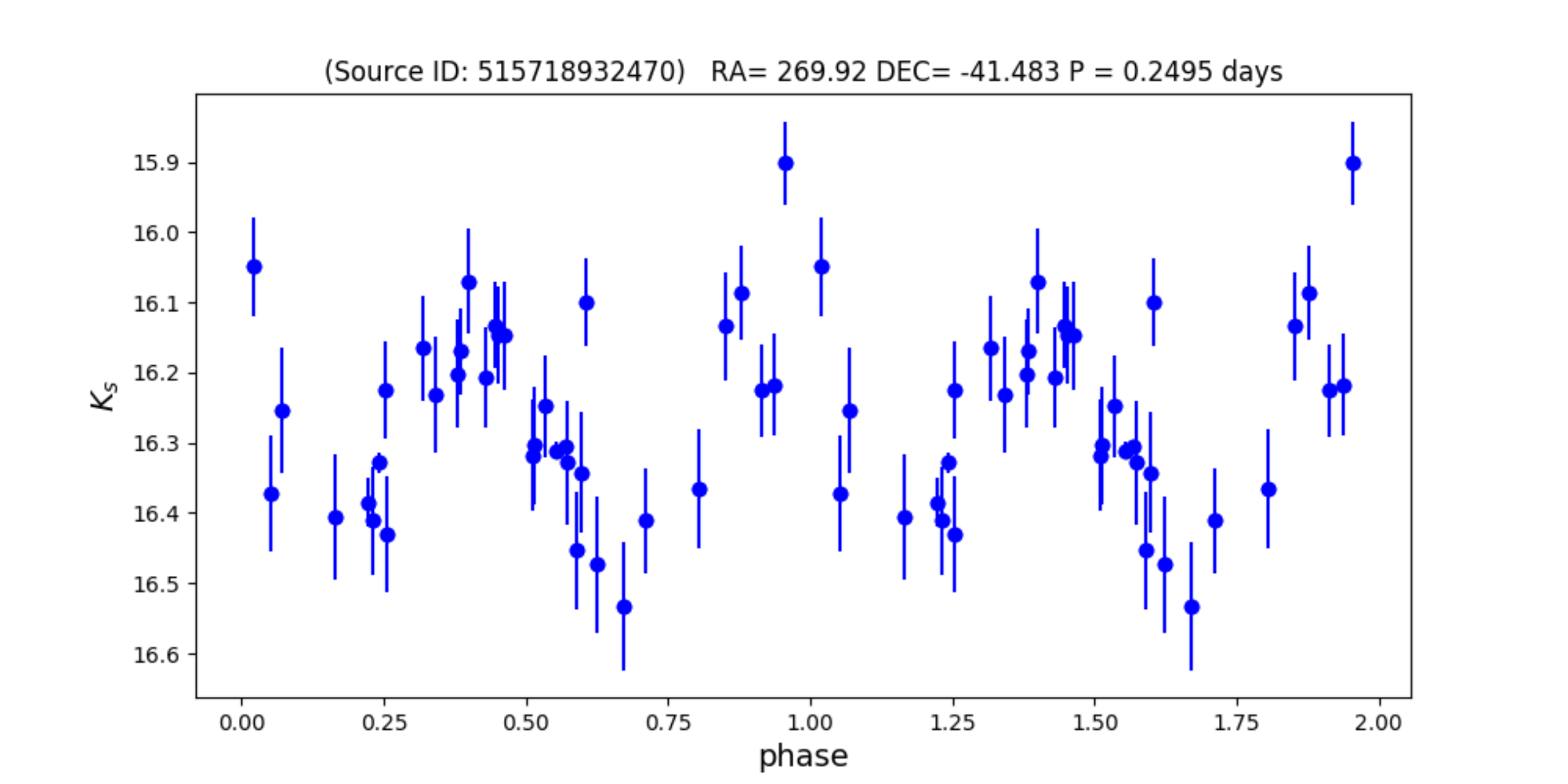} \\         
  \end{tabular}
  \includegraphics[width=.33\textwidth]{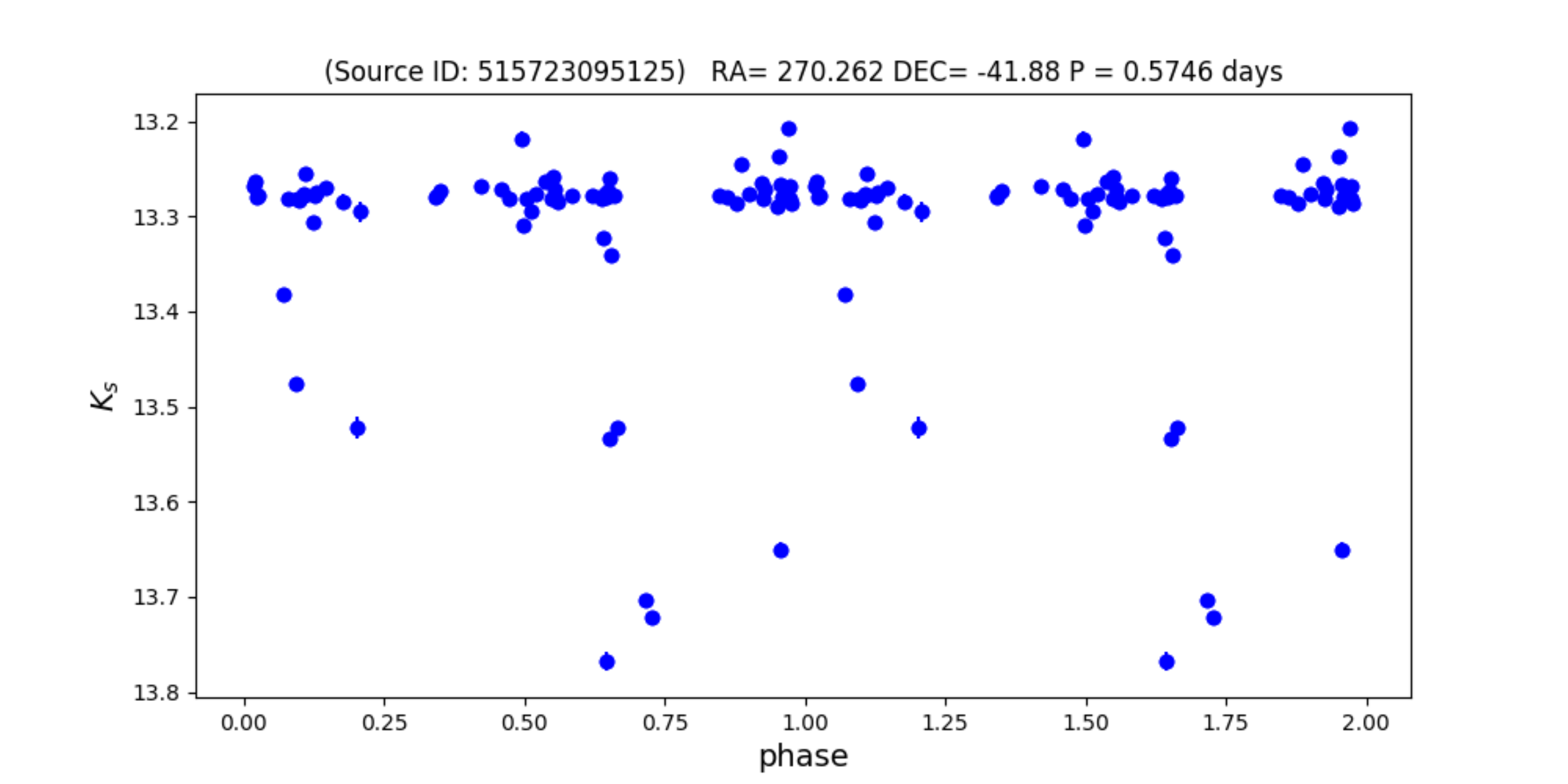} 
  \includegraphics[width=.33\textwidth]{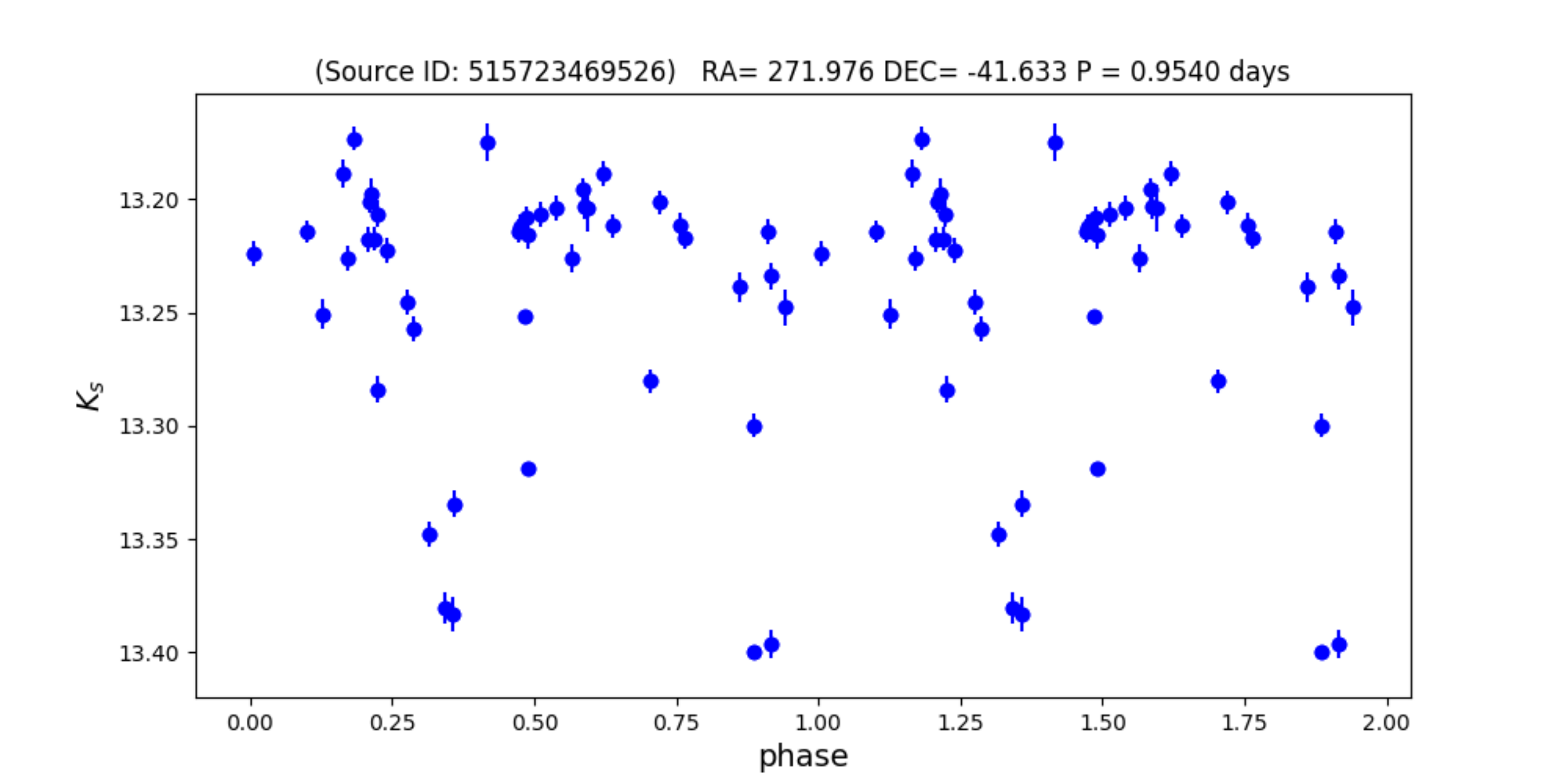}  
  \caption{BHB stars light curves corresponding to $2^{nd}$ category eclipsing binaries.}
  \label{fig:bhbB5}
\end{figure*}

\end{document}